\documentclass[UKenglish,a4paper]{scrbook}
\newcommand{\ismain}{1}

\usepackage{etex}
\reserveinserts{28}
\usepackage{geometry}
\usepackage[dutch,main=UKenglish]{babel}
\KOMAoptions{fontsize=11pt}

\SetSymbolFont{largesymbols}{normal}{OMX}{cmex}{m}{n}
\SetSymbolFont{largesymbols}{bold}  {OMX}{cmex}{m}{n}

\KOMAoptions{chapterprefix=true,numbers=noendperiod}

\usepackage{scrlayer-scrpage}
\clearpairofpagestyles
\ihead{\headmark}
\ohead{\pagemark}

\usepackage{xifthen}
\ifthenelse{\isundefined{\ismain}}{
  \newcommand{\ismain}{0}
}{}

\usepackage{caption}

\usepackage{amsmath,amsthm,amssymb}
\usepackage{mathtools}
\usepackage{xspace,enumerate,color,epsfig}
\usepackage{graphicx}
\graphicspath{{.}{./figures/}}
\usepackage{bm}
\usepackage{physics}
\usepackage{enumitem}

\usepackage{stmaryrd}
\usepackage{docmute}
\usepackage{keycommand}
\usepackage{multicol}
\usepackage{breakurl}

\usepackage{csquotes}
\usepackage[
  backend=bibtex,
  maxbibnames=99,
  style=numeric-comp,
  sorting=nyt,
  abbreviate=false,
  backref]{biblatex}
\bibliography{../bibliography}

\DeclareFieldFormat{postnote}{#1} 
\DeclareFieldFormat{multipostnote}{#1}

\DeclareFieldFormat{editortype}{\mkbibparens{#1}}
\DeclareDelimFormat{editortypedelim}{\addspace}

\renewbibmacro*{pageref}{}
\renewbibmacro*{finentry}{%
  \newunit\newblock
  \iflistundef{pageref}
    {}
    {\printtext{%
        $\uparrow$\addnbthinspace
        \printlist[pageref][-\value{listtotal}]{pageref}}}%
  \finentry}

\DeclareFieldFormat[article]{number}{\mkbibparens{#1}}
\renewbibmacro*{volume+number+eid}{%
  \printfield{volume}%
  \printfield{number}%
  \setunit{\addcomma\space}%
  \printfield{eid}}

\renewbibmacro*{issue+date}{%
  \setunit{\addcomma\space}
    \printfield{issue}%
    \setunit*{\addspace}%
    \usebibmacro{date}
  \newunit}

\newbibmacro*{related:eprint}[1]{%
  \entrydata{#1}{\usebibmacro{eprint}}}

\DeclareFieldFormat{eprint:hdl}{%
  \mkbibacro{HDL}\addcolon\space
  \ifhyperref
    {\href{http://hdl.handle.net/#1}{\nolinkurl{#1}}}
    {\nolinkurl{#1}}}

\renewbibmacro*{date}{%
  \printdate
  \iffieldundef{pubstate}
    {}
    {\setunit{\addcomma\space}%
      \printfield{pubstate}%
      \clearfield{pubstate}}}

\appto\biburlsetup{\Urlmuskip=0mu\relax}

\usepackage{multind}
\makeindex{default}
\makeindex{math}
\newcommand{\indexd}[1]{\index{default}{#1}}

\usepackage{hyperref}

\usepackage{tikzit}
\input{zx.tikzdefs}

\tikzstyle{white dot}=[inner sep=0mm, minimum size=2mm, draw=black, shape=circle, draw=black, fill=white]
\tikzstyle{gray dot}=[inner sep=0mm, minimum size=2mm, draw=black, shape=circle, draw=black, fill={rgb,255: red,191; green,191; blue,191}]
\tikzstyle{white phase dot}=[minimum size=5mm, font={\footnotesize}, shape=rectangle, rounded corners=2mm, inner sep=0.2mm, outer sep=-2mm, scale=0.8, tikzit shape=circle, draw=black, fill=white, tikzit draw=blue]
\tikzstyle{gray phase dot}=[minimum size=5mm, font={\footnotesize}, shape=rectangle, rounded corners=2mm, inner sep=0.2mm, outer sep=-2mm, scale=0.8, tikzit shape=circle, draw=black, fill={rgb,255: red,191; green,191; blue,191}, tikzit draw=blue]
\tikzstyle{small hadamard}=[fill=white, draw, inner sep=0.6mm, minimum height=1.5mm, minimum width=1.5mm, tikzit shape=rectangle]
\tikzstyle{small dot}=[inner sep=0.7mm, minimum width=0pt, minimum height=0pt, draw, shape=circle]
\tikzstyle{small white dot}=[small dot, fill=white]
\tikzstyle{small black dot}=[small dot, fill=black]
\tikzstyle{special dot}=[small white dot]
\tikzstyle{mbqc dot}=[small black dot]
\tikzstyle{mbqc input dot}=[small white dot]
\tikzstyle{mbqc output dot}=[small gray dot]
\tikzstyle{label}=[font={\footnotesize}, text height=1.5ex, text depth=0.25ex, yshift=0.5mm]
\tikzstyle{left label}=[label, anchor=east, xshift=1.5mm]
\tikzstyle{right label}=[label, anchor=west, xshift=-1.5mm]
\tikzstyle{inline text}=[text height=1.5ex, text depth=0.25ex, yshift=0.5mm]
\tikzstyle{small box}=[rectangle, inline text, fill=white, draw, minimum height=5mm, yshift=-0.5mm, minimum width=5mm, font={\small}]
\tikzstyle{medium box}=[rectangle, inline text, fill=white, draw, minimum height=5mm, yshift=-0.5mm, minimum width=10mm, font={\small}]
\tikzstyle{empty diagram}=[draw={gray!40!white}, dashed, shape=rectangle, minimum width=1cm, minimum height=1cm]
\tikzstyle{empty diagram small}=[draw={gray!50!white}, dashed, shape=rectangle, minimum width=0.6cm, minimum height=0.5cm]
\tikzstyle{box}=[shape=rectangle, text height=1.5ex, text depth=0.25ex, yshift=0.5mm, fill=white, draw=black, minimum height=5mm, yshift=-0.5mm, minimum width=5mm, font={\small}]
\tikzstyle{Z dot}=[white dot]
\tikzstyle{Z phase dot}=[white phase dot]
\tikzstyle{X dot}=[gray dot, tikzit fill={rgb,255: red,191; green,191; blue,191}]
\tikzstyle{X phase dot}=[gray phase dot, tikzit fill={rgb,255: red,191; green,191; blue,191}, tikzit draw=blue]
\tikzstyle{hadamard}=[small hadamard, tikzit shape=rectangle]
\tikzstyle{vertex}=[inner sep=0mm, minimum size=1mm, shape=circle, draw=black, fill=black]
\tikzstyle{vertex set}=[inner sep=0mm, minimum size=1mm, shape=circle, draw=black, fill=white, font={\footnotesize\boldmath}]

\tikzstyle{simple}=[-]
\tikzstyle{gs edge}=[-]
\tikzstyle{gs double edge}=[-, double, shorten <=-1mm, shorten >=-1mm, double distance=2pt]
\tikzstyle{gray edge}=[-, {gray!99!white}]
\tikzstyle{hadamard edge}=[-, color=gray, opacity=0.8, dashed, dash pattern=on 3pt off 1.5pt, thick]
\tikzstyle{brace edge}=[-, tikzit draw=blue, decorate, decoration={brace,amplitude=1mm,raise=-1mm}]
\tikzstyle{diredge}=[->]
\tikzstyle{highlight T}=[-, draw={rgb,255: red,8; green,0; blue,255}, very thick, shorten <=-0.5pt, shorten >=0.5pt]
\tikzstyle{dashed edge}=[-, dashed, dash pattern=on 2pt off 0.5pt, draw=black]

\usepackage{tikz}
\usepackage{tikz-cd}
\usetikzlibrary{arrows, matrix}
\usepackage{subcaption}

\makeatletter
\newcommand{\ChapterOutsidePart}{%
    \def\toclevel@chapter{-1}\def\toclevel@section{0}\def\toclevel@subsection{1}}
\newcommand{\ChapterInsidePart}{%
    \def\toclevel@chapter{0}\def\toclevel@section{1}\def\toclevel@subsection{2}}
\makeatother
\newcommand{\sectionfree}[1]{%
  \phantomsection
  \addcontentsline{toc}{section}{#1}
  \section*{#1}
}

\usepackage{array}
\makeatletter
\g@addto@macro{\endtabular}{\rowfont{}}
\makeatother
\newcommand{\rowfonttype}{}
\newcommand{\rowfont}[1]{
   \gdef\rowfonttype{#1}#1%
   }
\newcolumntype{L}{>{\rowfonttype}l}
\newcolumntype{C}{>{\rowfonttype}c}

\usepackage{algorithm}
\usepackage{algorithmicx}
\usepackage[noend]{algpseudocode}
\makeatletter
\algrenewcommand\ALG@beginalgorithmic{\small}
\makeatother

\usepackage{listings}
\DeclareFixedFont{\ttb}{T1}{txtt}{bx}{n}{10} 
\DeclareFixedFont{\ttm}{T1}{txtt}{m}{n}{10}  
\usepackage{xcolor}
\definecolor{deepblue}{rgb}{0,0,0}
\definecolor{deepred}{rgb}{0,0,0}
\definecolor{deepgreen}{rgb}{0.2,0.2,0.2}
\newcommand\pythonstyle{\lstset{
language=Python,
basicstyle=\ttm,
otherkeywords={self},             
keywordstyle=\ttb\color{deepblue},
emph={MyClass,__init__},          
emphstyle=\ttb\color{deepred},    
stringstyle=\color{deepgreen},
frame=tb,                         
showstringspaces=false,
commentstyle=\color{gray} %
}}

\lstnewenvironment{python}[1][]
{
\pythonstyle
\lstset{#1}
}
{}
\newcommand\pythoninline[1]{{\pythonstyle\lstinline!#1!}}

\usepackage{scalerel}
\newcommand\ssquare{\scaleobj{0.6}{\square}}

\newcommand{\fullcounter}{\thesection.\arabic{counter}}
\theoremstyle{definition}
\newcounter{counter}
\numberwithin{counter}{section}
\newtheorem{theorem}[counter]{Theorem}
\newtheorem*{theorem*}{Theorem}
\newtheorem{corollary}[counter]{Corollary}
\newtheorem{lemma}[counter]{Lemma}
\newtheorem{proposition}[counter]{Proposition}

\newtheorem{definition}[counter]{Definition}
\newtheorem{example}[counter]{Example}

\newtheorem{remark}[counter]{Remark}
\newtheorem{assumption}[counter]{Assumption}

\usepackage{color}

\usepackage[color]{changebar}

\makeatletter
\newcommand\etc{etc\@ifnextchar.{}{.\@}\xspace}
\newcommand\ie{i.e.\@\xspace}  
\newcommand\eg{e.g.\@\xspace}
\makeatother

\DeclarePairedDelimiter{\ceil}{\lceil}{\rceil}
\DeclarePairedDelimiter{\floor}{\lfloor}{\rfloor}
\DeclarePairedDelimiter{\inn}{\langle}{\rangle}

\newcommand\Define[1]{\textbf{#1}}
\DeclareMathOperator{\supp}{supp}

\newcommand{\bigovee}{\mathop{\vphantom{\sum}\mathchoice%
        {\vcenter{\hbox{\huge $\ovee$}}}%
        {\vcenter{\hbox{\Large $\ovee$}}}%
        {\ovee}{\ovee}}\displaylimits}

\newcommand{\R}{\mathbb{R}}
\newcommand{\C}{\mathbb{C}}
\newcommand{\N}{\mathbb{N}}

\newcommand{\sse}{\subseteq}

\newcommand{\rnk}{\text{rnk}\xspace}
\newcommand{\asrt}{\text{asrt}}
\newcommand{\id}{\text{id}}
\newcommand{\opp}{\text{op}}
\newcommand{\eff}{\text{Eff}}
\newcommand{\seff}{\text{SEff}}
\newcommand{\EJA}{\textbf{EJA}}
\newcommand{\OUS}{\textbf{OUS}}
\newcommand{\st}{\text{St}}
\newcommand{\sa}{\text{sa}}
\newcommand{\cl}[1]{\overline{#1}}
\newcommand{\im}[1]{\text{im}(#1)}
\newcommand{\mult}{\,\&\,}
\newcommand{\commu}{\,\lvert\,}

\newcommand{\XY}{\text{XY}\xspace}
\newcommand{\XZ}{\text{XZ}\xspace}
\newcommand{\YZ}{\text{YZ}\xspace}
\newcommand{\XYm}{\text{XY}\xspace}
\newcommand{\XZm}{\text{XZ}\xspace}
\newcommand{\YZm}{\text{YZ}\xspace}
\newcommand{\LOG}{labelled open graph\xspace}

\newcommand{\zxdiagram}{ZX-diagram\xspace}
\newcommand{\zxdiagrams}{ZX-diagram\xspace}
\newcommand{\ld}{\lambda}

\newcommand{\intf}[1]{\left\llbracket #1 \right\rrbracket} 
\newcommand{\odd}[2]{\textsf{Odd}_{#1}\left(#2\right)}

\newcommand{\pow}[1]{\ensuremath{\mathcal{P}\left( #1 \right)}}
\newcommand{\symd}{\mathbin{\Delta}\xspace}
\newcommand{\Symdi}[1]{\underset{\scriptstyle #1}{\scalebox{1.5}{$\symd$}}\,}

\newcommand{\pat}{\ensuremath{\mathfrak{P}}} 

\newcommand{\circuit}{\texttt{Circuit}\xspace}
\newcommand{\graph}{\texttt{Graph}\xspace}
\newcommand{\gate}{\texttt{Gate}\xspace}
\newcommand{\gadgetsimp}{\texttt{gadget-simp}\xspace}

\title{Quantum Theory from Principles, \\Quantum Software
from Diagrams}
\author{John van de Wetering}
\date{\today}

\begin{document}
\frontmatter   
\maketitle
\ChapterOutsidePart
\chapter{Acknowledgements}

I have had the pleasure to spend these last four years with many wonderful and interesting people, only some of whom I am able to thank here explicitly.

This thesis would not have existed without my promotor Bart Jacobs and copromotor Aleks Kissinger. I wish to thank Bart for having the faith in me to hire and support me, and for building an excellent group of researchers at the Radboud University. Aleks I want to thank for giving me the freedom to pursue my own interests in the first years of my PhD, which resulted in the first part of this thesis, while simultaneously pushing me to work on the ZX-calculus, which ended up as the second part of this thesis. 
This thesis could not have been written without the many hours I have spent in Aleks' office discussing research and asking questions and I could not wish for a better supervisor.

I gratefully acknowledge financial support from the ERC under the European Union's Seventh Framework Programme (FP7/2007-2013) / ERC grant n$^\text{o}$ 320571.

When I started my PhD I was lucky that there was already an established group of researchers in the Quantum Group Nijmegen of the Digital Security department. Sander Uijlen, Kenta Cho, Guillaume Allais and Bas and Bram Westerbaan made my stay that much more enjoyable and it was a sad thing to see them leave one by one. I especially wish to thank Bas and Bram Westerbaan who have answered many of my questions regarding operator algebras and quantum theory. Without their expertise this thesis would have been a lot shorter and less interesting.

Of course the Digital Security department is larger than this small group of quantum people, and I wish to thank all my colleagues there who have made it an enjoyable place to work.

Like most research, my work was not done in isolation. This thesis includes joint work with Bas and Bram Westerbaan, Aleks Kissinger, Ross Duncan, Simon Perdrix, Miriam Backens, Hector Miller-Bakewell, Giovanni de Felice and Leo Lobski. 
I have furthermore had the pleasure of collaborating with Sal Wolffs, Stach Kuijpers, Kenta Cho and Louis Lemonnier on work that did not make it into the thesis.

During my PhD I had the opportunity to travel to many places and meet many new people. In particular, QPL2019 was made extra enjoyable by an exploration of Los Angeles together with Joe Collins, Sean Tull, Cole Comfort and Richard East. I also especially wish to thank Dominic Horsman for taking me under his wing in Grenoble; Bob Coecke, for letting me stay in his shed in Oxford; and Ross Duncan for hiring me for a short stint at Cambridge Quantum Computing. 

My research has benefited from discussions with many people around the world. Carlo Maria Scandolo, John Selby, Stefano Gogioso, Titouan Carette, Quanlong Wang, Kang Feng Ng, Arianne Meijer-van de Griend, Sam Staton, Stan Gudder, Robert Furber, Niel de Beaudrap, Giulio Chiribella and Chris Heunen, and undoubtedly many others whose name I am forgetting: thank you for your time and wisdom.

A significant reason for why Nijmegen really felt like home were the many afternoons I spent as a student in my living-room away from home, de Noord-kantine. To all the people who were there (you know who you are!): thank you!

While a PhD is certainly good exercise for the mind, it is perhaps less good for the body. Fortunately, dancing is an especially good remedy for this issue. I wish to thank all the Nijmegen Lindy Hoppers for being awesome and \emph{gezellig} in quite a special way.

Besides these new friends in Nijmegen, some friends have been sticking around for a good while longer. Ch\'e, Joep en Ludo, thank you for still not being sick of me and the many good memories and LAN-parties. 

Mam, pap en Paul, bedankt dat je me al die jaren de vrijheid hebt gegeven om de richting op te gaan die ik wilde.

Lastly, Ema, thank you for having been so supportive and understanding during the long months it took to write this thesis.

\phantomsection
\addcontentsline{toc}{chapter}{Contents}
\tableofcontents

\if\ismain0 
\frontmatter
\ChapterOutsidePart

\microtypesetup{protrusion=false}
\pdfbookmark{\contentsname}{toc}
\tableofcontents
\microtypesetup{protrusion=true}

\fi 
\phantomsection
\chapter{Introduction}
This thesis deals with the question of how quantum theory can be recovered from first principles and how we can study quantum computation using diagrammatic methods. As these are quite disparate topics, this thesis is divided into two parts that can be read separately.

\sectionfree{Part A: quantum theory from principles}


In the early 20th century a series of discoveries was made that showed there was something wrong with the classical description of the universe. It took a number of decades for physicists to develop consistent mathematical theories that could explain these strange new phenomena. At first there were the theories of \emph{matrix mechanics} of Heisenberg and \emph{wave mechanics} of Schr\"odinger. These were later unified by von Neumann into the mathematical description of quantum theory as we know it today~\cite{neumann2013grundlagen}: Hilbert spaces, unitary maps, complex numbers, tensor products, and so on.

This raises the question \emph{why} these mathematical concepts should describe our reality so well. Why do we need to describe a system by a Hilbert space? Why do we need to use complex numbers instead of real numbers? Why are composite systems described by a tensor product?

It is instructive to compare the situation of quantum theory to that of Einstein's relativity. Unlike quantum theory, relativity was originally not based on much physical evidence. Instead, Einstein entertained a small number of physically compelling principles: the constancy of the speed of light, the invariance of the laws of physics on a chosen reference frame, and the equivalence of gravitational and inertial mass. Using these physical principles he worked through several thought experiments and managed to derive new mathematical laws of physics based on the mathematics of Lorentz, Minkowski and Riemann. This provides an answer to the question of \emph{why} we would need to use, for instance, Lorentz transformations for the physical laws: ``our chosen physical principles demand it''.

The goal of this part of the thesis is to explore possible principles from which the mathematics of quantum theory can be derived. 
As already mentioned, there is a philosophical reason for doing so: if we can derive quantum theory from a small set of physically compelling principles, then it shows our laws of nature could not have been different without breaking one of these principles, hence providing a satisfying answer to the question of why our universe `needs' quantum mechanics. 
Another related reason is that the search for principles allows us to see which parts of quantum theory are specific to it, and which are `generic' for a broad class of physical theories. For instance, the impossibility of cloning a quantum state has been shown to hold in basically any non-classical alternative~\cite{barrett2007information}, while the Tsirelson bound on the strength of quantum correlations holds in any physical theory satisfying the principle of \emph{information causality}~\cite{pawlowski2009information}.
Lastly, if quantum theory is derivable from some set of principles, then we know that any alternative or generalisation must abandon at least one of these principles, which simplifies the search for such alternatives.

Much work has been done this last century on deriving quantum theory from physically motivated principles, from von Neumann's seminal \emph{quantum logic} approach, to the more modern approach of using \emph{generalised probabilistic theories} (a short historic overview of which is given in Section~\ref{sec:history-of-principles}). In Part A of this thesis we add to these results two new approaches to deriving quantum theory from first principles. 

The first approach defines principles for how the process of \emph{sequential measurement} should behave. A crucial difference between classical and quantum theory is that in the latter case a measurement generally affects the state of the system. As a result, when doing a sequence of measurements, the order in which the measurements are performed is relevant.
Chapter~\ref{chap:seqprod} considers general physical theories that allow sequential measurement and then restricts the possibilities by assuming this process is well-behaved in certain ways; namely that ``compatible measurements should act classically''. We find that this is sufficient to recover standard quantum theory.
So even though quantum theory is more complicated than classical theory because measurements don't necessarily commute, it is special in that sequential measurements are still `nice' in certain ways.

The second approach takes an entirely different route. In quantum theory we can distinguish between `pure' processes that represent in a sense the true processes of nature, and `mixed' processes that arise from interactions with the classical world, for instance via measurements or noise. In Chapter~\ref{chap:effectus} we consider general physical theories where the subset of pure processes has certain properties one would expect to hold for the true processes of nature in a `nice' physical theory. Again we find that this forces such theories to be part of standard quantum theory. The crux here is that we define `pure process' in a different way than is usual, by using the abstract language of category theory.

Let us now outline in more detail the results of each of the chapters in Part A.

Chapter~\ref{chap:reconstruction-intro} presents the necessary preliminaries for this part of the thesis. We cover the basic mathematics of quantum theory, and recall the definitions and some important results regarding C$^*$-algebras, Euclidean Jordan algebras (EJAs), order unit spaces and generalised probabilistic theories (GPTs). We end the chapter with a brief history of results in the principled approach to quantum theory.

Then in Chapter~\ref{chap:seqprod} we present our first reconstruction of quantum theory. We consider the operation of \emph{sequential measurement}, where we first perform some measurement $a$ and then a measurement $b$, resulting in their \emph{sequential product} $a\mult b$. We find that a variation on the axioms for the sequential product of Gudder and Greechie~\cite{gudder2002sequential} can be operationally motivated. Using these axioms on the sequential product we show that the space of effects of a finite-dimensional single system must be a Euclidean Jordan algebra, a space generalising the space of observables of a quantum system~\cite{jordan1993algebraic}. We furthermore show that the only EJAs that compose in a \emph{locally tomographic} manner are the C$^*$-algebras, hence recovering the standard systems of quantum theory. In addition to recovering the space of observables of a quantum system, we also show how to recover the Born rule and the Schr\"odinger equation.

Chapter~\ref{chap:effectus} also presents a reconstruction of quantum theory, but based on wholly different assumptions. We cover the basics of \emph{effectus theory}, a new approach to categorical logic developed by Jacobs in 2015~\cite{jacobs2015new}. It generalises the convex structure of state and effect spaces that are required in a generalised probabilistic theory to allow arbitrary \emph{effect algebras}~\cite{foulis1994effect}. Since an effectus has much less structure than a GPT it naturally leads to a different viewpoint, and hence different ideas as to which notions are important. In particular, from effectus theory we get a new notion of \emph{pure} transformation, which is based on maps satisfying some particular universal properties. We succeed in reconstructing finite-dimensional quantum theory by postulating some reasonable assumptions on the set of pure transformations (namely, that they must form a \emph{dagger-category}).

The previous chapters introduce a variety of new assumptions for which it might not a priori be clear that they are actually true in quantum theory. In Chapter~\ref{chap:jordanalg} we study \emph{JBW-algebras}. These are to Euclidean Jordan algebras as what von Neumann algebras are to finite-dimensional C$^*$-algebras, and can be seen as a generalisation of infinite-dimensional quantum theory. We show that the assumptions regarding the sequential product and pure maps hold in the category of JBW-algebras with positive subunital maps. Along the way we find additional new structure in JBW-algebras: the existence of a division operation on effects, a `polar-decomposition-like' property, and a useful characterisation of when elements \emph{operator commute}.

Finally, in Chapter~\ref{chap:infinitedimension} we combine the results of the previous three chapters to present a (partial) reconstruction of infinite-dimensional quantum theory using a combination of assumptions from Chapters~\ref{chap:seqprod} and~\ref{chap:effectus}. Notably, we succeed in classifying the allowed sets of scalars in a \emph{$\omega$-effectus}, which allows us to make a reconstruction that a priori does not refer to the structure of the real numbers.


\sectionfree{Part B: quantum software from diagrams}

The idea that computers based on quantum-mechanical systems might outperform `classical' computers has been around for a number of decades now~\cite{feynman1982simulating}. It however took until very recently for technical advances to proceed to the point where quantum computers are actually powerful enough to perform computations that would take impractically long on a classical computer~\cite{arute2019quantum} (and even those computations are not necessarily useful as of yet). Since quantum computers are still very limited in the number of qubits they possess, and will remain so for the foreseeable future, it is necessary for computations to be performed as efficiently as possible. In this part of the thesis we find several ways to reduce the number of qubits and operations needed to implement a given quantum computation.

We approach quantum computation through a slightly unusual lens: the \emph{ZX-calculus}. The ZX-calculus is a language for reasoning about a class of diagrams called \emph{ZX-diagrams} developed by Coecke and Duncan in 2008~\cite{CD1,CD2}. These diagrams can represent any linear map between qubits, and in particular any computation done on a quantum computer. Their usefulness comes from the ability to graphically rewrite ZX-diagrams while preserving the linear map they represent. This allows us to reason about quantum computation in an efficient manner.

In Chapter~\ref{chap:zxcalculus} we introduce the basics of quantum computation and we present the ZX-calculus together with several useful rewrite rules.

Then in the next two chapters we study \emph{measurement-based quantum computation} (MBQC). This is a model of computation wherein some intricate resource state is prepared, and the computation proceeds by measuring qubits in a specific pattern (unlike the circuit model of quantum computation where the computation proceeds by applying unitary quantum gates to a simple input state). 
Chapter~\ref{chap:ms-mbqc} presents a new concrete model of MBQC that is deterministic and approximately universal, while only requiring measurements in the two Pauli bases X and Z. To our knowledge this was the first model to demonstrate these properties. We verify the correctness of our model using the ZX-calculus. This chapter doubles as a gentle introduction to the main concepts of MBQC.

In Chapter~\ref{chap:MBQC} we consider the most well-studied model of MBQC: the \emph{one-way model}~\cite{MBQC1}. In contrast to most work dealing with the one-way model, we do not restrict our measurements to a single measurement-plane, but instead allow measurements in all three principal planes of the Bloch sphere. This allows us to present a general set of rewrite strategies that transform measurement patterns in several useful ways, in particular reducing the number of qubits needed to implement a measurement pattern. We show how these rewrites preserve the existence of \emph{gflow}, a property that ensures that the measurement pattern is deterministically implementable~\cite{GFlow}. We end this chapter with an efficient algorithm that allows any measurement pattern with a gflow to be converted into a unitary quantum circuit.

In Chapter~\ref{chap:optimisation} we apply the results of Chapter~\ref{chap:MBQC} to the problem of quantum circuit optimisation. We introduce a rewrite strategy based on the ZX-calculus that reduces Clifford circuits to a new pseudo-normal form that has several desirable properties. In addition, this rewrite strategy results in an ancilla-free T-count optimiser for Clifford+T circuits that matches or outperforms every other existing method (at the moment of writing) at this task. Finally, we discuss how this algorithm can be used as a powerful circuit equality verifier. 

The results in this part of the thesis show that the ZX-calculus can be used to unify and improve several practical aspects of quantum computing: MBQC, circuit optimisation and circuit equality verification.
We expect the ZX-calculus to be useful in a much wider array of problems than just the ones discussed in this thesis. In Chapter~\ref{chap:future} we present some preliminary results in the problems of CNOT optimisation, circuit routing, Toffoli circuit optimisation, and quantum circuit simulation.

\sectionfree{Writing style}

Whenever a new term is defined, we print it in \textbf{bold}.
Some Propositions, Lemmas, and Theorems in this thesis are labelled with a specific reference, like for instance Proposition~\ref{prop:basic}. This denotes that I was not involved with originally proving the statement. For completeness sake, or when the proof was given for a different setting, we will still sometimes include a proof for these labelled statements.

Except for in this introduction, I will use `we' to denote myself, including possible coauthors, and the reader.

\sectionfree{Attribution \& main results}

This thesis is based on the following publications and preprints, some of which were joint work with other people: \cite{wetering-gflow,cliffsimp,pyzx,kissinger2019tcount,kissinger2017MBQC,wetering2018algebraic,wetering2018reconstruction,wetering2018characterisation,wetering2020commutativity,wetering2018sequential,westerbaan2018puremaps,wetering-effect-monoids}.
Besides the results presented in this thesis, I also worked on a number of other preprints and publications during my PhD that did not fit into this thesis~\cite{wetering2017quasistochastic,wetering2017ordering,zhphasefree,GraphicalFourier2019,Lemonnier2020hypergraph,Cho2020dichotomy,westerbaan2020sequential}.
The list below states the main results of each chapter and which material it was based on.

\begin{itemize}
  \item \textbf{Chapter~\ref{chap:reconstruction-intro}}: This is an introductory chapter to Part A of the thesis, and does not contain new results.

  \item \textbf{Chapter~\ref{chap:seqprod}}: We reconstruct quantum theory from assumptions on sequential measurement. Mathematically, we show that a finite-dimensional order unit space that is a \emph{sequential effect algebra}~\cite{gudder2002sequential} is a Euclidean Jordan algebra, and that the only such spaces which have a tensor product are C$^*$-algebras. 
  This chapter is based on the solo-author paper Ref.~\cite{wetering2018sequential}, but expanded to include more details and background. In particular, Sections~\ref{sec:bornrule},~\ref{sec:central-effects} and~\ref{sec:dynamics} include new material for the thesis.

  \item \textbf{Chapter~\ref{chap:effectus}}: We reconstruct quantum theory from assumptions on pure maps. Mathematically we show that a category of finite-dimensional order unit spaces that has suitably interacting \emph{filters} and \emph{compressions} must embed into the category of Euclidean Jordan algebras. 
  This chapter is based on the solo-author paper Ref.~\cite{wetering2018reconstruction}. Section~\ref{sec:effect-algebras} contains basic theory on effect algebras that can be found in for instance Ref.~\cite{foulis1994effect}. The proofs of Section~\ref{sec:filterscompressions} are adapted from Bas Westerbaan's PhD thesis~\cite{basthesis}.

  \item \textbf{Chapter~\ref{chap:jordanalg}}: We show that the category of JBW-algebras satisfies most of the assumptions outlined in the previous chapters. New results include the existence of filters and compressions in JBW-algebras, that the pure maps between JBW-algebras form a dagger-category, and that the unit interval forms a normal sequential effect algebra.
  These results generalise those for Euclidean Jordan algebras originally presented in Ref.~\cite{westerbaan2018puremaps} --- joint work with Bas and Bram Westerbaan --- and von Neumann algebras as presented in Bram Westerbaan's PhD thesis~\cite{bramthesis}. I am especially indebted to Bas and Bram Westerbaan for Sections~\ref{sec:division-filter} and~\ref{sec:diamond-dagger} as they helped prove most of the crucial results in these sections. Finally, Section~\ref{sec:JBW-SEA} appeared as Ref.~\cite{wetering2020commutativity}.
  For known results regarding JBW-algebras I cite Ref.~\cite{hanche1984jordan} where possible, with a few remaining results coming from Ref.~\cite{alfsen2012geometry}.

  \item \textbf{Chapter~\ref{chap:infinitedimension}}: The main result is a reconstruction of infinite-dimensional quantum theory from a combination of assumptions found in the previous chapters. New results include a characterisation of $\omega$-complete effect monoids, new sufficient conditions for a convex sequential effect algebra to be isomorphic to a JB-algebra, and a set of conditions for an $\omega$-effect-theory to embed into the category of JBW-algebras.
  Section~\ref{sec:effect-monoids} is based on Ref.~\cite{wetering-effect-monoids}, which is joint work with Bas and Bram Westerbaan. Section~\ref{sec:SEA} is based on parts of Refs.~\cite{wetering2018characterisation} and~\cite{wetering2018sequential} while Sections~\ref{sec:Jordan-from-effectus} and~\ref{sec:neumann-from-effectus} contain new material for this thesis.

  \item \textbf{Chapter~\ref{chap:zxcalculus}}: This is mostly introductory material to the rest of the thesis, except for a new diagrammatic proof of the Gottesman-Knill theorem that might be of independent interest.
  Sections~\ref{sec:intro-to-circuits}--\ref{sec:simple-derivations} are standard material on the ZX-calculus and quantum computation. Wherever possible, the proofs and notation follow Ref.~\cite{CKbook}. The exposition on phase gadgets presented in Section~\ref{sec:phasegadgets} is based on Ref.~\cite{kissinger2019tcount}, co-authored with Aleks Kissinger. Sections~\ref{sec:graph-like}--\ref{sec:graph-theoretic-simp} contain material from Ref.~\cite{cliffsimp} which is joint work with Ross Duncan, Aleks Kissinger and Simon Perdrix. The diagrammatic proof of the Gottesman-Knill theorem in Section~\ref{sec:graph-theoretic-simp} is new.

  \item \textbf{Chapter~\ref{chap:ms-mbqc}}: We introduce a new model for measurement-based quantum computation that was the first model to be deterministic, approximately universal, and only require measurements in the Pauli X and Z bases.
  This chapter is based on Ref.~\cite{kissinger2017MBQC} which is joint work with Aleks Kissinger.

  \item \textbf{Chapter~\ref{chap:MBQC}}: We find a set of rewrite rules for measurement patterns in the one-way model that allow us to remove all non-input qubits measured in a Clifford angle while preserving deterministic implementability. We find an efficient algorithm for transforming any measurement pattern with gflow into an ancilla-free quantum circuit.
  This chapter is based on Ref.~\cite{wetering-gflow} which is joint work with Miriam Backens, Hector Miller-Bakewell, Giovanni de Felice and Leo Lobski. That paper itself is a continuation of the results of Ref.~\cite{cliffsimp}, joint work with Ross Duncan, Simon Perdrix and Aleks Kissinger, which included an earlier version of the circuit extraction algorithm of Section~\ref{sec:circuit-extraction}. I can claim no credit for the crucial Lemma~\ref{lem:lc_gflow} which was proved by Miriam Backens, based on work by Simon Perdrix.

  \item \textbf{Chapter~\ref{chap:optimisation}}: We find a simplification routine for ZX-diagrams that allows us to reduce Clifford circuits to a new normal form with several desirable features, while simultaneously acting as a state-of-the-art T-count optimisation algorithm for ancilla-free circuits.
  Sections~\ref{sec:circuit-simplification} and~\ref{sec:clifford-T-optimisation}--\ref{sec:circuit-verification} are based on Ref.~\cite{kissinger2019tcount} while Section~\ref{sec:pyzx} is based on Ref.~\cite{pyzx} which both are joint work with Aleks Kissinger. The new normal form for Clifford circuits in Section~\ref{sec:clifford-circuit-optimisation} is from Ref.~\cite{cliffsimp}.

  \item \textbf{Chapter~\ref{chap:future}}: This chapter mostly contains preliminary results and speculation, and hence is not based on any concrete publications. The conjecture regarding hardness of general circuit extraction in Section~\ref{sec:extraction-improvements} resulted from discussions with Niel de Beaudrap. 
  The ideas behind Section~\ref{sec:circuit-routing} come from Aleks Kissinger and were further developed by Arianne Meijer-van de Griend in her Master's research~\cite{KissingerCNOT2019}. The hyperpivoting rule in Section~\ref{sec:ZH-calculus} was jointly discovered with Louis Lemonnier during his Masters' research~\cite{Lemonnier2020hypergraph}. The results regarding graphical proofs of certain Toffoli identities in Ref.~\cite{GraphicalFourier2019} are joint work with Aleks Kissinger and Stach Kuijpers.
\end{itemize}

\if\ismain0 
  \ChapterOutsidePart
  \addtocontents{toc}{\protect\addvspace{2.25em}}
   \cleardoublepage
    \begingroup
    \phantomsection
    \emergencystretch=1em\relax
    \printbibliography[heading=bibintoc]
    \endgroup

\fi 

\mainmatter 
\ChapterInsidePart

\if\ismain0 

\microtypesetup{protrusion=false}
\ChapterOutsidePart
\pdfbookmark{\contentsname}{toc}
\tableofcontents
\microtypesetup{protrusion=true}
\ChapterInsidePart

\fi 

\part{Quantum Theory from Principles}

\chapter{Reconstructions of quantum theory}\label{chap:reconstruction-intro}

This chapter contains the necessary preliminaries for Chapters~\ref{chap:seqprod}--\ref{chap:infinitedimension}. First, to remind the reader what it is we wish to reconstruct from first principles, we will recall the basic mathematical formulation of quantum theory in Section~\ref{sec:basic-mathematics}. Then we introduce two useful generalisations of quantum mechanical systems in Section~\ref{sec:operator-algebras}: C$^*$-algebras and Jordan algebras. We recall some fundamental results relating ordered vector spaces to these algebras in Section~\ref{sec:ordered-vector-space}. A useful framework for dealing with abstract physical systems is that of \emph{generalised probabilistic theories}, which is given in Section~\ref{sec:GPTs}. Finally, we end the chapter with a brief history of results regarding the derivation of quantum theory from first principles in Section~\ref{sec:history-of-principles}.

The mathematics of this chapter can be found in basically any quantum mechanics textbook that also deals with Jordan algebras, such as that of Alfsen and Shultz~\cite{alfsen2012state,alfsen2012geometry} or Landsman~\cite{landsman2012mathematical}.

\section{The mathematics of quantum mechanics}\label{sec:basic-mathematics}

In this section we will cover the basics of the mathematics of quantum theory: Hilbert spaces, unitarity and self-adjoint operators as observables.
We will assume the reader is familiar with undergraduate linear algebra, analysis and point-set topology, in particular being comfortable with complex numbers, the notion of (orthonormal) bases, eigenvectors, diagonalisation, norms, completeness of metric spaces, convergence and continuity.

\subsection{Unitary quantum mechanics}

The standard description of quantum theory is based on Hilbert spaces.

\begin{definition}
  Let $H$ be a vector space over $\mathbb{F}$ where $\mathbb{F}$ is either the complex numbers $\C$ or the real numbers $\R$. An \Define{inner product}\indexd{inner product} on $H$ is a map $\inn{\cdot,\cdot}:H\times H\rightarrow \mathbb{F}$ satisfying the following conditions for all $a,b,c\in H$ and $z\in \mathbb{F}$:
  \begin{itemize}
    \item Linearity: $\inn{a+zb,c} = \inn{a,c} + z\inn{b,c}$.
    \item Symmetry: $\inn{b,a} = \overline{\inn{a,b}}$, where $\overline{z}$ denotes the complex conjugation of the complex number $z$.
    \item Positivity: $\inn{a,a} \geq 0$.
    \item Definiteness: $\inn{a,a} = 0$ iff $a=0$.
  \end{itemize}
  Linearity and symmetry combine to make the inner product \Define{sesquilinear}: $\inn{a,c+zb} = \inn{a,c} + \overline{z}\inn{a,b}$. An inner product induces a norm on $H$ given by $\norm{a} := \sqrt{\inn{a,a}}$.\indexd{norm} We call~$H$ a (real or complex) \Define{Hilbert space}\indexd{Hilbert space} when it has an inner product and is complete in the induced norm of this inner product.
\end{definition}

\begin{remark}\label{rem:finite-dim-hilbert}
  In finite-dimensional vector spaces, any topology induced by a norm is complete, and hence any inner product makes a finite-dimensional vector space a Hilbert space. Any finite-dimensional complex Hilbert space is linearly isomorphic to $\C^n$ for some $n\in \N$ with the inner product of $v=\sum_i x_ie_i$ and $w=\sum_j y_je_j$ (where $e_i$ is the standard orthonormal basis) defined as $\inn{v,w} = \sum_i x_i\overline{y_j}$. A similar statement holds for finite-dimensional real Hilbert spaces.
\end{remark}

In (pure) quantum theory, a physical system is identified with a complex Hilbert space $H$. The states of this system correspond to unit vectors of the Hilbert space, up to a complex phase. I.e.~a state is a vector $v\in H$ satisfying $\inn{v,v} = 1$, with two normalised vectors $v,w\in H$ corresponding to the same physical state when $v = e^{i\alpha} w$ for some $\alpha\in \R$. 
We will usually denote a state with \Define{Dirac notation}\indexd{Dirac notation}: $\ket{\psi}$. We then denote the inner product of two quantum states $\ket{\psi}$, $\ket{\phi}$ by $\braket{\psi}{\phi}$. 

Because states are unit vectors up to complex phase, it is often helpful to represent states as 1-dimensional subspaces of a Hilbert space, or as the projectors corresponding to these 1-dimensional subspaces, as those are in 1-to-1 correspondence with the physical states of the system. For a state $\ket{\psi}$ we will denote by $\ketbra{\psi}{\psi}$ the projector that projects onto the 1-dimensional space $\{z\ket{\psi}~;~z\in \C\}$. I.e.~$\ketbra{\psi}{\psi} \ket{\phi} = \braket{\psi}{\phi} \ket{\psi}$.
These projectors are examples of bounded operators.

\begin{definition}
    Let $(V,\norm{\cdot}_V)$ and $(W,\norm{\cdot}_W)$ be normed vector spaces. Let $A\colon V\rightarrow W$ be a linear map. We say $A$ is \Define{bounded}\indexd{bounded linear map} when there is a $\lambda\in\R_{\geq 0}$ such that $\norm{Av}_W \leq \lambda \norm{v}_V$ for all $v\in V$. It is an \Define{isometry}\indexd{isometry} when $\norm{Av}_W = \norm{v}_V$ for all $v\in V$. We denote the set of bounded linear maps between $V$ and $W$ by $B(V,W)$, and we define the shorthand $B(V):=B(V,V)$.
\end{definition}

For a finite-dimensional Hilbert space $\C^n$ all linear maps are bounded and hence $B(\C^n)$ consists of all $n\times n$ complex matrices. We will therefore often write $M_n(\C) := B(\C^n)$ to denote this correspondence. Similarly, we will write $M_n(\R) := B(\R^n)$. 

We can associate an \emph{adjoint} to every bounded operator on a Hilbert space. This allows us to define a couple of useful classes of linear maps.

\begin{definition}
  Let $A:H\rightarrow K$ be a bounded linear map between (complex or real) Hilbert spaces. There is a unique linear map $A^*:K\rightarrow H$ that satisfies $\inn{Av,w} = \inn{v,A^*w}$ for all $v,w\in H$. We call this map the \Define{adjoint} of $A$. When $H=K$ and $A^*=A$ we say $A$ is \Define{self-adjoint}\indexd{self-adjoint}.
\end{definition}
It is easy to see that a bounded linear map of Hilbert spaces $A:H\rightarrow K$ is an isometry iff $A^*A = \id_H$. We say $A$ is a \Define{unitary}\indexd{unitary} when both $A$ and $A^*$ are isometries, and hence $A^*A = \id_H$ and $AA^* = \id_K$.

As quantum states correspond to normalised vectors, unitary maps send quantum states onto other quantum states. Unitary maps hence describe the possible dynamics of a quantum systems: the way quantum states can change through time.

For a given quantum system corresponding to a Hilbert space $H$ we describe the amount of energy of any given state by a self-adjoint map $A\in B(H)$ called the \Define{Hamiltonian}. The expectation value of the energy of a state $\ket{\psi}$ is given by $\bra{\psi}A\ket{\psi}$. The Hamiltonian describes the evolution of a state through time by the \Define{Schr\"odinger equation}: $\ket{\psi_t}:= e^{-itA}\ket{\psi}$. The linear map $e^{-itA}$ is unitary for all values of $t\in \R$, and hence $\ket{\psi_t}$ indeed remains a normalised vector, and hence a quantum state.
Instead of describing a state by a normalised vector $\ket{\psi}$ we can describe it by a projector $\ketbra{\psi}$ in which case the unitary evolution is given by $e^{-itH}\ketbra{\psi}e^{itH}$.

\subsection{Mixed-state quantum mechanics}\label{sec:mixed-state-quantum}

To complete the mathematical description of quantum mechanics, we need to include a notion of measurement.

We will view a measurement as asking a question about a system: we interact with the system in some manner to determine some property, and the outcome we get is the answer to our question. The most basic type of question would be `Is the system in the state $\ket{\psi}$?' When our system is actually in the state $\ket{\phi}$, the answer to this question has a probability $\lvert \braket{\psi}{\phi}\rvert^2$ of being `yes'. The formula giving this probability is known as the \Define{Born rule}\indexd{Born rule}, and the probability $\lvert \braket{\psi}{\phi}\rvert^2$ is sometimes called the \Define{transition probability}\indexd{transition probability} from $\phi$ to $\psi$.

The Born rule might look a bit arbitrary --- why is there an exponent of 2 there? --- but when we represent the states by their projectors, we can find a more elegant formula. Write $\rho = \ketbra{\phi}$ and $E=\ketbra{\psi}$. Then $\braket{\psi}{\phi}\rvert^2 = \tr(E\rho)$, where $\tr(\cdot)$ denotes the trace of the linear maps in the Hilbert space.
\begin{definition}
  Let $H$ be a (real or complex) Hilbert space, and fix an orthonormal basis $(e_i)$ of $H$. Let $A:H\rightarrow H$ be a bounded linear map. Then we define  \Define{trace}\indexd{trace} of $A$ as the (potentially infinite) number\footnote{The trace is only well-defined for \emph{trace class} operators. We do not make this distinction here as we are primarily interested in the finite-dimensional case where this is not an issue.}: $\tr(A) := \sum_i \inn{Ae_i, e_i}$.
\end{definition}

We used the change of notation to $E$ and $\rho$ deliberately, as it turns out that the most general type of measurement and state is not described by vectors on $H$.
\begin{definition}\label{def:positive-map-Hilbert-space}
  Let $A\in B(H)$ be a bounded operator on a Hilbert space. We say $A$ is \Define{positive} and write $A\geq 0$ when $\inn{Av,v}\geq 0$ for all $v\in H$. We extend this to a partial order on $B(H)$ by defining $A\leq B$ iff $B-A\geq 0$. We write $1\in B(H)$ for the identity: $1v = v$ for all $v\in H$. We say $A$ is an \Define{effect} when $0\leq A\leq 1$. We write $\eff(H) := [0,1]_{B(H)} := \{A\in B(H)~;~0\leq A\leq 1\}$.
\end{definition}

\begin{proposition}[{\cite{NielsenChuang}}]
  Let $H$ be a (real or complex) Hilbert space. Then the following statements are true.
  \begin{itemize}
    \item A positive map is self-adjoint.
    \item If $A\in B(H)$ is positive, then we can find a unique positive map $\sqrt{A}$ such that $\sqrt{A}^2 = A$.
    \item If $A\in B(H)$ is positive, then $\tr(A)\geq 0$.
    \item If $A\in B(H)$ is positive and $B\in B(H)$ is arbitrary, then $BAB^*$ is positive. In particular, if $B$ is positive, $\sqrt{B}A\sqrt{B}$ will again be positive.
  \end{itemize}
\end{proposition}

It turns out that the most general type of `questions' we can ask about states are given by effects. Any measurement with $k$ outcomes can be represented by a set of effects $E_1,E_2,\ldots, E_k$ satisfying $\sum_i E_i = 1$. Such a set of effects is called a \Define{POVM} (\emph{positive operator-valued measure})\indexd{POVM}. The probability that we observe outcome $i$ when the system is in a state $\rho$ is given by $\tr(E_i \rho)$. We note that this is indeed a probability: recall that the trace satisfies $\tr(ABC) = \tr(BCA)$, and hence $\tr(E_i \rho) = \tr(\sqrt{E_i}\sqrt{E_i}\rho) = \tr(\sqrt{E_i}\rho\sqrt{E_i}) \geq 0$ as $\sqrt{E_i}\rho\sqrt{E_i}$ is a positive map. Furthermore, $\sum_i \tr(E_i \rho) = \tr(\sum_i E_i \rho) = \tr(1\rho) =\tr(\rho) = 1$, and hence the probabilities of all the outcomes sum up to 1.

The state $\rho = \ketbra{\psi}$ is what we call a \Define{pure state}. It represents a state of maximal information. We however could also be in a situation where we are unsure in which state the system is. For instance, if we have a probability of $p_i$ to prepare the state $\ket{\psi_i}$ then our final prepared state is described by a probability distribution over the states $\ket{\psi_i}$, as $\rho = \sum_i p_i \ketbra{\psi_i}$. The resulting $\rho$ is an example of a density operator.

\begin{definition}
  Let $\rho \in B(H)$ be positive. We say $\rho$ is a \Define{density operator}\indexd{density operator} when $\tr(\rho) = 1$.
\end{definition}

The condition of having normalised trace replaces the condition of being a normalised state. We see furthermore that for any POVM $\{E_i\}$ that $\tr(E_i\rho)$ still forms a probability distribution. 

A measurement applied to a state will in general change the state. 
If the measured effect corresponds to a pure state $E=\ketbra{\psi}{\psi}$, then the state $\rho$ will be updated to $\ketbra{\psi}{\psi}$ itself. 
However, for more general measurement effects the update rule is more complicated, and depends on how the measurement process is actually implemented.
The most general type of state update can be described as follows~\cite[Chapter~III.2]{busch1996quantum}: if the outcome corresponding to the effect $E$ has been observed on a state $\rho$, then there exist operators $A_i\in B(H)$ satisfying $\sum_i A_i^*A_i = E$ such that the state $\rho$ has been updated to
\begin{equation}\label{eq:general-update-rule}
	\sum_i \frac{A_i\rho A_i^*}{\tr(E\rho)}.
\end{equation}
We divide here by $\tr(E\rho)$, the probability of observing the outcome associated to $E$, in order to preserve the normalisation of the state. Note that we can only observe $E$ when $\tr(E\rho) \neq 0$, so that this update rule is well-defined.

This update rule is so general that just knowing the effect $E$ barely gives any information on what the updated state $\rho$ will be. Indeed, the application to $\rho$ of any \emph{completely-positive map} (see next section) $\Phi$ satisfying $\Phi(1)=E$  can be described by some set of $A_i$ in this way. We can however gain a bit more insight into this generic update rule by viewing it as consisting of a 3-layered process: a coarse-graining, an actual update, and a unitary evolution. Let us consider this in more detail.

Write $\Phi_i(\rho) := A_i\rho A_i^*$. Then, ignoring the normalisation, the updated state is $\sum_i \Phi_i(\rho)$. We can then consider this outermost layer of the update rule as stating that we actually measured a more \emph{fine-grained} set of effects $\{A_1^*A_1,A_2^*A_2,\ldots\}$, one of which was actually observed resulting in the updated state $\Phi_i(\rho)$, but then we `forgot' the outcome of this more fine-grained measurement resulting in the more mixed state $\sum_i \Phi_i(\rho)$. This process of `forgetting' or `throwing on one heap' all the outcomes is called a \emph{coarse-graining}.
So let us now assume that no such coarse-graining happened in order to find what lies at the core of the state update rule of quantum theory. Hence assume that we have a single operator $A$ satisfying $A^*A=E$. Ignoring normalisation, the updated state is then $A\rho A^*$. We can take the \emph{polar decomposition} of $A$ in order to write it as $A=U\sqrt{A^*A}$ for some unitary $U$. Since $A^*A=E$ we can write this as $A=U\sqrt{E}$. The updated state is hence $U\sqrt{E}\rho \sqrt{E}U^*$.
We see then that the update rule consists of an update $\rho \mapsto \sqrt{E}\rho\sqrt{E}$, followed by a unitary evolution $\rho' \mapsto U\rho' U^*$.
A unitary evolution is reversible, and hence the `core part' of the update rule is the conjugation with $\sqrt{E}$.
Stripping away the coarse-graining and the unitary update we then arrive at what is known as the \emph{L\"uders update rule}:
\begin{equation}\label{eq:state-update-rule}
\rho \mapsto \frac{\sqrt{E}\rho\sqrt{E}}{\tr(E\rho)}
\end{equation}

Assume now that we are doing measurements in such a way that the state update is implemented by the L\"uders rule.
Suppose we have observed the effect $E_1$ on the state $\rho$, resulting in the updated state $\sqrt{E_1}\rho\sqrt{E_1}/\tr(E_1\rho)$. The probability that we now observe the outcome associated to an effect $E_2$ is given by:
\[\tr(E_2\frac{\sqrt{E_1}\rho\sqrt{E_1}}{\tr(E_1\rho)})\ =\  \frac{\tr(\sqrt{E_1}E_2\sqrt{E_1} \rho)}{\tr(E_1\rho)}\]
By using standard classical conditioning of probabilities we can then calculate the combined probability of observing first the outcome associated to $E_1$ and then that of $E_2$:
\[\frac{\tr(\sqrt{E_1}E_2\sqrt{E_1} \rho)}{\tr(E_1\rho)} \tr(E_1\rho) = \tr(\sqrt{E_1}E_2\sqrt{E_1} \rho).\]
We remark that $\sqrt{E_1}E_2\sqrt{E_1}$ is again an effect, and that it produces precisely the same statistics as first observing $E_1$ and then observing $E_2$. We hence are motivated to define the \Define{sequential product} $E_1\mult E_2 := \sqrt{E_1}E_2\sqrt{E_1}$~\cite{gudder2001sequential}.\indexd{sequential product!in quantum theory} 
This can be seen as an update rule for effects. Given POVMs $(E_i)_{i=1}^n$ and $(F_j)_{j=1}^m$, the POVM corresponding to the `sequential measurement' where $(E_i)$ is applied first followed by $(F_j)$, is then $(\sqrt{E_i}F_j\sqrt{E_i})_{i=j=1}^{n,m}$. As compositions of linear maps are generally not commutative, so that in general $E_1\mult E_2 \neq E_2\mult E_1$, this joint POVM might give different measurement statistics than the joint measurement $(\sqrt{F_j}E_i\sqrt{F_j})_{i=j=1}^{n,m}$. So, unlike in classical mechanics, the order in which we do measurements is important.

\subsection{Composite systems}\label{sec:intro-composite-systems}

Let $\rho$ be a density operator representing a state.
The unitary evolution of $\rho$ under a unitary $U$ is given by $U\rho\, U^*$. We see that $U\rho\, U^*\geq 0$, and $\tr(U\rho\, U^*) = \tr(\rho\, U^*U) = \tr(\rho 1) = 1$, so that this is again a density operator.

Just as density operators are a generalisation of pure states to allow for a lack of full information regarding the system, we can generalise unitary evolution to allow for transformations that include uncertainty.

\begin{definition}
  Let $H$ and $H'$ be (real or complex) Hilbert spaces. We say a linear map $\Phi:B(H)\rightarrow B(H')$ is \Define{positive} when for all $A\in B(H)$ with $A\geq 0$ we have $\Phi(A)\geq 0$ in $B(H')$. A positive map is \Define{trace-preserving} when furthermore for all positive $A$ we have $\tr(\Phi(A)) = \tr(A)$.
\end{definition}

A positive trace-preserving map sends density operators to density operators and hence seems like a good candidate for a more general type of transformation. This however misses an important point that we have so far not discussed: composite systems.

\begin{definition}
  Let $H_1$ and $H_2$ be (complex or real) Hilbert spaces. Their vector space tensor product has an inner product defined by setting $\inn{v_1\otimes v_2,w_1\otimes w_2} := \inn{v_1,w_1}\inn{v_2,w_2}$ and extending by linearity. We define the \Define{tensor product of Hilbert spaces} $H_1\otimes H_2$ to be the completion in the norm given by this inner product. This gives a bilinear map $\Phi:B(H_1)\times B(H_2)\rightarrow B(H_1\otimes H_2)$ defined by $\Phi(A_1,A_2)(v_1\otimes v_2) = (A_1v_1)\otimes (A_2v_2)$. We will simply write $A_1\otimes A_2:=\Phi(A_1,A_2)$ for the resulting linear map on $B(H_1\otimes H_2)$.
\end{definition}

Given two physical systems, described respectively by the Hilbert spaces $H_1$ and $H_2$, their \Define{composite system} consisting of both systems at once is described by the tensor product $H_1\otimes H_2$. If the spaces $H_i$ are in the states $\rho_i$ (describing the state by a density operator), then the state of the composite system is $\rho_1\otimes \rho_2$. These states are called \Define{separable}, as they describe non-interacting physical systems. We however also have states that are \Define{entangled}. These result from interactions between the two separate systems, and hence those systems can no longer be seen as truly separate. 
For instance, let $\ket{0},\ket{1}$ be the standard orthonormal basis of $\C^2$. Then the (unnormalised) state $\ket{\psi}$ on $\C^2 \otimes \C^2 \cong \C^{4}$ given by $\ket{\psi} := \ket{00}+\ket{11}$ is entangled.

The existence of entanglement explains why it is not sufficient for a map between operators of Hilbert spaces to be positive and trace-preserving: given such a map $\Phi:B(H)\rightarrow B(H')$ it should also be valid to apply this map to a part of a larger system, giving a map $\Phi\otimes \id_K: B(H\otimes K)\rightarrow B(H'\otimes K)$. Let for instance $H=H'=K=\C^2$ and take $\Phi$ to be the transpose map. Letting $\ket{\psi}$ be the entangled state above we have 
$\ketbra{\psi} = \ketbra{0}\otimes\ketbra{0} + \ketbra{0}{1}\otimes\ketbra{0}{1} + \ketbra{1}{0}\otimes \ketbra{1}{0} + \ketbra{1}\otimes\ketbra{1}$ 
and hence 
$\rho' = (\Phi\otimes \id)(\ketbra{\psi}) = \ketbra{0}\otimes\ketbra{0} + \ketbra{1}{0}\otimes\ketbra{0}{1} + \ketbra{0}{1}\otimes \ketbra{1}{0} + \ketbra{1}\otimes\ketbra{1}$. 
If we now let $v := -\ket{01} + \ket{10}$ in $\C^2\otimes \C^2$ then we can easily verify that $\inn{\rho' v, v} = -2$ and hence that $\rho'$ is not positive.
Hence, just sending positive maps to positive maps on your own system is not enough: the map must also preserve positivity when applied to part of a larger system.

\begin{definition}
	Let $\Phi:B(H)\rightarrow B(H')$ be a positive map. It is \Define{completely positive} when the maps $\Phi\otimes \id_n: B(H\otimes \C^n)\rightarrow B(H'\otimes \C^n)$ are positive for all~$n\in \N$.
\end{definition}

The physically realisable processes in quantum theory (at least in finite dimension) are precisely the completely positive trace-preserving maps.

Let $\Phi:M_n(\C)\rightarrow M_m(\C)$ be a completely positive map between finite-dimensional Hilbert spaces. Then there exist linear maps $V_i:\C^m\rightarrow \C^n$ for $i=1,\ldots,nm$ such that $\Phi(A) = \sum_{i=1}^{nm} V_i A V_i^*$ for all $A\in M_n(\C)$. If $\Phi$ is trace-preserving then we furthermore have $\sum_i V_iV_i^* = 1$.
This representation of $\Phi$ is called a \Define{Kraus decomposition}\indexd{Kraus decomposition} of $\Phi$, and the maps $V_i$ are called \Define{Kraus operators}. The minimal number of non-zero maps $V_i$ that are needed to represent $\Phi$ is called the \Define{Kraus-rank} of $\Phi$. In particular, the completely positive maps with Kraus-rank 1 are $\Phi(A) = VAV^*$ for some $V:\C^m\rightarrow \C^n$. If $\Phi$ is furthermore trace-preserving then we must have $VV^* = 1$, so that  $V^*$ is an isometry. Hence, when furthermore $n=m$, the only trace-preserving Kraus-rank 1 maps are given by a unitary $V$.

Another useful result concerning completely positive maps is \Define{Stinespring's dilation theorem}\indexd{Stinespring dilation}~\cite{stinespring1955positive}. In finite dimension, this states that for any trace-preserving completely positive map $\Phi: B(H)\rightarrow B(H')$ we can find a Hilbert space $K$ and an isometry $V:H\rightarrow H'\otimes K$ such that $\Phi(A) := (\id_{H'}\otimes \tr_K) (VAV^*)$.
When considering generalised probabilistic theories, or other abstracted versions of quantum theory, a Stinespring dilation of a map (or analogous constructions in their respective settings) is often called a \Define{purification}\indexd{purification}, as it `purifies' the mixed map $\Phi$ to a `pure' isometry $V$.

\section{Operator algebras}\label{sec:operator-algebras}

Our introduction to quantum theory in the previous section focused on the states of a system. Instead, we can focus on the sort of properties that can be measured of a system: the \Define{observables}\indexd{observable}\footnote{For physicists this shift in focus can be framed as a change from the Schr\"odinger picture to the Heisenberg picture, while a computer scientist might describe it as a change of description from state-transformers to predicate-transformers.}.

We already saw an example of an observable, the Hamiltonian, which observes the amount of energy in a system. Mathematically we identify observables with the self-adjoint maps of the Hilbert space.

Early on in the development of the mathematics of quantum theory it was realised that it would be useful to study spaces of observables abstractly as a particular type of algebra. These algebras are now known as \Define{operator algebras}\indexd{operator algebra}. The most well-known type of operator algebra is a C$^*$-algebra.

\begin{definition}
	Let $(\mathfrak{A},\norm{\cdot},\cdot, *)$ be a complex normed vector space with a bilinear associative operation $\cdot$ and a sesquilinear involution $*$ (\ie~$(a^*)^* = a$, $(a+b)^* = a^*+b^*$ and $(za)^* = \overline{z}a^*$ for all $a,b\in\mathfrak{A}$ and $z\in\C$). We say $\mathfrak{A}$ is a \Define{C$^*$-algebra}\indexd{C*-algebra} when it satisfies the following conditions.
	\begin{itemize}
		\item It is complete in the norm $\norm{\cdot}$.
		\item $(a\cdot b)^* = b^*\cdot a^*$ for all $a,b\in \mathfrak{A}$.
		\item $\norm{a^*\cdot a} = \norm{a}\norm{a^*}$.
	\end{itemize}
\end{definition}

\begin{example}
	Let $H$ be a complex Hilbert space. For a bounded operator $A\in B(H)$ define the \Define{operator norm}\indexd{operator norm} as $\norm{A} := \inf\{\lambda\in\R_{\geq 0}~;~\norm{Av}\leq \lambda\norm{v} \text{ for all } v\in H\}$. Then $B(H)$ with the operator norm, composition of linear maps, and adjoint forms a C$^*$-algebra. Furthermore, let $\mathfrak{A}\sse B(H)$ be any subspace of bounded linear maps that is closed in the operator norm and under taking adjoints. Then $\mathfrak{A}$ is a C$^*$-algebra.
\end{example}

The converse of the above result is also true.
\begin{theorem}[Gelfand-Naimark~\cite{gelfand1943inclusion}]
	Let $\mathfrak{A}$ be a C$^*$-algebra. Then there exists a complex Hilbert space $H$ and an injective linear map $\phi:\mathfrak{A}\rightarrow B(H)$ that is a {$*$-homomorphism}, \ie~$\phi(a\cdot b) = \phi(a)\phi(b)$ and $\phi(a^*)=\phi(a)^*$, and an isometry, $\norm{\phi(a)} = \norm{a}$.
\end{theorem}

In finite dimension, C$^*$-algebras are particularly simple.
\begin{theorem}[{\cite{kadison1997fundamentals}}]
	Let $\mathfrak{A}$ be a finite-dimensional C$^*$-algebra. Then there exist (unique) numbers $n_1,\ldots n_k\in \N_{>0}$ such that $\mathfrak{A}$ is isomorphic as a C$^*$-algebra to $M_{n_1}(\C)\oplus\cdots\oplus M_{n_k}(\C)$, where `$\oplus$' denotes the direct sum of algebras, consisting of the Cartesian product of the spaces with pointwise operations.
\end{theorem}

Interpreting matrix algebras as quantum systems, we can view the direct sum in the above theorem as a `classical mixture' of quantum systems, where we are allowed to prepare any of the quantum system $M_{n_i}(\C)$.
Hence, in finite-dimension, a C$^*$-algebra essentially describes a quantum system just like a Hilbert space would.

As remarked, observables correspond to self-adjoint operators. But a C$^*$-algebra necessarily contains non-self-adjoint, and hence `non-physical' elements. This is because of two reasons. First, for any $A\in B(H)$ self-adjoint we have $(iA)^* = -i A$ (where $i$ is the imaginary unit), and hence $iA$ is not self-adjoint when $A\neq 0$. In other words, the space of self-adjoint maps does not form a complex vector space. Second, the product~$AB$ of self-adjoint $A,B\in B(H)$ is self-adjoint if and only if $AB=BA$.

In order to get an algebra that consists solely of the `physical' operators, \ie~self-adjoint maps, we need to resolve these two problems. The first is easily solved by working with real vector spaces instead of complex ones, but for the second problem we will need a different algebra operation.

The crucial observation, made by Jordan and von Neumann, is that self-adjoint maps stay self-adjoint when you square them. As self-adjoint maps also form a real vector space, we can then define a binary operation
\[A*B := \frac12 ((A+B)^2 - A^2 - B^2) = \frac12 (AB+BA).\]
This product has quite different properties than the usual composition of linear maps. Like the usual composition, it is bilinear, but unlike composition, it is commutative (which is easily seen) and not associative. Let us demonstrate this last point with an example. Take
\[
 A= \begin{pmatrix}1&0\\0&0 \end{pmatrix}, \qquad
 B= \begin{pmatrix}0&0\\0&1 \end{pmatrix}, \qquad
 C= \begin{pmatrix}0&1\\1&0 \end{pmatrix}, \qquad
\]
then $(A*B)*C = 0$ while $A*(B*C) = \frac14 C$.

The product $*$ however does satisfy a weaker form of associativity. For any $A,B\in B(H)$ we have $A*(B*A^2) = (A*B)*A^2$ (note that $A^2 = A*A$). 

\begin{definition}
	Let $(E,*)$ be a vector space with a bilinear commutative product~$*$. We say $E$ is a \Define{Jordan algebra}\indexd{Jordan algebra} when $*$ satisfies the \Define{Jordan identity}\indexd{Jordan identity} $a*(b*(a*a)) = (a*b)*(a*a)$ for all $a,b\in E$.
\end{definition}

\begin{example}
	Let $(\mathfrak{A}, \cdot)$ be an associative algebra (such as a C$^*$-algebra). Then the product 
	\begin{equation}\label{eq:special-Jordan-product}
	a*b := \frac12 (a\cdot b + b\cdot a)
	\end{equation}
	makes $(\mathfrak{A}, *)$ a Jordan algebra. We will refer to this product as the \Define{special Jordan product}.
\end{example}

Our goal was to make an algebra of self-adjoint elements, but the above example shows that any C$^*$-algebra is a Jordan algebra. We will hence require additional restrictions.

\begin{definition}
	Let $(E,*)$ be a Jordan algebra over the real numbers. It is \Define{formally real} when for any finite set of elements $a_1,\ldots, a_n\in E$ the sum $\sum_i^n a_i*a_i = 0$ iff  $a_i = 0$ for all $i$. It is \Define{Euclidean} when it is a finite-dimensional real vector space that has an inner product satisfying $\inn{a*b,c} = \inn{b,a*c}$ for all $a,b,c\in E$.
\end{definition}

We will often abbreviate Euclidean Jordan algebra to EJA.

As it turns out, in finite dimension, being formally real and being Euclidean are equivalent, although showing this is highly non-trivial~\cite[Proposition VIII.4.2]{faraut1994analysis}.

\begin{example}
	Let $\mathfrak{A}$ be a C$^*$-algebra, and denote by $\mathfrak{A}_\sa$ its set of self-adjoint elements. Then $\mathfrak{A}_\sa$ equipped with the special Jordan product of Eq.~\eqref{eq:special-Jordan-product} is a formally real Jordan algebra. If $\mathfrak{A}$ is finite-dimensional, so that we can view it as a subspace of a $B(H)$ with $H$ finite-dimensional, then $\mathfrak{A}_\sa$ is a Euclidean Jordan algebra with inner product $\inn{a,b} := \tr(ab)$.
\end{example}

C$^*$-algebras are complex vector spaces, but we can also define a real analogue, which we can see as closed subspaces of real Hilbert spaces. The space of self-adjoint elements of real C$^*$-algebras, in particular $M_n(\R)_\sa$, also forms a formally real Jordan algebra. 

The \Define{quaternions}\indexd{quaternions} $\mathbb{H}$\index{math}{H@$\mathbb{H}$ (quaternions)} are a division algebra (\ie~a `non-commutative field') which can be seen as a 4-dimensional vector space over the real numbers with a basis $1,i,j,k$ satisfying the identities $i^2=j^2=k^2=ijk = -1$. For a quaternion $w=a+bi+cj+dk$ we define its conjugate as $\overline{w} := a-bi-cj-dk$. We can then define a quaternionic Hilbert space similar to how we defined a complex Hilbert space. The set of bounded maps on an $n$-dimensional quaternionic Hilbert space is then $M_n(\mathbb{H})$, the set of $n\times n$ quaternionic matrices. It turns out that the space of self-adjoint quaternionic matrices $M_n(\mathbb{H})_\sa$ is also a formally real (and Euclidean) Jordan algebra.

The \Define{octonions}\indexd{octonions} $\mathbb{O}$\index{math}{O@$\mathbb{O}$ (octonions)} form a division algebra, which has dimension 8 over the real numbers. Unlike the real, complex and quaternionic number systems, the octonions are not associative. We can also define an analogue to Hilbert spaces with octonions, which allows us to define $M_n(\mathbb{O})_\sa$. It turns out that this is a Jordan algebra if and only if $n\leq 3$.

Jordan, von Neumann and Wigner introduced (formally real) Jordan algebras (then called `r-number systems'), because they hoped it would lead to alternatives to quantum theory. They were however disappointed to learn that the above examples almost exhaust the possibilities. In fact, there is only one other type of finite-dimensional formally real Jordan algebra.

\begin{definition}\label{def:spin-factor}
  Let $n>1$ and let $H\cong \R^n$ be the real $n$-dimensional Hilbert space. Set $V_n=H\oplus \R$. Equip this space with the product $(v,r)*(w,s):= (sv+rw, \inn{v,w}+rs)$, so that for example the product of vectors $v,w\in H$ is the scalar $\inn{v,w}$. Then $(V_n,*)$ is a formally-real Jordan algebra that we call the $n$-dimensional \Define{spin factor}\indexd{spin factor}.
\end{definition}
We have already encountered a few examples of spin factors. For $M_2(\C)_\sa$ an orthonormal basis is given by the three Pauli matrices $\sigma_1,\sigma_2$ and $\sigma_3$ in combination with the identity matrix $I$. Letting $H$ be the restriction of $M_2(\C)_\sa$ to the linear span of the Pauli matrices and noting that $\sigma_i*\sigma_j = \delta_{ij}I$ we see that as a Jordan algebra we indeed have $M_2(\C)_\sa\cong H\oplus \R = V_3$. Similarly we also have $M_2(\R)_\sa \cong V_2$, $M_2(\mathbb{H})_\sa\cong V_5$ and $M_2(\mathbb{O})^{\text{sa}}\cong V_{9}$.

\begin{theorem}[{Jordan-von Neumann-Wigner~\cite{jordan1993algebraic}}]
\label{thm:Jordan-classification}
   Let $A$ be a finite-dimensional formally real (or equivantly, Euclidean) Jordan algebra. Then $A\cong A_1\oplus\cdots A_k$ where each $A_i$ is equal to one of the following non-isomorphic Jordan algebras.
  \begin{itemize}
    \item The set of real numbers $\R$.
    \item The algebras $M_n(F)_\sa$ where $F=\R$, $F=\C$ or $F=\mathbb{H}$ for $n\geq 3$.
    \item The spin factors $V_n$ for $n\geq 2$.
    \item The exceptional Jordan algebra $M_3(\mathbb{O})_\sa$.
  \end{itemize}
\end{theorem}

We revisit Euclidean Jordan algebras and (their generalisation) JBW-algebras in detail in Chapter~\ref{chap:jordanalg}.

\section{Ordered vector spaces}\label{sec:ordered-vector-space}

In the previous section we considered abstractions of the algebraic structure of $B(H)$. Let us now abstract its order structure.

\begin{definition}\label{def:orderedvectorspace}
  An \Define{ordered vector space}\indexd{ordered vector space} $(V,\leq)$ is a real vector space with a partial order $\leq$ such that for all $a,b,c\in V$:
  \begin{itemize}
    \item If $a \leq b$, then also $a+c \leq b+c$.
    \item If $a \leq b$, then also $\lambda a \leq \lambda b$ for $\lambda \in \R_{\geq 0}$. 
  \end{itemize}
  We call an element $a\in V$ \Define{positive}\indexd{positive element!in ordered vector space} when $a\geq 0$. A \Define{positive}\indexd{positive map!between ordered vector spaces} map $f:V\rightarrow W$ is a linear map between ordered vector spaces such that if $a\geq_V 0$, then $f(a)\geq_W 0$.
\end{definition}

\begin{example}
    The set of self-adjoint maps on a Hilbert space $B(H)_\sa$ is an ordered vector space with the order as defined in Definition~\ref{def:positive-map-Hilbert-space}.
\end{example}

\begin{example}
    Let $E$ be a formally real Jordan algebra. We say $a\in E$ is positive and write $a\geq 0$ when $\exists b: a=b*b$. We extend this to a partial order via $a\leq b$ iff $b-a\geq 0$. This makes $E$ into an ordered vector space (although showing that this is the case is actually quite non-trivial).
\end{example}

\begin{remark}
  In an ordered vector space $V$, $a\leq b$ if and only if $b-a \geq 0$. As a result the set of positive elements of an ordered vector space completely determines the partial order. Consequently, a positive map $f:V\rightarrow W$ is automatically \Define{monotone}\indexd{monotone map}: if $a\geq_V b$ then $f(a)\geq_W f(b)$.
\end{remark}

\begin{remark}
  The set of positive elements of an ordered vector space forms a \Define{cone}\indexd{cone}: a subset $C\sse V$ that is closed under addition and positive scalar multiplication. This cone is furthermore \Define{proper}, meaning that $C\cap (-C) = \{0\}$. Conversely, any proper cone determines a partial order that makes $V$ into an ordered vector space.
\end{remark}

The Koecher--Vinberg theorem is an important result that links the theory of ordered vector spaces to that of Jordan algebras. Before we state it we recall a few more definitions.

\begin{definition}\label{def:order-iso-homogeneous}
    Let $V$ be an ordered vector space. An \Define{order isomorphism}
    \indexd{order isomorphism}
    \indexd{isomorphism!order ---} 
    is a linear map $\Phi: V\rightarrow V$ such that $\Phi(a)\geq 0 \iff a\geq 0$ for all $a\in V$ (such a map is necessarily bijective). Suppose $V$ has a given topology (for instance, if $V$ is finite-dimensional, the unique topology compatible with the linear structure). 
    We say $V$ is \Define{homogeneous}\indexd{homogeneous cone} when for every $a,b\geq 0$ in the interior of the positive cone we can find an order isomorphism $\Phi:V\rightarrow V$ such that $\Phi(a)=b$ (\ie~when the order-automorphism group acts transitively on the interior of the positive cone). We say $V$ is \Define{self-dual}\indexd{self-dual inner product} when it has an inner product $\inn{\cdot,\cdot}$ such that $\inn{a,b}\geq 0$ for all $b\geq 0$ if and only if $a\geq 0$ (\ie~when the inner product determines the order).
\end{definition}

\begin{theorem}[Koecher--Vinberg~\cite{koecher1957positivitatsbereiche,vinberg1967theory}]
\label{thm:Koecher--Vinberg}
    Let $V$ be a finite-dimensional ordered vector space that is homogeneous and self-dual. Then $V$ is order-isomorphic to a formally real Jordan algebra. Conversely, any finite-dimensional formally real Jordan algebra is homogeneous and self-dual.
\end{theorem}

It turns out that for spaces that have a particularly simple positive cone, the requirement of self-duality is even superfluous.

\begin{definition}\label{def:strictly-convex}
    Let $C$ be a positive cone of an order unit space $V$. We call $F\subseteq C$ a \Define{face}\indexd{face} of $C$ if $F$ is a convex set such that whenever $\lambda a+(1-\lambda) b \in F$ with $0<\lambda<1$ for some $a,b\in C$, then $a,b \in F$. The face $\{\lambda p~;~ \lambda \in \R_{\geq 0}\}$ of $C$ defined by an extreme point $p\in C$ is called an \Define{extreme ray}\indexd{extreme ray}. A face of $C$ is called \Define{proper}\indexd{proper face} when it is non-empty and not equal to $C$. If the only proper faces of a cone are extreme rays the cone is \Define{strictly convex}\indexd{strictly convex cone}. 
\end{definition}

\begin{proposition}[\cite{ito2017p}]\label{prop:itochar}
     Let $V$ be a finite-dimensional ordered vector space with a strictly convex homogeneous positive cone, then $V$ is order-isomorphic to a spin factor, i.e.\ $V\cong H\oplus \R$ where $H$ is a real finite-dimensional Hilbert space with the order on $H\oplus \R$ given by $(v,t)\geq 0 \iff t\geq \sqrt{\inn{v,v}}$. Consequently, $V$ is self-dual\footnote{We attribute this result to Ref.~\cite{ito2017p}, but it was probably already known by Vinberg in 1967~\cite{vinberg1967theory}: strictly convex homogeneous cones are precisely those cones that are of \emph{rank} (in the sense of Vinberg) at most 2, and Vinberg gives a complete classification of non-self-dual homogeneous cones of rank 3, seemingly implying that he knows there are no non-self-dual homogeneous cones of rank 2. Nevertheless, as far as the author is aware, Vinberg has never formally proved this result.}.
\end{proposition}

Most of the time, we will require a bit more structure than just an order on a vector space.

\begin{definition}\label{def:orderunitspace}
  Let $V$ be an ordered vector space. For a positive $u\in V$ we write $[0,u]_V:=\{v\in V~;~0\leq v\leq u\}$. We call an element $u\in V$ an \Define{order unit}\indexd{order unit} when for all $a\in V$ we can find $n\in\N$ such that $-n u \leq a \leq n u$, or equivalently, when the \Define{unit interval}
  \indexd{unit interval}\index{math}{$[0,1]_V$ (unit interval in $V$)} 
  $[0,u]_V$ spans $V$. We call an order unit $u$ \Define{Archimedean}\indexd{Archimedean order unit} when $a\leq \frac1n u$ for all $n\in \N_{>0}$ implies $a\leq 0$.

  An ordered vector space $V$ is an \Define{order unit space} (OUS)
  \indexd{order unit space}%
  \index{math}{OUS (order unit space)}%
  when it has an Archimedean order unit. We will write $1\in V$ for the designated order unit of the order unit space $V$.
\end{definition}

\begin{remark}
  Some authors define an order unit space as any ordered vector space with an order unit, and our definition is then referred to as an Archimedean order unit space. Since we will usually have Archimedean order units we will refer to this weaker type of space as a `vector space with an order unit'.
\end{remark}

\begin{definition}\label{def:stateOUS}
  Let $(V,u)$ be a vector space with order unit $u$. An element $a\in V$ is an \Define{effect}\indexd{effect!in order unit space} when $0\leq a\leq u$ and hence $a\in [0,u]_V$. A \Define{state}\indexd{state!order unit space} is a positive map $\omega:V\rightarrow \R$ satisfying $\omega(u)=1$. We will denote the set of states of an order unit space by $\st(V)$.
\end{definition}

\begin{example}
    The set of self-adjoint maps of a Hilbert space $B(H)_\sa$ is an order unit space. The identity $1$ is an Archimedean order unit. If $H$ is finite-dimensional\footnote{In infinite dimension we need the further assumption that the state is normal.} all states $\omega:B(H)_\sa\rightarrow \R$ are given by a density operator $\rho$ via $\omega(A) = \tr(\rho A)$.
\end{example}

There are different equivalent ways to define an OUS.

\begin{proposition}[{\cite[Chapter 1]{alfsen2012state}}]\label{prop:OUSequivalentdefinitions}
  Let $V$ be a vector space with order unit $u$. The following are equivalent.
  \begin{enumerate}[label=\alph*)]
    \item $V$ is an order unit space (\ie~$u$ is Archimedean).
    \item The expression $\norm{a} := \inf\{r\in \R_{\geq 0}~;~ {-r u \leq a\leq ru}\}$ defines a norm, and the set of positive elements is closed in this norm.
    \item The set of states \Define{order-separate}\indexd{order separation} the effects: for all $a,b \in [0,u]_V$, if $\omega(a)\leq \omega(b)$ for all states $\omega$ then $a\leq b$.
  \end{enumerate}
\end{proposition}

\begin{definition}
    Let $(V,1)$ be an order unit space. The \Define{order-unit norm} is defined as $\norm{a} := \inf\{r\in \R_{\geq 0}~;~ {-r 1 \leq a\leq r1}\}$. We say $V$ is \Define{complete} when it is complete in the topology induced by the order-unit norm.
\end{definition}

Note that for $B(H)_\sa$ the order-unit norm coincides with the operator norm.

An important result in the theory of order unit spaces is Kadison's representation theorem. Before we state it we will give one more important class of order unit spaces.

\begin{definition}\label{def:CX}\index{math}{C(X)@$C(X)$}
  Let $X$ be a compact Hausdorff space. We denote by $C(X)$ the space of continuous functions $f:X\rightarrow \R$. This space has a partial order given by pointwise comparison. This partial order makes $C(X)$ into an OUS with the order unit given by the constant function $1(x) = 1$. Additionally, $C(X)$ is a commutative associative algebra by pointwise multiplication: $(f\cdot g)(x) := f(x)g(x)$. Note that if $f,g\geq 0$ that then $f\cdot g \geq 0$.
\end{definition}

\begin{theorem}[Kadison's representation theorem~\cite{kadison1951representation}]\label{thm:kadison}\indexd{Kadison's representation theorem}
     Let $V$ be a complete order unit space with a bilinear operation $\cdot$ that preserves positivity: $a\cdot b \geq 0$ when $a,b\geq 0$. Then there exists a compact Hausdorff space $X$ and a linear bijection $\Phi:V\cong C(X)$ that is both an order-isomorphism and an algebra-isomorphism.
\end{theorem}
Note that this theorem does not require the multiplication to be commutative nor associative, and that hence these properties follow for free.

\section{Generalised probabilistic theories}\label{sec:GPTs}

We will in this section present the basic concepts of \emph{generalised probabilistic theories} (GPTs). This name was coined by Barrett in 2007~\cite{barrett2007information}, describing a framework based on the work of Hardy~\cite{hardy2001quantum} (although similar ideas had been considered earlier, see for instance~Refs.~\cite{gunson1967algebraic,edwards1979mathematical,ludwig1985axiomatic})\footnote{Some authors use the term `operational probabilistic theory'~\cite{chiribella2010probabilistic,chiribella2011informational}. Although one could argue that there are differences, with operational probabilistic theories relying more on a graphical description and the existence of composite systems, we will here conflate operational probabilistic theories with GPTs.}.

Although there is no consensus on what exactly the mathematical description of a GPT should be, the main idea that unifies all versions is that ultimately any physical theory must describe what outcomes can be expected when an experiment is performed. An `experiment' in the GPT framework is divided into three parts. First, a given system is prepared in some state. 
Then, some transformation is applied to the system, potentially changing the state. 
And finally, the system is measured, giving a classical outcome. Every possible outcome of the measurement has a probability of occurring, and this gives a probability distribution of the measurement outcomes over the transformed input state.

As an illustration, the system could be a molecule, with the preparation stage preparing it in the ground state. The transformation could be the exposure of the molecule to a laser, and the measurement outcome could be whether we detect a photon emitted from the molecule.

Let us formalise these ideas a bit more. We will label the physical systems as $A,B,C,\ldots$ and we associate to each system $A$ a set of states $\st(A)$ that we can prepare the system in. A transformation $T:A\rightarrow B$ from system $A$ to $B$ is then a map $T:\st(A)\rightarrow \st(B)$ that transforms every given state into another one. Finally, we represent measurements by a collection of outcomes $a_1,a_2,\ldots, a_n$ that we call \emph{effects}, collected in the set $\eff(A)$. Given a state $\omega\in\st(A)$ and a measurement consisting of effects $a_i\in \eff(A)$ we have probabilities $\omega(a_i)\in[0,1]$ that tell us the likelihood of observing the outcome associated to the effect $a_i$ when the system $A$ is prepared in the state $\omega$. So for $a_1,\ldots, a_n$ to be a valid measurement, we should have $\sum_i \omega(a_i) = 1$ for all states $\omega\in\st(A)$.

\begin{remark}
  Following the literature on operator algebras we write $\omega(a)$ for the probability of observing $a$ when the system is in the state $\omega$.
  Note that in the literature on GPTs it is more common to write this the other way around: $a(\omega)$. This notation is useful in the Schr\"odinger picture\indexd{Schrodinger picture vs Heisenberg picture} where we view operations as modifying the state of the system.
  We will however focus more on the observables of a system, and hence it will be more convenient to adopt the Heisenberg picture wherein we view operations as modifying the effects.
\end{remark}

When we have a procedure to prepare one of $\omega_1,\omega_2\in \st(A)$, we can also prepare a \emph{mixture} by flipping a biased coin and preparing either $\omega_1$ or $\omega_2$. From the operational perspective, this results in a mixture of the measurement outcomes we will observe. For a given probability $t\in[0,1]$ we will call this mixed state $t\omega_1 + (1-t)\omega_2$. This structures makes the state space $\st(A)$ a convex set. For any effect $a\in \eff(A)$ we should then have $(t\omega_1 + (1-t)\omega_2)(a) = t\omega_1(a) + (1-t)\omega_2(a)$.
Analogously, we allow mixtures of effects $ta_1+(1-t)a_2$, and we require that $\omega(ta+(1-t)a_2) = t\omega(a) + (1-t)\omega(a)$.

When two states have exactly the same outcome probabilities for every possible measurement, then there is no way to physically detect any difference between the states. We say that these states are then \Define{operationally equivalent}.\indexd{operational equivalence} 
Similarly, two effects are operationally equivalent when they give the same outcome probabilities on every possible state they can be tested against.
A common assumption in the GPT framework, that we will make here as well, is that two states or effects that are operationally equivalent are in fact equal. So if $\omega_1,\omega_2\in\st(A)$ satisfy $\omega_1(a)=\omega_2(a)$ for all $a\in\eff(A)$ then $\omega_1=\omega_2$. Mathematically, we say that the effects \Define{separate} the states. Similarly, the states separate the effects: if we have $\omega(a_1)=\omega(a_2)$ for some $a_1,a_2\in\eff(A)$ for all $\omega\in\st(A)$, then $a_1=a_2$. 
We can now introduce a useful concept.

\begin{definition}\label{def:assocvectorspaceofeffects}
    Let $A$ be a system. It's \Define{associated vector space}\indexd{associated vector space} $V_A$ is defined to be the space of formal linear combinations $\sum_i \lambda_i a_i$ where $a_i \in \eff(A)$ and $\lambda_i \in \R$, modulo equality among all states:
    \[\sum_i \lambda_i a_i \sim \sum_j \mu_j a'_j \iff \sum_i \lambda_i \omega(a_i) = \sum_j \mu_j \omega(a'_j) \text{ for all } \omega\in \st(A) \]
\end{definition}

Because the states separate the effects, the effects of $A$ embed into its associated vector space: $\eff(A) \sse V_A$, and because the states act affinely on the effects, \ie~$\omega(ta_1+ta_2) = t\omega(a_1)+(1-t)\omega(a_2)$, this embedding preserves the convex structure of the effects. We hence consider the effect space $\eff(A)$ as simply a convex subset of the vector space $V_A$. This allows us to define expressions such as $a+b$ for effects $a,b\in\eff(A)$, where we take $a+b$ to be an element of $V_A$.

We will posit the existence of two `trivial' effects that we will call $0$ and $1$. The first is the effect that is never successful and thus has $\omega(0) = 0$ for all states $\omega$. The second is the opposite, always being successful: $\omega(1) = 1$. These effects always exist for any system, because we can just decide to make a measurement device that doesn't interact with the state, and simply always outputs ``success'' or ``fail''. Alternatively, we can interpret the effect $1$ as the effect that measures ``Does the system exist?''.

\begin{remark}
	The existence of an effect like $1$, a `deterministic effect', in every system is related to the theory obeying causality, or equivalently, not allowed signalling from the future~\cite{chiribella2010probabilistic,coecke2014terminality}.
	In categorical quantum mechanics the effect $1$ functions as a `discard' map that is interpreted as throwing away a system~\cite{CKbook}.
\end{remark}

The effect $0$ is interesting because it allows us to \emph{scale down} effects: $p e := pe + (1-p) 0$. This can be interpreted as doing the measurement $e$, but deciding with probability $1-p$ to throw away the outcome and returning false. In the vector space $V_A$, the effect $0$ corresponds to the zero vector. As a result we can extend the function $\omega: \eff(A)\rightarrow [0,1]$ for any $\omega\in\st(A)$ to a linear map $\omega: V_A\rightarrow \R$. Consequently, $V_A$ is in fact an ordered vector space: we set $v\geq w$ when $\omega(v)\geq \omega(w)$ for all $\omega\in\st(A)$. Furthermore, $1$ is an order unit of $V_A$. We remark then that the states $\omega\in\st(A)$ are also states of $V_A$ in the sense of Definition~\ref{def:stateOUS}.

Another assumption we will make in our version of the GPT framework is that effects allow \Define{negation}\indexd{negation of effect}\index{math}{$a^\perp$ (negation of effect)}. Given an effect $a$ we can consider its negation which returns true if and only if $A$ returns false. We will denote this effect as $a^\perp$, pronounced ``$a$ perp'', and its probabilities are given by $\omega(a^\perp) = 1-\omega(a)$. We will sometimes also refer to the negation as $1-a$.

\begin{remark}\label{rem:coarse-grain}
  The existence of negations of effects is closely related to the ability to `coarse-grain' measurements.\indexd{coarse-graining} Given a measurement of a state $\omega$ with $n$ outcomes defined by the effects $a_1,\ldots a_n$ we have a probability distribution with probabilities $\{\omega(a_k)\}$. A \Define{coarse-graining} of this measurement is the same measurement where we conflate some of the outcomes. E.g.~we can define a measurement with $n-1$ outcomes by identifying the outcomes of $a_{n-1}$ and $a_n$ to get an outcome with probability $\omega(a_{n-1}) + \omega(a_n)$. The resulting effect is then often denoted by $a_{n-1}+a_n$ so that $\omega(a_{n-1}+a_n) = \omega(a_{n-1}) + \omega(a_n)$.
  When we identify all the outcomes, except for $a_1$, then that outcome has a probability of $\sum_{k=2}^n \omega(a_k) = 1-\omega(a_1)$ and hence this `sum of effects' acts as the negation of $a_1$.
\end{remark}

Let us now summarise our version of the GPT framework.
\begin{assumption}[GPT framework]\label{assum:GPT}
  For every system $A$ we have an associated ordered vector space $V_A$ with order unit $1$ such that $\eff(A)\sse [0,1]_{V_A}$ is a convex subset containing both $0$ and $1$ and $a^\perp:= 1-a$ when $a\in \eff(A)$. The states $\st(A)$ are a convex subset of the states of $V_A$ that contains enough states to separate the elements of $V_A$, \ie~if $\omega(v)=\omega(w)$ for all $\omega\in\st(A)$ then $v=w$.
\end{assumption}

Almost all work that deals with GPTs furthermore assumes that the associated vector spaces are finite-dimensional. An operational assumption that guarantees this restriction (and in fact is equivalent to it) is the assumption of \Define{finite tomography}: that for every system $A$ we can find a finite set of effects $a_1,\ldots a_k$ such that $\omega_1(a_j) = \omega_2(a_j)$ for all $j=1,\ldots k$ if and only if $\omega_1=\omega_2$. If a system does not satisfy finite tomography it is impossible to (approximately) characterise a state with a finite number of experiments, and hence this assumption definitely makes sense from an operational viewpoint.

Let us discuss one more topic in the framework of GPTs: composite systems and local tomography.
When given two independent systems $A$ and $B$ we can consider them as parts of a composite system that we will denote by $A\otimes B$. If $A$ is in the state $\omega_A\in\st(A)$ and $B$ is in the state $\omega_B\in\st(B)$, then the system $A\otimes B$ is in the state $\omega_A\otimes \omega_B$. But $A\otimes B$ can also have states that do not arise in this manner (as is the case with entangled states in quantum theory). We similarly can make composites $a\otimes b$ of effects in $\eff(A)$ and $\eff(B)$. These composites need to satisfy $(ta_1 + (1-t)a_2)\otimes b = t(a_1\otimes b) + (1-t)(a_2\otimes b)$ for operational reasons. This leads to a bilinear map $V_A\times V_B\rightarrow V_{A\otimes B}$ similar to the bilinear map for tensor products of vector spaces.
We say the composite $A\otimes B$ is \Define{locally tomographic} when every state of $A\otimes B$ is fully characterised by local measurements on the subsystems $A$ and $B$, \ie~when for $\omega_1,\omega_2\in\st(A\otimes B)$ we have $\omega_1(a\otimes b) = \omega_2(a\otimes b)$ for all $a\in\eff(A)$ and $b\in\eff(B)$ if and only if $\omega_1=\omega_2$. If the systems satisfy finite tomography, so that the associated vector spaces are finite-dimensional, local tomography is equivalent to $\dim V_{A\otimes B} = \dim V_A ~\dim V_B$.
Interestingly, regular quantum theory (described by complex C$^*$-algebras) satisfies local tomography, whereas `real' quantum theory (where systems are real C$^*$-algebras) does not~\cite{barnum2014local}.

\section{A history of first principles for quantum theory}\label{sec:history-of-principles}

In order to put the results of this thesis into context, we will give a brief overview of previous work in the topic of first principles for quantum theory, highlighting the most commonly used types of assumptions.

\textbf{Foundational results} --- The starting point of the field of reconstructions of quantum theory can be considered to be von Neumann's seminal 1932 book \emph{Mathematical Foundations of Quantum Mechanics}~\cite{neumann2013grundlagen}, as from that point onwards it was clear which mathematics actually needed to be explained from first principles. 
In these early days, much work was done on axiomatic quantum theory in the hope that this would lead to some natural generalisation. It turned out however that such generalisations are elusive.
For instance, Wigner's theorem~\cite{wigner1931gruppentheorie} showed that the standard unitary maps naturally arise as symmetries of quantum states, while Stone's theorem on one-parameter unitary groups~\cite{stone1932one} showed that any unitary time-evolution must be implemented by a Hamiltonian, reconstructing the basic form of the Schr\"odinger equation.
Jordan, von Neumann and Wigner studied what later became known as Jordan algebras as a generalisation of the space of observables of a quantum system. To their disappointment they discovered that Jordan algebras are in fact very close to regular quantum theory indeed~\cite{jordan1993algebraic}, with almost all formally real Jordan algebras embedding into the set of bounded operators of a Hilbert space.
Another later result along this same line is Gleason's theorem~\cite{gleason1957measures} which showed that measures on the space of projections of a Hilbert space are necessarily characterised by a bounded operator, just like in the Born rule of quantum mechanics.

\textbf{Quantum logic} --- The need for a complete axiomatic reconstruction of quantum theory was outlined by Mackey in 1957~\cite{mackey1957quantum}. He wished to do so using the \emph{quantum logic} approach of von Neumann and Birkhoff~\cite{birkhoff1936logic} that takes the orthomodular poset $P(H)$ of projections on a Hilbert space $H$ as the central concept. 
This approach was already used by von Neumann in a set of unpublished notes from 1937~\cite{neumann1981continuous} to reconstruct quantum theory, albeit with a rather technical set of assumptions. 
Mackey motivates why a set of observables should be represented by an orthomodular poset, but stops short of recovering $P(H)$. Mackey's work was continued by Piron who in 1972 showed that any irreducible complete atomistic orthomodular lattice satisfying the \emph{covering property} (cf.~Section~\ref{sec:coverprop}) must be isomorphic to $P(H)$ where $H$ is a \emph{generalised} Hilbert space~\cite{piron1972survey}. This programme was essentially completed decades later by S\`oler when she showed that the only infinite-dimensional generalised Hilbert spaces are the Hilbert spaces over the reals, complex numbers or quaternions~\cite{soler1995characterization}. Hence, this approach essentially recovered the same spaces as the characterisation of Euclidean Jordan algebras (cf.~Theorem~\ref{thm:Jordan-classification}), albeit in infinite-dimension instead. For a more in-depth overview of the foundational work done in quantum logic we refer to the survey Ref.~\cite{coecke2000operational}.

\textbf{Early reconstructions} --- Contrary to the quantum logic approach that only considers `sharp' observables, \ie~projections, there is the operational approach where `fuzzy' observables (effects) are allowed, bringing us close to the GPT framework. 
Probably the first person to use a GPT-like framework for a reconstruction was Gunson in 1967~\cite{gunson1967algebraic}. 
His twelve axioms show that the space of observables is isomorphic to a $B(H)$ where $H$ is an infinite-dimensional real, complex or quaternionic Hilbert space. 
Of these twelve axioms, the first 6 essentially recover the GPT framework, the next 3 are algebraic in nature and have no clear physical interpretation, while the last 3 are related to the existence and properties of \emph{filters}, specific state transformations that project the state onto a certain `filtered' subspace (cf.~Section~\ref{sec:filterscompressions}). 
The importance of filters was further realised by Mielnik~\cite{mielnik1968geometry,mielnik1969theory} that considered more general physical systems based on properties of filters and \emph{transition probabilities}: the probability that a particle goes through a second filter if it successfully went through a first filter. He furthermore introduced the axiom of \emph{symmetry of transition probabilities} that says the probability stays the same if the order of the filters is interchanged.
Symmetry of transition probabilities has been used as an assumption in many subsequent reconstructions, such as in Refs.~\cite{fivel1994interference,landsman1997poisson,wilce2018royal,alfsen2012geometry} (and the derivation of this property will be a cornerstone in the reconstructions of Chapters~\ref{chap:seqprod} and~\ref{chap:effectus}).
The work of Gunson and Mielnik was continued by Guz in 1981~\cite{guz1981conditional} who was the first to realise the connection between the covering property of the lattice of observables and the well-behavedness of filters. He used the notion of \emph{pure states}, states that are convex-extreme in the state-space. To each pure state $\omega$ he associates a unique sharp measurement $p_\omega$ that represents `testing' for that pure state and this allowed him to define a `transition probability' between states as $\inn{\omega_1,\omega_2} := \omega_2(p_\omega)$. This approach has also been adopted by many other authors.

\textbf{Modern reconstructions} --- Most of the previously mentioned works have in common that they focus on (countably) infinite-dimensional spaces and that they only consider spaces in isolation, never dealing with composite systems and tensor products.
This stands in contrast to the modern approach, which could be said to have been initiated with Hardy's 2001 preprint \emph{Quantum Theory From Five Reasonable Axioms}~\cite{hardy2001quantum}. Hardy's main innovation was that he considered composite systems and subsystems in finite dimension, which allowed him to introduce the axiom of \emph{local tomography} (cf.~Section~\ref{sec:GPTs}). It is this latter axiom that `selects' complex Hilbert spaces over real or quaternionic ones (which were still possibilities in most previous approaches)~\cite{hardy2012limited}. He also used a \emph{pure transitivity} axiom. Such axioms state that the set of reversible transformations act transitively on the set of pure states, \ie~that for every pair of pure states $\omega_1$ and $\omega_2$ we can find a transformation $\Phi$ that has an inverse $\Phi^{-1}$ such that $\Phi(\omega_1) = \omega_2$. Versions of pure transitivity are assumed in most subsequent reconstructions, such as in Refs.~\cite{barnum2014higher,hardy2011reformulating,dakic2009quantum,masanes2014entanglement,krumm2017thermodynamics,chiribella2011informational,chiribella2016entanglement,tull2016reconstruction,selby2018reconstructing}, while local tomography is still one of the only compelling ways to distinguish real from complex quantum theory.

Since Hardy's work, many reconstructions by various authors have appeared (besides the ones mentioned above, also Refs.~\cite{clifton2003characterizing,masanes2011derivation,fivel2012derivation,hohn2017quantum,dakic2009quantum,niestegge2020simple,wetering2018reconstruction,wetering2018sequential,wilce2009four}). We will here only highlight what can be considered the most influential one: that of Chiribella, D'Ariano and Perinotti~\cite{chiribella2011informational,dariano2017firstprinciples}. This reconstruction motivates its axioms by considering them rules on information processing in physical systems. Besides versions of local tomography, a pure transitivity axiom, and an axiom related to the existence of filters, it introduces two new axioms. The first basically states that `purity is closed under composition'. Specifically, that the state resulting from measuring a pure effect on a part of a pure state on a composite system is still pure. The second new axiom states the possibility of `purification' of states, that any mixed state is the result of throwing away part of a pure composite state~\cite{chiribella2010probabilistic} (in quantum theory such purifications are given by Stinespring dilations; cf.~Section~\ref{sec:intro-composite-systems}). A version of purification has since been used in other reconstructions~\cite{tull2016reconstruction,selby2018reconstructing,barnum2017ruling}.

The number of reconstructions that have appeared in the last 20 years show that there is still little belief that the indisputable `right' set of principles has been discovered. 
However, the similarities in the assumptions and sometimes even the proofs in many of the approaches highlight that there are a few key mathematical features of (finite-dimensional) quantum theory. 
The properties of `pure' states (where `purity' can mean different things in different contexts; cf.~Section~\ref{sec:purity}), in particular that every mixed state can be `diagonalised' in terms of pure states, are a crucial part of the proof in many reconstructions. Similarly, pure transitivity, which ensures the existence of `enough' reversible transformations to map the pure states between each other, is often needed. 
Finally, the property of local tomography seems to be the only assumption so far that can distinguish between real and complex quantum theory.%
\footnote{Although it should be noted that Refs.~\cite{barnum2014higher,alfsen2012geometry} postulate an, arguably somewhat arbitrary, correspondence between dynamics and observables that also succeeds in distinguishing between real and complex quantum theory.}
In fact, by combining these three properties, diagonalisability, pure transitivity, and local tomography, quantum theory can essentially be derived~\cite{barnum2019strongly}.



\chapter{Sequential measurement}\label{chap:seqprod}

In this chapter we will show that, in finite dimension, quantum theory is the unique non-classical physical theory where sequential measurement is suitably well-behaved. Given two effects $a$ and $b$ we have an effect $a\mult b$ that corresponds to `measure $a$ and then $b$'. In classical theory the order of measurements is not important so that we have $a\mult b = b\mult a$ for all measurements $a$ and $b$. There is however no reason to assume that this would continue to hold for general physical theories, because a measurement will change the state of the system in a way that modifies the subsequent measurement probabilities (as is indeed the case in quantum theory). 
However, some effects $a$ and $b$ will be `compatible' in the sense that although $a$ changes the state, this does not affect the probability of observing $b$, and hence we can still have $a\mult b = b\mult a$ for such effects. Our assumptions regarding sequential measurement can be intuitively stated as `classical operations preserve compatibility of effects'.

Mathematically the results of this chapter state that a finite-dimensional order unit space equipped with a \emph{sequential product}~\cite{gudder2002sequential} that is continuous must be order-isomorphic to a Euclidean Jordan algebra, while the only such systems that have a locally tomographic composite with themselves are C$^*$-algebras.

With an eye on the extensive existing literature on reconstructions of quantum theory covered in Section~\ref{sec:history-of-principles}, it is worthwhile considering how the approach of this chapter is different from and adds to the literature. 
To start, this is the first reconstruction to use the concept of sequential measurement in a principled way, hence highlighting the importance of the structure of this operation to quantum theory.%
\footnote{The work of Niestegge~\cite{niestegge2020simple} reconstructs a part of quantum theory using assumptions on conditional probabilities, which are closely related to sequential measurement. He however only considers sharp measurements, and his assumptions are not satisfied by all quantum systems, such as the qubit system.}
It is furthermore noteworthy just how few different concepts we need to refer to in our assumptions: just effects, states and the sequential product. We require no specific assumptions regarding (reversible) transformations, pure states, information capacity, etc. In this sense, our reconstruction is `focused'.
Finally, as we will see in Chapter~\ref{chap:jordanalg}, the assumptions we have on the sequential product continue to hold in infinite dimension, in contrast to most (modern) reconstructions that employ assumptions that only hold in finite dimension. Hence, our assumptions could be considered more natural as they do not rely on the peculiarities of finite-dimensional systems. This naturality of the assumptions is further highlighted by the fact that (almost all) the assumptions we make on the sequential product were already considered in an axiomatic manner~\cite{gudder2002sequential}, making it seem more plausible that someone could have independently come up with our assumptions, without knowing a priori about quantum theory.

This chapter is structured as follows. We discuss our version of the GPT framework in Section~\ref{sec:gpt-framework}, which results in our systems being described by order unit spaces. Then we introduce the axioms of the sequential product in Section~\ref{sec:seqmeas}. We prove some basic results culminating in a spectral theorem and a proof of homogeneity of the space in Section~\ref{sec:basicresultsseqprod}. The most technical part of the proof is establishing the self-duality of the space, which is done in Section~\ref{sec:selfdual}. We discuss some consequences of self-duality with regards to the Born rule in Section~\ref{sec:bornrule}. As the space is homogeneous and self-dual we could use the Koecher--Vinberg theorem to conclude that our spaces are Euclidean Jordan algebras, but instead we will demonstrate an explicit construction of the Jordan product in Section~\ref{sec:jordanproduct}. In Section~\ref{sec:central-effects} we study central effects which are necessary for our results regarding locally tomographic composites in Section~\ref{sec:seqmeas-tensorproducts} which show that the only systems allowing such a composite are complex C$^*$-algebras. In Section~\ref{sec:dynamics} we sketch how our results imply the regular allowed dynamics of quantum theory.
We end with some concluding remarks in Section~\ref{sec:seqmeas-conclusion}.

\section{Framework}\label{sec:gpt-framework}

For this chapter we adopt the GPT framework as outlined in Section~\ref{sec:GPTs}, in particular Assumption~\ref{assum:GPT}. Hence we represent a physical system $A$ with a finite-dimensional real ordered vector space $V_A$ such that $\eff(A)\sse [0,1]_{V_A}$ and $\st(A)\sse \st(V_A)$. We will however require a bit more structure then that.

Imagine we have an ensemble of identical systems, each of which is prepared in the same state $\omega$. 
We can measure an effect $a$ on some of the states and an effect $b$ on some of the others. 
The probabilities of success are then given respectively by $\omega(a)$ and $\omega(b)$. 
When $\omega(a)+\omega(b)\leq 1$ for every possible state $\omega$, we wish to define the `statistical' effect $a+b\in V_A$ which can be interpreted as measuring ``$a$ or $b$ is true''. In other words, we expect $a+b \in \eff(A)$ is an effect whenever $\omega(a+b)\leq 1$ for all states $\omega\in \Omega$.  Similarly, if $\omega(b) - \omega(a) \geq 0$ for all states $\omega$ then we can consider the effect $b-a\in V_A$ which can be interpreted as measuring ``$b$ is true and $a$ is not true''.

\begin{assumption}\label{assum:effectspaceclosedunderaddition}
    For $a,b\in \eff(A)$, if $\omega(a)+\omega(b) \leq 1$ for all $\omega \in \st(A)$, then $a+b \in \eff(A)$. If $\omega(b)-\omega(a) \geq 0$ for all $\omega\in\st(A)$, then $b-a \in \eff(A)$.
\end{assumption}

\begin{remark}
	One could think that we can define the statistical effect by flipping a fair coin and based on that deciding whether to measure $a$ or $b$. However, this results in the effect $\frac12 a + \frac12 b$ instead of $a+b$. While the effect $\frac12 a + \frac12 b$ is guaranteed to exist by the Assumptions~\ref{assum:GPT} (since the effects are closed under convex combinations), the effect $a+b$ does not necessarily exist just based on these assumptions.
	Indeed, as will become clear from the next proposition, the existence of effects like $a+b$ implies that all mathematical effects necessarily correspond to physical effects, which is not the case when just using the Assumptions~\ref{assum:GPT}.
\end{remark}

\begin{proposition}\label{prop:effectspaceisfullunitinterval}
    Let $V_A$ be the vector space associated to a system $A$ satisfying Assumptions~\ref{assum:GPT} and~\ref{assum:effectspaceclosedunderaddition}. Then $\eff(A) = [0,1]_{V_A}$.
\end{proposition}
\begin{proof}
  The inclusion $\eff(A)\sse [0,1]_{V_A}$ is of course trivial. So let $0\leq c \leq 1$ in $V_A$, our goal is to show that $c$ is an effect, \ie~$c\in \eff(A)$. 

  The vector space $V_A$ is by definition spanned by linear combinations of effects of $A$, and hence $c = \sum_i \lambda_i a_i$ for $a_i\in \eff(A)$. We split this sum up into a positive and negative part based on whether $\lambda_i$ is positive or negative to get $c = \sum_{i} \lambda_i a_i - \sum_j \mu_j b_j$ where all $\lambda_i, \mu_j\geq 0$. 
  Let $\lambda = \sum_i \lambda_i$ so that $\frac{1}{\lambda} \sum_i \lambda_i a_i = \sum_i \frac{\lambda_i}{\lambda} p_i$ is a convex combination of effects, and hence lies in $\eff(A)$. By doing the same with the $b_i$ we see that we can write $c = \lambda a - \mu b$ where $a,b\in \eff(A)$, $a = \sum_i\frac{\lambda_i}{\lambda} a_i$ and $b = \sum_i\frac{\mu_i}{\mu}b_i$.
  We now make a case distinction based on whether $\lambda\leq \mu$ or $\lambda \geq \mu$.

  Suppose $\lambda\leq \mu$. Then $\frac{\lambda}{\mu}\leq 1$ and hence $\frac{\lambda}{\mu}a = \frac{\lambda}{\mu}a + (1-\frac{\lambda}{\mu}) 0\in\eff(A)$ so that $0\leq \frac1\mu c = \frac{\lambda}{\mu}a - b$, is a difference of effects. Assumption~\ref{assum:effectspaceclosedunderaddition} then implies that $\frac1\mu c \in \eff(A)$. If $\mu \leq 1$, then $\mu (\frac1\mu c) = c\in \eff(A)$ and we are done. If $\mu \geq 1$, then we write $\mu = n + \epsilon$ where $n\in\N$ and $0<\epsilon<1$ We note that $\sum_i^n \frac1\mu c = \frac{n}{\mu} c \leq \frac{\mu}{\mu} c = c\leq 1$ and hence by Assumption~\ref{assum:effectspaceclosedunderaddition} is an effect. Similarly $\frac{\epsilon}{\mu}c + \frac{n}{\mu}c = c \leq 1$ is also an effect, and we are done.

  The case where $\lambda \geq \mu$ is handled analogously.
\end{proof}

\begin{remark}\label{remark:norestrictionhypothesis}
    The property that $\eff(A)=[0,1]_{V_A}$, \ie~that the physically realisable effects exactly match the mathematically definable effects is a common assumption in GPTs known as the \Define{no-restriction hypothesis}~\cite{chiribella2011informational,janotta2013generalized}\indexd{no-restriction hypothesis}. Note that we have not shown that the states of the system satisfy the no-restriction hypothesis. The only thing that we currently have shown about the state space is that there are enough physical states to order-separate the effects. We revisit this issue for states in Section~\ref{sec:bornrule}.
\end{remark}

\begin{proposition}\label{prop:effectspaceisorderunitspace}
    Let $V_A$ be the vector space associated to a system $A$ satisfying Assumptions~\ref{assum:GPT} and~\ref{assum:effectspaceclosedunderaddition}. Then $V_A$ is an order unit space.
\end{proposition}
\begin{proof}
    We will show that the states order-separate the effects, which is sufficient by Proposition~\ref{prop:OUSequivalentdefinitions}. So suppose we have $a,b \in [0,1]_{V_A}$ such that $\omega(a)\leq \omega(b)$ for all $\omega\in \st(V)$. We need to show that $a\leq b$.

    By Proposition~\ref{prop:effectspaceisfullunitinterval} we have $a,b\in \eff(A)$. Note also that $\st(A)\sse\st(V_A)$ and hence $\omega(a)\leq\omega(b)$ for all $\omega\in\st(A)$ so that $\omega(b)-\omega(a) \geq 0$. By Assumption~\ref{assum:effectspaceclosedunderaddition} we then have $b-a\in\eff(A)$ and hence $b-a\geq 0$. But then $a\leq b$ and we are done.
\end{proof}

Based on the results of Propositions~\ref{prop:effectspaceisfullunitinterval} and \ref{prop:effectspaceisorderunitspace} we will identify a physical system with its associated order unit space $V_A$ and its effect space with $[0,1]_{V_A}$. We however still cannot know exactly which subset $\st(A)$ is of $\st(V_A)$.

\begin{remark}
 In the basic GPT framework of Section~\ref{sec:GPTs} we only require that effects are separated by states, instead of \emph{order}-separated. Separation of states is sufficient for the norm defined in Proposition~\ref{prop:OUSequivalentdefinitions} to be an actual norm. The order-separation is hence equivalent to requiring that the set of effects is closed in this norm, which is an assumption that is also made regularly in the literature on GPTs~\cite{chiribella2011informational}.
\end{remark}

\section{Sequential Measurement}\label{sec:seqmeas}

\index{math}{$a\mult b$ (sequential measurement)} Let $a$ and $b$ denote two effects that can be measured on the same system. Their \Define{sequential product}\indexd{sequential product} is the effect that is implemented by first measuring $a$ and then measuring $b$. The sequential product, that we will denote by $a\mult b$, is considered successful when $a$ and $b$ are both successful. Hence, letting $\omega_a$ denote the state that results from an observation of $a$ after a preparation of $\omega$, we have $\omega(a\mult b) = \omega_a(b)\omega(a)$. The expression $a\mult b$ can be read as ``we measure $a$ \emph{and then} we measure $b$.''

The sequential product gives us a map $\&: \eff(A)\times \eff(A)\rightarrow \eff(A)$ that takes two effects and produces a new one. In the following paragraphs we will motivate the conditions we require of this map. The reader not interested in the physical motivation of our assumptions is welcome to skip to Definition~\ref{def:seqprod}.

Since both the measurements of $a$ and $b$ can influence the system in some non-trivial way, we wouldn't expect the outcome probabilities of $a\mult b$ to be the same as those of $b\mult a$, the measurement that is implemented by reversing the order of measurement. For some measurements however, the order might not be important. When this is the case we will call the measurements \Define{compatible}\indexd{compatible effects}. Following Gudder and Greechie~\cite{gudder2002sequential} we will argue that when compatible measurements are considered, the sequential product should act in a `classical' way.

Suppose that we have an ensemble of identical states and that we measure the effect $a$ on all of them. Let $b$ and $c$ now be measurements so that their disjunction $b+c$ exists, and split the ensemble into two. Measure $b$ on the first set and $c$ on the second. The complete process is now described by the effect $a\mult (b+c)$. The same situation however can equivalently be described as splitting the ensemble into two and then measuring $a\& b$ on the first group, and $a\& c$ on the second group. This measurement is described by $a\mult b + a\mult c$. As a result we should have $a\mult(b+c)=a\mult b + a\mult c$. 
Crucially, we have no reason to expect the same property to hold in the first argument (that is: $(b+c)\mult a = b\mult a + c\mult a$) because $b+c$ is a measurement that only exists in a statistical sense. The expression $(b+c)\mult a$ hence does not make sense (in general).

When effects $a$ and $b$ are compatible we will write $a\commu b$ as a shorthand. By definition we have $a\commu b$ when $\omega(a\mult b)=\omega(b\mult a)$ for all states $\omega$, but since states separate effects this is only true when $a\mult b = b\mult a$. 
As the order of measurement of compatible effects is not relevant it makes sense to view the measurements as being performed at the same time. This is captured by the equality $a\mult(b\mult c) = (a\mult b)\mult c$, \ie~measuring $a$ and then $b$ \emph{and then} $c$ should be the same as measuring $a$ and $b$ `at the same time' and only then measuring $c$. 

\begin{remark}\indexd{associativity in sequential product}
    It would seem to be more natural to require associativity of the sequential product \emph{in all cases}, and not just for compatible measurements, as was pointed out in Ref.~\cite{gudder2001sequential}. It however turns out that quantum theory does not satisfy this assumption of associativity. In fact, we will see that in combination with the other assumptions we will make, the only systems satisfying associativity of the sequential product are classical (i.e.~commutative, see Proposition~\ref{prop:assoc-is-commutative}). To the authors knowledge there is still no satisfying interpretation of the expression `$(a\mult b)\mult c$' when $a$ and $b$ are not compatible.
\end{remark}

If we have an effect $a$, its negation $a^\perp$ can be physically implemented in the same way, since the negation is merely a change of classical description of its output. We therefore expect $a^\perp \commu b$ to hold whenever $a\commu b$. Similarly, if $a$ is compatible with $b$ and with $c$, then $a$ should be compatible with $b+c$ (if it is defined).

Suppose $a\mult b = 0$. This states that it is impossible for both effects $a$ and $b$ to be observed on a given state. It seems reasonable to expect that this situation does not change when we interchange the order of execution. We will therefore require $b\mult a =0$ whenever $a\mult b=0$. In that case we will call $a$ and $b$ \Define{orthogonal}\indexd{orthogonal effect}. Note that we must of course have $0\mult a = 0$ (and relatedly $1\mult a = a$).

\begin{remark}
	This assumption that $a\mult b = 0$ iff $b\mult a = 0$ holds for the L\"uders update rule of quantum theory, but does not hold for other more general types of updates.
	We still however feel warranted in using this condition as it is a quite natural assumption for a hypothetical physical theory and one that someone working in the framework we have set up who is not aware of quantum theory could still reasonably come up with independently (for instance by adopting it from the classical framework).
	Regardless, in the interest of generality we try to avoid this assumption as much as possible in the early derivations of this chapter and the results of Section~\ref{sec:basicresultsseqprod} do not require it.
\end{remark}

We will need one additional assumption. Any physical measurement is noisy, but as the amount of noise is reduced, the measurement statistics should converge to the value of the idealised measurement. We capture this property by requiring the sequential product to be continuous: if $a_n\rightarrow a$ then $a_n\mult b \rightarrow a\mult b$ for all effects $b$. The continuity of the sequential product map in the second argument will follow automatically, as it turns out to be linear in that argument.

Our framework and all the assumptions regarding the sequential product are summarised in the following definition.
\begin{definition}\label{def:seqprod}
  Let $V$ be an order unit space with a function ${\&:E\times E\rightarrow E}$ where $E=[0,1]_V$ is its set of effects. We write $a\commu b$ when $a\mult b = b\mult a$ and say that $a$ and $b$ are \Define{compatible}\indexd{compatible effects}. We call $\&$ a \Define{sequential product}\indexd{sequential product!on order unit space} if it satisfies the following conditions for all $a,b,c\in E$.
  \begin{enumerate}[label=({S}\theenumi), ref=S\theenumi]
        \item \label{ax:add} Additivity: $a\mult (b+c) = a\mult b+ a\mult c$ whenever $b+c\leq 1$.
        \item \label{ax:cont} Continuity: The map $a\mapsto a\mult b$ is continuous in the norm.
        \item \label{ax:unit} Unit: $1\mult a = a$.
        \item \label{ax:orth} Compatibility of orthogonal effects: If $a\mult b = 0$ then also $b\mult a =0$.
        \item \label{ax:assoc} Associativity of compatible effects: If $a\commu b$ then $a\mult (b\mult c) = (a\mult b)\mult c$.
        \item \label{ax:compadd} Additivity of compatible effects: If $a\commu b$ then $a \commu b^\perp$. If furthermore $a\commu c$ and $b+c\leq 1$, then $a\commu (b+c)$.
    \end{enumerate}
    We call an order unit space with a sequential product a \Define{sequential effect space}\indexd{sequential effect space}.
\end{definition}

\begin{remark}\index{sequential effect algebra}
    The properties for the sequential product are close to those required in a \emph{sequential effect algebra} as introduced by Gudder and Greechie \cite{gudder2002sequential} (cf.~Definition~\ref{def:normal-SEA}) and studied in \cite{gudder2018convex,gudder2005uniqueness,jun2009remarks,wetering2018characterisation}. The only difference is that we add the requirement of continuity~\ref{ax:cont}, while they have a further axiom stating that if $a\commu b$ and $a\commu c$, then also $a\commu (b\mult c)$. We will require this further axiom in Chapter~\ref{chap:infinitedimension}.
\end{remark}


\begin{example}
	Let $X$ be a compact Hausdorff space. Denote by $C(X)$ the space of continuous functions $f:X\rightarrow \R$. Then $C(X)$ is an order unit space with unit the constant function $1(x) = 1$. Its unit interval consists of the continuous functions that restrict to $f:X\rightarrow [0,1]$. We can then define a (commutative) sequential product on the unit interval by pointwise multiplication: $(f\mult g)(x) = f(x)g(x)$.
\end{example}

\begin{example}
Let $V = B(H)_\sa$, the set of self-adjoint operators on a complex Hilbert space $H$ (representing a quantum system). Given two effects $a$ and $b$, the operation $a\mult b:= \sqrt{a}b\sqrt{a}$ is a sequential product and we have $a\commu b$ if and only if $ab=ba$~\cite{gudder2002sequentially}.
\end{example}

\begin{remark}
As discussed in Section~\ref{sec:mixed-state-quantum} the L\"uders rule $(a,b)\mapsto \sqrt{a}b\sqrt{a}$ is not the only possibility of an update rule for effects.
Another set of update rules that satisfies the axioms of a sequential product is given by $(a,b)\mapsto u_a \sqrt{a}b\sqrt{a} u_a^\dagger$ where $u_a$ is a particular unitary commuting with $a$. This map can be interpreted as the instrument that implements `measure $a$, wait for a while, and then measure $b$', where in between the two measurements, the system is dynamically evolving in the basis of $a$.
In fact, this is the most general form of update rule for $B(H)_\sa$ that satisfies the axioms~\cite{weihua2009uniqueness}.
Hence, the most general form of update rule~\eqref{eq:general-update-rule} is not compatible with our axioms. It is still not entirely clear why this should be the case.
A hint is given by considering the action of the effect $1$. According to our axioms, observing $1$ should not affect the state, while with~\eqref{eq:general-update-rule} the state could be changed in an almost arbitrary way. Hence, our axioms require a more strict correspondence between the effect and its action on the state then the general update rule implies.

While the axioms of Definition~\ref{def:seqprod} do not uniquely pick out the L\"uders update rule as special, some variations on the properties of the sequential product have been proposed that do characterise this update rule~\cite{gudder2008characterization,westerbaan2016universal,wetering2018characterisation}.
\end{remark}

Our aim now is to study finite-dimensional sequential effect spaces and show that these correspond to quantum-like systems (we will study a similar structure in infinite dimension in Chapter~\ref{chap:infinitedimension}). Before we do so however, it is interesting to note that the assumption that the underlying space has an Archimedean order unit is necessary, as otherwise some more pathological spaces satisfy our assumptions.
\begin{example}\label{ex:weird-seq-prod}
  Let $V$ be an ordered vector space and let $R$ be the space of linear functions $f:V\rightarrow V$. For $f,g\in R$ we set $f\leq g$ when $f(v)\leq g(v)$ for all $v\geq 0$ in~$V$, making $R$ into an ordered vector space. Let $E := [0,\id]_R := \{f\in R~;~ 0\leq f\leq \id\}$. Then the regular composition of linear maps $f\circ g$ is a bilinear associative product on $E$ that satisfies all the axioms of a sequential product when $f\circ g = 0$ implies $g\circ f = 0$.

  In particular, let $V=\R^2$ be equipped with the order determined by $(a,b) > 0$ iff  $a+b > 0$ and define
    $R$ and $E$ as above.
    Of course $R$ is just the space of $2\times 2$ real matrices.
    With some straightforward but tedious calculation it can be verified
    that
\begin{equation*}
    A  :=  \begin{pmatrix}
        a&b\\
        c&d
    \end{pmatrix} \in E \quad \iff\quad  A=0 \ \text{or}\ A=\id\  \text{or}\ 1>a+c=b+d>0.
\end{equation*}
    Define a map $\tau: E\rightarrow [0,1]$ by $\tau(\begin{psmallmatrix}a&b\\c&d\end{psmallmatrix}) = a+c=b+d$. Then it is straightforward to check that $\tau$ is monotone ($A\leq B \implies \tau(A)\leq \tau(B)$), multiplicative ($\tau(A\cdot B) = \tau(A)\tau(B)$), and $A=0$ iff~$\tau(A)=0$. As a result $A\cdot B = 0$ iff $A=0$ or $B=0$. Hence~$E$ satisfies $A\cdot B = 0$ iff $B\cdot A = 0$, so that the regular composition of matrices in $E$ indeed satisfies all the assumptions of a sequential product. We remark that $E$ is the set of effects of a 3-dimensional vector space with a separating (but not order-separating) set of states, and that the product $\circ$ is non-commutative and associative.
\end{example}

\section{Basic results}\label{sec:basicresultsseqprod}

Unless otherwise stated, we will let $V$ denote a finite-dimensional sequential effect space, $E=[0,1]_V$ its set of effects and $\&: E\times E\rightarrow E$ a sequential product. For $a\in E$ we let $a^\perp=1-a$ denote its \Define{complement}\indexd{complement} which by virtue of $a$ lying in the unit interval of $V$ is also an effect.

\begin{proposition}[{cf.~\cite{gudder2002sequential}}]\label{prop:basic}
    Let $a,b,c \in E$.
    \begin{multicols}{2}
    \begin{enumerate}[label=\alph*), ref=\fullcounter.\alph*)]
        \item $a\mult 1=1\mult a = a$.
        \item \label{prop:unitzero} $a\mult 0 = 0\mult a = 0$.
        \item \label{prop:decreasing} $a\mult b \leq a$.
        \item \label{prop:orderpreserve} If $a\leq b$, then $c\mult a\leq c\mult b$.
        \item \label{prop:compmult} If $a\commu b$, $a\commu c$ and $b\commu c$, then $a\commu (b\mult c)$.
    \end{enumerate}
    \end{multicols}
\end{proposition}
\begin{proof}~
    \begin{enumerate}[label=\alph*)]
        \item We of course have $a\commu a$ and by \ref{ax:compadd} we have $a\commu a^\perp$. Using \ref{ax:compadd} again we then see that $a\commu (a+a^\perp)$. As $a+a^\perp =1$, then $a\commu 1$ so that by \ref{ax:unit} $1\mult a = a\mult 1 = a$. 
        \item By the previous point $a\commu 1$ and hence also $a\commu 1^\perp = 0$ so that it remains to show that $a\mult 0 = 0$. This follows by \ref{ax:add} as $a\mult 0 = a\mult(0+0) = a\mult 0 + a\mult 0$.
        \item By the previous point and \ref{ax:add} $a = a\mult 1 = a\mult (b+b^\perp) = a\mult b + a\mult b^\perp$ so that $a-a\mult b = a\mult b^\perp\geq 0$ and hence indeed $a\mult b \leq a$.
        \item We have $b-a \geq 0$ so by \ref{ax:add} we have $c\mult b = c\mult (b-a + a) = c\mult (b-a) + c\mult a$. Hence $c\mult(b-a) = c\mult b - c\mult a$. Since the left-hand side is greater than zero, the right-hand side must be as well.
        \item Using axiom \ref{ax:assoc} repeatedly:
    $a\mult (b\mult c) = (a\mult b)\mult c = (b\mult a)\mult c = b\mult (a\mult c) = b\mult (c\mult a) = (b\mult c)\mult a$.\qedhere
    \end{enumerate}
\end{proof}

\begin{proposition} \label{prop:linearity}
    Let $a,b\in E$ and let $q$ be any rational number between zero and one, and $\lambda$ any real number between zero and one.
    \begin{multicols}{2}
    \begin{enumerate}[label=\alph*),ref=\fullcounter.\alph*)]
        \item $a\mult (qb) = q(a\mult b)$.
        \item $a\mult (\lambda b) = \lambda(a\mult b)$.
        \item $(\lambda a)\mult b = a\mult (\lambda b) = \lambda(a\mult b)$.
        \item \label{prop:commumult} If $a\commu b$, then $a\commu \lambda b$.
    \end{enumerate}
    \end{multicols}
\end{proposition}
\begin{proof}~
  \begin{enumerate}[label=\alph*)]
      \item Of course $a\mult b = a\mult (n \frac1n b) = n (a\mult (\frac1n b))$ by \ref{ax:add}. Dividing by $n$ gives $a\mult(\frac1n b) = \frac1n (a\mult b)$. By summing this equation multiple times we see that we get $a\mult (q b) = q(a\mult b)$ for any rational $0\leq q\leq 1$.

      \item Let $q_i$ be an increasing sequence of positive rational numbers that converges to $\lambda$. Using the order-unit norm of $V$ we compute
      \begin{align*}
      \norm{\lambda (a\mult b) - a\mult (\lambda b)} &= \norm{(\lambda - q_i)(a\mult b) + q_i(a\mult b) - a\mult (\lambda b)} \\
      &= \norm{(\lambda-q_i)(a\mult b) - a\mult ((\lambda - q_i)b)}.
      \end{align*}
      Note that $(\lambda - q_i)b\leq (\lambda-q_i)\norm{b}1$ and hence, using Proposition \ref{prop:orderpreserve}, we have $\norm{a\mult ((\lambda - q_i)b)} \leq \norm{a}\norm{(\lambda -q_i)b} = (\lambda - q_i)\norm{a}\norm{b}$. 
      But then:
      $$\norm{\lambda (a\mult b) - a\mult (\lambda b)} \leq 2(\lambda - q_i)\norm{a}\norm{b}.$$
      This expression indeed vanishes as $i$ increases so that $\lambda (a\mult b) = a\mult (\lambda b)$.

      \item Clearly $\frac{1}{n}a\commu \frac{1}{n}a$ so that by \ref{ax:compadd} $\frac{1}{n}a\commu a$. 
      In the same way we also get $qa\commu a$ and $qa^\perp \commu a^\perp$ for any rational $0\leq q\leq 1$. 
      Using the rule $a\commu b \implies a\commu b^\perp$ from \ref{ax:compadd} 
      we then also get $qa^\perp \commu a$ so that $a\commu (qa+qa^\perp)=q1$, 
      and hence also $b\commu qb$. 
      Then $(q1)\mult b = b\mult (q1) = q(b\mult 1) = qb$ so that also $(qa)\mult b = (a\mult (q1))\mult b = a\mult((q1)\mult b)) = a\mult qb = q(a\mult b)$. 
      Now let $\lambda\in[0,1]$ be a real number and let $q_i$ be a sequence of rational numbers converging to $\lambda$ so that also $q_i a \rightarrow \lambda a$ and $q_i(a\mult b) \rightarrow \lambda(a\mult b)$. Then $q_i(a\mult b) = (q_ia)\mult b \rightarrow (\lambda a)\mult b$ by \ref{ax:cont}. We conclude that $(\lambda a)\mult b = \lambda(a\mult b) = a\mult(\lambda b)$.

      \item Using the previous point we calculate: $a\mult (\lambda b) = \lambda (a\mult b) = \lambda (b\mult a) = (\lambda b)\mult a$. \qedhere
  \end{enumerate}
\end{proof}

As a result of this proposition, the \Define{left-product map}\index{math}{La@$L_a$ (sequential product map)} $L_a:E\rightarrow E$ for $a\in E$ given by $L_a(b) = a\mult b$ can be extended by linearity to the entirety of $V$ by $L_a(\lambda b-\mu c) = \lambda L_a(b) - \mu L_a(c)$. Similarly we can define the sequential product for any positive $a\in V$ by rescaling: $a\mult b := \norm{a} ((\frac{1}{\norm{a}} a)\mult b)$.
Note that all $L_a:V\rightarrow V$ are positive maps and that by \ref{ax:assoc} we have $a\commu b \iff L_aL_b = L_bL_a$.

\begin{definition}\label{def:sharp-effect}
    An effect $p\in E$ is called \Define{sharp}\indexd{sharp effect} when the only effect below both $p$ and $p^\perp$ is the zero effect, i.e\ when $b\leq p$ and $b\leq p^\perp$ implies $b=0$.
\end{definition}

When $V=B(H)_\sa$ the sharp effects are precisely the projections. This should be clear considering the following proposition.

\begin{proposition}[{\cite{gudder2002sequential}}]\label{prop:sharpness}
    Let $a\in E$ be an effect, $a$ is sharp if and only if $a\mult a^\perp = 0$ if and only if $a\mult a = a$.
\end{proposition}
\begin{proof} 
  Note that $a = a\mult 1 = a\mult (a+a^\perp) = a\mult a + a\mult a^\perp$ and hence $a\mult a^\perp =0$ iff $a\mult a = a$.

  Let us assume $a$ is sharp. By \ref{ax:compadd} we have $a\commu a^\perp$ so that $a\mult a^\perp = a^\perp \mult a$. By \ref{prop:decreasing} we have $a\mult a^\perp \leq a$ and $a\mult a^\perp = a^\perp \mult a \leq a^\perp$. As $a\mult a^\perp$ is then below both $a$ and $a^\perp$ we must have $a\mult a^\perp = 0$ by assumption of sharpness. 
  Conversely, suppose $a\mult a^\perp = 0$ and let $b\leq a$ and $b\leq a^\perp$. Then by \ref{prop:decreasing} we get $a\mult b \leq a\mult a^\perp =0$, and similarly we get $a^\perp \mult b = 0$. Then using \ref{ax:orth} we conclude that $b = b\mult 1 = b\mult (a+a^\perp) = b\mult a + b\mult a^\perp = 0 + 0 = 0$.
\end{proof}

Let us now introduce the notion of orthogonal effects which was hinted at in \ref{ax:orth}:
\begin{definition}
    We call two effects $a$ and $b$ \Define{orthogonal}\indexd{orthogonal effect!in sequential effect space} when $a\mult b = 0$.
\end{definition}
Of course by \ref{ax:orth} orthogonality is a symmetric relation, and we note that therefore orthogonal effects are also compatible.

\begin{definition}
    Let $a\in E$ be an effect. We define the powers of $a$ inductively to be $a^0 := 1$ and $a^n := a\mult a^{n-1}$. We define the \Define{classical algebra of $a$}\indexd{classical algebra} to be the linear space $C(a)$\index{math}{C(a)@$C(a)$ (classical algebra)} spanned by all the powers of $a$ and $a^\perp$.
\end{definition}

\begin{proposition}\label{prop:polyspace}
    Let $a\in E$ be an effect. Then $C(a)$ is a commutative sequential effect space.
\end{proposition}
\begin{proof}
    $C(a)$ inherits the order structure from $V$ in the obvious way, and as $1\in C(a)$ every state of $V$ also restricts to a state on $C(a)$ so that $C(a)$ is an order unit space. 
    Of course $a\commu a$ and $a\commu a^\perp$ and thus by Proposition~\ref{prop:compmult} we have $a^n\commu a^m$ and $a^n \commu (a^\perp)^m$ for all $n$ and $m$. Because of \ref{ax:compadd} and Proposition \ref{prop:commumult} linear combinations of compatible effects are also compatible and hence all effects of $C(a)$ are compatible.
\end{proof}

The next result uses Kadison's representation theorem for order unit spaces (Theorem~\ref{thm:kadison}).

\begin{proposition}
    Let $a\in E$ be an effect. Then there is an $n\in \N$ such that $C(a)$ is both order-isomorphic and algebra-isomorphic (interpreting $\&$ as the product) to $\R^n$.
\end{proposition}
\begin{proof}
    The sequential product is linear in the second argument. Since $C(a)$ is a commutative sequential effect space by Proposition \ref{prop:polyspace}, its product is also linear in the first argument, and hence this operation is bilinear. It obviously preserves positivity, and since $C(a)\sse V$ is finite-dimensional, it is complete, so that Kadison's theorem applies and $C(a) \cong C(X)$ for some compact Hausdorff space $X$. As $C(X)$ then also has to be finite-dimensional, we see that $X$ is finite. The only finite Hausdorff spaces are discrete and hence we conclude that $C(X) \cong \R^n$ for some $n\in \N$.
\end{proof}

\begin{corollary}
    Let $a\in E$ be an effect. There exists a set of orthogonal non-zero sharp effects $p_i$ compatible with $a$ and positive $\lambda_i\in \R$ such that $a = \sum_i \lambda_i p_i$.
\end{corollary}
\begin{proof}
    By the previous proposition $C(a) \cong \R^n$ and this space is obviously spanned by orthogonal sharp effects, hence we can find the desired $p_i$ and $\lambda_i$. By construction $p_i\in C(a)$ so that they are compatible with $a$.
\end{proof}
\noindent We will refer to a decomposition of $a$ in the above sense as a \Define{spectral decomposition}\indexd{spectral decomposition} of $a$. 

We can now show why the lack of associativity of the sequential product is necessary for non-commutative, and hence non-classical, sequential products.

\begin{proposition}\label{prop:assoc-is-commutative}\indexd{associativity in sequential product}
	Let $V$ be a finite-dimensional sequential effect space where the sequential product is associative. Then the sequential product is commutative.
\end{proposition}
\begin{proof}
	Assume $\&$ is associative. Let $a$ be any effect and let $p$ be sharp. Of course $p^\perp\mult a\leq p^\perp$ is orthogonal to $p$ and hence $0 = (p^\perp\mult a)\mult p = p^\perp\mult (a\mult p)$. But then $p^\perp \commu a\mult p$, and hence $p\commu a\mult p$. Similarly, we also get $p\commu a\mult p^\perp$. As a result $p\commu (a\mult p^\perp + a\mult p) = a$. As $a$ was arbitrary we see that sharp elements are compatible with every effect and as any effect can be written as a linear combination of sharp effects, this shows that the sequential product is commutative.
\end{proof}

The existence of spectral decompositions is also enough to show that the space must be homogeneous (Definition~\ref{def:order-iso-homogeneous}), \ie~that for every pair of internal positive elements $a,b\in V$ there exists an order isomorphism $\Phi$ such that $\Phi(a)=b$. Note that $a\in V$ is an internal positive element iff there is an $\epsilon\in\R_{>0}$ such that  $\epsilon 1 \leq a$. Given a spectral decomposition $a=\sum_i\lambda_i p_i$ of such an element, it is easy to see that necessarily $\sum_i p_i = 1$.

\begin{definition}
  Let $a$ be an internal positive element so that $\epsilon 1\leq a$ for some $\epsilon \in \R_{>0}$, and let $a=\sum_i\lambda_i p_i$ with all $\lambda_i > 0$ and $p_i\neq 0$ be a spectral decomposition. We define the \Define{inverse} of $a$ with respect to this decomposition as $a^{-1} := \sum_i \lambda_i^{-1} p_i$. 
\end{definition}

Note that the name of inverse is chosen well as indeed $a\mult a^{-1} = \sum_{i,j} \lambda_i \lambda_j^{-1} p_i \mult p_j = \sum_i \lambda_i \lambda_i^{-1} p_i = \sum_i p_i = 1$.

\begin{proposition}\label{prop:homogen}
    Let $V$ be a finite-dimensional sequential effect space. Then $V$ is homogeneous (cf.~Definition~\ref{def:order-iso-homogeneous}).
\end{proposition}
\begin{proof}
    Let $a$ be an arbitrary internal positive element with spectral decomposition $a=\sum_i \lambda_i p_i$ and inverse $a^{-1}=\sum_i \lambda_i^{-1} p_i$. The sequential product map $L_a(b) := a\mult b$ is positive and since $a^{-1}\commu a$ it also has a positive inverse $L_{a^{-1}}$ (using~\ref{ax:assoc}): $a^{-1}\mult (a\mult b) = (a^{-1}\mult a)\mult b = 1\mult b = b$. The map $L_a$ is therefore an order isomorphism when $a$ lies in the interior of the positive cone. Now, for $a$ and $b$ in the interior we define $\Phi: V\rightarrow V$ by $\Phi = L_bL_{a^{-1}}$. As this is a composition of order isomorphisms, it is also an order isomorphism and of course $\Phi(a) = b\mult (a^{-1}\mult a) = b\mult 1 = b$ as desired.
\end{proof}

\section{Proof of self-duality}\label{sec:selfdual}
With homogeneity of $V$ now established, we set our sights on proving self-duality (cf.~Definition~\ref{def:order-iso-homogeneous}). We do this in a few steps. First we study the lattice of sharp effects in Section~\ref{sec:subsharpeffects}. We then consider properties of the \emph{atoms} of this lattice in Section~\ref{sec:atomeffect}. Then in Section~\ref{sec:coverprop} we establish that this lattice has the \emph{covering property} as defined in Ref.~\cite{alfsen2012state}. The covering property has as a consequence that for every sharp effect $p$ there is a unique number $r$ called the \emph{rank} of $p$ such that we can write $p=\sum_{i=1}^r p_i$ where the $p_i$ are atomic and orthogonal. We then define the rank of a space as the rank of the unit effect. The existence of well-defined ranks of sharp effects allows us to reduce the question of self-duality to that of self-duality in spaces of rank 2. This problem is in turn solved by appealing to the classification result of Ref.~\cite{ito2017p} that homogeneous spaces of rank 2 are always self-dual, which is done in Section~\ref{sec:subselfdual}.

\subsection{The lattice of sharp effects}\label{sec:subsharpeffects}

First, we establish some results regarding sharp effects.

\begin{proposition}\label{prop:sharpprop} Let $a\in E$ be any effect and let $p\in E$ be sharp.
    \begin{enumerate}[label=\alph*), ref=\fullcounter.\alph*)]
        \item \label{prop:belowsharp} $a\leq p$ if and only if $p\mult a = a\mult p = a$ if and only if $p^\perp \mult a = 0$.
        \item \label{prop:abovesharp} $p\leq a$ if and only if $p\mult a = a\mult p = p$.
    \end{enumerate}
\end{proposition}
\begin{proof}~
    \begin{enumerate}[label=\alph*)]
        \item Suppose $a\leq p$ with $p$ sharp. Then $p^\perp \mult a \leq p^\perp \mult p = 0$ by Proposition \ref{prop:sharpness} and \ref{prop:decreasing}. Hence $a\commu p^\perp$ and $a\commu p$ so that $a = a\mult (p+p^\perp) = a\mult p + a\mult p^\perp = a\mult p = p\mult a$. For the other direction we note that $a=p\mult a \leq p$ by \ref{prop:decreasing}.

        \item Suppose $p\leq a$ with $p$ sharp, then $a^\perp \leq p^\perp$ with $p^\perp$ sharp so that by the previous point $a\commu p$ and $p\mult a^\perp = 0$ so that $p=p\mult(a+a^\perp) = p\mult a$. \qedhere
    \end{enumerate}
\end{proof}

\begin{definition}\index{math}{aceil@$\ceil{a}$ (ceiling of element)}
\index{math}{afloor@$\floor{a}$ (floor of element)}

    For an effect $a\in E$ we let $\ceil{a}$ denote the smallest sharp element above $a$, called the \Define{ceiling}\indexd{ceiling!in sequential effect space} of $a$ and we let $\floor{a}$ denote the largest sharp element below $a$, called the \Define{floor}\indexd{floor!in sequential effect space} of $a$.
\end{definition}

A priori, the floor and effect of an effect do not have to exist. In our setting however, they always do.
\begin{proposition}
    The ceiling and the floor exist for any $a$. Moreover, writing $a=\sum_i \lambda_i p_i$ with $1\geq \lambda_i> 0$ and the $p_i$ sharp and orthogonal, then $\ceil{a} = \sum_i p_i$ and $\floor{a}=\ceil{a^\perp}^\perp$.
\end{proposition}
\begin{proof}
    Write $a=\sum_i \lambda_i p_i$. Of course $\sum_i p_i$ is an upper bound of $a$. Suppose $a\leq r$ for some sharp $r$. Then $\lambda_i p_i \leq r$, so by Proposition \ref{prop:belowsharp} $r\mult (\lambda_i p_i) = \lambda_i p_i$. But as $r\mult (\lambda_i p_i) = \lambda_i (r\mult p_i)$ this reduces to $r\mult p_i = p_i$. Hence $r\mult \sum_i p_i = \sum_i r\mult p_i = \sum_i p_i$ so that by Proposition~\ref{prop:belowsharp} $\sum_i p_i\leq r$ so that $\sum_i p_i$ is indeed the least upper bound. The other statement now follows because $a\leq b \iff b^\perp \leq a^\perp$.
\end{proof}

As a corollary of the above we also see that $\ceil{\lambda a} = \ceil{a}$ when $1\geq \lambda > 0$ and that $a$ is sharp if and only if $\ceil{a}=a$ or $\floor{a}=a$. We also note that $a\leq b$ implies that $\ceil{a}\leq \ceil{b}$.

\begin{proposition}\label{prop:lattice}
    The sharp effects form an ortholattice: for two sharp effects $q$ and $p$, their least upper bound $p\vee q$ and greatest lower bound $p\wedge q$ exist and the following relation holds: $(p\vee q)^\perp = p^\perp \wedge q^\perp$.
\end{proposition}
\begin{proof}
    We claim that $p\vee q = \ceil{\frac12(p+q)}$. Note that $p\leq p+q$ and thus that $\frac12 p \leq \frac12(p+q)$ so that $p = \ceil{p} = \ceil{\frac12p} \leq \ceil{\frac12(p+q)}$. Similarly we also have $q\leq \ceil{\frac12(p+q)}$ and thus this is an upper bound. Suppose now that $p\leq a$ and $q\leq a$ for some $a$. Then also $p = \floor{p} \leq \floor{a}$ and $q=\floor{q}\leq\floor{a}$ and hence $\frac12(p+q)\leq \frac12(\floor{a}+\floor{a}) = \floor{a}$. Taking the ceiling on both sides then shows that $\ceil{\frac12(p+q)}\leq \ceil{\floor{a}} = \floor{a} \leq a$ so that indeed $p\vee q = \ceil{\frac12(p+q)}$.

    To find $p\wedge q$ we note that $(\cdot)^\perp$ is an order-anti-isomorphism, and thus that it interchanges joins with meets: $(p\vee q)^\perp = p^\perp\wedge q^\perp$.
\end{proof}

\begin{proposition}[{\cite{gudder2002sequential}}]\label{prop:sharplattice} Let $a\in E$ be any effect and let $p\in E$ be sharp.
    \begin{enumerate}[label=\alph*), ref=\fullcounter.\alph*)]
        \item \label{prop:meetsharp} $p\mult a = 0$ if and only if $p+a\leq 1$ in which case $p+a = p\vee a$. When $p\mult a = 0$ their sum $p+a$ is sharp if and only if $a$ is also sharp.
        \item \label{prop:joinsharp} If both $a$ and $p$ are sharp and $a\commu p$, then $p\mult a$ is sharp and $p\wedge a = p\mult a$.
    \end{enumerate}
\end{proposition}
\begin{proof}~
    \begin{enumerate}[label=\alph*)]
        \item $p\mult a = 0$ if and only if $p^\perp\mult a = a$ which by Proposition \ref{prop:abovesharp} is true if and only if $a\leq p^\perp = 1-p$ so that indeed $p+a\leq 1$. That $p+a$ is an upper bound of $p$ and $a$ is obvious. Suppose now that $b$ is also an upper bound so that $p\leq b$ and $a\leq b$. 
        We then calculate using Proposition \ref{prop:abovesharp} $p = p\mult b = p\mult(b-a + a) = p\mult(b-a) + p\mult a = p\mult(b-a)$ so that again by Proposition~\ref{prop:abovesharp} $p\leq b-a$. Hence $p+a \leq b$ and since $b$ was arbitrary indeed $p+a = p\wedge a$.

        Now to show $p+a$ is sharp if and only if both $p$ and $a$ are sharp: since $p\mult a = 0$ we have $p\commu a$ and thus also $p\commu p+a$ and $a\commu p+a$ by \ref{ax:compadd}. We calculate $(p+a)\mult (p+a) = p\mult p + 2 p\mult a + a\mult a = p + a\mult a = (p+a) + (a-a\mult a)$. We therefore have $(p+a)\mult (p+a) = p+a$ if and only if $a-a\mult a = 0$ which proves the result by Proposition \ref{prop:sharpness}.

        \item As $p\commu a$ we also have $a\commu a\mult p$ and $p\commu a\mult p$ by Proposition~\ref{prop:compmult}. We calculate:
        \begin{align*}
        (p\mult a)\mult (p\mult a) &= (p\mult a)\mult(a\mult p) =p\mult(a\mult(a\mult p)) \\
        &= p\mult (a\mult p) = p\mult (p\mult a) = p\mult a.
        \end{align*}
        Hence $p\mult a$ is sharp. It is a lower bound of $p$ and $a$ by \ref{prop:decreasing}. Suppose $b\leq p, a$ is also a lower bound. We need to show that $b \leq p\mult a$. We calculate $p\mult a = p\mult (a-b + b) = p\mult (a-b) + p\mult b = p\mult(a-b) + b \geq b$, where $p\mult b = b$ due to Proposition \ref{prop:belowsharp}.\qedhere
    \end{enumerate}
\end{proof}

\begin{lemma} \label{lem:ceilzero}
    Let $a,b\in E$ with $b\mult a = 0$. Then $b\mult \ceil{a} = 0$.
\end{lemma}
\begin{proof}
    Write $a= \sum_i \lambda_i p_i$ with $p_i \neq 0$ and $\lambda_i>0$. If $b\mult a = 0 = \sum_i \lambda_i b\mult p_i$, then we must have $b\mult p_i=0$ for all $p_i$. Since $\ceil{a}=\sum_i p_i$ the claim follows.
\end{proof}

\begin{lemma}\label{lem:ceilceil}
    Let $p\in E$ be sharp and $a\in E$ arbitrary. Then $\ceil{p\mult a} = \ceil{p\mult \ceil{a}}$.
\end{lemma}
\begin{proof}
    Of course $p\mult a \leq p\mult \ceil{a}$ and hence $\ceil{p\mult a}\leq \ceil{p\mult \ceil{a}}$ so that it remains to prove the other inequality. Because $p\mult a \leq p$ we also have $\ceil{p\mult a}\leq \ceil{p}=p$ and hence $\ceil{p\mult a}^\perp \commu p$.
    Now because $p\mult a \leq \ceil{p\mult a}$ we can use Proposition \ref{prop:belowsharp} to write $0=\ceil{p\mult a}^\perp \mult (p\mult a) = (\ceil{p\mult a}^\perp\mult p)\mult a = (\ceil{p\mult a}^\perp \mult p)\mult \ceil{a} = \ceil{p\mult a}^\perp\mult (p\mult \ceil{a})$ where we have used Lemma \ref{lem:ceilzero} to replace $a$ with $\ceil{a}$. Since then $\ceil{p\mult a}^\perp\mult (p\mult \ceil{a}) = 0$ we use \ref{prop:belowsharp} again to conclude $p\mult \ceil{a}\leq \ceil{p\mult a}$ so that indeed $\ceil{p\mult \ceil{a}}\leq \ceil{p\mult a}$.
\end{proof}

\subsection{Atomic effects}\label{sec:atomeffect}

\begin{definition}
    An effect $p\in E$ is \Define{atomic}\indexd{atomic effect} when $p$ is a nonzero sharp effect and if for all $a\in E$ with $a\leq p$ we have $a=\lambda p$ for some $\lambda\in[0,1]$.
\end{definition}

The name atomic comes from the fact that in the lattice of sharp effects, the atomic effects are the smallest nonzero elements. We call a lattice itself \Define{atomistic}\indexd{atomistic lattice} when every element can be written as a supremum of (possibly an infinite number of) atoms. This holds for the lattice of sharp elements in our setting:

\begin{proposition}\label{prop:sharpeffectssumofatomic}
    Every sharp effect can be written as a sum of orthogonal atomic effects.
\end{proposition}
\begin{proof}
    Let $p$ be sharp. If $p=0$ or $p$ is atomic we are already done, so suppose this is not the case. Then we can find $0\leq a\leq p$ such that $a\neq \lambda p$ for any $\lambda\in [0,1]$. Write $a= \sum_i \lambda_i q_i$ where the $q_i\neq 0$ are sharp and orthogonal and $\lambda_i>0$. Then $\lambda_i q_i\leq p$ and thus also $\ceil{\lambda_i q_i} = q_i\leq \ceil{p}=p$. If all the $q_i$ are equal to $p$, then $a$ is a multiple of $p$, so at least one of the $q_i$ is strictly smaller than $p$. If $q_i$ and $p-q_i$ are now both atomic we are done. If for instance $q_i$ is not atomic, we can find a $a'\leq q_i$ with $a'\neq \lambda q_i$ for all $\lambda\in[0,1]$ and repeat the argument. In this way we get a sequence of nonzero orthogonal sharp effects that sum up to $p$. As the space is finite-dimensional and orthogonal effects are linearly independent this process must stop after a finite number of steps in which case we are left with atomic effects.
\end{proof}
\begin{corollary}\label{cor:spectralatomic}
    Every $a\in V$ can be written as $a=\sum_i \lambda_i p_i$ where the $p_i$ are orthogonal atomic effects.
\end{corollary}
\begin{proof}
    For every $a\in V$ we can find a spectral decomposition in terms of orthogonal sharp effects. The previous proposition shows that these sharp effects can be further decomposed into atomic effects.
\end{proof}

\noindent Recall that the norm of an element $a$ in an order unit space is the smallest number $r$ such that $-r 1\leq a\leq r 1$.

\begin{lemma} \label{lem:atomicnorm}
    A non-zero effect $p$ is atomic if and only if we have $p\mult a = \norm{p\mult a}p$ for all $a\in E$.
\end{lemma}
\begin{proof}
    First note that any non-zero sharp effect $q$ satisfies $q = q\mult q \leq q \mult(\norm{q} 1) = \norm{q} q\mult 1 = \norm{q} q$ so that $\norm{q}\geq 1$. But since also $q\leq 1$ we must have $\norm{q}\leq 1$. For an arbitrary effect $a$ with spectral decomposition $a=\sum_i \lambda_i q_i$ we then get $\norm{a} = \sup_i \lambda_i$. Hence, if $\norm{a}=1$ we also have $\norm{a^2} = 1$.
    
    Suppose $p$ is atomic. Because $0\leq p\mult a \leq p$ we must have $p\mult a = \lambda p$ for some $0\leq \lambda \leq 1$ so that $\norm{p\mult a} = \lambda \norm{p} = \lambda$ because $p$ is sharp.

    Conversely, we first note that $p= p\mult \ceil{p} = \norm{p\mult \ceil{p}} p = \norm{p}p$ so that necessarily $\norm{p}=1$ (since $p\neq 0$). Then $p^2 = p\mult p = \norm{p^2} p = p$ so that $p$ is sharp. Let $q\leq p$ be non-zero and sharp. Then $\norm{q}=1$ and hence $q=p\mult q = \norm{p\mult q}p=p$. Hence, there are no non-zero sharp effects strictly below $p$. Now let $a\leq p$ be arbitrary with spectral decomposition $a=\sum_i \lambda_i q_i$. Then $\lambda_iq_i\leq p$ so that $\ceil{\lambda_i q_i} = q_i\leq \ceil{p}=p$ and hence $q_i=p$. We conclude that $a=\lambda p$. As $a$ was arbitrary, $p$ is indeed atomic.
\end{proof}
\begin{corollary}
    The set of atomic effects is closed in the norm topology.
\end{corollary}
\begin{proof}
    Let $p_n\rightarrow p$ be a norm-converging set of atomic effects $p_n$. We need to show that $p$ is also atomic. Note first of all that since $\norm{p_n} = 1$ for all $n$ that also $\norm{p} = 1$ and hence $p\neq 0$.
    Furthermore, by the previous lemma we have $p_n\mult a = \norm{p_n \mult a} p_n$ for all effects $a$. By continuity of $\&$ (i.e.\ axiom \ref{ax:cont}) we have $p_n\mult a \rightarrow p\mult a$ so that $p\mult a = \lim p_n \mult a = \lim \norm{p_n\mult a}p_n = \norm{p\mult a} p$. Using the previous lemma again we conclude that $p$ is indeed atomic.
\end{proof}

\begin{proposition}\label{prop:atompreservation}
    Let $a\in E$ be arbitrary and $p\in E$ be atomic. Then $a\mult p$ is proportional to an atomic effect, \ie~$a\mult p = \lambda q$ where $q$ is atomic and $\lambda\in[0,1]$.
\end{proposition}
\begin{proof}
  Let $\Phi$ be an order isomorphism and suppose $0\leq a\leq \Phi(p)$. Then $0\leq \Phi^{-1}(a)\leq p$ so that $\Phi^{-1}(a) = \lambda p$ and hence $a= \lambda \Phi(p)$. This shows that $\Phi(p)$ is proportional to an atomic effect.
  If $a$ is invertible then $L_a: V\rightarrow V$ given by $L_a(b):= a\mult b$ is an order isomorphism (cf.~Proposition~\ref{prop:homogen}), so that $L_a(p)$ must be proportional to an atomic effect. 

  Suppose now that $a$ is not necessarily invertible. If $a\mult p = 0$ we are already done, so assume that $a\mult p \neq 0$.
  Define $a_n = a+\frac{1}{n}1$, so that $a_n$ is invertible and the sequence $a_n$ converges to $a$. Set $q_n = (a_n\mult p)/\norm{a_n \mult p}$. Then all the $q_n$ are atomic. By the continuity condition \ref{ax:cont} we have $\lim_n a_n\mult p = a\mult p$ so that also $\lim_n \norm{a_n \mult p} = \norm{a\mult p}\neq 0$. The sequence $q_n$ is therefore also convergent and since the set of atomic effects is closed by the previous corollary we conclude that $\lim_n q_n = (a\mult p)/\norm{a\mult p}$ is atomic.
\end{proof}

\subsection{The Covering Property}\label{sec:coverprop}

At this point we know that the set of sharp effects forms an atomic lattice, but in fact we can show that it has the much stronger \emph{covering property} that allows us to attach a \emph{rank} to each sharp effect: the number of atomic effects needed to make the effect. 

\begin{definition}
  Let $L$ be an atomistic lattice. For $p,q\in L$ we say $p$ \Define{covers} $q$ when $q\neq p$, $q\leq p$ and for any $r$ with $q\leq r\leq p$ we have $r=q$ or $r=p$ (in other words: $p$ is the smallest element above $q$). We say $L$ has the \Define{covering property}\indexd{covering property} when for any $q\in L$ atomic and $p\in L$ arbitrary, either $q\vee p = p$ or $q\vee p$ covers $p$.
\end{definition}

To prove this property for the lattice of sharp effects in a sequential effect space we will adapt some results from Alfsen and Shultz~\cite{alfsen2012geometry} that were proven in a slightly different setting.

\begin{lemma}[{cf.~\cite[Lemma 8.9]{alfsen2012geometry}}]\label{lem:wedgeminus}
    Let $p,q\in E$ be sharp with $q\leq p$. Then $p-q = p\wedge q^\perp$.
\end{lemma}
\begin{proof}
    We note that $q\commu p$ so that it is easily seen that $p-q$ is sharp. Since also $p\commu q^\perp$ we conclude using Proposition~\ref{prop:joinsharp} that indeed $p-q = p\mult q^\perp = p\wedge q^\perp$.
\end{proof}

\begin{lemma}[{cf.~\cite[Theorem 8.32]{alfsen2012geometry}}]
    Let $p\in E$ be sharp and $a\in E$ arbitrary, then $\ceil{p\mult a} = (\ceil{a}\vee p^\perp)\wedge p$.
\end{lemma}
\begin{proof}
    Because $\ceil{p\mult a} = \ceil{p\mult \ceil{a}}$ by Lemma~\ref{lem:ceilceil} it suffices to prove this for sharp $a$. We prove the equality by showing that an inequality holds in both directions.

    Since $p^\perp \leq a\vee p^\perp$ we have $p^\perp \commu (a\vee p^\perp)$ by Proposition~\ref{prop:belowsharp} so that in turn $p\commu (a\vee p^\perp)$ by \ref{ax:compadd}.
    We proceed by using \ref{ax:assoc}: 
    $$(a\vee p^\perp)\mult (p\mult a) = ((a\vee p^\perp)\mult p)\mult a = p\mult ((a\vee p^\perp)\mult a) = p\mult a,$$
    where in the last step $(a\vee p^\perp)\mult a = a$ because $a\vee p^\perp \geq a$. Therefore $p\mult a \leq a\vee p^\perp$ which implies that $\ceil{p\mult a} \leq a\vee p^\perp$. Since also $p\mult a\leq p$ and therefore $\ceil{p\mult a}\leq p$ we conclude that $\ceil{p\mult a}\leq (a\vee p^\perp)\wedge p$ as desired.

    Now for the converse direction: we obviously have $p^\perp\mult (p\mult a) = (p^\perp \mult p)\mult a = 0$ by \ref{ax:compadd} and \ref{ax:assoc} so that by Lemma~\ref{lem:ceilzero} $p^\perp\mult \ceil{p\mult a}=0$. Then $p\commu \ceil{p\mult a}^\perp$ and by Proposition~\ref{prop:joinsharp} we have $\ceil{p\mult a}^\perp\mult p = \ceil{p\mult a}^\perp\wedge p$. 
    Since $p\mult a \leq \ceil{p\mult a}$ we calculate using Proposition~\ref{prop:belowsharp}: 
    $$0=\ceil{p\mult a}^\perp\mult (p\mult a) = (\ceil{p\mult a}^\perp \mult p)\mult a = (\ceil{p\mult a}^\perp \wedge p)\mult a$$
    so that $a\leq (\ceil{p\mult a}^\perp\wedge p)^\perp = \ceil{p\mult a}\vee p^\perp$
    by Proposition~\ref{prop:lattice}. 
    Then of course also $a\vee p^\perp \leq \ceil{p\mult a}\vee p^\perp$ and by noting that $\ceil{p\mult a}$ and $p^\perp$ are orthogonal and using Proposition~\ref{prop:meetsharp}: $\ceil{p\mult a}\vee p^\perp = \ceil{p\mult a}+p^\perp$. Bringing the $p^\perp$ to the other side we then have $(a\vee p^\perp) - p^\perp \leq \ceil{p\mult a}$. Finally, we have $a\vee p^\perp - p^\perp = (a\vee p^\perp)\wedge p$ because of Lemma~\ref{lem:wedgeminus} (which applies because $p^\perp\leq a\vee p^\perp$). Hence indeed $(a\vee p^\perp)\wedge p \leq \ceil{p\mult a}$ as desired.
\end{proof}

\begin{proposition}[{cf.~\cite[Proposition 9.7]{alfsen2012geometry}}]\label{prop:coverprop}
  For $q$ atomic and $p$ sharp, the expression $(q\vee p)\wedge p^\perp = (q\vee p)-p$ is either zero or atomic. Consequently, the lattice of sharp effects has the covering property.
\end{proposition}
\begin{proof}
  Let us first demonstrate how $(q\vee p)\wedge p^\perp = (q\vee p)-p$ being zero or atomic implies the covering property. Suppose $p\leq r \leq q\vee p$. Subtracting $p$ gives $0\leq r-p \leq (q\vee p) - p$. Hence, as $r-p$ is sharp and $(q\vee p) -p$ is atomic we must have $r-p = 0$ or $r-p = (q\vee p) - p$ so that indeed $r=p$ or $r=q\vee p$.

  The previous lemma gives $(q\vee p)\wedge p^\perp = \ceil{p^\perp \mult q}$, while Proposition~\ref{prop:atompreservation} shows that $p^\perp\mult q$ is proportional to an atom. Hence $p^\perp\mult q = 0$, in which case $(q\vee p)\wedge p^\perp = 0$, or $p^\perp\mult q \neq 0$ in which case $\ceil{p^\perp \mult q} = (q\vee p)\wedge p^\perp$ is an atom. The equality $(q\vee p)\wedge p^\perp = (q\vee p)-p$ follows directly from Lemma~\ref{lem:wedgeminus}.
\end{proof}

\begin{definition}
    Let $p$ be sharp and let $p_i$ be a collection of atomic orthogonal effects such that $p=\sum_i^n p_i$. The minimal size of such a collection is called the \Define{rank}\indexd{rank!of effect} of $p$. We define the rank of a sequential effect space to be the rank of the unit effect $1$.
\end{definition}

The covering property has as a consequence the following `dimension theorem':

\begin{proposition}[{cf.~\cite[Proposition 1.66]{alfsen2012state}}]\label{prop:rankofsharpeffects}
    Write $p=\sum_i^n p_i$ where the $p_i$ are orthogonal and atomic. Then $n=\rnk~ p$, i.e.\ all ways of writing $p$ as a sum of atomic effects require an equal number of atomic effects. Furthermore, when $q\leq p$ we have $\rnk~ q \leq \rnk~ p$ and if also $\rnk~ q = \rnk~ p$ then necessarily $q=p$.
\end{proposition}
\begin{proof}
  Note that since all the $p_i$ are orthogonal that we have $p_i\vee p_j = p_i + p_j$ when $i\neq j$ by Proposition~\ref{prop:sharplattice}.
    Let $p^\prime = p_1\vee\ldots\vee p_{n-1}$. Then $p^\prime\vee p_n = p$ and by the covering property there is no sharp effect strictly between $p^\prime$ and $p$. Suppose now $q\leq p$ is atomic and suppose that $q$ is not below $p^\prime$. Then $p^\prime \vee q$ must be strictly greater than $p^\prime$, but since this must also lie below $p$ we conclude that $p^\prime \vee q = p$.

    Let $p = \sum_j^r q_j = q_1\vee\ldots\vee q_r$ where $r:=\rnk ~ p$ is the minimal number of terms needed to write $p$ as a sum of atomic effects. We must then of course have $r\leq n$. Let $q=q_2\vee\ldots\vee q_r$, so that $q$ lies strictly below $p$ (as $q_1\leq p$ but not $q_1\leq q$). It then follows that there must be a $p_i$ such that $p_i$ does not lie below $q$ as well, since otherwise $p = p_1\vee\ldots\vee p_n \leq q < p$. Without loss of generality let this $p_i$ be $p_1$. By the previous paragraph we must have $p_1\vee q = p_1\vee q_2\ldots\vee q_r = p$. This shows that $q_1$ can be replaced with $p_1$ in this decomposition of $p$. We can do the same with $q_2,\ldots, q_r$ until we are left with the equation $p_1\vee \ldots\vee p_r = p$. Suppose $n>r$, then because $p_n$ is orthogonal to all the other $p_i$'s we have in particular $p_n\leq p_1^\perp \wedge \ldots \wedge p_r^\perp = (p_1\vee \ldots \vee p_r)^\perp = p^\perp$. Since also $p_n\leq p$ we have $p_n\leq p\wedge p^\perp = 0$ which contradicts $p_n\neq 0$. We therefore have $n=r$.

    Now suppose $q=\sum_j^s q_j\leq p=\sum_i^r p_i$ where $s=\rnk~ q$. Since $p-q$ is sharp we can write $p-q = \sum_k^t v_k$ for some atoms $v_k$. Then because $p = \sum_j^s q_j + \sum_k^t v_k$ we must by the above argument have $s+t = r$ so that indeed $\rnk~ q\leq \rnk~ p$. When $\rnk~ q = \rnk~ p$ we must have $t=0$ so that indeed $p-q = 0$.
\end{proof}
\begin{corollary}\label{cor:rank}
    Let $p\neq q$ be two atomic sharp effects and suppose $0\leq a\leq p\vee q$. Then $a=\lambda_1 r_1 + \lambda_2 r_2$ where the $r_i$ are orthogonal and atomic and $r_1 + r_2 = p\vee q$.
\end{corollary}
\begin{proof}
    By Proposition \ref{prop:coverprop} $(p\vee q) - p$ is atomic so that $p\vee q$ can be written as the sum of two atomic sharp effects. The previous proposition consequently gives $\rnk~ p\vee q = 2$. Suppose $0\leq a \leq p\vee q$. Let $a=\sum_i^n \lambda_i r_i$ be a spectral decomposition of $a$ with the $r_i$ orthogonal and atomic. Of course $\ceil{a} \leq p\vee q$ so that by the previous proposition we must have $\rnk \ceil{a} \leq 2$. Since also by the previous proposition $\rnk~\sum_i^n r_i = n$ we see that we must have $n=2$ and thus that $a$ is as desired.
\end{proof}

\subsection{Self-duality}\label{sec:subselfdual}

In this section we will apply the characterisation theorem of strictly convex homogeneous cones of Proposition~\ref{prop:itochar} to show sequential effect spaces must be self-dual.

\begin{definition}\label{def:orderideal}
    Let $p\neq q$ be a pair of atomic effects. We define the \Define{order ideal}\indexd{order ideal} generated by $p$ and $q$ as $V_{p\vee q}:= \{v\in V~;~\exists n: -n~p\vee q \leq v \leq n ~p\vee q\}$.
\end{definition}
$V_{p\vee q}$ is an order unit space with order unit $p\vee q$. If we have $a,b\in [0,1]_{V_{p\vee q}}$ then $a\mult b \leq a \leq p\vee q$ so that the sequential product of $V$ restricts to $V_{p\vee q}$. Hence, $V_{p\vee q}$ is also a sequential effect space so that by Proposition~\ref{prop:homogen} we see that this space has a homogeneous positive cone, while the results of the previous section show that the sharp effects in $V_{p\vee q}$ have the covering property. Furthermore, $V_{p\vee q}$ has rank 2 and since for any atom $r\in V_{p\vee q}$ we have $r+r^\perp = 1$ we see that $r^\perp$ must also be an atom.

\begin{lemma}\label{lem:order-ideal-strictly-convex}
    Let $p\neq q$ be a pair of atomic effects. The positive cone of $V_{p\vee q}$ is strictly convex (cf.~Definition~\ref{def:strictly-convex}).
\end{lemma}
\begin{proof}
    Let $F$ be a proper face of the positive cone of $V_{p\vee q}$. We must show that it is an extreme ray, or equivalently, that it contains a unique atom. Let $a\in F$ and let $\lambda\in \R_{>0}$. Then we can write $a=\lambda (\lambda^{-1} a) + \lambda^\perp 0$, so that also $\lambda^{-1} a \in F$ and hence $F$ is closed under positive scalar multiplication, so that $F$ is completely determined by the effects it contains. As $F$ is already closed under convex combinations we see that it is now also closed under sums. Let $a\in F$ be an effect. By Corollary~\ref{cor:rank} we can write $a=\lambda r + \mu r^\perp$ for some $\lambda,\mu\geq 0$ and $r$ atomic. Suppose both $\lambda,\mu >0$. Then we must have $r,r^\perp \in F$ so that $1 = r+r^\perp\in F$. For any atomic $s$ we have $1=s+s^\perp$ so that then also $s,s^\perp\in F$. As $s$ is arbitrary, $F$ must then be the entire positive cone, contradicting the assumption that $F$ is proper. Hence, we must have had $a=\lambda r$ for some atomic $r$. If there were some other atomic $s \in F$, then we can consider $a' = \frac{1}{2}(r + s)$. We know that $a'$ can't be atomic so we can write it as $a=\lambda t + \mu t^\perp$ for some atomic $t$ with $\lambda, \mu > 0$ which by our previous argument would contradict the properness of $F$. We conclude that $F$ indeed contains a unique atom so that it is an extreme ray.
\end{proof}

\begin{corollary}
    Let $p\neq q$ be a pair of atomic effects. Then $V_{p\vee q}$ is isomorphic to a spin factor (cf.~Definition~\ref{def:spin-factor}).
\end{corollary}
\begin{proof}
    Combine the previous lemma with Proposition~\ref{prop:itochar}.
\end{proof}

Recall that a state on an order unit space is a positive linear map $\omega: V\rightarrow \R$ such that $\omega(1) = 1$. For an atomic effect $p$ in a spin factor (or any Euclidean Jordan algebra) there exists a unique state $\omega_p$ such that $\omega_p(p)=1$. Indeed, any state $\omega$ on an Euclidean Jordan algebra (EJA) is determined by some effect $\rho$ via the EJAs inner product: $\omega(a) = \inn{a,\rho}$. The only state which has $\omega(p) =1$ must then have $\rho = p$. EJAs furthermore satisfy \Define{symmetry of transition probabilities}\indexd{symmetry of transition probabilities} \cite{alfsen2012geometry}: $\omega_p(q)=\omega_q(p)$ for any two atomic effects $p$ and $q$. This again easily follows by the correspondence of states and effects via the inner product: $\omega_p(q) = \inn{q,p} = \inn{p,q} = \omega_q(p)$.

\begin{proposition}\label{prop:uniquestate}
     For any atomic $p\in E$ there is a unique state $\omega_p$ satisfying $\omega_p(p)=1$. For any pair of atomic effects $p,q\in E$ these states satisfy $\omega_p(q)=\omega_q(p)$.
\end{proposition}
\begin{proof}
    The states separate the effects in an order unit space (Proposition~\ref{prop:OUSequivalentdefinitions}) so that for $p$ we can find a state $\omega$ such that $\omega(p)\neq 0$. Let $\omega_p(a) := \omega(p\mult a)/(\omega(p))$. Then $\omega_p$ is a state and $\omega_p(p)=1$. Suppose there is another state $\omega'$ such that $\omega'(p)=1$. Let $q\neq p$ be any other atomic effect (if there is no atomic $q\neq p$ then $V\cong \R$ and we are already done) and look at the restrictions of the states $\omega_p$ and $\omega'$ to the space $V_{p\vee q}$. These restriction maps are still states as $\omega_p(p\vee q)\geq \omega_p(p)=1$ (and similarly for $\omega'$). Because states with the property $\omega(p) = 1$ are unique on spin factors we see that the states $\omega'$ and $\omega_p$ are equal on $V_{p\vee q}$ and hence in particular $\omega_p(q)=\omega'(q)$. Since $q$ was arbitrary and the atomic effects span $V$ we conclude that $\omega_p=\omega'$ so that $\omega_p$ is indeed unique.

    For any two atomic $p$ and $q$ their unique states $\omega_p$ and $\omega_q$ when restricted to the spin factor $V_{p\vee q}$ are still the unique states satisfying $\omega_p(p) = 1$ and $\omega_q(q) = 1$. As spin factor satisfy symmetry of transition probabilities we then see that indeed $\omega_p(q)=\omega_q(p)$.
\end{proof}

\begin{proposition}\label{prop:atomicprod}
    Let $p,q\in E$ be atomic. Then $p\mult q = \omega_p(q) p$. Consequently, $p$ and $q$ are orthogonal if and only if $\omega_p(q) = \omega_q(p) = 0$.
\end{proposition}
\begin{proof}
  Because $p$ is atomic we have $p\mult q = \lambda p$ for some $\lambda \geq 0$. Let $\omega^\prime(a) = \omega_p(p\mult a)$. Then $\omega^\prime(p) = \omega_p(p\mult p) = \omega_p(p) = 1$, so that by the uniqueness of $\omega_p$ we have $\omega^\prime = \omega_p$. We then see that $\omega_p(q) = \omega^\prime(q) = \omega_p(p\mult q) = \omega_p(\lambda p) = \lambda \omega_p(p) = \lambda$.  Hence indeed $p\mult q = \omega_p(q) p$.
\end{proof}

\begin{proposition}\label{prop:SESselfdual}
    There exists an inner product $\inn{\cdot,\cdot}$ on $V$ such that the positive cone is self-dual with respect to this inner product.
\end{proposition}
\begin{proof}
    For atomic $p$ and $q$ we set $\inn{p,q}:= \omega_p(q)=\omega_q(p)=\inn{q,p}$. We can then extend it by linearity to arbitrary $a=\sum_i \lambda_i p_i$ and $b=\sum_j \mu_j q_j$ in $V$ by $\inn{a,b}:= \sum_{i,j} \lambda_i\mu_j \inn{p_i,q_j}$. For this to be well-defined, the inner product must be independent of the choice of spectral decomposition of $a$ and $b$. So suppose $b=\sum_k \mu'_k q'_k$ is a different spectral decomposition. Then 
    $$\sum_{i,j} \lambda_i\mu_j \inn{p_i,q_j} = \sum_{i} \lambda_i \omega_{p_i}(\sum_j \mu_j q_j) = \sum_i \lambda_i\omega_{p_i}(b) = \sum_i\lambda_i \omega_{p_i}(\sum_k \mu'_k q'_k),$$
    as desired. The well-definedness in the first argument follows via commutativity of the expression:
    $$\inn{a,b} = \sum_{i,j} \lambda_i \mu_j\omega_{p_i}(q_j) = \sum_{j,i} \mu_j \lambda_i\omega_{q_j}(p_i) = \inn{b,a}$$
    We see that $\inn{a,a} = \sum_{i,j} \lambda_i \lambda_j \omega_{p_i}(p_j) = \sum_i \lambda_i^2$ since $p_i$ and $p_j$ are orthogonal when $i\neq j$ and $\omega_{p_i}(p_i)=1$. We conclude that $\inn{a,a}\geq 0$ and that it is only equal to zero when $a=0$ so that $\inn{\cdot,\cdot}$ is indeed an inner product.

    If $a$ and $b$ are positive elements then we can write them as $a=\sum_i \lambda_i p_i$ and $b=\sum_j \mu_j q_j$ where all the $\lambda_i$ and $\mu_j$ are greater than zero. But then $\inn{a,b}\geq 0$ because $\omega_{p_i}(q_j)\geq 0$ for all $i$ and $j$. Conversely, if $a=\sum_i \lambda_i p_i$ with $\lambda_i$ not necessarily positive and $\inn{a,b}\geq 0$ for all $b\geq 0$, then by taking $b=p_j$ we see that $0\leq \inn{a,p_j} = \lambda_j$. Hence we must have $\lambda_j\geq 0$ for all $j$ and thus $a\geq 0$.
\end{proof}

\begin{remark}
    Since we have now shown that finite-dimensional sequential effect spaces are both homogeneous (Proposition~\ref{prop:homogen}) and self-dual (Proposition~\ref{prop:SESselfdual}), we could use the Koecher--Vinberg theorem (Theorem~\ref{thm:Koecher--Vinberg})\indexd{Koecher--Vinberg theorem} to show that these spaces are order-isomorphic to Euclidean Jordan algebras. For completeness sake we will explicitly construct a Jordan product from the sequential product in Section~\ref{sec:jordanproduct}.
\end{remark}


\section{The Born rule}\label{sec:bornrule}

Now that we have seen that the sequential effect spaces are self-dual we can recover the familiar Born rule of quantum mechanics.

Recall that the states of a quantum mechanical system represented by the matrix algebra $M_n(\C)$ are the density operators\indexd{density operator} $\rho \in M_n(\C)$ (\ie~positive matrices with $\tr(\rho)=1$). A measurement is represented by a POVM\indexd{POVM} $\{E_i\}$ where the $E_i\in M_n(\C)$ are a set of effects satisfying $\sum_i E_i = 1$.

The probability of observing the outcome $i$ when measuring $\rho$ with the POVM $\{E_i\}$ is then given by the Born rule: $P(\rho\lvert i) = \tr(\rho E_i)$.\indexd{Born rule} A more convenient form of the Born rule is the equality $\tr(\rho E_i) = \tr(\sqrt{\rho}E_i\sqrt{\rho})$. The expression $\sqrt{\rho}E_i\sqrt{\rho}$ corresponds to the standard sequential product $\rho\mult E_i$ on $M_n(\C)_\sa$. Hence, the Born rule can also be presented as $P(\rho\lvert i) = \tr(\rho \mult E_i)$.
We will see that a similar rule can be derived for sequential effect spaces.

\begin{remark}
    The expression $\tr(\sqrt{\rho}E_i\sqrt{\rho})$ might seem a bit foreign. Recall that we are working in the Heisenberg picture, in which our primary concern is the effects instead of the states. This is why we chose to represent this expression in such a way to highlight the linearity in the effects instead of the more well-known expression $\tr(\sqrt{E_i}\rho\sqrt{E_i})$ which highlights the linearity in the states.
\end{remark}


As before, we will let $V$ be a finite-dimensional sequential effect space.

\begin{definition}
    Let $a\in V$. The \Define{trace}\indexd{trace (sequential effect space)} of $a$ is $\tr(a):= \inn{a,1}$. A \Define{density operator}\indexd{density operator!on sequential effect space} on $V$ is an element $a\in V$ with $a\geq 0$ and $\tr(a) = 1$.
\end{definition}

\begin{proposition}\label{prop:statesaredensityoperators}
    Let $\omega:V\rightarrow \R$ be a state. Then there exists a density operator $\rho \in V$ such that $\omega(a) = \inn{\rho,a}$ for all $a\in V$. Conversely, any density operator defines a state on $V$ in this manner.
\end{proposition}
\begin{proof}
    Let $V^* := \{f:V\rightarrow \R \text{ linear}\}$ denote the \Define{dual space}\indexd{dual vector space} of $V$.
    The inner product on $V$ gives a map $\Phi:V\rightarrow V^*$ defined by $\Phi(v)(w) = \inn{v,w}$.
    Note that if $\Phi(v) = \Phi(v')$, then $\inn{v-v',w} = 0$ for all $w\in V$ and in particular for $w=v-v'$ so that $v=v'$.
    Hence, $\Phi$ is injective and because $V$ is finite-dimensional, $\dim V^* = \dim V$ so that $\Phi$ is necessarily also bijective.

    For each state $\omega:V\rightarrow \R$ we can then find a unique $\rho\in V$ such that $\Phi(\rho)=\omega$. By definition of $\Phi$ we have $\omega(a) = \inn{\rho,a}$ for all $a\in V$. 
    As $\omega$ is positive, we have $0\leq \omega(a) = \inn{\rho, a}$ for all $a$ positive and hence by self-duality of $V$ we have $\rho \geq 0$. Furthermore $1=\omega(1) = \inn{\rho, 1} = \tr(\rho)$, and hence $\rho$ is indeed a density operator.

    Conversely it is clear how any $\rho\geq 0$ with $\tr(\rho) = 1$ defines a state $\omega_\rho(a) = \inn{\rho,a}$.
\end{proof}

\begin{lemma}\label{lem:sequentialproductadjointon1}
    Let $L_a:V\rightarrow V$ denote the sequential product map of $a$, and let $L_a^*:V\rightarrow V$ denote its adjoint with respect to the inner product. Then $L_a^*(1) = a$.
\end{lemma}
\begin{proof}
    We note that the classical algebra $C(a)$ of $a$ is isomorphic to $\R^n$ and that the sequential product there is the standard coordinatewise product. The inner product is also the standard inner product on $\R^n$ so that $L_a^*=L_a$ when the maps are restricted to this associative algebra. Since $1 \in C(a)$ we indeed have $L_a^*(1)=L_a(1) = a$.
\end{proof}

We can now prove a Born rule for sequential effect spaces.

\begin{proposition}\label{prop:bornrule}
    Let $\omega:V\rightarrow \R$ be a state. Then there exists a density operator $\rho\in V$ such that $\omega(a) = \tr(\rho\mult a)$ for all $a\in V$. Conversely any density operator defines a state in this way.
\end{proposition}
\begin{proof}
    By Proposition~\ref{prop:statesaredensityoperators} there is a density operator $\rho\in V$ such that $\omega(a) = \inn{\rho,a}$ for all $a$. 
    Using Lemma~\ref{lem:sequentialproductadjointon1} we calculate: $\omega(a) = \inn{\rho, a} = \inn{a,\rho} = \inn{a,L_\rho^*(1)} = \inn{L_\rho(a),1} = \inn{\rho\mult a, 1} = \tr(\rho\mult a)$.
\end{proof}

Given a system $A$ we associated an order unit space $V_A$ to it, and we know that $\st(A)\sse \st(V_A)$. We however do not yet know whether every state of $V_A$ corresponds to an actual state of $A$, \ie~whether the states satisfy the no-restriction hypothesis\indexd{no-restriction hypothesis} (cf.~Remark~\ref{remark:norestrictionhypothesis}). This will require an additional assumption. This assumption has no bearing on the results outside this section, we merely include it to show how one could get all the mathematical states to be physical states.

Given a state $\omega \in \st(A)$ and some effect $a\in \eff(A)$ such that $\omega(a)\neq 0$ there should be some state $\omega_a$ that results from observing $a$ on $\omega$. Any effect $b$ measured on this state has (by definition) the same probabilities as $a\mult b$ measured on $\omega$, and hence the state should satisfy $\omega_a(b) = \omega(a\mult b)$. But in fact, this state is not normalized as $\omega_a(1) = \omega(a\mult 1) = \omega(a)$ does not have to be equal to 1. In order to fix this we define the state $\omega_a$ to be $\omega_a(b) := \frac{1}{\omega(a)}\omega(a\mult b)$. Dividing by the probability $\omega(a)$ captures the fact that we have ``post-selected'' for $a$ being true on $\omega$.

\begin{assumption}\label{assum:statesclosedundersequentialproduct}
    Suppose $\omega \in \st(A)$ and $a \in \eff(A)$ such that $\omega(a)\neq 0$. Then there is a state $\omega_a\in \st(A)$ satisfying $\omega_a(b) = \frac{1}{\omega(a)}\omega(a\mult b)$ for all $b\in \eff(A)$.
\end{assumption}

Note that this assumption is merely saying that the mathematical state $\omega_a$ on $V$ defined by $\omega_a(b) := \frac{1}{\omega(a)}\omega(a\mult b)$ is in fact a physical state in our framework.

\begin{proposition}\label{prop:allstatesarephysical}
    Let $A$ be a physical system satisfying Assumption~\ref{assum:statesclosedundersequentialproduct} (in addition to the Assumptions~\ref{assum:GPT} and~\ref{assum:effectspaceclosedunderaddition} that we made before). Then $\st(A) = \st(V_A)$.
\end{proposition}
\begin{proof}
  Let $\omega\in \st(V_A)$. We need to show that $\omega \in \st(A)$.
 By Proposition~\ref{prop:statesaredensityoperators} there is a density operator $\rho \in V_A$ such that $\omega(a) = \inn{\rho, a}$ for all $a\in V$. 
 Write $\rho = \sum_i \lambda_i p_i$ where the $p_i$ are atomic. 
 Note that $\lambda_i\geq 0$ and $1=\tr(\rho) = \sum_i \lambda_i \tr(p_i) = \sum_i \lambda_i$, so that the $\lambda_i$ form a probability distribution. Hence $\rho$ is a convex combination of the $p_i$ so that $\omega = \sum_i \lambda_i \omega_{p_i}$ is a convex combination of the $\omega_{p_i}$ that are the unique states satisfying $\omega_{p_i}(p_i) = 1$.  By Assumption~\ref{assum:GPT} $\st(A)$ is a convex set, so it remains to show that $\omega_{p_i}\in \st(A)$.

  Let $p\in [0,1]_{V_A}$ be atomic. By Proposition~\ref{prop:effectspaceisfullunitinterval} $[0,1]_{V_A} = \eff(A)$, and since the states separate the effects by Assumption~\ref{assum:GPT}, we can find a state $\omega\in\st(A)$ so that $\omega(p) \neq 0$. Then Assumption~\ref{assum:statesclosedundersequentialproduct} implies there is a state $\omega_p\in \st(A)$ such that $\omega_p(p) = \frac{1}{\omega(p)} \omega(p\mult p) = \frac{\omega(p)}{\omega(p)} = 1$. But such states are unique by Proposition~\ref{prop:uniquestate} and hence we are done.
\end{proof}

Combining Propositions~\ref{prop:effectspaceisfullunitinterval}, \ref{prop:bornrule} and \ref{prop:allstatesarephysical} we see that for a system $A$ we can find a self-dual order unit space $V_A$ such that $\eff(A) = \eff(V_A)$, $\st(A) = \st(V_A)$ and there is a one-to-one correspondence between states $\omega\in\st(A)$ and density operators $\rho_\omega \in V_A$ so that the Born rule holds: $\omega(a) = \tr(\rho_\omega \mult a)$ for all $a\in \eff(a)$.

\section{The Jordan product}\label{sec:jordanproduct}
We wish to show that the order unit spaces we have been working with are actually Euclidean Jordan algebras. To do this we must construct the Jordan product, and show that it behaves well with respect to the inner product.

We will use the construction of the Jordan product from the work of Alfsen and Shultz \cite{alfsen2012geometry}, but then adapted to our setting. In this section we will again let $V$ be a finite-dimensional sequential effect space, which by the previous sections is self-dual and homogeneous.

\begin{proposition}
    Let $p$ be atomic and let $a,b\in V$ be arbitrary.
    \begin{enumerate}[label=\alph*),ref=\fullcounter.\alph*)]
        \item $p\mult a = \omega_p(a) p = \inn{p,a} p$.
        \item \label{prop:atomadjoint} ${\inn{p\mult a, b} = \inn{a,p\mult b}}$.
     \end{enumerate} 
\end{proposition}
\begin{proof}~
    \begin{enumerate}[label=\alph*)]
        \item By Proposition \ref{prop:atomicprod} this is true when $a$ is atomic. Writing $a=\sum_i \lambda_i q_i$ where the $q_i$ are atomic we then get by linearity $p\mult a = \sum_i \lambda_i p\mult q_i = \sum_i \lambda_i \omega_p(q_i) p = \omega_p(\sum_i\lambda_i q_i) p = \omega_p(a) p$. Since $\omega_p(q_i) = \inn{p,q_i}$ the second equality follows in a similar way.
        \item Follows easily from the previous point as: 
        $$\inn{p\mult a, b} = \inn{\inn{p,a}p,b} = \inn{p,a}\inn{p,b} = \inn{a,p}\inn{p,b} = \inn{a,\inn{p,b}p} = \inn{a,p\mult b}.\qedhere$$
    \end{enumerate}
\end{proof}

\begin{lemma}\label{lem:restrict}
    Let $p$ and $q$ be atomic and set $p' = p\vee q - p$. Then $p^\perp \mult q = p' \mult q$.
\end{lemma}
\begin{proof}
    First note that $p^\perp = 1-p = 1-p\vee q + p\vee q -p = (p\vee q)^\perp + p'$ and hence that $p' \leq p^\perp$ so that $p' \commu p^\perp$ by Proposition \ref{prop:belowsharp}. As $p\commu p\vee q$ we also have $p^\perp \commu p\vee q$ so that $p^\perp\mult (p\vee q) = (p\vee q)\mult p^\perp = (p\vee q) \mult ((p\vee q)^\perp + p') = p'$ (where in the last step we used $p' \leq p\vee q$). As also $q\leq p\vee q$ we calculate $p^\perp \mult q = p^\perp \mult ((p\vee q)\mult q) = (p^\perp \mult (p\vee q))\mult q = p' \mult q$.
\end{proof}

Recall that $L_a: V\rightarrow V$ is given by $L_a(b) = a\mult b$.

\begin{definition}
    Let $p\in V$ be an atomic effect and let $b\in V$ be arbitrary. We define their \Define{Jordan product}\indexd{Jordan product!in sequential effect space} as $p*b = \frac{1}{2}(\id + L_p - L_{p^\perp}) b$.
\end{definition}

\begin{lemma}\label{lem:atomiccommute}
    \cite[Lemma 9.29]{alfsen2012geometry}: Let $p$ and $q$ be atomic effects.
    \begin{enumerate}[label=\alph*),ref=\fullcounter.\alph*)]
        \item $p*q = q*p$ and $p*p = p$.
        \item When $p\mult q = 0$ we have $p*q=0$ and $p*(q*b)=q*(p*b)$ for all $b\in V$.
    \end{enumerate}
\end{lemma}
\begin{proof}~
    \begin{enumerate}[label=\alph*)]
        \item $p*p =p$ follows immediately from $p\mult p = p$ and $p^\perp \mult p =0$.
        Now let $p$ and $q$ be atomic. If $p=q$ we of course have $p*q=q*p$, so suppose also that $p\neq q$. Following Lemma~\ref{lem:restrict} define $p' = p\vee q - p$. By Proposition~\ref{prop:coverprop} $p'$ is atomic and hence by Proposition~\ref{prop:atomicprod} we have $p\mult q = \omega_p(q) p = \inn{p,q} p$ and $p' \mult q = \inn{p',q} p'$. Note that hence $p^\perp \mult q = p' \mult q = \inn{p',q}p'$ by Lemma \ref{lem:restrict}. We calculate:
        \begin{align*}
        2 (p*q) &= q + \inn{p,q}p - \inn{p', q} p' \\
        &= q + \inn{p,q}p - \inn{p', q} (p\vee q - p) \\
        &= q + (\inn{p,q}p + \inn{p', q})p - \inn{p^\prime, q} (p\vee q) \\
        &= q + \inn{p\vee q,q}p + \inn{p\vee q - p, q} (p\vee q) \\
        &= q+p + (1-\inn{p,q})(p\vee q).
        \end{align*}
        This final expression is symmetric in $p$ and $q$ and hence $p*q = q*p$.

        \item When $p\mult q = 0$ we have $q\leq p^\perp$ so that $p^\perp \mult q = q$ which indeed gives $p*q = {\frac12(q + p\mult q - p^\perp \mult q)} = \frac12(q-q) = 0$. Furthermore, because $p\mult q = 0$ we have $p\commu q, q^\perp$ and $q\commu p^\perp$, and hence the maps $L_p, L_{p^\perp}, L_q$ and $L_{q^\perp}$ all mutually commute so that the maps $b\mapsto p*b$ and $b\mapsto q*b$ commute as well. \qedhere
    \end{enumerate}
\end{proof}
Using these results we extend the Jordan product to the entire space.
\begin{definition}
    Let $a,b\in V$ be arbitrary. Let $a=\sum_i \lambda_i p_i$ and $b=\sum_j \mu_j q_j$ be spectral decompositions with the $p_i$ and $q_j$ atomic. Define their Jordan product as $a*b = \sum_{i,j} \lambda_i \mu_j (p_i*q_j)$. We write $T_a:V\rightarrow V$ for the map that sends $b$ to $a*b$.
\end{definition}

\begin{proposition}\label{prop:seqprodisjordan}
    Let $V$ be a finite-dimensional sequential effect space. The Jordan product $*$ defined above makes $V$ a Jordan algebra.
\end{proposition}
\begin{proof}
  First we check that $*$ is independent of how $a$ and $b$ are represented as a linear combination of atoms, so that $*$ is indeed well-defined.
  Write $a$ as a spectral decomposition into atomic effects $a=\sum_i \lambda_i p_i$. Then $a*b = \sum_i \lambda_i p_i*b$ so that $a*b$ is independent of how $b$ is represented as a sum of atomic sharp effects. Using the previous lemma we easily see that $a*b = b*a$ and hence it is also independent of how $a$ is represented. Bilinearity then follows from linearity in the second argument.
  
  It now remains to show that the Jordan identity $a*(a^2*b) = a^2*(a*b)$ holds (where $a^2:= a*a$).
    Write $a = \sum_i \lambda_i p_i$ with all the $p_i$ orthogonal atoms so that $p_i*p_j = 0$. Then $a*a = \sum_{i,j} \lambda_i\lambda_j p_i*p_j = \sum_i \lambda_i^2 p_i$. We now calculate:

    $${}\qquad\qquad a*(a^2*b) = \sum_{i,j} \lambda_i\lambda_j^2 p_i*(p_j*b) \stackrel{\ref{lem:atomiccommute}.b)}{=} \sum_{i,j} \lambda_i\lambda_j^2 p_j*(p_i*b) = a^2*(a*b). \qquad\qquad {}\qedhere$$
\end{proof}

At this point we do not yet know whether $V$ is a \emph{Euclidean} Jordan algebra. As we already know that $V$ is a real Hilbert space, it remains to prove the identity $\inn{a*b,c} = \inn{b,a*c}$. This comes down to showing that $T_a = T_a^*$ where $T_a^*$ is the Hilbert space adjoint of $T_a$. By linearity it suffices to prove this for $a$ atomic, for which we have $T_a = \frac12(\id + L_a - L_{a^\perp})$. If we can therefore show that $L_p^* = L_p$ for any sharp $p$, we are done.

We know that $a$ has an inverse if and only if $L_a$ is invertible and in which case $L_a^{-1} = L_{a^{-1}}$. For these maps we have $(L_a^*)^{-1}=L_{a^{-1}}^*$. Note furthermore that $L_a$ and $L_b$ commute if and only if $L_a^*$ and $L_b^*$ commute and that $L_0^* = L_0 = 0$.



\begin{proposition}\label{prop:orderisometry}
    Let $\Phi:V\rightarrow V$ be a unital order-isomorphism. Then $\Phi^*=\Phi^{-1}$.
\end{proposition}
\begin{proof}
  Any order-isomorphism $\Phi$ sends an atom $p$ to a $\Phi(p)$ that is proportional to an atom (cf.~Proposition~\ref{prop:atompreservation}). When it is furthermore unital, it also preserves sharpness of effects, so that $\Phi(p)$ is also atomic and hence $(\omega_p \Phi^{-1})(\Phi(p)) = \omega_p(p)=1$. By Proposition~\ref{prop:uniquestate} states with this property are unique so that necessarily $\omega_{\Phi(p)} = \omega_p \Phi^{-1}$. We conclude that $\inn{\Phi(p),\Phi(q)} = \omega_{\Phi(p)} \Phi(q) = \omega_p \Phi^{-1}\Phi(q) = \omega_p(q) = \inn{p,q}$ for atomic $p$ and $q$ so that $\Phi$ preserves the inner product so that indeed $\Phi^*=\Phi^{-1}$.
\end{proof}

\begin{lemma}\label{lem:orderisoexists}
    Let $a$ be invertible. There exists a unital order-isomorphism $\Phi$ that commutes with $L_a$ and $L_a^*$ and satisfies $L_a^*=L_a \Phi$.
\end{lemma}
\begin{proof}
    Since $a$ is invertible, $L_a$ and $L_a^*$ are also invertible. Define $\Phi := L_a^{-1} L_a^*$. Then $\Phi$ is an order isomorphism and using Lemma~\ref{lem:sequentialproductadjointon1} we have $\Phi(1) = L_a^{-1}L_a^*(1)= L_a^{-1}(a) = 1$ so that it is unital. By definition we have $L_a \Phi = L_a^*$. 

    For the commutativity we note that for all $b$ and $c$: $\inn{L_a^*b,c} = \inn{L_a\Phi b,c} = \inn{\Phi b, L_a^* c}= \inn{\Phi b, L_a \Phi c} = \inn{\Phi^{-1}L_a^*\Phi b, c}$  so that $\Phi^{-1}L_a^*\Phi = L_a^*$ which shows that $\Phi$ commutes with $L_a^*$ and therefore with $L_a$.
\end{proof}

\begin{lemma}\label{lem:uniformcont}
  Denote by $B(V)$ the set of bounded linear maps on $V$ (which by finite-dimensionality of $V$ are all the linear maps). 
  The map $L:E\rightarrow B(V)$ that sends an effect to its sequential product map is continuous (where $B(V)$ is equipped with the operator-norm topology).
\end{lemma} 
\begin{proof}
    For a fixed $b$ the map $a\mapsto a\mult b$ is continuous by \ref{ax:cont}. Furthermore, the map $b\mapsto a\mult b$ is continuous because it is linear and bounded. With a standard argument it can then be shown that $(a,b)\mapsto a\mult b$ is jointly continuous. Because $V$ is finite-dimensional the space of effects $E$ is compact. The map $\&: E\times E\rightarrow V$ is therefore \emph{uniformly} continuous.

    Now for every $\epsilon>0$ we need to find a $\delta>0$ such that $\norm{a_1-a_2}\leq \delta$ implies $\norm{L_{a_1} - L_{a_2}}\leq \epsilon$. Recall that $\norm{L_{a_1}-L_{a_2}} = \sup_{\norm{b}\leq 1} \norm{L_{a_1}(b) - L_{a_2}(b)}$ and hence it suffices to find a $\delta>0$ such that $\norm{L_{a_1}(b) - L_{a_2}(b)}<\epsilon$ for all $b\in E$ whenever $\norm{a_1-a_2}< \delta$, but this is simply the uniform continuity of $\&: E\times E\rightarrow V$.
\end{proof}

\begin{proposition}
    For any $a\in V$ positive, $L_a$ and $L_a^*$ commute.
\end{proposition}
\begin{proof}
    Suppose first that $a$ is invertible. By Lemma \ref{lem:orderisoexists} we then have $L_a = L_a^*\Phi = \Phi L_a^*$. But then $L_aL_a^* = L_a \Phi L_a = \Phi L_a L_a = L_a^* L_a$. Now if $a\in E$ is not invertible then there is a sequence of invertible effects $a_n$ that converge to $a$. By the previous lemma we then also see that $L_{a_n}$ converges to $L_a$ in the operator norm, and similarly $L_{a_n}^*$ converges to $L_a^*$. Therefore $0 = L_{a_n}L_{a_n}^* - L_{a_n}^*L_{a_n}$ converges to $L_aL_a^* - L_a^*L_a$.
\end{proof}

\begin{proposition}
  For any positive $a$ let $U_a = L_aL_a^*$ be its \Define{quadratic product}\indexd{quadratic product!in sequential effect space}. Then
  \begin{enumerate}[label=\alph*)]
    \item $U_a^2 = U_{a^2}$,
    \item $U_p = L_p$ for any sharp $p$, and hence $L_p^* = L_p$, and
    \item $T_a^* = T_a$ for any $a\in V$.
  \end{enumerate}
\end{proposition}
\begin{proof}~
\begin{enumerate}[label=\alph*)]
    \item $U_a^2 = L_aL_a^* L_a L_a^** = L_a L_a L_a^* L_a^* = L_{a^2} L_{a^2}^* = U_{a^2}$.
    \item We have $U_p^2 = U_{p^2} = U_p$. Furthermore $L_pU_p = L_pL_pL_p^* = L_pL_p^* = U_p$ and similarly $U_pL_p^* = U_p$.
    Write $A = U_p - L_p$ and note that $A^* = U_p - L_p^*$. Hence:
    \[AA^* = (U_p - L_p)(U_p - L_p^*) = U_p^2 - U_pL_p^* - L_pU_p + L_pL_p^* = U_p - U_p - U_p + U_p = 0.\]
    But then for arbitrary $a\in V$ we calculate $\inn{A^*a,A^*a} = \inn{AA^*a,a} = 0$ so that $A^*a = 0$ and hence $0=A^*=U_p - L_p^*$, so that $L_p = (L_p^*)^* = U_p^* = U_p$.
    \item By the previous point, for any sharp $p$: $T_p^* = \frac12(\id^* + L_p^* - L_{p^\perp}^*) = \frac12(\id + L_p - L_{p^\perp})$. Now write $a=\sum_i \lambda_i p_i$ where the $p_i$ are atomic. Then $T_a = \sum_i \lambda_i T_{p_i}$, and hence $T_a^* = \sum_i \lambda_i T_{p_i}^* = \sum_i \lambda_i T_{p_i} = T_a$. \qedhere
  \end{enumerate}
\end{proof}

\begin{theorem}\label{thm:seqprodisjordan}
  Let $V$ be a finite-dimensional sequential effect space. Then $V$ is a Euclidean Jordan algebra.
\end{theorem}
\begin{proof}
  By Proposition~\ref{prop:seqprodisjordan} it is a Jordan algebra, and by the previous proposition the Jordan algebra maps are self-adjoint so that $V$ is indeed Euclidean.
\end{proof}

We note that the converse is also true: any Euclidean Jordan algebra is a finite-dimensional sequential effect space (cf.~Theorem~\ref{thm:JBW-seqprod}).

\section{Central effects}\label{sec:central-effects}

Before we go on to study composite systems, we first need to know a bit more out the structure of a single sequential effect space, and how it decomposes into factors.

\begin{definition}
  We call a sharp effect \Define{central}\indexd{central effect} when it is compatible with all effects. We say a central effect is \Define{minimal}\indexd{minimal central effect} if it is non-zero and there is no non-zero central effect beneath it.
\end{definition}

\begin{proposition}
    Let $c_1$ and $c_2$ be central effects.
    \begin{enumerate}[label=\alph*)]
        \item $c_1^\perp$ is central.
        \item If $c_1$ and $c_2$ are orthogonal, then $c_1+c_2$ is central.
        \item $c_1\mult c_2$ is central.
        \item If $c_1\leq c_2$, then $c_2-c_1$ is central.
        \item If $c_1$ is minimal, then either $c_1\leq c_2$, or $c_1$ is orthogonal to $c_2$.
        \item If $c_1\neq c_2$ are both minimal, then $c_1$ and $c_2$ are orthogonal.
    \end{enumerate}
\end{proposition}
\begin{proof}We let $a$ denote an arbitrary effect.
    \begin{enumerate}[label=\alph*)]
        \item As $a\commu c_1$ for all $a$, also $a\commu c_1^\perp$. Since $c_1$ is sharp, $c_1^\perp$ is also sharp.
        \item Since $a\commu c_1, c_2$, we also have $a\commu c_1+c_2$. The sum of two orthogonal sharp effects is again sharp.
        \item Similar to the previous point, but using Proposition~\ref{prop:compmult}.
        \item We have $c_2-c_1 = c_2\mult c_1^\perp$, and hence it is central by the previous points.
        \item By point c), $c_1\mult c_2$ is central and of course $c_1\mult c_2\leq c_1$. Since $c_1$ is minimal, we must then have $c_1\mult c_2 = c_1$, or $c_1\mult c_2 =0$, giving the desired result.
        \item Direct consequence of the previous point by minimality of $c_2$. \qedhere
    \end{enumerate}
\end{proof}

\begin{proposition}
    Every central effect is equal to the sum of the minimal central effects below it. In particular, $1$ is equal to the sum of all minimal central effects.
\end{proposition}
\begin{proof}
    Let $c$ be central. If it is minimal then we are done. If it is not minimal, then there is a non-zero central effect $c_1 < c$. We now have two orthogonal central effects $c_1$ and $c-c_1$. By repeating this procedure these central effects we get a sequence of orthogonal central elements below $c$. As orthogonal elements are linearly independent, by finite-dimensionality, this procedure has to terminate and hence we have $c=\sum_i c_i$ where all the $c_i$ are minimal.

    Now suppose $c^\prime \leq c$ is minimal. Then $c^\prime = c^\prime \mult c = \sum_i c^\prime \mult c_i$. Each of these summands is either zero or equal to $c_i$ by the previous proposition, and hence $c^\prime = c_i$ for some $i$. As a result, all the minimal central elements below $c$ are already included in the set $\{c_i\}$.
\end{proof}

\begin{proposition}\label{prop:splitovercentraleffects}
    Let $\{c_i\}$ be a collection of central effects that sums to $1$, and let $a$ be an arbitrary effect. Then there are unique $a_i\leq c_i$ such that $a=\sum_i a_i$. Furthermore, for any other $b=\sum_i b_i$ with $b_i\leq c_i$ we get $a\mult b = \sum_i a_i\mult b_i$.
\end{proposition}
\begin{proof}
    Letting $a_i = c_i\mult a$ we get $a = a\mult 1 = a\mult (\sum_i c_i) = \sum_i a\mult c_i = \sum_i c_i\mult a$. For uniqueness suppose now that $a=\sum_i a_i^\prime$ is another decomposition with $a_i^\prime \leq c_i$. Then $a_j = c_j \mult a = c_j \mult (\sum_i a_i^\prime) = a_j^\prime$, because $c_i\perp c_j$ when $i\neq j$. Now let $b=\sum_i b_i$ with $b_i\leq c_i$. Then $a\mult b = \sum_i a\mult b_i = \sum_i a\mult (c_i \mult b_i) = \sum_i (a\mult c_i)\mult b_i = \sum_i a_i\mult b_i$.
\end{proof}

\begin{lemma}
  Let $c$ be a central effect and let $p$ be an atom. Then either $p\leq c$ or $p\perp c$.
\end{lemma}
\begin{proof}
  Since $c\commu p$ we have $c\mult p = p\mult c = \lambda p$ for some $\lambda$. But also $\lambda p = c\mult p = c\mult(c\mult p) = \lambda (c\mult p) = \lambda^2 p$. Hence, either $\lambda = 0$ or $\lambda = 1$. In the first case $p\perp c$, and in the second case $c\mult p = p$ so that $p\leq c$.
\end{proof}

\begin{proposition}\label{prop:incompatibleatomscentral}
  Let $c$ be a central effect and suppose $p\leq c$ is an atom. Suppose $q$ is an atom incompatible with $p$, then also $q\leq c$.
\end{proposition}
\begin{proof}
  We prove the contrapositive. Suppose $q$ is not below $c$. Then by the previous lemma $q\mult c = 0$ and hence $q\mult p = q\mult (c\mult p) = (q\mult c)\mult p = 0\mult p = 0$ so that $q\commu p$.
\end{proof}

\begin{definition}
  We call $V$ \Define{simple}\indexd{simple sequential effect space} if the only central effects are $0$ and $1$.
\end{definition}

\begin{definition}
  Let $V$ and $W$ be sequential effect spaces. We define their \Define{direct sum}\indexd{direct sum!of sequential effect spaces} $V\oplus W$ to be the direct sum of the vector spaces with the order and sequential product defined coordinatewise. We call a sequential effect space \Define{indecomposable}\indexd{indecomposable sequential effect space} if it is not isomorphic to $V\oplus W$ for $V,W\neq\{0\}$.
\end{definition}

\begin{proposition}
  A space is simple if and only if it is indecomposable.
\end{proposition}
\begin{proof}
  We will show that $V$ is decomposable if and only if $V$ is not simple.

  Suppose $V$ is not simple. Then there is a central effect $c\neq 0,1$. Let $V_1 := c\mult V := \{c\mult a~;~ a\in V\} $ and similarly $V_2 = c^\perp \mult V$. By Proposition~\ref{prop:splitovercentraleffects} any $a\in V$ can be written as $a=a_1+a_2$ where $a_i\in V_i$. Also by that proposition, when we have $b=b_1+b_2$ split in the same way, then $a\mult b = a_1\mult b_1 + a_2\mult b_2$, and hence the sequential product separates over $V_1$ and $V_2$. Hence $V\cong V_1\oplus V_2$. 

  For the other direction, suppose $V=V_1\oplus V_2$, then it is straightforward to verify that $(1,0)\in V_1\oplus V_2$ is central, and hence that $V$ is not simple.
\end{proof}

\begin{proposition}\label{prop:uniquesimpledecomposition}
  Let $V$ be a finite-dimensional sequential effect space. There exist unique (up to permutation) simple sequential effect spaces $V_1,\ldots V_n$ such that $V\cong V_1\oplus\ldots\oplus V_n$.
\end{proposition}
\begin{proof}
  Let $c_1,\ldots,c_n$ be the set of minimal non-zero central effects in $V$ and set $V_i = c_i\mult V$. By Proposition~\ref{prop:splitovercentraleffects} we then indeed get $V=V_1\oplus\ldots\oplus V_n$. Since the $c_i$ are minimal, each of the $V_i$ must be simple.

  Suppose there is another decomposition $V\cong W_1\oplus\ldots\oplus W_k$ into simple sequential effect spaces $W_j$. The $1_{W_j}$ are minimal central effects, and hence they must correspond to the $c_i$ so that there is some permutation $\delta$ such that $V_i = c_i\mult V = 1_{W_{\delta(i)}} \mult V = W_{\delta(i)}$.
\end{proof}



We will need the following proposition. Unfortunately the author does not know of a simple proof of this fact using the sequential product. However, because $V$ is a Euclidean Jordan algebra, we can resort to the extensive literature on that subject.

\begin{proposition}\label{prop:simplecornerissimple}
    Suppose $V$ is simple and let $p\in V$ be any sharp effect. Then $p\mult V$ is again simple.
\end{proposition}
\begin{proof}
    As $V$ is a Euclidean Jordan algebra, it is also a JBW-algebra (Proposition~\ref{prop:EJA-is-JBW}) and hence Proposition 5.2.17 of Ref.~\cite{hanche1984jordan} applies that states exactly this result.
\end{proof}

\section{Composite systems}\label{sec:seqmeas-tensorproducts}
\indexd{tensor product}

Definition \ref{def:seqprod} only concerns single systems, but a physical theory must also describe how multiple systems combine into a larger one. Let $A$ and $B$ denote a pair of systems where the effects come from order unit spaces $V_A$ and $V_B$. The composite system $A\otimes B$ is then represented by some space $V_{A\otimes B}$. Given states $\omega_A:V_A\rightarrow \R$ and $\omega_B:V_B\rightarrow \R$ and effects $a\in V_A$ and $b\in V_B$, we represent their composites on the system $A\otimes B$ by $\omega_A\otimes \omega_B$ and $a\otimes b$ (note that a priori the symbol `$\otimes$' does not have to be related to the regular tensor product of vector spaces). Since these are product states and effects, they represent non-interacting systems. The probability of the outcome $a\otimes b$ being observed on $\omega_A\otimes \omega_B$ is therefore given by $(\omega_A\otimes \omega_B)(a\otimes b) = \omega_A(a)\omega_B(b)$. Similarly, when we have effects $a_1$ and $a_2$ on $A$ and effects $b_1$ and $b_2$ on $B$, the independence of their composites should be respected by the sequential product: $(a_1\otimes b_1)\mult (a_2\otimes b_2)=(a_1\mult a_2)\otimes (b_1\mult b_2)$. 

We will also assume that our composites satisfy local tomography\indexd{local tomography}. Recall from Section~\ref{sec:GPTs} that local tomography demands that a state defined on a composite system is completely determined by measurements on each separate component (the \emph{local} measurements). In other words, given states $\omega_1,\omega_2\in\st(A\otimes B)$ we only have $\omega_1(a\otimes b)=\omega_2(a\otimes b)$ for all $a\in \eff(A)$ and $b\in \eff(B)$ when $\omega_1=\omega_2$. As our systems satisfy the no-restriction hypothesis, this requirement reduces to the equality~$\dim(V_{A\otimes B}) = \dim(V_A)\dim(V_B)$~\cite{barrett2007information}.

\begin{definition}\label{def:localtomographiccomposite}
  Let $V$ and $W$ be finite-dimensional sequential effect spaces. We say that $V\otimes W$ is a \Define{locally tomographic composite}\indexd{locally tomographic composite} when there is a bilinear map $\otimes: V\times W\rightarrow V\otimes W$ and
  \begin{enumerate}[label=\alph*)]
    \item $\dim(V\otimes W) = \dim(V)\dim(W)$,
    \item for all $a_1,a_2\in V$, $b_1,b_2 \in W$: $(a_1\otimes b_1)\mult (a_2\otimes b_2)=(a_1\mult a_2)\otimes (b_1\mult b_2)$,
    \item for all states $\omega_1:V\rightarrow \R$, $\omega_2:W\rightarrow \R$, there is a state $\omega:V\otimes W\rightarrow \R$ satisfying for all $a\in V$ and $b\in W$: $\omega(a\otimes b) = \omega_1(a)\omega_2(b)$. We denote this state by $\omega_1\otimes \omega_2$.
  \end{enumerate}
\end{definition}

For the remainder of this section we will let $V$ and $W$ denote finite-dimensional sequential effect spaces, and $V\otimes W$ a locally tomographic composite of them.

\begin{lemma}\label{lem:tensorisbijective}
    The bilinear map $V\times W\rightarrow V\otimes W$ given by $(a,b)\mapsto a\otimes b$ is bijective.
\end{lemma}
\begin{proof}
    Let $\{p_i\}$ be a basis of atomic effects of $V$ and similarly let $\{q_j\}$ be a basis of atomic effects in $W$. We will show that $\{p_i \otimes q_j\}$ is linearly independent in $V\otimes W$ and hence the dimension of the image of the tensor product map is $\dim(V) \dim(W)$. Since $\dim(V\otimes W) = \dim(V) \dim(W)$ the map must then be bijective. 

    Reasoning towards a contradiction, suppose that the set $\{p_i \otimes q_j\}$ is linearly dependent.
    Without loss of generality we can then write $p_1\otimes q_1 = \sum_{i,j} \lambda_{ij} p_i\otimes q_j$ where the $\lambda_{ij}$ are some real numbers and the sum goes over all $i,j$ except $i=j=1$. 
    Let $\omega_1$ and $\omega_2$ be arbitrary states on $V$ respectively $W$ and apply the map $\omega_1\otimes \omega_2$ to both sides to get
    \[\omega_1(p_1)\omega_2(q_1) = \sum_{i,j} \lambda_{ij} \omega_1(p_i) \omega_2(q_j) = \omega_1(p_1)\left(\sum_j \lambda_{1j} \omega_2(q_j)\right) + \sum_{i>1} \omega_1(p_i) \left(\sum_j \lambda_{ij}\omega_2(q_j)\right).\]
    Rewrite this to
    \[\omega_1(p_1)\left(\omega_2(q_1)-\sum_j \lambda_{1j} \omega_2(q_j)\right) = \omega_1\left(\sum_{i>1} p_i \sum_j \lambda_{ij}\omega_2(q_j)\right).\]
    This holds for all states $\omega_1$ and since states separate the effects, we hence have
    \[p_1\left(\omega_2(q_1)-\sum_j \lambda_{1j} \omega_2(q_j)\right) = \sum_{i>1} p_i \sum_j \lambda_{ij}\omega_2(q_j).\]
    By assumption the $p_i$ are linearly independent and so this can only hold when $\sum_j \lambda_{ij} \omega_2(q_j) = \omega_2(\sum_j \lambda_{ij} q_j) =  0$ for all $i>1$. 
    This must again hold for all states $\omega_2$ so that $\sum_j \lambda_{ij} q_j = 0$. 
    Since the $q_j$ are also linearly independent this shows that $\lambda_{ij} = 0$ when $i>1$. By interchanging the role of $p_1$ and $q_1$ we also get $\lambda_{ij} = 0$ when $j>1$, so that the only nonzero value could be $\lambda_{11}$, which finishes the contradiction.
\end{proof}

\begin{lemma}\label{lem:localtomographyspanning}
  Let $c\in V\otimes W$ be arbitrary. Then $c = \sum_i \lambda_i a_i\otimes b_i$ for some $a_i\in V$, $b_i\in W$ and $\lambda_i \in \R$.
\end{lemma}
\begin{proof}
  By the previous proposition, the `pure tensors' $a\otimes b$ span $V\otimes W$ and hence we can find a basis of $V\otimes W$ that consists of pure tensors $a_i\otimes b_i$.
\end{proof}

\begin{proposition}
  The following are true.
  \begin{enumerate}[label=\alph*),ref=\fullcounter.\alph*)]
    \item \label{prop:atomictensor} $p\otimes q\in V\otimes W$ is atomic when $p\in V$ and $q\in W$ are atomic.
    \item \label{prop:classicaltensor}$c\otimes d$ is central in $V\otimes W$ when $c\in V$ and $d\in W$ are central.
  \end{enumerate}
\end{proposition}
\begin{proof}~
  \begin{enumerate}[label=\alph*)]

    \item Because $(p\otimes q)\mult (p\otimes q) = (p\mult p)\otimes (q\mult q) = p\otimes q$, we see $p\otimes q$ is sharp. Let $c = \sum_i \lambda_i a_i\otimes b_i$ be an arbitrary element of $V\otimes W$, then using Lemma~\ref{lem:atomicnorm} $(p\otimes q)\mult c = \sum_i \lambda_i (p\mult a_i)\otimes (q\mult b_i) = \sum_i \lambda_i \norm{p\mult a_i}\norm{q\mult b_i} (p\otimes q) = \mu (p\otimes q)$ for some $\mu\in \R$. Now suppose $0\leq c \leq p\otimes q$. Since $p\otimes q$ is sharp we get $c = (p\otimes q)\mult c = \mu (p\otimes q)$, and hence $p\otimes q$ is indeed atomic.

    \item $c\commu a$ for all $a \in V$ and $d\commu b$ for all $b\in W$, and hence $c\otimes d\commu a\otimes b$. Consequently $c\otimes d$ is compatible with all linear combinations of pure tensors, which span the entirety of $V\otimes W$. \qedhere
  \end{enumerate}
\end{proof}

\begin{lemma}\label{lem:nonzeroatompairs}
    Suppose $V$ is simple and let $p,q\in V$ be atoms. Then there is an atom $r\in V$ such that $r\mult p \neq 0$ and $r\mult q \neq 0$.
\end{lemma}
\begin{proof}
    If $p\mult q \neq 0$ then we are done (pick $r=p$), so assume that $p$ and $q$ are orthogonal and hence $p\commu q$. Consider $W:=(p + q)\mult V$. If the only atoms in $W$ are $p$ and $q$ then both $p$ and $q$ are central in $W$ and hence $W$ would not be simple, which contradicts Proposition~\ref{prop:simplecornerissimple}. Hence, there is an atom $r\in W$ with $r\neq p,q$. If $r\mult p = 0$, then $p+r \leq 1_W = p+q$.
    As $W$ has rank 2 by Corollary~\ref{cor:rank} this is only possible when $r=q$, a contradiction. Hence $r\mult p \neq 0$ and by symmetry also $r\mult q \neq 0$.
\end{proof}

\begin{proposition}\label{prop:tensordirectsum}
    Decompose $V = E_1\oplus\ldots \oplus E_n$ and $W = F_1\oplus \ldots \oplus F_m$ with the $E_i$ and $F_j$ simple as in Proposition~\ref{prop:uniquesimpledecomposition}. Pick $1\leq k \leq n$ and $1\leq l\leq m$ and let $p_1,\ldots, p_r$ be a maximal collection of orthogonal atomic effects in $E_k$, and $q_1,\ldots, q_s$ a maximal collection of orthogonal atoms in $F_l$. Then $(p_i\otimes q_j)_{i=1,j=1}^{r,s}$ belong to the same simple summand in $V\otimes W$ and they form a maximal collection of orthogonal non-zero atomic effects in this summand.
\end{proposition}
\begin{proof}
    For each pair $p_i$ and $p_j$ pick an atom $r_{ij}$ such that $r_{ij}\mult p_i \neq 0$ and $r_{ij}\mult p_j\neq 0$, which exists by Lemma~\ref{lem:nonzeroatompairs}. Similarly, for every $q_k$ and $q_l$ pick an atom $r'_{kl}$ such that $r'_{kl}\mult q_k \neq 0$ and $r'_{kl}\mult q_l \neq 0$.

    By Proposition~\ref{prop:atomictensor} $r_{ij}\otimes r'_{kl}$ and $p_i\otimes q_k$ are atomic for all $i,j,k,l$. By construction we of course have $0\neq (r_{ij}\mult p_i)\otimes (r'_{kl}\mult q_k)=(r_{ij}\otimes r'_{kl})\mult (p_i\otimes q_k)$ and similarly $(r_{ij}\otimes r'_{kl})\mult (p_j\otimes q_l) \neq 0$. Then by Proposition~\ref{prop:incompatibleatomscentral}, $r_{ij}\otimes r'_{kl}$ must be below the same minimal central effect as both $p_i\otimes q_k$ and $p_j\otimes q_l$. Doing this with all possible pairings we see that indeed all the $p_i\otimes q_j$ are below the same minimal central element in $V\otimes W$ and hence belong to the same simple summand.

    Since $\sum_i p_i=1_{E_k}$, this sum is a central effect. The same holds for $\sum_j q_j=1_{F_l}$. Their tensor product $1_{E_k}\otimes 1_{F_l} = \sum_{i,j} p_i\otimes q_j$ is then also central by Proposition~\ref{prop:classicaltensor}. Since the only nonzero central effect in a simple summand is the identity, this expression must be equal to the identity of this summand. As a result the set $(p_i\otimes q_j)_{i,j}$ is indeed maximal.
\end{proof}
Using this proposition we conclude that for each of the simple summands $E$ of $V$ and $F$ of $W$ there must exist a summand in $V\otimes W$ which has rank $(\rnk~ E)(\rnk~ F)$. Because the tensor product map is injective this factor must have dimension at least $\dim E~ \dim F$, and since the total dimension of the factors must equal $\dim (V\otimes W) = \dim(V)\dim(W)$ the dimension must be exactly equal to $\dim E~\dim F$.
Now when $V$ is a sequential effect space for which a locally tomographic composite $V\otimes V$ exists, the above must in particular be true when $E=F$ so that for all simple factors $E$ of $V$ there must exist a simple factor with rank $(\rnk E)^2$ and dimension $(\dim E)^2$.

To show that this is quite a restrictive property we will use the classification theorem for Euclidean Jordan algebras presented in Theorem~\ref{thm:Jordan-classification}, which allows us to prove the following proposition.

\begin{proposition}\label{prop:simpleEJAsquare}
    Let $E$ be a simple Euclidean Jordan algebra of rank $r$ and dimension $N$. There exists a simple Euclidean Jordan algebra of rank $r^2$ and dimension $N^2$ if and only if $E=M_r(\C)_\sa$.
\end{proposition}
\begin{proof}
    If $E=M_r(\C)_\sa$ is a complex matrix algebra then $\dim E = N=r^2$ and the property is true because $M_{r^2}(\C)_\sa$ is a simple EJA of rank $r^2$ and dimension $r^4 = N^2$. It remains to check that every other simple EJA does not satisfy the conditions.
    If $E= M_3(\mathbb{O})_\sa$ is the 3-dimensional octonion Hilbert space, then $r=3$ and $N=27$. The highest-dimensional simple EJA of rank $9$ is the quaternionic $M_9(\mathbb{H})_\sa$ which has dimension $9\cdot(2\cdot 9-1)=153<27^2 = 729$. Hence, there is no simple EJA with the right dimension and rank.
    If $E=M_r(\mathbb{H})_\sa$ is quaternionic, then $N = r(2r-1)$. The highest-dimensional simple EJA of rank $r^2$ is also quaternionic so that its dimension is $r^2(2r^2-1)$. It is easy to check that $N^2 = r^2(2r-1)^2 > r^2(2r^2-1)$ when $r>1$ so that again, $E$ does not satisfy the property.
    If $E=M_r(\R)_\sa$, then by dimension counting we can again see that there does not exist an EJA with rank $r^2$ and dimension $N^2$.
    If $E$ is a spin factor then its rank is 2. The rank 4 simple EJAs have dimension 10, 16 and 28. The only one of these which is a square is 16, and hence in this case $E$ must have dimension 4 and hence be the spin factor $S_3$. This spin factor is isomorphic to $M_2(\C)_\sa$.
\end{proof}

\begin{theorem}\label{thm:seqprodlocalcomp}
    Suppose $V$ is a finite-dimensional sequential effect space which has a locally tomographic composite with itself. Then there exists a C$^*$-algebra $\mathfrak{A}$ such that $V\cong \mathfrak{A}_\sa$.
\end{theorem}
\begin{proof}
    Decompose $V$ into simple sequential effect spaces $V_k$ using Proposition~\ref{prop:uniquesimpledecomposition}: $V=V_1\oplus\cdots\oplus V_n$.
    As a result of Proposition~\ref{prop:tensordirectsum} for each summand $V_i$ of $V$ there must exist a simple sequential effect space of rank $(\rnk V_i)^2$ and dimension $(\dim V_i)^2$. 
    By Theorem~\ref{thm:seqprodisjordan} all the $V_i$ are simple Euclidean Jordan algebras, hence by Proposition~\ref{prop:simpleEJAsquare} this is only possible if $V_i=M_{r_i}(\C)_\sa$ for some $r_i\in \N$. 
    Hence $V = M_{r_1}(\C)_\sa \oplus\cdots\oplus M_{r_n}(\C)_\sa = \mathfrak{A}_\sa$ where $\mathfrak{A}$ is the C$^*$-algebra $\mathfrak{A} = M_{r_1}(\C) \oplus\cdots\oplus M_{r_n}(\C)$.
\end{proof}

This theorem completes our reconstruction of quantum theory. It shows that any physical theory which satisfies the conditions we specified with regard to sequential measurements and composite systems must be represented by a complex C$^*$-algebra. Conversely, any finite-dimensional complex C$^*$-algebra satisfies our assumptions.  

\begin{remark}
	The more interesting part of this derivation is that sequential product spaces correspond to Jordan algebras. The result that Euclidean Jordan algebras that have a locally tomographic composite correspond to complex C$^*$-algebras has been shown in many different ways before. To the authors knowledge the first time a result like this was shown was by Hanche-Olsen~\cite{hanche1985jb}. Local tomography has been used explicitly as an axiom to force this property in several reconstructions such as in Refs.~\cite{selby2018reconstructing,masanes2014entanglement} and we will use it for the same purpose for a different reconstruction in Chapter~\ref{chap:effectus}.
\end{remark}

\section{Dynamics of the theory}\label{sec:dynamics}

We have so far focused on the states and effects of a physical system, but an important part of any physical theory is of course the set of allowed dynamics. In this section we will sketch how our assumptions are also sufficient to retrieve the standard dynamics of quantum theory: trace-preserving completely positive maps, unitary conjugation, and the Schr\"odinger equation.

Any state-transformation $T:A\rightarrow B$ of a system $A$ to $B$, which we will assume to be convex for the same reasons we assumed states to be convex, gives rise to a map of the effect spaces $T:\eff(B)\rightarrow \eff(A)$ via $\omega(T(a)) := T(\omega)(a)$. We then necessarily have $T(0) = 0$ as $\omega(T(0)) = T(\omega)(0) = 0$ for all $\omega$ and similarly $T(1)=1$. Since we have $\eff(A)=[0,1]_{V_A}$ and transformations preserve the convex structure of the effect space we see that $T$ extends to a positive unital linear map $T:V_B\rightarrow V_A$. 
Theorem~\ref{thm:seqprodlocalcomp} shows that we can represent systems by (the self-adjoint part) of complex C$^*$-algebras: $V_A=\mathfrak{A}_\sa$, $V_B=\mathfrak{B}_\sa$. The linear map $T:\mathfrak{B}_\sa\rightarrow\mathfrak{A}_\sa$ is easily extended to the entirety of the C$^*$-algebra by setting $T(a+ib) = T(a)+iT(b)$. 
Hence, transformations correspond to positive unital maps between C$^*$-algebras. 
As discussed in Section~\ref{sec:intro-composite-systems}, to deal with composite systems, $T$ must furthermore be completely positive. Hence, we have retrieved the standard type of dynamics that are allowed in quantum theory. Note that the reason we get unital maps instead of trace-preserving maps is because we are dealing with effect-transformations. The adjoint of the map $T$ gives a state-transformation that is trace-preserving instead of unital.

In order to retrieve the unitary structure of quantum theory we consider reversible transformations. A transformation $T:A\rightarrow A$ is \Define{reversible}\indexd{reversible transformation} when it has an inverse $T^{-1}:A\rightarrow A$ so that $T\circ T^{-1} = \id = T^{-1}\circ T$.
Hence in our setting reversible transformations are precisely those unital completely positive maps $T:\mathfrak{A}\rightarrow\mathfrak{A}$ that have a completely-positive two-sided inverse $T^{-1}$. 
Such maps are in particular unital order-isomorphisms, and hence Jordan-isomorphisms (cf.~Proposition~\ref{prop:orderiso-is-Jordan}). As a result, $T$ maps (minimal) central effects to (minimal) central effects. This means that $T$ permutes the simple factors $\mathfrak{A} = \mathfrak{A}_1\oplus\cdots\oplus\mathfrak{A}_n$ in some manner (note that $T$ can only interchange isomorphic factors). Composing $T$ with the map $P$ that undoes this permutation of the simple factors we get a completely-positive invertible map $P\circ T$ that restricts to each $\mathfrak{A}_k\cong M_{r_k}(\C)$ separately. The unital completely-positive invertible maps $f:M_{r_k}(\C)\rightarrow M_{r_k}(\C)$ have been classified~\cite{nayak2006invertible} and are all of the form $f(M) = UMU^*$ for some unitary $U\in M_{r_k}(\C)$.
We conclude that reversible transformations consist of a permutation of the simple factors followed by a unitary conjugation on each of the separate factors. In particular, if the system is `maximally non-classical', having no non-central effects, then the only possible reversible transformations are unitary conjugations.

Finally, to reconstruct the Schr\"odinger equation, let us consider a time-evolution of the system. We represent the time-evolution of the system $A$ by a family of reversible transformations $\{\Phi_t:A\rightarrow A~;~ t\in \R\}$ that maps an initial state $\omega\in \st(A)$ to the state $\omega_t:=\Phi_t(\omega)$ that represents the same state after some time $t$ has passed. Of course $\Phi_0 = \id$ and $\Phi_t\circ\Phi_s = \Phi_{t+s} = \Phi_s\circ \Phi_t$, as first letting $s$ units of time pass and then $t$ is the same as letting $t+s$ units pass at once. Consequently, we also have $\Phi_{-t} = \Phi^{-1}_t$ so that $t\mapsto \Phi_t$ is a group homomorphism from $\R$ to the set of reversible transformations of $A$.
As each of the $\Phi_t$ is a reversible transformation, it must consist of a permutation of the simple factors of the C$^*$-algebra underlying $A$ followed by a unitary conjugation. However, since we have $\Phi_t = (\Phi_{t/n})^n$ for every $n$, the permutation of $\Phi_t$ must be trivial. Hence the time evolution is implemented by unitary conjugation on each separate factor of $A$. Let us then assume that the C$^*$-algebra of $A$ is simple and hence equal to $M_n(\C)$ for some $n$. Then for each $t$ we have $\Phi_t(M) = U_tMU_t^*$ for some unitary $U_t$.
If we now make the final reasonable assumption that when $t\rightarrow 0$ we have $\Phi_t \rightarrow \id$ (in the strong operator topology), \ie~that the time-evolution is continuous, or in other words, that small changes in time lead to small changes in the state, we have the conditions for Stone's theorem on one-parameter unitary groups~\cite{stone1932one}. As a consequence of this theorem there must be some self-adjoint matrix $H\in M_n(\C)$ such that $\Phi_t(M) = e^{it H} M e^{-it H}$, which is the form of the Schr\"odinger equation for a Hamiltonian $H$.

\section{Conclusion}\label{sec:seqmeas-conclusion}

We have shown that we can reconstruct quantum theory from assumptions on the behaviour of sequential measurement. We first considered an isolated system and showed that its effect space must be isomorphic to that of a Euclidean Jordan algebra. Then we showed that the only systems that additionally have locally tomographic composites with themselves must be complex C$^*$-algebras. The Born rule and the Schr\"odinger equation were derived as consequences from these results.

An interesting question is of course to ask whether our assumptions can be weakened. Example~\ref{ex:weird-seq-prod} shows that we need the effect spaces to be order unit spaces --- just having an ordered vector space separated by states is not enough. Of the axioms of the sequential product in Definition~\ref{def:seqprod}, the orthogonality axiom \ref{ax:orth} is perhaps the least well-motivated from an operational standpoint. We remark however that most of the results of Section~\ref{sec:basicresultsseqprod} do not need this axiom. In particular, the proof of the spectral theorem and the homogeneity of the space can be derived without it. It is unclear whether the rest of our results continue to hold without the use of \ref{ax:orth}. It is also not clear whether any of the other axioms of Definition~\ref{def:seqprod} can be dropped.

We dealt in this chapter with finite-dimensional systems. In Chapter~\ref{chap:infinitedimension} we consider infinite-dimensional systems. In particular, in Section~\ref{sec:SEA} we consider a slightly different variant of the sequential product that allows us to derive many of the same results in infinite dimension. We identify in particular two additional assumptions that are sufficient for the space to be isomorphic to a JB-algebra (a generalisation of EJAs that also allows infinite-dimensional algebras).

\chapter{Pure processes}\label{chap:effectus}

In the previous chapter we used the framework of generalised probabilistic theories. This framework however has a hidden assumption: the representation of probabilities with real numbers. This allowed us to work with convex state and effect spaces, and in turn allowed us to derive that it was sufficient to consider the order unit spaces associated to physical systems.

In this chapter we start an investigation in what can be said about the basic structure of state and effect spaces if we do not consider probabilities to take the form of real numbers. The result will be a structure known as an \emph{effect algebra} which was originally introduced in 1994 by Foulis and Bennett~\cite{foulis1994effect}. An effect algebra is an axiomatisation of the structure present in the space of effects of a (quantum) physical system. This axiomatisation involves a partial addition, negation and existence of a $0$ and $1$ effect (abstract versions of~Assumptions~\ref{assum:GPT} and~\ref{assum:effectspaceclosedunderaddition}).

Effect algebras are used in a relatively new branch of categorical logic known as \emph{effectus theory}\index{effectus}, which was initiated by Jacobs in 2014~\cite{jacobs2015new}. An effectus is essentially a generalisation of the type of GPT we described in Section~\ref{sec:GPTs}, and can describe deterministic, probabilistic, and quantum models of logic or even more exotic alternatives~\cite{kentathesis}, either in finite or infinite dimension~\cite{basthesis}.

In this chapter we consider a weaker form of an effectus that we call an \emph{effect theory}. We use this alternative as it is sufficient for our purposes and easier to define. As many of the concepts we use were originally framed in the language of effectus theory, we adapt them where necessary. No previous knowledge on effectus theory is assumed, as this chapter is a self-contained exposition of the parts of effectus theory that we will need in this thesis.

This chapter consists basically of two parts.
The first part (Sections~\ref{sec:category-theory}--\ref{sec:filterscompressions}) recalls the basic concepts of effect algebras and effect theories and introduces some important structure that can be present in an effect theory: filters, compressions and sharp effects. These concepts will be essential in this chapter and the remainder of this part of the thesis. We mostly adapt these results from Refs.~\cite{kentathesis,cho2015introduction,basthesis}.
The second part (Sections~\ref{sec:purity}--\ref{sec:monoidalPETs}) presents a reconstruction of quantum theory complementary to that of Chapter~\ref{chap:seqprod}. Whereas Chapter~\ref{chap:seqprod} reconstructed the effect and state spaces of quantum theory from assumptions on how sequential measurement should behave, in this chapter the notion of \emph{purity} is central. We study what it means for a state, effect or transformation to be pure, and we find suitable properties that the set of pure maps should satisfy in a physical theory.

Although the framework of effect theories allows for more general probabilities than just real numbers, for the reconstruction in this chapter we do still require the structure of real numbers. In Chapter~\ref{chap:infinitedimension} we will combine the assumptions of this chapter with those of Chapter~\ref{chap:seqprod} to reconstruct infinite-dimensional quantum theory without a priori assuming the presence of real numbers.

\section{A primer on category theory}\label{sec:category-theory}

Some of the results in this chapter will be framed in the language of category theory, so let us recall some of the basic notions that will be relevant for us before we continue. A \Define{category}\indexd{category} $\mathbf{C}$ consists of a set of \Define{objects} and a set of \Define{morphisms}\indexd{morphism!in category theory}. A morphism $f:A\rightarrow B$ goes from an object $A$ to an object $B$. Given morphisms $f:A\rightarrow B$ and $g:B\rightarrow C$ we can compose them to get a morphism $g\circ f:A\rightarrow C$. Composition is associative. Every object $A$ has a unique \Define{identity} morphism $\id_A:A\rightarrow A$ satisfying $f\circ \id_A = f$ for all $f:A\rightarrow B$ and $\id_A\circ g = g$ for all $g:B\rightarrow A$.

Let $f:A\rightarrow B$ be a morphism in a category. We say $f$ is an \Define{isomorphism} when there exists a (necessarily unique) morphism $g:B\rightarrow A$ so that $g\circ f= \id_A$ and $f\circ g = \id_B$. When there is an isomorphism between two objects $A$ and $B$ we say they are \Define{isomorphic} and write $A\cong B$.
We say $f$ is \Define{epic}\indexd{epic morphism} when for every pair of morphisms $g_1,g_2:B\rightarrow C$ the equality $g_1\circ f = g_2\circ f$ implies $g_1=g_2$. Dually, $f$ is \Define{monic}\indexd{monic morphism} when an equality $f\circ g_1 = f\circ g_2$ for morphisms $g_1,g_2:C\rightarrow A$ implies $g_1=g_2$.

Given categories $\mathbf{C}$ and $\mathbf{D}$ a \Define{functor}\indexd{functor} $F:\mathbf{C}\rightarrow\mathbf{D}$ maps each object $A$ of $\mathbf{C}$ to an object $F(A)$ of $\mathbf{D}$, and each morphism $f:A\rightarrow B$ in $\mathbf{C}$ to a morphism $F(f):F(A)\rightarrow F(B)$ in $\mathbf{D}$ and is required to satisfy $F(\id_A)=\id_{F(A)}$ and $F(f\circ g) = F(f)\circ F(g)$. We say a functor is \Define{faithful}\indexd{faithful functor} when $F(f) = F(g)$ implies $f=g$.

Given a category $\mathbf{C}$ we denote by $\mathbf{C}^\opp$ the category with the same objects but where each morphism $f^\opp:A\rightarrow B$ is a morphism $f:B\rightarrow A$ in $\mathbf{C}$. Composition is defined as $g^\opp\circ^\opp f^\opp = (f\circ g)^\opp$. 

A \Define{monoidal category}\indexd{monoidal category} is intuitively a category with tensor products. Specifically, for every pair of objects $A$ and $B$ we have an object $A\otimes B$ and for every pair of morphisms $f_i:A_i\rightarrow B_i$ we have a morphism $f_1\otimes f_2:A_1\otimes A_2\rightarrow B_1\otimes B_2$. These tensor products are required to satisfy $(f_1\otimes f_2)\circ (g_1\otimes g_2) = (f_1\circ g_1)\otimes (f_2\circ g_2)$. Furthermore, we have a special object $I$ called the \Define{monoidal unit} that satisfies $A\otimes I\cong A$ for all objects $A$. A monoidal category is required to satisfy certain `coherence axioms' that will not be relevant to us~\cite{mac2013categories}.

\section{Effect algebras}\label{sec:effect-algebras}

We split up the definition of an effect algebra into two parts, first introducing a more primitive notion.

\begin{definition}[{\cite{jacobs2012coreflections}}]\label{def:partialcommutativemonoid}
    A \Define{partial commutative monoid}\indexd{partial commutative monoid} $(P, \ovee, 0)$ is a set $P$
together with a distinguished element~$0 \in P$ and
        a partial binary operation~$\ovee$
    such that for all~$a,b,c \in P$ --- writing~$a \perp b$ whenever~$a \ovee b$ is defined
    --- the following conditions hold.
\begin{itemize}
\item Commutativity: Suppose $a\perp b$. Then $b \perp a$ and $a\ovee b = b \ovee a$.
\item Zero: $a\perp 0$ and $a\ovee 0 = a$.
\item Associativity: Suppose $a\perp b$ and $(a\ovee b)\perp c$. Then
    $b\perp c$, 
$a\perp (b \ovee c)$, and $(a\ovee b) \ovee c = a\ovee (b\ovee c)$.
\end{itemize}
When $a\perp b$ we say that $a$ and $b$ are \Define{summable}\indexd{summable elements (effect algebra)}. When we write an equation like `$a\ovee b = c$' we mean that both $a\perp b$, so that $a\ovee b$ is indeed defined, and $a\ovee b = c$.\index{math}{$a\ovee b$ (sum in effect algebra)}\index{math}{$a\perp b$ (summable in effect algebra)}
\end{definition}

\begin{remark}
  What we call ``summable'' is in the literature often referred to as ``orthogonal''. As this does not correspond with the standard notion of orthogonality in quantum theory for non-projections, we do not use this term in this context.
\end{remark}

\begin{definition}[{\cite{foulis1994effect}}]\label{def:effectalgebra}
    An \Define{effect algebra}\indexd{effect algebra}
   $(E, \ovee, 0, (\ )^\perp)$ is a partial commutative monoid with an additional
    (total) unary operation $a \mapsto a^\perp$ called the \Define{complement},
    \indexd{complement (effect algebra)}
    \index{math}{$a^\perp$ (complement in effect algebra)}
    that satisfies the following properties (writing $1:=0^\perp$).
\begin{itemize}
\item Given~$a\in E$, the complement
    $a^\perp$ is the unique element with $a \ovee a^\perp = 1$.
\item If $a\perp 1$ for some~$a\in E$, then $a=0$.
\end{itemize}
\end{definition}
Note that as $a \ovee a^\perp = 1$ we have by uniqueness of complements $(a^\perp)^\perp = a$.

\begin{example}
  Let $V$ be an order unit space, and let $E=[0,1]_V$ be its unit interval. Then $E$ is an effect algebra with $a^\perp := 1-a$ and $a\perp b$ iff $a+ b\leq 1$ and then $a\ovee b := a+b$.
\end{example}

\begin{definition}
  Let $E$ and $F$ be effect algebras. A \Define{morphism}\indexd{morphism! between effect algebras} between $E$ and $F$ is a map $f:E\rightarrow F$ that satisfies for all $a,b \in E$:
  \begin{itemize}
    \item $f(1) = 1$.
    \item If $a\perp b$, then $f(a)\perp f(b)$ and $f(a\ovee b) = f(a)\ovee f(b)$.
  \end{itemize}
  We denote the category of effect algebras with morphisms between them by \textbf{EA}.
\end{definition}

\begin{proposition}
  Let $f:E\rightarrow F$ be a morphism between effect algebras. Then $f(0) = 0$ and $f(a)^\perp = f(a^\perp)$.
\end{proposition}
\begin{proof}
  As $1\perp 0$ we have $f(1)\perp f(0)$. But as $f(1)=1$ we then have $f(0)\perp 1$ and hence $f(0) = 0$.
  For the second point note that $1=f(1) = f(a\ovee a^\perp) = f(a) \ovee f(a^\perp)$, and so by uniqueness of complements $f(a)^\perp = f(a^\perp)$.
\end{proof}

\begin{proposition}\label{prop:effectalgebracancellative}
  Let $E$ be an effect algebra. Then addition is cancellative: for $a,b,c \in E$, if $a\ovee c = b\ovee c$, then $a=b$.
\end{proposition}
\begin{proof}
  We of course have $(a\ovee c)\ovee (a\ovee c)^\perp = 1$, and hence $a\ovee (c\ovee (a\ovee c)^\perp) = 1$. By uniqueness of complements we then have $a^\perp = c\ovee(a\ovee c)^\perp$. Similarly $b^\perp = c\ovee(b\ovee c)^\perp$. But since $a\ovee c = b\ovee c$, we have $(a\ovee c)^\perp = (b\ovee c)^\perp$, and hence also $b^\perp = c\ovee (b\ovee c)^\perp = c\ovee (a\ovee c)^\perp = a^\perp$ and thus indeed $a=b$.
\end{proof}

\begin{definition}
  Let $E$ be an effect algebra with $a,b\in E$. We write $a\leq b$ iff there exists a $c\in E$ with $a\ovee c = b$.
\end{definition}
  

\begin{proposition}\label{lem:effectalgebraorder}
  Let $E$ be an effect algebra, and $a,b \in E$. 
  \begin{enumerate}[label=\alph*)]
    \item If $a\ovee b = 0$, then $a=b=0$.
    \item $\leq$ is a partial order.
    \item $a\leq b$ iff $b^\perp \leq a^\perp$.
    \item $a$ and $b$ are summable iff $a\leq b^\perp$ iff $b\leq a^\perp$.
    \item If $a\leq b$, then there is a unique $c\in E$, such that $a\ovee c = b$.
  \end{enumerate}
\end{proposition}
\begin{proof}~
  \begin{enumerate}[label=\alph*)]
    \item If $a\ovee b = 0$, then $1\perp a\ovee b$ and hence by associativity also $1\perp a$, so that $a=0$. Similarly $b=0$.
    \item Reflexivity and transitivity follow from respectively the existence of $0\in E$, and the associativity of addition. It remains to show anti-symmetry, \ie~that $a\leq b$ and $b\leq a$ implies $a=b$. So suppose $a\leq b$ and $b\leq a$. By definition there exist $c_1,c_2 \in E$ such that $a\ovee c_1 = b$ and $b\ovee c_2 = a$. Then $a\ovee 0 = a = b\ovee c_2 = a\ovee (c_1\ovee c_2)$, and hence by cancellativity of addition $0 = c_1\ovee c_2$. By the previous point then $c_1=c_2 = 0$ and thus indeed $a=b$.

    \item If $a\leq b$ then by definition $a\ovee c = b$ for some $c \in E$. Hence $b^\perp = (a\ovee c)^\perp$. We note that $a\ovee(c\ovee(a\ovee c)^\perp) = 1$, and hence by uniqueness of complements $a^\perp = c\ovee (a\ovee c)^\perp = c\ovee b^\perp$. By definition then $b^\perp \leq a^\perp$. The same argument works in the opposite direction.
    
    \item Suppose $a$ and $b$ are summable. Then $a\ovee b$ is defined, and hence $a\ovee (b\ovee (a\ovee b)^\perp) = 1$, so that $a^\perp = b\ovee (a\ovee b)^\perp$, and thus indeed $b \leq a^\perp$. The previous point then also gives $a = (a^\perp)^\perp \leq b^\perp$. Now suppose $a\leq b^\perp$. Then $a\ovee c = b^\perp$ for some $c$. Hence, $1=b\ovee b^\perp = b\ovee (a\ovee c) = (b\ovee a)\ovee c$ so that $b$ and $a$ are indeed summable.
    
    \item Follows directly from the definition of $\leq$ and the cancellativity of addition. \qedhere
  \end{enumerate}
\end{proof}

\begin{definition}
  Let $E$ be an effect algebra, and $a,b \in E$ with $a\leq b$. We denote the unique element $c\in E$ such that $a\ovee c = b$ by $b\ominus a$.
\end{definition}

All the results of this section, \eg the definition and properties of the partial order on an effect algebra, and the definition of $b\ominus a$ will be used without further reference throughout this chapter.

It might seem that the axioms of an effect algebra are somewhat arbitrary, and that we might instead want to work with some slightly other definition. The following remark shows that the definition of an effect algebra is in fact quite ``canonical''.

\begin{remark}
  By the previous results an effect algebra is a \Define{bounded poset}\indexd{bounded posets}: a partially ordered set with a minimal and maximal element ($0$ respectively $1$). In Ref.~\cite{mayet1995classes} it was shown that any bounded poset $P$ can be embedded into a orthomodular poset $K(P)$. This is known as the \Define{Kalmbach extension}\indexd{Kalmbach extension}, named after the results of Ref.~\cite{kalmbach1977orthomodular}. It was then later shown~\cite{harding2004remarks} that this extends to a functor from the category of bounded posets to the category of orthomodular posets, and that this functor is in fact left adjoint to the forgetful functor going in the opposite direction. This adjunction gives rise to the \Define{Kalmbach monad} on the category of bounded posets. The Eilenberg-Moore category for the Kalmbach monad is isomorphic to the category of effect algebras, and hence effect algebras form algebras in the category of bounded posets.
\end{remark}

\section{Effect theories}

In Section~\ref{sec:GPTs} we introduced a general framework for describing physical theories based on real probabilities. In this section we introduce a generalisation that we call \emph{effect theories}.

We will assume that we have some kind of special system $I$ that denotes the `empty' or `trivial' system. A transformation $\omega:I\rightarrow A$ is then a procedure that creates a system from nothing, i.e.\ it is a preparation. We will call such transformations \Define{states}\indexd{state!in effect theory}. They represent the different ways in which a system $A$ can be prepared. We define the \Define{state space} of $A$ to be St$(A):=\{\omega: I\rightarrow A\}$\index{math}{St$(A)$ (state space)}.

Dually, the transformations $a:A\rightarrow I$ are the ways in which the system $A$ can be destroyed. We call these transformations \Define{effects}\indexd{effect!in effect theory} and they model the different types of measurements you can perform on a system $A$. We define the \Define{effect space} of $A$ to be $\eff(A):=\{a:A\rightarrow I\}$.

The composition of a state with an effect results in a transformation $a\circ\omega: I\rightarrow I$. Such a transformation from the unit system to itself we will call a \Define{scalar}\indexd{scalar!in effect theory}. 

The framework of Section~\ref{sec:GPTs} is a special case of this. In that setting, the states and effects have a convex structure, and the scalars are real numbers from the unit interval. The scalar $a\circ \omega$ is then the probability that a measurement $a$ returns true on a state $\omega$. 
Instead of real numbers and convex structures, our assumption is going to be that the effect space of a system forms an effect algebra.

\begin{definition}
  An \Define{effect theory}\indexd{effect theory} is a category $\mathbf{E}$ with a designated object $I$ that we call the \Define{trivial system}\indexd{trivial system (effect theory)} such that for all objects $A$, its \Define{effect space}\indexd{effect space (effect theory)} $\eff(A):=\{a\colon A\rightarrow I\}$\index{math}{Eff(A)@$\eff(A)$ (effect space)} has a special element $0$, a complement operation $(\cdot)^\perp$ and a binary operation $\ovee$ making it into an effect algebra, such that every map $g:B\rightarrow A$ preserves its addition: $0\circ g = 0$ and $(a\ovee b)\circ g = (a\circ g)\ovee(b\circ g)$ whenever $a\ovee b: A\rightarrow I$ is defined as an effect.
\end{definition}

As the effects of an object form an effect algebra, we have for each object a special effect $1$. This corresponds to the `truth' effect that always holds on any state.

\begin{remark}\label{rem:effect-vs-effectus}
  Our definition of an effect theory is related to that of an \emph{effectus in partial form}~\cite{cho2015introduction}. In fact, any such effectus is an effect theory. While an effect theory only defines addition of some effects (using the effect algebra structure), an effectus furthermore defines the sum of some morphisms, as all the homsets are required to be partial commutative monoids (cf.~Definition~\ref{def:partialcommutativemonoid}). In addition, an effectus always has coproducts, allowing the definition of a classical disjunction of physical systems.
  All the examples we will see of effect theories in this thesis are also effectuses, and essentially all results of this thesis could be reframed for effectuses.
  The only reason we work with effect theories in this thesis instead of effectuses is because they are easier to define and sufficient for our purposes.
\end{remark}

We will also need to be able to describe composite systems, i.e.\ given independent systems $A$ and $B$, it should be possible to describe them as forming one bigger system $A\otimes B$. The structure we wish to use for this is that of a monoidal category.

\begin{definition}\label{def:monoidal-effect-theory}
    We call an effect theory $(C,I)$ \Define{monoidal}\indexd{monoidal effect theory}\indexd{effect theory!monoidal structure} when $C$ is a monoidal category such that the trivial system $I$ is also the monoidal unit and such that the tensor product preserves the effect algebra structure: $1\otimes 1 = 1$ and for all effects $a$, $0\otimes a = 0$ and whenever $b_1\ovee b_2$ is defined, $(b_1\otimes a)\ovee(b_2\otimes a)$ is also defined and equal to $(b_1\ovee b_2)\otimes a$, and similarly $(a\otimes b_1)\ovee (a\otimes b_2) = a\otimes (b_1\ovee b_2)$.
\end{definition}

In order to highlight our interpretation of an effect theory as an abstract description of a physical theory we will often call a morphism in an effect theory a \emph{transformation}
We end this section with a few observations regarding transformations in an effect theory. Note that we take composition to bind stronger than addition, so that $a_1\circ f \ovee b$ should be interpreted as $(a_1\circ f)\ovee b$.

\begin{proposition}\label{prop:effect-theory-basic} The following are true in any effect theory.
    \begin{enumerate}
        \item Let $a,b \in \eff(A)$ be effects such that $a\leq b$, and let $f:B\rightarrow A$ be some transformation. Then $a\circ f \leq b\circ f$.
        \item Let $\Theta: A\rightarrow B$ be an isomorphism, i.e.\ a transformation that has a 2-sided inverse $\Theta^{-1}:B\rightarrow A$. Then $1\circ \Theta = 1$.\indexd{isomorphism!in effect theory}
    \end{enumerate}
\end{proposition}
\begin{proof}~
    \begin{enumerate}
    \item By definition, $a\leq b$ if and only if there is some $c\in \eff(A)$ such that $a\ovee c = b$. Maps in an effect theory preserve addition and hence $b\circ f = (a\ovee c)\circ f = a\circ f \ovee c\circ f$ so that indeed $a\circ f \leq b\circ f$.
    \item Let $1\circ \Theta = a$. We need to show that $a^\perp = 0$. Of course $a\circ \Theta^{-1} = 1$ and hence $1\circ \Theta^{-1} = (a\ovee a^\perp)\circ \Theta^{-1} = a\circ \Theta^{-1} \ovee a^\perp \circ \Theta^{-1} = 1 \ovee a^\perp \circ \Theta^{-1}$. Since the only effect summable with $1$ is $0$, we get $a^\perp\circ \Theta^{-1} = 0$. As a result $0 = 0\circ \Theta = a^\perp\circ\Theta^{-1}\circ \Theta = a^\perp$ and we are done. \qedhere
    \end{enumerate}
\end{proof}

\section{Operational effect theories}

Before we continue with abstract effect theories, we first relate them to the more familiar GPT framework.
To do that we need to consider the category of order unit spaces. Recall that a linear map $f:V\rightarrow W$ between ordered vector spaces is positive when $f(v)\geq 0$ whenever $v\geq 0$, or equivalently when it is monotone: $v\leq w\implies f(v)\leq f(w)$.

\begin{definition}
    We call a positive linear map $f:V\rightarrow W$ between order unit spaces \Define{sub-unital}\indexd{sub-unital map!between order unit spaces} when $f(1)\leq 1$. We denote the category of order unit spaces with positive sub-unital maps by \OUS\index{math}{OUS@\OUS (category of order unit spaces)}.
\end{definition}

The category $\OUS^\opp$ is an effect theory. The states of this category correspond to positive sub-unital maps $\omega: V\rightarrow \R$ in \OUS, while the effects are positive sub-unital maps $\hat{a}:\R \rightarrow V$. These effects are completely determined by $\hat{a}(1) =: a \in [0,1]_V$ and hence the effects correspond to $a\in [0,1]_V = \{v\in V~;~0\leq v \leq 1\}$.

\begin{remark}\indexd{Schrodinger picture vs Heisenberg picture}
	The reason that $\OUS^\opp$ is an effect theory instead of $\OUS$, is because we framed the language of effect theories in the Schr\"odinger picture where states take a central role, while order unit spaces describe a space of effects and hence are framed in the Heisenberg picture. The `Schr\"odinger counterpart' to order unit spaces are called \emph{base norm spaces}~\cite{alfsen2012geometry}. These will not play a significant role in this thesis.
	The switching of the direction of morphisms when viewing it from the perspective of the effects is a theme that will repeat itself multiple times throughout this chapter. It is an artifact of morphisms in an effect theory being naturally interpreted as state transformers, while we wish to deal with them as effect/predicate transformers.
\end{remark}

The category \textbf{OUS}$^\opp$ is important, because it is in a sense the most general example of what we call an \emph{operational effect theory}. Such categories roughly correspond to GPTs. Before we give the definition of an operational effect theory, we must introduce a few other concepts.
\begin{definition}\label{def:tomographies}
    We say an effect theory satisfies
    \begin{itemize}
        \item \Define{order-separation by unital states}\indexd{order separation!in effect theories} when for every system $A$ and effects $a,b\in \eff(A)$ we have $a\leq b$ whenever $ a\circ \omega \leq b\circ \omega $ for all $\omega\in $ St$(A)$ with $1\circ\omega = 1$;
        \item \Define{local tomography}\indexd{local tomography!in effect theory}\indexd{tomography (effect theory)!local tomography} when the effects separate the transformations, \ie~if for all $f,g: B\rightarrow A$ we have $f=g$ whenever $a\circ f = a\circ g$ for all $a\in \eff(A)$;
        \item \Define{tomography}\indexd{tomography (effect theory)} when it is monoidal and the effects monoidally separate the transformations, \ie~if for all $f,g: B\rightarrow A$ we have $f=g$ whenever $a\circ (f\otimes \id_C) = a\circ (g\otimes \id_C)$ for all systems $C$ and $a\in \eff(A\otimes C)$.
    \end{itemize}
\end{definition}

\noindent Local tomography tells us that transformations are completely determined by what they do on effects.%
\footnote{What we call `local tomography' here might also be called `process tomography'. It is in general not equivalent to the notion of local tomography we adopted in the previous chapter. However, in the presence of composites which allow for maximally entangled states it is equivalent to local tomography via an argument involving process-state duality (also known as the Choi-Jamio\l{}kowski isomorphism)~\cite[Theorem~7.41]{CKbook}.}
When monoidal structure is present in the theory, there are more ways to let a transformation interact with an effect, and hence we weaken the definition to (non-local) tomography.
The states order-separating the effects essentially tells us that there can be no `infinitesimal' effects. These properties allow us to relate effect theories to order unit spaces.

\begin{definition}\label{def:convexeffectalgebra}
    Let $E$ be an effect algebra. We say $E$ is \Define{convex} if
    there exists an action $\cdot: [0,1]\times E\rightarrow E$,
    where $[0,1]$ is the standard real unit interval, satisfying
    the following axioms for all $a,b\in E$ and~$\lambda, \mu \in
    [0,1]$:
    \begin{itemize}
        \item $\lambda\cdot (\mu\cdot a) = (\lambda\mu)\cdot a$.
    \item If $\lambda+\mu \leq 1$, then $\lambda \cdot a \perp
    \mu \cdot a$ and $\lambda \cdot a \ovee \mu \cdot a =
    (\lambda+\mu)\cdot a$.
        \item If $a\perp b$, then $\lambda\cdot a \perp \lambda \cdot b$ and $\lambda\cdot(a\ovee b) = \lambda\cdot a \ovee \lambda \cdot b$.
        \item $1\cdot a = a$.
    \end{itemize}
\end{definition}

When the scalars $\eff(I)$ of an effect theory are the real numbers (with the regular addition, and where composition corresponds to multiplication of real numbers) all the effect spaces $\eff(A)$ are convex effect algebras, because for every effect $a:A\rightarrow I$ and scalar $s:I\rightarrow I$ we get a new effect $s\circ a$. This action of the real unit interval of $\eff(A)$ is straightforwardly verified to satisfy the conditions required of a convex effect algebra.

\begin{proposition}[{\cite{gudder1998representation}}]\label{prop:convextotalisation}
    Let $E$ be a convex effect algebra. Then there exists an ordered vector space $V$ with (possibly non-Archimedean) order unit $1$ such that $E\cong [0,1]_V$. 
\end{proposition}
\begin{proof}
  We take the set $T(E)$ of formal linear combinations of elements in $E$ modulo the evident equalities in $E$ analogous to Definition~\ref{def:assocvectorspaceofeffects}. It is then straightforward, but tedious to show that $T(E)$ has the desired properties. See~\cite{gudder1998representation} for the details.
\end{proof}

\begin{proposition}\label{prop:effectsformorderunitspace}
    Let $\mathbf{E}$ be an effect theory where the scalars are the real unit interval, and where the unital states order-separate the effects. Then for all systems $A$ of $\mathbf{E}$, there exists an order unit space $V_A$, such that $\eff(A) \cong [0,1]_{V_A}$.
\end{proposition}
\begin{proof}
  As every effect space $\eff(A)$ is a convex effect algebra, we can associate an ordered vector space $V_A$ with order unit $1$ such that $\eff(A) \cong [0,1]_{V_A}$. This space is ordered by the set of states of $A$, and by assumption the states order-separate $[0,1]_V = E$, so that by Proposition~\ref{prop:OUSequivalentdefinitions}, $V$ is an order unit space.
\end{proof}

The order unit space $V_A$ associated to $A$ has its effects associated one-to-one with those of $A$, but (like we observed in Remark~\ref{remark:norestrictionhypothesis}) this is not necessarily true for the states. Any state $\omega:I\rightarrow A$ can be mapped to a state $\omega^*:V_A\rightarrow \R$ (see Theorem~\ref{theor:opefftheor} below), but not every state on $V_A$ necessarily comes from some state in the effect theory.

\begin{definition}
    Let $A$ be a system in an effect theory with real scalars where the unital states order-separate the effects and let $V_A$ be the OUS such that $[0,1]_{V_A} \cong \eff(A)$.
    \begin{itemize}
    \item We say $A$ is \Define{finite-dimensional}\indexd{finite-dimensional (effect theory)} when $V_A$ is.
    \item We say $A$ is \Define{state-closed}\indexd{state-closed system} when the collection of unital states St$_1(A)$ is closed as a subset of $V_A^*$ (with respect to the topology induced by the norm of $V_A$). 
    \item We call $A$ \Define{scalar-like}\indexd{scalar-like system} when $\eff(A) \cong [0,1]$.
    \end{itemize}
\end{definition}

\begin{definition}\label{def:OET}
    A (monoidal) \Define{operational effect theory} (OET)
    \indexd{operational effect theory} 
    \indexd{effect theory!operational ---}
    \index{math}{OET (operational effect theory)} is a (monoidal) effect theory satisfying the following additional properties.
    \begin{enumerate}
        \item The set of scalars is the real unit interval: $\eff(I)=[0,1]$.
        \item The unital states order-separate the effects.
        \item All systems are finite-dimensional.
        \item All systems are state-closed.
        \item Every scalar-like system is isomorphic to the trivial system $I$.
    \end{enumerate}
\end{definition}
Note that we do not require an OET to satisfy (local) tomography. As a result, this last condition is needed to prevent the case where we have multiple copies of the trivial system that are taken to be non-isomorphic. This property can hence be seen as a restricted form of tomography.

\begin{remark}
  The definition of an OET brings us very close to the GPT framework we adopted in Section~\ref{sec:gpt-framework}.
  In both cases the scalars are of course the real numbers. 
  As in Assumption~\ref{assum:effectspaceclosedunderaddition} we require order-separation of the effects which is related to the operational equivalence assumption on GPTs~\cite{chiribella2011informational}. As the effect spaces are effect algebras, we have a causal effect $1$ and furthermore, we have for every effect $a$ a negation $a^\perp$. As in the previous chapter, we are assuming finite-dimensionality of the systems. The assumption of the closure of the state space was not needed in the previous chapter, but is still a standard background assumption in the literature on GPTs~\cite{chiribella2011informational,barnum2014higher}.
\end{remark}

If an OET satisfies local tomography, then all the information in the theory is captured by the structure of the effects:

\begin{theorem}\label{theor:opefftheor}
    Let $\mathbb{E}$ be an operational effect theory. Then there is a functor $F: \mathbb{E}\rightarrow \OUS^\opp$ such that $\eff(A)\cong [0,1]_{F(A)}$. Furthermore, this functor is faithful if and only if $\mathbb{E}$ satisfies local tomography.
\end{theorem}
\begin{proof}
    For a map $f:A\rightarrow B$ we get a map of effect algebras $f^*:\eff(B)\rightarrow \eff(A)$ via $f^*(b) := b\circ f$. By definition of an effect theory, $f^*$ preserves the addition. Furthermore, for any scalar $s:I\rightarrow I$, we see that $f^*(s\circ b) = (s\circ b)\circ f = s\circ (b\circ f) = s\circ f^*(b)$, so that $f^*$ also preserves the convex action induced by the scalar composition.

    For every system $A$ we have the order unit space $V_A$ associated to $\eff(A)$ (see Proposition~\ref{prop:effectsformorderunitspace}). Now, every element $v\in V_A$ can be written as ${v=\lambda a - \mu b}$ where $\lambda, \mu \in \R^+$ and $a,b\in [0,1]_{V_A}$, and hence a linear map on $V_A$ is determined by what it does on the unit interval (and similarly for $V_B$). It is then straightforward to check that $f^*:\eff(B)\rightarrow \eff(A)$ extends to a positive linear sub-unital map $f^*:V_B\rightarrow V_A$.

    Hence, we can define the functor as:
    $$F(A) := V_A \quad \text{and}\quad F(f:A\rightarrow B) := (f^*:V_B\rightarrow V_A).$$
    For the details we refer to Ref.~\cite{jacobs2012coreflections}.
    Unfolding the definitions we see that the faithfulness of this functor exactly corresponds to $\mathbf{E}$ satisfying local tomography.
\end{proof}

\section{Filters and compressions}\label{sec:filterscompressions}

An important class of maps present in some effect theories are \emph{filters} and \emph{compressions}. These are maps satisfying certain universal properties that can intuitively be thought of as post-selecting on an effect, respectively embedding a subsystem into a larger system.

Let us consider a system $A$ in an effect theory and let $B$ be a `subsystem' of $A$, i.e.\ some system with an embedding map $\pi: B\rightarrow A$. 
If we have some state $\omega:I\rightarrow B$ on this subsystem, then we can view it as a state on the whole system by `forgetting' it was actually defined on the subsystem: $\pi\circ \omega: I\rightarrow A$. Now let $a:A\rightarrow I$ be the effect that `witnesses' whether a state came from the subsystem $B$. That is, it is the smallest effect such that $a\circ \pi\circ\omega = 1$ for all states $\omega:I\rightarrow B$. This is the case when $a\circ \pi = 1\circ \pi$. Conversely, we can ask for a given effect $a$ on $A$ which subsystem it witnesses. I.e.~what is the subsystem of $A$ where $a$ is true? We will call the map that maps the subsystem onto $A$ a \emph{compression} for $a$ as it `compresses' the system onto the part where $a$ is true.

\begin{definition}\label{def:compression}
    Let $a: A\rightarrow I$ be an effect in an effect theory. We call a system $A_a$ a \Define{compression system}\indexd{compression system} for $a$ when there is a map $\pi_a:A_a\rightarrow A$ such that $1\circ \pi_a = a\circ \pi_a$ that is \Define{final} with this property: whenever $f:B\rightarrow A$ is such that $1\circ f = a\circ f$ then there is a unique $\cl{f}:B\rightarrow A_a$ such that the following diagram commutes:
    \[\begin{tikzcd}[ampersand replacement = \&]
    A_a \arrow{r}{\pi_a}\& A \\
    B \arrow[dotted]{u}{\overline{f}}\arrow{ru}[swap]{f}\&
    \end{tikzcd}\] 
    The map $\pi_a$ is called a \Define{compression}\indexd{compression} for $a$, and as it satisfies a universal property is unique up to unique isomorphism\footnote{The term `compression' should not be confused with the maps of the same name in Ref.~\cite{alfsen2012geometry}. The compressions from that paper correspond to what we will later call `assert maps'. What we call a compression is sometimes also known as an `encoding' or `embedding'~\cite{chiribella2011informational}.}. We say an effect theory \Define{has compressions} when every effect has a compression.
\end{definition}

The universal property of compressions tells us that the system $A_a$ is the `largest' system such that there is a map $\pi:A_a\rightarrow A$ with $1\circ\pi = a\circ \pi$, and hence this is the system that we should see as the subsystem that $a$ witnesses.
As the map goes from the smaller system to the larger system it might not be apparent why we call these maps `compressions'. However, from the viewpoint of the effect spaces it is a map $\pi^*:A\rightarrow A_a$ that does compress the effects into the smaller space.

\begin{example}\label{ex:compression-quantum}
  Let $B(H)$ be the set of bounded operators on a (complex) Hilbert space and let $A\in B(H)$ be an effect. Denote by $P$ the largest projection (idempotent effect) below $A$, \ie~$P$ projects to the eigenspace of $A$ of eigenvalue $1$. Denote this space by $K\sse H$. Then a compression of $A$ (in the opposite category of C$^*$-algebras with positive sub-unital maps) is the map $\pi_A:B(H)\rightarrow B(K)$ given by $\pi_A(B) = PBP$. 
\end{example}


While a compression can be seen as a map that `forgets' that a state came from a subsystem, a filter is the opposite, describing how a state can be `filtered' to fit inside a subsystem.

\begin{definition}\label{def:filter}
    Let $a:A\rightarrow I$ be an effect in an effect theory. A \Define{filter}\indexd{filter} for $a$ is a map $\xi_a:A\rightarrow A^a$ such that $1\circ\xi_a\leq a$ which is \Define{initial} for this property: for any map $f:A\rightarrow B$ which satisfies $1\circ f\leq a$ there is a unique $\cl{f}:A^a\rightarrow B$ such that the following diagram commutes:
    \[\begin{tikzcd}[ampersand replacement = \&]
    A^a \arrow[dotted,swap]{d}{\cl{f}}\&\arrow[swap]{l}{\xi_a} A\arrow{dl}{f} \\
     B\&
    \end{tikzcd}\]
    We say an effect theory \Define{has filters} when every effect has a filter.
\end{definition}

\noindent The interpretation of the filter $\xi_a$ is that it represents a non-destructive measurement of $a$ which got a positive outcome, i.e.\ it is a post-selection for $a$. The space $A_a$ is the subsystem where $a$ has a nonzero probability of being true.

\begin{example}\label{ex:filter-quantum}
  Again, let $B(H)$ be the set of bounded operators on a (complex) Hilbert space and let $A\in B(H)$ be an effect. Let $K\sse H$ be $K=(\ker A)^\perp$, \ie~the closure of the eigenspaces of $A$ of non-zero eigenvalue. Then a filter for $A$ (in the opposite category of C$^*$-algebras with positive sub-unital maps) is the map $\xi_A: B(K)\rightarrow B(H)$ given by $\xi_A(B) = \sqrt{A}B\sqrt{A}$.

\end{example}

\begin{remark}
    In effectus theory, compressions are called \emph{comprehensions} and filters are called \emph{quotients}. They arise in a wide variety of settings~\cite{cho2015quotient}. Quotients and comprehensions have an interesting categorical significance relating to fibered category theory. We refer to Ref.~\cite{kentathesis} for a full account of this relation.
\end{remark}

\subsection{Sharpness}

A compression $\pi_a: A_a\rightarrow A$ tells us that $A_a$ is a subsystem of $A$. We also know that $1\circ\pi_a = a\circ\pi_a$, but $a$ might not be the smallest effect with this property. This is because an effect can be \emph{fuzzy}, meaning that it does not make a sharp distinction between where it holds true, and where it does not. In contrast, if $a$ \emph{is} the smallest effect with this property, then we call it \emph{sharp}, since it sharply delineates its subspace.

\begin{definition}
    Let $f:A\rightarrow B$ be a transformation in an effect theory. The \Define{image}\indexd{image (effect theory)} of $f$, when it exists, is the smallest effect $a:B\rightarrow I$ such that $a\circ f = 1\circ f$, i.e.\ if $b:B\rightarrow I$ is also such that $b\circ f = 1\circ f$, then $a\leq b$. We denote the image of $f$ by $\im{f}$. We say an effect theory \Define{has images} when all the transformations have an image.
\end{definition}

By the discussion above, we would call an effect sharp when $\im{\pi_a}=a$. Instead we will use the slightly more general formulation used in Ref.~\cite{basthesis} that turns out to be equivalent (cf.~Proposition~\ref{prop:floorceiling}).

\begin{definition}\label{def:effectus-sharp}
    Let $a:A\rightarrow I$ be an effect. We call $a$ \Define{sharp}\indexd{sharp effect!in effect theory} when there is some transformation $f:B\rightarrow A$ such that $\im{f} = a$.
\end{definition}

\noindent In this chapter we will denote arbitrary effects by $a,b,c$ and sharp effects by $p,q,r$. We defined sharp effects in Chapter~\ref{chap:seqprod} to be those effects $p$ which satisfied $p\wedge p^\perp = 0$. In Proposition~\ref{prop:sharpadd} we will see that in certain well-behaved effect theories sharp effects have this property, that we call `ortho-sharpness', as well. In general effect theories, it is unclear whether the converse direction holds. However, in quantum theory (or more generally, in Euclidean Jordan algebras), the two notions of sharpness coincide, and hence our reconstruction establishes their equivalence.

\subsection{Properties of filters and compressions}

Let us now study some basic consequences of the existence of filters, compressions and images. These results can be found in Refs.~\cite{cho2015introduction,basthesis}, but for completeness we will give their proofs.

First, the following results are immediate from the universal properties of filters and compressions.

\begin{proposition} \label{prop:quotcompr}
  Let $a:A\rightarrow I$ be an effect and let $\pi_a:A_a\rightarrow A$ be a compression for $a$ and $\xi_a:A\rightarrow A^a$ a filter. 
  \begin{itemize}
    \item Let $\Theta:B\rightarrow A_a$ be an isomorphism. Then $\pi\circ \Theta$ is also a compression for $a$. Conversely, for any two compressions $\pi$ and $\pi^\prime$ for $a$ there is an isomorphism $\Theta$ such that $\pi^\prime = \pi\circ \Theta$.
    \item Let $\Theta:A_a\rightarrow B$ be an isomorphism. Then $\Theta\circ \xi$ is also a filter for $a$. Conversely, for any two filters $\xi$ and $\xi^\prime$ for $a$ there exists an isomorphism $\Theta$ such that $\xi^\prime = \Theta\circ \xi$.
  \end{itemize}
\end{proposition}

\begin{proposition}
  Isomorphisms are filters and compressions for the truth effect $1$.
\end{proposition}
\begin{proof}
  The identity is a filter and compression for the $1$ effect. Hence by the previous proposition the desired result follows.
\end{proof}

In the previous section we described filters as a post-selection of an effect. By the propositions above we see that it is actually more accurate to describe them as a measurement post-selection (the filter) followed by some post-processing in the form of a reversible transformation (the isomorphism). Similarly, a compression allows some preprocessing in the form of an isomorphism.

\begin{lemma}\label{lem:imageofcomposedmaps}
  Let $f$ and $g$ be composable maps and suppose $\im{f\circ g}$ and $\im{f}$ exist. Then $\im{f\circ g} \leq \im{f}$. Furthermore, if $g$ is an isomorphism, then $\im{f\circ g} = \im{f}$.
\end{lemma}
\begin{proof}
  We of course have $1\circ (f\circ g) = (1\circ f)\circ g = (\im{f}\circ f)\circ g = \im{f}\circ (f\circ g)$, and hence $\im{f}\leq \im{f\circ g}$.

  If $g$ is an isomorphism, then we furthermore have $\im{f} = \im{(f\circ g)\circ g^{-1}} \leq \im{f\circ g} \leq \im{f}$, and hence $\im{f} = \im{f\circ g}$.
\end{proof}

This lemma makes the following well-defined, as any two compressions for the same effect are related by an isomorphism.

\begin{definition}
Let $a$ be an effect and let $\pi$ be a compression for $a$. The \Define{floor}\indexd{floor!in effect theory} of $a$ is defined as $\floor{a}:= \im{\pi}$. The \Define{ceiling}\indexd{ceiling!in effect theory} is defined as the De Morgan dual: $\ceil{a} = \floor{a^\perp}^\perp$.
\end{definition}

The ceiling and floor are related to the ceiling and floor of Chapter~\ref{chap:seqprod}, but are a priori not equal. This is due to the notion of sharpness in an effectus not being equal to that of sharpness as defined in the previous chapter. For the particular effect theories we are interested in however (cf.~section~\ref{sec:PET}) they will coincide.

Before we prove some properties of floors and ceilings in the next proposition, let us remark on a useful computational tool involving images. The image $\im{f}$ of a transformation satisfies $\im{f}\circ f = 1\circ f$. Hence $1\circ f = (\im{f}\ovee \im{f}^\perp)\circ f = \im{f}\circ f \ovee \im{f}^\perp\circ f = 1\circ f \ovee \im{f}^\perp \circ f$. Cancelling $1\circ f$ on both sides gives $\im{f}^\perp \circ f = 0$. In fact, $\im{f}^\perp$ is the largest effect for which $f$ is zero: if $a\circ f = 0$, then $a\leq \im{f}^\perp$. We will use this alternative characterisation of the image without further reference.  Note that as $a\circ \pi_a = 1\circ \pi_a$ we have for the same reasons $a^\perp\circ \pi_a = 0$.

\begin{proposition}[{\cite[203IV]{basthesis}}]\label{prop:floorceiling}
  In an effect theory with images and compressions, the following are true for any effects $a\leq b$ and composable map $f$.
  \begin{multicols}{2}
  \begin{enumerate}[label=\alph*)]
    \item $\floor{a}\leq a \leq \ceil{a}$.
    \item $\floor{\floor{a}}=\floor{a}$.
    \item $\floor{b}\leq \floor{a}$ and $\ceil{b}\leq \ceil{a}$.
    \item $\ceil{a\circ f} = \ceil{\ceil{a}\circ f}$.
    \item $\ceil{a}\circ f = 0 \iff a\circ f = 0$.
    \item $a$ is sharp if and only if $\floor{a}=a$.
  \end{enumerate}
  \end{multicols}
\end{proposition}
\begin{proof}
  Let $a, b: A\rightarrow I$ be effects, and let $\pi_a: A_a \rightarrow A$ be a compression for $a$, and $\pi_b:A_b \rightarrow A$ a compression for $b$.
  \begin{enumerate}[label=\alph*)]
    \item Of course $1\circ \pi_a = a\circ \pi_a$, and hence $\floor{a} := \im{\pi_a} \leq a$. Hence also $\floor{a^\perp} \leq a^\perp$, and thus $\ceil{a} := \floor{a^\perp}^\perp \geq (a^\perp)^\perp = a$.
    
    \item We will show that $\pi_a$ is a compression for $\floor{a}$, and hence $\pi_{\floor{a}} = \pi_a \circ\Theta$ for some isomorphism $\Theta$. The result then follows using Lemma~\ref{lem:imageofcomposedmaps}, because $\floor{\floor{a}} := \im{\pi_{\floor{a}}} = \im{\pi_a\circ \Theta} = \im{\pi_a} = \floor{a}$.

    Note first that $\floor{a}\circ \pi_a = \im{\pi_a}\circ \pi_a = 1\circ \pi_a$. Now let $f:B\rightarrow A$ be some map with $\floor{a}\circ f = 1\circ f$. As $\floor{a}\leq a$, we then also have $a\circ f = 1\circ f$, and hence by the universal property of $\pi_a$ there is a unique $\cl{f}$ with $f=\pi_a \circ \cl{f}$. Hence, $\pi_a$ is also a compression for $\floor{a}$.

    \item We have $1\circ \pi_b = b\circ \pi_b \leq a\circ \pi_b \leq 1\circ \pi_b$, and hence $a\circ \pi_b = 1\circ \pi_b$. Hence $\pi_b = \pi_a \circ \cl{\pi_b}$ for a unique $\cl{\pi_b}$. With Lemma~\ref{lem:imageofcomposedmaps} we calculate $\floor{b} := \im{\pi_b} = \im{\pi_a\circ\cl{\pi_b}} \leq \im{\pi_a} = \floor{a}$. To show $\ceil{b}\leq \ceil{a}$, we note that as $b\leq a$, we have $a^\perp \leq b^\perp$ and hence $\floor{a^\perp}\leq \floor{b^\perp}$. Then: $\ceil{b}:=\floor{b^\perp}^\perp \leq \floor{a^\perp}^\perp = \ceil{a}$.

    \item First note that since $\ceil{a}\circ f \geq a\circ f$, we have by point c): $\ceil{\ceil{a}\circ f} \geq \ceil{a\circ f}$. It remains to show the inequality in the other direction.

    Because $a\circ (f\circ \pi_{(a\circ f)^\perp}) = 0$, there must be an $h$ with $f\circ \pi_{(a\circ f)^\perp} = \pi_{a^\perp}\circ h$. By point b) there must be some isomorphism $\Theta$ such that $\pi_{a^\perp} = \pi_{\floor{a^\perp}} \circ \Theta = \pi_{\ceil{a}^\perp} \circ \Theta$. We then calculate:
    $$\ceil{a}\circ f\circ \pi_{(a\circ f)^\perp} = \ceil{a}\circ \pi_{a^\perp}\circ h = \ceil{a} \circ \pi_{\ceil{a}^\perp} \circ \Theta\circ h = 0.$$
    Hence $\ceil{a}\circ f \leq \im{\pi_{(a\circ f)^\perp}}^\perp = \floor{(a\circ f)^\perp}^\perp = \ceil{a\circ f}$. Using points c) and b): $\ceil{\ceil{a}\circ f}\leq \ceil{\ceil{a\circ f}} = \ceil{a\circ f}$. 

    \item Of course if $\ceil{a}\circ f = 0$, then $a\circ f \leq \ceil{a}\circ f = 0$. For the other direction, we remark that $\ceil{0} = \floor{1}^\perp = 1^\perp = 0$, so that by the previous point: $0 = \ceil{0} = \ceil{a\circ f} = \ceil{\ceil{a}\circ f}$, and hence $\ceil{a}\circ f \leq \ceil{\ceil{a}\circ f} = 0$.

    \item If $\floor{a} = a$, then $a=\im{\pi_a}$, and hence $a$ is sharp. Now suppose $a$ is sharp, and hence is the image of some map $f$: $a=\im{f}$. Then by the universal property of $\pi_a$, there is some $\cl{f}$ such that $f=\pi_a \circ \cl{f}$. We then calculate using Lemma~\ref{lem:imageofcomposedmaps} $a=\im{f} = \im{\pi_a\circ \cl{f}} \leq \im{\pi_a} = \floor{a}$. As $\floor{a}\leq a$ by point a), we are done. \qedhere
  \end{enumerate}
\end{proof}

\noindent Note that this proposition implies that the floor of $a$ is the largest sharp predicate below $a$ (by combining points c) and f)), so that the name is indeed well-chosen. It is not clear whether $\ceil{a}$ will always be sharp in an effect theory. For this to hold we need the implication `$a$ is sharp iff $a^\perp$ is sharp'. This is a quite reasonable assumption: as discussed, when $p$ is a sharp effect, there is some subsystem where $p$ is `true' and hence it stands to reason that there should also be some subsystem where $p$ is `false', corresponding to the subsystem of $p^\perp$. Interestingly, we do not know of any effect theory containing compressions and images where $\ceil{a}$ is not always sharp. We will assume this implication on the sharpness of complements starting in Section~\ref{sec:diamond-effect-theory}.
First we will prove some more results that require the existence of all filters, compressions and images.

\begin{definition}
  A transformation $f:A\rightarrow B$ in an effect theory is \Define{unital}\indexd{unital map!in an effect theory} when $1\circ f = 1$. It is \Define{faithful}\indexd{faithful transformation} when $a\circ f = 0$ implies $a=0$. This is easily seen to be equivalent to $\im{f}=1$.
\end{definition}

\begin{proposition}[{\cite{cho2015introduction}}]\label{prop:faithfulfilters}
  In an effect theory with filters and compressions the following are true.
  \begin{enumerate}[label=\alph*)]
    \item Let $\xi_a$ be a filter for $a$, then $1\circ \xi_a = a$.
    \item Filters are epic and faithful.
    \item Compressions are monic and unital.
    \item A composition of filters is again a filter.
  \end{enumerate}
\end{proposition}
\begin{proof}~
  \begin{enumerate}[label=\alph*)]
    \item Since $1\circ a = a \leq a$, we can use the universal property of $\xi_a$ to find an $\cl{a}:A_a\rightarrow I$ with $\cl{a}\circ \xi_a = a$. But then $a=\cl{a}\circ\xi_a \leq 1\circ \xi_a \leq a$, and hence necessarily $1\circ \xi_a = a$.
    \item Let $g_1,g_2:A^a\rightarrow B$, such that $f := g_1\circ \xi_a = g_2\circ \xi_a$. We see that $1\circ f = 1\circ g_1\circ \xi_a \leq 1\circ \xi_a = a$ and hence there is a unique $\cl{f}$ such that $f=\cl{f}\circ \xi_a$. But then necessarily $g_1 = \cl{f} = g_2$. So $\xi_a$ is indeed epic.

    Now suppose $b:A_a\rightarrow I$ satisfies $b\circ \xi_a = 0$. Since also $0\circ \xi_a = 0$ and $\xi_a$ is epic we conclude that $b=0$. Hence, $\xi_a$ is indeed faithful.

    \item Showing $\pi_a$ is monic follows along similar lines to the argument in the previous point. It remains to show that $\pi_a$ is unital. 
    By the universal property of $\xi_{1\circ \pi_a}$ there is an $f$ such that $\pi_a = f\circ \xi_{1\circ \pi_a}$. Then $1\circ f\circ \xi_{1\circ \pi_a} = 1\circ \pi_a = a\circ \pi_a = a\circ f\circ \xi_{1\circ \pi_a}$. 
    As filters are epic, we then have $1\circ f = a\circ f$. Hence by the universal property of $\pi_a$ there exists a $\cl{f}$ such that $f = \pi_a\circ \cl{f}$. We then calculate $\pi_a = f\circ \xi_{1\circ \pi_a} = \pi_a\circ\cl{f}\circ \xi_{1\circ\pi_a}$. 
    Since $\pi_a$ we then have $\cl{f}\circ \xi_{1\circ \pi_a} = \id$. Now $1\circ \pi_a = 1\circ \xi_{1\circ \pi_a} \geq (1\circ \cl{f})\circ \xi_{1\circ \pi_a} = 1\circ \id = 1$, so that indeed $1\circ \pi_a = 1$.

    \item Let $\xi_a$ and $\xi_b$ be filters for $a$ respectively $b$. We claim that $\xi_a\circ\xi_b$ is a filter for $a\circ\xi_b$. To this end we let $\xi$ be a filter for $a\circ\xi_b$. Then there is a unique $g$ such that $\xi_a\circ\xi_b = g\circ \xi$, which we need to show is an isomorphism. As $1\circ\xi = a\circ\xi_b \leq 1\circ\xi_b=b$ we have $\xi = h_1\circ\xi_b$ for a unique $h_1$. Because $1\circ h_1\circ \xi_b = 1\circ\xi = a\circ \xi_b$ we have $1\circ h_1 = a$ because $\xi_b$ is epic and hence $h_1 = h_2\circ \xi_a$. Then
    \[g\circ h_2\circ \xi_a\circ \xi_b = g\circ \xi = \xi_a\circ \xi_b\quad \text{and} \quad h_2\circ g \circ \xi = h_2 \circ \xi_a\circ \xi_b = h_1\circ \xi_b = \xi\]
  so that because $\xi_a\circ \xi_b$ and $\xi$ are epic we have $g\circ h_2 = \id$ and $h_2\circ g = \id$.\qedhere
  \end{enumerate} 
\end{proof}

We end this section with a proposition that relates the structure of effects of a filter space to the bigger space.
\begin{proposition}\label{prop:filterisotodownset}
    Let $a:A\rightarrow I$ be an effect and let $\xi_a:A^a\rightarrow A$ be a filter for $a$. The set of effects of $A^a$ is isomorphic as an effect algebra to the \Define{downset}\indexd{downset} of $a$: $\eff(A^a)\cong \{b \in \eff(A)~;~ b\leq a\}$.
\end{proposition}
\begin{proof}
  Let $E= \{b\in \eff(A)~;~b\leq a\}$ be the downset of $a$ in $\eff(A)$. Note that for an effect $c:A^a\rightarrow I$ the map $c\circ \xi_a: A\rightarrow I$ is an effect in $A$. Moreover $c\circ \xi_a \leq 1\circ \xi_a = a$, so that $c\circ \xi_a\in E$. We then let the map $f: \eff(A^a) \rightarrow E$ be $f(c) := c\circ \xi_a$. This map is of course additive.

  For any $b\in E$ we have $1\circ b = b \leq a$, and hence by the universal property of $\xi_a$ there is a unique $\cl{b}:A^a\rightarrow I$ with $\cl{b}\circ \xi_a = b$. We define the map $g: E\rightarrow \eff(A^a)$ by $g(b) = \cl{b}$. Using the uniqueness of the $\cl{b}$ we can show that this map is also additive.

  We of course have for any $b\in E$: $g(f(b)) = \cl{b}\circ \xi_b = b$, so that $g\circ f = \id$. For any $c\in \eff(A^a)$ we have $f(g(c)) = \cl{c\circ \xi_a}$. This is the unique effect such that $\cl{c\circ \xi_a}\circ \xi_a = c\circ \xi_a$. But as $c$ also has this property, we must have $\cl{c\circ \xi_a} = c$. As a result also $f\circ g = \id$ and hence $g=f^{-1}$, and we are done.
\end{proof}


\subsection{\texorpdfstring{$\diamond$}{Diamond}-effect theories}\label{sec:diamond-effect-theory}

We will now study some further consequences of the existence and interplay of filters, compressions and images. While we will not need most of these results in this chapter, they will help to simplify some arguments in Chapter~\ref{chap:jordanalg}.

\begin{definition}[{\cite[Section 3.5]{basthesis}}]\label{def:diamond-effect-theory}
  We call an effect theory a \Define{$\diamond$-effect theory}\indexd{diamond-effect-theory@$\diamond$-effect-theory} when it has images, filters and compressions and an effect $p$ is sharp iff $p^\perp$ is sharp.
\end{definition}

The diamond $\diamond$ refers to the `possibilistic' structure that is present in such effect theories:\index{math}{diamond@$\diamond$ (effect theory)}
\begin{definition}
  Let $A$ be a system in a $\diamond$-effect theory. Let SEff$(A)$ be its poset of sharp effects. For any transformation $f:A\rightarrow B$ we define 
  \[f^\diamond:\text{SEff}(B)\rightarrow \text{SEff}(A) \qquad \text{and} \qquad f_\diamond:\text{SEff}(A)\rightarrow \text{SEff}(B)\]
  by $f^\diamond(p) := \ceil{p \circ f}$ for $p\in \text{SEff}(B)$ and $f_\diamond(p) := \im{f\circ\pi_p}$ for $p\in \text{SEff}(A)$ where $\pi_p$ is any compression for $p$.
\end{definition}

Note that in a $\diamond$-effect theory, for any effect $a$, $\ceil{a} := \floor{a^\perp}^\perp$ is sharp by assumption, and hence the maps $f^\diamond$ are indeed maps between the sharp effects. To prove some results regarding $f^\diamond$ and $f_\diamond$ it will be useful to introduce a third such map: $f^{\ssquare}(p) := (f^\diamond(p^\perp))^\perp$.

\begin{definition}\label{def:galois-connection}
  Let $P$ and $Q$ be posets with functions $f:P\rightarrow Q$ and $g:Q\rightarrow P$. When $f(a)\leq b \iff a\leq b$ we say $f$ and $g$ form a \Define{Galois connection}\indexd{Galois connection} and we say $f$ is \Define{left Galois adjoint} to $g$.
\end{definition}

\begin{proposition}[{\cite[Proposition~207]{basthesis}}]\label{prop:galois-properties}
  Let $f:A\rightarrow B$ and $g:B\rightarrow C$ be transformations in a $\diamond$-effect theory and let $p\in \text{SEff}(B)$ and $q\in \text{SEff}(A)$. Then the following are true.
  \begin{multicols}{2}
  \begin{enumerate}[label=\alph*)]
    \item $f^\diamond$ and $f^{\ssquare}$ are monotone.
    \item $f^\diamond(p)\leq q^\perp \iff f_\diamond(q)\leq p^\perp$.
    \item $f_\diamond$ and $f^{\ssquare}$ form a Galois connection.
    \item $f_\diamond$ is monotone.
    \item $f_\diamond\circ f^{\ssquare}\circ f_\diamond = f_\diamond$.
    \item $(\id)^\diamond = (\id)_\diamond = (\id)^{\ssquare} = \id$.
    \item $(f\circ g)^\diamond = g^\diamond \circ f^\diamond$, $(f\circ g)^{\ssquare} = g^{\ssquare} \circ f^{\ssquare}$.
    \item $(f\circ g)_\diamond = f_\diamond \circ g_\diamond$.
  \end{enumerate}
  \end{multicols}
\end{proposition}
\begin{proof}~
  \begin{enumerate}[label=\alph*)]
    \item Suppose $p\leq q$. Then $p\circ f\leq q\circ f$ and hence $f^\diamond(p) = \ceil{p\circ f} \leq \ceil{q\circ f} = f^\diamond(q)$. Also $q^\perp \leq p^\perp$ and hence $\ceil{q^\perp \circ f} \leq \ceil{p^\perp\circ f}$. Taking complements again gives $f^{\ssquare}(p)\leq f^{\ssquare}(q)$.
    
    \item Suppose $f^\diamond(p)\leq q^\perp$. Then $p\circ f \leq \ceil{p\circ f} = f^\diamond(p) \leq q^\perp = \im{\pi_q}^\perp$ and hence $p\circ f \circ \pi_q = 0$ so that $p\leq \im{f\circ \pi_q}^\perp$. But then $f_\diamond(q) = \im{f\circ \pi_q} \leq p^\perp$.
    
    \item Suppose $f_\diamond(q)\leq p$. We need to show $q\leq f^{\ssquare}(p)$. The previous point gives $f_\diamond(q)\leq p$ iff $f^\diamond(q^\perp) \leq p^\perp$ and hence $p\leq f^\diamond(q^\perp)^\perp =: f^{\ssquare}(q)$.
    
    \item As $f_\diamond(q)\leq f_\diamond(q)$ the previous point gives $q\leq f^{\ssquare}(f_\diamond(q))$. Now suppose $p\leq q$. Then $p\leq q\leq f^{\ssquare}(f_\diamond(q))$ so that again by the previous point $f_\diamond(p)\leq f_\diamond(q)$.

    \item We have $f_\diamond(f^{\ssquare}(p))\leq p$. Leting $p:= f_\diamond(q)$ we get $f_\diamond f^{\ssquare} f_\diamond (q) \leq f_\diamond(q)$. We also have $q\leq f^{\ssquare} f_\diamond (q)$ and hence applying the monotone $f_\diamond$ to both sides gives the other inequality.

    \item For sharp $p$ we have $\ceil{p} = p = \im{\pi_p}$. The statements then follow easily.
    \item By Proposition~\ref{prop:floorceiling}: $(f\circ g)^\diamond(p) = \ceil{p\circ f\circ g} = \ceil{\ceil{p\circ f}\circ g} = g^\diamond(f^\diamond(p))$. Furthermore $(f\circ g)^{\ssquare}(p) = (f\circ g)^\diamond(p^\perp)^\perp = g^\diamond(f^\diamond(p^\perp))^\perp = g^\diamond(f^{\ssquare}(p)^\perp)^\perp = g^{\ssquare}(f^{\ssquare}(p))$.
    \item $(f\circ g)_\diamond$ is left Galois adjoint to $(f\circ g)^{\ssquare}$. As Galois adjoints are unique it suffices to show that $f_\diamond\circ g_\diamond$ is also left Galois adjoint to $(f\circ g)^{\ssquare}$. We calculate:
    \[f_\diamond(g_\diamond(p))\leq q \iff g_\diamond(p) \leq f^{\ssquare}(q) \iff p \leq g^{\ssquare}(f^{\ssquare}(q)) = (f\circ g)^{\ssquare}(q). \qedhere\]
  \end{enumerate}
\end{proof}

\begin{proposition}[{\cite[208IX]{basthesis}}]\label{prop:sharpadd}
  Let $A$ be a system in a $\diamond$-effect theory.
  \begin{enumerate}[label=\alph*)]
    \item If $p\in \eff(A)$ is sharp, then it is also \Define{ortho-sharp}\indexd{ortho-sharp}\indexd{sharp effect!ortho-sharp}: for any $a\in\eff(A)$ with $a\leq p$ and $a\leq p^\perp$ we have $a=0$. In other words: $p\wedge p^\perp = 0$.
    \item Every two sharp effects $p,q\in \eff(A)$ have a sharp infimum $p\wedge q$.

    \item $\seff(A)$ is an ortholattice and a sub-effect-algebra of $\eff(A)$. In particular sums and differences of sharp effects are again sharp.
  \end{enumerate}
\end{proposition}
\begin{proof}~
  \begin{enumerate}[label=\alph*)]
    \item Let $p$ be sharp, and suppose $a\leq p,p^\perp$. Then also $\ceil{a}\leq p,p^\perp$. But then $1\circ \pi_{\ceil{a}} = \ceil{a}\circ \pi_{\ceil{a}} \leq p\circ \pi_{\ceil{a}}$ so that by the universal property of $\pi_p$ there is a map $h$ such that $\pi_{\ceil{a}} = \pi_p\circ h$. But then $1\circ \pi_{\ceil{a}} = \ceil{a}\circ\pi_{\ceil{a}} = \ceil{a} \circ \pi_p \circ h \leq p^\perp \circ \pi_p\circ h = 0$, so that $a\leq \ceil{a} = \im{\pi_{\ceil{a}}} = 0$.

    \item Let $p,q \in \seff(A)$. We claim that $p\wedge q = (\pi_p)_\diamond(\pi_p^{\ssquare}(q))$. First of all, as $q\leq q$, we have $\pi_p^{\ssquare}(q)\leq \pi_p^{\ssquare}(q)$ and thus $(\pi_p)_\diamond(\pi_p^{\ssquare}(q)) \leq q$. Second, $(\pi_p)_\diamond(\pi_p^{\ssquare}(q)) \leq (\pi_p)_\diamond(1) = \im{\pi_p} = p$, so it is indeed a lower bound. Now let $r$ be any sharp element with $r\leq p$ and $r\leq q$. 
    Then as in the previous point $\pi_r = \pi_p \circ h$ for some map $h$. Now using that $(\pi_p)_\diamond = (\pi_p)_\diamond\circ (\pi_p)^{\ssquare} \circ (\pi_p)_\diamond$; cf.~Proposition~\ref{prop:galois-properties}.e):
    \[(\pi_r)_\diamond = (\pi_p)_\diamond \circ h_\diamond = (\pi_p)_\diamond\circ (\pi_p)^{\ssquare}\circ (\pi_p)_\diamond \circ h_\diamond = (\pi_p)_\diamond\circ (\pi_p)^{\ssquare}\circ (\pi_r)_\diamond\]
    and thus $r = (\pi_r)_\diamond(1) = (\pi_p)_\diamond\circ (\pi_p)^{\ssquare}(r) \leq (\pi_p)_\diamond\circ (\pi_p)^{\ssquare}(q)$.
    Now for a general $a\leq p,q$, we will also have $\ceil{a}\leq p,q$ and hence $a\leq \ceil{a} \leq p\wedge q$.

    \item By assumption $\seff(A)$ is closed under complements, so that $p\vee q = (p^\perp\wedge q^\perp)^\perp$ makes $\seff(A)$ into an ortholattice. Of course $1$ and $0$ are in $\seff(A)$, so to be a sub-effect-algebra it suffices to show that it is closed under addition. Suppose $p\perp q$. Then $p\wedge q \leq p\wedge p^\perp = 0$, by the first point. By Ref.~\cite[177I]{basthesis} then $p\ovee q = p\vee q$ which is again sharp. %
    \qedhere
  \end{enumerate}
\end{proof}

So far we have not required any type of coherence between filters and compressions. For the next results we however do need to know a bit more about their interplay.
 
\begin{definition}\label{def:compatible-filters-compressions}
  Let $\mathbb{E}$ be an effect theory with filters and compressions. We say the filters and compressions are \Define{compatible}\indexd{compatible filters and compressions} when for every compression $\pi_p$ of a sharp effect $p$ there exists a filter $\xi_p$ of $p$ such that $\xi_p\circ \pi_p = \id$.
\end{definition}
In the next section we will introduce a dagger structure on pure maps which in particular will entail that $\pi_p^\dagger$ is a filter with $\pi_p^\dagger \circ \pi_p = \id$.

The following results seem to require this compatibility assumption in order to be proven. We will use these in Section~\ref{sec:polar-decomposition}.

\begin{proposition}[{\cite[211XI]{basthesis}}]\label{prop:composition-of-compressions}
  In a $\diamond$-effect-theory with compatible filters and compressions the composition of two compressions is again a compression.
\end{proposition}
\begin{proof}
  Let $\pi_p$ and $\pi_q$ be compressions for sharp effects $p$ respectively $q$. 
  We will show that $\pi_p\circ \pi_q$ is a compression for $\im{\pi_p \circ \pi_q}$. 
  To this end let $f$ be any map with $\im{\pi_p\circ \pi_q}\circ f = 1\circ f$. 
  As $\im{\pi_p\circ \pi_q} \leq \im{\pi_p} = p$ we also have $p\circ f = 1\circ f$ and hence $f = \pi_p\circ g_1$ for a unique $g_1$. 
  Let $\xi_p$ be a filter for $p$ such that $\xi_p\circ \pi_p = \id$. 
  Then $q\circ \xi_p\circ\pi_p\circ \pi_q = q\circ \pi_q = 1\circ\pi_q = 1\circ \pi_p \circ \pi_q$ and hence $q\circ \xi_p \geq \im{\pi_p\circ \pi_q}$ so that
  \[q\circ g_1 = q\circ \xi_p\circ \pi_p \circ g_1 = q\circ \xi_p \circ f \geq \im{\pi_p\circ \pi_q} \circ f = 1\circ f = 1\circ g_1.\]
  Hence there is a unique $g_2$ such that $g_1 = \pi_q\circ g_2$ which gives $f = \pi_p\circ g_1 = (\pi_p\circ \pi_q) \circ g_2$. Uniqueness of $g_2$ with the property that $f=(\pi_p\circ \pi_q)\circ g_2$ follows because $\pi_p\circ \pi_q$ is monic.
\end{proof}

\begin{proposition}[{\cite[Proposition~212III]{basthesis}}]\label{prop:effect-decomposition-of-maps}
    Let $f:A\rightarrow B$ be a transformation in a $\diamond$-effect-theory with compatible filters and compressions.
    \begin{enumerate}[label=\alph*)]
        \item There is a unique unital $g$ such that $f = g\circ\xi_{1\circ f}$.
        \item There is a unique faithful $g'$ such that $f=\pi_{\im{f}}\circ g'$.
        \item There is a unique unital and faithful $\cl{f}: A^{\ceil{1\circ f}} \rightarrow A_{\im{f}}$ such that $f = \pi_{\im{f}}\circ \cl{f}\circ \xi_{1\circ f}$.
    \end{enumerate}
\end{proposition}
\begin{proof}~
    \begin{enumerate}[label=\alph*)]
        \item By the universal property of $\xi_{1\circ f}$ there is a unique map $g$ such that $f = g\circ \xi_{1\circ f}$. To show that $g$ is unital note that $1\circ g\circ \xi_{1\circ f} = 1\circ f = 1\circ \xi_{1\circ f}$. As filters are epic, we have $1\circ g = 1$.

        \item By the universal property of $\pi_{\im{f}}$, there is a unique $g$ with $f=\pi_{\im{f}}\circ g$. We claim that $g$ is faithful. So suppose $a\circ g = 0$ for some effect $a$. We need to show that $a=0$. Let $\xi_{\im{f}}$ be a filter for $\im{f}$ such that $\xi_{\im{f}}\circ \pi_{\im{f}} = \id$. Then $0 = a\circ g = a\circ \xi_{\im{f}}\circ \pi_{\im{f}}\circ g = a\circ \xi_{\im{f}}\circ f$ and so $a\circ \xi_{\im{f}} \leq \im{f}^\perp$. Hence $a = a\circ \xi_{\im{f}}\circ \pi_{\im{f}} \leq \im{f}^\perp\circ \pi_{\im{f}} = 0$.

        \item First write $f = \pi_{\im{f}}\circ g'$ where $g'$ is faithful. Using $1\circ\pi_{\im{f}}=1$ we calculate $1\circ g' = 1\circ \pi_{\im{f}}\circ g' = 1\circ f$ and hence there is a unique unital map $g$ with $g' = g\circ \xi_{1\circ f}$. As $g'$ is faithful we have $\im{g'} = 1$, and so we calculate: $1=\im{g'} = \im{g\circ \xi_{1\circ f}}\leq \im{g}$ so that also $\im{g} = 1$ and hence $g$ is faithful. This shows the existence of a suitable map. Uniqueness follows because filters are epic and compressions are monic. \qedhere
    \end{enumerate}
\end{proof}

\section{Purity}\label{sec:purity}

As mentioned in the introduction of this chapter, the axioms of this chapter's reconstruction mostly concern the specific definition of purity from effectus theory. In order to appreciate this definition, let us first take a broader look at the concept of purity in quantum physics.

In quite a variety of topics in quantum information theory, a notion of purity has proved fruitful~\cite{devetak2005distillation,brandao2013resource}. While there is consensus about which states should be considered pure, when talking about pure maps, the situation is more muddled. There is a variety of different definitions in play that each have their own benefits and drawbacks.
In this section we will review those different definitions, but first let us consider some properties that would be desirable or expected of an intuitive definition of purity.

First of all, what does it mean to say that a map is `pure'? In a way, saying that a map is pure is saying that it is `fundamental' in some way. 
This can mean multiple things. It could mean that every other map can be made in some way using pure maps, and thus that the pure maps are the basic building blocks of the theory. 
It could also mean that the pure maps are the only transformations that are part of the fundamental theory, other transformations merely reflecting our ignorance of these `true' dynamics. 
For instance, in standard `pure' quantum mechanics, the systems are Hilbert spaces, while the only allowed transformations are unitaries. Since this is the fundamental level of the physical theory, all these unitaries can be considered pure. Transitioning to the more general framework of C*-algebras, we also have the liberty to describe classical systems and interactions that do not seem to warrant being called pure, such as the action of throwing away a system by the partial trace map.

We will take the view that a map is pure when it is somehow fundamental to the theory. We will now argue that those maps should form a dagger-category.

\begin{definition}[{\cite{selinger2007dagger}}]
    A \Define{dagger-category}\indexd{dagger-category} $C$ is a category equipped with an involutive endofunctor $(\cdot)^\dagger:C\rightarrow C^{\text{op}}$ that is the identity on objects: $A^\dagger = A$. In other words, there is an operation $\dagger$ that sends every map $f:A\rightarrow B$ to some map $f^\dagger: B\rightarrow A$ such that $(f^\dagger)^\dagger = f$, $\id^\dagger = \id$ and $(f\circ g)^\dagger = g^\dagger\circ f^\dagger$.
\end{definition}

Saying that the pure maps should form a dagger category is in fact stating three different things:
\begin{itemize}
    \item \emph{The identity map is pure.} Every physical theory should be able to describe the act of not changing a system.
    \item \emph{The composition of pure maps is pure.} If we describe a fundamental set of transformations, then when two transformations could possibly happen after one another, i.e.\ when they are composable, this combined transformation should also be describable in this fundamental theory and hence be pure.
    \item \emph{The time-reverse of a pure map is pure.} We consider the dagger action as describing the reversal of the arrow of time. Saying that the pure maps have a dagger is then akin to saying that for every fundamental operation, the reversed operational is also fundamental.
\end{itemize}

If we also wish to describe composite systems, then there is an obvious additional requirement for pure maps that a composite of pure maps should again be pure. In this case the pure maps form a monoidal dagger category\footnote{In a monoidal dagger category it is common to require that the dagger preserves the monoidal structure. We will not need this additional requirement for now, but we will use it in Chapter~\ref{chap:infinitedimension}.}.

Now let us go over the definitions of purity found in the literature and see how they compare. 
Probably the most well-known is that of \Define{atomicity}\indexd{atomic map} and the related notion of \Define{convex-extremality}\indexd{convex-extreme map} used extensively in generalised probabilistic theories~\cite{barrett2007information,chiribella2010probabilistic,chiribella2011informational}. A map $f$ is atomic when any decomposition $f=g_1+g_2$ implies $g_i = \lambda_i f$ for some $\lambda_i\in [0,1]$. The maps $f$, $g_1$ and $g_2$ here are then required to be `sub-causal'. For causal maps one can consider convex-extremality. A map $f$ is convex extreme when any decomposition $f=\lambda g_1 + (1-\lambda)g_2$ for $0<\lambda<1$ implies that $g_1=g_2=f$. If we take atomicity to be our definition of pure, an immediate problem arises. Consider the C*-algebra $M_n(\C)\oplus M_n(\C)$. The identity can then be written as $\id = \id_1+\id_2$, and hence it is not atomic and thus not pure. Taking convex-extremality as our definition of purity leads to a more subtle problem: the dagger of convex-extreme maps does not have to be sub-unital.

Other definitions of purity are those given by leaks~\cite{selby2018reconstructing}, orthogonal factorizations~\cite{cunningham2017purity} or dilations~\cite{tull2019phdthesis}. Without going into the details, these definitions of purity are in general not closed under a dagger operation. These definitions also require the existence of composite systems in order to be stated. On a conceptual level there is then the issue that the purity of a map using these definitions can only be established by considering external systems so that purity does not seem to be an inherent property of the system and its dynamics.

It should be noted that all these definitions of purity were specifically designed to be applicable to finite-dimensional systems. When considering for instance infinite-dimensional von Neumann algebras, it is no longer clear that these definitions serve their intended purpose.

\subsection{Pure effect theories}\label{sec:PET}

Let us give the definition of pure maps in effect theories.

\begin{definition}[{\cite[Definition~201II]{basthesis}}]\label{def:pure}
    A map $f:A\rightarrow B$ in an effect theory is \Define{pure}\indexd{pure map} when $f=\pi\circ\xi$ where $\xi$ is a filter and $\pi$ is a compression.
\end{definition}

To motivate this definition, let us see what the action of a pure map would be on a pure state in the sense of quantum theory, \ie~a state of maximal information.
Letting $\xi_a$ be a filter for an effect $a$, the (unnormalised) state $\xi_a\circ \omega$ corresponds to the state $\omega$ on which a measurement and post-selection of the effect $a$ is applied. Post-selecting to an effect increases the amount of information we have about $\omega$ and hence $\xi_a\circ\omega$ should remain a state of maximal information. In this sense, $\xi_a$ preserves the property of states being pure.
Similarly, if $\pi_b$ is a compression for an effect $b$, then $\pi_b\circ \omega$ is simply $\omega$ seen as a state on a bigger system, and hence this preserves the purity of the state.
Our definition of pure maps then consists of maps that send pure states (\ie~states of maximal information) to pure states\footnote{As further evidence to the naturalness of this definition of purity, Theorem~171VII of Ref.~\cite{basthesis} shows that a normal completely positive map between von Neumann algebras is pure if and only if its \emph{Paschke dilation}, i.e.~its purification (cf.~Section~\ref{sec:intro-composite-systems}), is surjective.}.

In Chapter~\ref{chap:jordanalg} we will see that contrary to the definitions of purity discussed before, this notion will be well-behaved, even in the quite general setting of JBW-algebras.

It is not a priori clear that the pure maps of Definition~\ref{def:pure} are closed under composition. In particular, it is not clear whether a composition `in the wrong order' $\xi\circ \pi$ is pure. In our definition of a pure effect theory below we have the assumption that the pure maps form a (dagger-)category, and hence we simply assert the closure of pure maps under composition.

In a general effect theory, filters and compressions are not required to interact in a meaningful way. Let us therefore discuss a bit more how filters and compressions of sharp effects \emph{should} interact.
Let $\pi_p:A_p\rightarrow A$ be a compression for a sharp effect $p$. Let $\omega:I\rightarrow A_p$ be a state defined on the system $A_p$ where $p$ holds with certainty.
The state $\pi_p\circ\omega$ is then the same state on $A$ where we have forgotten that $p$ holds. Now consider the time-reverse $\pi_p^\dagger$ of $\pi_p$ (which we will assume exists, since we take $\pi_p$ to be a pure map). As the compression $\pi_p$ forgets that the effect $p$ holds for $\omega$, the map $\pi_p^\dagger$ should `remember it'. In other words: it is a post-selection for $p$ and hence is a filter for $p$. Now consider $\pi_p^\dagger\circ \pi_p$. This is a post-selection to $p$ after we already knew that the effect holds (since $\pi_p$ is an embedding from the space where $p$ holds with certainty). Hence we should have $\pi_p^\dagger \circ \pi_p = \id$. These considerations give rise to the assumptions we will require. In the parlance of dagger-categories: compressions for sharp effects should be isometries whose adjoint is a filter for the same effect.

\begin{definition}\label{def:PET}
    A (monoidal) \Define{pure effect theory}\indexd{pure effect theory} (PET)\index{math}{PET (pure effect theory)} is a (monoidal) effect theory satisfying the following properties.
    \begin{enumerate}[label=({P}\theenumi), ref=P\theenumi]
        \item \label{pet:filtcompr} All effects have filters and compressions.
        \item \label{pet:dagger} The pure maps form a (monoidal) dagger-category.
        \item \label{pet:images} All maps have images.
        \item \label{pet:sharpnegation} The negation of a sharp effect is sharp (if $p$ is sharp, then $p^\perp$ is sharp)
        \item \label{pet:sharpadjoint} Compressions of sharp effects are adjoint to its filters (if $\pi_p$ is a compression for sharp $p$, then $\pi_p^\dagger$ is a filter for $p$, and vice versa).
        \item \label{pet:sharpisometry} Compressions of sharp effects are isometries ($\pi_p^\dagger\circ\pi_p = \id$ for sharp $p$).
    \end{enumerate}
\end{definition}

\begin{remark}
  The combination of \ref{pet:filtcompr}, \ref{pet:images}, and \ref{pet:sharpnegation} makes a PET into a $\diamond$-effect-theory. By \ref{pet:sharpadjoint} and \ref{pet:sharpisometry} for every compression $\pi_p$ we have a filter $\xi_p$ (namely $\xi_p=\pi_p^\dagger$) such that $\xi_p \circ \pi_p = \id$ and hence filters and compressions are compatible as in Definition~\ref{def:compatible-filters-compressions}.
  The properties we assume for a PET are a subset of what is called a \emph{$\dagger$-effectus} in Ref.~\cite{basthesis}. In the next chapter we will show that JBW-algebras satisfy these stronger conditions and hence that JBW-algebras in particular form a PET.
\end{remark}

\begin{remark}
    Several of these properties are closely related to more familiar categorical definitions. It is shown in Ref.~\cite{basthesis} that an effect theory has all compressions if and only if it has all \Define{kernels}\indexd{kernel} (a compression of $q$ is a kernel of $q^\perp$). It has all \Define{cokernels}\indexd{cokernel} if and only if all maps have an image and every sharp effect has a filter (a filter is the cokernel of a compression). The assumptions~\ref{pet:sharpadjoint} and \ref{pet:sharpisometry} can equivalently be stated as ``all kernels are dagger-kernels, and the dagger of a kernel is a cokernel'', and hence the subcategory of pure maps of a PET is a \Define{dagger kernel category}~\cite{heunen2010quantum}\indexd{dagger kernel category}.
\end{remark}

\noindent Now that we have the definition of a PET, we can state the main results of this chapter: the reconstruction of quantum theory from assumptions on purity.
\begin{theorem*}[\textbf{\ref{theor:OETEJA}}]
    Let $\mathbb{E}$ be an operational pure effect theory. Then there exists a functor into the opposite category of Euclidean Jordan algebras and positive sub-unital maps $F:\mathbb{E}\rightarrow \EJA_{\text{psu}}^{\text{op}}$ such that $[0,1]_{F(A)} \cong \eff(A)$. Furthermore, this functor is faithful if and only if $\mathbb{E}$ satisfies local tomography.
\end{theorem*}

\begin{theorem*}[\textbf{\ref{theor:compositealgebras}}]
    Let $\mathbb{E}$ be a monoidal operational pure effect theory. Then the above functor restricts either to the category of real C*-algebras or to the category of complex C*-algebras. 
\end{theorem*}

In this last theorem we expect that $F$ is faithful if and only if $\mathbb{E}$ satisfies tomography, but showing this requires establishing that $F$ is monoidal, which is currently an open question. For more discussion regarding this we refer to Section~\ref{sec:monoidalPETs}.

\subsection{Properties of pure maps}\label{sec:firsttwo}


Let $p$ be a sharp effect and let $\pi_p$ be a compression for it. By \ref{pet:sharpadjoint} the map $\pi_p^\dagger$ is a filter for $p$ and by \ref{pet:sharpisometry} we have $\pi_p^\dagger\circ \pi_p = \id$.

\begin{definition}
  Let $p:A\rightarrow I$ be a sharp effect in a PET and let $\pi_p: A_p\rightarrow A$ be a compression of $p$. The \Define{assert map}\indexd{assert map} $\asrt_p:A\rightarrow A$ of $p$ is then $\asrt_p := \pi_p\circ \pi_p^\dagger$. 
\end{definition}

\begin{remark}
    The definition of the assert map does not depend on the choice of compression. If $\pi_p^\prime$ is also a compression for $p$ then $\pi_p^\prime = \pi_p\circ\Theta_1$ for some isomorphism $\Theta_1$ by Proposition~\ref{prop:quotcompr}, and similarly since $(\pi_p^\prime)^\dagger$ is a filter by \ref{pet:sharpadjoint} we have $(\pi_p^\prime)^\dagger = \Theta_2\circ\pi_p^\dagger$ for some other isomorphism $\Theta_2$. Now $\id = (\pi_p^\prime)^\dagger\circ \pi_p^\prime = \Theta_2 \circ \pi_p^\dagger \circ \pi_p \circ \Theta_1 = \Theta_2\circ \Theta_1$ so that $\Theta_2 = \Theta_1^{-1}$. As a result $\pi_p^\prime \circ (\pi_p^\prime)^\dagger = \pi_p\circ \Theta_1\circ \Theta_1^{-1} \circ \pi_p^\dagger = \pi_p \circ \pi_p^\dagger = \asrt_p$.
\end{remark}
\begin{remark}
    Since $\xi_p^\dagger$ is also a compression for $p$ we could have defined the assert map equally well as $\asrt_p = \xi_p^\dagger \circ \xi_p$ for any filter of $p$.
\end{remark}

\begin{example}
	Let $B(H)$ be the set of bounded operators on a (complex) Hilbert space and let $A\in B(H)$ be an effect. Then the assert map $\asrt_A:B(H)\rightarrow B(H)$ (in the opposite category of C$^*$-algebras with positive sub-unital maps) is given by $\asrt_A(B) = \sqrt{A}B\sqrt{A}$. I.e.~it is the sequential product map of $A$. The name `assert' comes from the fact that it asserts the effect $A$ to be true.
\end{example}


\begin{proposition}[{\cite[Section 3.6]{basthesis}}] \label{prop:assertmaps}
   Let $p$ be a sharp effect and let $f$ be any composable map in a PET. The following are true:
  \begin{multicols}{2}
  \begin{enumerate}[label=\alph*)]
    \item $\asrt_p\circ \asrt_p = \asrt_p$.
    \item $\im{\asrt_p} = p$.
    \item $1\circ \asrt_p = p$.
    \item $\im{f}\leq p \iff \asrt_p\circ f = f$.
    \item $1\circ f \leq p \iff f\circ \asrt_p = f$.
  \end{enumerate}
  \end{multicols}
\end{proposition}
\begin{proof} ~
  \begin{enumerate}[label=\alph*)]
    \item $\asrt_p\circ \asrt_p = \pi_p\circ\pi_p^\dagger\circ\pi_p\circ\pi_p^\dagger = \pi_p\circ \id\circ \pi_p^\dagger = \asrt_p$.

    \item Note that $p\circ \asrt_p = p\circ \pi_p \circ \pi_p^\dagger = 1\circ \pi_p\circ \pi_p^\dagger = 1\circ \asrt_p$ so that $p\geq \im{\asrt_p}$. Conversely, suppose $q\leq \im{\asrt_p}^\perp$. Then $0 = q\circ \asrt_p = q\circ \pi_p\circ \pi_p^\dagger$. Because $\pi_p^\dagger$ is a filter, it is faithful by Proposition~\ref{prop:faithfulfilters}. As a result $q\circ \pi_p =0$ so that $q\leq \im{\pi_p}^\perp = p^\perp$. Taking $q=\im{\asrt_p}^\perp$ we then have $p \leq \im{\asrt_p}$. 

    \item By Proposition~\ref{prop:faithfulfilters}, first c) and then a): $1\circ \asrt_p = 1\circ \pi_p \circ \pi_p^\dagger = 1\circ \pi_p^\dagger = p$.

    \item If $\asrt_p\circ f = f$, then $\im{f} = {\im{\asrt_p\circ f}} \leq \im{\asrt_p} = p$. Conversely, if $\im{f}\leq p$, then $p\circ f = 1\circ f$ so that by the universal property of $\pi_p$ we have $f = \pi_p\circ \cl{f}$ for some $\cl{f}$. Now $\cl{f} = \id\circ \cl{f} = \pi_p^\dagger\circ \pi_p \circ \cl{f} = \pi_p^\dagger \circ f$ so that $f = \pi_p \circ \cl{f} = \pi_p\circ \pi_p^\dagger\circ f = \asrt_p\circ f$.
    \item Suppose $f\circ \asrt_p = f$. Then $1\circ f = (1\circ f)\circ \asrt_p \leq 1\circ \asrt_p = p$. Conversely, if $1\circ f \leq p$, then by the universal property of $\pi_p^\dagger$ we have $f = \cl{f}\circ \pi_p^\dagger$ for some $\cl{f}$. Now $\cl{f} = \cl{f}\circ \id = \cl{f} \circ \pi_p^\dagger \circ \pi_p = f \circ \pi_p$ so that $f = \cl{f}\circ \pi_p^\dagger = f\circ \pi_p\circ \pi_p^\dagger = f\circ \asrt_p$. \qedhere
  \end{enumerate}
\end{proof}

\noindent Recall that in an effect algebra the addition operation is a partial operation. This will allow us to talk about orthogonality.
\begin{definition}
  Let $p,q\in$ $\eff(A)$ be sharp effects. We call them \Define{orthogonal}\indexd{orthogonal effect!in effect theory} when $p$ and $q$ are summable. That is, when $p+q$ is defined and therefore $p+q\leq 1$. We call two arbitrary effects (not necessarily sharp) orthogonal when their ceilings are orthogonal.
\end{definition}
Note that for non-sharp effects, being summable is weaker than being orthogonal, as for instance $\frac12 p$ is always summable with itself (assuming we have a scalar acting like $\frac12$).
\begin{proposition}\label{prop:orthog}
  Let $p,q\in\seff(A)$ be sharp effects in a PET. Then $p$ and $q$ are orthogonal if and only if $q\circ \asrt_p = 0$.
\end{proposition}
\begin{proof}
    We have $\im{\asrt_p} = p$, so that for any $q\leq p^\perp =\im{\asrt_p}^\perp$ we have $q\circ \asrt_p \leq \im{\asrt_p}^\perp\circ \asrt_p = 0$. If $q\circ \asrt_p = 0$ then $q\leq \im{\asrt_p}^\perp =p^\perp$.
\end{proof}

\noindent There is a lot more that can be done in this abstract situation of effect theories with filters and compressions, some of which we will explore in Chapter~\ref{chap:infinitedimension}. For more results concerning effectus theory we refer the interested reader to Refs.~\cite{cho2015introduction,basthesis,kentathesis}. We will now switch to the more concrete setting of operational effect theories.

\section{From operational PETs to Jordan algebras}\label{sec:opeffect}

In this section we will study PETs in a more familiar convex setting by working with operational PETs. The goal is to show that in this setting effect spaces correspond to Euclidean Jordan algebras, and hence most of the structure of quantum theory is recovered. 
On a surface level the proof is structured much the same way as in Chapter~\ref{chap:seqprod}:
we first derive a diagonalisation theorem in Section~\ref{sec:diagonalisation} and then in Section~\ref{sec:duality} we construct a self-dual inner product on the effect spaces. In combination with some further technical arguments presented in Section~\ref{sec:structureoffaces}, this will show that the systems correspond to EJAs.

For the duration of this section we will assume that we have fixed some operational PET, and that $A$ is a system therein. Furthermore $V$ will denote the order unit space associated to the system: $\eff(A) \cong [0,1]_V$. Sharpness, ceilings and floors in $V$ are all defined by the same notion on effects in $A$.

\subsection{Diagonalisation}\label{sec:diagonalisation}
We will construct a diagonalisation in terms of sharp effects using properties of the ceiling of effects. That is: we will show that for any $v\in V$ we can find a collection of eigenvalues $\lambda_i\in \R$ and a set of sharp effects $p_i\in V$ that are all orthogonal to each other such that $v=\sum_i \lambda_i p_i$. The following proposition collects the needed properties of the ceiling established in Section~\ref{sec:firsttwo}:
\begin{proposition}~\label{prop:ceilings}
  Let $a,b\in \eff(A)$  and $f:B\rightarrow A$.
  \begin{itemize}
    \item If $a\leq b$ then $\ceil{a}\leq \ceil{b}$ and $\floor{a}\leq \floor{b}$.
    \item Let $\lambda\neq 0$ be a scalar. Then $\ceil{\lambda \circ a}=\ceil{a}$.
    \item $a\circ f = 0$ iff $\ceil{a}\circ f = 0$.
  \end{itemize}
\end{proposition}
\begin{proof}
  All the points are straight from Proposition~\ref{prop:floorceiling} except for $\lambda \neq 0$ implying that $\ceil{\lambda \circ a}=\ceil{a}$. This follows from the point in Proposition~\ref{prop:floorceiling} that $\ceil{a\circ f}=\ceil{\ceil{a}\circ f}$ by letting $f:=a$ and $a:=\lambda$ and observing that in our current setting $\ceil{\lambda}=1$ when $\lambda\neq 0$.
\end{proof}

Recall that the norm on a order-unit space is given by $\norm{v}:=\inf\{r>0~;~-r1\leq v\leq r1\}$. Let $C := \{v\in V~;~ v\geq 0\}$ be the positive cone of $V$ and denote its interior with respect to the order-unit-norm topology by $C^\circ$. We have $v\in C^\circ$ iff there is an $\epsilon>0$ such that $\epsilon 1 \leq v$. Denote the boundary of the cone by $\partial C = C\backslash C^\circ$.

\begin{lemma} \label{lem:var}
  Let $v\in C$. Then 
    \begin{enumerate}[label=\alph*)]
      \item $\norm{v}1 - v \in \partial C$,
      \item if $\norm{v}<1$, then $1-v=v^\perp \in C^\circ$,
      \item if $v$ is sharp and $\norm{v}<1$, then $v=0$,
      \item if $\norm{v}<1$ and $v\leq p$ where $p$ is sharp, then $\ceil{p-v}=p$,
      \item if $\norm{v}<1$ and $\ceil{v}\perp p$ where $p$ is sharp, then $\floor{p+v} = p$.
    \end{enumerate}
\end{lemma}
\begin{proof} ~
  \begin{enumerate}[label=\alph*)]
  \item As $v\leq \norm{v}1$, if $\norm{v}1-v \not\in\partial C$ then $v-\norm{v}1\in C^\circ$ so that there must be an $\epsilon>0$ such that $\epsilon 1 \leq v-\norm{v}1$ which means that $v-(\norm{v}+\epsilon)1 \geq 0$ contradicting the defining property of the order-unit norm. 
  \item Of course $0\leq v\leq \norm{v}1$ and hence $0\leq 1-\norm{v}1 \leq 1-v=v^\perp$. Since $1-\norm{v}1 = (1-\norm{v})1>0$ we conclude that $v^\perp \in C^\circ$.
  \item Let $v$ be sharp with $\norm{v}<1$. Then by the previous point $v^\perp \in C^\circ$ so that $\epsilon 1\leq v^\perp$ for some $\epsilon>0$. Then $1=\ceil{1}=\ceil{\epsilon 1} \leq \ceil{v^\perp} = v^\perp$ because $v^\perp$ is sharp. But then $v^\perp = 1$ so indeed $v=0$.
  
  \item Let $v\leq p$ with $\norm{v}<1$, and let $q:=\ceil{p-v}\leq p$.
  Then we can write $p=q+r$ where $r:=p-q$ is a sharp effect by Proposition~\ref{prop:sharpadd}.c). 
  Now $p-v\leq \ceil{p-v}=q$ so that $r=p-q\leq v$. Taking the norm on both sides gives $\norm{r} \leq \norm{v}<1$ so that by the previous point $r=0$. So indeed $p=q+0 = \ceil{p-v}$.
    \item As $\ceil{v}\perp p$ we have $\ceil{v}+p\leq 1$ so that also $v+p\leq 1$ is indeed an effect. Furthermore as $v\leq \ceil{v}\leq p^\perp$ and $\norm{v}<1$ the previous point applies and thus $\ceil{p^\perp - v} = \ceil{p^\perp}$. We then calculate: $\floor{p+v} = \ceil{(p+v)^\perp}^\perp = \ceil{1-p-v}^\perp = \ceil{p^\perp - v}^\perp = \ceil{p^\perp}^\perp = \floor{p}$ as desired. \qedhere
  \end{enumerate}
\end{proof}
We need to show that the order unit spaces we deal with are `finite-rank' in a suitable way. This turns out to follow from the next lemma.

\begin{lemma}\label{lem:finitedimrank}
    Let $0\leq v\leq 1$ in $V$. We have $v\in C^\circ \iff \ceil{v}=1$.
\end{lemma}
\begin{proof}
  When $v\in C^\circ$ we have $\epsilon 1\leq v$ and hence $\ceil{v}\geq \ceil{\epsilon 1} = 1$. It remains to verify the converse direction.

  Let $0\leq v\leq 1$ and suppose $\ceil{v}=1$. We want to show that there exists an $\epsilon>0$ such that $\epsilon 1\leq v$. 
  Define $f:\st_1(A)\rightarrow [0,1]$ by $f(\omega)=v\circ \omega$ where St$_1(A)$ denotes the set of unital states on $A$. By assumption St$_1(A)\subset V_A^*$ is closed in $V_A^*$. As it is also bounded, and $V_A^*$ is a finite-dimensional vector space we conclude that St$_1(A)$ is compact. Therefore the image of $f$ will be some compact subset of $[0,1]$. In particular, there is an $\omega\in $ St$_1(A)$ that achieves the minimum of $f$. Suppose that $v\circ \omega = 0$. Then $\ceil{v}\circ\omega = 1\circ\omega = 0$ by the last point of Proposition~\ref{prop:ceilings}. This is a contradiction as $1\circ \omega = 1$, so we must have $v\circ \omega=:\epsilon > 0$. But as $\omega$ is the minimum value we must have $v\circ \omega \geq \epsilon = (\epsilon 1)\circ \omega$ for all $\omega\in $ St$_1(A)$. Since the states order-separate the effects we then get $v\geq \epsilon 1$ which shows that $v\in C^\circ$.
\end{proof}

\begin{remark}
    This lemma explicitly requires the state-space to be closed, and $V$ to be finite-dimensional. Many of the following results can be proved without using finite-dimensionality or the closure of state-space if one simply assumes the consequence of this lemma as an additional axiom.
\end{remark}

\begin{proposition} \label{prop:ceilint}
    Let $v\in [0,1]_V$ with $\norm{v}=1$. Then $\floor{v}\neq 0$.
\end{proposition}
\begin{proof}
  By Lemma~\ref{lem:var} $\norm{v}1-v \not\in C^\circ$ and hence by Lemma~\ref{lem:finitedimrank} $\ceil{\norm{v}1-v} \neq 1$. Supposing now that $\norm{v}=1$ we immediately get $\floor{v}^\perp = \ceil{v^\perp} = \ceil{1-v} = \ceil{\norm{v}1 -v} \neq 1$ and hence $\floor{v}\neq 0$.
\end{proof}



\begin{lemma}\label{lem:orthindep}
  Let $\{p_i\}$ be a finite set of non-zero orthogonal sharp effects in $V$. Then they are linearly independent.
\end{lemma}
\begin{proof}
  Reasoning towards contradiction, assume that there is a non-trivial linear combination of the orthogonal sharp effects. Then without loss of generality $p_1 = \sum_{i>1} \lambda_i p_i$. Since all the $p_i$ are orthogonal we note that $p_j\circ \asrt_{p_i} = 0$ when $i\neq j$ by Proposition~\ref{prop:orthog}, so that $p_1 = p_1\circ\asrt_{p_1} = \sum_{i>1}\lambda_i p_i\circ \asrt_{p_1} = 0$, a contradiction.
\end{proof}

\noindent We can now prove our diagonalisation theorem.

\begin{proposition}\label{prop:diag}
  Let $v\in[0,1]_V$. There is a $k\in \N$ and a strictly decreasing sequence of scalars $\lambda_1>\ldots>\lambda_k>0$ such that $v=\sum_{i=1}^k \lambda_i p_i$ where the $p_i$ are non-zero orthogonal sharp effects.
\end{proposition}
\begin{proof}
If $v=0$ the result is trivial, so assume that $v\neq 0$. 
Let $v^\prime = \norm{v}^{-1} v$ so that $\norm{v^\prime} = 1$. Write $v^\prime = \floor{v^\prime} + w$ for $w=v^\prime - \floor{v^\prime}$. By Proposition~\ref{prop:ceilint} $\floor{v^\prime} \neq 0$. 
Note that $\floor{w}+\floor{v^\prime}\leq v^\prime$ implying that $\floor{\floor{w}+\floor{v^\prime}} = \floor{w}+\floor{v^\prime}\leq \floor{v^\prime}$ by Proposition~\ref{prop:sharpadd}.c), and hence $\floor{w} = 0$. But then $\norm{w}\neq 1$ by Proposition~\ref{prop:ceilint}. 
Now since $w=v^\prime-\floor{v^\prime}\leq 1 - \floor{v^\prime}$ we can write $\ceil{w}\leq \ceil{1-\floor{v^\prime}} = \ceil{\floor{v^\prime}^\perp} = \floor{\floor{v^\prime}}^\perp = \floor{v^\prime}^\perp$ so that $\ceil{w} \perp \floor{v^\prime}$.

Now using $v = \norm{v}v^\prime$ we see that we can write $v = \lambda_1 p + w$ where $p$ is a sharp effect, $\lambda_1 = \norm{v}$ and $w$ is an effect orthogonal to $p$ with $\norm{w}<\norm{v}$. We can now repeat this procedure for $w$. This has to end at some point for if it would not then we get an infinite sequence of orthogonal sharp effects $\{p_i\}_{i=1}^\infty$ which by Lemma~\ref{lem:orthindep} can only be the case when the space is infinite-dimensional.
\end{proof}

\noindent This diagonalisation is unique in the following sense:

\begin{proposition}
  Let $v = \sum_{i=1}^k \lambda_i p_i$ and $v=\sum_{j=1}^l \mu_j q_j$ where $\lambda_i>\lambda_j>0$ and $\mu_i>\mu_j>0$ for $i<j$ and all the $p_i$ and $q_i$ are sharp and non-zero with all the $p_i$ being orthogonal and all the $q_i$ being orthogonal. Then $k=l$, $\lambda_i=\mu_i$ and $p_i=q_i$ for all $i$.
\end{proposition}
\begin{proof}
  Note first that if $v=0$ that then $k=l=0$ and hence we are done. So assume that $v\neq 0$. It is clear that $\norm{v}=\lambda_1=\mu_1$. Consider $v^\prime = \lambda_1^{-1}v$. Now $\floor{v^\prime}=\floor{p_1 + \sum_{i>1}\lambda_1^{-1}\lambda_i p_i} = p_1$ by Lemma~\ref{lem:var}.e). But similarly, using the other decomposition we also get $\floor{v^\prime} = q_1$, so that $p_1=q_1$. Now we can consider $v_2 = v-\lambda_1p_1$ and continue the procedure.
\end{proof}

We can diagonalise an arbitrary positive element $a$ by first diagonalising the effect $\norm{a}^{-1}a$ and then rescaling. Now for an arbitrary element $a$ (not necessarily positive) we have $-n1\leq a \leq n1$ for some $n$, so that $0\leq a+ n1\leq 2n1$. This gives us a diagonalisation $a+n1 = \sum_i\lambda_i p_i$, so that $a = \sum_i\lambda_i p_i - n1 = \sum_i(\lambda_i-n)p_i + n(1-\sum_i p_i)$. As a corollary:
\begin{proposition}
  Any vector $v\in V$ can be written as $v= v^+ - v^-$ where $v^+, v^-\geq 0$ are orthogonal.
\end{proposition}

We can get a more fine-grained diagonalisation than just the one in terms of sharp effects.

\begin{definition}
  We call a non-zero sharp effect $p$ \Define{atomic}\indexd{atomic effect} when for all $q$ with $0\leq q\leq p$ we have $q=\lambda p$. Or in other words when $\downarrow p \cong [0,1]$ where $\downarrow p$ denotes the \Define{downset}\indexd{downset} of $p$.
\end{definition}

\begin{proposition}\label{prop:sharprep}
  Each non-zero sharp effect can be written as a sum of atomic effects.
\end{proposition}
\begin{proof}
  Analogous to the proof of Proposition~\ref{prop:sharpeffectssumofatomic}.
\end{proof}

\begin{corollary}\label{cor:atomicspectrum}
    Any $v\in V$ can be written as $v=\sum_i \lambda_i p_i$ where $\lambda_i\in \R$ and the $p_i$ are atomic and orthogonal.
\end{corollary}
\begin{proof}
    First diagonalise $v$ in terms of sharp effects, and then write every sharp effect as a sum of atomic effects.
\end{proof}



\subsection{Duality}\label{sec:duality}
The goal of this section is to find a self-dual inner product on $V$. We do this by establishing a duality between pure states and effects.

The following proposition establishes that our definition of purity coincides with atomicity when considering states and effects. This correspondence does not hold for arbitrary maps, as there are pure maps that aren't atomic. Note that this is the only result for which we need the assumption that scalar-like systems are isomorphic to the trivial system in Definition~\ref{def:OET}.

\begin{proposition} \label{prop:purestate}
  An effect $q:A\rightarrow I$ is pure if and only if its corresponding effect $q\in [0,1]_V$ is proportional to an atomic effect. A unital state is pure if and only if its image is atomic.
\end{proposition}
\begin{proof}
    When $q=0$ this is trivial, so assume that $q\neq 0$.

  Suppose $q:A\rightarrow I$ is pure. Then we have $q=\pi_s\circ \xi_t$ for some filter $\xi_t: A\rightarrow B$ and compression $\pi_s:B\rightarrow I$. Note that $q = 1\circ q = 1\circ \pi_s\circ \xi_t = 1\circ \xi_t = t$, so that $t=q$.
  The image $\im{q}$ is a scalar and must be sharp. As $q\neq 0$ we must then have $\im{q}=1$ so that $1=\im{\pi_s\circ\xi_{t}}\leq \im{\pi_s} = s$ and hence $s=1$. Compressions for the unit effect are isomorphisms, and hence $\pi_s\circ \xi_t = \xi_t^\prime$ is still a filter for $t=q$. 
  The filter has type $\xi_t^\prime:A\rightarrow I$ so that $I\cong A^q$ and hence by Proposition~\ref{prop:filterisotodownset} $[0,1]\cong \eff(I) \cong \eff(A^q) \cong \,\downarrow q$. As a result $q$ is indeed proportional to an atomic effect. 
  
  For the converse direction, suppose $q$ is proportional to an atomic effect. 
  We use Proposition~\ref{prop:effect-decomposition-of-maps}.a) to write an effect $q$ as $q=g\circ\xi_{q}$ where $g$ is unital and $\xi_{q}:A\rightarrow A^q$. Our goal is to show that $g$ is an isomorphism so that $q$ is indeed pure. 
  As $g:A^q\rightarrow I$ is unital, we have $g = \id_I\circ g = 1_I\circ g = 1_{A_q}$ and since $\downarrow q \cong [0,1]$ by assumption, Proposition~\ref{prop:filterisotodownset} gives us $\eff(A^q)\cong \downarrow q \cong [0,1] \cong \eff(I)$ so that $A^q$ is a scalar-like system. Hence by definition of an operational effect theory, there is an isomorphism $\Theta:I\rightarrow A_q$. Isomorphisms are unital and hence $g\circ\Theta = 1_{A^q}\circ\Theta = 1_I = \id_I$, so that $g$ is the inverse of $\Theta$, and hence is an isomorphism. We conclude that $q$ is a composition of a filter and an isomorphism, and hence is pure.
  

  Using an analogous argument, when we have a pure state $\omega=\pi_s\circ \xi_t$ we must have $t=1$ so that $\xi_1$ is an isomorphism and then $\im{\omega}=\im{\pi_s}=s$ where $s$ must be proportional to an atomic effect because $[0,1]\cong \eff(I)\cong \eff(\{A\lvert s\}) \cong \downarrow s$.
\end{proof}

\begin{proposition} \label{prop:atomicstate}
  Let $q$ be an atomic effect. There exists a unique unital state $\omega_q$ such that $\im{\omega_q}=q$. This state is pure and given by $\omega_q:=q^\dagger$.
\end{proposition}
\begin{proof}
  Since $q$ is atomic, it is pure by Proposition~\ref{prop:purestate} and hence $\omega_q:=q^\dagger$ exists and is a pure state. From the proof of Proposition~\ref{prop:purestate} we also see that $q$ is a filter for $q$ and hence $\omega_q = q^\dagger$ is a compression for $q$ by~\ref{pet:sharpadjoint}. Hence $\im{\omega_q}= q$. Furthermore, since compressions are unital (Proposition~\ref{prop:faithfulfilters}), $\omega_q$ is as well.

  For uniqueness we note that any unital state $\omega$ can be written as $\pi_{\im{\omega}}\circ \cl{\omega}$ where $\cl{\omega}$ is also a unital state and $\pi_{\im{\omega}}$ is a compression for $\im{\omega}$. Here $\cl{\omega}$ is a map to the object $A_{\im{\omega}} \cong A^{\im{\omega}}$ which has $\eff(A_{\im{\omega}}) \cong \downarrow \im{\omega}$ by Proposition~\ref{prop:filterisotodownset}. If $\im{\omega}=q$ is atomic then the effect space will be the real numbers so that $A_{\im{\omega}}$ is a scalar-like system. Hence, using the same argumentation as in the proof of Proposition~\ref{prop:purestate}, $\cl{\omega}$ will be the unique unital state on this system. This means that any state with $\im{\omega}=q$ will be equal to $\pi_q$.
\end{proof}

\noindent We now have a correspondence between pure states and pure effects. A pure state $\omega$ has an atomic image $q=\im{\omega}$ that is a pure effect. It is the unique pure effect such that $q\circ\omega = 1$. In turn $\omega$ is the unique pure state for $q$ such that $q\circ\omega = 1$. We want this correspondence to satisfy the following property:

\begin{definition}
    We will say an operational PET has \Define{symmetry of transition probabilities}\indexd{symmetry of transition probabilities} \cite{alfsen2012geometry} when for any two atomic effects $p$ and $q$ on the same system we have $q\circ \omega_p= p\circ \omega_q$ where $\omega_p$ is the unique pure unital state with $\im{\omega_p}=p$.
\end{definition}

\noindent This is easy to show in the following case:
\begin{proposition}\label{prop:puredistinguish}
  Let $p$ and $q$ be atomic effects. We have $p\circ \omega_q = 0 \iff q\circ \omega_p = 0 \iff p\perp q$.
\end{proposition}
\begin{proof}
  $p\circ \omega_q=0 \iff p\leq \im{\omega_q}^\perp = q^\perp \iff p\perp q \iff q\perp p \iff q\leq \im{\omega_p}^\perp \iff q\circ \omega_p =0$.
\end{proof}

\noindent The general case is a bit harder to show, and we need the following lemma.

First, note that all scalars are pure maps as they are filters for themselves, and hence the dagger is defined for scalars.
\begin{lemma}\label{lem:scalarselfadjoint}
    Let $s: I\rightarrow I$ be a scalar, then $s^\dagger = s$.
\end{lemma}
\begin{proof}
    A scalar $s:I\rightarrow I$ of course corresponds to some number $s\in[0,1]$, and composition of scalars $s\circ t$ is then equal to their product $st$. We then have $(st)^\dagger = (s\circ t)^\dagger = t^\dagger \circ s^\dagger = s^\dagger t^\dagger$, so that the dagger preserves multiplication. 
    Note that for real numbers between $0$ and $1$ we have $s\leq t \iff \exists r\in[0,1]: s = rt$. As a consequence we get $s\leq t \iff s = rt \iff s^\dagger = r^\dagger t^\dagger \iff s^\dagger \leq t^\dagger$ so that the dagger is also an order-isomorphism for the unit-interval.
    Now suppose that $s\leq s^\dagger$, then by taking the dagger on both sides we get $s^\dagger \leq s$ so that $s=s^\dagger$. This of course also holds when we start with $s^\dagger \leq s$. Since the unit interval is totally ordered, one of these cases must be true and we are done.
\end{proof}

\begin{proposition}\label{prop:symoftrans}
    An operational PET has symmetry of transition probabilities.
\end{proposition}
\begin{proof}
    Let $p$ and $q$ be atomic effects. By Proposition~\ref{prop:atomicstate} we have $\omega_p = p^\dagger$ and $\omega_q = q^\dagger$. 
    Note that $q\circ \omega_p$ is a scalar and hence by the previous lemma $q\circ \omega_p = (q\circ \omega_p)^\dagger$. But also $(q\circ \omega_p)^\dagger = \omega_p^\dagger \circ q^\dagger = p\circ \omega_q$ so that $q\circ \omega_p = p\circ \omega_q$ as desired.
\end{proof}

\begin{definition}
    Let $v,w \in V$ be arbitrary vectors and write them as $v=\sum_i \lambda_i p_i$ and $w=\sum_j \mu_j q_j$, where the $p_i$ are orthogonal atoms and the same for the $q_j$. Such a decomposition can always be found by Corollary~\ref{cor:atomicspectrum}. Define the \Define{inner product}\indexd{inner product} of $v$ and $w$ to be $\inn{v,w} := \sum_{i,j} \lambda_i \mu_j (q_j\circ\omega_{p_i})$.
\end{definition}
\begin{proposition}
    The inner product defined above is indeed an inner product: well-defined, symmetric, bilinear and $\inn{v,v}\geq 0$ with $\inn{v,v}=0$ iff $v=0$.
\end{proposition}
\begin{proof}
    First note that with $v$ and $w$ as defined above, 
    $$\inn{v,w} = \sum_{i,j} \lambda_i \mu_j (q_j\circ\omega_{p_i}) = \sum_i \lambda_i (\sum_j \mu_j q_j)\circ\omega_{p_i} = \sum_i \lambda_i (w\circ\omega_{p_i})$$ 
    so that the inner product is independent of the representation of $w$ in terms of atomic effects and linear in the second argument. 
    Second, due to symmetry of transition probabilities $q_j\circ\omega_{p_i} = p_i\circ\omega_{q_j}$ and hence $\inn{v,w} = \inn{w,v}$ so that it is also independent of the representation of $v$ and linear in the first argument.

    Lastly, we have $\inn{v,v} = \sum_{i,j} \lambda_i \lambda_j p_j\circ\omega_{p_i} = \sum_i \lambda_i^2 p_i\circ \omega_{p_i} = \sum_i \lambda_i^2\geq 0$ due to Proposition~\ref{prop:puredistinguish}, and hence $\inn{\cdot,\cdot}$ indeed forms an inner product.
\end{proof}
\begin{proposition}
    The inner-product makes $V$ self-dual\indexd{inner product!self-dual ---}\indexd{self-dual}, \ie~$v\geq 0$ iff for all $w\geq 0$ we have $\inn{v,w}\geq 0$.
\end{proposition}
\begin{proof}
    If $v\geq 0$ then we can write $v=\sum_i \lambda_i p_i$ with the $p_i$ atomic and orthogonal and $\lambda_i \geq 0$ for all $i$. It then easily follows that $\inn{v,w}\geq 0$ if $w$ is also positive. For the other direction, suppose $\inn{v,w}\geq 0$ for all positive $w$, then in particular $\inn{v,p_i} = \lambda_i \geq 0$, so that $v$ is indeed positive.
\end{proof}
\begin{corollary} \label{cor:purestateconvex}
    Let $\omega\in $ St$_1(A)$ be a unital state, then $\omega = \sum_i \lambda_i \omega_{p_i}$ with $\lambda_i\geq 0$, $\sum_i\lambda_i = 1$ and the $\omega_{p_i}$ being pure states.
\end{corollary}
\begin{proof}
    The inner product defines a linear map $f:V\rightarrow V^*$ by $f(v)(w):= \inn{v,w}$. This map is an injection, so that due to finite-dimensionality it is a bijection. In particular, we can find for every $\omega \in V^*$ an element $v\in V$ such that $f(v)=\omega$ and hence $\omega(w) = \inn{v,w}$. Since $\omega(w)\geq 0$ for all $w\geq 0$ we must have $v\geq 0$. By expanding $v$ in terms of atomic effects we then get the desired result.
\end{proof}
\begin{corollary}\label{cor:stateconvexpure}
    A unital state on $A$ is pure if and only if it is convex extremal in St$_1(A)$.
\end{corollary}

\subsection{Structure of faces}\label{sec:structureoffaces}
At this point we could try to explicitly construct a Jordan product as in Section~\ref{sec:jordanproduct}. which would require results analogous to those of Section~\ref{sec:coverprop}. Proving these results in our current setting is however not straightforward, and hence we will take a different route. 

The combination of a self-dual inner product and assert maps brings us close to the setting of Alfsen and Shultz's work in Ref.~\cite{alfsen2012geometry}. In particular, by making the appropriate translations we can use their results to prove the main theorem of this chapter: that systems in an operational PET correspond to Euclidean Jordan algebras. 
Their proof, Theorem 9.33 of Ref.~\cite{alfsen2012geometry}, relies on a type of operators they also call \emph{compressions}. Their compressions will turn out to be our assert maps. To distinguish our compressions from theirs, we will call their compressions \emph{AS-compressions}.

\begin{definition}\label{def:AS-compression}
    Let $P:A\rightarrow A$ be a map in an effect theory. We call it an \Define{AS-compression}\indexd{AS-compression} when $P$ is idempotent and it is \Define{bicomplemented}\indexd{bicomplement}, \ie~when there exists an idempotent map $Q:A\rightarrow A$ such that for all effects $a$ and states $\omega$ the following implications hold:
    \begin{multicols}{2}
    \begin{itemize}
        \item $a\circ P = a \iff a\circ Q = 0$.
        \item $a\circ Q = a \iff a\circ P =0$.
        \item $P\circ \omega = \omega \iff Q\circ \omega = 0$.
        \item $Q\circ \omega = \omega \iff P\circ \omega =0$.
    \end{itemize}
    \end{multicols}
\end{definition}

\begin{proposition}\label{prop:gencompr}
    In a PET where the effects separate the states (i.e.\ where $\omega=\omega^\prime$ when $p\circ \omega = p\circ \omega^\prime$ for all effects $p$) the assert map $\asrt_p$ of a sharp effect $p$ is an AS-compression with bicomplement $\asrt_{p^\perp}$.
\end{proposition}
\begin{proof}
    In Proposition~\ref{prop:assertmaps} it was already shown that assert maps of sharp effects are idempotent so it remains to show that they are bicomplemented. The bicomplement of $\asrt_p$ will turn out to be $\asrt_{p^\perp}$.

    We have $a=a\circ \asrt_p \iff a\leq p=(p^\perp)^\perp \iff a\circ \asrt_{p^\perp}=0$ by an application of Propositions \ref{prop:assertmaps} and \ref{prop:orthog}. Obviously the same holds with $p$ and $p^\perp$ interchanged.

    Let $\omega$ be a state. Suppose $\asrt_p\circ \omega = \omega$. Then $a\circ \asrt_{p^\perp}\circ \omega = a\circ\asrt_{p^\perp}\circ \asrt_p\circ\omega \leq p^\perp\circ \asrt_p\circ \omega = 0\circ\omega = 0$ for all effects $a$. Because the effects separate the states we can conclude that $\asrt_{p^\perp}\circ \omega = 0$. Conversely, suppose $\asrt_{p^\perp}\circ \omega = 0$. 
    Then $0=1\circ \asrt_{p^\perp}\circ \omega = p^\perp\circ \omega$ so that $1\circ \omega = (p+p^\perp)\circ \omega = p\circ \omega$. Hence $\im{\omega}\leq p$ so that by Proposition \ref{prop:assertmaps} $\asrt_p\circ \omega = \omega$. The other direction we get by interchanging $p$ and $p^\perp$. So $\asrt_p$ and $\asrt_{p^\perp}$ are indeed bicomplemented.
\end{proof}

The results of Ref.~\cite{alfsen2012geometry} require a few concepts from convex geometry, in particular the notion of a face (Definition~\ref{def:strictly-convex}). Let us recall the definition of a face along with several new properties.

\begin{definition}
    Let $W$ be a real vector space and let $K\subseteq W$ be a convex subset. A \Define{face}\indexd{face} $F$ of $K$ is a convex subset such that $\lambda x + (1-\lambda) y\in F$ with $0<\lambda<1$ implies that $x, y \in F$. Any extreme point  $p\in K$ forms a face $\{p\}$. A face $F$ is called \Define{norm exposed}\indexd{norm exposed face} when there exists a bounded affine positive functional $f:K\rightarrow \R^+$ such that $f(a)=0\iff a\in F$. A face $F$ is called \Define{projective}\indexd{projective face} when there exists an AS-compression $P:W\rightarrow W$ such that for all $x\in K$ we have $P(x)=x \iff x\in F$.
\end{definition}

\begin{lemma}\label{lem:normexposed}
  Let $A$ be a system in an operational PET. Any norm-exposed face of St$_1(A)$ is projective.
\end{lemma}
\begin{proof}
  Let $V$ be the vector space associated to $A$. Let $F\subseteq $ St$_1(A)$ be a norm-exposed face. That means that there is a positive affine functional $f:$ St$_1(A)\rightarrow \R_{\geq 0}$ with $f(\omega)=0\iff \omega\in F$. As St$_1(A)$ is the state space of an order unit space $V$, it forms a \Define{base} of the \Define{base norm space}\indexd{base norm space} $V^*$~\cite[Theorem 1.19]{alfsen2012state}. As a result, its span is the entire positive cone of $V^*$~\cite[Definition 1.10]{alfsen2012state}, and $f$ extends uniquely to a positive linear map $f:V^*\rightarrow \R$~\cite[Proposition 1.11]{alfsen2012state}. In finite dimension we of course have $V\cong V^*$ so that there must be a $q\geq 0$ in $V$ such that $\forall \omega\in$ St$(A): f(\omega) = \omega(q)$. We can rescale $q$ without changing the zero set, so we can take $q$ to be an effect. By Proposition~\ref{prop:floorceiling} $\omega(q)=0\iff \omega(\ceil{q})=0$, so $q$ can be replaced by a sharp effect without changing the zero set. Now $\omega \in F \iff \omega(q)=0 \iff $ $\im{\omega}\leq q^\perp \iff \asrt_{q^\perp}\circ \omega = \omega$. Since assert maps of sharp effects are AS-compressions we see that $F$ is indeed projective.
\end{proof}

\begin{corollary} \label{cor:comp-assert}
  Let $A$ be a system in an operational PET. Any AS-compression is an assert map.
\end{corollary}
\begin{proof}
  If we have a AS-compression $P$ with complement $Q$ then we can construct $f:$ St$_1(A)\rightarrow \R^+$ by $f(\omega) = \norm{Q(\omega)}$ which is affine (this is a standard result for the norm on base norm spaces in order separation with an order unit space \cite{alfsen2012geometry}). Now obviously $f(v)=0 \iff Q(v)=0 \iff P(v)=v$, so we see that the projective face generated by $P$ is also norm exposed. But then by the previous proposition it is also the projective face of some assert map which necessarily must have the same projective unit. Because the projective unit determines the compression uniquely we see that $P$ must be equal to this assert map.
\end{proof}

\begin{lemma}\label{lem:compressionpure}
  In an operational PET, the AS-compression of a convex extremal state is proportional to a convex extremal state.
\end{lemma}
\begin{proof}
    By Corollary~\ref{cor:stateconvexpure} convex extremal states are precisely the pure states.
  The only AS-compressions are the assert maps of sharp effects, and assert maps are pure maps. By \ref{pet:dagger} the composition of pure maps is again pure, so that an AS-compression sends a convex extremal state to a pure state, which again by Corollary~\ref{cor:stateconvexpure} must be convex extremal.
\end{proof}

\begin{theorem} \label{theor:OETEJA}
  Let $\mathbb{E}$ be an operational PET. Then there exists a functor ${F:\mathbb{E}\rightarrow \EJA^\opp_{\text{psu}}}$ such that for any system $A$ the effect space $\eff(A)$ is isomorphic to the unit interval of its corresponding EJA: $\eff(A)\cong [0,1]_{F(A)}$. Furthermore, this functor is faithful if and only if $\mathbb{E}$ is locally tomographic.
\end{theorem}
\begin{proof}
    By Theorem~\ref{theor:opefftheor} we have a functor $F:\mathbb{E}\rightarrow \OUS^\opp$ that is faithful when $\mathbb{E}$ satisfies local tomography, and hence it suffices to prove that the order unit spaces in the image are actually EJAs. We show this by establishing all the conditions of Theorem 9.33 of \cite{alfsen2012geometry}. This theorem states that a state-space is isomorphic to that of a Jordan algebra when the following conditions are met:
  \begin{itemize}
      \item Every norm exposed face is projective (Lemma \ref{lem:normexposed}).
      \item The convex extremal points span the space (Corollary \ref{cor:purestateconvex}).
      \item It satisfies symmetry of transition probabilities (Proposition \ref{prop:symoftrans}).
      \item AS-compressions preserve convex extremal states (Lemma \ref{lem:compressionpure}).
  \end{itemize}
  Hence, we can indeed conclude that our systems are Euclidean Jordan algebras.
\end{proof}

We have now shown that systems in an operational PET correspond to Euclidean Jordan algebras. Conversely, in Chapter~\ref{chap:jordanalg} we will show that the opposite category of EJAs with positive sub-unital maps is an operational PET. Hence, we have characterized operational PETs. 

\section{Monoidal Operational PETs}\label{sec:monoidalPETs}

In this section we will see what additional restrictions are imposed on the systems by adding a monoidal structure to the theory that respects the structure of pure maps.

The prototypical example of a monoidal operational PET is \textbf{CStar}$^\opp_{\text{cpsu}}$, the opposite category of complex finite-dimensional C$^*$-algebras with completely positive sub-unital maps. Another example is \textbf{RStar}$^\opp_{\text{cpsu}}$ of \emph{real} finite-dimensional C$^*$-algebras with completely positive sub-unital maps. A real finite-dimensional C$^*$-algebra is a direct sum of the real matrix algebras $M_n(\R)$. The aim of this section is to prove that these are in fact the only two possibilities. 

\begin{remark}
  Many of the results in this section are very similar to those of Section~\ref{sec:seqmeas-tensorproducts}. In those cases we will omit the proof and give the appropriate reference.
\end{remark}

In this section we will fix a monoidal operational PET $\mathbf{E}$, and we will let $V$ and $W$ denote the order unit spaces associated to a pair of objects in $\mathbf{E}$. By the results of the previous section, these are in fact EJAs. The monoidal structure of the PET lifts to a bilinear map $V\times W\rightarrow V\otimes W$ where $V\otimes W$ is another EJA corresponding to some system of $\mathbf{E}$. In particular, if we have $v\geq 0$ in $V$ and $w\geq 0$ in $W$ then $v\otimes w\geq 0$ in $V\otimes W$.
Note that a priori $V\otimes W$ is not necessarily related to the regular vector space tensor product. 

\begin{proposition} \label{prop:compositeproperties}
    In a monoidal operational PET the following are true.
    \begin{enumerate}[label=\alph*)]
        \item A composite of pure maps is again pure.
        \item A composite of normalised states is again a normalised state.
        \item A composite of atomic effects is again an atomic effect.
        \item For atomic effects $p$ and $q$ we have $\omega_p\otimes \omega_q = \omega_{p\otimes q}$.
        \item If $q_1\perp p_1$ and $q_2\perp p_2$ are atomic orthogonal effects, then $q_1\otimes q_2 \perp p_1\otimes p_2$.
    \end{enumerate}
\end{proposition}
\begin{proof}~
    \begin{enumerate}[label=\alph*)]
        \item By definition of a monoidal PET.
        \item Given two normalized states $\omega_1$ and $\omega_2$ we see that $1\circ(\omega_1\otimes \omega_2) = (1\otimes 1)\circ(\omega_1\otimes \omega_2) = (1\circ\omega_1) \otimes (1\circ \omega_2) = 1\otimes 1 = 1$.
        \item Let $p$ and $q$ be atomic effects. By Proposition \ref{prop:purestate} this is equivalent to them being sharp and pure. By the previous point we know that $p\otimes q$ is also pure and hence $p\otimes q$ must be proportional to an atom: $p\otimes q = \lambda r$. We calculate $1 = (p\otimes q)\circ (\omega_p\otimes \omega_q) = (\lambda r)\circ (\omega_p\otimes \omega_q) \leq (\lambda 1)\circ (\omega_p\otimes \omega_q) = \lambda$, and hence $\lambda=1$ and we are done.
        \item We know that $p\otimes q$ is atomic, and we know that $\omega_p\otimes \omega_q$ is a unital pure state (since it is a composite of pure unital states). We of course have $(p\otimes q)\circ (\omega_p\otimes \omega_q) = 1$, but by Proposition \ref{prop:atomicstate}, the state $\omega_{p\otimes q}$ is the unique state with this property and hence $\omega_{p\otimes q} = \omega_p\otimes \omega_q$.
        \item By Proposition \ref{prop:puredistinguish}, atomic effects $p$ and $q$ are orthogonal if and only if $q\circ \omega_p = 0$. So supposing that $q_1\perp p_1$ and $q_2\perp p_2$ we calculate $(q_1\otimes q_2)\circ\omega_{p_1\otimes p_2} = (q_1\otimes q_2)\circ (\omega_{p_1}\otimes \omega_{p_2}) = (q_1\circ \omega_{p_1})\otimes (q_2\circ \omega_{p_2}) = 0\otimes 0 = 0$. \qedhere
    \end{enumerate}
\end{proof}

\begin{definition}
    The \Define{rank}\indexd{rank!of Euclidean Jordan algebra} of an EJA $V$, denoted by $\rnk~V$, is equal to the maximal size of any set of orthogonal atomic effects in $V$.
\end{definition}
\noindent The rank of the $n\times n$ matrix algebra $M_n(\mathbb{F})_\sa$ is equal to $n$. The rank of a spin factor is always equal to 2. Note that the size of a set of of orthogonal atomic effects $\{p_i\}$ is equal to the rank of the space if and only if $\sum_i p_i = 1$.

\begin{proposition}\label{prop:rankcomposite}
    Rank is preserved by compositing: $\rnk~V\otimes W = \rnk~V \rnk~W$. We also have $\dim~V\otimes W \geq \dim~V \dim~W$.
\end{proposition}
\begin{proof}
    Let $\{p_i\}$ be a maximal set of orthogonal atomic effects in $V$, and let $\{q_j\}$ be a maximal set of orthogonal atomic effects in $W$. By maximality we must have $\sum_i p_i = 1_V$ and $\sum_j q_j = 1_W$. By Proposition \ref{prop:compositeproperties} the set $\{p_i\otimes q_j\}$ also consists of orthogonal atomic effects. Furthermore $\sum_{i,j} p_i\otimes q_j = (\sum_i p_i)\otimes (\sum_j q_j) = 1_V\otimes 1_W = 1_{V\otimes W}$ so that this set must also be maximal.

    For the second part let $\{p_i\}$ be a basis of atomic effects of $V$ and similarly let $\{q_j\}$ be a basis of atomic effects in $W$. Suppose that $\dim~V\otimes W < \dim~V \dim~W$, then $\{p_i \otimes q_j\}$ must be linearly dependent in $V\otimes W$. In the same way as in the proof of Lemma~\ref{lem:tensorisbijective} this can be shown to lead to a contradiction, and hence we must have $\dim~V\otimes W \geq \dim~V \dim~W$.
\end{proof}

\begin{proposition}\label{prop:simplecomposite}
    Let $V$ and $W$ be simple. Then their composite $V\otimes W$ is also simple.
\end{proposition}
\begin{proof}
    We know that the composite $V\otimes W$ has rank $(\rnk~V) (\rnk~W)$, so if we can show that $V\otimes W$ must contain a simple factor of this rank than we are done.
    Writing $p\mult a := a\circ \asrt_p$ for a sharp effect $p\in V$ and any effect $a\in V$ we can use Proposition~\ref{prop:simplecornerissimple}, Lemma~\ref{lem:nonzeroatompairs} and finally the proof of Proposition~\ref{prop:tensordirectsum} to prove this result.
\end{proof}

\begin{proposition}\label{prop:simpleEJAcomplexreal}
    Let $V$ be simple. Then $V=M_n(\mathbb{F})_\sa$ with $\mathbb{F}=\R$ or $\mathbb{F}=\C$.
\end{proposition}
\begin{proof}
    This can be shown by a simple case distinction and dimension counting argument as in Proposition~\ref{prop:simpleEJAsquare}. We will work out one specific case, the other ones follow similarly. Suppose $V=M_n(\mathbb{H})_\sa$ for $n\geq 2$. By Propositions \ref{prop:rankcomposite} and \ref{prop:simplecomposite} we then know that $V\otimes V$ must be a simple EJA with rank $n^2$ and $\dim(V\otimes V)\geq \dim(V)^2$. The simple EJA of rank $n^2$ with the highest dimension is $M_{n^2}(\mathbb{H})_\sa$. When $n>1$ however, the dimension of this space is still lower than $\dim(M_n(\mathbb{H})_\sa)^2$, and hence such an EJA does not exist.

    Note that the spin factors $S_2$ and $S_3$ do allow the right sort of composites, but that these are in fact isomorphic to respectively real and complex matrix algebras.
\end{proof}

\begin{proposition}\label{prop:complexrealexclusion}
    Let $V$ and $W$ both be simple. Then both $V$ and $W$ are real matrix algebras, or both of them are complex matrix algebras.
\end{proposition}
\begin{proof}
    By the previous proposition we know that they both must be real or complex, so that the only thing we need to show is that it cannot be that $V$ is complex while $W$ is real.

    Let $V=M_n(\C)_\sa$ and $W=M_m(\R)_\sa$. By dimension counting and the previous proposition we know that $V\otimes W = M_{nm}(\C)_\sa$. Let $\omega$ be a pure state on $V$. The identity map on $W$ is of course pure, so that $\omega\otimes \id: M_{nm}(\C)_\sa \rightarrow M_m(\R)_\sa$ is also a pure map. However it follows from Examples~\ref{ex:compression-quantum} and~\ref{ex:filter-quantum} that the compression systems for any complex matrix algebra are again complex matrix algebras. Hence, a pure map $f: M_k(\C)_\sa\rightarrow W$ (which is a composition of a filter and a compression) must have $W\cong M_l(\C)_\sa$ for some $l\in\N$. As a real and complex matrix algebra are only isomorphic in the trivial case, the desired result follows.
\end{proof}

\begin{theorem}\label{theor:compositealgebras}
    Let $\mathbb{E}$ be an monoidal operational PET. Then there is a functor $F:\mathbb{E}\rightarrow \mathbb{D}$ where $\mathbb{D} = \textbf{CStar}^\opp_{\text{cpsu}}$ or $\mathbb{D} = \textbf{RStar}^\opp_{\text{cpsu}}$. This functor preserves the effect space of objects: $\eff(A)\cong [0,1]_{F(A)}$. 
\end{theorem}
\begin{proof}
By Theorem~\ref{theor:OETEJA} we have a functor into $\EJA_{\text{psu}}^\opp$, so it suffices to show that this functor restricts in the correct manner.

Let $A$ be a system in a monoidal operational PET. We know that $\eff(A)$ is isomorphic to the unit interval of a Euclidean Jordan algebra. Let $c$ be a minimal central element of this EJA, that hence corresponds to a simple factor of the algebra. The compression space associated to $c$ is then a simple algebra, and hence by Proposition~\ref{prop:simpleEJAcomplexreal} it must be a real or complex matrix algebra. let $d$ be another minimal central element. Then its associated simple factor is also a real or complex matrix algebra and furthermore this algebra is associated to some system in the PET. By Proposition~\ref{prop:complexrealexclusion} we can then conclude that either both the simple factors associated to $c$ and $d$ are real, or they are complex. We conclude that $\eff(A)$ is indeed isomorphic to the unit interval of a real or complex C$^*$-algebra. That all the systems of $\mathbb{E}$ must be real or complex follows similarly.
Hence the functor of Theorem~\ref{theor:OETEJA} restricts to the category of real or complex C$^*$-algebras. That all the maps must be completely positive follows because positivity must be preserved even on composite systems.
\end{proof}

\begin{remark}
  This theorem does not mention the faithfulness of this functor. If $\mathbb{E}$ satisfies local tomography, then in the same way as in Theorem~\ref{theor:OETEJA} we see that it is faithful, but we expect that faithfulness already holds if $\mathbb{E}$ merely satisfies tomography. To prove this we would need to show that the functor $F$ is strongly monoidal, which does not seem trivial. Analogously to Theorem~\ref{thm:seqprodlocalcomp} we also expect that if $\mathbb{E}$ satisfies local tomography that the effect spaces of $\mathbb{E}$ must then be isomorphic to complex C$^*$-algebras, but this again requires knowing more about the interaction of the functor with the monoidal structure.
\end{remark}

\begin{remark}
The opposite category of von Neumann algebras with completely positive normal sub-unital maps $\textbf{vNA}^\opp_{\text{cpsu}}$  is also a monoidal PET~\cite{bramthesis,basthesis}, but since we require all systems to be finite-dimensional, it is not an operational PET. Finding suitable conditions under which we retrieve $\textbf{vNA}^\opp_{\text{cpsu}}$ (or the bigger category of JBW-algebras and positive maps) is discussed in Chapter~\ref{chap:infinitedimension}.
\end{remark}

\chapter{The category of JBW-algebras}\label{chap:jordanalg}

In the preceding two chapters we studied physical theories satisfying properties related to sequential measurement and purity. We established that these properties forced the systems in these physical theories to correspond to Euclidean Jordan algebras (or C$^*$-algebras when composite systems are also required). In this chapter we will consider the converse direction: we will show that Euclidean Jordan algebras satisfy the assumptions outlined in Chapters~\ref{chap:seqprod} and~\ref{chap:effectus}.

In fact, we will consider a larger class of Jordan algebras known as \emph{JBW-algebras} that also include infinite-dimensional algebras. JBW-algebras are to EJAs what von Neumann algebras are to finite-dimensional C$^*$-algebras.

We will first introduce the basic theory of Jordan algebras in Section~\ref{sec:Jordanalg}. Then in Section~\ref{sec:Jordan-operator-alg} we introduce JBW-algebras. Sections~\ref{sec:JBW-floor-ceiling}--\ref{sec:division-filter} establish that JBW-algebras have compressions and filters as defined in Section~\ref{sec:filterscompressions}.
Section~\ref{sec:diamond-dagger} shows that the pure maps in a JBW-algebra form a dagger-category, proving that JBW-algebras are an example of a PET (cf.~Definition~\ref{def:PET}). In fact, we find that the pure maps satisfy some stronger conditions that we will use as new axioms for a reconstruction in Chapter~\ref{chap:infinitedimension}.
Finally, in Section~\ref{sec:JBW-SEA} we establish that the set of effects of a JBW-algebra has a sequential product satisfying the conditions outlined in Section~\ref{sec:seqmeas}.

As the theory of Jordan and JBW-algebras is intricate and technical, we will give appropriate references where needed, instead of reproving known results. Our standard reference is the excellent book \emph{Jordan operator algebras} by Hanche-Olsen and St\"ormer~\cite{hanche1984jordan}. Other books containing most of the material we need are Refs.~\cite{mccrimmon2006taste,alfsen2012geometry}.
While we endeavored to make this chapter as accessible as possible, some basic familiarity with C$^*$-algebras or von Neumann algebras should help the reader's understanding.

The title of this chapter is a reference to the PhD thesis of Abraham Westerbaan, \emph{The category of von Neumann algebras}~\cite{bramthesis}, as this chapter generalises many of his results. Wherever possible we tried to follow his line of argument, although in quite some cases the proofs become considerably harder in our setting.

\section{Jordan algebras}\label{sec:Jordanalg}

Let us recall some of the basic definitions regarding Jordan algebras that we introduced in Section~\ref{sec:operator-algebras}.

\begin{definition}
    A \Define{Jordan algebra}\indexd{Jordan algebra} $(E,*,1)$ over a field $\mathbb{F}$ is a vector space over $\mathbb{F}$ equipped with a unital commutative bilinear operation $*: E\times E\rightarrow E$
    that satisfies the \Define{Jordan identity}\indexd{Jordan identity}: 
    \[(a*b)*(a*a) = a*(b*(a*a))\]
    We will refer to this operation as the \Define{Jordan product}\indexd{Jordan product} of the algebra.
    A linear map $f:E\rightarrow F$ between Jordan algebras is a \Define{(unital) Jordan homomorphism}\indexd{homomorphism!Jordan ---} when $f(a*b) = f(a)*f(b)$ (and $f(1)=1$).
\end{definition}

\begin{remark}
    The Jordan product does not have to be associative. The Jordan identity can be seen as a weaker form of associativity that only applies to certain combinations of the product.
\end{remark}

\begin{example}
    Let $(A,\cdot,1)$ be any associative unital algebra over a field $\mathbb{F}$, \ie~a vector space over $\mathbb{F}$ with $\cdot:A\times A\rightarrow A$ bilinear and satisfying $a\cdot(b\cdot c) = (a\cdot b)\cdot c$. Assume $\mathbb{F}$ is of characteristic different than $2$ so that $2$ has an inverse in $\mathbb{F}$.
    Then the operation $*:A\times A\rightarrow A$ defined by
    \[a*b := \frac12 (a\cdot b + b\cdot a)\]
    makes $(A,*,1)$ into a Jordan algebra. We will refer to this operation as the \Define{special Jordan product}\indexd{special Jordan product} of the algebra.
    Any Jordan algebra that embeds into an associative algebra equipped with this Jordan product is called \Define{special}\indexd{special Jordan algebra}\indexd{Jordan algebra!special ---}\indexd{special Jordan algebra}.
\end{example}

\begin{assumption}
    For the remainder of this section we will assume that all Jordan algebras are over some field that is not of characteristic 2. This is needed so that $-1\neq 1$. The theory of Jordan algebras over a field of characteristic 2 is subtly different, and out of scope for this thesis.
\end{assumption}

To proceed we need some basic algebraic properties
of Jordan algebras, which are most conveniently expressed
with some additional notation.

\begin{definition}
Let~$E$ be a Jordan algebra.
\begin{itemize}
\item
We write $a^0:=1$,\quad $a^1:= a$,\quad  $a^2:= a*a$,\quad  
        $a^3:= a*a^2$, \dots .
Note that since~$*$ is not associative
it's not a priori clear whether equations like  $a^4 = a^2 * a^2$ hold.
\item We write $J(a)\sse E$ for the Jordan algebra generated by $a$ (which will turn out to consist of all polynomials in $a$).
\item
Given~$a\in E$ we write $T_a: E\rightarrow E$ for the linear operator $T_a(b) := a*b$. We call these operators \Define{product maps}\indexd{product map!in Jordan algebras}\index{math}{Ta@$T_a$ (product map of Jordan algebra)}.
\item 
Given two linear maps $S,T\colon E\to E$
we write $[S,T]:= ST-TS$ for the commutator of~$S$ and~$T$.
\end{itemize}
\end{definition}

Note that because the Jordan product is bilinear, we have $T_{a+\lambda b} = T_a + \lambda T_b$. This allows us to derive the \Define{linearised Jordan equations}\indexd{linearised Jordan equations}:

\begin{lemma}[{\cite[Section 2.4.2]{hanche1984jordan}}]\label{lem:jordan-equations}
Given a Jordan algebra~$E$,
and $a,b,c\in E$, we have
\begin{enumerate}[label=\alph*), ref=\fullcounter.\alph*]
\item \label{eq:jordan1}
$[T_a,T_{a^2}] = 0$
\item \label{eq:jordan2}
$[T_b,T_{a^2}] = 2[T_{a*b},T_{a}]$ 
\item \label{eq:jordan3}
$[T_a,T_{b*c}] + [T_b,T_{c*a}] + [T_c,T_{a*b}] = 0$;
\end{enumerate}
\end{lemma}
\begin{proof}~
\begin{enumerate}[label=\alph*)]
    \item The first equation, $[T_a,T_{a^2}]=0$, is just a reformulation of the Jordan identity:
        \begin{equation*}
                T_aT_{a^2}b\,=\, a*(b*a^2) \ =\  (a*b)*a^2
            \,=\, T_{a^2}T_a b.
        \end{equation*}
    
    \item Take the equality $[T_d,T_{d^2}] = 0$ and let $d=a\pm b$: $[T_{a\pm b},T_{(a\pm b)^2}]~=~0$. After expanding the terms using linearity we are left with
    $$[T_a, T_{a^2}] \pm [T_b, T_{b^2}] \pm \left([T_b, T_{a^2}] + 2 [T_a, T_{ab}]\right) +\left([T_a, T_{b^2}] + 2 [T_b, T_{ab}]\right) = 0.$$
    Subtracting the equation for $d=a+b$ from the equation for $d=a-b$ and dividing the result by 2 (here we use that the field is not of characteristic 2) we have the desired equation (as $[T_a,T_{a^2}] = [T_b,T_{b^2}]=0$).

    \item Take the equation of the previous point and replace $a$ by $a\pm c$. Using the same trick as before we arrive at the desired equation. \qedhere
\end{enumerate}
\end{proof}

\begin{lemma}[{\cite[Section 2.4.4]{hanche1984jordan}}]
    Given a Jordan algebra~$E$ and $a,b,c \in E$ we have
    \begin{equation}\label{eq:jordan-normalise}
        T_{a*(b*c)} = T_aT_{b*c} + T_b T_{c*a} + T_c T_{a*b} - T_b T_a T_c - 
    T_c T_a T_b.
    \end{equation}
\end{lemma}
\begin{proof}
    Apply an element $d$ to both sides of the equation \eqref{eq:jordan3} and bring all the negative terms to the right to get:
    $$a((bc)d) + b((ac)d) + c((ab)d) = (bc)(ad) + (ac)(bd) + (ab)(cd).$$
    Note that for clarity we are simply writing `$ab$' instead of `$a*b$'.
    Observe that the right-hand side of this equation is invariant under an interchange of $a$ and $d$ so that the left-hand side must be as well. This leads to the equality
    \begin{align*}
    a((bc)d) + b((ac)d) + c((ab)d) &= d((bc)a) + b((dc)a) + c((db)a) \\
    &= ((bc)a)d + b(a(cd)) + c(a(bd))
    \end{align*}
    where we have used the commutativity of the product to move $d$ to the end of all the terms in the last equality.
    Translating this back into multiplication operators, using that this equality holds for all $d$, and bringing some terms to the other side then gives the desired equation.
\end{proof}

We will refer to equation \eqref{eq:jordan-normalise} as the \Define{normalisation equation}\indexd{normalisation equation (Jordan algebra)}. Using this equation we can reduce a product map of an arbitrary expression into a polynomial of product maps containing at most two terms. For instance:

\begin{proposition}\label{prop:Jordan-normalise-powers}
  Let $E$ be a Jordan algebra and let $a\in E$. For any $n\in \N$, $T_{a^n}$ can be written as a polynomial in $T_{a^2}$ and $T_a$.
\end{proposition}
\begin{proof}
  We prove by induction. It is obviously true for $n=1,2$. Suppose it is true for all $k\leq n$. Then by Eq.~\eqref{eq:jordan-normalise} $T_{a^{n+1}} = T_{a*(a*a^{n-1})} = T_a T_{a^n} + T_a T_{a^n} + T_{a^{n-1}} T_{a^2} - T_a^2T_{a^{n-1}} - T_{a^{n-1}} T_a^2$. Expanding each of the $T_{a^n}$ and $T_{a^{n-1}}$ as polynomials of $T_a$ and $T_{a^2}$ finishes the proof.
\end{proof}

\subsection{Triple and quadratic product}

Calculating with the Jordan product is not always straightforward. In many situations it turns out to be easier to work with a related operation that can be defined on any Jordan algebra.

\begin{definition}
    Let $E$ be a Jordan algebra, $a,b,c\in E$. We define its \Define{triple product}\indexd{triple product (Jordan algebra)} as 
    \begin{equation}
        \{a,b,c\} := (a*b)*c + (c*b)*a - (a*c)*b.
    \end{equation}
    This is linear in all three arguments, and in particular in $b$, so that we can define the \Define{triple product map} of $a$ and $c$ as $Q_{a,c}b = \{a,b,c\}$, i.e.~$Q_{a,c} := T_aT_c + T_cT_a - T_{a*c}$.\index{math}{Qab@$Q_{a,b}$ (triple product map)}
    We define the \Define{quadratic product}\indexd{quadratic product!of a Jordan algebra} of $a$ as $Q_a := Q_{a,a}$\index{math}{Qa@$Q_a$ (quadratic product)} so that $Q_a b = \{a,b,a\}$. Note that $Q_a b = 2a*(a*b) - a^2 * b$.
\end{definition}

\begin{example}
    Let $(A,\cdot,1)$ be an associative algebra and let $*$ denote the special Jordan product. Then $\{a,b,c\} = \frac12 (a\cdot b\cdot c + c\cdot b\cdot a)$ and $Q_a c = \{a,c,a\} = a\cdot c\cdot a$
\end{example}

\begin{remark}
  Recall that the sequential product in a C$^*$-algebra was defined for effects as $a\mult b = \sqrt{a}b\sqrt{b}$. The form of the quadratic product in a special Jordan algebra above hence shows that $a\mult b = Q_{\sqrt{a}}b$ in the special case of a C$^*$-algebra. We will see in Section~\ref{sec:JBW-SEA} that this operation is also a sequential product in an arbitrary JBW-algebra.
\end{remark}

It is easily verified that $Q_a 1 = a^2$. The quadratic product satisfies several other useful identities.

\begin{proposition}
    Let $E$ be a Jordan algebra, $a\in E$. Then $Q_{a^2} = Q_a^2$.
\end{proposition}
\begin{proof}
        The equation $Q_{a^2} = 2L_{a^2}^2 - L_{a^2*a^2}$ can be normalised to $L_{a^2}^2 + 4 L_a^4 - 4 L_{a^2}L_a^2$ using equation~\eqref{eq:jordan-normalise} repeatedly. This is easily shown to be equal to $Q_a^2$ by usage of Eq.~\eqref{eq:jordan1}.
\end{proof}

The next identity is known as \Define{the fundamental identity of quadratic Jordan algebras}\indexd{fundamental identity} (or just `fundamental identity' for short).
\begin{theorem}
    Let $E$ be a Jordan algebra and $a,b \in E$. Then $Q_{Q_a b} = Q_aQ_bQ_a$.
\end{theorem}

Although this equation might seem quite arbitrary, it is in fact of fundamental importance to the theory of Jordan algebras. Note that this equation is evidently true in a special Jordan algebra:
$Q_{Q_a b} c = (Q_a b)\cdot c \cdot (Q_a b) = (a\cdot b\cdot a)\cdot c \cdot (a\cdot b\cdot a)$, while $Q_aQ_bQ_a c = a\cdot(b\cdot(a\cdot c\cdot a)\cdot b)\cdot a$. 

There is unfortunately no easy proof of the fundamental identity in an arbitrary Jordan algebra. Many textbooks~\cite{hanche1984jordan,alfsen2012geometry,chu2011jordan,mccrimmon2006taste} prove the fundamental identity as a consequence of \Define{MacDonald's theorem}\indexd{MacDonald's theorem}. This theorem states that any polynomial identity of Jordan products in 3 variables that is linear in at least one variable, is true for any Jordan algebra if and only if it is true for any special Jordan algebra. For real Jordan algebras methods from analysis can be used to prove it~\cite{faraut1994analysis}.

A more straightforward, though lengthy, algebraic proof using an automated program to do the dozens of necessary rewrite steps is found in Ref.~\cite{wetering2018algebraic}.

\subsection{Operator commutativity}\label{sec:Jordan-commutativity}

In this section we will collect and prove some results regarding operator commutativity in general Jordan algebras.

\begin{definition}\label{def:Jordan-operator-commute}
    Let $E$ be a Jordan algebra. We say $a,b\in E$ \Define{operator commute} when their Jordan product maps $T_a,T_b:E\rightarrow E$ commute, or equivalently when $a*(c*b) = (a*c)*b$ for all $c\in A$. We write $a\commu b$ to denote that $a$ and $b$ operator commute. An element is \Define{central}\indexd{central element} when it operator commutes with every element of the algebra.
\end{definition}

\begin{definition}
    Let $E$ be a Jordan algebra, and let $S\sse E$ be some subset. We write $S'$ for the \Define{commutator} of $S$, defined as $S' := \{a\in E~;~ \forall s\in S: a\commu s\}$. 
\end{definition}

The following shows that a Jordan algebra is always \Define{power associative}.\indexd{power associative}

\begin{proposition}
  For any $a\in E$ and $n,m\in \N$, $a^n$ and $a^m$ operator commute and $a^n*a^m = a^{n+m}$, and hence $a^n*(b*a^m)=(a^n*b)*a^m$ for all $b\in E$.
\end{proposition}
\begin{proof}
  By Proposition~\ref{prop:Jordan-normalise-powers} we can reduce both $T_{a^n}$ and $T_{a^m}$ to polynomials in $T_{a^2}$ and $T_a$, which commute by Eq.~\eqref{eq:jordan1}.
\end{proof}

\begin{corollary}
    For any $a\in E$, the Jordan algebra $J(a)$ generated by $a$ is associative.
\end{corollary}

For general Jordan algebras, it could be that $a$ and $b$ operator commute, while $a^2$ does not operator commute with $b$ (see for instance~Ref.~\cite[Remark 2.5.2]{hanche1984jordan}).
Using the decomposition of $T_{a^n}$ into $T_{a^2}$ and $T_a$ we do get the following.

\begin{lemma}\label{lem:Ja-prime}
  For any $a,b\in E$, $b\in J(a)'$ iff $b\commu a,a^2$.
\end{lemma}

Slightly more involved is the following result that gives a sufficient condition for $a$ and $b$ to generate an associative Jordan algebra.

\begin{proposition}\label{prop:simple-operator-commutation}
  Let $E$ be a Jordan algebra with $a,b\in E$ and suppose $a\commu b, b^2$ and $b\commu a, a^2$. Then $a$ and $b$ generate an associative Jordan algebra of mutually operator commuting elements.
\end{proposition}
\begin{proof}
  By repeatedly applying Eq.~\eqref{eq:jordan-normalise} any $T_p$ where $p$ is a polynomial in $a$ and $b$ can be reduced to a polynomial in $T_a$, $T_{a^2}, T_b, T_{b^2}$ and $T_{a*b}$. It hence remains to show that $a^2\commu b^2$, and that $a*b$ operator commutes with $a, a^2, b$ and $b^2$. Eq.~\eqref{eq:jordan2} already gives $a*b\commu a,b$. With the same equation, but now taking $b:=b^2$, we see that $b^2 \commu a^2 \iff a*b^2\commu a$. Applying Eq.~\eqref{eq:jordan-normalise} to $T_{a*b^2}$ we see that it reduces to a polynomial in $T_b, T_a, T_{a*b}$ and $T_{b^2}$ and since $T_a$ commutes with them all, it commutes with $T_{a*b^2}$, and hence $a^2\commu b^2$.

    Taking Eq.~\eqref{eq:jordan3} with $a:= a^2, b:= a, c:= b$ we get $[T_{a^2}, T_{a*b}] = - [T_a, T_{b*a^2}] - [T_b, T_{a^3}]$. As $b\commu a,a^2$ we also have $b\commu a^3$, and hence this last term disappears. Normalising $T_{b*a^2}$ we see that $[T_a, T_{b*a^2}] = 0$ and hence indeed $[T_{a^2}, T_{a*b}] = 0$. Showing that $b^2\commu a*b$ follows entirely analogously.
\end{proof}

The situation regarding operator commutativity will turn out to be more streamlined in JBW-algebras, although proving that requires quite some setup. We revisit the topic of operator commutativity in Sections~\ref{sec:JBW-commutativity} and~\ref{sec:JBW-commutation-revisited}.

\section{Jordan operator algebras}\label{sec:Jordan-operator-alg}

The previous section concerned itself with the abstract theory of general Jordan algebras. For the remainder of this chapter we will work with a more concrete structure, where the field is the real numbers and the Jordan algebra is equipped with a norm. We recall a few definitions regarding normed spaces.

\begin{definition}
  A (real or complex) \Define{Banach space}\indexd{Banach space} $(V,\norm{\cdot})$ is a (real or complex) vector space $V$ equipped with a norm $\norm{\cdot}$ such that $V$ is complete in the topology induced by the norm. A linear map $f:V\rightarrow W$ between Banach spaces $(V,\norm{\cdot}_V)$ and $(W,\norm{\cdot}_W)$ is an \Define{isometry}\indexd{isometry} when $\norm{f(v)}_W = \norm{v}_V$ for all $v\in V$. We say $V$ and $W$ are \Define{isometrically isomorphic} when there is an isometry $f:V\rightarrow W$ that is a bijection.
\end{definition}

\begin{definition}\label{def:JB-algebra}
  Let $(A,*,1,\norm{\cdot})$ be a real Banach space $(A,\norm{\cdot})$ that is also a Jordan algebra $(A,*,1)$. The space $A$ is a \Define{JB-algebra}\indexd{JB-algebra} (Jordan-Banach) if the Jordan product $*$ satisfies for all $a,b\in A$:
  \begin{enumerate}[label=\alph*)]
    \item $\norm{a*b} \leq \norm{a}\norm{b}$.
    \item $\norm{a^2} = \norm{a}^2$.
    \item $\norm{a^2} \leq \norm{a^2 + b^2}$.
  \end{enumerate}
\end{definition}

\begin{remark}
	In Ref.~\cite{hanche1984jordan} JB-algebras are not required to have a unit. In this thesis we will only deal with unital JB-algebras.
\end{remark}

JB-algebras can be seen as a generalisation of Euclidean Jordan algebras that also allows infinite-dimensional algebras. In fact, in finite dimension, JB-algebras are precisely EJAs.

\begin{proposition}\label{prop:EJA-is-JB}
  A Euclidean Jordan algebra is a JB-algebra. Conversely, any finite-dimensional JB-algebra is a Euclidean Jordan algebra.
\end{proposition}
\begin{proof}
  EJAs are precisely formally real finite-dimensional Jordan algebras by Ref.~\cite[Proposition VIII.4.2]{faraut1994analysis}, and as shown in Ref.~\cite[Corollary 3.1.7]{hanche1984jordan}, any finite-dimensional formally real Jordan algebra is a JB-algebra. Conversely, by \cite[Corollary 3.3.8]{hanche1984jordan}, any JB-algebra is formally real, and hence any finite-dimensional JB-algebra is an EJA.
\end{proof}

Definition~\ref{def:JB-algebra} establishes a JB-algebra as a real Jordan algebra with a suitably interacting complete norm. We could have equivalently defined it as a complete order unit space with a suitably interacting Jordan algebra structure, as the following proposition makes clear.

\begin{proposition}[{\cite[Proposition 3.1.6]{hanche1984jordan}}]\label{prop:JB-define-as-OUS}
  Let $A$ be a JB-algebra. Then $A$ is an order unit space complete in the order-unit norm, and for all $a\in A$:
  \begin{equation}\label{eq:Jordan-OUS}
    -1 \leq a \leq 1 \implies 0\leq a^2\leq 1.
  \end{equation}
  Conversely, any complete order unit space with a Jordan product satisfying Eq.~\eqref{eq:Jordan-OUS} is a JB-algebra.
\end{proposition}

\begin{remark}
  As a JB-algebra is an order unit space, it comes with a partial order~$\leq$. The positive elements $a\geq 0$ in a JB-algebra precisely correspond with the squares: \[a\geq 0 \iff \exists b: a=b^2 := b*b.\]
\end{remark}

\begin{example}
	Let $\mathfrak{A}$ be a unital C$^*$-algebra. Then the set of self-adjoint elements $\mathfrak{A}_{\sa}$ forms a JB-algebra with the special Jordan product $a*b := \frac12(ab+ba)$ and the regular C$^*$-norm. The positive elements of $\mathfrak{A}_{\sa}$ (in the JB-algebra sense) are precisely the positive elements of $\mathfrak{A}$ (in the C$^*$-algebra sense).
\end{example}

\begin{definition}
	Let $A$ be a JB-algebra. We say $A$ is a \Define{JC-algebra}\indexd{JC-algebra} when there exists a C$^*$-algebra $\mathfrak{A}$ so that $A$ is isometrically isomorphic to a norm-closed subset of~$\mathfrak{A}_{\sa}$.
\end{definition}

The isometry $\phi: A\rightarrow \mathfrak{A}_{\sa}$ mapping a JC-algebra into its C$^*$-algebra is necessarily a Jordan homomorphism~\cite{wright1978isometries}. When studying JC-algebras as a Jordan algebra we can then hence without loss of generality assume it be a Jordan subalgebra of $\mathfrak{A}_{\sa}$.

\begin{definition}
	Let $f:A\rightarrow B$ be a linear map between JB-algebras. 
  It is \Define{positive}\indexd{positive map!between JB-algebras} when $f(a)\geq 0$ if $a\geq 0$. 
  It is \Define{unital}\indexd{unital map!between JB-algebras} when $f(1)=1$, and \Define{sub-unital}\indexd{sub-unital map!between JB-algebras} when $f(1)\leq 1$. 
  We denote by \textbf{JB}$_{\text{psu}}$ the category of JB-algebras with positive sub-unital maps.
	\index{math}{JBpsu@\textbf{JB}$_\text{psu}$} 
  The \Define{states}\indexd{state!state of JB-algebra} of $A$ are positive unital maps $f:A\rightarrow \R$ . We denote the set of states of $A$ by~$\st(A)$.
\end{definition}

Note that a Jordan homomorphism $f$ between JB-algebras is always positive, since for a positive $a:=b^2$ we have $f(a) = f(b^2) = f(b)^2 \geq 0$. The Jordan-product maps $T_a:A\rightarrow A$ are in general not positive. Fortunately, the quadratic maps are better behaved.
\begin{proposition}[{\cite[Proposition 3.3.6]{hanche1984jordan}}]
	Let $A$ be a JB-algebra and let $a\in A$ be an arbitrary element (not necessarily positive). Then $Q_a:A\rightarrow A$ is a positive map.
\end{proposition}

\begin{proposition}\label{prop:JB-is-effect-theory}
    \textbf{JB}$_{\text{psu}}^\opp$ is an effect theory.%
    \footnote{It is in fact also an effectus, which can be easily shown using the \emph{grounded biproduct} construction of an effectus; cf.~Ref.~\cite[Section~7.1]{kentathesis}. The same holds for the later categories of JB- and JBW-algebras we define.}
\end{proposition}
\begin{proof}
    JB-algebras are order unit spaces and hence \textbf{JB}$_{\text{psu}}^\opp$ is a full subcategory of $\OUS^\opp$ that we already saw was an effect theory.
\end{proof}

The effects of a JB-algebra $A$ in the effect theory \textbf{JB}$_{\text{psu}}^\opp$ are precisely the elements of $[0,1]_A$. We will hence refer to elements $a\in A$ satisfying $0\leq a\leq 1$ as \Define{effects}\indexd{effect!in JB-algebra}.

Isomorphisms in effect theories are often useful, hence let us remove some subtlety regarding different types of isomorphisms.

\begin{proposition}\label{prop:orderiso-is-Jordan}
  A morphism $\Phi$ in \textbf{JB}$_{\text{psu}}^\opp$ has a two-sided inverse $\Phi^{-1}$ (and hence is an isomorphism in the sense of an effect theory) if and only if $\Phi$ is a unital Jordan-isomorphism.
\end{proposition}
\begin{proof}
  If $\Phi$ is a unital Jordan-isomorphism, then both $\Phi$ and $\Phi^{-1}$ are positive and sub-unital and hence lie in \textbf{JB}$_{\text{psu}}^\opp$. 
  Conversely, if $\Phi$ and $\Phi^{-1}$ lie in \textbf{JB}$_{\text{psu}}^\opp$ and are each others inverses, then they are unital (Proposition~\ref{prop:effect-theory-basic}) and since both $\Phi$ and $\Phi^{-1}$ are monotone (being positive), $\Phi$ is an order isomorphism. Theorem 2.80 of Ref.~\cite{alfsen2012geometry} shows that unital order-isomorphisms between JB-algebras are precisely Jordan isomorphisms.
\end{proof}

\subsection{JBW-algebras}

This chapter deals with JBW-algebras, which are a class of JB-algebras that have more structure. They relate to JB-algebras in a analogous manner to how von Neumann algebras (W$^*$-algebras) related to C$^*$-algebras, hence the `W' in `JBW'.
The reason we will resort to using JBW-algebras instead of JB-algebras is because their order structure is richer. Before we continue let us therefore recall a few order-theoretic concepts.


\begin{definition}\label{def:directedcompleteness}
    Let $P$ be a partially ordered set (such as an order unit space). We call a subset $D\sse P$ is \Define{directed}\indexd{directed set} when for any two elements $x_1,x_2$ we can find a third element $x\in D$ such that $x_1\leq x$ and $x_2\leq x$. A subset $D\sse P$ is \Define{bounded}\indexd{bounded set} when it has an upper bound, \ie~when there is a $p\in P$ such that $x\leq p$ for all $x\in D$. We say $P$ is \Define{(bounded) directed-complete}\indexd{directed complete} when any non-empty (bounded) directed set has a supremum.
\end{definition}

\begin{definition}\label{def:normality}
    Let $P$ and $Q$ be partially ordered sets. A monotone map $f\colon P\rightarrow Q$ is called \Define{normal}\indexd{normal!map between posets} when it preserves suprema of bounded directed sets, \ie~when $f(\bigvee D) = \bigvee f(D)$ for any bounded directed set $D\sse P$ which has a supremum $\bigvee D$ in $P$.
\end{definition}

\begin{definition}
    Let $A$ be a JB-algebra, and let $S\sse \st(A)$ be a subset of the states of $A$. We say $S$ is \Define{separating}\indexd{separating set of states} when $\omega(a) = \omega(b)$ for all $\omega \in S$ implies that $a=b$ for any $a,b \in A$.
\end{definition}

\begin{definition}\label{def:JBW-algebra}
  \index{math}{JBWpsu@\textbf{JBW}$_{\text{psu}}$}
  \index{math}{JBWnpsu@\textbf{JBW}$_{\text{npsu}}$}
  A JB-algebra $A$ is a \Define{JBW-algebra}\indexd{JBW-algebra} when it is bounded directed-complete and has a separating set of normal states. We denote by \textbf{JBW}$_{\text{psu}}$ the category of JBW-algebras with positive sub-unital maps, and by \textbf{JBW}$_{\text{npsu}}$ for the wide subcategory of normal positive sub-unital maps.
\end{definition}

\begin{proposition}
    Both \textbf{JBW}$_{\text{psu}}^\opp$ and \textbf{JBW}$_{\text{npsu}}^\opp$ are effect theories.
\end{proposition}
\begin{proof}
    That \textbf{JBW}$_{\text{psu}}^\opp$ is an effect theory follows analogously to Proposition~\ref{prop:JB-is-effect-theory}. That \textbf{JBW}$_{\text{npsu}}^\opp$ is an effect theory follows because all positive sub-unital maps ${\hat{q}:\R\rightarrow A}$ are necessarily normal, and hence the set of effects coincides with that of \textbf{JBW}$_{\text{psu}}^\opp$.
\end{proof}

\begin{proposition}\label{prop:EJA-is-JBW}
  Any Euclidean Jordan algebra is a JBW-algebra.
\end{proposition}
\begin{proof}
  By Proposition~\ref{prop:EJA-is-JB} EJAs are precisely finite-dimensional JB-algebras. Theorem 4.4.16 of Ref.~\cite{hanche1984jordan} establishes that JBW-algebras are precisely those JB-algebras $A$ that have a \Define{predual}\indexd{predual}, \ie~a space $A_0$ such that $A_0^* \cong A$. As an EJA $A$ is finite-dimensional we of course have $A\cong (A^*)^*$ and hence $A$ has a predual, making it a JBW-algebra.
\end{proof}

\begin{example}
	Let $\mathfrak{A}$ be a \Define{von Neumann algebra}\indexd{von Neumann algebra}, \ie~a $C^*$-algebra that is bounded directed-complete and has a separating set of normal states~\cite{kadison1956operator}. Then its set of self-adjoint elements $\mathfrak{A}_{\sa}$ is a JBW-algebra with the special Jordan product.%
	\footnote{What we call a `von Neumann algebra' is sometimes also called a `W$^*$-algebra', whereas the term `von Neumann algebra' is reserved for concrete algebras represented on a Hilbert space. We will not make this distinction and refer to both by the term `von Neumann algebra'.}
\end{example}

\begin{definition}
    Let $A$ be a JBW-algebra, denote by $V$ the vector space spanned by its normal states. The \Define{weak} topology\indexd{weak topology} of $A$ is the $\sigma(A,V)$ topology, \ie~it is the weakest topology that makes every map in $V$ continuous. Concretely, a net $(a_\alpha)$ converges weakly to $a$ if $(\omega(a_\alpha))$ converges in $\C$ to $\omega(a)$ for every normal state $\omega$.
    The \Define{strong} topology\indexd{strong topology} is the locally convex topology defined by the semi-norms $a\mapsto \sqrt{\omega(a^2)}$ for all normal states $\omega$. Concretely, a net $(a_\alpha)$ converges strongly to $a$ if $\sqrt{\omega( (a_\alpha-a)^2)} \rightarrow 0$ for all normal states $\omega$.
\end{definition}

\begin{remark}
	We use the names of weak and strong topology on JBW-algebras as in Ref.~\cite{hanche1984jordan}. In Ref.~\cite{alfsen2012geometry} these are called respectively $\sigma$-weak and $\sigma$-strong. In the literature on von Neumann algebras the corresponding topologies are called ultraweak and ultrastrong~\cite{bramthesis}.
\end{remark}

\begin{definition}
	A JBW-algebra $A$ is a \Define{JW-algebra}\indexd{JW-algebra} when it is Jordan-isomorphic to an ultraweakly closed subset of the self-adjoint elements of a von Neumann algebra.
\end{definition}

We collect below a few results regarding the weak and strong topology that we will use throughout the chapter without further reference. Recall that for $a\in A$, $T_a$ denotes the Jordan product map $T_a(b) = a*b$ while $Q_a$ denotes the quadratic map $Q_a(b) = 2a*(a*b) - a^2*b$.

\begin{proposition}
	Let $A$ be a JBW-algebra and let $a\in A$ be an arbitrary element.
	\begin{enumerate}[label=\alph*)]
		\item Norm convergence implies strong convergence, and strong convergence implies weak convergence~\cite[Remark 4.1.3]{hanche1984jordan}.
		\item Let $D\sse A$ be a bounded directed subset. Then the net $(a)_{a\in D}$ converges strongly and weakly to $\bigvee D$~\cite[Remark 4.1.3]{hanche1984jordan}.
		\item The operators $T_a$ and $Q_a$ are weakly continuous~\cite[Corollary 4.1.6]{hanche1984jordan}.
		\item The operators $T_a$ and $Q_a$ are strongly continuous~\cite[Lemma 4.1.8]{hanche1984jordan}.
		\item The Jordan product is jointly strongly continuous on bounded subsets~\cite[Lemma 4.1.9]{hanche1984jordan}.
		\item A normal state is strongly and weakly continuous~\cite[Corollary 4.5.4]{hanche1984jordan}.
		\item A Jordan homomorphism $\phi: A\rightarrow B$ between JBW-algebras $A$ and $B$ is normal if and only if it is weakly continuous~\cite[Remark 4.5.6]{hanche1984jordan}.
	\end{enumerate}
\end{proposition}


For positive maps, normality and weak continuity coincide:

\begin{proposition}\label{prop:normal-iff-weak-cont}
	Let $f:A\rightarrow B$ be a positive map between JBW-algebras $A$ and $B$. Then the following are equivalent:
	\begin{enumerate}[label=\alph*)]
		\item $f$ is weakly continuous.
		\item $f$ is weakly continuous on $[0,1]_A$.
		\item $f$ is normal.
		\item $\omega\circ f$ is normal for every normal state $\omega$.
	\end{enumerate}
\end{proposition}
\begin{proof}
	a) to b) and c) to d) are trivial. For b) to c) we need to show that $f(\bigvee D) = \bigvee f(D)$ for a bounded directed subset $D\sse A$. But note that since $D$ is bounded, we can rescale and translate it so that it lies inside $[0,1]_A$. As directed sets converge weakly to their suprema, the desired result follows.
	For d) to a) we note that $\omega\circ f$ is normal if and only if it is weakly continuous. The rest of the proof is then simply unpacking definitions.
\end{proof}

We can use this result to define pointwise weak limits of positive maps. Note that we make no distinction in notation for limits: if $(a_\alpha)$ converges strongly/weakly/in the norm to $a$ then we write $a=\lim_\alpha a_\alpha$ for the strong/weak/norm limit. If the meaning of `$\lim_\alpha$' is unclear from context, we will specify exactly which limit is meant.

\begin{proposition}\label{prop:weak-pointwise-limit}
	Let $f_n:A\rightarrow B$ be a sequence of positive maps between JBW-algebras $A$ and $B$ such that $f_n(a)$ converges weakly for all $a\in A$. 
  Then the map $f$ defined as the pointwise weak limit $f(a):=\lim_n f_n(a)$ is a positive map. 
  Furthermore, if the convergence is uniform on $[0,1]_A$ and all the $f_n$ are normal, then $f$ is normal as well.
\end{proposition}
\begin{proof}
	Since addition and scalar multiplication is weakly continuous, the map $f$ is easily seen to be linear. Suppose $a\in A$ is positive. We need to show that $f(a) = \lim_n f_n(a)$ is positive. But as each $f_n(a)$ is positive, we see that $\omega(f(a)) = \lim_n \omega(f_n(a)) \geq 0$ for every normal (and hence weakly continuous) state $\omega$. As the normal states order-separate the elements, we indeed have $f(a)\geq 0$.

	Now suppose all the $f_n$ are normal and the convergence is uniform on $[0,1]_A$. By Proposition~\ref{prop:normal-iff-weak-cont}, the $f_n$ are weakly continuous. But then $f$ restricted to $[0,1]_A$ is the uniform limit of weakly continuous functions, and hence is weakly continuous on $[0,1]_A$. Again by Proposition~\ref{prop:normal-iff-weak-cont}, $f$ is normal. 
\end{proof}

\begin{definition}[{\cite[Definition~4.5.9]{hanche1984jordan}}]\label{prop:JBW-subalgebras}
	Let $A$ be a JBW-algebra and let $B\sse A$ be a Jordan subalgebra. We say $B$ is a \Define{JBW-subalgebra}\indexd{JBW-algebra!subalgebra} if $B$ is norm-closed and monotone-closed (\ie~closed under suprema of bounded directed sets).
\end{definition}
Clearly, a JBW-subalgebra is a JBW-algebra itself. 
By \cite[Proposition 4.5.10]{hanche1984jordan} a norm-closed Jordan subalgebra of a JBW-algebra is a JBW-subalgebra if and only if it is weakly closed.

\begin{definition}
	Let $a\in A$ be an arbitrary element of a JBW-algebra $A$. We denote by $W(a)$ the weak closure of the Jordan algebra $J(a)$.
\end{definition}

\begin{proposition}[{\cite[Remark 4.1.10]{hanche1984jordan}}]
	Let $A$ be a JBW-algebra and $a\in A$ arbitrary. Then $W(a)$ is an associative JBW-algebra, and hence a JBW-subalgebra of $A$.
\end{proposition}
\begin{proof}
	$W(a)$ is the weak closure of the associative Jordan algebra $J(a)$. That $W(a)$ is a Jordan algebra and associative then both follow from the weak continuity of the Jordan product. As norm convergence implies weak convergence, $W(a)$ is both norm-closed and weakly closed and hence is a JBW-algebra.
\end{proof}

The following technical result will be of tremendous use.

\begin{proposition}[{\cite[Corollary 2.56]{alfsen2012geometry}}]
	The unit ball of a JBW-algebra is weakly compact.
\end{proposition}

\begin{corollary}\label{cor:weakly-Cauchy}
	Any bounded sequence in a JBW-algebra converges weakly if and only if it is weakly Cauchy. That is, if we have a sequence of elements $a_1,a_2,\ldots$ such that there is some $c\in \R$ with $\norm{a_k}\leq c$ for all $k$, such that $\omega(a_j - a_k) \rightarrow 0$ as $j,k\rightarrow \infty$ for all normal states $\omega$, then $a_1,a_2,\ldots$ converges weakly to some $a$.
\end{corollary}
We will use this fact freely in the remainder of this chapter.


\subsection{Idempotents and the Peirce decomposition}

One of the benefits of using JBW-algebras over JB-algebras is that they have an abundance of idempotents.

\begin{definition}
	Let $E$ be a Jordan algebra. We call an element $p\in E$ \Define{idempotent}\indexd{idempotent!in Jordan algebra} when $p*p = p$.
\end{definition}

\begin{proposition}[{\cite[Proposition~4.2.3]{hanche1984jordan}}]\label{prop:JBW-idempotents-dense}
	Let $A$ be a JBW-algebra. The linear span of the idempotents of $A$ lies norm-dense in $A$.
\end{proposition}

Note that if $p$ is an idempotent that then $1-p$ is also an idempotent. We denote this idempotent by $p^\perp$.
Furthermore $Q_p 1 = p^2 = p$ and hence $Q_p p^\perp = 0$. Lastly, as $Q_{a^2} = Q_a^2$ for any $a$ we have $Q_p^2 = Q_p$.

Idempotents interact well with operator commutation.

\begin{proposition}[{\cite[Lemma 2.5.5 and Lemma 5.2.5]{hanche1984jordan}}]\label{prop:operator-commutation}
  Let $A$ be a Jordan algebra, $p\in A$ an idempotent, and $a\in A$ arbitrary.  Then the following are equivalent:
  \begin{itemize}
    \item $a$ and $p$ operator commute.
    \item $T_p a = Q_p a$.
    \item $a = (Q_p + Q_{p^\perp}) a$.
    \item $a$ and $p$ generate an associative subalgebra of $A$.
  \end{itemize}
  If furthermore $A$ is a JBW-algebra, then the above are equivalent to $p$ operator commuting with all elements of $W(a)$.
\end{proposition}

\begin{lemma}\label{lem:projection-zero}
	Let $p, a\in E$ be elements of a Jordan algebra with $p$ idempotent. Then $Q_p a = 0$ if and only if $p*a = 0$, and in that case $p$ and $a$ operator commute.
\end{lemma}
\begin{proof}
	$Q_p a = 0 \iff p*a = 0$ is Ref.~\cite[Remark 4.1.14]{hanche1984jordan}. Operator commutation then follows by Proposition~\ref{prop:operator-commutation}.
\end{proof}

We can now state some useful calculation rules involving idempotents.
\begin{proposition}\label{prop:idempotent-rules}
	Let $A$ be a JB-algebra, and let $p\in A$ be an idempotent, and $a\in A$ arbitrary.
	\begin{enumerate}[label=\alph*)]
		\item Suppose $0\leq a \leq p$. Then $0 = Q_{p^\perp} a = Q_a p^\perp = p^\perp*a$, $Q_p a = p*a = a$, and $Q_a p = a^2$.
		\item Suppose $p\leq a \leq 1$. Then $Q_p a = p$ and $Q_a p = p$.
	\end{enumerate}
\end{proposition}
\begin{proof}~
	\begin{enumerate}[label=\alph*)]
		\item Using the positivity of $Q_{p^\perp}$ we calculate $0 = p*p^\perp = Q_{p^\perp} p \geq Q_{p^\perp} a \geq 0$, and thus $Q_{p^\perp} a = 0$. Then by the previous lemma $p^\perp$ operator commutes with $a$ and $p^\perp * a = 0$, so that $p*a = a$ and hence $Q_p a = a$. Using the fundamental identity we see that $Q_{p^\perp} Q_a Q_{p^\perp} = Q_{Q_{p^\perp} a} = Q_0 = 0$. Hence, since $p^\perp$ and $a$ operator commute, we calculate 
		$Q_a p^\perp = Q_a Q_{p^\perp} Q_{p^\perp} 1 = Q_{p^\perp} Q_a Q_{p^\perp} 1 = 0$. Finally, $a^2 = Q_a 1 = Q_a (p+p^\perp) = Q_a p$.

		\item If $p\leq a \leq 1$ then $0\leq a^\perp \leq p^\perp$ and hence by the previous point $0 = Q_{p^{\perp\perp}} a^\perp = Q_p (1-a) = Q_p 1 - Q_p a = p - Q_p a$, so that indeed $Q_p a = p$. As $p$ and $a$ operator commute we furthermore calculate, again using the fundamental identity, $Q_a p = Q_a Q_p Q_p 1 = Q_p Q_a Q_p 1 = Q_{Q_p a} 1 = Q_p 1 = p$\qedhere
	\end{enumerate}
\end{proof}

Another useful result regarding an idempotent $p$ in a Jordan algebra $E$ is that they `split-up' the algebra into three parts corresponding to the eigenvalues of $T_p$, known as its \Define{Peirce decomposition}\indexd{Peirce decomposition}.

\begin{proposition}[{\cite[Lemma~2.6.3]{hanche1984jordan}}]
	Let $E$ be a Jordan algebra and $p\in E$ an idempotent. Then $E=E_1\oplus E_{\frac12} \oplus E_0$  (where `$\oplus$' denotes direct sum of vector spaces, not of Jordan algebras) such that for all $a\in E_k$ we have $p*a = k a$. Furthermore, $E_1$ and $E_0$ are Jordan subalgebras and $E_1 = Q_p(E)$ while $E_0 = Q_{p^\perp}(E)$.
\end{proposition}

We will introduce some special notation for the `1-eigenspace' of an idempotent: $E_p := Q_p(E)$. For a JBW-algebra, this subalgebra is actually a JBW-subalgebra.\index{math}{Ap@$A_p$ (Peirce decomposition)}

\begin{proposition}[{\cite[Lemma 4.1.13]{hanche1984jordan}}]\label{prop:peirce}
    Let $p\in A$ be a non-zero idempotent in a JBW-algebra $A$. Then $A_p$ is a JBW-subalgebra of $A$. Furthermore, if $a\in A_p$ and $b\in A$ with $0\leq b\leq a$, then $b\in A_p$.
\end{proposition}

\subsection{Operator commutativity in JBW-algebras}\label{sec:JBW-commutativity}

We extend the results of Section~\ref{sec:Jordan-commutativity} regarding operator commutativity to JBW-algebras. Recall that we write $S'$ for the set of elements that operator commutate with every element of $S$.

\begin{lemma}[{\cite[Lemma 1]{jacobson1989operator}}]\label{lem:idempotent-commutator}
	Let $E$ be a Jordan algebra, and let $p\in E$ be idempotent. Then $\{p\}'$ is a subalgebra of $E$.
\end{lemma}

\begin{proposition}
	Let $S\sse A$ be a Jordan subalgebra of the JBW-algebra $A$. Then $S'$ is a JBW-subalgebra.
\end{proposition}
\begin{proof}
	Let $a_\alpha \in S$ be a net that converges weakly to some $a$ (not necessarily in $S$). Suppose $b\in S'$. Then for any $c\in A$, $T_bT_a c = b*(c*a) = \lim_\alpha b*(c*a_\alpha) = \lim_\alpha T_bT_{a_\alpha}c = \lim_\alpha T_{a_\alpha}T_b c = \lim_\alpha (b*c)*a_\alpha = (b*c)*a = T_aT_bc$. So any $b\in S'$ also commutes with everything in the weak closure of $S$. Without loss of generality we may hence assume that $S$ is a JBW-subalgebra. Furthermore, in a similar way we can show that the limit of any weakly convergent net $b_\alpha \in S'$ also commutes with all elements of $S$ so that $S'$ is weakly closed. It then remains to show that $S'$ is a Jordan algebra.

	As $S$ is a JBW-subalgebra, Proposition~\ref{prop:JBW-idempotents-dense} shows that the span of idempotents lies (norm) dense in $S$ and hence $b\in S'$ iff $b\commu p$ for every idempotent $p\in S$. As a result $S' = \bigcap_{p^2=p\in S}\ \{p\}'$. But as each $\{p\}'$ is a Jordan algebra by Lemma~\ref{lem:idempotent-commutator}, $S'$ is the intersection of Jordan algebras and hence is a Jordan algebra itself. 
\end{proof}

\begin{proposition}\label{prop:commuting-elements-span-associative-algebra}
	Let $A$ be a JBW-algebra, and $a,b\in A$ arbitrary. If $b\commu a$ and $b\commu a^2$, then $b^2\commu a$ and there is an associative subalgebra $B$ consisting of mutually operator commuting elements such that $W(a),W(b)\sse B$.
\end{proposition}
\begin{proof}
	If $b\commu a$ and $b\commu a^2$, then $b\in J(a)'$ (Lemma~\ref{lem:Ja-prime}). By weak continuity of the Jordan product then also $b \in W(a)'$. By the previous proposition $W(a)'$ is a Jordan algebra, and hence also $b^2 \in W(a)'$, so that in particular $b^2\commu a$. Then by Proposition~\ref{prop:simple-operator-commutation}, $a$ and $b$ generate an associative Jordan algebra $S$ of mutually operator commuting elements. Let $B$ be weak closure of $S$. Then $B$ has the desired properties.
\end{proof}

In Section~\ref{sec:JBW-commutation-revisited} we will prove a stronger version of this statement, namely that $a\commu b$ iff $a$ and $b$ generate an associative subalgebra. 
But whereas Proposition~\ref{prop:commuting-elements-span-associative-algebra} is relatively straightforward to prove, the results of Section~\ref{sec:JBW-commutation-revisited} require a lot more knowledge regarding the global structure of JBW-algebras.

\section{Floors, ceilings and compressions}\label{sec:JBW-floor-ceiling}

Having covered some of the basic material of the theory of JBW-algebras we will now work towards proving some new results. Our ultimate aim is to show that all the structure surrounding pure maps discussed in Chapter~\ref{chap:effectus} is present in JBW-algebras. To start, in this section we will establish the existence of compressions and images, which follows relatively straightforward using standard techniques.

\begin{definition}\label{def:JBWfloorceiling}
    Let $A$ be a JBW-algebra and let $a\in A$ be an effect. We define the \Define{floor}\indexd{floor!in JBW-algebra} of $a$ as $\floor{a} := \bigwedge_n a^n$. The \Define{ceiling}\indexd{ceiling!in JBW-algebra} of $a$ is defined as $\ceil{a} := \floor{a^\perp}^\perp$.
\end{definition}

Note first of all that $\floor{a}$ (and hence $\ceil{a}$) indeed exist, as for effects $a$, the sequence $a^n$ is decreasing, and hence has an infimum. Note furthermore that $\ceil{a^\perp}^\perp = \floor{a}$.

\begin{proposition}\label{prop:JBWfloorceiling}
    Let $a\in A$ be an effect in a JBW-algebra $A$. The floor and ceiling are both idempotents. Furthermore, $\floor{a}$ is the largest idempotent below $a$ and $\ceil{a}$ is the smallest idempotent above $a$.
\end{proposition}
\begin{proof}
    By the weak continuity of the Jordan product we get $a^k*\floor{a} = T_{a^k} (\wedge_n a^n) = \wedge_n T_{a^k} a^n = \wedge_n a^{n+k} = \wedge_n a^n$. Hence $\floor{a}*\floor{a} = \floor{a}*\wedge_n a^n = \wedge_n a^n *\floor{a} = \wedge_n a^n = \floor{a}$, so that it is indeed a idempotent. The same of course holds for $\floor{a^\perp}$ and since the complement of an idempotent is again an idempotent we see that $\ceil{a} = \floor{a^\perp}^\perp$ is also an idempotent.

    Suppose $p$ is a idempotent with $p\leq a$. Then by Proposition~\ref{prop:idempotent-rules} $p*a = p$ and $p$ and $a$ operator commute. Hence $p*a^n = p*(a*a^{n-1}) = (p*a)*a^{n-1} = p$ by induction. Then by weak continuity of the Jordan product $p*\floor{a} = p*\wedge_n a^n = \wedge_n p*a^n = \wedge_n p = p$ so that $p\leq \floor{a}$. So $\floor{a}$ is indeed the largest idempotent below $a$.

    Now if $p\geq a$ is an idempotent, then we note that $p^\perp \leq a^\perp$ and hence by the previous paragraph $p^\perp \leq \floor{a^\perp}$ so that also $p\geq \floor{a^\perp}^\perp = \ceil{a}$.
\end{proof}

\begin{remark}
	In the literature on von Neumann algebras and JBW-algebras, the smallest idempotent above an element $a$ is usually referred to as the \Define{range projection}\indexd{range projection} of the element and denoted as $r(a)$. We use the name ceiling to link it to the corresponding notion in Chapters~\ref{chap:seqprod} and~\ref{chap:effectus}.
\end{remark}

\indexd{Cauchy-Schwarz inequality}
Recall that the Cauchy-Schwarz inequality states that for any real-valued pre-inner-product $\inn{\cdot,\cdot}$, \ie~a bilinear map which has $\inn{a,a}\geq 0$ for all $a$, but does not necessarily satisfy $\inn{a,a}=0\implies a=0$, we have $\lvert \inn{a,b}\rvert^2 \leq \inn{a,a}\inn{b,b}$.
The following is then an easy consequence, but essential in many of our proofs:

\begin{lemma}\label{lem:JBWCauchySchwarz}
    Let $\omega:A\rightarrow \R$ be a positive linear map. Then $\inn{a,b}_\omega := \omega(a*b)$ defines a pre-inner-product and hence satisfies the Cauchy-Schwarz inequality $\lvert \inn{a,b}_\omega\rvert^2 \leq \inn{a,a}_\omega
    \inn{b,b}_\omega$, i.e.~$\lvert \omega(a*b)\rvert^2 \leq \omega(a^2)\omega(b^2)$.
\end{lemma}

\begin{lemma}\label{lem:JBWpreserveimage}
    Let~$g: A\rightarrow B$ be a positive linear map between JBW-algebras such that $g(p)=g(1)$ for some idempotent $p$. Then $g(Q_pa)= g(a)$ for all $a \in A$.
\end{lemma}
\begin{proof}
    Suppose the lemma holds for positive linear maps $\omega: A\rightarrow \R$. Then if we let $\omega': B\rightarrow \R$ be any positive linear map, we know that $\omega := \omega'\circ g:A\rightarrow \R$ satisfies the property, and hence $\omega'\circ g(Q_p a) = \omega(Q_p a) = \omega(a) = \omega'\circ g(a)$. Now since $\omega'$ is arbitrary, and the states of $B$ separate the elements, we conclude that $g(Q_p a) = g(a)$.

    It remains to show that the property holds for positive linear $\omega: A\rightarrow \R$. Note that if $\omega(p) = \omega(1)$ that then $\omega(p^\perp) = 0$.
    Using the Cauchy-Schwarz inequality of the previous lemma with $a:= p^\perp$ and $b:= a$ we calculate
    \[\lvert \omega(p^\perp*a) \rvert^2 =  \lvert \inn{p^\perp,a}_{\omega}\rvert^2 \leq \inn{p^\perp,p^\perp}_{\omega}\inn{a,a}_{\omega} = \omega(p^\perp) \omega(a^2) = 0.\] 
    As a result $\omega(p^\perp * a) = 0$ for all $a\in A$ and hence $\omega(a) = \omega(1*a) = \omega((p+p^\perp)*a) = \omega(p*a)$. Unfolding the definition of $Q_p$ we then easily get $\omega(Q_p a) = \omega(a)$.
\end{proof}

\begin{lemma}\label{lem:JBWfloor}
    Let $g: A\rightarrow B$ be a normal positive linear map between JBW-algebras
    such that $g(a)=g(1)$ for some effect $a\in A$.
    Then~$g(\floor{a})=g(1)$.
\end{lemma}
\begin{proof}
    We will show the result for normal states $\omega: A\rightarrow \R$. The general result then follows as in the previous lemma by the separation of the elements by the normal states.

    So let $\omega:A\rightarrow \R$ be a normal state. We need to show that if $\omega(a)=\omega(1)=1$ that then $\omega(\floor{a})=1$.

    Recall that $\floor{a} := \wedge_n a^n$, and hence by the normality of $\omega$ we get $\omega(\floor{a}) = \wedge_n \omega(a^n)$, so that it suffices to show that $\omega(a^n) = 1$ for all $n$. 
    Using the Cauchy-Schwarz inequality of Lemma~\ref{lem:JBWCauchySchwarz} with $b:= 1$, we get $\lvert \omega(a)\rvert^2 \leq \omega(a^2) \omega(1^2) = \omega(a^2)$. Since by assumption $\omega(a) = \omega(1) = 1$ we have $\omega(a^2) \geq \lvert \omega(a)\rvert^2 = 1$. By Eq.~\eqref{eq:Jordan-OUS} $a^2\leq 1$ and hence also $\omega(a^2)\leq 1$, so that indeed $\omega(a^2) = 1$. Repeating this argument we get $\omega(a^{2^k}) = 1$ for all $k$, and hence also $\omega(a^n) = 1$ for all $n$.
\end{proof}

\begin{corollary}\label{cor:JBW-ceil-zero}
	Let $g: A\rightarrow B$ be a normal positive linear map between JBW-algebras, such that $g(a) = 0$ for some effect $a$. Then $g(\ceil{a}) = 0$.
\end{corollary}
\begin{proof}
	Suppose that $g(a) = 0$ so that $g(a^\perp) = g(1)$. Then $g(\floor{a^\perp}) = g(1)$, so that $g(\ceil{a}) = g(\floor{a^\perp}^\perp) = 0$.
\end{proof}

We can now show that the first special type of structure we defined for effect theories is present in JBW-algebras: compressions.
Because we must take the \emph{opposite} category of JBW-algebras to make an effect theory, we flip the direction of the arrows in the proposition below.

\begin{proposition}\label{prop:JBWcorner}
    Let $a$ be an effect of $A$. Define $\pi_a: A\rightarrow A_{\floor{a}}$ to be $\pi_a(b) = Q_{\floor{a}} b$ restricted to the space $A_{\floor{a}}$ (cf.~Proposition~\ref{prop:peirce}).
    Then~$\pi_a$ is a compression for $a$, \ie~$\pi_a(1) = \pi_a(a)$ and for any map $g: A\rightarrow B$ with $g(1) = g(a)$ there is a unique map $\cl{g}:A_{\floor{a}}\rightarrow B$ such that $g=\cl{g}\circ \pi_a$. Furthermore, if $g$ is normal, then $\cl{g}$ is also normal.
    We will refer to $\pi_a$ as the \Define{standard compression}\indexd{standard compression (JBW-algebra)} for $a$.\indexd{compression!standard --- (JBW-algebra)}
\end{proposition}
\begin{proof}
    First of all we have $\pi_a(1) = Q_{\floor{a}} 1 = \floor{a} = Q_{\floor{a}} a = \pi_a(a)$. Now suppose $g:A\rightarrow B$ is a positive sub-unital map such that $g(a)=g(1)$. We must show that there is a unique $\cl{g}:A_{\floor{a}} \rightarrow B$ such that $\cl{g}\circ \pi_a= g$.

    Define $\cl{g}:A_{\floor{a}}\rightarrow B$ as the restriction of $g$ to $A_{\floor{a}}$. To prove that $\cl{g}\circ \pi_a= g$, we need to show that~$g(b)=g(Q_{\floor{a}}b)$ for all $b$.

    By Lemma~\ref{lem:JBWfloor} $g(\floor{a})=g(1)$ and hence by Lemma~\ref{lem:JBWpreserveimage} $g(Q_{\floor{a}} b) = g(b)$, so that $\cl{g}$ indeed satisfies the required condition.

    For uniqueness of $\cl{g}$ suppose we have a $h: A_{\floor{a}} \rightarrow B$ such that $h\circ \pi_a = g = \cl{g}\circ \pi_a$. Let $b\in A_{\floor{a}}$ be arbitrary. Then $\pi_a(b) = b$, so that $h(b) = h(\pi_a(b)) =
    \cl{g}(\pi_a(b)) = \cl{g}(b)$, as desired.

    If $g$ is normal, then $\cl{g}$, being a restriction of $g$, will also be normal.
\end{proof}

\begin{remark}\label{remark:normal-compression}
	The property of being a compression in principle depends on which effect theory we are considering, \textbf{JBW}$_{\text{psu}}^\opp$ or \textbf{JBW}$_{\text{npsu}}^\opp$. But as the standard compression is normal, and the unique map $\cl{g}$ appearing in the universal property is normal whenever $g$ is, the standard compression is a compression in both categories. Furthermore, as all compressions for an effect are connected by an isomorphism, and isomorphisms are always normal, compressions coincide in both categories.
\end{remark}

Images also exist for maps between JBW-algebras, as long as we restrict ourselves to normal maps.

\begin{proposition}[{\cite[Lemma 4.2.8]{hanche1984jordan}}]
    The set of idempotents in a JBW-algebra form a complete lattice. Hence, the supremum and infimum of any set of idempotents exists and is a idempotent again.%
    \footnote{There is a subtlety here that the supremum taken in the set of idempotents could a priori be different from the supremum taken in the JBW-algebra itself. Fortunately, these two different notions of suprema coincide.}
\end{proposition}

\begin{proposition}\label{prop:normal-has-image}
	Let $f:A\rightarrow B$ be a normal positive map between JBW-algebras. Then $f$ has an image, \ie~a smallest effect $p\in A$ such that $f(p) = f(1)$, or equivalently, $f(p^\perp) = 0$. We denote the image of $f$ by $\im{f}$.
\end{proposition}
\begin{proof}
	Let $\im{f} := \bigwedge\{p ~;~ f(p)=f(1), p^2 = p\}$. Then by normality of $f$: 
  \[f(\im{f}) = f\left(\bigwedge_{f(p)=f(1)} p\right) = \bigwedge_{f(p)=f(1)} f(p) = \bigwedge_{f(p)=f(1)} f(1) = f(1),\]
  so $\im{f}$ satisfies the required equality. It is also by definition the smallest such element among the idempotents. Suppose $f(a) = f(1)$ for an effect $a$. By Lemma~\ref{lem:JBWfloor} then $f(\floor{a}) = f(1)$ and hence $\im{f}\leq\floor{a}\leq a$, so that $\im{f}$ is also the smallest among all effects.
\end{proof}

\begin{remark}
  This proposition shows that images of normal maps are always idempotents. Hence, the sharp effects of Definition~\ref{def:effectus-sharp} are idempotents in the setting of Jordan algebras. Conversely for any idempotent $p$ we have $\im{\pi_p} = \floor{p} = p$ so that they are also sharp.  
\end{remark}

We now know that \textbf{JBW}$_{\text{npsu}}^\opp$ is an effect theory with compressions and images, and hence we can start using some results from Chapter~\ref{chap:effectus}. Note that as the \emph{opposite} category forms an effect theory, we will have to change the order of composition in the translation.

\begin{lemma}\label{lem:JBW-image-of-composed}
	Let $f$ and $g$ be composable normal positive sub-unital maps between JBW-algebras. Then $\im{f\circ g} \leq \im{g}$.
\end{lemma}
\begin{proof}
	This is Lemma~\ref{lem:imageofcomposedmaps}, but with flipped order of the maps.
\end{proof}

\begin{lemma}\label{lem:JBW-ceil-in-ceil}
	Let $f:A\rightarrow B$ be a normal positive sub-unital map between JBW-algebras. Then for any effect $a\in A$: $\ceil{f(a)} = \ceil{f(\ceil{a})}$
\end{lemma}
\begin{proof}
	This is Proposition~\ref{prop:floorceiling}.d).
\end{proof}

\section{Spectral theorem}\label{sec:spectral-theorem}

In this section we will recall a spectral theorem for JBW-algebras that will prove invaluable in the remainder of this chapter. Before we state this spectral theorem we will need to know more about associative JBW-algebras, \ie~algebras where the Jordan product satisfies $a*(b*c) = (a*b)*c$ for all $a$, $b$ and $c$.

\subsection{Basically disconnected spaces}\label{sec:basically-disconnected}

We can relate associative JB(W)-algebras to topological spaces using Kadison's representation theorem.

\begin{proposition}
    Let $A$ be an associative JB-algebra. Then $A\cong C(X)$ where $X$ is a compact Hausdorff space.
\end{proposition}
\begin{proof}
    $A$ is a complete order unit space where the positive elements correspond to the squares (as defined by the Jordan product). Hence using the associativity and commutativity of the Jordan product, for any two positive elements $a^2, b^2 \in A$, their product $a^2*b^2 = a*((a*b)*b) = a*(b*(a*b)) = (a*b)*(a*b)$ is again a square and hence positive. As a result we can use Kadison's representation theorem (Theorem~\ref{thm:kadison}) to conclude that $A\cong C(X)$ for some compact Hausdorff space $X$.
\end{proof}

When $A$ is an associative JBW-algebra, the compact Hausdorff space is of course going to be of a more specific type.

\begin{definition}
  Let $X$ be a compact Hausdorff space. We call $V\subseteq X$ a \Define{zero set}\indexd{zero set} if a continuous function $f:X\rightarrow \R$ exists such that $V=f^{-1}(0)$. Similarly we call $U\subseteq X$ a \Define{cozero set}\indexd{cozero set} when $U=f^{-1}((0,\infty))$ for some continuous $f$. We call $X$ \Define{basically disconnected}\indexd{basically disconnected} when the interior of all zero sets is closed or, equivalently, when the closure of a cozero set is again open. We will call sets that are both open and closed \Define{clopen}\indexd{clopen set}. 
\end{definition}

We will now show that if $C(X)$ is a JBW-algebra, then $X$ is basically disconnected (but note that the converse is false; cf~Section~\ref{sec:JBW-structure}). For the remainder of this section let $C(X)$ be a JBW-algebra.

\begin{lemma}\label{lem:sharpclopen}
	A $p\in C(X)$ is idempotent iff it is the characteristic function of a clopen set.
\end{lemma}
\begin{proof}
	Let $p:X\rightarrow \R$ be a continuous function such that $p^2=p$, \ie~$\forall x\in X: p(x)^2 = p(x)$. This implies that $p(x)=0$ or $p(x)=1$. Then $p = \chi_S$, the characteristic function of some $S\subseteq X$. Let $U\subseteq [0,1]$ be any open subset of the unit interval. Since $p$ is continuous $p^{-1}(U)$ must be open. This set is either $\emptyset, X, S, X\backslash S$ depending on whether $U$ contains $1$ or $0$ or both. We conclude that both $S$ and $X\backslash S$ must be open, so that $S$ is indeed clopen.
\end{proof}

\begin{lemma}\label{lem:imclopen}
	For an effect $a:X\rightarrow [0,1]$ its ceiling is $\ceil{a}=\chi_S$ where $S=\cl{a^{-1}((0,1])}$ (where $\cl{A}$ denotes the closure of a set $A$). Consequently, $\cl{a^{-1}((0,1])}$ is clopen for all effects $a$.
\end{lemma}
\begin{proof}
	Fix an effect $a:X\rightarrow [0,1]$. Write im$(a) = a^{-1}(0,\infty) = a^{-1}((0,1])$ for the open set where $a$ is nonzero. The ceiling $\ceil{a}$ of the effect $a$ must be the characteristic function $\chi_S$ for some clopen $S$. As $a\leq \chi_S$ we must have im$(a)\subseteq S$ so that $\cl{\text{im}(a)}\subseteq \cl{S}=S$. Let $T = S/ \cl{\text{im}(a)}$. Note that as $T$ is the difference of an open set and a closed one, that $T$ is open. Towards contradiction, assume it is non-empty and pick an $x\in T$. As $X/ T$ and $\{x\}$ are disjoint closed sets we can use to Urysohn's lemma to find a continuous $f:X\rightarrow [0,1]$ such that $f(x)=1$ and $f(X/ T)=\{0\}$. 
	Then by construction $Q_{\sqrt{f}}a = fa = 0$ and hence by Corollary~\ref{cor:JBW-ceil-zero} $f\chi_S = f\ceil{a}=0$. But as $(f\chi_S)(x) = 1$, this is a contradiction. We conclude that $T$ must have been empty and thus that $S=\overline{\text{im}(p)}$ so that $\overline{\text{im}(a)}$ is clopen.
\end{proof}
\begin{corollary}\label{cor:zerosetint}
	The interior of all zero sets, that is $f^{-1}(0)$ for some continuous function $f:X\rightarrow \R$, is clopen. In other words: $X$ is basically disconnected.
\end{corollary}
\begin{proof}
	Let $S=f^{-1}(0)$ be the zero set of some continuous function. Without loss of generality we may assume that $f$ is an effect (as post-composing $f$ with the absolute value function and rescaling it preserves its zero set). As $X\backslash S = f^{-1}( (\infty, 0))$ we see by the previous lemma that $\cl{X\backslash S} = X\backslash S^\circ$ is clopen. Its complement $S^\circ$ is then clopen as well.
\end{proof}

We have now seen that if $A$ is an associative JBW-algebra, then $A=C(X)$ where $X$ is a basically disconnected compact Hausdorff space. In fact, $X$ satisfies a stronger property that we explore in Section~\ref{sec:JBW-structure}. 

We need one more result regarding basically disconnected spaces.

\begin{proposition}\label{prop:CXsupremumofsimpleelements}
	Let $X$ be basically disconnected and let $E=[0,1]_{C(X)}$. Then any $a\in E$ is the supremum and norm limit of an increasing sequence of elements of $E$ of the form $\sum_{k=1}^n \lambda_i p_i$ where $\lambda_i>0$ and the $p_i$ are orthogonal idempotents.
\end{proposition}
\begin{proof}
	Let $a:X\rightarrow [0,1]$ be an element of $E$. Write $p_n^k$ for the characteristic function of the clopen set $\cl{a^{-1}((\frac{k}{n},1])}$, so that we have the implication $p_n^k(x)=1 \implies a(x)\geq \frac{k}{n}$. Of course $p_n^k\geq p_n^{k+1}$. 
  Define $q_n:= \sum_{k=1}^n \frac{1}{n}p_n^k$. Then $q_n$ only takes values $\frac{l}{n}$ for some $0\leq l\leq n$ and we see that $q_n(x)=\frac{l}{n} \iff p_n^l(x)=1 \implies a(x)\geq \frac{l}{n}$ so that $q_n\leq a$.

	In general $q_n$ and $q_m$ might not have an obvious order relation, but if we consider $q_n$ and $q_{2n}$ then it is straightforward to check that the latter will always be larger. The sequence $(q_{2^k})$ is therefore an increasing sequence that has $a$ as an upper bound. We will show that $\norm{a-q_{2^n}}\leq 2^{1-n}$ so that this sequence indeed converges to $a$. As it is an increasing sequence, $a$ will then also be its supremum.

	For every $x\in X$ there exists a $0\leq l\leq 2^n$ such that $\frac{l}{2^n}\leq a(x)< \frac{l+1}{2^n}$. We then always have either $q_{2^n}(x)=\frac{l-1}{2^n}$ or $q_{2^n}(x)=\frac{l}{2^n}$. In both cases $a(x)-q_{2^n}(x) \leq \frac{2}{2^n}=2^{1-n}$. Since this bound does not depend on $x$ or $l$, we indeed have $\norm{a-q_{2^n}} \leq 2^{1-n}$.

	Finally, we have to show that $q_n$ can be written as a linear combination of orthogonal sharp effects. This is easily done by observing that $q_n = \sum_{k=1}^n \frac{1}{n} p_n^k$ can equivalently be written as $q_n = \sum_{k=1}^n \frac{k}{n} (p_n^k - p_n^{k+1})$.
\end{proof}

\subsection{Spectral theorem}

Recall that for an element $a$ of a JBW-algebra $A$ we defined $W(a)$ to be the JBW-algebra generated by $a$ and that $W(a)$ is associative.
By the results of the previous section we can then associative a compact Hausdorff space $X_a$ to each $a$ such that $W(a) \cong C(X_a)$. Note that this isomorphism is both an order-isomorphism (and hence is normal) and an algebra isomorphism.

The element $a$ itself then corresponds to some function $\hat{a}:X_a\rightarrow \R$. For any continuous function $f:\R\rightarrow \R$ we can then define a new element $f(a)\in W(a)$ as the function $\widehat{f(a)}:X_a\rightarrow \R$ given by $\widehat{f(a)}(x) := f(\hat{a}(x))$. In particular, if $a$ is positive, then we can define a unique positive square root $\sqrt{a}\in W(a)$ as the function $\widehat{\sqrt{a}}:X_a\rightarrow \R_{\geq 0}$ given by $\widehat{\sqrt{a}}(x) := \sqrt{\hat{a}(x)}$.

\begin{definition}\index{math}{a@$\sqrt{a}$ (in JBW-algebra)}
	Let $a\in A$ be a positive element of a JBW-algebra. We define its \Define{square root}\indexd{square root (JBW-algebra)} as the unique positive element $\sqrt{a}$ in $W(a)$ such that $\sqrt{a}^2 = a$.
\end{definition}

For an effect $0\leq a\leq 1$ we now see that $a^2 = Q_{\sqrt{a}} a \leq Q_{\sqrt{a}} 1 = a$. We will use the fact that $a^2\leq a$ for effects without further reference in the rest of this chapter.

Recall that if an element $b$ operator commutes with $a$ and $a^2$, then it operator commutes with all of $W(a)$ (Proposition~\ref{prop:commuting-elements-span-associative-algebra}). In particular, $b$ will operator commute with $f(a)$ for any continuous function $f$.

The floor of an element $a$ was defined as $\floor{a}=\wedge_n a^n$ and hence is an element of $W(a)$, as it is weakly closed. Similarly, $\ceil{a}\in W(a)$. We saw in Lemma~\ref{lem:imclopen} that $\ceil{a}$ is then the characteristic function of $\cl{X_a\backslash C}$ where $C = a^{-1}(\{0\})$. It then immediately follows that the ceiling of $a$ and $a^2$ coincide.
\begin{lemma}
	Let $a\in A$ be any element of a JBW-algebra. Then $\ceil{a^2} = \ceil{a}$.
\end{lemma}

We so far only defined the ceiling for effects. For arbitrary positive $a$ we can define an analogous ceiling by $\ceil{a} := \ceil{\frac{a}{\norm{a}}}$. By the above argument, this definition coincides with the regular one when $a$ is an effect. For arbitrary $a\in A$ (not necessarily positive) we define $\ceil{a} := \ceil{a^2}$, which by the previous lemma also coincides with the regular definition for positive $a$.

\subsection{Approximate pseudo-inverses}

In this section we will see how to use the spectral theorem to define (pseudo-)inverses to elements in a JBW-algebra.

\begin{definition}
	Let $a\in A$ be an element of a JBW-algebra. We say $a$ is \Define{pseudo-invertible}\indexd{pseudo-inverse (JBW-algebra)} when there exists a element $b\in W(a)$ such that $a*b = \ceil{a}$. When furthermore $\ceil{a}=1$ (and hence $a*b=1$), we say $a$ is \Define{invertible}\indexd{inverse (JBW-algebra)}.
\end{definition}

When $a$ is invertible, the element $b$ with $a*b=1$ is unique. We denote this element by $a^{-1}$ and call this the \Define{inverse} of $a$. If $a$ is merely pseudo-invertible the element $b$ with $a*b = \ceil{a}$ is not necessarily unique, but there is a unique one that additionally satisfies $\ceil{a}*b = b$.

\begin{lemma}
	Let $a$ be pseudo-invertible. Then there is a unique $b\in W(a)$ such that $a*b = \ceil{a}$ and $\ceil{a}* b = b$.
\end{lemma}
\begin{proof}
	Let $b$ be such that $a*b = \ceil{a}$. Let $b' = \ceil{a}*b$. Then $a*b' = a*(b*\ceil{a}) = (a*b)*\ceil{a} = \ceil{a}^2 = \ceil{a}$, so that $b'$ satisfies the required properties. For uniqueness, suppose that $b''$ also satisfies $a*b'' = \ceil{a}$ and $\ceil{a}*b'' = b''$. Then $b'' = \ceil{a}*b'' = (b'*a)*b'' = b'*(a*b'') = b'*\ceil{a} = b'$.
\end{proof}

We will denote the unique $b$ with $a*b=\ceil{a}$ and $\ceil{a}*b = b$ by $a^{-1}$ and call it the \Define{pseudo-inverse} of $a$. As we use the same notation for inverses and pseudo-inverses, we will always specify which is meant.

Let us demonstrate more clearly how (pseudo-)inverses are defined.
For an example of an invertible element, let $f:X_a\rightarrow \R$ be a positive continuous function with $f(x) > \epsilon$ for all $x\in X_a$ for some fixed $\epsilon$. Then $f$ has an inverse $f^{-1}$ defined pointwise by $f^{-1}(x) := f(x)^{-1}$. 
A pseudo-invertible element that is not invertible can be constructed in the following way. Let $p_1,p_2,\ldots,p_n$ be a finite set of orthogonal idempotents with $\sum_k p_k \neq 1$, and let $\lambda_1,\ldots,\lambda_n$ be strictly positive real numbers. Define $a = \sum_{k=1}^n \lambda_k p_k$. Its ceiling is then $\ceil{a} = \sum_{k=1}^n p_k$, and the unique pseudo-inverse satisfying $\ceil{a}*a^{-1} = a^{-1}$ is given by $a^{-1}:= \sum_{k=1}^n \lambda_k^{-1} p_k$.

(Pseudo-)inverses are a powerful tool, but they do not exist for all elements of a JBW-algebra. There does however exist a closely related structure originally described in the context of von Neumann algebras~\cite{bramthesis}.

\begin{definition}
  Let $a\in A$ be an element in a JBW-algebra. An \Define{approximate pseudo-inverse} for $a$ is a sequence $t_1,t_2,\ldots$ in $W(a)$ such that $t_1*a, t_2*a, \ldots$ is a sequence of orthogonal idempotents and $\sum_i t_i*a = \ceil{a}$ (where we define this infinite sum as the weak limit of the finite sums).
\end{definition}

\begin{remark}
  When $p_1,p_2,\ldots$ are orthogonal idempotents (like the $t_i*a$) above, their weak (and strong) sum $\sum_i p_i$ exists and is equal to their supremum, cf.~\cite[Remark 4.2.9]{hanche1984jordan}.
\end{remark}


\begin{proposition}[{cf.~\cite[Theorem 80IV]{bramthesis}}]\label{prop:approxpseudoinverse}
  Let $A$ be a JBW-algebra, and let $a\in A$ be positive. Then $a$ has an approximate pseudo-inverse.
\end{proposition}
\begin{proof}
  We restrict our attention to $W(a)$ and hence we may assume that the JBW-algebra is a $C(X)$ where $X$ is a basically disconnected space. For an element $a:X\rightarrow \R$ let $a^+:X\rightarrow [0,\infty)$ be the function $a^+(x) = a(x)$ when $a(x)\geq 0$ and $a^+(x) = 0$. Similarly we define $a^-(x) = -a(x)$ when $a(x)\leq 0$ and $a^-(x) = 0$ otherwise. Note that $a^+$ and $a^-$ are clearly continuous so that $a^+,a^-\in C(X)$, that $a = a^+ - a^-$, that $a^+,a^-\geq 0$ and that $Q_{\sqrt{a^+}} a^- = a^+*a^- = 0$. Hence, by Lemma~\ref{lem:JBW-ceil-in-ceil}, also $Q_{\sqrt{a^+}} \ceil{a^-} = 0$. Repeating the argument we get $\ceil{a^+}*\ceil{a^-} = 0$. Note that furthermore $a*\ceil{a^+} = a^+$.

  Now let $a\geq 0$ and define $a_n := a-\frac1n$ for $n\geq 1$. Then both $a_n$ and $a_n^+$ are increasing sequences that converge in norm to $a$. We also have $\bigvee_n \ceil{a_n^+} = \ceil{a}$. 
  Write $p_n := \ceil{a_n^+}$ so that $p_1\leq p_2\leq \ldots$ converges weakly to $\ceil{a}$. Define $q_0 := p_1$ and $q_n := p_{n+1}-p_n$ for $n\geq 1$ so that all the $q_n$ are orthogonal and $\sum_{i=0}^k q_i = p_{k+1}$. Hence the weak limit $\sum_i q_i$ exists and is equal to $\ceil{a}$.

  We calculate $(a-\frac1n)*p_n = a_n*\ceil{a_n^+} = a_n^+ \geq 0$ so that $a*p_n \geq \frac1n p_n$. Then we easily verify that $\frac{1}{n+1}q_n \leq a*q_n \leq \frac1n q_n$ and thus $q_n = \ceil{a*q_n}$.
  Hence, $a*q_n$ has a pseudo-inverse $0\leq t_n\leq q_n = \ceil{a*q_n}$ satisfying $t_n*a = (t_n*q_n)*a = t_n*(a*q_n) = q_n$.
  Finally, $\lim_k\sum_{i=0}^k t_n*a = \lim_k \sum_{i=0}^k q_n = \ceil{a}$ as desired.
\end{proof}

\section{Division and filters}\label{sec:division-filter}

In this section we will show that effects in JBW-algebras have filters. We will do this by proving that JBW-algebras allow a kind of division operation on its elements.

Before we do so, we will need to prove some additional properties of the triple product.
Recall that $Q_{a,b} := T_aT_b + T_bT_a - T_{a*b}$, and hence that $Q_{a,b}$ is linear in both $a$ and $b$ and that $Q_{a,b}=Q_{b,a}$. Also recall that $Q_a := Q_{a,a}$.

\begin{lemma}\label{lem:linearizedfundamentalequality}
    Let $A$ be a Jordan algebra, and let $a,b,c \in A$ be arbitrary. Then $Q_{Q_a b, Q_a c} = Q_a Q_{b,c} Q_a$.
\end{lemma}
\begin{proof}
    Recall the fundamental equality: $Q_{Q_a d} = Q_a Q_d Q_a$. Take $d=b+c$ and expand both sides using linearity to get $Q_{Q_a b} + Q_{Q_a c} + 2 Q_{Q_a b, Q_a c} = Q_a Q_b Q_a + Q_a Q_c Q_a + 2 Q_a Q_{b,c} Q_a$. Cancelling the terms corresponding to the fundamental equality on both sides and dividing by two then gives the desired expression.
\end{proof}

\begin{lemma}\label{lem:commutingtripleproduct}
    Let $A$ be a Jordan algebra, and let $a,b,c \in A$. Suppose $a$ and $b$ operator commute with $c$. Then $Q_{a,b} c = T_{a*b} c$.
\end{lemma}
\begin{proof}
    $Q_{a,b} c := a*(b*c) + b*(a*c) - (a*b)*c = T_aT_c b + T_b T_c a - (a*b)*c = T_cT_a b + T_c T_b a - (a*b)*c = c*(a*b) + c*(b*a) - c*(a*b) = (a*b)*c = T_{a*b}c$
\end{proof}

\begin{lemma}\label{lem:JBW-Cauchy-Schwarz-triple}
  Let $a,b \in A$ be arbitrary, and let $c\in A$ be positive. Let $\omega:A\rightarrow \R$ be a positive linear map. Then $\lvert \omega(Q_{a,b} c) \rvert^2 \leq \omega(Q_a c)~\omega(Q_b c)$.
\end{lemma}
\begin{proof}
  It is easily verified that $\inn{a,b}_\omega := \omega(Q_{a,b} c)$ defines a pre-inner-product on $A$ (this is where we need the positivity of $c$). The desired inequality is then simply the Cauchy-Schwarz inequality.
\end{proof}

\begin{lemma}\label{lem:quadraticweakconverge}
  Let $a_1,a_2,\ldots$ be a bounded sequence that strongly converges to $a\in A$. Let $b\in A$ be arbitrary. Then $Q_{a_n} b$ converges weakly to $Q_a b$ and uniformly so for $0\leq b\leq 1$.
\end{lemma}
\begin{proof}
  First, we will assume $b$ is positive.
  It is easily verified using the linearity of the triple product that $Q_a b - Q_{a_n} b = Q_{a-a_n} b + 2 Q_{a-a_n, a_n} b$. We will show that each of these two terms on the right converge weakly to zero. The first term readily follows because $0\leq Q_{a-a_n} b \leq \norm{b} (a-a_n)^2$.
  For the second term we observe, using Lemma~\ref{lem:JBW-Cauchy-Schwarz-triple}, that for any state $\omega$
  $$\lvert \omega(Q_{a-a_n, a_n}b)\rvert^2 \leq \omega(Q_{a-a_n}b) \omega(Q_{a_n} b).$$
  As the sequence $a_n$ is bounded, $\omega(Q_{a_n} b)\leq \norm{b}\omega(a_n^2)$ is also bounded and hence, since $\omega(Q_{a-a_n}b)\leq \norm{b}\omega((a-a_n)^2)$ vanishes, we conclude that $\omega(Q_{a-a_n, a_n}b)$ vanishes. Since $\omega$ was arbitrary, $Q_{a-a_n, a_n} b$ indeed converges weakly to zero, so that we conclude that $Q_{a_n} b$ converges weakly to $Q_a b$. Since the only dependence on $b$ in the bounds was via the quantity $\norm{b}$, the convergence is uniform for $0\leq b\leq 1$.

  Now if $b$ is arbitrary, we simply write $b = b^+ - b^-$ where $b^+$ and $b^-$ are positive, and we use the weak continuity of addition to get the desired result.
\end{proof}

We can now start proving the existence of a division operation on JBW-algebras.

\begin{proposition}\label{prop:JBW-division}
    Let $a,b \in A$ be effects with $a\leq b^2$ and let $t_1,t_2,\ldots$ denote an approximate pseudo-inverse for $b$. Then $Q_{\sum_{i=1}^N t_i} a$ converges weakly to an effect $c\in A$ satisfying $Q_b c = a$. This convergence is uniform in $a$. We denote this $c$ by $a/b^2$.\indexd{division (JBW-algebra)}\index{math}{a/b@$a/b$}
\end{proposition}
\begin{proof}
    Let $t_1,t_2,\ldots$ be an approximate pseudo-inverse for $b$. Recall from Proposition~\ref{prop:approxpseudoinverse} that the $t_i$ are all orthogonal and lie in $W(b)$ and hence operator commute with $b$, $b^2$ and $\sqrt{b}$.

    Define $s_N := \sum_{i=1}^N t_i$. Then all the $s_N$ also operator commute with $b$, $b^2$ and $\sqrt{b}$.

    Define $a_N := Q_{s_N}a$. We will show that $a_N$ weakly converges to an element $c$ that has the desired property.
    First we remark that $a_N\geq 0$ for all $N$. Using the operator commutation of $s_N$ and $\sqrt{b}$ we calculate
    $$a_N = Q_{s_N}a \leq Q_{s_N} b^2 = Q_{s_N} Q_b 1 = Q_{\sqrt{b}}Q_{s_N}Q_{\sqrt{b}} 1 = Q_{Q_{\sqrt{b}} s_N} 1 \stackrel{\eqref{lem:commutingtripleproduct}}{=} b^2*s_n \leq \ceil{b}.$$
    Hence, the sequence $a_N$ is bounded.

    To show that $a_N$ converges weakly it then suffices to show that it is weakly Cauchy (Corollary~\ref{cor:weakly-Cauchy}).
	Let $M>N$. Using linearity of the triple product we easily verify that $Q_{s_M} - Q_{s_N} = Q_{s_M-s_N} + 2 Q_{s_M-s_N, s_N}$, and hence $a_M-a_N = Q_{s_M} a - Q_{s_N} a = Q_{s_M-s_N} a  + 2Q_{s_M-s_N, s_N} a$.
    The first term is positive and is bounded by $Q_{s_M-s_N} b^2$ which indeed weakly vanishes as $M$ and $N$ become large. 
    For the second term we note that for any state $\omega$:
    $$\lvert \omega(Q_{s_M-s_N, s_N} a)\rvert^2 \stackrel{\eqref{lem:JBW-Cauchy-Schwarz-triple}}{\leq} \omega(Q_{s_M-s_N} a)^2 \omega(Q_{s_N} a)^2.$$
    As $Q_{s_N} a = a_N\leq 1$ is bounded, and $\omega(Q_{s_M-s_N}a) \leq \omega(Q_{s_M-s_N} b^2)$ vanishes, we see that $\omega(Q_{s_M-s_N, s_N} a)$ also vanishes as $M$ and $N$ become large. Since $\omega$ was arbitrary we conclude that the second term weakly tends to zero, and hence $a_N$ forms a weak Cauchy sequence, so that it weakly converges. As the bounds we found did not make reference to the $a_N$, this convergence is uniform in $a$.

    Now let $c$ denote the weak limit of the sequence $a_N$. We claim that $Q_b c = a$. Define $b_N = s_N*b$ so that the $b_N$ are bounded and converge strongly to $\ceil{b}$. We see that $Q_b Q_{s_N} = Q_{\sqrt{b}}Q_{s_N}Q_{\sqrt{b}} = Q_{Q_{\sqrt{b}}s_N} = Q_{b_N}$ using Lemma~\ref{lem:commutingtripleproduct}. Now using the weak continuity of $Q_b$:
    $$Q_b c =  Q_b \lim_N a_N = \lim_N Q_b a_N = \lim_N Q_b Q_{s_N} a = \lim_N Q_{b_N} a \stackrel{\ref{lem:quadraticweakconverge}}{=}Q_{\ceil{b}} a = a. \qedhere$$ 
\end{proof}

\begin{proposition}\label{prop:divisionunique}
    Let $a,b,c$ be effects with $a,c\leq \ceil{b}$. Then $Q_b a = Q_b c$ iff $a=c$.
\end{proposition}
\begin{proof}
    Obviously if $a=c$ then $Q_b a = Q_b c$. For the other direction, let $a,c\leq \ceil{b}$ and suppose that $Q_b a = Q_b c$. Let $t_1,t_2,\ldots$ be an approximate pseudo-inverse for $b$. Analogously to the previous proposition, letting $b_N := \sum_{i=1}^N t_i*b$, we have $Q_{\sum_{i=1}^Nt_i} Q_b a = Q_{b_N} a$ and similarly $Q_{\sum_{i=1}^N t_i} Q_b c = Q_{b_N} c$. As $b_N$ strongly converges to $\ceil{b}$, we calculate (using Lemma~\ref{lem:quadraticweakconverge}) $a = Q_{\ceil{b}} a = \lim_n Q_{b_N} a = \lim_n Q_{b_N} c =  Q_{\ceil{b}} c = c$ (where we have used that $Q_{\ceil{b}} a = a$ since $a\leq \ceil{b}$). 
\end{proof}

By the previous proposition, the element $c$ satisfying $Q_b c = a$ for a pair $a\leq b^2$ is unique, given that we pick $c\leq \ceil{b}$. We can extend the definition of such elements to the entire positive cone:

\begin{proposition}
    Let $a,b\in A$ be positive, satisfying $a\leq \lambda b^2$ for some $\lambda > 0$. Then there is a unique positive $c\leq \lambda \ceil{b}$ satisfying $Q_{b} c = a$.
\end{proposition}
\begin{proof}
    If $a\leq \lambda b^2$, then $\frac{a}{\lambda \norm{b^2}} \leq \frac{1}{\norm{b^2}} b^2 = (b/\norm{b})^2$, and these are both effects. So there exists a unique effect $c'\leq \ceil{b/\norm{b}} = \ceil{b}$ satisfying $Q_{\frac{1}{\norm{b}} b} c' = \frac{1}{\lambda \norm{b^2}} a$. Multiplying both sides by $\lambda\norm{b^2}$ then gives $Q_b \lambda c' = a$ and hence setting $c=\lambda c'$ gives the correct positive element. Uniqueness easily follows by uniqueness of $c'$.
\end{proof}

\begin{definition}
    Let $a,b\in A$ be positive with $a\leq \lambda b$ for some $\lambda > 0$. We denote by $a/b$ the unique positive $c\in A$ with $c\leq \lambda \ceil{b}$ that satisfies $Q_{\sqrt{b}} c = a$.
\end{definition}

\begin{proposition}
    Let $a\leq \lambda b$ and $c\leq \lambda' b$. Then $(a+c)/b = a/b + c/b$. Furthermore, for any $\mu > 0$, $(\mu a)/b = \mu (a/b)$.
\end{proposition}
\begin{proof}
    Note first that $a+c\leq (\lambda+\lambda')b$, and hence $(a+c)/b$ is indeed defined. The defining property of $(a+c)/b$ is that $Q_{\sqrt{b}} ((a+c)/b) = a+c$. But we obviously also have $Q_{\sqrt{b}} (a/b + c/b) = a+c$, and hence by uniqueness: $(a+c)/b = a/b + c/b$. We prove similarly that $(\mu a)/b = \mu (a/b)$.
\end{proof}

\begin{lemma}
    Let $b\geq 0$, and let $a = a_1 - a_2$ with $a_1, a_2 \geq 0$, such that $a_i \leq \lambda_i b$ for some $\lambda_i$. Then there is a unique $c = c_1 - c_2$ with $0 \leq c_i \leq \lambda_i\ceil{b}$ such that $Q_{\sqrt{b}} c = a$. We write $a/b$ for this unique element.
\end{lemma}
\begin{proof}
    Let $a = a_1 - a_2$ satisfy the conditions as specified. Let $c_i = a_i/b$ and set $c = c_1- c_2$. Then of course $Q_{\sqrt{b}} c = a$. Uniqueness follows by the uniqueness and linearity of the division.
\end{proof}

\begin{proposition}\label{prop:Q-injective}
  Let $b\in A$ be positive. Then $Q_b$ restricted to $A_{\ceil{b}}$ is injective.
\end{proposition}
\begin{proof}
  Suppose $Q_b a = Q_b c$ for $a,c \in A_{\ceil{b}}$. By the previous lemma, there are unique $a', c' \in A_{\ceil{b}}$ such that $Q_b a' = Q_b a$, $Q_b c' = Q_b c$. Hence, by uniqueness, $a=a'=c'=c$.
\end{proof}

Using division we can prove the existence of filters.

\begin{proposition}\label{prop:JBW-filters}
    Let $a\in A$ be an effect. Define~$\xi_a: A_{\ceil{a}} \rightarrow A$ to
    be the map $\xi_a(b) := Q_{\sqrt{a}} b$.
    Then $\xi_a$ is a filter for $a$. I.e.~$f(1)\leq a$ and for any map $f:B\rightarrow A$ with $f(1)\leq a$, there exists a unique map $\cl{f}: B\rightarrow A_{\ceil{a}}$ satisfying $f = \xi_a\circ \cl{f}$. Furthermore, if $f$ is normal, then so is $\cl{f}$.
    We will call the map $\xi_a$ the \Define{standard filter}\indexd{standard filter (JBW-algebra)} for $a$.
\end{proposition}
\begin{proof}
    Clearly~$\xi_a(1)=a$. We need to show that this map is final with respect to this property.
    To this end, assume~$f: B\rightarrow A$ is any positive sub-unital linear map with~$f(1)\leq a$.
    We need to find a unique~$\cl{f}:B\rightarrow A_{\ceil{a}}$ satisfying $\xi_a\circ \cl{f}=f$.

    For any $0\leq b\leq 1$ we have $f(b)\leq f(1) \leq a$. Any element $b\in B$ can be written as $b=b_1-b_2$ where $0\leq b_i\leq \norm{b_i}1$, so that $f(b_i)\leq \norm{b_i}a$. Hence, by the previous results we can define the map $\cl{f}:B\rightarrow A_{\ceil{a}}$ by
    $\cl{f}(b)= f(b)/a$. Again by the previous propositions, this map is linear and positive and $\cl{f}(1) \leq \ceil{a}$ so that it is also sub-unital. Furthermore
    $(\xi_a\circ \cl{f})(b) = Q_{\sqrt{a}} (f(b)/a) = f(b)$, so that is satisfies the required condition.

    Uniqueness follows by injectivity of $Q_a$ restricted to $A_{\ceil{a}}$ (Proposition~\ref{prop:Q-injective}).

    Suppose $f$ is normal. Then $f$ is weakly continuous, and as the unit interval of $B$ is weakly compact, $f$ restricted to $[0,1]_B$ is uniformly weakly continuous. For $0\leq b\leq 1$ we recall from Proposition~\ref{prop:JBW-division} that $\cl{f}(b) = f(b)/a$ is defined as the weak limit $\cl{f}(b) = \lim_N Q_{\sum_{i=1}^N t_i} f(b)$, where $t_1,t_2,\ldots$ forms an approximate pseudo-inverse for $\sqrt{a}$. Since each $Q_{\sum_{i=1}^N t_i}$ is normal, and the weak limit is uniform in $b$ and hence by the uniform continuity of $f$ also in $f(b)$, we conclude using Proposition~\ref{prop:normal-iff-weak-cont} that $\cl{f}$ is weakly continuous and hence normal.
\end{proof}

\begin{remark}
	As in Remark~\ref{remark:normal-compression}, the notion of filter could a priori be different for the category of normal positive sub-unital maps versus all positive sub-unital maps. But as the standard filter is normal, and the unique map $\cl{f}$ for the universal property is normal when $f$ is, filters in both categories coincide.
\end{remark}

\section{Diamond and dagger}\label{sec:diamond-dagger}

The previous results can be combined to show that JBW-algebras form a $\diamond$-effect-theory (Definition~\ref{def:diamond-effect-theory}).

\begin{theorem}\indexd{diamond-effect-theory@$\diamond$-effect-theory}
  The category \textbf{JBW}$_{\text{npsu}}^\opp$ is a $\diamond$-effect-theory.
\end{theorem}
\begin{proof}
  By Proposition~\ref{prop:JBWcorner} every effect has a compression, while Proposition~\ref{prop:JBW-filters} shows that filters exist for every effect. The existence of images is shown in Proposition~\ref{prop:normal-has-image}. Finally, an effect $q$ is sharp if and only if it is idempotent, and hence $q$ is sharp iff $q^\perp$ is sharp.
\end{proof}

Furthermore, it is easily seen that for the standard filter $\xi_p$ and compression $\pi_p$ of a sharp effect $p$ that $\xi_p\circ \pi_p = \id$, and hence filters and compressions are compatible as in Definition~\ref{def:compatible-filters-compressions}. All the results of Section~\ref{sec:diamond-effect-theory} are therefore applicable to \textbf{JBW}$_{\text{npsu}}^\opp$.

For instance, let us restate Proposition~\ref{prop:effect-decomposition-of-maps}.c) in the language of JBW-algebras. Recall that a map $f$ is faithful when $f(a)=0$ implies $a=0$ for all effects $a$, or equivalently when $\im{f} = 1$.
\begin{proposition}\label{prop:JBW-decomposition-of-maps}
    Let $f:A\rightarrow B$ be a normal positive sub-unital map between JBW-algebras. 
    Then there is a unique unital and faithful map $\cl{f}: A_{\im{f}} \rightarrow A_{\ceil{f(1)}}$ such that $f = \xi_{f(1)}\circ \cl{f}\circ \pi_{\im{f}}$.
\end{proposition}
\begin{corollary}\label{cor:pure-decomp}
	Let $f:A\rightarrow B$ be a pure map, \ie~$f=\xi\circ\pi$ for a compression $\pi$ and filter $\xi$. Then the unique map $\cl{f}$ above is an isomorphism. If $f$ is faithful, then $f$ is a filter and if $f$ is unital, then $f$ is a compression.
\end{corollary}
\begin{proof}
	As $\pi(1) = 1$ we have $f(1) = \xi(1)$ and hence $\xi = \xi_{f(1)}\circ \Theta_1$ for an isomorphism $\Theta_1$. Similarly, as $\im{\xi} = 1$ we have $\im{f} = \im{\pi}$ so that $\pi = \Theta_2\circ \pi_{\im{f}}$. Putting it together we have $f = \xi_{f(1)}\circ\Theta_1\circ\Theta_2\circ \pi_{\im{f}}$. The uniqueness of $\cl{f}$ then makes it equal to the isomorphism $\Theta_1\circ\Theta_2$.

	If $f$ is faithful, then $\pi_{\im{f}} = \pi_1 = \id$ so that $f=\xi\circ\cl{f}$ and hence it is a filter. Similarly, if $f(1)=1$ we have $\xi_{f(1)} = \xi_1 = \id$, making $f$ a compression.
\end{proof}

We will now work towards establishing the remaining properties required of a pure effect theory. We will do this in several stages. First, we will establish that the pure maps are closed under composition and hence form a category. To do this we need to derive an analogue to \emph{polar decompositions} as found in von Neumann algebras (cf.~Section~\ref{sec:polar-decomposition}). Then we will establish some new results regarding the $\diamond$-structure in Section~\ref{sec:unique-diamond}. This will allow us to define a dagger structure on the category of pure maps in Section~\ref{sec:dagger-pure-maps} via an abstract argument found in Bas Westerbaan's PhD thesis~\cite[215III]{basthesis}.




Before we continue, we need to know more about the $\diamond$-structure of the quadratic maps. As the directions with regard to the $\diamond$-structure are flipped, let us state explicitly the definitions in a JBW-algebra for convenience.
\begin{definition}
	Let $f:A\rightarrow B$ be a normal positive map between JBW-algebras. We define $f^\diamond(p) := \ceil{f(p)}$ for idempotents $p\in A$, and $f_\diamond(q) := \im{\pi_q \circ f} = \im{Q_q \circ f}$ for idempotents $q\in B$.
\end{definition}
Adapting Proposition~\ref{prop:galois-properties}, we see that $(f\circ g)^\diamond = f^\diamond \circ g^\diamond$ while $(f\circ g)_\diamond = g_\diamond\circ f_\diamond$ for maps $f$ and $g$ between JBW-algebras.

\begin{lemma}\label{lem:quadratic-is-contraposed}
	Let $A$ be a JBW-algebra and $a\in A$ arbitrary. Let $p,q\in A$ be idempotents. Then $Q_a p \leq q^\perp \iff Q_a q \leq p^\perp$.
\end{lemma}
\begin{proof}
	Suppose $Q_a p \leq q^\perp$. Then $Q_q Q_a p = 0$ and hence $Q_a Q_q Q_a p = Q_{Q_a q} p = 0$. Precomposing with $Q_p$ then gives $0 = Q_p Q_{Q_a q} Q_p 1 = Q_{Q_p Q_a q} 1 = 0$ so that $Q_p Q_a q = 0$ and hence $Q_a q \leq p^\perp$. Since the situation is symmetric in $p$ and $q$ the other direction follows analogously.	
\end{proof}


\begin{proposition}\label{prop:Q-diamond-self-adjoint}
	Let $A$ be a JBW-algebra and $a\in A$ arbitrary. Then $(Q_a)_\diamond = Q_a^\diamond$.
\end{proposition}
\begin{proof}
	By the previous lemma $(Q_a)^\diamond$ is left Galois adjoint to $(Q_a)^{\ssquare}$ (Definition~\ref{def:galois-connection}). Since $(Q_a)_\diamond$ is also left adjoint to $(Q_a)^{\ssquare}$ the result follows by uniqueness of these adjoints.
\end{proof}

\begin{corollary}\label{cor:image-of-quadratic}
	Let $A$ be a JBW-algebra with $a\in A$. Then $\im{Q_a} = \ceil{a} = \ceil{Q_a1}$.
\end{corollary}
\begin{proof}
	$\im{Q_a} = (Q_a)_\diamond(1) = Q_a^\diamond(1) = \ceil{Q_a 1} = \ceil{a^2} = \ceil{a}$.
\end{proof}

\subsection{Polar decomposition}\label{sec:polar-decomposition}

In this section we will generalise the notion of polar decompositions as present in von Neumann algebras to the setting of JBW-algebras. The \Define{polar decomposition}\indexd{polar decomposition} of $a\in \mathfrak{A}$ where $\mathfrak{A}$ is a von Neumann algebra is the decomposition $a=up$ where $p = \sqrt{a^*a}$ is positive and $u$ is a \emph{partial isometry}\indexd{partial isometry}\indexd{isometry!partial ---} satisfying $uu^* = \ceil{aa^*}$ and $u^*u = \ceil{a^*a}$.

Of course, non-self-adjoint elements (including most partial isometries) do not exist in a JBW-algebra. So in order to generalize the concept of a polar decomposition to JBW-algebras we instead will consider the maps $\Phi, \Phi^*:\mathfrak{A}_{\sa}\rightarrow \mathfrak{A}_{\sa}$ given by $\Phi(b) := ubu^*$ and $\Phi^*(b) := u^*bu$ that a decomposition induces, in particular for the polar decomposition of a product of self-adjoint $a$ and $b$.

\begin{definition}
	Let $A$ be a JBW-algebra with $a,b \in A$. A \Define{polar pair}\indexd{polar pair} for $a$ and $b$ is a pair of maps $\Phi, \Phi^*:A\rightarrow A$ such that:
	\begin{itemize}
		\item $Q_bQ_a = \Phi Q_{\sqrt{Q_ab^2}}$ and $Q_aQ_b = Q_{\sqrt{Q_a b^2}} \Phi^*$.
		\item $\Phi(1) = \ceil{Q_b a^2}$ and $\Phi^*(1) = \ceil{Q_a b^2}$.
		\item $\Phi^*\Phi = Q_{\ceil{Q_a b^2}}$ and $\Phi\Phi^* = Q_{\ceil{Q_b a^2}}$.
	\end{itemize}
\end{definition}

Note that the `star' in $\Phi^*$ has no special significance, it is just to remind the reader what the origin is of these maps. To motivate the definition, let us show the existence of polar pairs in JW-algebras.

\begin{proposition}\label{prop:polardecomp-JW}
	Let $A$ be a JW-algebra with $a, b\in A$. Then $a$ and $b$ have a polar pair.
\end{proposition}
\begin{proof}
	Let $\mathfrak{A}$ denote the von Neumann algebra that $A$ acts on. Let $u\in \mathfrak{A}$ be the partial isometry associated to the polar decomposition of $ba$, \ie~which satisfies $ba = u\sqrt{(ba)^*ba} = u\sqrt{ab^2a}$, $uu^* = \ceil{(ba)(ba)^*} = \ceil{ba^2b}$ and $u^*u = \ceil{(ba)^*(ba)} = \ceil{ab^2a}$. 
	Then for any $c\in\mathfrak{A}$ set $\Phi(c) = ucu^*$ and $\Phi^*(c) = u^*cu$. Note that $\Phi Q_{\sqrt{Q_a b^2}}~c = u\sqrt{ab^2a}~c\sqrt{ab^2a}u^* = bac(ba)^* = bacab = Q_bQ_a c$. The other required properties are also easily checked.
\end{proof}

Now we will prove that polar pairs in fact exist for all JBW-algebras. Proving this is much more involved.

\begin{proposition}\label{prop:JBW-polardecomp}
    Let $A$ be a JBW-algebra with $a, b\in A$. Then $a$ and $b$ have a polar pair.
\end{proposition}
\begin{proof}
    We would wish to take $\Phi = Q_bQ_aQ_{(Q_a b^2)^{-1/2}}$ and $\Phi^* = Q_{(Q_a b^2)^{-1/2}}Q_aQ_b$, but as $Q_a b^2$ does not have to have a pseudo-inverse, we will instead build $\Phi$ and $\Phi^*$ as weak limits involving the approximate pseudo-inverse of $Q_a b^2$.

    Let $t_1, t_2,\ldots$ denote the elements of an approximate pseudo-inverse of $Q_a b^2$. To recall, $t_n*(Q_a b^2) = \ceil{t_n}$ and all the $\ceil{t_n}$ are orthogonal with $\sum_n \ceil{t_n} = \ceil{Q_a b^2}$. The $t_n$ belong to $W(Q_a b^2)$ and hence they operator commute with $\ceil{Q_a b^2}$, $Q_a b^2$ and $\sqrt{Q_a b^2}$. The same holds for the $\sqrt{t_n}$. As a result $Q_{\sqrt{t_n}} (Q_a b^2) = \sqrt{t_n}^2*(Q_a b^2) = \ceil{t_n}$.

    Define $s_n := \sum_{i=1}^n t_i$. Then the $s_n$ are also in the bicommutant of $Q_a b^2$ and we note that $\ceil{s_n}$ is an increasing sequence with supremum $\ceil{Q_a b^2}$, and hence is also strongly convergent. As the $t_n$ are orthogonal we have $\sqrt{s_n} = \sum_{i=1}^n \sqrt{t_i}$. Note that if $n\leq m$ then $\sqrt{s_n}*\sqrt{s_m} = s_n$. Furthermore, $Q_{s_n} Q_{Q_a b^2} = Q_{\sqrt{s_n}} Q_{Q_a b^2} Q_{\sqrt{s_n}} = Q_{Q_{\sqrt{s_n}} (Q_a b^2)} = Q_{\ceil{s_n}}$.

    Define $\Phi_n = Q_bQ_aQ_{\sqrt{s_n}}$ and $\Phi_n^* = Q_{\sqrt{s_n}} Q_a Q_b$. We of course have $\Phi_n^*(1) = Q_{\sqrt{s_n}} Q_a b^2 = \ceil{s_n}$. The expression for $\Phi_n(1)$ does not simplify as easily, and hence we will give it a new name. Set $r_n := \Phi_n(1) = Q_bQ_a s_n$.

    For the remainder of this proof we will let $n,m \in \N_{>0}$ with $n\leq m$. We calculate:
    \begin{equation}\label{eq:polardecomp-1}
    \Phi_n^*\Phi_m = Q_{\sqrt{s_n}} Q_{Q_a b^2} Q_{\sqrt{s_m}} = Q_{\sqrt{s_n}}Q_{\sqrt{s_m}} Q_{Q_a b^2} = Q_{s_n} Q_{Q_a b^2} = Q_{\ceil{s_n}}.
    \end{equation}
    Similarly we also calculate:
    \begin{equation}\label{eq:polardecomp-2}
    \Phi_n \Phi_m^* = Q_b Q_a Q_{\sqrt{s_n}} Q_{\sqrt{s_m}} Q_a Q_b = Q_b Q_a Q_{s_n} Q_a Q_b = Q_{Q_bQ_a s_n} = Q_{r_n}
    \end{equation}

    Before we proceed, we need to know more about the $r_n$. First of all, they are idempotent:
    \begin{align*}
    r_n^2 &= Q_{r_n}1 = Q_bQ_aQ_{s_n} Q_aQ_b 1 = Q_bQ_aQ_{\sqrt{s_n}} Q_{\sqrt{s_n}} (Q_a b^2) \\
    &= Q_bQ_aQ_{\sqrt{s_n}} \ceil{s_n} = Q_b Q_a s_n = r_n.
    \end{align*}
    Second, $r_1,r_2,\ldots$ forms an increasing sequence:
    \begin{align*}
    Q_{r_n} r_m &= Q_bQ_aQ_{s_n}Q_aQ_bQ_bQ_a s_m = Q_bQ_aQ_{s_n} Q_{Q_ab^2} s_m = Q_bQ_a Q_{\ceil{s_n}} s_m \\
    &= Q_bQ_a s_n = r_n.
    \end{align*}
    As a result $r=\vee_n r_n$ exists, is an idempotent, and is also the strong limit of the sequence $r_1,r_2,\ldots$.

    We wish to define, for any $c\in A$, $\Phi(c)$ and $\Phi^*(c)$ as the weak pointwise limits of $\Phi_n(c)$ and $\Phi_n^*(c)$. As we also want $\Phi$ and $\Phi^*$ to be normal, these pointwise limits must be uniform when restricted to the unit interval (see Proposition~\ref{prop:weak-pointwise-limit}). For $\Phi^*$ this follows as in the proof of Proposition~\ref{prop:JBW-filters} (with $f:=Q_aQ_b$ and $a:=Q_a b^2$). For $\Phi$ we have to use a slightly more involved argument.

    First, let $c\in A$ be an effect. As $\Phi_n(c) \leq \Phi_n(1) = r_n\leq 1$, it is a bounded sequence, and hence for $\Phi(c) = \lim_n \Phi_n(c)$ to exist, it suffices to show that it is weakly Cauchy.
    In other words, we need to show that $\Phi_m(c) - \Phi_n(c) = Q_bQ_a(Q_{\sqrt{s_m}}-Q_{\sqrt{s_n}})c$ vanishes weakly as $m$ and $n$ become large. Our proof is analogous to that of Proposition~\ref{prop:JBW-division}. Recall from that proof that for any $d$ and $e$, $Q_d - Q_e = Q_{d-e} + 2 Q_{d-e, e}$, and hence we have
    \[\Phi_m(c) - \Phi_n(c) = Q_bQ_aQ_{\sqrt{s_m}-\sqrt{s_n}} c + 2 Q_bQ_aQ_{\sqrt{s_m}-\sqrt{s_n},\sqrt{s_n}} c.\]
    The first term is positive and bounded by $Q_bQ_aQ_{\sqrt{s_m}-\sqrt{s_n}} 1 = Q_bQ_a (s_m - s_n) = r_m - r_n$ and thus vanishes as the $r_n$ weakly go to $r$. For the second term we will require another Cauchy-Schwarz inequality.
    
    Let $\omega$ be any normal state. Define $\inn{d,e}_\omega := \omega(Q_bQ_aQ_{d,e}\, c)$. This is a pre-inner-product (bilinearity and symmetry are clear and it is positive semi-definite because $c\geq 0$) and hence we get a Cauchy-Schwarz inequality:
    \[\lvert \omega(Q_bQ_aQ_{d,e} c)\rvert^2 = \lvert \inn{d,e}_\omega\rvert^2 \leq \inn{d,d}_\omega\inn{e,e}_\omega = \omega(Q_bQ_aQ_d c)~\omega(Q_bQ_aQ_e c).\]
    Applying this inequality with $d:=\sqrt{s_m}-\sqrt{s_n}$ and $e:=\sqrt{s_n}$ gives the quantity we want to see vanish on the left-hand side, while the right-hand side becomes $\omega(Q_bQ_aQ_{\sqrt{s_m}-\sqrt{s_n}} c)\, \omega(Q_bQ_a Q_{\sqrt{s_n}} c)$. This right-hand side vanishes as the latter term is bounded ($0\leq \omega(Q_bQ_aQ_{\sqrt{s_n}}c) \leq \omega(Q_bQ_a s_n) = \omega(r_n)\leq 1$), while the first term is positive and bounded by $\omega(r_m-r_n)$. We conclude that $\Phi_n(c)$ converges weakly, and as our bounds did not involve $c$, it converges uniformly for all effects. The map is then easily extended to all $c\in A$, and by Proposition~\ref{prop:weak-pointwise-limit} $\Phi$ is normal.

    From Eq.~\eqref{eq:polardecomp-1} we had $\Phi_n^*\Phi_m = Q_{\ceil{s_n}}$, and hence, using the weak continuity of $\Phi_n^*$, for any $c\in A$: $(\Phi_n^*\Phi)(c) = \lim_m \Phi_n^* \Phi_m(c) = \lim_m Q_{\ceil{s_n}}c = Q_{\ceil{s_n}}c$. Hence $(\Phi^*\Phi)(c) = \lim_n \Phi_n^*\Phi(c) = \lim_n Q_{\ceil{s_n}}c = Q_{\ceil{Q_a b^2}}c$, using Lemma~\ref{lem:quadraticweakconverge} as $\ceil{s_n}\rightarrow \ceil{Q_a b^2}$ strongly. So indeed $\Phi^*\Phi = Q_{\ceil{Q_a b^2}}$.
    By a similar argument, but using Eq.~\eqref{eq:polardecomp-2}, we get $\Phi\Phi^* = Q_r$.
    Additionally, as $\Phi_n^*(1) = \ceil{s_n}$ and $\Phi_n(1) = r_n$ we also easily calculate $\Phi^*(1) = \ceil{Q_a b^2}$ and $\Phi(1) = r$. We then need to show that $r=\ceil{Q_b a^2}$, but before we do that we establish some of the other equalities first.

    We have $\Phi_n Q_{\sqrt{Q_a b^2}} = Q_bQ_a Q_{\ceil{s_n}}$ and  $Q_{\sqrt{Q_a b^2}} \Phi_n^* = Q_{\ceil{s_n}} Q_a Q_b$. Hence, taking weak limits we get 
    \begin{equation}\label{eq:polardecomp-3}
    \Phi Q_{\sqrt{Q_a b^2}} = Q_bQ_a Q_{\ceil{Q_a b^2}} \ \ \text{and} \ \ Q_{\sqrt{Q_a b^2}}\Phi^* = Q_{\ceil{Q_a b^2}} Q_a Q_b = Q_a Q_b.
    \end{equation} 
    For the first equality we would like to be able to drop the $Q_{\ceil{Q_a b^2}}$. To do this we note that \[\im{Q_b Q_a} = (Q_b Q_a)_\diamond(1) = (Q_a)_\diamond (Q_b)_\diamond(1) = Q_a^\diamond Q_b^\diamond(1) = \ceil{Q_a b^2}\]
    and so by Lemma~\ref{lem:JBWpreserveimage} $(Q_bQ_a)Q_{\ceil{Q_a b^2}} = Q_b Q_a$.

    Using Eqs.~\eqref{eq:polardecomp-3} we calculate:
    \[Q_{Q_b a^2} = Q_bQ_aQ_aQ_b = \Phi Q_{\sqrt{Q_a b^2}} Q_{\sqrt{Q_a b^2}} \Phi^*.\]
    Hence, by plugging $1$ into both sides and taking ceilings:
    \[\ceil{Q_b a^2} = \ceil{\Phi Q_{Q_a b^2} \Phi^*(1)} = \ceil{\Phi Q_{Q_a b^2} \ceil{Q_a b^2}} = \ceil{\Phi (Q_a b^2)^2} \leq \ceil{\Phi(1)} = r.\]

    It now remains to prove the other inequality $r\leq \ceil{Q_b a^2}$. To do this note that $r=\im{Q_r} = \im{\Phi\Phi^*} \leq \im{\Phi^*}$ (these images exist since $\Phi$ and $\Phi^*$ are normal, see Proposition~\ref{prop:normal-has-image}), so if we show $\im{\Phi^*} = \ceil{Q_b a^2}$ we will be done.

    Let $s$ be an idempotent. Note first that if $Q_a Q_b s = 0$, then $\Phi_n^*(s) = 0$ for all $n$ and hence $\Phi^*(s) = 0$. Conversely, if $\Phi^*(s) = 0$, then $Q_{\sqrt{Q_a b^2}}\Phi^*(s) = Q_a Q_b s = 0$. Hence:
    \begin{align*}
    \im{\Phi^*} \leq s^\perp &\iff \Phi^*(s) = 0 \\
    &\iff Q_a Q_b s = 0 \\
    &\iff Q_b^\diamond(s) = \ceil{Q_b s} \leq \im{Q_a}^\perp = \ceil{a}^\perp \\
    &\iff (Q_b)_\diamond(\ceil{a}) \leq s^\perp.
    \end{align*}
    Note that $(Q_b)_\diamond(\ceil{a}) = Q_b^\diamond(\ceil{a}) = \ceil{Q_b \ceil{a}} = \ceil{Q_b a^2}$. Taking $s=(\im{\Phi^*})^\perp$ we get $\ceil{Q_b a} \leq \im{\Phi^*}$, and taking $s=\ceil{Q_b a}^\perp$ we get $\im{\Phi^*} \leq \ceil{Q_b a}$. Hence, $r\leq \im{\Phi^*} = \ceil{Q_b a}$ as desired.
\end{proof}

\begin{corollary}\label{cor:polardecomp-iso}
	Let $A$ be a JBW-algebra with $a,b\in A$ and let $\Phi$ and $\Phi^*$ be a polar pair of $a$ and $b$. Then the restriction $\Phi: A_{\ceil{Q_a b^2}} \rightarrow A_{\ceil{Q_b a^2}}$ is an order-isomorphism with inverse $\Phi^*$.
\end{corollary}
\begin{proof}
	As $\Phi(1) = \ceil{Q_b a^2}$ it obviously restricts to a map $\Phi:A_{\ceil{Q_a b^2}}\rightarrow A_{\ceil{Q_b a^2}}$, and similarly $\Phi^*$ restricts to $\Phi^*: A_{\ceil{Q_b a^2}}\rightarrow A_{\ceil{Q_a b^2}}$. As $\Phi\Phi^* = Q_{\ceil{Q_b a^2}}$ and $\Phi^*\Phi = Q_{\ceil{Q_a b^2}}$, they are each others inverses when restricted to these spaces.
\end{proof}

\begin{lemma}
	Let $\Phi, \Phi^*$ be a polar pair for $a$ and $b$. Then $\im{\Phi} = \Phi^*(1)$ and $\im{\Phi^*} = \Phi(1)$.
\end{lemma}
\begin{proof}
	As $\Phi^*\Phi = Q_{\ceil{Q_a b^2}}$ we have $\Phi^*(1) = \ceil{Q_a b^2} = \im{\Phi^*\Phi} \leq \im{\Phi}$. For the other direction note that $\Phi(\ceil{Q_a b^2}) = \Phi(\Phi^*(1)) = (\Phi\Phi^*)(1) = \ceil{Q_b a^2} = \Phi(1)$, and hence $\ceil{Q_a b^2}\geq \im{\Phi}$. That $\im{\Phi^*} = \Phi(1)$ follows similarly.
\end{proof}

\begin{proposition}\label{prop:pure-composition}
	Let $\xi_a: A_{\ceil{a}}\rightarrow A$, $\xi_a =
	Q_{\sqrt{a}}$ be the standard filter of an
	effect $a$ and $\pi_p: A \rightarrow E_p$, $\pi_p = Q_p$ be the standard compression of an idempotent effect $p$.
	Then $\pi_p \circ \xi_a = \xi_b\circ \Phi\circ \pi_q$ where
	$b$ and $q$ are some effects and $\Phi$ is an isomorphism.
	In other words: $\pi_p\circ \xi_a$ is pure (cf.~Definition~\ref{def:pure}).
\end{proposition}
\begin{proof}
	Let $f = \pi_p \circ \xi_a = Q_p Q_{\sqrt{a}}$. Then $f(1) = Q_p a$ and $\im{f} = \ceil{Q_{\sqrt{a}} p}$. Hence, by Proposition~\ref{prop:JBW-decomposition-of-maps} there is a unique unital and faithful map $\cl{f}: A_{\ceil{Q_{\sqrt{a}p}}} \rightarrow A_{\ceil{Q_p a}}$ such that $f = \xi_{Q_p a}\circ \cl{f}\circ \pi_{\ceil{Q_{\sqrt{a}}p}}$. It remains to show that $\cl{f}$ is an isomorphism.

	Let $\Phi,\Phi^*$ be a polar pair for $\sqrt{a}$ and $p$. By Corollary~\ref{cor:polardecomp-iso}, $\Phi^*$ restricts to an isomorphism $\Phi^*:A_{\ceil{Q_{\sqrt{a}p}}} \rightarrow A_{\ceil{Q_p a}}$. So in particular it is unital and faithful. If we can then show that $f = \xi_{Q_p a}\circ \Phi^*\circ \pi_{\ceil{Q_{\sqrt{a}}p}}$, then by the uniqueness of $\cl{f}$ we have $\cl{f} = \Phi^*$ and we are done.

	Expand the definitions to get $\xi_{Q_p a}\circ \Phi^*\circ \pi_{\ceil{Q_{\sqrt{a}}p}} = Q_{\sqrt{Q_p a}} \Phi^* Q_{\ceil{Q_{\sqrt{a}}p}}$. As $\im{\Phi^*} = \Phi(1) = \ceil{Q_{\sqrt{a}}p}$ we have by Lemma~\ref{lem:JBWpreserveimage} $\Phi^* Q_{\ceil{Q_{\sqrt{a}}p}} = \Phi^*$. Furthermore, by the definition of a polar pair $Q_{\sqrt{Q_p a}} \Phi^* = Q_p Q_{\sqrt{a}}$, and hence indeed 
	\[\xi_{Q_p a}\circ \Phi^*\circ \pi_{\ceil{Q_{\sqrt{a}}p}} = Q_{\sqrt{Q_p a}} \Phi^* Q_{\ceil{Q_{\sqrt{a}}p}} = Q_p Q_{\sqrt{a}} = f. \qedhere\]
\end{proof}

Though not strictly a corollary, we can reuse the proof of the previous proposition, for the following:

\begin{corollary}
    Let $A$ be a JBW-algebra and $a\in A$ with $-1\leq a \leq 1$. Then $Q_a$ is pure.
\end{corollary}
\begin{proof}
    It is clear that $Q_{\sqrt{a^2}}$ is a pure map as it is equal to $\pi_{\ceil{a}}\circ \xi_{a^2}$. Let $\Phi$ and $\Phi^*$ be a polar pair of $1$ and $a$. Then $Q_a = Q_1Q_a = \Phi Q_{\sqrt{a^2}} = Q_{\sqrt{a^2}} \Phi^*$. In the previous proof let $f:=Q_a$. Then in the same way we can show that $\cl{f} = \Phi^*$ is an isomorphism. 
\end{proof}

\begin{theorem}\label{thm:JBW-pure-composition}
	Pure maps are closed under composition in \textbf{JBW}$_{\text{psu}}^\opp$.
\end{theorem}
\begin{proof}
	Given a pure map $f=\xi\circ \pi$ we can write $\xi = \xi_q\circ \Theta_1$ and $\pi = \Theta_2\circ \pi_p$ where $\xi_q$ is the standard filter for $q$, $\pi_p$ is the standard compression for $p$ and the $\Theta_i$ are isomorphisms. Hence any pure map can be written as $f=\xi_q\circ \Theta\circ \pi_p$ where $\Theta=\Theta_1\circ \Theta_2$ is an isomorphism.

	So for $i=1,2$ let $f_i = \xi_{q_i}\circ\Theta_i\circ \pi_{p_i}$ be pure maps. We need to show that $f_1\circ f_2$ can again be written in this form. By Proposition~\ref{prop:pure-composition} $\pi_{p_1}\circ \xi_{q_2} = \xi_a \circ \Phi \circ \pi_b$ where $a$ and $b$ are effects and $\Phi$ is an isomorphism. Note that $\xi' = \xi_{q_1}\circ \Theta_1$ is a filter and $\pi' = \Theta_2\circ \pi_{p_2}$ is a compression. Hence:
 	\[f_1\circ f_2 = \xi'\circ \xi_a \circ \Phi \circ \pi_b \circ \pi'.\]
 	By Proposition~\ref{prop:faithfulfilters} the composition of filters is again a filter and by Proposition~\ref{prop:composition-of-compressions} the composition of compressions is a compression. Hence, indeed $f_1\circ f_2 = \xi''\circ \Phi\circ \pi''$ for some filter $\xi''$ and compression $\pi''$.
\end{proof}

\subsection{Unique diamond-positivity}\label{sec:unique-diamond}

We now know that pure maps in JBW-algebras are closed under composition. However, the definition of a pure effect theory required the pure maps to form a dagger category. As constructing this dagger manually is tedious, we will take a round-about approach to finding it, in the process establishing a new property that will turn out to be a useful axiom in Chapter~\ref{chap:infinitedimension}. This property is the uniqueness of $\diamond$-positive maps.

\begin{definition}[{\cite[Definition~206II]{basthesis}}]
	We say a pure map $f$ is \Define{$\diamond$-self-adjoint} when $f^\diamond = f_\diamond$. It is \Define{$\diamond$-positive} when $f=g\circ g$ where $g$ is $\diamond$-self-adjoint.
\end{definition}

We already saw that $Q_a$ is $\diamond$-self-adjoint for arbitrary $a$ (Proposition~\ref{prop:Q-diamond-self-adjoint}. When $a$ is positive then $Q_a = Q_{\sqrt{a}}^2$ and hence $Q_a$ is $\diamond$-positive. These are in fact the only $\diamond$-positive maps.

\begin{theorem}\label{thm:unique-diamond-positivity}
	Let $f:A\rightarrow A$ be a $\diamond$-positive map on a JBW-algebra $A$. Then $f=Q_{\sqrt{f(1)}}$.
\end{theorem}

An analogous statement was proven for completely positive normal sub-unital $\diamond$-positive maps between von Neumann algebras in Ref.~\cite{bramthesis}. Our proof follows along the same lines, but some steps, particularly Lemma~\ref{lem:diamond-positive-lemma} will require more work to prove.

\begin{lemma}\label{lem:triple-quadratic-is-zero}
  Let $a,b,c \in A$ be arbitrary. Then $Q_a Q_b Q_c 1 = 0$ if and only if $Q_c Q_b Q_a 1 = 0$.
\end{lemma}
\begin{proof}
  Suppose $Q_a Q_b Q_c 1 = 0$. Then also $Q_c Q_b Q_a Q_a Q_b Q_c 1 = 0$, and hence by the fundamental equality $Q_{Q_cQ_b a^2} 1 = 0$. As a result $Q_c Q_b a^2 = Q_c Q_b Q_a 1 = 0$. The other direction is shown similarly.
\end{proof}

\begin{lemma}\label{lem:diamond-positive-lemma}
  Let $\Theta: A\rightarrow A$ be a unital Jordan homomorphism, $a\in A$ an effect with $\ceil{a} = 1$, and $p\in A$ an idempotent. Suppose that $\ceil{Q_a \Theta(p)} \leq p$ and $\ceil{Q_a \Theta(p^\perp)} \leq p^\perp$. Then $p$ and $a$ operator commute and $\Theta(p) = p$.
\end{lemma}
\begin{proof}
  We remark that the assumptions are symmetric in $p$ and $p^\perp$, and hence any equation we prove involving $p$ and $p^\perp$ is also true with the roles of $p$ and $p^\perp$ interchanged.

  Apply $\ceil{Q_{p^\perp}\cdot}$ to both sides of $p \geq \ceil{Q_a \Theta(p)}$ to get 
  $$0 = \ceil{Q_{p^\perp} p} \geq \ceil{Q_{p^\perp} \ceil{Q_a \Theta(p)}} \stackrel{\ref{lem:JBW-ceil-in-ceil}}{=} \ceil{Q_{p^\perp} Q_a \Theta(p)}.$$
  Hence $0 = Q_{p^\perp} Q_a Q_{\Theta(p)} 1$, as $\Theta(p)^2 = \Theta(p)$. By Lemma~\ref{lem:triple-quadratic-is-zero} we then also have $Q_{\Theta(p)} Q_a Q_{p^\perp} 1 = 0$ and since these are all positive maps, we have in fact $Q_{\Theta(p)} Q_a Q_{p^\perp}=0$.
  By symmetry of $p$ and $p^\perp$, then also
  $Q_{\Theta(p^\perp)} Q_a Q_p = 0$.


  Note that $\id = Q_1 = Q_{\Theta(p)+\Theta(p^\perp)} = Q_{\Theta(p)} + Q_{\Theta(p^\perp)} + 2 Q_{\Theta(p),\Theta(p^\perp)}$. We hence calculate, using the linearised fundamental equality (Lemma~\ref{lem:linearizedfundamentalequality}):
  \begin{align}\label{eq:proof-diamond-positive-lemma-0}
  Q_a Q_a Q_{p^\perp} &= Q_a \id Q_a Q_{p^\perp} \nonumber \\
  &= Q_a Q_{\Theta(p)}Q_aQ_{p^\perp} + Q_a Q_{\Theta(p^\perp)}Q_aQ_{p^\perp} + 2 Q_a Q_{\Theta(p),\Theta(p^\perp)}Q_aQ_{p^\perp} \nonumber\\
  &= 0 + Q_a Q_{\Theta(p^\perp)}Q_aQ_{p^\perp} + 2 Q_{Q_a\Theta(p), Q_a\Theta(p^\perp)}Q_{p^\perp}.
  \end{align}
  We claim that the last term in this equation is also zero. To see this, note that as $0\leq Q_a\Theta(p) \leq \ceil{Q_a\Theta(p)} \leq p$ and $0\leq Q_a\Theta(p^\perp)\leq p^\perp$, both $Q_a\Theta(p)$ and $Q_a\Theta(p^\perp)$ operator commute with both $p$ and $p^\perp$ so that $Q_{Q_a\Theta(p), Q_a\Theta(p^\perp)}$ commutes with $Q_{p^\perp}$ and hence:
  \begin{align*}
  Q_{Q_a\Theta(p), Q_a\Theta(p^\perp)}Q_{p^\perp} &= 
  Q_{Q_a\Theta(p), Q_a\Theta(p^\perp)}Q_{p^\perp}Q_{p^\perp} =
  Q_{p^\perp} Q_{Q_a\Theta(p), Q_a\Theta(p^\perp)}Q_{p^\perp} \\
  &\stackrel{\ref{lem:linearizedfundamentalequality}}{=} Q_{Q_{p^\perp}Q_a\Theta(p), Q_{p^\perp}Q_a\Theta(p^\perp)} = Q_{0, Q_{p^\perp}Q_a\Theta(p^\perp)} = 0.
  \end{align*}
  Hence, Eq.~\eqref{eq:proof-diamond-positive-lemma-0} reduces to $Q_aQ_a Q_{p^\perp} = Q_a Q_{\Theta(p^\perp)}Q_a Q_{p^\perp}$.
  As $\ceil{a} = 1$, $Q_a$ is injective (Proposition~\ref{prop:Q-injective}), and hence $Q_aQ_{p^\perp} = Q_{\Theta(p^\perp)}Q_aQ_{p^\perp}$. By symmetry in $p$ and $p^\perp$, we then also have:
  \begin{equation}\label{eq:proof-diamond-positive-lemma-1}
    Q_aQ_p = Q_{\Theta(p)} Q_a Q_p
  \end{equation}

  We will now use a similar argument with the goal of showing that $Q_{\Theta(p)}Q_a = Q_{\Theta(p)} Q_aQ_p$, which together with Eq.~\eqref{eq:proof-diamond-positive-lemma-1} would give $Q_{\Theta(p)} Q_a = Q_aQ_p$.

  So let us expand $Q_1$ again, but now using $p$ and $p^\perp$:
  \begin{align*}
    Q_aQ_{\Theta(p)}Q_a &= \id Q_aQ_{\Theta(p)}Q_a = Q_{p^\perp}Q_aQ_{\Theta(p)}Q_a + Q_{p}Q_aQ_{\Theta(p)}Q_a + 2 Q_{p,p^\perp} Q_a Q_{\Theta(p)}Q_a \\
    &= 0 + Q_{p}Q_aQ_{\Theta(p)}Q_a + 2 Q_{p,p^\perp} Q_a Q_{\Theta(p)}Q_a.
  \end{align*}
  Now again, since $Q_a \Theta(p)$ operator commutes with $p$ and $p^\perp$, so that $\sqrt{Q_a \Theta(p)}$ also operator commutes with $p$ and $p^\perp$, we calculate:
  \begin{align*}
  Q_{p,p^\perp} Q_a Q_{\Theta(p)}Q_a &= Q_{p,p^\perp} Q_{Q_a \Theta(p)} = Q_{\sqrt{Q_a\Theta(p)}} Q_{p,p^\perp} Q_{\sqrt{Q_a\Theta(p)}} \\
  &= Q_{Q_{\sqrt{Q_a\Theta(p)}}p, Q_{\sqrt{Q_a\Theta(p)}}p^\perp} \stackrel{\ref{prop:idempotent-rules}}{=} Q_{Q_{\sqrt{Q_a\Theta(p)}}p, 0} =  0.
  \end{align*}
  Here the second to last equality uses Proposition~\ref{prop:idempotent-rules} with $\sqrt{Q_a\Theta(p)}\leq p$.

  As a result, we have $Q_a Q_{\Theta(p)}Q_a = Q_pQ_aQ_{\Theta(p)}Q_a = Q_p Q_{Q_a\Theta(p)} = Q_{Q_a\Theta(p)} Q_p = Q_aQ_{\Theta(p)}Q_a Q_p$. By again canceling the first $Q_a$ we are left with $Q_{\Theta(p)}Q_a = Q_{\Theta(p)}Q_a Q_p$. But this right-hand side agrees with that of \eqref{eq:proof-diamond-positive-lemma-1}, and hence:
  \begin{equation}\label{eq:proof-diamond-positive-lemma-2}
  Q_{\Theta(p)}Q_a = Q_a Q_p.
  \end{equation}
  We assumed that $\ceil{a} =1 $ and hence also $\ceil{a^2}$ = 1. Now:
  \begin{align*}
  p &\geq \ceil{Q_a \Theta(p)} = \ceil{Q_a Q_{\Theta(p)} 1} = \ceil{Q_a Q_{\Theta(p)}\ceil{a^2}} \stackrel{\ref{lem:JBW-ceil-in-ceil}}{=} \ceil{Q_aQ_{\Theta(p)}Q_a 1} \\
  &\stackrel{\eqref{eq:proof-diamond-positive-lemma-2}}{=} \ceil{Q_a Q_a Q_p 1} = \ceil{Q_{a^2} p}
  \end{align*}
  Doing the analogous argument with $p^\perp$ we see that we arrive at the inequalities:
  \begin{equation*}
    \ceil{Q_{a^2} p} \leq p \qquad \qquad \ceil{Q_{a^2}p^\perp} \leq p^\perp
  \end{equation*}
  We can now repeat our proof from the start, but with $a^2$ instead of $a$, and~$\id$ instead of $\Theta$. The analogous version of Equation~\eqref{eq:proof-diamond-positive-lemma-2} is then $Q_p Q_{a^2} = Q_{a^2} Q_p$, and hence $p$ operator commutes with $a^2$. By Proposition~\ref{prop:operator-commutation}, $p$ then operator commutes with everything in $W(a^2)$, including $\sqrt{a^2} = a$.
  Using this operator commutation, we can rewrite \eqref{eq:proof-diamond-positive-lemma-2} to $Q_{\Theta(p)} Q_a = Q_p Q_a$. Then $Q_{Q_a \Theta(p)} = Q_a Q_{\Theta(p)}Q_a = Q_a Q_p Q_a = Q_{Q_a p}$, and hence $(Q_a\Theta(p))^2 = Q_{Q_a\Theta(p)} 1 = Q_{Q_a p} 1 = (Q_a p)^2$. As $Q_a\Theta(p)$ and $Q_a p$ are both positive, we must then have $Q_a p = Q_a \Theta(p)$. By the injectivity of $Q_a$ we indeed get $p=\Theta(p)$ as desired.
\end{proof}

\begin{corollary}
	Let $a\in A$ be an effect with $\ceil{a} = 1$, and let $\Theta:A\rightarrow A$ be a unital Jordan homomorphism such that $\ceil{Q_a \Theta(p)} \leq p$ for all idempotents $p\in A$. Then $a$ is central and $\Theta = \id$.
\end{corollary}

\begin{proposition}[{cf.~\cite[Proposition~104VII]{bramthesis}}]
	Let $a,b \in A$ be effects with $\ceil{a} = \ceil{b} = 1$, and suppose there is a normal unital Jordan homomorphism $\Theta:A\rightarrow A$ such that $\ceil{Q_a p} = \ceil{Q_b \Theta(p)}$ for all idempotents $p\in A$. Then $\Theta=\id$.
\end{proposition}
\begin{proof}
	Let $p$ be any idempotent that operator commutes with $a$, so that $Q_p a^2 = Q_pQ_a1 = Q_aQ_p1 = Q_a p$. We then calculate:
	\[p = \ceil{Q_p1} = \ceil{Q_p\ceil{a^2}} = \ceil{Q_p a^2} = \ceil{Q_a p} = \ceil{Q_b\Theta(p)}\]
	As $p^\perp$ also operator commutes with $a$ we also have $p^\perp = \ceil{Q_b\Theta(p^\perp)}$ and hence Lemma~\ref{lem:diamond-positive-lemma} applies and we see that $\Theta(p) = p$ and that $p$ operator commutes with $b$. As linear combinations of idempotents lie norm-dense in $W(a)$, $b$ also operator commutes with $a$ and $a^2$ and $\Theta(a) = a$.

	By Proposition~\ref{prop:commuting-elements-span-associative-algebra}, $a$ and $b$ span an associative JBW-algebra, and hence, analogous to the proof of Proposition~\ref{prop:approxpseudoinverse}, we can find a series of orthogonal idempotents $p_1,p_2,\ldots$ that operator commute with both $a$ and $b$ and satisfy $\sum_k p_k = \ceil{a} = 1$ such that $p_k*a$ and $p_k*b$ are pseudo-invertible for all $k$.
    Let $s_n = \sum_{k=1}^n p_n$, then the $s_n$ are idempotents that operator commute with $a$ and strongly converge to $1$. Furthermore, $s_n*a$ and $s_n*b$ are pseudo-invertible for all $n$.
	
	As the $s_n$ operator commutes with $a$ we have $\Theta(s_n) = s_n$ so that $\Theta$ restricts to a map $\Theta: A_{s_n} \rightarrow A_{s_n}$. Suppose that $\Theta$ is equal to the identity on each of these restrictions. As the sequence $s_n$ strongly converges to $1$, by Lemma~\ref{lem:quadraticweakconverge} we have for any $c\in A$: $\Theta(c)= \Theta(Q_1 c) =  \Theta(\lim_n Q_{s_n} c) = \lim_n \Theta(Q_{s_n} c) = \lim_n Q_{s_n} c = Q_1 c = c$ (since $\Theta$ is assumed to be normal, it is weakly continuous), so that indeed $\Theta = \id$. It thus suffices to show that $\Theta$ is the identity on each $A_{s_n}$. Because $s_n*a$ and $s_n*b$ are pseudo-invertible in $A$, $a$ and $b$ are invertible when restricted to $A_{s_n}$. Without loss of generality we may therefore assume now that $a$ and $b$ are invertible.

	Because $a$ and $b$ span an algebra of mutually operator commuting elements, $b$ also operator commutes with $a^{-1}$ and hence $Q_{a^{-1}} Q_b = Q_{a^{-1}*b}$.
	Now, let $p$ be any idempotent. We compute:
	\begin{align*}
	\ceil{Q_{a^{-1}*b} \Theta(p)} &= \ceil{Q_{a^{-1}} Q_b \Theta(p)} = \ceil{Q_{a^{-1}} \ceil{Q_b \Theta(p)}} \\
	&= \ceil{Q_{a^{-1}} \ceil{Q_a p}} = \ceil{Q_{a^{-1}} Q_a p} = \ceil{p} = p
	\end{align*}
	Hence, by the previous corollary, $\Theta=\id$ (and $a^{-1}*b$ is central).
\end{proof}

\begin{proposition}[{cf.~\cite[Proposition~104IX]{bramthesis}}]\label{prop:faithful-diamond-positive}
	Let $f:A\rightarrow A$ be a faithful $\diamond$-positive map. Then $f=Q_{\sqrt{a}}$ where $a := f(1)$.
\end{proposition}
\begin{proof}
	Since $f$ is faithful and pure, it is a filter (Corollary~\ref{cor:pure-decomp}), and hence it is of the form $f=Q_{\sqrt{a}}\Theta$ for some isomorphism $\Theta:A\rightarrow A$. It remains to show that $\Theta=\id$. 

	Because $f$ is faithful, $1=\im{f}$, and because it is $\diamond$-self-adjoint: $\ceil{\sqrt{a}} = \ceil{f(1)} = f^\diamond(1) = f_\diamond(1) = \im{f} = 1$. Hence, to show that $\Theta=\id$, it suffices by the previous proposition to show that there is some effect $b$ with $\ceil{b}= 1$ and $f^\diamond(p) := \ceil{Q_{\sqrt{a}}\Theta(p)} = \ceil{Q_b p}$ for all idempotents $p$.

	By definition of $\diamond$-positivity, there exists a $\diamond$-self-adjoint map $g:A\rightarrow A$ with $f=g\circ g$. Since $1=\im{f} = f_\diamond(1) = g_\diamond(g_\diamond(1)) \leq g_\diamond(1) = \im{g}$ we must have $\im{g} = 1$. Since $g$ is then faithful and pure, it is also a filter. As $\xi := Q_{\sqrt{g(1)}}$ is also a filter of $g(1)$, there is an isomorphism $\Phi:A\rightarrow A$ with $g = \xi\circ \Phi$. We calculate $\xi^\diamond\circ \Phi^\diamond = g^\diamond = g_\diamond = \Phi_\diamond\circ \xi_\diamond = (\Phi^\diamond)^{-1} \xi^\diamond$. Bringing $(\Phi^\diamond)^{-1}$ to the other side then gives $\Phi^\diamond\circ \xi^\diamond \circ \Phi^\diamond = \xi^\diamond$.
	As a result $f^\diamond = (g\circ g)^\diamond = \xi^\diamond\circ \Phi^\diamond\circ \xi^\diamond \circ \Phi^\diamond = \xi^\diamond \circ \xi^\diamond = (\xi\circ\xi)^\diamond$.
	For an arbitrary idempotent $p$ we then see that $\ceil{Q_{\sqrt{a}}\Theta(p)} = f^\diamond(p) = (\xi\circ \xi)^\diamond(p) = \ceil{Q_{\xi(1)}p}$. By the previous proposition then indeed $\Theta = \id$, and hence $f = Q_{\sqrt{a}}$.
\end{proof}

\begin{lemma}\label{lem:compression-adjoint-filter}
	Let $p\in A$ be an idempotent. Then its standard compression and standard filter are $\diamond$-adjoint: $\xi_p^\diamond = (\pi_p)_\diamond$.
\end{lemma}
\begin{proof}
	Let $q\in A_p$ be an idempotent. Then $q\leq p$ and hence $Q_p q = q$ and $\pi_q\circ \pi_p = \pi_q$. Hence $\xi_p^\diamond(q) = \ceil{Q_p q} = q$ and $(\pi_p)_\diamond(q) = \im{\pi_q\circ \pi_p} = \im{\pi_q} = q$.
\end{proof}

\begin{proof}[Proof of Theorem~\ref{thm:unique-diamond-positivity}]
	We need to show that each $\diamond$-positive map $g$ satisfies $g = Q_{\sqrt{g(1)}}$.

	First, let $f$ be a $\diamond$-self-adjoint map. Then in particular $\im{f} = \ceil{f(1)}$. Using the universal properties of compressions and filters, we factor $f$ as $f = \xi_{\im{f}}\circ \cl{f}\circ \pi_{\im{f}}$ where $\cl{f}:A_{\im{f}}\rightarrow A_{\im{f}}$. Note that by definition $\xi_{\im{f}}$ is just the inclusion map into $A$. We also see that $\cl{f} = \pi_{\im{f}}\circ f \circ \xi_{\im{f}}$ (since $\pi_{\im{f}}\circ \xi_{\im{f}} = \id$) so that $\cl{f}$ is a composition of pure maps and hence also pure by Theorem~\ref{thm:JBW-pure-composition}.

	We see that $\cl{f}(1) = \pi_{\im{f}}(f(\im{f})) = f(\im{f})=f(1)$ and furthermore
	\[\cl{f}^\diamond = (\pi_{\im{f}}\circ f\circ \xi_{\im{f}})^\diamond = \pi_{\im{f}}^\diamond \circ f^\diamond \circ \xi_{\im{f}}^\diamond \stackrel{\eqref{lem:compression-adjoint-filter}}{=} (\xi_{\im{f}})_\diamond \circ f_\diamond \circ (\pi_{\im{f}})_\diamond 
	= \cl{f}_\diamond.\] 

	Note that $\im{f^2} = f^\diamond(f_\diamond(1)) = f^\diamond(\im{f}) = f^\diamond(1) = f_\diamond(1)=\im{f}$. For a $\diamond$-self-adjoint $f$ we then get $\cl{f}^2 = \pi_{\im{f}}\circ f \circ \xi_{\im{f}}\circ \pi_{\im{f}}\circ f \circ \xi_{\im{f}} = \pi_{\im{f}}\circ f^2 \circ \xi_{\im{f}} = \cl{f^2}$. 

	Now let $g=f\circ f$ be $\diamond$-positive. Then $\cl{g}=\cl{f^2}=\cl{f}^2$ is also $\diamond$-positive. Since $\cl{g}$ is faithful, we have $\cl{g} = Q_{\sqrt{a}}$ where $a = \cl{g}(1)=g(1)$ by Proposition~\ref{prop:faithful-diamond-positive}. Since $\im{g} = \ceil{g(1)} = \ceil{a}$ we can then calculate $g=\xi_{\im{g}} \circ Q_{\sqrt{p}} \circ \pi_{\im{g}} = Q_{\sqrt{p}}Q_{\ceil{p}} = Q_{\sqrt{p}}$.
\end{proof}

\subsection{A dagger on pure maps}\label{sec:dagger-pure-maps}

Finally, let us see how these results help us define a dagger $\dagger$ on pure maps. Any pure map can be written as a composition of a standard filter, a standard compression, and an isomorphism, and hence a dagger is completely determined by its action on these types of maps. Our goal will be to make $\diamond$-self-adjoint maps also $\dagger$-self-adjoint. 

We will map isomorphisms $\Theta$ to their inverse: $\Theta^\dagger = \Theta^{-1}$. Recall that for compressions $\pi_a = \pi_{\floor{a}}$ and hence it suffices to define the action of the dagger on standard compressions for idempotents. We will set $\pi_p^\dagger = \xi_p$ for idempotents $p$. This determines the action of the dagger on filters of idempotents: $\xi_p^\dagger = \pi_p$. The only type of map left then is a filter for a non-idempotent. To define a dagger for this map we first split it up into two parts: $\xi_a = Q_{\sqrt{a}} \xi_{\ceil{a}}$. As $Q_{\sqrt{a}}$ is $\diamond$-self-adjoint we will define $Q_{\sqrt{a}}^\dagger = Q_{\sqrt{a}}$, and hence we arrive at $\xi_a^\dagger = \pi_{\ceil{a}}Q_{\sqrt{a}}$.

Although this in principle defines a dagger, we still need to check that it \emph{actually} results in a dagger structure. Specifically, that $(f^\dagger)^\dagger = f$ and $(f\circ g)^\dagger = g^\dagger\circ f^\dagger$. Since proving this is quite lengthy, we will refer to a result from Bas Westerbaan's PhD thesis~\cite{basthesis}. Before we state that result we recall some further definitions.

\begin{definition}[{cf.~\cite[Definition~211II]{basthesis}}]
	A $\diamond$-effect-theory is an \Define{\&-effect-theory} when pure maps are closed under composition and when for each effect $a\in\eff(A)$ there is a unique $\diamond$-positive map $\asrt_a:A\rightarrow A$ such that $1\circ\asrt_a = a$. This map is called the \Define{assert map of $a$}. We write $a\mult b := b\circ \asrt_a$ and $a^2 = a\mult a$.
\end{definition}

\begin{proposition}
	The category \textbf{JBW}$_{\text{npsu}}^\opp$ is an \&-effect-theory.
\end{proposition}
\begin{proof}
	Closure of pure maps under composition is Theorem~\ref{thm:JBW-pure-composition}, while uniqueness of $\diamond$-positive maps is Theorem~\ref{thm:unique-diamond-positivity}.
\end{proof}

Recall that a map $f$ in a dagger-category is \Define{$\dagger$-positive} when $f=g^\dagger\circ g$ for some map $g$.

\begin{definition}[{cf.~\cite[Definition~215I]{basthesis}}]\label{def:dagger-effect-theory}
	An \&-effect-theory is a \Define{$\dagger$-effect-theory} when
	\begin{itemize}
		\item the pure maps form a dagger-category such that $\asrt_a^\dagger = \asrt_a$ for all effects $a$ and $f^\dagger$ is $\diamond$-adjoint to $f$ for all pure maps $f$;
		\item for every $\dagger$-positive map $f$, there is a unique $\dagger$-positive $g$ with $g\circ g = f$;
		\item $\diamond$-positive maps are $\dagger$-positive.
	\end{itemize}
\end{definition}

\begin{theorem}[{\cite[Theorem~215III]{basthesis}}]\label{thm:dagger-effectus}
	A \&-effect theory is a $\dagger$-effect-theory if and only if
	\begin{itemize}
		\item for every effect $a$ there is a unique effect $b$ with $a=b^2$;
		\item for all effects $a$ and $b$ we have $\asrt_{a\mult b}^2 = \asrt_a\circ\asrt_b^2\circ \asrt_a$;
		\item a filter for a sharp effect $\xi$ is a \Define{sharp map}\indexd{sharp map}: $\ceil{p\circ \xi} = p\circ \xi$ for all sharp $p$.
	\end{itemize}
\end{theorem}

\begin{remark}
	The proof of Theorem~\ref{thm:dagger-effectus} as stated in Theorem~215III of Ref.~\cite{basthesis} assumes the structure of an effectus, not just an effect theory. The proof however does not require any of the special structure present in an effectus beyond that which is already present in an effect theory, and hence continues to apply in our setting. Alternatively, it is not too hard to show that \textbf{JBW}$_{\text{npsu}}^\opp$ is actually an effectus itself, so that we are still warranted in using it.
\end{remark}

\begin{theorem}\label{thm:JBW-dagger}
	The category \textbf{JBW}$_{\text{npsu}}^\opp$ is a $\dagger$-effect-theory.
\end{theorem}
\begin{proof}
	We already know that \textbf{JBW}$_{\text{npsu}}^\opp$ is a \&-effect-theory so it remains to show that the three conditions of Theorem~\ref{thm:dagger-effectus} hold.

	First note that $\asrt_a = Q_{\sqrt{a}}$ and hence $a\mult b = Q_{\sqrt{a}} b$ so that $b^2 = b\mult b = Q_{\sqrt{b}} b = b*b$.
	Hence the first point of Theorem~\ref{thm:dagger-effectus} follows because every effect in a JBW-algebra has a unique positive square root.

	Unfolding the definition of the assert map, we see that the second point reduces to proving $Q_{\sqrt{Q_{\sqrt{a}} b}}^2 = Q_{\sqrt{a}} Q_{\sqrt{b}}^2 Q_{\sqrt{a}}$, but this is just the fundamental identity.

	Finally, the third point follows because the standard filter $\xi_p:A_p\rightarrow A$ of a sharp (\ie~idempotent) effect $p$ is simply an embedding satisfying $\xi_p(q) = q$.
\end{proof}

\begin{proposition}
	The category \textbf{JBW}$_{\text{npsu}}^\opp$ is a PET (cf.~Definition~\ref{def:PET}).
\end{proposition}
\begin{proof}
	We already have \ref{pet:filtcompr}, \ref{pet:images} and \ref{pet:sharpnegation}. That the pure maps form a dagger-category (\ref{pet:dagger}) follows from Theorem~\ref{thm:JBW-dagger}. By Ref.~\cite[Proposition~216VII]{basthesis} we have $\xi_p^\dagger = \pi_p$ for any sharp $p$ in a $\dagger$-effectus, and hence \ref{pet:sharpadjoint} holds. Finally, we have $\pi_p\circ \pi_p^\dagger = \pi_p \circ \xi_p = \id$ so that \ref{pet:sharpisometry} holds.
\end{proof}

\begin{remark}
	As discussed in Ref.~\cite[Section~3.8.3]{basthesis}, any $\dagger$-effectus is a \emph{pointed homological category}~\cite{grandis1992categorical}\indexd{ homological category (pointed)}. Hence, results about exact sequences such as the well-known \indexd{snake lemma}\emph{Snake Lemma} hold in \textbf{JBW}$_{\text{npsu}}^\opp$. As far as the author is aware, this structure has not been considered before for JBW-algebras. We leave its implications for future work.
\end{remark}

\section{Sequential products}\label{sec:JBW-SEA}
In the previous section we found the structures of Chapter~\ref{chap:effectus} which are present in JBW-algebras. Now we set our sights on the structure of sequential measurement studied in Chapter~\ref{chap:seqprod}.
We will establish that the unit interval of a JBW-algebra has a sequential product given by $a\mult b := Q_{\sqrt{a}}b$ that satisfies all the assumptions of Definition~\ref{def:seqprod}. Before we are able to do so however, we need to establish some more results regarding operator commutation in JBW-algebras and for that we will need to know more about the global structure of a JBW-algebra.

\subsection{Structure theorem}\label{sec:JBW-structure}

Recall that special Jordan algebras are those that embed into an associative algebra. The counterpart to those algebras are the \emph{exceptional} Jordan algebras.

\begin{definition}
	Let $A$ be a JB-algebra. We call $A$ \Define{purely exceptional}
  \indexd{purely exceptional JB-algebra} 
  \indexd{JB-algebra!purely exceptional ---}
  when any Jordan homomorphism $\phi:A\rightarrow \mathfrak{A}_{\sa}$ onto a C$^*$-algebra $\mathfrak{A}$ is necessarily zero.
\end{definition}

\begin{theorem}[{\cite[Theorem 7.2.7]{hanche1984jordan}}]\label{thm:JBW-decomposition}
	Let $A$ be a JBW-algebra. Then there is a unique decomposition $A=A_{\text{sp}} \oplus A_{\text{ex}}$ where $A_{\text{sp}}$ is a JW-algebra and $A_{\text{ex}}$ is a purely exceptional JBW-algebra.
\end{theorem}

We saw in Section~\ref{sec:spectral-theorem} that if $C(X)$ is a JBW-algebra, that $X$ is basically disconnected.
In fact, $X$ actually satisfies a stronger property: 

\begin{definition}
	A compact Hausdorff space $X$ is \Define{extremally disconnected}\indexd{extremally disconnected} when the closure of any open set is again open (and hence clopen). If $C(X)$ is additionally separated by normal states, then $X$ is called \Define{hyperstonean}\indexd{hyperstonean space}.
\end{definition}

The space $X$ is hyperstonean if and only if $C(X)$ is an associative JBW-algebra. Since the self-adjoint part of a commutative von Neumann algebra is also isomorphic to $C(X)$ with $X$ hyperstonean, there is a correspondence between commutative von Neumann algebras and associative JBW-algebras.

\begin{example}[{\cite{shultz1979normed}}]
	Let $X$ be a hyperstonean space and let $E = M_3(\mathbb{O})_{\sa}$ denote the exceptional Albert algebra of self-adjoint octonion matrices. Denote by $C(X,E)$ the set of continuous functions $f:X\rightarrow E$. Then $C(X,E)$ is a purely exceptional JBW-algebra with the Jordan product given pointwise by $(f*g)(x) = f(x)*g(x)$.
\end{example}

The above example is actually the only type of purely exceptional JBW-algebra, as the following result by Shultz shows.

\begin{theorem}[{\cite{shultz1979normed}}]\label{thm:purely-exceptional-char}
	Let $A$ be a purely exceptional JBW-algebra. Then there exists a hyperstonean space $X$, such that $A\cong C(X,M_3(\mathbb{O}))$.
\end{theorem}

Since any JBW-algebra splits up into a direct sum of a JW-algebra and an algebra of the form $C(X,M_3(\mathbb{O})_{\sa})$, many questions regarding JBW-algebras can be settled by studying von Neumann algebras and Euclidean Jordan algebras (of which $M_3(\mathbb{O})_{\sa}$ is an example). This is not the most elegant way to prove something, but for some results we do not know of any other way to prove them, such as for some results in the next section.

\subsection{Operator commutation revisited}\label{sec:JBW-commutation-revisited}

Recall that an element $a$ in a C$^*$-algebra is called \Define{normal}\indexd{normal!element of C$^*$-algebra} when $aa^* = a^*a$ (not to be confused with the definition of `normal' as a suprema preserving map, which will not play a role in this section). The following classic result regarding normal elements will be used several times throughout this section.

\begin{theorem}[Fuglede-Putnam-Rosenblum]
  Let $m,n,a\in \mathfrak{A}$ be elements of a C$^*$-algebra, with $m$ and $n$ normal and $ma = an$. Then $m^*a = an^*$.
\end{theorem}

\begin{proposition}\label{prop:JW-operator-commutation}
  Let $A\sse \mathfrak{A}$ be a JW-algebra acting on the von Neumann algebra $\mathfrak{A}$. Elements of $A$ operator commute if and only if they commute as elements of $\mathfrak{A}$: $T_aT_b = T_bT_a \iff ab=ba$.
\end{proposition}
\begin{proof}
  This result is proven for JC-algebras in \cite[Lemma 5.1]{hanche1983structure} using representation theory. We present here a more algebraic proof based on Gudder's proof of a similar result for operators on Hilbert spaces~\cite{gudder2002sequentially}.
  Obviously, when $ab=ba$, we have $T_aT_b = T_bT_a$. So let us prove the converse direction.
  For $a,b,c\in A$ we easily calculate:
  \begin{equation}\label{eq:proof-operator-commutation-in-vNA}
    (T_aT_b-T_bT_a)c = \frac14 ((ab-ba)c - c(ab-ba))
  \end{equation}
  Hence, when $a$ and $b$ operator commute we have $(ab-ba)c = c(ab-ba)$ for all $c\in A$. 
  In particular, by taking $c=a$ and $c=b$ we get
  \[2aba = ba^2 + a^2b \quad \text{and} \quad 2bab = ab^2 + b^2a.\]
  Now multiply the first equation by $b$ on the right and the second with $a$ on the left:
  \[2abab = ba^2b +a^2b^2 \qquad 2abab = a^2b^2 + ab^2a\]
  As the left-hand sides now agree, we can combine the equations to get $ba^2b = ab^2a$. 
  This equation shows that $ab$ is normal: $(ab)(ab)^* = ab^2a = ba^2b = (ab)^*(ab)$, and hence by the Fuglede-Putnam-Rosenblum theorem, since $(ab)a = a(ba)$ we also have $ba^2 = (ab)^*a = a(ba)^* = a^2b$ and so $b$ and $a^2$ commute.

  Recall that we had the equation $2aba = ba^2 + a^2b$. Using the operator commutation we get $aba = a^2b$.
  Apply the approximate pseudoinverse of $a$ on $aba = a^2b$ to get $\ceil{a}ba = ab$. As $b$ commutes with $a^2$, it commutes with $\ceil{a^2} = \ceil{a}$, and hence $ab=\ceil{a}ba = b\ceil{a}a= ba$, and we are done.
\end{proof}

\begin{proposition}\label{prop:JW-quadratic-commutation}
    Let $A\sse \mathfrak{A}$ be a JW-algebra acting on the von Neumann algebra~$\mathfrak{A}$. Let $a,b\in A$ with at least one of $a$ and $b$ positive. Then the following are equivalent.
    \begin{enumerate}[label=\alph*)]
        \item $Q_aQ_b =Q_bQ_a$.
        \item $Q_ab^2 = Q_b a^2$.
        \item $a$ and $b$ operator commute.
    \end{enumerate}
\end{proposition}
\begin{proof}
    a) to b) is trivial. For c) to a) we note that $a$ and $b$ operator commute if $ab=ba$ in $\mathfrak{A}$, and hence also $a$ and $a^2$ operator commute with $b$ and $b^2$. The result then follows by the definition of $Q_a$ and $Q_b$ in terms of $T_a, T_{a^2}, T_b$ and $T_{b^2}$. It remains to prove b) to c). Suppose $Q_ab^2 = Q_ba^2$. Written in terms of the associative product of $\mathfrak{A}$ this becomes $ab^2a = ba^2b$. Hence, the product $ab$ is normal. Without loss of generality, assume that $a$ is positive. Since $(ab)a = a(ba)$, by the Fuglede-Putnam-Rosenblum theorem: $ba^2 = (ab)^*a = a(ba)^* = a^2 b$, so that $b$ and $a^2$ commute. By positivity of $a$, $\sqrt{a^2} = a$. Since $\sqrt{a^2}$ lies in the bicommutant of $a^2$ we then see that $b$ also commutes with $a$ in $\mathfrak{A}$ and hence they operator commute in $A$.
\end{proof}

\begin{remark}
	Of course a) implies b) and c) implies a) regardless of positivity of the elements $a$ and $b$, but for the other implications, the requirement that at least one of $a$ and $b$ is positive is necessary. Take for instance any non-commuting $a$ and $b$ satisfying $a^2=b^2 = 1$ such as $a=\ket{+}\bra{+}-\ket{-}\bra{-}$ and $b=\ket{0}\bra{0}-\ket{1}\bra{1}$. Then b) holds, but a) and c) do not. Keeping $b$ the same, but letting $a=\ket{0}\bra{1}+\ket{1}\bra{0}$ we get a) but not c). 
\end{remark}

For our next results we will need the following powerful theorem.
\begin{theorem}[Shirshov-Cohn theorem for JBW-algebras]
	A JBW-algebra generated by two elements (and possibly the unit) is a JW-algebra.
\end{theorem}
\begin{proof}
	Theorem 7.2.5 of Ref.~\cite{hanche1984jordan} proves the analogous result for JB-algebras, but the proof directly translates to JBW-algebras.
\end{proof}

\begin{proposition}\label{prop:JBW-two-elements-associative}
	Let $A$ be a JBW-algebra, and suppose $a,b \in A$ either operator commute or that at least one is positive and they satisfy $Q_ab^2 = Q_ba^2$. Then the JBW-algebra spanned by $a$ and $b$ (and possibly the unit) is associative.
\end{proposition}
\begin{proof}
	Let $B$ denote the JBW-algebra spanned by $a$ and $b$. By the Shirshov-Cohn theorem, $B$ is a JW-algebra. Let $\mathfrak{B}$ denote the von Neumann algebra $B$ acts on.

	If $a$ and $b$ operator commute in $A$, then they of course also operator commute in $B$ and hence by Proposition~\ref{prop:JW-operator-commutation} $ab=ba$ in $\mathfrak{B}$. Similarly, but by Proposition~\ref{prop:JW-quadratic-commutation}, if one of $a$ and $b$ is positive and they satisfy $Q_ab^2 = Q_ba^2$, they operator commute (inside of $B$), and hence also $ab=ba$ in $\mathfrak{B}$.

	In both cases we then also have $a^2b = ba^2$ and hence $a^2$ operator commutes with $b$ when restricted to $B$. By Proposition~\ref{prop:commuting-elements-span-associative-algebra}, there is an associative JBW-subalgebra $B'$ of $B$ containing both $a$ and $b$. But as $B$ is already the smallest JBW-algebra generated by $a$ and $b$ we necessarily have $B' = B$.
\end{proof}

\begin{corollary}\label{cor:JBW-assoc-quadratic}
	Let $a,b\in A$ be positive elements in a JBW-algebra. If $Q_ab^2 = Q_ba^2$, then $Q_ab^2 = a^2*b^2$.
\end{corollary}
\begin{proof}
	By the previous proposition, $a$ and $b$ span an associative algebra, and the statement is true for associative Jordan algebras.
\end{proof}

\begin{proposition}
    Let $E$ be a Euclidean Jordan algebra with $a,b \in E$ where at least one of $a$ and $b$ is positive. Then the following are equivalent.
    \begin{enumerate}[label=\alph*)]
    	\item $Q_aQ_b =Q_bQ_a$.
        \item $Q_ab^2 = Q_b a^2$.
        \item $b$ and $b^2$ operator commute with $a$ and $a^2$.
    \end{enumerate}
\end{proposition}
\begin{proof}
	a) to b) and c) to a) are trivial, hence it suffices to prove b) to c). Nevertheless, it will be useful to first prove b) to a).

	So assume that $Q_ab^2 = Q_ba^2$. 
	Since $E$ is a Euclidean Jordan algebra, it is a real Hilbert space, and hence the space of bounded operators on $E$, $B(E)$, is a (real) C$^*$-algebra. Recall that $Q_a$ for any $a\in E$ is a self-adjoint operator and hence, if $a$ is positive, $Q_a = Q_{\sqrt{a}}^2$ is a positive operator in the Hilbert space sense, \ie~positive in $B(E)$.
	By the fundamental equality we have $Q_aQ_b^2Q_a = Q_{Q_ab^2} = Q_{Q_ba^2} = Q_bQ_a^2Q_b$ so that $Q_aQ_b$ is normal as an element of $B(E)$. Hence, analogously to the proof of Proposition~\ref{prop:JW-quadratic-commutation}, we can use the Fuglede-Putnam-Rosenblum theorem to conclude that $Q_a Q_b = Q_bQ_a$ which proves b) to a).

	To prove b) to c) we will use the identity $T_a = Q_{a+1} - Q_a - \id$ that holds for any element $a$ in a Jordan algebra. Remark that since $Q_ab^2 = Q_ba^2$, Proposition~\ref{prop:JBW-two-elements-associative} shows that $a$, $b$ and $1$ span an associative JBW-algebra $B$, which in this case is an EJA. 
	Then $a+1$ also lies in $B$, and hence $Q_{a+1}b^2 = (a+1)^2*b^2 = Q_b (a+1)^2$. Using b) to a) we then see that $Q_{a+1}Q_b = Q_bQ_{a+1}$. As also $Q_aQ_b = Q_bQ_a$ and $T_a = Q_{a+1} - Q_a - \id$ we then necessarily have $T_aQ_b = Q_bT_a$. Similarly we can show $T_aQ_{b+1} = Q_{b+1}T_a$ and hence $T_aT_b = T_bT_a$. As $a^2$ is also part of the same associative JBW-algebra, we can repeat the argument with $a^2$ instead of $a$ to see that $T_{a^2}T_b = T_bT_{a^2}$. Similarly, we can take $b^2$ instead of $b$.
\end{proof}

\begin{lemma}\label{lem:exceptional-quadratic-commutation}
	Let $A$ be a purely exceptional JBW-algebra with $a,b \in A$ where at least one of $a$ and $b$ is positive. Then the following are equivalent.
	\begin{enumerate}[label=\alph*)]
    	\item $Q_aQ_b =Q_bQ_a$.
        \item $Q_ab^2 = Q_b a^2$.
        \item $b$ and $b^2$ operator commute with $a$ and $a^2$.
    \end{enumerate}
\end{lemma}
\begin{proof}
	a) to b) and c) to a) are trivial, so only b) to c) remains. Hence, suppose that $Q_ab^2 = Q_ba^2$.

	Since $A$ is a purely exceptional, $A\cong C(X,E)$ where $X$ is a hyperstonean space, and $E=M_3(\mathbb{O})_{\text{sa}}$ is the exceptional Albert algebra and hence an EJA. Let $f,g:X\rightarrow E$ denote the functions corresponding to $a$ and $b$. Then for every $x\in X$, $Q_{f(x)}g(x)^2 = (Q_f g^2)(x) = (Q_g f^2)(x) = Q_{g(x)} f(x)^2$. As these are elements of an EJA, we use the previous proposition to see that $T_{g(x)}$ and $T_{g(x)^2}$ operator commute with $T_{f(x)}$ and $T_{f(x)^2}$ for all $x\in X$. We wish to conclude from this that $T_g$ and $T_{g^2}$ operator commute with $T_f$ and $T_{f^2}$. So let $h:X\rightarrow E$ be any other function, then we should have for every $x\in X$, $(T_fT_gh)(x) = (T_gT_fh)(x)$ (and similarly for $f^2$ and $g^2$), but as $(T_fT_gh)(x) = T_{f(x)}T_{g(x)}h(x)$ this directly follows.
\end{proof}

\begin{proposition}\label{prop:JBW-quadratic-commute}
	Let $A$ be a JBW-algebra with $a,b \in A$ where at least one of $a$ and $b$ is positive. Then the following are equivalent.
	\begin{enumerate}[label=\alph*)]
    	\item $Q_aQ_b =Q_bQ_a$.
        \item $Q_ab^2 = Q_b a^2$.
        \item $b$ and $b^2$ operator commute with $a$ and $a^2$.
    \end{enumerate}
\end{proposition}
\begin{proof}
	Write $A=A_1\oplus A_2$ where $A_1$ is a JW-algebra, and $A_2$ is a purely exceptional JBW-algebra. If $Q_ab^2 = Q_ba^2$, then this equation also holds with $a$ and $b$ restricted to $A_1$ or $A_2$. By Proposition~\ref{prop:JW-quadratic-commutation} and Lemma~\ref{lem:exceptional-quadratic-commutation} the desired result then follows.
\end{proof}

\begin{theorem}\label{thm:JBW-assoc-iff-commute}
    Let $A$ be a JBW-algebra and $a,b\in A$ arbitrary. Then the following are equivalent.
    \begin{enumerate}[label=\alph*)]
        \item $a$ and $b$ operator commute.
        \item $a$ and $b$ generate an associative JBW-algebra.
        \item $a$ and $b$ generate an associative JBW-algebra of mutually operator commuting elements.
        \item $a$ and $a^2$ operator commute with $b$ and $b^2$.
    \end{enumerate}
    If at least one of $a$ and $b$ is positive then these statements are furthermore equivalent to $Q_a b^2 = Q_b a^2$.
\end{theorem}
\begin{proof}
	a) to b) follows by Proposition~\ref{prop:JBW-two-elements-associative}. c) to d) follows because $a^2$ and $b^2$ are part of the associative algebra of mutually operator commuting elements. d) to a) is of course trivial. It remains to show b) to c).

    So suppose $a$ and $b$  generate an associative JBW-algebra $B$. Let $c,d\in B$ be positive. By associativity $Q_c d^2 = c^2*d^2 = Q_d c^2$, and hence by Proposition~\ref{prop:JBW-quadratic-commute} $c$ and $d$ operator commute. As the positive elements span $B$, we are done.

    Now suppose one of $a$ and $b$ is positive. If they span an associative algebra, then of course $Q_a b^2 = a^2*b^2 = Q_b a^2$. For the converse direction we again refer to Proposition~\ref{prop:JBW-quadratic-commute}.
\end{proof}

\subsection{The sequential product}

We can now finally show the main result of this section: that the unit interval of a JBW-algebra has a sequential product as in Definition~\ref{def:seqprod}. At the same time we will also establish two extra properties that are required in a \emph{normal sequential effect algebra} (see Definition~\ref{def:normal-SEA}).

\begin{lemma}[{\cite[Lemma~1.26]{alfsen2012geometry}}]\label{lem:quadratic-is-zero}
	Let $a$ and $b$ be positive elements in a JB-algebra. Then $Q_a b =0$ iff $Q_b a = 0$, and in that case $a*b = 0$.
\end{lemma}

\begin{theorem}\label{thm:JBW-seqprod}
	Let $A$ be a JBW-algebra. Define the operation $a\mult b := Q_{\sqrt{a}}b$. Then $\&$ satisfies all the axioms of Definition~\ref{def:seqprod}:
	\begin{enumerate}[label=\alph*)]
		\item $a\mult (b+c) = a\mult b + a\mult c$.
		\item The map $a\mapsto a\mult b$ is continuous in the norm.
		\item $1\mult a = a$.
		\item If $a\mult b = 0$, then also $b\mult a = 0$.
		\item If $a\mult b = b\mult a$, then $a\mult (b\mult c) = (a\mult b)\mult c$.
		\item If $a\mult b = b\mult a$, then $a\mult b^\perp = b^\perp \mult a$, and if also $a\mult c = c\mult a$, then $a\mult (b+c) = (b+c)\mult a$.
	\end{enumerate}
	Furthermore, it also satisfies some additional properties:
	\begin{enumerate}[resume, label=\alph*)]
		\item for any directed set of effects $S$ we have $a\mult \bigvee S = \bigvee_{b\in S} a\mult b$ and if $a\mult b = b\mult a$ for all $b\in S$ then $a\mult \bigvee S = \bigvee S \mult a$.
		\item If $a\mult b = b\mult a$ and $a\mult c = c\mult a$, then $a\mult (b\mult c) = (b\mult c)\mult a$.
	\end{enumerate}
\end{theorem}
\begin{proof}
	Points a), b) and c) are trivial.
	
	For d), if $a\mult b = 0 = Q_{\sqrt{a}} b$, then $Q_b \sqrt{a} = 0$ by Lemma~\ref{lem:quadratic-is-zero}. Hence also $Q_b a^2 \leq Q_b a \leq Q_b \sqrt{a} = 0$. Applying Lemma~\ref{lem:quadratic-is-zero} again gives $Q_a b = 0$, so that also $Q_a b^2 = 0$. But then $Q_a b^2 = 0 = Q_b a^2$, so by Proposition~\ref{prop:JBW-quadratic-commute}, $a$ operator commutes with $b$ and $b^2$. By Proposition~\ref{prop:commuting-elements-span-associative-algebra}, there is then an associative JBW-subalgebra $B$ of mutually operator commuting elements containing both $a$ and $b$. This algebra necessarily also contains $\sqrt{b}$, and hence $\sqrt{b}$ and $a$ operator commute. But then $b\mult a = Q_{\sqrt{b}} \sqrt{a}^2 = a*b = 0$ by Corollary~\ref{cor:JBW-assoc-quadratic} and Lemma~\ref{lem:quadratic-is-zero}.

	Note that if $a\mult b = b\mult a$, then by definition $Q_{\sqrt{a}} \sqrt{b}^2 = Q_{\sqrt{b}}\sqrt{a}^2$, so that by Proposition~\ref{prop:JBW-quadratic-commute}, $Q_{\sqrt{a}}Q_{\sqrt{b}} = Q_{\sqrt{b}}Q_{\sqrt{a}}$, and by Corollary~\ref{cor:JBW-assoc-quadratic} $a\mult b = Q_{\sqrt{a}}\sqrt{b}^2 = \sqrt{a}^2*\sqrt{b}^2 = a*b$. Furthermore, by the same argument as in the previous paragraph, $Q_{a^{1/4}}$ commutes with $Q_{\sqrt{b}}$, and hence $a^{1/4}$ and $\sqrt{b}$ generate an associative JBW-algebra by Proposition~\ref{prop:JBW-two-elements-associative}.

	For point e) suppose that $a\mult b = b\mult a$. As the JBW-algebra spanned by $a^{1/4}$ and $\sqrt{b}$ is associative we easily verify that $Q_{a^{1/4}}\sqrt{b} = \sqrt{Q_{\sqrt{a}}b}$ and then calculate:
	\[a\mult (b\mult c) = Q_{\sqrt{a}} Q_{\sqrt{b}} c = Q_{a^{1/4}}Q_{\sqrt{b}} Q_{a^{1/4}} c = Q_{Q_{a^{1/4}}\sqrt{b}} c = Q_{\sqrt{Q_{\sqrt{a}}b}} c = (a\mult b)\mult c.\]

	For point f) suppose that $a\mult b = b\mult a$. As $\sqrt{a}$ and $\sqrt{b}$ span an associative JBW-algebra containing $1$, we must have $a\mult b^\perp = Q_{\sqrt{a}}b^\perp = a*b^\perp = Q_{\sqrt{b^\perp}} a = b^\perp \mult a$. Suppose now that also $a\mult c = c\mult a$.
	Then $a$ operator commutes with $b$ and $c$ and hence with $b+c$. By Theorem~\ref{thm:JBW-assoc-iff-commute}, $a$ and $b+c$ then generate an associative algebra, and hence $(b+c)\mult a = Q_{\sqrt{b+c}} a = (b+c)*a = Q_{\sqrt{a}}(b+c) = a\mult (b+c)$ as desired.

  For g) the first point follows from the normality of $Q_{\sqrt{a}}$. For the second point suppose that $a\mult b = b\mult a$ for all $b\in S$. Then $a$ operator commutes with $b$ for all $b$ in $S$, and since the Jordan product is weakly continuous, $a$ also operator commutes with $\bigvee S$. Then $a$ and $\bigvee S$ generate an associative algebra and thus indeed $a\mult \bigvee S = \bigvee S \mult a$.

	Finally, for h) assume that $a\mult b = b \mult a$ and $a\mult c = c\mult a$, then $[Q_{\sqrt{a}},Q_{\sqrt{b}}] = 0$ and $[Q_{\sqrt{a}},Q_{\sqrt{c}}] = 0$, so that also 
	\[Q_{Q_{\sqrt{b}}c} Q_{\sqrt{a}} = Q_{\sqrt{b}}Q_{\sqrt{c}}^2Q_{\sqrt{b}} Q_{\sqrt{a}} = Q_{\sqrt{a}} Q_{\sqrt{b}}Q_{\sqrt{c}}^2Q_{\sqrt{b}} = Q_{\sqrt{a}} Q_{Q_{\sqrt{b}}c}.\]
	Hence $[Q_{\sqrt{Q_{\sqrt{b}}c}}, Q_{\sqrt{a}}] = 0$ so that indeed $a\mult (b\mult c) = (b\mult c)\mult a$.
\end{proof}

We conclude that any JBW-algebra is a sequential effect space as defined in Definition~\ref{def:seqprod}. Hence, together with Theorem~\ref{thm:seqprodisjordan} we see that a finite-dimensional order unit space is a sequential effect space if and only if it is a Euclidean Jordan algebra.

\chapter{Reconstructing infinite-dimensional quantum theory}\label{chap:infinitedimension}

In Chapters~\ref{chap:seqprod} and~\ref{chap:effectus} we found two ways to recover quantum theory using a set of natural assumptions, related to respectively sequential measurement and pure maps.

The process of recovering quantum theory crucially relied on two background assumptions. The first was that we were assuming that probabilities take the form of real numbers; an assumption we started to challenge in Chapter~\ref{chap:effectus} but ultimately had to use to complete the reconstruction. The second was that once we had this convex structure of the real numbers, we required our effect spaces to be finite-dimensional. In Chapter~\ref{chap:jordanalg} we found that all our assumptions regarding sequential measurement and purity hold in the infinite-dimensional setting of JBW-algebras, so it stands to reason that we should be able to use these assumptions when we drop the requirement of finite-dimensionality.

In this chapter we will combine and augment the requirements regarding sequential measurement and purity of Chapter~\ref{chap:seqprod} and~\ref{chap:effectus} in order to get a reconstruction of quantum theory that includes infinite-dimensional systems without a priori assuming our probabilities are real numbers. Unlike in those previous chapters, we can no longer claim that the set of assumptions we require are all physically reasonable. Instead, we view the results of this chapter as a first step towards finding reasonable assumptions from which infinite-dimensional quantum theory can be recovered. We expect subsequent approaches to succeed with a more elegant or smaller set of assumptions. 

We also remark that the reconstruction in this chapter is not completely `finished': we will show that our assumptions force every effect space $\eff(A)$  to be represented by a JW-algebra $V_A$ such that $\eff(A)\cong [0,1]_{V_A}$. Recall that a JW-algebra is a JBW-algebra that embeds into the self-adjoint part of a von Neumann algebra $\mathfrak{A}_\sa$, and hence our effect spaces satisfy $\eff(A)\xhookrightarrow{} [0,1]_{\mathfrak{A}}$. In a full reconstruction we would expect our effect spaces to be equal to the entire set of effects: $\eff(A) \cong [0,1]_{\mathfrak{A}}$. We leave a derivation of this requirement to future work.

The primary additional tool we require for the reconstruction is one that already proved useful in the study of JBW-algebras: the existence of suprema of directed sets. 
Recall that a subset $D$ of a partially ordered set is directed when for any $a,b\in D$ there exists $c\in D$ such that $a,b\leq c$, and that a partially ordered set is directed complete when any directed subset has a supremum (cf.~Definition~\ref{def:directedcompleteness}).
In this chapter we will primarily interested in the countable `version' of directed completeness:

\begin{definition}
    Let $P$ be a partially ordered set.
    It is~\Define{$\omega$-complete}\indexd{omega-complete@$\omega$-complete}
    if suprema of countable directed sets exist, or equivalently if any increasing sequence $a_1\leq a_2\leq \ldots$ in $P$ has a supremum.
\end{definition}

\begin{definition}
	An \Define{$\omega$-effect-theory} is an effect theory where every effect space is $\omega$-complete.
\end{definition}

This chapter is divided into three parts. First, we study the scalars $\eff(I)$ of an $\omega$-effect theory in Section~\ref{sec:effect-monoids}. We will see that the structure of the real numbers naturally appears in these scalars.
Second, in Section~\ref{sec:SEA} we study convex normal sequential effect algebras and find suitable conditions under which these are JB-algebras.
Third, we combine these two ingredients with some further assumptions to get the structure of a JB-algebra in Section~\ref{sec:Jordan-from-effectus}, and we add a tensor product to restrict these furthermore to JW-algebras in Section~\ref{sec:neumann-from-effectus}.

\section{Effect monoids}\label{sec:effect-monoids}

The effect space $\eff(I)$ of the trivial object $I$, like for any other system in an effect theory, is an effect algebra. But additionally, since $a,b\in \eff(I)$ are morphisms $a,b:I\rightarrow I$, they can be composed using the regular composition $a\circ b:I\rightarrow I$. This gives a total multiplication operation that makes $\eff(I)$ into an \emph{effect monoid}.

\begin{definition}[{\cite{jacobs2015new}}]\label{def:effectmonoid}
    An \Define{effect monoid}\indexd{effect monoid} is an effect algebra $(M,\ovee, 0, ^\perp,
    \,\cdot\,)$ with an additional (total) binary operation $\,\cdot\,$,
such that the following conditions hold 
    for all~$a,b,c \in M$.
\begin{itemize}
\item Unit: $a\cdot 1 = a = 1\cdot a$.
\item Distributivity: if $b\perp c$, then $a\cdot b\perp  a\cdot c$,\quad
    $b\cdot a\perp c\cdot a$,
        $$a\cdot (b\ovee c) \ =\  (a\cdot b) \ovee (a\cdot c),\ \text{and}\ 
(b\ovee c)\cdot a \ =\  (b\cdot a) \ovee (c\cdot a).$$
Or, in other words,
    the operation~$\,\cdot\,$ is bi-additive.
\item Associativity: $a \cdot (b\cdot c) = (a \cdot b) \cdot c$.
\end{itemize}
We call an effect monoid~$M$ \Define{commutative} if $a\cdot b = b \cdot a$
    for all~$a,b \in M$;
an element $p$ of~$M$  \Define{idempotent}
whenever~$p^2 := p \cdot p = p$;
elements~$a$, $b$  of~$M$ \Define{orthogonal}
when $a\cdot b = b\cdot a = 0$;
and we denote the set of idempotents of $M$ by~$P(M)$.
\end{definition}

\begin{remark}
The category of effect algebras is symmetric monoidal~\cite{jacobs2012coreflections}. The monoids in the category
of effect algebras resulting from this tensor product are precisely the effect monoids, hence the name.
\end{remark}

\begin{example}\label{ex:booleanalgebra}
	A Boolean algebra  $(B,0,1,\wedge,\vee,(\ )^\perp)$,
    is an effect algebra with the partial addition defined by 
    $x\perp y \iff x\wedge y = 0$ and in that case  $x\ovee y = x\vee y$. The complement is just the standard Boolean orthocomplement, $(\ )^\perp$.
	It is a commutative effect monoid with
    multiplication defined by $x \cdot y= x \wedge y$.
\end{example}

\begin{example}
	Let $X$ be a compact Hausdorff space. Then its unit interval $[0,1]_{C(X)}$ is a commutative effect monoid with pointwise addition and multiplication.
\end{example}

There exist non-commutative effect monoids, see for instance Refs.~\cite[Ex.~4.3.9]{kentathesis} and~\cite[Cor.~51]{basmaster}.

In this section we will be primarily interested in effect monoids that form the set of scalars of an $\omega$-effect-theory, and hence whose underlying effect algebra is $\omega$-complete. Examples of $\omega$-complete effect monoids are $\omega$-complete Boolean algebras, and the unit intervals of $C(X)$ where $X$ is a basically disconnected compact Hausdorff space.

The main result of this section is that these are in a sense the only possibilities for an $\omega$-complete effect monoid. Before we can state this result formally, we need some more definitions.

\begin{definition}
    Let $M$ and $N$ be effect monoids. A \Define{morphism}\indexd{morphism!between effect monoids} from $M$ to $N$ is
    an effect algebra morphism $f:M\rightarrow N$ that additionally preserves multiplication: $f(a\cdot b) = f(a)\cdot f(b)$ for all $a,b\in M$.
    A morphism is an \Define{embedding}\indexd{embedding (effect monoid)}\indexd{morphism!embedding of effect monoids} when it is also \Define{order reflecting}\indexd{order reflecting map}:
    if $f(a)\leq f(b)$ then $a\leq b$.
    Observe that an embedding is automatically injective,
    and that any 
    isomorphism (i.e.~a bijective morphism whose inverse is a morphism too) 
    is an embedding.
    Conversely, any surjective embedding is an isomorphism.
    We say $M$ and $N$ are \Define{isomorphic}\indexd{isomorphism!of effect monoids} and write $M\cong N$ when 
    there exists a surjective embedding (and hence an isomorphism) from $M$ to $N$.
\end{definition}

\begin{example}
    Given two effect algebras/monoids~$E_1$ and $E_2$ we define
    their \Define{direct sum}\indexd{direct sum!of effect algebras} $E_1\oplus E_2$ as the Cartesian
    product with pointwise operations. This is again an effect
    algebra/monoid.
\end{example}

We can now state the main result of this section:

\begin{theorem}\label{thm:omega-complete-classification}
    Let $M$ be an $\omega$-complete effect monoid.
    Then $M$ embeds into $M_1\oplus M_2$, where
    $M_1$ is an $\omega$-complete Boolean algebra,
    and $M_2 =[0,1]_{C(X)}$, 
    where $X$ is a basically disconnected compact Hausdorff space.
\end{theorem}

Intuitively, this theorem shows that an $\omega$-complete effect monoid splits up into a `sharp' part that is a Boolean algebra, and a `convex' part described by a compact Hausdorff space. 
In the setting where there are no zero-divisors (\ie~when $a\cdot b = 0$ implies $a=0$ or $b=0$) this picture further simplifies.

\begin{theorem}\label{thm:no-zero-divisors}
    Let $M$ be an $\omega$-complete effect monoid with no non-trivial
    zero divisors.
    Then either $M=\{0\}$, $M=\{0,1\}$ or $M\cong [0,1]$.
\end{theorem}

Hence, in an $\omega$-effect-theory where the scalars have no non-trivial zero divisors, the scalars are either trivial ($\{0\}$), sharp ($\{0,1\}$) or regular probabilities ($[0,1]$).

We can also frame this result through a different lens.

\begin{definition}
  We say an effect monoid $M$ is \Define{reducible}\indexd{reducible effect monoid} when $M\cong M_1\oplus M_2$ for some effect monoids $M_1\neq \{0\}$ and $M_2\neq \{0\}$. If no such decomposition exists we say $M$ is \Define{irreducible}\indexd{irreducible effect monoid}.
\end{definition}

\begin{theorem}\label{thm:irreducible-effect-monoid}
  Let $M$ be an irreducible $\omega$-complete effect monoid. Then $M=\{0\}$, $M=\{0,1\}$ or $M=[0,1]$.
\end{theorem}

The rest of this section is dedicated to proving Theorems~\ref{thm:omega-complete-classification}--\ref{thm:irreducible-effect-monoid}. 
This is rather technical and lengthy and will have no further bearing on the rest of this chapter. A reader not interested in the details can continue to Section~\ref{sec:SEA}.

\subsection{Basic results}\label{sec:basicresults}

We do not assume any commutativity of the product in an effect
monoid. Nevertheless, some commutativity comes for free.

\begin{lemma}\label{lem:ssperpcommute}
    For any~$a \in M$ in an effect monoid~$M$,
        we have~$a\cdot a^\perp = a^\perp \cdot a$.
\end{lemma}
\begin{proof}
    $a^2 \ovee  (a^\perp
    \cdot a) = (a\ovee a^\perp) \cdot a  = 1\cdot a = a = a\cdot 1 = a \cdot (a\ovee a^\perp) = a^2 \ovee  (a
    \cdot a^\perp)$. Cancelling $a^2$ on both sides gives the desired
    equality.
\end{proof}

\begin{lemma}
    \label{lem:idempotentiff}
    An element $p\in M$ is an idempotent if and only if $p\cdot p^\perp = 0$.
\end{lemma}
\begin{proof}
    $p = p\cdot 1 = p\cdot (p\ovee p^\perp) = p^2 \ovee p\cdot p^\perp$. Hence $p=p^2$ if and only if $p\cdot p^\perp = 0$.
\end{proof}

\begin{lemma}\label{lem:preserveunderidempotent}
    For~$a, p \in M$ with~$p^2=p$, we have
    \begin{equation*}
            p \cdot a \ =\  a \quad \iff \quad
            a \cdot p \ =\  a \quad \iff \quad
            a \ \leq\  p.
    \end{equation*}
\end{lemma}
\begin{proof}
Assume $a\leq p$.
Then $a\cdot p^\perp \leq p\cdot p^\perp = 0$, so
    that~$a \cdot p^\perp = 0$. Similarly $p^\perp \cdot a = 0$.
     Hence~$a=a\cdot 1=a\cdot (p\ovee p^\perp) = a\cdot p \ovee a \cdot p^\perp
        = a\cdot p$. Similarly~$p \cdot a = a$.

Now assume~$p \cdot a = a$.
Then immediately~$a = p\cdot a \leq p \cdot 1 = p$.
    The final implication (that~$a \cdot p = a \implies a \leq p$) is proven
    similarly.
\end{proof}

\begin{definition}
	Let~$M$ be an effect monoid and let~$p \in M$ be some idempotent.
    The subset~$pM := \{p \cdot a; \ a\in M\}$
        is called the \Define{left corner} by~$p$
        and is an effect monoid with~$(p\cdot a)^\perp := p \cdot a^\perp$
            and all other operations inherited from~$M$.
\end{definition}

\begin{lemma}\label{lem:idempotentscommute}
    Let $M$ be an effect monoid with idempotent~$p \in M$.
Then~$p\cdot a = a \cdot p$ for any~$a \in M$.
\end{lemma}
\begin{proof}
Clearly~$p\cdot a \leq p \cdot 1 = p$
    and so by Lemma~\ref{lem:preserveunderidempotent}
    $p \cdot a \cdot p = a \cdot p$.
    Similarly~$a \cdot p \leq p$ and so~$p \cdot a \cdot p = p \cdot a$.
    Thus~$p\cdot a = p\cdot a\cdot p = a \cdot p$, as desired.
\end{proof}
\begin{corollary}\label{cor:corneriso}
    Let~$M$ be an effect monoid with idempotent~${p \in M}$.
    The map~$a \mapsto (p \cdot a, p^\perp \cdot a)$
    is an isomorphism~$M \cong pM \oplus p^\perp M$.
\end{corollary}

The following two lemmas are simple observations
that will be used several times.
\begin{lemma}
\label{lem:forcing}
Let~$a\leq b$ be elements of an effect algebra~$E$.
If~$ b\ovee b'\leq a\ovee a'$ for some~$a'\leq b'$ from~$E$,
    then~$a=b$ (and~$a'=b'$).
\end{lemma}
\begin{proof}
    Since~$a\leq a'$ and~$b\leq b'$,
    we have~$a\ovee a'\leq b\ovee b'$,
    and so~$a\ovee a' = b\ovee b'$.
Then  $0 = (b\ovee b') \ominus (a\ovee a')=
    (b\ominus a)\ovee (b'\ominus a')$,
    yielding $b\ominus a= 0$
    and $b'\ominus a'=0$,
so~$b=a$ and~$b'=a'$.
\end{proof}

\begin{lemma}\label{lem:summableunderidempotent}
Let~$p$ be an idempotent from an effect monoid~$M$,
    and let~$a,b\leq p$ be elements below~$p$.
If~$a\ovee b$ exists,
then~$a\ovee b\leq p$.
\end{lemma}
\begin{proof}
    Since $a\leq p$, we have $a \cdot p^\perp = 0$,
    and similarly, $b\cdot p^\perp=0$.
    But then $(a\ovee b)\cdot p^\perp
    = 0$, and hence $(a \ovee b)\cdot p = a\ovee b$. By
    Lemma~\ref{lem:preserveunderidempotent} we then have $a\ovee b\leq
    p$.
\end{proof}

We defined directed set to mean upwards directed. Using the fact
that $a \mapsto a^\perp$ is an order anti-isomorphism, an $\omega$-complete
effect algebra also has all countable infima of downwards directed (or `filtered') 
sets.

Recall that given an element~$a$ of an
ordered group~$G$
a subset~$S$ of~$G$ has a supremum $\bigvee S$ in~$G$
if and only if $\bigvee_{s\in S} a+s$ exists,
which follows immediately from 
the observation that~$a+(\ )\colon G\to G$ is an order isomorphism.
For effect algebras
the situation is a bit more complicated,
and we only have the implications
mentioned in the lemma below.
We will see in Corollary~\ref{cor:sumomegaem}
that the situation is simpler
for $\omega$-complete effect monoids.
\begin{lemma}\label{lem:additionisnormal}
Let~$x$ be an element and $S$ a non-empty
subset of an effect algebra~$E$.
    If $S\subseteq [0,x^\perp]_E$, then
\begin{alignat*}{3}
    \text{$\bigvee S$ exists}
    \quad&\implies\quad&
    x\ovee \bigvee S &\,=\,
    \bigvee x\ovee S\text{, and}
    \\
    \text{$\bigwedge x\ovee S$ exists}
    \quad&\implies\quad&
x\ovee \bigwedge S &\,=\,
    \bigwedge x\ovee S \text{.}
    \\
    \intertext{Here ``$=$'' means also that the 
    sums, suprema and infima on either side exist.
    Similarly, if $S\subseteq [x,1]_E$, then}
    \text{$\bigwedge_{s\in S} s\ominus x$ exists}
    \quad&\implies\quad&
\bigl(\bigvee S\bigr) \ominus x
     &\,=\, 
    \bigvee_{s\in S} s \ominus x\text{, and}
    \\
    \text{$\bigwedge S$ exists}
    \quad&\implies\quad&
\bigl(\bigwedge S\bigr) \ominus x
     &\,=\, 
    \bigwedge_{s\in S} s \ominus x\text{.}
    \\
\intertext{Moreover, if $S\subseteq[0,x]_E$, then}
    \text{$\bigvee S$ exists}
    \quad&\implies\quad&
x \ominus \bigvee S
     &\,=\,
    \bigwedge x\ominus S\text{, and}
    \\
    \text{$\bigvee x \ominus S$ exists}
    \quad&\implies\quad&
x \ominus \bigwedge S
     &\,=\, 
    \bigvee x\ominus S\text{.}
\end{alignat*}
\end{lemma}
\begin{proof}
    Note that $a \mapsto x\ovee a$
    gives an order isomorphism~$[0,x^\perp]_E\to [x,1]_E$
    with inverse~$a \mapsto a \ominus x$.
    Hence~$x\ovee(\ )$ preserves and reflects
    all infima and suprema \emph{restricted to}~$[0,x^\perp]_E$ and~$[x,1]_E$.
Given elements $a\leq b$ from~$E$,
and a subset~$S$ of the interval~$[a,b]_E$,
it is clear
that any supremum (infimum)
    of~$S$ in~$E$ will be the supremum (infimum) of~$S$ in~$[a,b]_E$ too
    (using here that~$S$ is non-empty).
The converse does not always hold,
    but when $S$ has a supremum in $[a,1]_E$,
    then this is the supremum in~$E$ too
    (and when~$S$ has an infimum in $[0,b]_E$,
    then this is the infimum in~$E$ too).
These considerations yield the first four equations.
For the latter two
we just add the observation
    that $x\ominus (\ )$
    gives an order anti-isomorphism~$[0,x]_E\to[0,x]_E$.
\end{proof}

We can now prove a few basic yet useful facts of $\omega$-complete effect monoids. 
These lemmas deal with elements that are summable with
    themselves: elements $a$ such that $a\perp a$ which means that
    $a\ovee a$ is defined. For $n\in \N$ we will use the notation $na
    = a\ovee \ldots \ovee a$ for the $n$-fold sum of $a$ with itself
    (when it is defined).
We study these self-summable elements
    to  be able to define a ``$\frac{1}{2}$''
    in some effect monoids later on.

\begin{lemma}\label{lem:selfsummable}
    For any~$a \in M$ in an effect monoid~$M$,
    the element~$a \cdot a^\perp$ is summable with itself.
\end{lemma}
\begin{proof}
Since $1=1\cdot 1 = (a\ovee a^\perp)\cdot(a\ovee a^\perp)
    = a\cdot a\ovee a\cdot a^\perp \ovee a^\perp \cdot a \ovee 
    a^\perp\cdot a^\perp$,
    and $a\cdot a^\perp = a^\perp \cdot a$ by Lemma~\ref{lem:ssperpcommute},
    we see that~$a\cdot a^\perp\ovee a\cdot a^\perp$  indeed exists.
\end{proof}

\begin{lemma}\label{lem:archemedeanomegadirectedcomplete}\label{lem:asummablepowerszero}\label{lem:nonilpotents}
    Let $a$ be an element of an $\omega$-complete effect monoid $M$.
    \begin{enumerate}[label=\alph*)]
        \item If $na$ exists for all $n$, then $a=0$.
        \item If $a^2 = 0$, then $a=0$.
        \item If $a\perp a$, then $\bigwedge_n a^n = 0$.
    \end{enumerate}
\end{lemma}
\begin{proof}
    For point a), we have $a\ovee \bigvee_n na
    =\bigvee_n a\ovee na
    = \bigvee_n (n+1)a = \bigvee_n na$,
    and so~$a=0$.

    For point b), since $a^2 = 0$ we have $a = a\cdot 1 = a\cdot (a\ovee a^\perp) = a\cdot a^\perp$, and hence (because of
    Lemma~\ref{lem:selfsummable}) $a$ is summable with itself. But
    furthermore~$(a\ovee a)^2 = 4a^2 = 0$, and so $(a\ovee a)^2=0$.
    Continuing in this fashion,
        we see that~$2^n a$ exists for every~$n \in \N$
        and~$(2^n a)^2 = 0$.
        Hence, for any~$m \in \N$
        the sum~$m a$ exists 
    so that by the previous point
    $a = 0$.

    For point c), write~$b := \bigwedge_n a^n$.
    As $(2a)^n = 2^n a^n$ and $b\leq a^n$ we see that
    $2^n b$ is defined. But this is true for all $n$, and so again by the point a), $b=0$.
\end{proof}

\subsection{Floors, ceilings and division}\label{sec:floorceiling}

In this section we will see that any $\omega$-complete effect monoid has
\emph{floors} and \emph{ceilings}, \ie~respectively the largest idempotent
below an element and the smallest idempotent above an element.
We will also construct a `division': for $a\leq b$ we will find 
an element $a/b$ such that $(a/b)\cdot b = a$. Although the same structure can be found in JBW-algebras (cf.~Sections~\ref{sec:JBW-floor-ceiling}, \ref{sec:division-filter}), the proofs will be quite different.

Using ceilings and division we will show that multiplication in a
$\omega$-complete effect monoid is always normal, i.e.~that 
$b\cdot \bigvee S = \bigvee b\cdot S$
for non-empty~$S$ for which~$\bigvee S$ exists. This technical result will be frequently used in the remainder of the proof of the characterisation theorem.

\begin{definition}
    Let~$(x_i)_{i \in I}$ be a (potentially infinite)
    family of elements from an effect algebra~$E$.
    We say that the sum~$\bigovee_{i \in I} x_i$ exists
        if for every finite subset~$S \subseteq I$
        the sum~$\bigovee_{i \in S} x_i$ exists
        and the supremum~$\bigvee_{\text{finite }S \subseteq I}
        \bigovee_{i \in S} x_i$ exists
            as well.
    In that case we write $\bigovee_{i \in I} x_i :=
        \bigvee_{\text{finite }S \subseteq I}
        \bigovee_{i \in S} x_i$.
\end{definition}
\begin{lemma}
    \label{lem:proto-ceilfloor}
Given~$a \in M$ for an effect monoid~$M$, we have
\begin{equation*}
    \textstyle
    (a^N)^\perp\ =\ a^\perp\ \ovee\  a^\perp \cdot a\ \ovee\  a^\perp \cdot a^2
    \ \ovee\  \dotsb\ \ovee\  a^\perp \cdot a^{N-1}
\end{equation*}
for every natural number~$N$.
\end{lemma}
\begin{proof}
We simply compute:
    \begin{equation*}
\begin{alignedat}{3}
{}\qquad\quad\ \  1 \ &=\ a^\perp \,\ovee\,a
\\
    &=\  a^\perp \,\ovee\,(a^\perp\ovee a)\cdot a
    \qquad\qquad\ \ \,=\  a^\perp\,\ovee\,a^\perp \cdot a \,\ovee\,a^2
\\
    &=\  a^\perp \,\ovee\,a^\perp \cdot a \,\ovee\,(a^\perp\ovee a) \cdot a^2
    \ =\  a^\perp\,\ovee\,a^\perp\cdot a
    \,\ovee\, a^\perp\cdot  a^2\,\ovee\,
    a^3
\\
    &\ \vdots
\\
    &=\ \Bigl(\,\bigovee_{n=0}^{N-1} a^\perp \cdot a^n\,\Bigr)\,\ovee\, a^N \qquad\qquad\qquad\qquad\qquad\qquad\qquad\qquad\qquad\qedhere
\end{alignedat}
\end{equation*}
\end{proof}
\begin{corollary}
    The sum~$\bigovee_{n=0}^\infty a^\perp \cdot a^n$ exists
        for any element~$a$ 
        in an~$\omega$-complete effect monoid~$M$.
\end{corollary}
\begin{definition}
Given an element~$a$ of an $\omega$-complete
effect monoid~$M$
    \begin{equation*}
        \ceil{a}\ :=\ \bigovee_{n=0}^\infty a\cdot (a^\perp)^n
        \qquad\text{and}\qquad
        \floor{a}\ := \ \bigwedge_{n=0}^\infty a^n
    \end{equation*}
    are called the \Define{ceiling}\indexd{ceiling!in effect monoid} of a and the \Define{floor}\indexd{floor!in effect monoid}
    of~$a$, respectively.
\end{definition}
We list
some basic properties of~$\ceil{a}$
and~$\floor{a}$ in Proposition~\ref{prop:ceilfloor},
after we have made the observations
necessary to establish them.

\begin{lemma}
    \label{lem:infaperpan}
    Let $M$ be an $\omega$-complete effect monoid and let $a\in M$ be arbitrary.
    Then $\bigwedge_n a^\perp\cdot  a^n = 0$.
\end{lemma}
\begin{proof}
Define $b := \bigwedge_n a^\perp \cdot a^n$.
Since~$a$ and~$a^\perp$ commute by Lemma~\ref{lem:ssperpcommute},
    we compute
\begin{equation*}
1\ =\ 1^n\ =\ (a^\perp \ovee a)^n
    \ =\ \bigovee_{k=0}^n \binom{n}{k}\bigl(\, (a^\perp)^k\cdot a^{n-k}
    \,\bigr),
\end{equation*}
where $\binom{n}{k}$ represents a binomial coefficient.
We in particular see
    that the sum $\binom{n}{1} (a^\perp \cdot a^{n-1})= 
    n(a^\perp \cdot a^{n-1})$ exists.
Because $b\leq a^\perp \cdot a^{n-1}$,
the $n$-fold sum~$nb$ exists too for all $n$ and hence~$b=0$
    by Lemma~\ref{lem:archemedeanomegadirectedcomplete}.
\end{proof}
\begin{lemma}
\label{lem:flooraa}
    We have~$\floor{a}=\floor{a}\cdot a = a \cdot \floor{a}$
for any element~$a$ of an $\omega$-complete
effect monoid~$M$.
\end{lemma}
\begin{proof}
Using Lemmas~\ref{lem:ssperpcommute} and~\ref{lem:infaperpan} we compute $\floor{a} \cdot a^\perp
    = (\bigwedge_n a^n) \cdot a^\perp
\leq \bigwedge_n  a^n \cdot a^\perp 
= \bigwedge_n a^\perp \cdot a^n 
= 0$,
    and so~$\floor{a}\cdot a=\floor{a}$.
The other identity
    has a similar proof.
\end{proof}


\begin{lemma}
\label{lem:zerodivsum}
Given elements $a, b_1, b_2,\dotsc$
of an $\omega$-complete effect monoid~$M$
such that~$\bigovee_n b_n$ exists,
and~$a\cdot b_n=0$ for all~$n\in \N$,
we have~$a \cdot \bigovee_n b_n=0$.
\end{lemma}
\begin{proof}
    Writing~$s_N = \bigovee_{n=1}^N b_n$,
    we have~$s_1\leq s_2\leq \dotsb$
    and~$a\cdot s_n=0$ for all~$n$.
    Since~$s_n=(a\ovee a^\perp)\cdot s_n
    = a \cdot s_n \ovee a^\perp \cdot s_n = a^\perp \cdot s_n$
    for all~$n \in \N$,
    we have
\begin{equation*}
    \textstyle \bigvee_n s_n \ = \ \bigvee_n a^\perp \cdot s_n
    \ \leq\  a^\perp \cdot \bigvee_n s_n
    \ \leq\  \bigvee_n s_n,
\end{equation*}
which implies that~$a^\perp\cdot \bigvee_ns_n= \bigvee_n s_n$,
and thus~$a\cdot \bigovee_n b_n = a \cdot \bigvee_n s_n=0$.
\end{proof}
\begin{proposition}
\label{prop:ceilprod}
Let $a$ and~$b$ be elements of an $\omega$-complete effect monoid~$M$. Then
$a\cdot b\,=\,0\implies a\cdot\ceil{b}\,=\,0$.
\end{proposition}
\begin{proof}
If~$a\cdot b=0$,
then also $a\cdot b\cdot (b^\perp)^n=0$
    for all~$n$.
Hence by Lemma~\ref{lem:zerodivsum}
$a\cdot \ceil{b}= a\cdot \bigovee_{n=1}^\infty b\cdot (b^\perp)^n
    =0$.
\end{proof}
\begin{proposition}
\label{prop:ceilfloor}
Let~$a$ be an element of an $\omega$-complete
effect monoid~$M$.
\begin{enumerate}[label=\alph*)]
\item
    \label{prop:ceilfloor-floor}
The floor~$\floor{a}$ of~$a$
        is an idempotent with~$\floor{a}\leq a$.
        In fact, $\floor{a}$
        is the greatest idempotent below~$a$.
\item
    \label{prop:ceilfloor-ceil}
The ceiling~$\ceil{a}$ of~$a$
is the least idempotent above~$a$.
\item
    \label{prop:ceilfloor-duality}
    We have
        $\ceil{a}^\perp = \floor{a^\perp}$
        and
        $\floor{a}^\perp = \ceil{a^\perp}$.
\end{enumerate}
\end{proposition}
\begin{proof}
Point~\ref{prop:ceilfloor-duality}
    follows from Lemma~\ref{lem:proto-ceilfloor}.
Concerning point \ref{prop:ceilfloor-floor}:
Since $\floor{a} \cdot a^\perp = 0$
    (by Lemma~\ref{lem:flooraa})
    we have $\floor{a}\cdot \ceil{a^\perp}=0$
    by Proposition~\ref{prop:ceilprod},
    and so $\floor{a}\cdot \floor{a}^\perp=0$
    because
     $\floor{a}^\perp = \ceil{a^\perp}$
     by point~\ref{prop:ceilfloor-duality}. Hence $\floor{a}$ is an idempotent.
Also, 
since~$\floor{a}=\bigwedge_n a^n$,
we clearly have~$\floor{a}\leq a$.
Now,
if~$s$ is an idempotent in~$M$
with~$s\leq a$,
then~$s=s^n\leq a^n$,
and so~$s\leq \bigwedge_n a^n= \floor{a}$.
So that~$\floor{a}$ is indeed the greatest idempotent below~$a$.
Point~\ref{prop:ceilfloor-ceil}
now follows easily from~\ref{prop:ceilfloor-floor},
    since~$\ceil{\,\cdot\,}$ is the dual of~$\floor{\,\cdot\,}$
    under the order anti-isomorphism $(\,\cdot\,)^\perp$.
\end{proof}
\begin{lemma}\label{lem:sumofceil}
$\ceil{a\ovee b}=\ceil{a}\vee \ceil{b}$
for all summable
    elements $a$ and~$b$ of an $\omega$-complete effect monoid~$M$
    (that is,
    $\ceil{a\ovee b}$ is the supremum of~$\ceil{a}$ and~$\ceil{b}$).
\end{lemma}
\begin{proof}
Since~$\ceil{a\ovee b}\geq a\ovee b \geq a$,
    we have $\ceil{a\ovee b } \geq \ceil{a}$,
    and similarly, $\ceil{a\ovee b}\geq \ceil{b}$.
Let~$u$ be an upper bound of~$\ceil{a}$ and~$\ceil{b}$;
    we claim that~$\ceil{a\ovee b}\leq u$.
    Since~$\ceil{a}\leq u$ and~$\ceil{b}\leq u$, 
    we have $a\leq \ceil{a}\leq \floor{u}$ and
    $b\leq \ceil{b}\leq \floor{u}$,
    and so~$a\ovee b\leq \floor{u}$ by Lemma~\ref{lem:summableunderidempotent}.
    Hence~$\ceil{a\ovee b}\leq \floor{u}\leq u$.
\end{proof}
Any $\omega$-complete effect monoid is a \emph{lattice effect algebra}~\cite{rievcanova2000generalization}:
\begin{theorem}\label{thm:latticeemon}
Let $a,b\in M$ be elements of an $\omega$-complete
effect monoid~$M$. Then $a$ and $b$ have an infimum $a\wedge b$ given by
$$
    a\wedge b\,=\, \bigovee_{n=1}^\infty a_n\cdot b_n
    \ \text{where}\ 
    \left[\ 
    \begin{alignedat}{3}
        a_1\,&=\, a \quad&& a_{n+1} \,&=\, a_n\cdot b_n^\perp \\
        b_1\,&=\, b && b_{n+1} \,&=\, a_n^\perp \cdot b_n
    \end{alignedat}\right.
$$
Consequently, $a$ and $b$ also have a supremum given by $a\vee b= (a^\perp \wedge b^\perp)^\perp$.
\end{theorem}
\begin{proof}
First we need to show
    that the sum~$\bigovee_{n=1}^N a_n \cdot b_n$ exists for every~$N$.
    In fact,
    we'll show that $a\ominus \bigovee_{n=1}^N a_n\cdot b_n = a_{N+1}$
    for all~$N$,
    by induction.
Indeed, for~$N=1$
we have~$a\ominus a\cdot b=a\cdot b^\perp = a_2$,
and if
$a\ominus \bigovee_{n=1}^N a_n\cdot b_n = a_{N+1}$
for some~$N$,
then~$a_{N+2} = a_{N+1} \cdot b_{N+1}^\perp
    = a_{N+1} \ominus a_{N+1}\cdot  b_{N+1}
    = (a\ominus \bigovee_{n=1}^N a_n\cdot b_n)\ominus a_{N+1}\cdot b_{N+1}
    = a\ominus \bigovee_{n=1}^{N+1} a_n \cdot b_n$.
    Hence,
    $\bigovee_{n=1}^\infty a_n \cdot b_n$ exists
    and moreover
    $$
    a\ =\ \bigwedge_{m=1}^\infty a_m \ \ovee\ 
    \bigovee_{n=1}^\infty a_n \cdot b_n.
    $$
By a similar reasoning, we get
    $$
    b\ =\ \bigwedge_{m=1}^\infty b_m \ \ovee\ 
    \bigovee_{n=1}^\infty a_n \cdot b_n.
    $$
Already writing~$a\wedge b :=
    \bigovee_{n=1}^\infty a_n \cdot b_n$,
    we know at this point that
    $a\wedge b \leq a$ and~$a\wedge b\leq b$.
    It remains to show that~$a\wedge b$
    defined above is the greatest lower bound of~$a$ and~$b$.
    So let~$\ell\in M$ with $\ell\leq a$ and~$\ell \leq b$ 
    be given; we must show that~$\ell\leq a\wedge b$.


    First, we observe that~$\bigl(\bigwedge_n a_n\bigr)\cdot
    \bigl(\bigwedge_m b_m\bigr)=0$.
    Indeed, we have
    $\bigl(\bigwedge_n a_n\bigr)\cdot
    \bigl(\bigwedge_m b_m\bigr)
    \leq \bigwedge_n a_n\cdot b_n$,
    and
    $\bigwedge_n a_n \cdot b_n = 0$
    because~$\bigovee_{n=1}^\infty a_n\cdot b_n$ exists (see Lemma~\ref{lem:archemedeanomegadirectedcomplete}).
    By Proposition~\ref{prop:ceilprod}
    it follows that $\bigl(\bigwedge_n a_n\bigr)\cdot
    \ceil{\bigwedge_m b_m} = 0$.
    Hence, writing~$p = \ceil{\bigwedge_m b_m}$,
    we have $p\cdot \bigwedge_n a_n = \bigl(\bigwedge_n a_n\bigr)\cdot p
    = 0$ 
    using Lemma~\ref{lem:idempotentscommute}.
    Observing that $\bigwedge_n b_n\leq p$ and using Lemma~\ref{lem:preserveunderidempotent} we also have $p^\perp \cdot \bigwedge_n b_n = 0$.
    We then calculate  $p\cdot a
    =p\cdot \bigl(\,\bigwedge_n a_n \,\ovee\, a\wedge b\,\bigr) 
    = p\cdot (a\wedge b)$ and similarly
    $p^\perp \cdot b= p^\perp \cdot (a\wedge b)$.

    Returning to the problem of whether~$\ell\leq a\wedge b$,
    we have
   \begin{equation*}
\ell \ =\  
p\cdot \ell\,\ovee\, p^\perp\cdot  \ell
\ \leq\  p\cdot a\,\ovee\, p^\perp\cdot  b
    \ =\ p\cdot (a\wedge b) \,\ovee\, p^\perp\cdot (a\wedge b)
     \ = \ a\wedge b.
   \end{equation*}
    As $l$ was arbitrary, $a\wedge b$ is indeed the infimum of~$a$ and~$b$.
\end{proof}
The presence of finite infima and suprema in $\omega$-complete
effect monoids prevents certain subtleties
around the existence of arbitrary suprema and infima.
\begin{corollary}
\label{cor:intervalsuprema}
Let~$a\leq b$ be elements of an $\omega$-complete
effect monoid~$M$,
    and let $S$ be a non-empty subset of~$[a,b]_M$.

    Then~$S$ has a supremum (infimum) in~$M$
    if and only if~$S$ has a supremum (infimum) in~$[a,b]_M$,
    and these suprema (infima) coincide.
\end{corollary}
\begin{proof}
It is clear that if~$S$ has a supremum in~$M$, then
    this is also the supremum in~$[a,b]_M$.
For the converse,
suppose that~$S$
has a supremum~$\bigvee S$ in $[a,b]_M$,
and let~$u$ be an upper bound for~$S$ in~$M$;
in order to show that~$\bigvee S$ is the supremum of~$S$
in~$M$ too,
we must prove that~$\bigvee S\leq u$.
    Note that~$b\wedge u$
    is an upper bound for~$S$.
    Indeed, given~$s\in S\subseteq [a,b]_M$
    we have~$s\leq b$ and~$s\leq u$ so that $s\leq b\wedge u$.
    Moreover, one easily sees
    that $b\wedge u \in [a,b]_M$
    using the fact that~$S$ is non-empty.
    Hence~$b\wedge u$ is an upper bound
    of~$S$ in $[a,b]_M$, and so
    $\bigvee S\leq b\wedge u\leq u$,
    making~$\bigvee S$ the supremum of~$S$ in~$M$.
An analogous proof works for proving infima of $S$ are preserved.
\end{proof}
\begin{corollary}
\label{cor:sumomegaem}
Given an element~$a$
and a non-empty subset~$S$ 
of an $\omega$-complete
effect monoid~$M$
such that $a\ovee s$ exists for all~$s\in S$,
\begin{itemize}
\item
the supremum~$\bigvee S$ exists
iff $\bigvee a\ovee S$ exists,
and in that case $a\ovee \bigvee S=\bigvee a\ovee S$;
\item
the infimum~$\bigwedge S$ exists
iff $\bigwedge a\ovee S$ exists,
and in that case $a\ovee \bigwedge S=\bigwedge a\ovee S$.
\end{itemize}
\end{corollary}
\begin{proof}
The map~$a\ovee(\ )\colon [0,a^\perp]_M\to [a,1]_M$,
being an order isomorphism,
preserves and reflects suprema and infima.
    Now apply Corollary~\ref{cor:intervalsuprema}.
\end{proof}

Now that we know more about the existence of suprema and infima, we set our sights on proving that multiplication interacts with suprema and infima as desired, namely that it preserves them. To do this we introduce a partial division operation.
\begin{definition}
Given elements~$a\leq b$ of an $\omega$-complete
effect monoid, set
\begin{equation*}
    a/b \ := \ \bigovee_{n=0}^\infty a \cdot (b^\perp)^n.
\end{equation*}
    Note that the sum exists,
    because $\bigovee_{n=0}^N
    a\cdot (b^\perp)^n 
    \leq {\bigovee_{n=0}^\infty b \cdot (b^\perp)^n}
    =\ceil{b}$ for all~$N$.
\end{definition}
\begin{lemma}
\label{lem:div}
Let $b$ be an element of an $\omega$-complete
effect monoid~$M$.
\begin{enumerate}[label=\alph*)]
\item
\label{lem:div-adiva}
$b/b = \ceil{b}$.
\item
\label{lem:div-additive}
$(a_1\ovee a_2)/ b = a_1/b\,\ovee a_2/b$
for all summable~$a_1,a_2\in M$ with $a_1\ovee a_2\leq b$.
\item
\label{lem:div-abdivb}
$(a\cdot b)/b = a\cdot \ceil{b}$
for all~$a\in M$.
\item
\label{lem:div-adivbb}
$(a/b)\cdot b = a$
for all~$a\in M$ with~$a\leq b$.
\item
    \label{lem:div-mb}
        $\{ a \cdot b~;~ a \in M \} =  Mb  = [0,b]_M = \{a\in M~;~a \leq b\}$.
\item
    \label{lem:div-iso}
        The maps $a\mapsto a\cdot b\colon M\ceil{b} \to M b$ and $a\mapsto b\cdot a\colon \ceil{b}M \to b M$
are order isomorphisms.
\end{enumerate}
\end{lemma}
\begin{proof}
    Points~\ref{lem:div-adiva} and~\ref{lem:div-additive} are easy,
and left to the reader.
Concerning~\ref{lem:div-abdivb},
first note that
\begin{equation*}
    (a\cdot b)/b\ =\ 
    \bigovee_{n=0}^\infty a \cdot b \cdot (b^\perp)^n
\ \leq\ 
    a \cdot \bigovee_{n=0}^\infty b \cdot (b^\perp)^n
    \ = \ a\cdot \ceil{b}.
\end{equation*}
Thus $(a \cdot b)/b\leq a\cdot \ceil{b}$.
    Since similarly $(a^\perp \cdot b)/b\leq a^\perp\cdot \ceil{b}$,
    we get, using~\ref{lem:div-adiva} and~\ref{lem:div-additive}:
\begin{equation*}
\ceil{b}\ = \ 
b/b\ = \ 
    (a\cdot b)/b\,\ovee\,(a^\perp \cdot b)/b
    \ \leq\ a\cdot \ceil{b}\,\ovee\,a^\perp \cdot\ceil{b}
    \ =\ \ceil{b}.
\end{equation*}
Hence, $(a\cdot b)/b = a\cdot \ceil{b}$ byLemma~\ref{lem:forcing}).
For point~\ref{lem:div-adivbb},
note that given~$a,b \in M$ with~$a \leq b$
we have
    $a=a\cdot \ceil{b}$ (by Lemma~\ref{lem:preserveunderidempotent}, 
    since $a\leq b\leq \ceil{b}$,) and so
    \begin{alignat*}{3}
        a 
         \ =\ a\cdot \ceil{b} 
        & \ =\ (a\cdot b)/b &\qquad& \text{by point~\ref{lem:div-abdivb}}\\
        & \ =\  \bigovee_{n=0}^\infty  a\cdot b\cdot (b^\perp)^n \\
        & \ =\  \bigovee_{n=0}^\infty a\cdot (b^\perp)^n \cdot b 
                    && \text{by Lemma~\ref{lem:ssperpcommute}} \\
        & \ \leq\  \Bigl(\bigovee_{n=0}^\infty a\cdot (b^\perp)^n\Bigr) \cdot b  \ =\ (a/b)\cdot b.
    \end{alignat*}
    Since similarly $b\ominus a\leq ((b\ominus a)/b)\cdot b$,
    we get
\begin{equation*}
    b\ =\ a\ovee (b\ominus a) 
    \ \leq\ 
    (a/b)\cdot b\,\ovee\,((b\ominus a)/b)\cdot b\ =\ (b/b)\cdot b =\ceil{b}\cdot b\ =\ b,
\end{equation*}
which forces~$a=(a/b)\cdot b$.
For point~\ref{lem:div-mb},
    note that~$Mb, bM\subseteq [0,b]_M$
    since~$b\cdot a, a\cdot b\leq b$ for all~$a\in M$,
    and~$[0,b]_M\subseteq Mb$,
    because~$a=(a/b)\cdot b$ for any~$a\in [0,b]_M$
    by point~\ref{lem:div-adivbb}.

Finally, 
concerning
    point~\ref{lem:div-iso}:
    the maps 
     $a\mapsto a\cdot b\colon M \ceil{b}
    \to M b$
and  $a\mapsto a/b\colon M b
    \to M \ceil{b}$
    are clearly order preserving,
    and each other's inverse
    by points~\ref{lem:div-abdivb} and~\ref{lem:div-adivbb},
    and thus order isomorphisms.
    The proof that~$a\mapsto b\cdot a\colon M \ceil{b} \to b M$
    is an order isomorphism follows along entirely similar lines,
    but involves $b\backslash a$ defined by 
    $b\backslash a:= \bigovee_n (b^\perp)^n \cdot a$
    and uses the fact that~$bM = [0,b]_M = M b$.
\end{proof}

Finally, we can prove that multiplication is indeed normal:
\begin{theorem}\label{thm:multisnormal}
    \label{prop:bsupinf}
    Let $b$ and~$b'$ be elements
    of an $\omega$-complete
    effect monoid~$M$,
    and let $S\subseteq M$ be any 
    (potentially uncountable or non-directed) non-empty subset.
    \begin{itemize}
        \item \label{prop:bsupinf-supp}
            If $\bigvee S$ exists, then
            so does $\bigvee_{s\in S} b\cdot s \cdot b'$,
            and $b\cdot (\bigvee S ) \cdot b'
            = \bigvee_{s\in S} b\cdot s\cdot b'$.
        \item
            \label{prop:bsupinf-inf}
        If $\bigwedge S$ exists, then
            so does $\bigwedge_{s\in S} b\cdot s \cdot b'$,
            and $b\cdot (\bigwedge S) \cdot b' = \bigwedge_{s\in S} 
            b\cdot s\cdot b'$.
    \end{itemize}
\end{theorem}
\begin{proof}
Suppose that~$\bigvee S$ exists.
    We will prove that~$b\cdot \bigvee S=\bigvee_{s\in S} b\cdot s$,
and leave the remainder to the reader.
    Note that  $b\cdot(\ )\colon [0,\ceil{b}]_M\to [0,b]_M$,
    being an order isomorphism by Lemma~\ref{lem:div}.\ref{lem:div-iso},
    preserves suprema and infima.
    The set~$S$ need, however, not be part of $[0,\ceil{b}]_M$,
    so we consider instead of~$b$
    the element $b':=  b\ovee \ceil{b}^\perp$,
    for which~$\ceil{b'}=\ceil{b}\vee \ceil{b}^\perp =1$
    by Lemma~\ref{lem:sumofceil}.
    We then get an order isomorphism~${b'\cdot(\ )\colon M\to [0,b']_M}$,
    which preserves suprema,
    so that~$b'\cdot \bigvee S$
is the supremum of $b'\cdot S$ in $[0,b']_M$,
    and hence in~$M$, by Corollary~\ref{cor:intervalsuprema}.
    Then
\begin{alignat*}{3}
    (b\ovee \ceil{b}^\perp)\cdot \bigvee S 
    \ &=\ \bigvee_{s\in S} (b\ovee \ceil{b}^\perp)\cdot s \\
    \ &\leq\ \bigvee_{s\in S} b\cdot  s 
    \ \ovee\ \bigvee_{s'\in S} \ceil{b}^\perp \cdot   s' \\
    \ &\leq\ b\cdot \bigvee S \ \ovee\ \ceil{b}^\perp\cdot \bigvee S \\
    \ & = \ (b\ovee \ceil{b}^\perp)\cdot\bigvee S
\end{alignat*}
forces $\bigvee_{s\in S} b\cdot s= b\cdot \bigvee S$
    by Lemma~\ref{lem:forcing}.
\end{proof}

\subsection{Boolean algebras, halves and convexity}\label{sec:boolhalvesconvexity}
We are ready to study the two important types of idempotents
    in an effect monoid: those that are \emph{Boolean}
    and those that are \emph{halvable}.

\begin{definition}
We say that an element~$a$ of an effect monoid~$M$
    is \Define{Boolean} when each~$b\leq a$
    is idempotent. We say an effect monoid is Boolean
    when $1$ is Boolean.
\end{definition}

\begin{proposition}\label{prop:booleanalgebra}
    The set of idempotents~$P(M)$ of an effect monoid~$M$
        is a Boolean algebra.
        Thus an effect monoid is Boolean iff it is a Boolean algebra.
\end{proposition}
\begin{proof}
    First we will show that in fact~$p\cdot q = p\wedge q$
        for~$p,q \in P(M)$, where the infimum~$\wedge$ is taken in~$M$.
    Using Lemma~\ref{lem:idempotentscommute},
    we see~$(p \cdot q)^2 = p\cdot q\cdot p\cdot q = p\cdot
    p\cdot q\cdot q = p\cdot q$ and so~$p \cdot q$ is an idempotent.
    Let~$r \leq p, q$.
    Then~$r \cdot p = r$ and~$r \cdot q = r$
        so that~$r \cdot p \cdot q = r$, and
        hence~$r \leq p \cdot q$ by~Lemma~\ref{lem:preserveunderidempotent},
        which shows~$p \cdot q = p \wedge q$.
     As the complement is an order anti-isomorphism,
        we find~$p\vee q = (p^\perp \wedge q^\perp)^\perp$ and
    hence $P(M)$ is an ortholattice.
    It remains to show that
    it satisfies distributivity: $p\wedge(q\vee r) = (p\wedge q)\vee
    (p\wedge r)$.
    By uniqueness of complements,
     it is easily shown that~$p\vee q
     = p \ovee (p^\perp \cdot q) = q \ovee ( p\cdot q^\perp)$.
    The remainder is a straightforward exercise in writing out
    the expressions $p\wedge(q\vee r)$ and $(p\wedge q)\vee
        (p\wedge r)$ and noting that they are equal.
\end{proof}

\begin{proposition}\label{prop:completelattice}
  The set of idempotents~$P(M)$ of an $\omega$-complete effect monoid~$M$
        is an $\omega$-complete Boolean algebra.
\end{proposition}
\begin{proof}
    Let~$A \subseteq P(M)$ be a countable subset. Pick an enumeration of 
    its elements $p_1,p_2,\ldots$. Let $q_n$ be iteratively defined
    as $q_1 = p_1$ and $q_n = q_{n-1}\vee p_n$.
    Then the $q_n$ form an increasing sequence and hence it has
    a supremum $q$. We claim that $q$ is also the supremum of $A$.
    Of course $q\geq q_n \geq p_n$ and hence $q$ is an upper bound.
    Suppose that $r\geq p_n$ for all $n$. Then $r\geq q_1$, and hence by induction
    if $r\geq q_n$ then $r\geq q_n \vee p_n = q_{n+1}$. Hence also $r\geq q$.
\end{proof}

\begin{proposition}\label{prop:sharpeffectmonoidisbooleanalgebra}
    Let $M$ be an $\omega$-complete Boolean effect monoid. 
    Then $M$ is an $\omega$-complete Boolean algebra.
\end{proposition}
\begin{proof}
    By Propositions~\ref{prop:booleanalgebra} and \ref{prop:completelattice} 
    $P(M)$ is an $\omega$-complete Boolean algebra. 
    But by assumption every element of $M$ is an idempotent, and hence $M=P(M)$.
\end{proof}

The counterpart to the Boolean effect monoids, are the \emph{halvable effect monoids}
\begin{definition}
We say that an element~$a$ of an effect algebra~$E$
    is \Define{halvable}
    when~$a=b\ovee b$ for some~$b\in E$. We say an effect algebra
    is halvable when $1$ is halvable.
\end{definition}

A halvable effect monoid actually has much more structure then might be apparent: it is a convex effect algebra. Recall from Definition~\ref{def:convexeffectalgebra} that a convex effect algebra has an action of the real unit interval $\lambda \mapsto \lambda\cdot a$ satisfying certain axioms.
In a convex effect monoid we will usually write
the convex action without any symbol in order to distinguish it 
from the multiplication coming from the monoid structure. 
So if $\lambda \in [0,1]$ is a real number and $a,b \in M$ is 
a convex effect monoid, then we write $\lambda(a\cdot b)$. 
Note that a priori it is not clear whether 
$\lambda (a\cdot b) = (\lambda a)\cdot b = a \cdot (\lambda b)$.

\begin{proposition}\label{prop:dircompleteisconvex}
    Let $M$ be a halvable $\omega$-complete effect monoid.
        Then $M$ is convex.
\end{proposition}
\begin{proof}
    Pick any~$a \in M$ with~$a \ovee a = 1$.
    Let $q = \frac{m}{2^n}$ be a dyadic rational number with $0\leq m\leq
    2^n$. We define a corresponding element~$\cl{q} \in M$  by
        $\cl{q} = m a^n$, which is easily seen to be independent
        of the choice of~$m$ and~$n$.
    This yield an action~$(q, s) \mapsto \cl{q} \cdot s$ that satisfies
        all axioms of Definition~\ref{def:convexeffectalgebra}
        restricted to dyadic rationals.
    
    Assume~$\lambda \in (0,1]$.
    Pick a strictly increasing
        sequence~${0\leq q_1< q_2<\ldots}$ of dyadic rationals
        with~$\sup q_i = \lambda$.

        We will define~$\cl\lambda \in M$ by~$\bigvee_i \cl {q_i}$,
            but first we have to show that it is independent
                of the choice of the sequence
                    and that it coincides with the definition
                    just given for dyadic rationals.
        So assume~$0 \leq p_1 < p_2 < \ldots$
                is any other sequence of dyadic rationals
                with~$\sup p_i=\lambda$.
    For any~$p_i$ we can find a~$q_j$ with~$p_i \leq q_j$,
        so~$\cl {p_i} \leq \cl {q_j}$, hence~$\bigvee_i \cl{p_i}
                \leq \bigvee_j \cl{q_j}$.
        As the situation is symmetric between the sequences,
            we also have~$\bigvee_j \cl{q_j} \leq \bigvee_i \cl{p_i}$
            and so~$\bigvee_j \cl{q_j} = \bigvee_i \cl{p_i}$.
        Hence~$\cl \lambda$ is independent of the choice of sequence.
    Next, assume~$\lambda = q$ is a non-zero dyadic rational.
    Pick~$m$ with~$2^{-m} \leq q$.
    Then~$q_n = q - {2^{-(m+n)}}$
        is a sequence of dyadic rationals with~$\sup q_n = q$.
    We have~$\bigvee_n \cl {q - 2^{-(m+n)}}
                    = \cl q \ominus \bigwedge_n \cl{2^{-(m+n)}}
                    = \cl q \ominus \bigwedge_n a^{m+n}
                    = \cl q $,
                    (where in the last step we used Lemma~\ref{lem:asummablepowerszero}) so both definitions of~$\cl q$ coincide.
        As a result we are indeed justified
            to define~$\overline\lambda = \bigvee_i \cl {q_i}$.
 We can then define an action by $(\lambda, s)\mapsto \cl{\lambda}\cdot s$.
 As both addition and multiplication preserve suprema 
 by Theorem~\ref{thm:multisnormal}, it is straightforward to show 
 that this action indeed satisfies all the axioms of a convex action.
\end{proof}

\subsection{Embedding theorems}\label{sec:embeddingtheorem}

In this section we will show that $\omega$-complete effect monoids always embed into a direct sum of a Boolean algebra and a convex effect monoid.

Some of the results in our section require us to prove the existence of certain maximal elements. Recall that for a partially ordered set $P$ an element $a\in P$ is \Define{maximal}\indexd{maximal element (poset)} when for any $b\in P$ with $a\leq b$ we have $a=b$. To find such maximal elements we use (either explicitly or implicitly) the well-known Zorn's lemma, a statement that is equivalent to the Axiom of Choice. 
\begin{theorem}[Zorn's Lemma]\label{zornslemma}\indexd{Zorn's Lemma} Let $P$ be any partially ordered set. A subset $D\sse P$ is called a \Define{chain}\indexd{chain (poset)} when it is totally ordered. If all chains in $P$ have a supremum, then $P$ has at least one maximal element.
\end{theorem}

\begin{lemma}
\label{lem:ceilhalveable}
    Let $M$ be an $\omega$-complete effect monoid, and let $a$ be a halvable element.
    Then the ceiling~$\ceil{a}$
    is halvable as well.
\end{lemma}
\begin{proof}
Write~$a =b\ovee b$.
We compute
\begin{multline*}
\textstyle
    \ceil{a} \ \equiv\ 
    \bigovee_n a  \cdot  (a^\perp)^n \ =\  
    \bigovee_n (b\ovee b) \cdot (a^\perp)^n
    \ =\  
\textstyle
    \bigl(\bigovee_n b \cdot (a^\perp)^n\bigr) \,\ovee\, \bigl(
    \bigovee_n b \cdot (a^\perp)^n \bigr),
\end{multline*}
    and hence $\ceil{a}$ is indeed halvable.
\end{proof}

\begin{proposition}\label{prop:maxcollectionbooleanhalveable}
Each $\omega$-complete effect monoid~$M$
has a subset~$E \subseteq M$ 
\begin{itemize}
\item
that is a maximal collection of non-zero orthogonal idempotents,
and
\item
where each element of~$E$ is either halvable or Boolean.
\end{itemize}
\end{proposition}
\begin{proof}
Let~$H$ be a maximal collection
of non-zero orthogonal halvable idempotents of~$M$,
and let~$E$ be a maximal collection
of non-zero orthogonal idempotents of~$M$
that extends~$H$.
(Such sets~$E$ and~$H$ exist by Zorn's Lemma).
By definition, $E$ is a maximal collection
of non-zero orthogonal idempotents of~$M$,
so the only thing to prove
is that each~$p\in E\backslash H$ is Boolean.
Hence, let~$a$ be an element of~$M$ below
some~$p\in E\backslash H$;
we must show that~$a$ is an idempotent.
    Note that~$2(a^\perp \cdot a)= a^\perp \cdot a \ovee a^\perp\cdot a$ 
    (which exists by e.g.~Lemma~\ref{lem:selfsummable}) is halvable,
    and $2(a^\perp\cdot a) \leq p$,
    because $2(a^\perp \cdot a)\cdot p = 2(a^\perp \cdot a \cdot p)
     = 2(a^\perp\cdot a)$.
Then the idempotent $\ceil{2a\cdot a^\perp} \leq p$,
    which is halvable by Lemma~\ref{lem:ceilhalveable},
is orthogonal to all~$h\in H$
    (since it is below~$p$)
and must therefore be zero (since it would contradict the maximality of~$H$ otherwise).
We then have
    $a^\perp\cdot a=0$ since $a^\perp \cdot a \leq \ceil{2a^\perp\cdot a}=0$,
    and so~$a$ is an idempotent by Lemma~\ref{lem:idempotentiff}.
So~$p$ is indeed Boolean.
\end{proof}

Note that the only idempotent that is both Boolean and halvable is zero, and hence
each element in the above set is either Boolean \emph{or} halvable.

    

\begin{proposition}
Given a maximal orthogonal collection of non-zero idempotents~$E$
of an $\omega$-complete effect monoid~$M$,
the map
$$
    a\mapsto (a\cdot p)_p\colon M\longrightarrow \bigoplus_{p\in E} pM
$$
is an embedding of effect monoids.
\end{proposition}
\begin{proof}
    The map obviously maps $1$ to $1$, and preserves addition. 
    Hence it also preserves the complement and the order.
    By Lemma~\ref{lem:idempotentscommute} we have 
    $(a\cdot p)\cdot (b\cdot p) = (a\cdot b)\cdot (p\cdot p) =(a\cdot b)\cdot p$,
    and so the map also preserves the multiplication.
    It remains to show that the map is order reflecting. 
    Note that if we had~$\bigovee E = 1$,
    then for any $a$ by Theorem~\ref{thm:multisnormal} $a = a\cdot 1 = a\cdot \bigovee_{p\in E} p 
    = \bigovee_{p\in E} a\cdot p$, and hence
    if $a\cdot p \leq b\cdot p$ for all $p\in E$ we have
    $a = \bigovee_{p\in E} a\cdot p \leq \bigovee_{p\in E} b\cdot p = b$, which proves that it is indeed order reflecting.

    So let us prove that $\bigovee E = 1$. Suppose~$u$ is an upper bound for~$E$.
    We must show that~$u=1$.
    Note that~$\floor{u}$ is an upper bound for~$E$ too,
    since~$E$ contains only idempotents.
    It follows that the idempotent~$\floor{u}^\perp = \ceil{u^\perp}$ 
    is orthogonal
    to all~$p\in E$,
    which is impossible (by maximality of~$E$) unless
    $\ceil{u^\perp}= 0$.
    Hence~$u^\perp \leq \ceil{u^\perp}=0$, so that indeed $u=1$.
\end{proof}

\begin{theorem}\label{thm:omegacompleteembed}
    Let $M$ be an $\omega$-complete effect monoid. Then there exist
    $\omega$-complete effect monoids $M_1$ and $M_2$ where
    $M_1$ is convex, and $M_2$ is an $\omega$-complete Boolean algebra
    such that $M$ embeds into $M_1\oplus M_2$.
\end{theorem}
\begin{proof}
    Let $E=H\cup B$ be a maximal collection of non-zero orthogonal idempotents of
    Proposition~\ref{prop:maxcollectionbooleanhalveable} such that the idempotents
    $p\in H$ are halvable, while the $q\in B$ are Boolean.

    Let $M_1 \equiv \bigoplus_{p\in H} pM$ and $M_2 \equiv \bigoplus_{q\in B} qM$.
    It is easy to see that $M_1$ is then again halvable and $M_2$ is Boolean.
    By Propositions~\ref{prop:dircompleteisconvex} and~\ref{prop:sharpeffectmonoidisbooleanalgebra} $M_1$ is convex while $M_2$ is an $\omega$-complete Boolean algebra.
    By the previous proposition $M$ embeds into 
    $\bigoplus_{p\in E}pM \cong M_1\oplus M_2$.
\end{proof}

One might be tempted to think that the above result could be strengthened to an isomorphism. The following example shows that this is not the case:

\begin{example}
    Let $X_1$ and $X_2$ be uncountably infinite sets, and let $A$ be the set of all pairs of functions 
    $$A := \{(f_1:X_1\rightarrow [0,1], f_2: X_2\rightarrow \{0,1\})\}.$$
    Let $S_0, S_1\subseteq A$ be subsets where both functions are unequal to 0 respectively 1 only at a countable number of spots:
    \begin{align*}
    S_0 &:= \{(f_1,f_2)~;~ \text{both } \{x_1\in X_1~;~f_1(x_1)\neq 0\} \text{ and } \{x_2\in X_2~;~f_2(x_2)\neq 0\} \text{ countable}\} \\
    S_1 &:= \{(f_1,f_2)~;~ \text{both }\{x_1\in X_1~;~f_1(x_2)\neq 1\} \text{ and } \{x_2\in X_2~;~f_2(x_2)\neq 1\} \text{ countable}\}
    \end{align*}
    Finally, define $M = S_0\cup S_1$. It is then straightforward to check that $M$ is an $\omega$-complete effect monoid. It is easy to see that $M$ has no maximal halvable idempotent, and hence for any $M_1$ halvable and $M_2$ Boolean, necessarily ${M\neq M_1\oplus M_2}$.
\end{example}

Though the embedding is not an isomorphism for $\omega$-complete effect monoids, if the effect monoid was directed complete, then we would have an isomorphism. We direct the interested reader to Ref.~\cite{wetering-effect-monoids}.

\subsection{Convexity and order unit spaces}\label{sec:convexeffectalgebrasandOUS}

Recall from Proposition~\ref{prop:convextotalisation} that every convex effect algebra is the unit interval of an ordered vector space.
We wish to use Kadison's representation theorem (Theorem~\ref{thm:kadison}) to conclude that any $\omega$-complete convex effect monoid is of the form~$[0,1]_{C(X)}$ for some compact Hausdorff space. This will require some setup.

\begin{lemma}\label{lem:completenessousemod}
    Let $V$ be an ordered vector space with order unit $1$. Its unit interval is $\omega$-complete
    if and only if $V$ is \Define{bounded $\omega$-directed-complete}\indexd{omega-complete@$\omega$-complete!--- order unit space};
    that is if every bounded increasing sequence in~$V$
    has a supremum.
\end{lemma}
\begin{proof}
Since any subset of the unit interval is obviously bounded, any
    bounded $\omega$-directed-complete vector space
    has an $\omega$-complete unit interval.
For the other direction, assume~$S \subseteq V$
    is some countable directed subset with upper bound~$b \in V$.
    Pick any~$a \in S$ and define~$S' = \{v \in S; a \leq v\}$.
    It is then sufficient to show that~$S'$ has a supremum.
    Pick any~$n \in \N$, $n\neq 0$ with~$-n \cdot 1 \leq a,b  \leq n\cdot 1$.
Then clearly~$\{\frac{1}{n} (s-a); {s\in S'}\} \subseteq [0,1]_V$
    has a supremum and hence so does~$S'$ as~$v \mapsto nv + a$
    is an order isomorphism.
\end{proof}

\begin{lemma}\label{lem:boundeddircomplisarchous}
    A bounded $\omega$-directed-complete ordered vector space with order unit is Archimedean, and hence is an order unit space.
\end{lemma}
\begin{proof}
Assume~$V$ is bounded~$\omega$-directed-complete.
    Let~$v \in V$ be given with~$v \leq \frac1n 1$ for all~$n \in \N$.
As~$a \mapsto -a$ is an order anti-isomorphism,
    all bounded directed subsets of~$V$ have an infimum too,
    so~$v \leq \bigwedge_n \frac1n 1 = 0$
    as desired.
\end{proof}

\begin{definition}
  Let $M$ be a convex effect monoid. We say its multiplication is \Define{homogeneous}\indexd{homogeneous multiplication} when for any $a,b\in M$ and $\lambda\in[0,1]$ we have $\lambda(a\cdot b) = (\lambda a)\cdot b = a\cdot (\lambda b)$.
\end{definition}

\begin{proposition}\label{prop:multiplicationisbilinear}
    Let $M$ be an $\omega$-complete convex effect monoid.
    Then the multiplication is homogeneous.
\end{proposition}
\begin{proof}
    We will only show $\lambda(a\cdot b) = a\cdot(\lambda b)$. The other equality follows similarly.

    Clearly~$n(a\cdot (\frac1n b)) = a\cdot b = n \frac1n (a\cdot b)$
        and so~$a\cdot (\frac1n b) = \frac1n (a \cdot b)$.
    Hence~$m (a \cdot (\frac 1n b )) = a\cdot (\frac mn b)
            = \frac mn (a \cdot b)$ for any~$m \leq n$.
    We have now shown the desired equality for rational~$\lambda$.
    To prove the general case, 
        let~$a,b \in M$ and~$\lambda \in [0,1]$ be given.
    Pick a sequence~$0 \leq q_1 < q_2 < \ldots $ of rationals
        with~$\bigvee_n q_n = \lambda$.
    Let~$V$ be the ordered vector space with~$M \cong [0,1]_V$ (as a convex effect algebra),
        which exists due to Proposition~\ref{prop:convextotalisation}, and is an order unit space due to Lemma~\ref{lem:boundeddircomplisarchous}.
    Note that the multiplication of~$M$ is only defined on~$[0,1]_V$.
    For~$a \in[0,1]_V$ we have~$a \leq \| a \| 1$
        and so~$\| a \cdot b \| \leq \| a \| \| b \|$ for any~$b \in [0,1]_V$.
        Hence
        \begin{align*}
        \norm{\lambda (a\cdot b) - a\cdot (\lambda b)}
            &\ =\  \norm{(\lambda - q_i)(a\cdot b) + a\cdot (q_i b) - a\cdot (\lambda b)} \\
        &\ =\  \norm{(\lambda-q_i)(a\cdot b) - a\cdot ((\lambda - q_i)b)} \\
        &\ \leq\  (\lambda-q_i)\norm{a\cdot b} + (\lambda-q_i)\norm{a}\norm{b}.
        \end{align*}
        The right-hand side vanishes
            as $i\rightarrow \infty$.
            Thus $\norm{\lambda(a\cdot b) - a\cdot (\lambda b)}=0$.
            Since $V$ is Archimedean by Lemma \ref{lem:boundeddircomplisarchous}, $\norm{\cdot}$
            is a proper norm so that then $\lambda(a\cdot b) - a\cdot (\lambda b) = 0$.
\end{proof}

\begin{proposition}[{\cite[Theorem 46]{basmaster}}]\label{prop:convexeffmonoidextendsbilinear}
Let $M$ be a convex effect monoid with homogeneous multiplication, and let $V$ be the OUS such that ~$M \cong [0,1]_V$.
Then the multiplication of~$M$ extends
    to a bilinear, unital, associative and positivity preserving multiplication
        on~$V$. 
\end{proposition}

\begin{proposition}[{\cite[Lemma 1.2]{wright1972measures}}]\label{prop:orderunitcompleteness}
    Let $V$ be a bounded $\omega$-directed-complete OUS. Then $V$ is complete in its norm. 
\end{proposition}

\begin{theorem}\label{thm:convexextremallydisconnected}
    Let $M$ be a convex $\omega$-complete effect monoid.
    Then $M \cong [0,1]_{C(X)}$ for some basically
        disconnected Hausdorff space~$X$.
\end{theorem}
\begin{proof}
    By Proposition~\ref{prop:convextotalisation}
            there is an OUS~$V$
            such that~$M \cong [0,1]_V$ as a convex effect algebra.
    $V$ is bounded~$\omega$-directed-complete
        by Lemma~\ref{lem:completenessousemod}
        and thus complete by Proposition~\ref{prop:orderunitcompleteness}
        and Archimedean by Lemma~\ref{lem:boundeddircomplisarchous}.
The multiplication on~$M$ is homogeneous
    by Proposition~\ref{prop:multiplicationisbilinear}
    and so it extends to
    a bilinear, unital, associative and positive product on~$V$
    by Proposition~\ref{prop:convexeffmonoidextendsbilinear}.
Then by Theorem~\ref{thm:kadison}
    there exists a compact Hausdorff space~$X$
    such that~$V \cong C(X)$.
    As~$C(X)$ is bounded~$\omega$-directed-complete
     iff~$X$ is basically disconnected, the result follows.
\end{proof}

\subsection{Characterisation of \texorpdfstring{$\omega$}{omega}-directed-complete effect monoids}\label{sec:maintheorems}

Finally, we can establish the results we set out to prove.

\begin{theorem*}[\ref{thm:omega-complete-classification}]
    Let $M$ be an $\omega$-complete effect monoid.
    Then there exists a basically disconnected compact Hausdorff space $X$,
    and an $\omega$-complete Boolean algebra $B$ such that 
    $M$ embeds into $[0,1]_{C(X)} \oplus B$.
\end{theorem*}
\begin{proof}
    By Theorem~\ref{thm:omegacompleteembed} there exist
    $\omega$-complete effect monoids $M_1$ and $M_2$
    such that $M$ embeds into $M_1\oplus M_2$, where $M_1$ is convex and 
    $M_2$ is an $\omega$-complete Boolean algebra.
    By Theorem~\ref{thm:convexextremallydisconnected} $M_1=[0,1]_{C(X)}$
    for a basically disconnected compact Hausdorff space $X$.
\end{proof}

\begin{theorem*}[\ref{thm:no-zero-divisors}]
    Let $M$ be an $\omega$-complete effect monoid with no non-trivial
    zero divisors.
    Then either $M=\{0\}$, $M=\{0,1\}$ or $M\cong [0,1]$.
\end{theorem*}
\begin{proof}
    Assume that $M\neq \{0,1\}$ and $M\neq \{0\}$. 
    We remark first that for any idempotent $p\in M$ we have $p\cdot p^\perp = 0$,
    and hence by the lack of non-trivial zero divisors we must have $p=0$ or $p=1$.
    Since $M\neq \{0,1\}$, there is an $s\in M$ such that $s\neq 0,1$, and hence
    we must have $s\cdot s^\perp \neq 0$. 
    By Lemma~\ref{lem:selfsummable} we then
    have an element $2(s\cdot s^\perp)$ that is halvable.
    Hence by Lemma~\ref{lem:ceilhalveable} 
    $\ceil{2s\cdot s^\perp}$ is also halvable.
    As this ceiling is an idempotent it must be equal to $1$ or to $0$. 
    If it were zero then 
    $2s\cdot s^\perp \leq \ceil{2s\cdot s^\perp} = 0$, 
    which contradicts $s\cdot s^\perp \neq 0$.
    So $1=\ceil{2s\cdot s^\perp}$ is halvable. 
    By Proposition~\ref{prop:dircompleteisconvex}, $M$ is then convex.
    Hence, by Theorem~\ref{thm:convexextremallydisconnected}~$M = [0,1]_{C(X)}$
        for some basically disconnected $X$.
    We will show that $X$ has a single element, which will complete the proof.

    As idempotents of~$[0,1]_{C(X)}$ correspond to clopens of~$X$,
        there are only two clopens in~$X$, namely~$X$ and~$\emptyset$.
    Reasoning towards contradiction,
        assume there are~$x,y \in X$ with~$x \neq y$.
    By Urysohn's lemma we can find~$f \in C(X)$
        with~$0 \leq f \leq 1$, $f(x) = 0$ and~$f(y) = 1$.
        $U_x = f^{-1}([0,\frac13))$ and~$U_y= f^{-1}((\frac23,1])$
        are two open sets with disjoint closure.
    Using Urysohn's lemma again, we can find~$g \in C(X)$
        with~$g(\overline{U_x}) = \{0\}$ and $g(\overline{U_y}) = \{1\}$.
    As~$X$ is basically disconnected,
        we know~$\overline{\supp g}$ is clopen.
    We cannot have~$\overline{\supp g} = \emptyset$
        as~$y \in U_y \subseteq \overline{\supp g}$.
        Hence~$\overline{\supp g} = X$.
        But then~$x \in U_x \subseteq X - \overline{\supp g} = \emptyset$, a contradiction.
        Hence~$X$ has only one point and so~$M \cong [0,1]$.
\end{proof}

\begin{theorem*}[\ref{thm:irreducible-effect-monoid}]
  Let $M$ be an irreducible $\omega$-complete effect monoid. Then $M=\{0\}$, $M=\{0,1\}$ or $M=[0,1]$.
\end{theorem*}
\begin{proof}
  By the previous theorem it suffices to show that $M$ has no non-trivial zero divisors.
  By Corollary~\ref{cor:corneriso}, any idempotent $p\in M$ with $p\neq 0,1$ would make $M$ reducible. Hence, the only idempotents in $M$ are $0$ and $1$. Now suppose $a\cdot b = 0$. By Proposition~\ref{prop:ceilprod} then $a\cdot \ceil{b} = 0$. If $\ceil{b} = 1$, this implies $a=0$. Otherwise necessarily $\ceil{b}=0$ so that $b\leq \ceil{b} = 0$. Hence, $M$ has no non-trivial zero divisors.
\end{proof}

\section{Convex normal sequential effect algebras}\label{sec:SEA}

The axioms of the sequential product in Definition~\ref{def:seqprod} were based on that of a sequential effect algebra (SEA), introduced by Gudder and Greechie~\cite{gudder2002sequential}. For this chapter we will be interested in \emph{$\omega$-SEAs}, that additionally have countable suprema that interact well with the sequential product.

\begin{definition}\label{def:normal-SEA}
    A \Define{sequential effect algebra} (SEA)~\cite{gudder2002sequential}
    \indexd{sequential effect algebra}
    \indexd{effect algebra!sequential effect algebra}
    \index{math}{SEA (sequential effect algebra)} 
    $(E,\ovee ,0,(\ )^\perp,\mult)$ is an
     effect algebra with an additional (total)
    binary operation~$\mult$,
        called the \Define{sequential product}\indexd{sequential product}\indexd{sequential product!on effect algebra},
        satisfying the axioms listed below.
        We say~$a$ and~$b$ are \Define{compatible}\indexd{compatible effects},
            writing~$a \commu b$,
            whenever~$a \mult b = b \mult a$.\index{math}{ab@$a\commu b$ (compatible effects)}
    \begin{enumerate}
            \item $a\mult (b\ovee c) = a\mult b \ovee a \mult c$.
            \item $1\mult a = a$.
            \item $a\mult b = 0 \implies b\mult a =0$.
        \item If $a\commu b$, then $a\commu b^\perp$ and $a\mult
                (b\mult c) = (a\mult b)\mult c$ for all $c$.
        \item If $c\commu a$ and $c\commu b$ then also $c\commu
        a\mult b$ and if furthermore $a\perp  b$, then $c\commu a\ovee
        b$.
    \end{enumerate}
    A SEA~$E$ is a \Define{$\omega$-SEA}
        provided~$E$ is $\omega$-complete and
    \begin{enumerate}[resume]
            \item For any increasing sequence $s_1\leq s_2\leq \ldots$ we have
            $a\mult \bigvee_n s_n = \bigvee_n (a\mult s_n)$ and 
            if $a\commu s_n$ for all $n$, then $a\commu \bigvee_n s_n$.
    \end{enumerate}
    We say~$E$ is \Define{commutative}\indexd{commutative sequential effect algebra} whenever~$a \commu b$ for
        all~$a,b \in E$.
    We say~$a,b \in E$ are \Define{orthogonal}\indexd{orthogonal effect!in sequential effect algebra},
        provided~$a \mult b =0$. An element $p\in E$ is \Define{idempotent}\indexd{idempotent!in sequential effect algebra} when $p^2 := p\mult p = p$ or equivalently, when $p\mult p^\perp = 0$.
\end{definition}

\begin{remark}
    The definition of a sequential effect algebra is very similar to that of Definition~\ref{def:seqprod}, but then specified for effect algebras. The differences are that it does not include the axiom of continuity \ref{ax:cont}, and that it has the new axiom stating that $a\commu (b\mult c)$ when $a\commu b,c$. The axiom of `normality', \ie~that is preserves suprema, will take over the role of continuity in the results of this section.
\end{remark}

We will be primarily interested in SEAs $E$ that are also convex effect algebras (cf.~Definition~\ref{def:convexeffectalgebra}). Recall from Proposition~\ref{prop:convextotalisation} that we can then find an ordered vector space $V$ such that $E\cong [0,1]_V$. If furthermore $E$ is $\omega$-complete (such as when $E$ is an $\omega$-SEA), then $V$ is an order unit space (cf.~Lemma~\ref{lem:boundeddircomplisarchous}).

\begin{example}
	The unit interval of a JBW-algebra is a directed-complete convex effect algebra. The operation $a\mult b:= Q_{\sqrt{a}} b$ makes this unit interval a convex $\omega$-SEA (cf.~Theorem~\ref{thm:JBW-seqprod}).
\end{example}

The following two propositions contain similar results to those found in Section~\ref{sec:basicresultsseqprod}.
\begin{proposition}
    Let $E$ be a SEA and let $a,b,c \in E$.
    \begin{itemize}
        \item $a\mult 0 = 0\mult a = 0$ and $a\mult 1=1\mult a = a$.
        \item $a\mult b \leq a$.
        \item If $a\leq b$ then $c\mult a\leq c\mult b$.
    \end{itemize}
\end{proposition}
\begin{proof}
    See Proposition~\ref{prop:basic}.
\end{proof}

\begin{proposition}\label{prop:SEAlinearity}
    Let $E$ be a convex $\omega$-SEA. Let $a,b \in E$ and let $\lambda\in[0,1]$. Then:
    \begin{enumerate}[label=\alph*)]
        \item $a\mult (\lambda b) = \lambda (a\mult b)$.
        \item $(\lambda a)\mult b = a\mult (\lambda b) = \lambda (a\mult b)$ and if $a\commu b$ then $a\commu \lambda b$.
    \end{enumerate}
\end{proposition}
\begin{proof}~
    \begin{enumerate}[label=\alph*)]
        \item See Proposition~\ref{prop:linearity} a) and b).
        \item Clearly $\frac{1}{n}a\commu \frac{1}{n}a$ so that by summing terms together $\frac{1}{n}a\commu a$. In the same way we also see that $qa\commu a$ and $qa^\perp \commu a^\perp$ for any rational $0\leq q\leq 1$. Then also $qa^\perp \commu a$ and hence $a\commu (qa+qa^\perp)=q1$ so that $(q1)\mult a = a\mult (q1) = q(a\mult 1) = qa$. 
        Now let $\lambda\in[0,1]$ be a real number and let $q_i$ be an increasing set of rational numbers that converges to $\lambda$, then it is straightforward to show that $\vee_i q_i 1 = \lambda 1$ and because $a\commu q_i 1$ we indeed have $a\commu \vee_i q_i1=\lambda 1$. As a result $(\lambda a)\mult b = (a\mult (\lambda 1))\mult b = a\mult ((\lambda 1)\mult b) = a \mult (\lambda b) = \lambda (a\mult b)$. 

        Now if $a\commu b$ then we get $a\mult (\lambda b) = \lambda (a\mult b) = \lambda (b\mult a) = (\lambda b)\mult a$ so that indeed $a\commu \lambda b$. \qedhere
    \end{enumerate}
\end{proof}

Let $V$ be the order unit space associated to a convex $\omega$-SEA $E$ via $E\cong [0,1]_V$. Let $v\in V$ be arbitrary and write $v=\lambda_1b_1 - \lambda_2b_2$ where $\lambda_1,\lambda_2\in \R_{\geq 0}$ and $b_1,b_2\in E$. Then we define $a\mult v := \lambda_1 a\mult b_1 - \lambda_2 a\mult b_2$. This is well-defined because $a\mult (\lambda b) = \lambda (a\mult b)$ shows that the product is linear in the second argument.

Similarly, because $(\lambda a)\mult b = \lambda (a\mult b)$, we can also define the sequential product for any $v> 0$ in the associated OUS of $E$ by $v\mult w:= \norm{v} (\norm{v}^{-1}v)\mult w$. We will therefore use the sequential product for all elements in the positive cone of the order unit space associated to a convex $\omega$-SEA without further reference to this fact.

Analogous to the definition in Jordan algebras, we define the commutant and bicommutant of a set of elements (cf.~Definition~\ref{def:Jordan-operator-commute}).

\begin{definition}
    Let $S\subseteq E$ be a subset of elements of a SEA $E$. We define the \Define{commutant} of $S$ to be $S^\prime := \{a\in E~;~ \forall s\in S: s\commu a\}$, the set of elements in $E$ that are compatible with every element in $S$. Similarly the \Define{bicommutant} $S^{\prime\prime}:=(S^\prime)^\prime$ of $S$ is the set of elements in $E$ that are compatible with every element in $S^\prime$.
\end{definition}

\begin{lemma}
    Let $S\subseteq E$ be a subset of elements of a SEA $E$. Then $S'$ is an effect sub-algebra. If $E$ is an $\omega$-SEA, then so is $S'$ and if $E$ is a convex $\omega$-SEA then so is $S'$.
\end{lemma}
\begin{proof}
    If we have $a,b\in S'$ such that $a\ovee b$ is defined in $E$ then for all $s\in S$ we have $a\commu s$ and $b\commu s$ so that $(a\ovee b)\commu s$ and $(a\mult b)\commu s$ as well which means that $S'$ is closed under addition and multiplication. It is similarly also closed under taking the complement. Obviously $1$ and $0$ commute with every element so that $0,1\in S'$ so that $S'$ is indeed a sub-SEA.

    Suppose now that $E$ is a $\omega$-SEA and let $a_1\leq a_2\leq \ldots$ be an increasing sequence in $S'$. This has a supremum $\bigvee_n a_n$ in $E$. Since for all $s\in S$ we have $s\commu a_n$ we also have $s\commu \bigvee_n a_n$ by an axiom of $\omega$-SEAs. Since $s$ was arbitrary we have $\bigvee_n a_n \in S'$ and since $S' \sse E$, $\bigvee_n a_n$ is also the supremum inside $S'$. Hence $S'$ is an $\omega$-complete effect algebra, and since the sequential product is inherited from $E$ it is also an $\omega$-SEA.

    Finally, suppose $E$ is a convex $\omega$-SEA. By Proposition~\ref{prop:SEAlinearity} when $a\commu b$ we also have $\lambda a\commu b$ and hence we see that $S'$ is closed under scalar multiplication and is thus also convex.
\end{proof}

\begin{lemma}\label{lem:bicommutant}
    Let $S\subseteq E$ be a set of mutually compatible elements in a SEA $E$. Then $S''$ is a commutative sequential effect sub-algebra of $E$. If $E$ is a convex $\omega$-SEA, then $S''$ is a convex $\omega$-SEA as well.
\end{lemma}
\begin{proof}
    By the previous lemma we already know that $S''$ is a (convex $\omega$-)SEA, so the only thing left to prove is that it is commutative. Since $S$ consists of mutually compatible elements we see that $S\subseteq S'$ and hence all elements of $S''$ are compatible with all elements of $S$ so that $S''\subseteq S'$. As the elements of $S''$ by definition are compatible with everything in $S$ we see that everything in $S''$ must be mutually compatible.
\end{proof}

\begin{lemma}\label{lem:commkadisonSEA}
	Let $E$ be a commutative convex $\omega$-SEA. Then there exists a basically disconnected compact Hausdorff space $X$ such that $E$ is isomorphic as a sequential effect algebra to the unit interval of $C(X)$.
\end{lemma}
\begin{proof}
	$E$ is isomorphic to the unit interval of some bounded $\omega$-directed-complete order unit space $V$. By Proposition~\ref{prop:orderunitcompleteness} $V$ is complete in its norm. The sequential product is linear in its second argument, but since it is assumed to be commutative it is also linear in its first argument. It obviously preserves positivity. Hence Theorem~\ref{thm:kadison} applies, and we see that there must be some compact Hausdorff space $X$ such that $V\cong C(X)$. As $V$ is bounded $\omega$-directed-complete, $X$ must be basically disconnected (cf.~Section~\ref{sec:basically-disconnected}).
\end{proof}

\begin{definition}
    Let $a\in E$ be an element of a (convex $\omega$-)SEA. We denote the commutative sub-algebra generated by $a$ by $C(a):=\{a\}^{\prime\prime}$.
\end{definition}
When $a\commu b$ we get $a^\perp\commu b$, $a^n\commu b$ and also $(\lambda a^n+\mu a^m)\commu b$ for scalars $\lambda,\mu\in [0,1]$. Consequently, $C(a)$ contains (among other effects) all polynomials in $a$ and $a^\perp$ that are contained in $E$.

\begin{proposition}\label{prop:spectraltheorem}
	Let $E$ be a convex $\omega$-SEA and let $a\in E$. Then $a$ can be written as the supremum and norm limit of an increasing sequence of effects of the form $\sum_i \lambda_i p_i$ where $\lambda_i>0$ and the $p_i$ are orthogonal idempotent effects that are compatible with $a$. Furthermore, there is a unique effect $\sqrt{a}$ that satisfies $\sqrt{a}^2 = a$.
\end{proposition}
\begin{proof}
	$C(a)$ is a commutative convex $\omega$-SEA by Lemma~\ref{lem:bicommutant}. By Lemma~\ref{lem:commkadisonSEA} we then see that $C(a)$ is isomorphic to the unit interval of some $C(X)$ where $X$ is basically disconnected. The claims then follows by Proposition~\ref{prop:CXsupremumofsimpleelements} and the remarks in Section~\ref{sec:spectral-theorem}.
\end{proof}
\begin{corollary}
	Let $E$ be a convex $\omega$-SEA and let $V$ be its associated order unit space. Let $v\in V$ be an arbitrary element, then $v$ can be written as the norm limit of elements of the form $\sum_i \lambda_i p_i$ where $\lambda_i\in \R$ and $p_i\in E$ are orthogonal idempotents.
\end{corollary}
\begin{proof}
	If $v\geq 0$ we can rescale it such that $\norm{v}\leq 1$ and use the previous theorem. Otherwise we know that $v+\norm{v}1\geq 0$ so that we get $v+\norm{v}1 = \lim a_n$ which we rewrite to $v = \lim (a_n - \norm{v}1)$. Each $a_n$ is of the form $a_n = \sum_i \lambda_{n,i} p_{n,i}$ so we can write $a_n - \norm{v}1 = \sum_i (\lambda_{n,i} - \norm{v})p_{n,i} - \norm{v} (1-\sum_i p_{n,i})$.
\end{proof}

It will be useful for future reference to have a name for elements of the form $\sum_i \lambda_i p_i$.
\begin{definition}\label{def:simpleeffect}
	Let $a$ be an effect in a convex $\omega$-SEA $E$. We call $a$ \Define{simple}\indexd{simple element} when  $a=\sum_{i=1}^n \lambda_i p_i$ for some $n\in \N$ with all $0<\lambda_i\leq 1$ and $p_i\neq 0$ idempotent and orthogonal. We denote the subset of simple effects of $E$ by $E_0:=\{a\in E~; a \text{ simple}\}$ and similarly we denote by $V_0\subseteq V$ the order unit space spanned by $E_0$.
\end{definition}
Proposition~\ref{prop:spectraltheorem} shows that the set of simple effects $E_0$ is dense in $E$ (with respect to the topology induced by the norm) and similarly that $V_0$ is dense in $V$. 

Let us end this section by noting that the order unit space of a convex $\omega$-SEA is also homogeneous, as was the case in finite dimension (cf.~Proposition~\ref{prop:homogen}).

\begin{proposition}\label{prop:homogen-infinite}
	Let $V$ be the order unit space associated to a convex $\omega$-SEA $E$. Then $V$ is homogeneous, i.e.~for every $a,b\in V$ in the interior of the positive cone there is an order isomorphism $\Phi: V\rightarrow V$ such that $\Phi(a) = b$.
\end{proposition}
\begin{proof}
	As $a\in V$ lies in the interior of the positive cone there is an $\epsilon\in\R_{>0}$ such that $a\geq \epsilon 1$. Hence, considering the commutative algebra $C(a)$ that corresponds to the space of continuous functions of some space $X$ we see that $a$ corresponds to a function $a:X\rightarrow \R$ with $a(x)\geq \epsilon$ for all $x\in X$. Hence, we can define the function $a^{-1}:X\rightarrow \R$ by $a^{-1}(x) = a(x)^{-1}$. This $a^{-1}$ is then also a positive element of $V$ and we have $a\mult a^{-1} = a^{-1}\mult a = 1$. Hence, the multiplication operator $L_a:V\rightarrow V$ defined by $L_a(b) = a\mult b$ has an inverse $L_a^{-1} = L_{a^{-1}}$. As both $L_a$ and $L_{a^{-1}}$ are positive, this means that $L_a$ is an order isomorphism. 

	Now for any $b\in V$ in the interior of the positive cone $L_b$ is also an order isomorphism, and setting $\Phi:=L_b\circ L_{a^{-1}}$ gives us an order isomorphism satisfying $\Phi(a) = b$.
\end{proof}

\subsection{From sequential effect algebras to JB-algebras}

In this section we will introduce two additional properties that force a convex $\omega$-SEA to have a Jordan algebra structure. 
Analogously to the previous chapters, we call a map $\omega:E\rightarrow [0,1]$ for a convex SEA $E$ a state when $\omega$ is linear and $\omega(1)=1$.

\begin{definition}\label{def:SEA-compressible}
    We say the sequential product of a convex $\omega$-SEA $E$ is \Define{compressible}\indexd{sequential product!compressible ---} when for all idempotent effects $p\in E$ the following implication holds for all states $\omega:E\rightarrow [0,1]$: if $\omega(p) = 1$, then  $\omega(p\mult a) = \omega(a)$ for all $a\in E$.
\end{definition}

What this property says is that if an effect $p$ already holds with certainty on a state $\omega$, then measuring $p$ does not effect the probabilities of other effects holding in the state $\omega$. This property holds for the sequential product on JBW-algebras by Lemma~\ref{lem:JBWpreserveimage}. Later we will see that this property is related to the SEA having compressions, hence the name.
Furthermore, the multiplication maps $L_p(a) = p\mult a$ are AS-compressions (cf.~Definition~\ref{def:AS-compression}) iff the sequential product is compressible (all the conditions in Definition~\ref{def:AS-compression} always hold for a pair $L_p$ and $L_{p^\perp}$ except for the implication $\omega\circ L_{p^\perp} = 0 \implies \omega\circ L_p = \omega$, which is equivalent to it being compressible).

The second property is a weaker version of the fundamental identity of quadratic Jordan algebras.
\begin{definition}\label{def:SEA-quadratic}
    We say the sequential product of a SEA $E$ is \Define{quadratic}\indexd{sequential product!quadratic} when for any two idempotents $p,q\in E$ we have $q\mult(p\mult q) = (q\mult p)^2$.
\end{definition}

For the remainder of the section we will let $E$ be a convex $\omega$-SEA with a compressive and quadratic sequential product and let $V$ be its associated order unit space.

\begin{remark}
  As far as the author is aware, there is no known example of a convex sequential effect algebra that is not compressible, nor one that is not quadratic. Hence, it might be that these properties hold for all convex SEAs.
\end{remark}

\begin{lemma}
    Let $a,b\in E$ with $a\commu b$. Then $a^2 - b^2 = (a+b)\mult (a-b)$.
\end{lemma}
\begin{proof}
    We simply calculate $(a+b)\mult (a-b) = (a+b)\mult a - (a+b) \mult b = a\mult (a+b) - b\mult (a+b) = a\mult a + a\mult b - b\mult a - b\mult b = a\mult a - b\mult b$.
\end{proof}
\begin{lemma}
    For $p,q \in E$ idempotent, $(q\mult p)^2 - (q\mult p^\perp)^2 = q\mult p - q\mult p^\perp$.
\end{lemma}
\begin{proof}
    As $q\mult p \leq q$ we have $q\mult p \commu q$ and as $q\mult p^\perp = q - q\mult p$ we then also have $q\mult p \commu q\mult p^\perp$. Using the previous lemma with $a:=q\mult p$ and $b:=q\mult p^\perp$ we get $(q\mult p)^2 - (q\mult p^\perp)^2 = (q\mult p + q\mult p^\perp)\mult (q\mult p - q\mult p^\perp) = q\mult (q\mult (p-p^\perp)) = q\mult p - q\mult p^\perp$
\end{proof}

\begin{lemma}
    Let $q\in E$ be idempotent, and let $\omega:V\rightarrow \R$ be a state. Then if $\omega(q)=0$ we have $\omega(p\mult q) = \omega(p^\perp \mult q)$ for any idempotent $p\in E$.
\end{lemma}
\begin{proof}
  Suppose $\omega(q)=0$. Then of course $\omega(q^\perp)=1$.  Hence, by the compressive property it is sufficient to show that $q^\perp\mult(p\mult q) = q^\perp\mult(p^\perp\mult q)$, since then 
  $$\omega(p\mult q) = \omega(q^\perp\mult(p\mult q)) = \omega(q^\perp\mult(p^\perp \mult q)) = \omega(p^\perp \mult q).$$

  From the previous lemma we have $$(q^\perp \mult p)^2 - (q^\perp\mult p^\perp)^2 = q^\perp \mult p - q^\perp \mult p^\perp.$$
  Bringing the negative terms to the other side and using the quadratic property to write $(q^\perp \mult p)^2 = q^\perp\mult(p \mult q^\perp)$ (and similarly with $p$ replaced with $p^\perp$) we get 
  $$q^\perp\mult(p\mult q^\perp) + q^\perp\mult p^\perp = q^\perp\mult(p^\perp\mult q^\perp) + q^\perp\mult p.$$
  Now subtract $q^\perp\mult(p\mult q^\perp) + q^\perp\mult(p^\perp\mult q^\perp)$ from both sides of this equality to get
  $$ q^\perp\mult p^\perp - q^\perp\mult(p^\perp\mult q^\perp) = q^\perp\mult p - q^\perp\mult(p\mult q^\perp).$$ 
  By linearity this can easily be rewritten to $q^\perp\mult(p\mult q) = q^\perp\mult(p^\perp\mult q)$ as desired. \qedhere


\end{proof}

\begin{lemma}\label{lem:SEAquadratic-orderderivation}
    Let $a\in V^+$, and let $\omega:V\rightarrow \R$ be a state. Then if $\omega(a)=0$ we have $\omega(p\mult a) = \omega(p^\perp \mult a)$ for any idempotent $p\in E$.
\end{lemma}
\begin{proof}
  We first show the result for simple $a$. So suppose $a=\sum_i \lambda_i q_i$. Since $\omega(a)=0$ we of course also have $\omega(q_i)=0$ and hence by the previous lemma $\omega(p\mult q_i) = \omega(p^\perp \mult q_i)$. Then by linearity $\omega(p\mult a) = \sum_i\lambda_i\omega(p\mult q_i) = \sum_i \lambda_i\omega(p^\perp\mult q_i) = \omega(p^\perp \mult a)$. 

  Now suppose $a\geq 0$ is arbitrary and satisfies $\omega(a)=0$.
  By Proposition~\ref{prop:spectraltheorem} we have $a=\lim_n a_n$ where the $a_n\leq a$ are positive, simple and converge in the norm to $a$. Hence $0\leq \omega(a_n) \leq \omega(a) = 0$ and thus $\omega(p\mult a_n) = \omega(p^\perp \mult a_n)$.
  Write $f:V\rightarrow \R$ for the norm-continuous map $f(v) := \omega(p\mult v - p^\perp \mult v)$ so that $f(a_n)=0$. By continuity then also $f(a) = 0$, which gives the desired result.
\end{proof}

\begin{definition}
	Let $W$ be an order unit space, and let $\delta: W\rightarrow W$ be a bounded linear map. We call $\delta$ an \Define{order derivation}\indexd{order derivation} when $e^{t\delta} := \sum_{n=0}^\infty \frac{(t\delta)^n}{n!}$ is an order isomorphism for all $t\in \R$.
\end{definition}

\begin{proposition}[{\cite[Proposition 1.108]{alfsen2012state}}]
	Let $W$ be a complete order unit space, and let $\delta: W\rightarrow W$ be a bounded linear map. Then $\delta$ is an order derivation if and only if for all $a\in W^+$ and states $\omega:W\rightarrow \R$ the following implication holds: $\omega(a) = 0 \implies \omega(\delta (a)) = 0$.
\end{proposition}

\begin{proposition}
	Let $p\in E$ be idempotent. Then the map $D_p:V\rightarrow V$ given by $D_p(v) = p\mult v - p^\perp \mult v$ is an order derivation.
\end{proposition}
\begin{proof}
	Immediate consequence of Lemma~\ref{lem:SEAquadratic-orderderivation} and the previous proposition.
\end{proof}

We introduce some additional notation. For any $a\in V^+$ we write $L_a:V\rightarrow V$ for the map $L_a(v) = a\mult v$. For any state $\omega:V\rightarrow \R$ and positive map $f:V\rightarrow V$ we write $f^*\omega$ for the state given by $f^*\omega(v) = \omega(f(v))$. Finally, for linear maps $f,g:V\rightarrow V$ we write $[f,g]:V\rightarrow V$ for their commutator $[f,g] = f\circ g - g\circ f$.

\begin{proposition}[{\cite[Proposition 1.114]{alfsen2012state}}]\label{prop:orderderivationcommutator}
	Let $W$ be a complete order unit space with order derivations $\delta_1, \delta_2:W\rightarrow W$. Then $[\delta_1,\delta_2]$ is also an order derivation.
\end{proposition}

\begin{definition}
	Let $p\in E$ be idempotent. Define $T_p:V\rightarrow V$ by 
  $$T_p := \frac12 (\id + D_p) = \frac12 (\id + L_p - L_{p^\perp}).$$
\end{definition}

\begin{proposition}[{\cite[Theorem 9.48]{alfsen2012geometry}}]\label{prop:SEA-sharpJordancommute}
	Let $p,q \in E$ be idempotent. Then $T_p q = T_q p$.
\end{proposition}
\begin{proof}
	We reformulate the proof of Ref.~\cite[Theorem 9.48]{alfsen2012geometry} for completeness sake.

	Because $T_p 1 = p$ and $T_q 1 = q$ we see that the statement is equivalent to $[T_p, T_q] 1 = 0$. This again is easily seen to be equivalent to $[D_p, D_q] 1 = 0$, since $\id$ commutes with all maps.

	By Proposition~\ref{prop:orderderivationcommutator} $[D_p,D_q]$ is an order derivation, hence $e^{t[D_p,D_q]} 1$ must be a positive operator for all $t\in \R$. Supposing that $[D_p, D_q]^2 1 = 0$, then we would have for all $t\in \R$: $0\leq e^{t[D_p,D_q]}1 = 1 + t[D_p,D_q]1$ as the higher order terms disappear. But obviously this can only hold for all $t$ iff $[D_p, D_q]1=0$, which is the desired result. Hence, it suffices for us to prove that $[D_p,D_q]^2 1 = 0$.

	Define $E_p := D_p^2 = L_p + L_{p^\perp}$, and similarly $E_q := D_q^2$. We then easily calculate:
	\begin{align*}
	[D_p,D_q]^2 1 &= D_pD_qD_pD_q 1 + D_qD_pD_qD_p 1 - D_pD_q^2 D_p 1 - D_qD_p^2D_p 1 \\
	&= D_pD_qD_pD_q 1 + D_qD_pD_qD_p 1 - D_pE_q D_p 1 - D_qE_pD_p 1 \\
	&= D_pD_qD_pD_q 1 + D_qD_pD_qD_p 1 - D_pE_q D_p E_q1 - D_qE_pD_p E_p 1 \\
	&= D_p(D_qD_pD_q-E_q D_p E_q) 1 + D_q(D_pD_qD_p - E_pD_p E_p)1
	\end{align*}
	where going to the third line we used $E_p 1 = 1$ and $E_q 1 = 1$. It hence suffices to prove that $D_qD_pD_q = E_q D_p E_q$ (by symmetry between $p$ and $q$, the second one then also follows).

	We note that for any $a\in V^+$ and state $\omega$ we have $L_{q^\perp}^*\omega(L_q a) = \omega(L_{q^\perp} L_q a) = 0$. Hence by applying Lemma~\ref{lem:SEAquadratic-orderderivation} with $a:=L_q a$ and $\omega:=L_{q^\perp}^*\omega$ we get $L_{q^\perp}^*\omega(D_p L_q a) = 0$. But seeing as $a$ and $\omega$ are arbitrary, this implies $L_{q^\perp} D_p L_q = 0$. Similarly we also get $L_q D_p L_{q^\perp} = 0$. By expanding the definition of $D_q$ we then see that $D_q D_p D_q = L_q D_p L_q + L_{q^\perp} D_p L_{q^\perp}$ and similarly by expanding $E_q = L_q + L_{q^\perp}$ we have $E_q D_p E_q = L_q D_p L_q + L_{q^\perp} D_p L_{q^\perp}$ so that indeed	$D_q D_p D_q = E_q D_p E_q$ as desired.
\end{proof}

\begin{lemma}
	Let $a=\sum_{i=1}^n \lambda_i p_i = \sum_{j=1}^m \mu_j q_j$ be two decompositions of a simple element $a\in V$. Then the maps $\sum_i^n \lambda_i T_{p_i}$ and $\sum_j^m \mu_j T_{q_j}$ coincide.
\end{lemma}
\begin{proof}
	Let $r\in E$ be idempotent. Then $\sum_i^n \lambda_i T_{p_i} r = \sum_i^n \lambda_i T_r p_i = T_r (\sum_i^n \lambda_i p_i) = T_r a = T_r (\sum_j^m \mu_j q_j) = \sum_j^m \mu_j T_{q_j} r$. Since the linear span of the idempotents $r$ lie dense in $V$, the two maps must indeed coincide.
\end{proof}

This lemma ensures that the following is well-defined.
\begin{definition}
	Let $a\in V_0$ be simple with a decomposition $a=\sum_i^n\lambda_i p_i$. Then we define $T_a := \sum_i^n \lambda_i T_{p_i}$. We write $a*b := T_a b$.
\end{definition}

\begin{proposition}
	The operation $*$ is a Jordan product on $V_0$. Furthermore, $a*a = a^2$, so that if $-1\leq a \leq 1$, then $0\leq a*a\leq 1$.
\end{proposition}
\begin{proof}
	Let $a,b \in V_0$ be simple with decompositions $a=\sum_i \lambda_i p_i$ and $b=\sum_j \mu_j q_j$. Then $a*b := T_a b = \sum_{i,j} \lambda_i\mu_j T_{p_i} q_j = \sum_{i,j} \lambda_i\mu_j T_{q_j} p_i = b*a$ by Proposition~\ref{prop:SEA-sharpJordancommute}. Since the operation $*$ is obviously linear in the second argument, it is then also linear in the first argument.

	To show that $a*a = a^2 = \sum_i \lambda_i^2 p_i$ we remark that $T_{p_i} p_j = \delta_{ij} p_i$, from which it easily follows. Note furthermore that because $p_i\commu p_j$ for all $i$ and $j$ that $T_{p_i} T_{p_j} = T_{p_j}T_{p_i}$. Recall that the Jordan identity can be cast as $[T_{a*a}, T_a] = 0$. This easily follows from the commutation of the $T_{p_i}$ and $T_{p_j}$.
\end{proof}

\begin{theorem}\label{thm:normalSEAisJB}
	Let $E$ be a convex $\omega$-SEA and suppose its sequential product is comprehensive and quadratic. Then it is the unit interval of a bounded $\omega$-directed-complete JB-algebra.
\end{theorem}
\begin{proof}
	Recall that a norm-complete order unit space $V$ is a JB-algebra iff it has a Jordan product $*$ where for any $a\in V$ with $-1\leq a \leq 1$ we have $0\leq a*a \leq 1$ (Proposition~\ref{prop:JB-define-as-OUS}). By the previous proposition, the space of simple elements $V_0$ has a Jordan product satisfying this condition. Since the simple elements lie dense in $V$, we can extend the product by continuity to the whole space. For the details we refer to \cite[Theorem 9.43]{alfsen2012geometry}.
\end{proof}

\begin{remark}
  It is unclear whether the theorem above continues to hold when the assumptions of the sequential product being comprehensive and quadratic are dropped. However it is not possible to drop the requirement of $\omega$-completeness: Example~\ref{ex:weird-seq-prod} is a convex SEA that is not $\omega$-complete. As its only idempotents are $0$ and $1$, it is trivially comprehensive and quadratic, but it is obviously not equal to the unit interval of a Jordan algebra.
\end{remark}

The property of being quadratic does not seem like a very natural condition.
It turns out that in a more restricted setting we can `trade it in' for a more natural set of conditions.

\begin{definition}
	Let $E$ be a SEA. We say it is \Define{atomic} when beneath every sharp effect there is an atomic effect.
\end{definition}

\begin{definition}
	Let $E$ be a convex $\omega$-SEA and let $V$ be its associated order unit space. We say $E$ has \Define{finite bits} when for every pair of atoms $p,q\in E$ their order ideal (cf.~Definition~\ref{def:orderideal}) $V_{p,q}$ is finite-dimensional.
\end{definition}

\begin{theorem}
	Let $E$ be an atomic convex $\omega$-SEA with finite bits and a norm-continuous sequential product. Then $E$ is the unit interval of an atomic bounded $\omega$-directed-complete JB-algebra.
\end{theorem}
\begin{proof}
	Because the sequential product is norm-continuous the results of Section~\ref{sec:atomeffect} continue to hold for $E$. In particular, for an atom $p$ and an arbitrary effect $a$, the effect $a\mult p$ is proportional to an atom (Proposition~\ref{prop:atompreservation}).
	As a result, the lattice of sharp effects of $E$ has the finite covering property (Section~\ref{sec:coverprop}), so that (via Corollary~\ref{cor:rank} and Lemma~\ref{lem:order-ideal-strictly-convex}) the order ideal $V_{p,q}$ of any pair of atoms $p,q\in E$ is strictly convex. As $V_{p,q}$ is also a convex $\omega$-SEA, it is homogeneous (Proposition~\ref{prop:homogen-infinite}).
	By the assumption of finite bits it is then a homogeneous strictly convex finite-dimensional space, and hence by Proposition~\ref{prop:itochar} it is order-isomorphic to a spin-factor. 
	Hence, following the results of Section~\ref{sec:subselfdual} we see that $E$ satisfies symmetry of transition probabilities.

	Let $E_f$ denote the SEA generated by finite linear combinations of the atoms of $E$ and let $V_f$ be its associated OUS. By symmetry of transition probabilities we can define a self-dual inner product on $V_f$. Then in much the same way as in Section~\ref{sec:jordanproduct} we can show $V_f$ is a Jordan algebra. Now letting $V_0$ denote the norm-closure of $V_f$ we can extend the Jordan product by continuity so that $V_0$ becomes a JB-algebra (for the details see Ref.~\cite[Prop.~9.30]{alfsen2012geometry}). It then remains to extend this Jordan product to the entire space $V$. This process is described in Theorem~9.38 of~\cite{alfsen2012geometry}.
\end{proof}

\begin{remark}
	The requirement of `finite bits' is needed because it is unknown whether the characterisation of Proposition~\ref{prop:itochar} continues to hold in infinite dimension. If it does hold, then this requirement is not needed. Alternatively, we can replace this requirement by the assumption that all order ideals $V_{p,q}$ are order-isomorphic to (possibly infinite-dimensional) spin-factors. This is similar to the \emph{Hilbert ball property} of~\cite[Definition~9.9]{alfsen2012geometry}.
	With this assumption, the requirement that the sequential product be norm-continuous is also no longer necessary.
\end{remark}

\section{From effect theories to Jordan algebras}\label{sec:Jordan-from-effectus}

We will now see how we can combine the results from Section~\ref{sec:effect-monoids} and~\ref{sec:SEA} to reconstruct infinite-dimensional quantum theory.

First, we remark that if we want to represent any non-classical effects in an $\omega$-effect theory that we \emph{need} to use real numbers.

\begin{definition}
  Let $\mathbf{E}$ be an effect theory. We say its scalars are \Define{trivial} when $\eff(I) = \{0\}$ or $\eff(I) = \{0,1\}$. We say its scalars are \Define{reducible} when $\eff(I) = M_1\oplus M_2$ for some non-zero effect monoids $M_1$ and $M_2$. If the scalars are not reducible we say the scalars are \Define{irreducible}.
\end{definition}

\begin{theorem}\label{thm:effect-theory-scalars}
    Let $\mathbf{E}$ be an $\omega$-effect theory where the scalars are irreducible and where the states separate the effects (\ie~$a\circ \omega=b\circ \omega$ for all $\omega\in\st(A)$ implies $a=b$). Then exactly one of the following is true:
    \begin{enumerate}
        \item The only scalar is $0=1$, and for every system $A$, $\eff(A) = \{0\}$.
        \item There are exactly two scalars $0$ and $1$, and for every system $A$ there is a Boolean algebra $B_A$ and an effect algebra morphism $\phi_A:\eff(A)\rightarrow B_A$ that is injective.
        \item The scalars are the real unit interval, and for every system $A$, $\eff(A) = [0,1]_{V_A}$ for some complete bounded $\omega$-directed-complete order unit space $V_A$.
    \end{enumerate}
\end{theorem}
\begin{proof}
    The scalars $M = \eff(I)$ of an effect theory form an effect monoid, and hence in an $\omega$-effect theory they form an $\omega$-complete effect monoid. Since the scalars are irreducible, by Theorem~\ref{thm:irreducible-effect-monoid} we must have $M=\{0\}$, $M=\{0,1\}$ or $M=[0,1]$.

    Suppose $M=\{0\}$. Let $A$ be any system and let $p\in \eff(A)$ be an effect. We have $p = 1\cdot p = 0\cdot p = 0$, since $0=1$ in $M$.

    Suppose $M = \{0,1\}$. Let $A$ be any system. Let $B_A := P(\st(A))$, the powerset of states of $A$. Define $\phi_A:\eff(A)\rightarrow B_A$ by $\phi_A(a) := \{\omega\in \st(A)~;~a\circ\omega=1\}$. Then it is straightforward to verify that $\phi_A$ is an effect algebra morphism. It is injective because the states separate the effects.

    Finally, if $M=[0,1]$, then for any system $A$, its effect space $\eff(A)$ is a convex effect algebra. By the results of Section~\ref{sec:convexeffectalgebrasandOUS} it is then isomorphic to the unit interval of a bounded $\omega$-directed-complete (and hence norm-complete) order unit space.
\end{proof}

Hence, we see that if the scalars in an $\omega$-effect-theory are trivial, that then all effect spaces embed into Boolean algebras. Hence, for the purpose of presenting non-classical theories, these are not interesting. The theorem furthermore shows that the only possible set of non-trivial irreducible scalars is the real unit interval $[0,1]$ and that this automatically leads to a representation of effects as elements of a complete order unit space.

Now let us define the type of effect theory we will use to reconstruct infinite-dimensional quantum theory. We combine the assumptions of a pure effect theory (Definition~\ref{def:PET}), sequential effect space (Definition~\ref{def:seqprod}), together with some of the new structure we found was present in JBW-algebras (Theorem~\ref{thm:unique-diamond-positivity}).
Recall that a map $f:A\rightarrow A$ in a dagger-category is $\dagger$-positive when there exists a map $g:A\rightarrow B$ such that $f = g^\dagger \circ g$.

\begin{definition}\label{def:sequential-effect-theory}
  A (monoidal) \Define{sequential} $\omega$-effect theory is a (monoidal) $\omega$-effect theory additionally satisfying the following conditions.
  \begin{enumerate}
    \item Every effect has a filter and a compression.
    \item Every pure map has an image.
    \item The pure maps form a (monoidal) dagger category. In the monoidal setting the maps satisfy in particular $(f\otimes g)^\dagger = f^\dagger\otimes g^\dagger$.
    \item Every pure map $f$ is $\diamond$-adjoint to $f^\dagger$.
    \item For every effect $a\in \eff(A)$ there is a unique $\dagger$-positive map $\asrt_a: A\rightarrow A$ satisfying $1\circ \asrt_a = a$ called the \Define{assert map} of $a$.
    \item For every system $A$, the operation $\&: \eff(A)\times \eff(A)\rightarrow \eff(A)$ given by $a\& b := b\circ \asrt_a$ is a normal sequential product, making $\eff(A)$ into an $\omega$-SEA.
  \end{enumerate}
\end{definition}

\begin{remark}
	Points 1--3 correspond to \ref{pet:filtcompr}--\ref{pet:images} in a PET. Point 4 takes over the role of \ref{pet:sharpadjoint} and \ref{pet:sharpisometry}. Point 6 is a variation on the axioms of the sequential product in Definition~\ref{def:seqprod}. Point 5 is a new assumption. We remark that the uniqueness condition can be framed as the implication $1\circ f^\dagger\circ f = 1\circ g^\dagger\circ g \implies f^\dagger\circ f = g^\dagger\circ g$ for any pure $f$ and $g$. In this sense it is a weaker version of the \emph{CPM axiom}~\cite{coecke2010environment} for an \emph{environment structure} (\ie~choice of pure maps) which states that $1\circ f^\dagger\circ f = 1\circ g^\dagger\circ g \implies f = g$.
\end{remark}

Denote by $\omega\text{-}\textbf{JB-alg}_{\text{psu}}$ the full subcategory of $\textbf{JB-alg}_{\text{psu}}$ consisting of the bounded $\omega$-directed-complete JB-algebras.
Our goal of this section will be to prove the following theorem.

\begin{theorem}\label{thm:JB-embedding}
  Let $\mathbb{E}$ be a sequential $\omega$-effect theory with non-trivial irreducible scalars. Then there is a functor $F:\mathbb{E}\rightarrow \omega\text{-}\textbf{JB-alg}_{\text{psu}}^\opp$ satisfying $F(\eff(A))\cong [0,1]_{F(A)}$. This functor is faithful if and only if $\mathbb{E}$ satisfies local tomography.
\end{theorem}

For the remainder of the section, we will let $\mathbb{E}$ be a sequential $\omega$-effect theory with non-trivial irreducible scalars, and we let $A$ denote an arbitrary system in $\mathbb{E}$. By Theorem~\ref{thm:effect-theory-scalars}, the scalars of $\mathbb{E}$ are equal to the real unit interval $[0,1]$, and we can associate an $\omega$-complete order unit space $V_A$ to $A$ such that $\eff(A)\cong [0,1]_{V_A}$. It then remains to show that these order unit spaces are JB-algebras. By assumption, $\eff(A)$ is a convex $\omega$-SEA, and hence by Theorem~\ref{thm:normalSEAisJB} it suffices to show that the sequential product given by the assert maps is comprehensive and quadratic. 

For any effect $a$ we will write $a^2 := a\mult a := a\circ \asrt_a$.
\begin{lemma}
  For all effects $a\in \eff(A)$ we have
	$\asrt_a^2 = \asrt_{a^2}$.
\end{lemma}
\begin{proof}
	By definition $\asrt_a$ is $\dagger$-positive, and hence so is $\asrt_a^2$. We have $1\circ \asrt_a^2 = a\circ \asrt_a = a^2 = 1\circ \asrt_{a^2}$ so that by the uniqueness of $\dagger$-positive maps: $\asrt_a^2 = \asrt_{a^2}$.
\end{proof}

\begin{lemma}
  For any effect $a\in \eff(A)$ there is a unique $\sqrt{a}\in \eff(A)$ with $\sqrt{a}^2 = a$.
\end{lemma}
\begin{proof}
By assumption, $\eff(A)$ is an $\omega$-SEA. As it is also convex, we can use the spectral theorem of Proposition~\ref{prop:spectraltheorem} to conclude that for every $a\in \eff(A)$ there is a unique effect $\sqrt{a}\in\eff(A)$ satisfying $\sqrt{a}^2 := \sqrt{a}\mult \sqrt{a}:=\sqrt{a}\circ\asrt_{\sqrt{a}} = a$.
\end{proof}

\begin{lemma}
	The assert maps $\asrt_a$ are $\diamond$-positive, \ie~there exists a $\diamond$-self-adjoint map $f$ such that $\asrt_a = f\circ f$.
\end{lemma}
\begin{proof}
  By assumption any pure map $f$ is $\diamond$-adjoint to $f^\dagger$.
	Assert maps are $\dagger$-positive and so in particular are $\dagger$-self-adjoint, so that they are also $\diamond$-self-adjoint. The desired result now follows because $\asrt_a = \asrt_{\sqrt{a}}^2$.
\end{proof}

\begin{proposition}
  An effect $p\in \eff(A)$ is sharp in the sense of an effect theory if and only if $p^2 = p$. As a result, $p$ is sharp iff $p^\perp$ is sharp.
\end{proposition}
\begin{proof}
  If $p^2=p$ then by definition we have $p\circ \asrt_p = 1\circ \asrt_p$ and hence $p\geq \im{\asrt_p} = (\asrt_p)_\diamond(1) = (\asrt_p)^\diamond(1) = \ceil{1\circ\asrt_p} = \ceil{p} \geq p$ so that $p=\ceil{p}$, and hence $p$ is sharp.

  Conversely, if $p=\ceil{p}$, then $p = \im{\asrt_p}$ so that $p^2 = p\circ \asrt_p = 1\circ \asrt_p = p$.
\end{proof}

\begin{corollary}
  A sequential $\omega$-effect-theory is a $\diamond$-effect-theory (Definition~\ref{def:diamond-effect-theory}).
\end{corollary}

The next lemma shows that the assert maps of sharp effects are of the same form as in Section~\ref{sec:firsttwo}.

\begin{lemma}[{\cite[211VII]{basthesis}}]\label{lem:compatible-filter-compression}
  For any compression $\pi_p$ of a sharp effect $p$ there exists a filter $\xi_p$ of $p$ such that $\xi_p \circ \pi_p = \id$ and $\pi_p\circ \xi_p = \asrt_p$.
\end{lemma}
\begin{proof}
  Let $p$ be a sharp effect. As $\asrt_p$ is pure we have $\asrt_p = \pi\circ\xi$ for some compression $\pi$ and filter $\xi$. Now, $\pi\circ \xi = \asrt_p = \asrt_p\circ\asrt_p = \pi\circ \xi\circ\pi\circ \xi$ so that $\xi\circ\pi = \id$ (cf.~Proposition~\ref{prop:faithfulfilters}). We see that $1\circ\xi = 1\circ\pi\circ\xi = 1\circ\asrt_p = p$ so that $\xi$ is a filter for $p$. Furthermore $p=\im{\asrt_p} =\im{\pi\circ\xi}\leq \im{\pi}$ and $\im{\pi} = \im{\pi\circ\xi\circ\pi} = \im{\asrt_p\circ\pi} \leq \im{\asrt_p} = p$ so that $\im{\pi} = p$ and hence $\pi$ is a compression for $p$.

  Now let $\pi'$ be a another compression for $p$. Then $\pi' = \pi\circ\Theta$ for some isomorphism $\Theta$. Define $\xi' := \Theta^{-1}\circ\xi$. Then $\pi'\circ\xi' = \pi\circ\xi = \asrt_p$ and $\xi'\circ\pi' = \id$.
\end{proof}
The above lemma implies that $\mathbb{E}$ has compatible filters and compressions as defined in Definition~\ref{def:compatible-filters-compressions}.

\begin{proposition}[{\cite[216VII]{basthesis}}]\label{prop:dagger-of-compression}
	Let $p$ be a sharp effect. Then $\pi_p^\dagger = \xi_p$, where $\pi_p$ and $\xi_p$ form a pair of a compression and a filter of $p$ with $\xi_p\circ \pi_p = \id$ and $\pi_p\circ\xi_p = \asrt_p$.
\end{proposition}
\begin{proof}
	By assumption $\pi_p$ is $\diamond$-adjoint to $\pi_p^\dagger$. Hence $\ceil{1\circ \pi_p^\dagger} = (\pi_p^\dagger)^\diamond(1) = (\pi_p)_\diamond(1) = \im{\pi_p} = p$. Similarly we calculate $\im{\xi_p^\dagger} = p$. By the universal property of filters respectively compressions there are then unique maps $h$ and $g$ such that $\pi_p^\dagger = h\circ\xi_p$ and $\xi_p^\dagger = \pi_p\circ g$. Using $\xi_p\circ\pi_p = \id$ twice we calculate $\id = \id^\dagger = \pi_p^\dagger \circ \xi_p^\dagger = h\circ\xi_p \circ \pi_p \circ g = h\circ g$. As a result $1=1\circ \id = 1\circ h\circ g \leq 1\circ g$ so that $g$ is unital, and hence $1\circ \xi_p^\dagger = 1\circ\pi_p\circ g = 1$.

  By uniqueness of $\dagger$-positive maps we have $\xi_p^\dagger\circ \xi_p = \asrt_{1\circ \xi_p^\dagger\circ \xi_p} = \asrt_{1\circ \xi_p} = \asrt_p = \pi_p\circ \xi_p$. Because $\xi_p$ is epic we conclude that indeed $\xi_p^\dagger = \pi_p$.
\end{proof}

\begin{corollary}\label{cor:dagger-of-iso}
	Let $\Theta$ be an isomorphism. Then $\Theta^\dagger = \Theta^{-1}$.
\end{corollary}
\begin{proof}
	$\Theta$ is a filter for $1$, which is sharp, and $\Theta^{-1}$ is a compression for $1$. As $\Theta^{-1}\circ \Theta = \id$ and $\Theta\circ \Theta^{-1} = \id = \asrt_1$, they satisfy the conditions of the previous proposition.
\end{proof}

\begin{lemma}[{\cite[212III]{basthesis}}]\label{lem:pure-decomposition}
	Every pure map $f$ factors as $f=\pi_{\im{f}}\circ \Theta\circ \xi_{\ceil{1\circ f}}\circ \asrt_{1\circ f}$ where $\Theta$ is an isomorphism.
\end{lemma}
\begin{proof}
  Combining Proposition~\ref{prop:effect-decomposition-of-maps} and Corollary~\ref{cor:pure-decomp} we see that $f=\pi_{\im{f}}\circ\Theta\circ \xi_{1\circ f}$. It hence remains to show that $\xi_{\ceil{1\circ f}}\circ \asrt_{1\circ f}$ is a filter for $1\circ f$. 

  Write $a=1\circ f$. Note first of all that $1\circ (\xi_{\ceil{a}}\circ \asrt_a) = \ceil{a}\circ\asrt_a = 1\circ \asrt_{\ceil{a}}\circ \asrt_a = 1\circ \asrt_a = a$ so that it remains to show that $\xi_{\ceil{a}}\circ \asrt_a$ is a filter. We see that $\im{\xi_{\ceil{a}}\circ \asrt_a} = (\xi_{\ceil{a}})_\diamond((\asrt_a)_\diamond(1)) = (\xi_{\ceil{a}})_\diamond(\ceil{a}) = \im{\xi_{\ceil{a}}\circ\pi_{\ceil{a}}} = \im{\id} = 1$.
  Being a composition of pure maps, $\xi_{\ceil{a}}\circ \asrt_a$ is a pure map itself, and hence is equal to $\pi\circ\xi$ for some compression $\pi$ and filter $\xi$. Now we calculate $1=\im{\xi_{\ceil{a}}\circ \asrt_a} = \im{\pi\circ\xi} \leq \im{\pi}$ so that $\im{\pi}=1$ and hence $\pi$ is an isomorphism. We conclude that $\xi_{\ceil{a}}\circ \asrt_a$ is a filter.
\end{proof}

\begin{proposition}[{\cite[216XIII]{basthesis}}]
	Let $a$ and $b$ be arbitrary effects on the same system. Then $\asrt_{a\& b}^2 = \asrt_a \circ \asrt_b^2 \circ \asrt_a$.
\end{proposition}
\begin{proof}
  Note that $1\circ \asrt_b\circ\asrt_a = a\mult b$ and 
  $$\im{\asrt_b\circ \asrt_a} = (\asrt_b)_\diamond(\ceil{a}) = (\asrt_b)^\diamond(\ceil{a}) = \ceil{b\mult \ceil{a}} \stackrel{\ref{prop:floorceiling}.d)}{=} \ceil{b\mult a}.$$ 
  Using Lemma~\ref{lem:pure-decomposition} we then get 
  $\asrt_b\circ\asrt_a = \pi_{\ceil{b\circ a}}\circ \Theta\circ\xi_{\ceil{a\mult b}}\circ \asrt_{a\mult b}$
   for some isomorphism $\Theta$. Applying the dagger to both sides and using Proposition~\ref{prop:dagger-of-compression} and Corollary~\ref{cor:dagger-of-iso} gives us:
   \begin{align*}
   \asrt_a\circ\asrt_b &= (\asrt_b\circ \asrt_a)^\dagger 
   = \asrt_{a\mult b}\circ \xi_{\ceil{a\mult b}}^\dagger \circ \Theta^\dagger \circ \pi_{\ceil{b\circ a}}^\dagger \\
   &= \asrt_{a\mult b}\circ \pi_{\ceil{a\mult b}} \circ \Theta^{-1} \circ \xi_{\ceil{b\circ a}}.
   \end{align*}
   We calculate:
   \begin{align*}
    \asrt&_a\circ\asrt_b^2\circ \asrt_a \\
    &=\asrt_{a\mult b}\circ \pi_{\ceil{a\mult b}} \circ \Theta^{-1} \circ \xi_{\ceil{b\circ a}}\circ \pi_{\ceil{b\circ a}}\circ \Theta\circ\xi_{\ceil{a\mult b}}\circ \asrt_{a\mult b} \\
    &= \asrt_{a\mult b}\circ \pi_{\ceil{a\mult b}}\circ\xi_{\ceil{a\mult b}}\circ \asrt_{a\mult b} & \text{Lem.~\ref{lem:compatible-filter-compression}}\\
    &= \asrt_{a\mult b}\circ \asrt_{\ceil{a\mult b}}\circ \asrt_{a\mult b} \\
    &=\asrt_{a\mult b}\circ \asrt_{a\mult b} = \asrt_{(a\mult b)^2}. & \text{Prop.~\ref{prop:assertmaps}.e)}
   \end{align*}
   And hence we are done.
\end{proof}

Now we can conclude that the sequential product in $\eff(A)$ is quadratic (cf.~Definition~\ref{def:SEA-quadratic})

\begin{corollary}\label{cor:assert-is-quadratic}
	Let $p$ and $q$ be sharp effects. Then $(p\& q)^2 = p\&(q\& p)$.
\end{corollary}
\begin{proof}
	Just plug $1$ into the expression of the previous proposition and use $\asrt_q^2 = \asrt_q$ for sharp $q$.
\end{proof}

\begin{proof}[Proof of Theorem~\ref{thm:JB-embedding}]
  $\mathbb{E}$ is an $\omega$-effect-theory with irreducible scalars, and hence the scalars are equal to $\{0\}$, $\{0,1\}$ or $[0,1]$. Since we further assume that the scalars are non-trivial we then necessarily have $\eff(I)=[0,1]$. Following Theorem~\ref{theor:opefftheor} we see that there is then a functor $F:\mathbb{E}\rightarrow \OUS^\opp$ with $\eff(A) \cong [0,1]_{F(A)}$ which is faithful if and only if $\mathbb{E}$ satisfies local tomography. It remains to show that these order unit spaces are actually bounded $\omega$-directed-complete JB-algebras.

  For every system $A$ we know that $\eff(A)\cong [0,1]_{V_A}$ has a normal sequential product. This sequential product is quadratic by Corollary~\ref{cor:assert-is-quadratic}. Furthermore, for any state $\omega$, if $p\circ \omega = 1$, then $\im{\omega}\leq p$ and hence $\asrt_p\circ \omega = \omega$ by Proposition~\ref{prop:assertmaps}.d). As a result, $(p\mult a)\circ \omega := a\circ\asrt_p\circ\omega = a\circ \omega$, so that the sequential product is compressible (Definition~\ref{def:SEA-compressible}). Hence by Theorem~\ref{thm:normalSEAisJB}, the desired result follows.
\end{proof}

It is currently not clear whether the converse also holds. Namely, whether the category $\omega\text{-}\textbf{JB-alg}_{\text{psu}}^\opp$ is a sequential $\omega$-effect-theory. We have however seen in Chapter~\ref{chap:jordanalg} that the category of JBW-algebras does indeed satisfy all the assumptions. Let us therefore strengthen some of the conditions used in a sequential $\omega$-effect-theory in order to get a more natural correspondence.

\begin{definition}\label{def:sequential-complete-effect-theory}
  Let $\mathbb{E}$ be a (monoidal) sequential $\omega$-effect theory. We say it is \Define{complete} when it satisfies the following additional requirements.
  \begin{enumerate}
    \item Every effect space is directed-complete.
    \item Every map is normal.
    \item The states separate the effects.
  \end{enumerate}
\end{definition}

\begin{theorem}\label{thm:JBW-embedding}
  Let $\mathbb{E}$ be a complete sequential $\omega$-effect theory with non-trivial irreducible scalars. Then there is a functor $F:\mathbb{E}\rightarrow \textbf{JBW-alg}_{\text{npsu}}^\opp$ satisfying $F(\eff(A))\cong [0,1]_{F(A)}$. This functor is faithful if and only if $\mathbb{E}$ satisfies local tomography.
\end{theorem}
\begin{proof}
  By Theorem~\ref{thm:JB-embedding} we already know that there exists an appropriate functor $F$ mapping each system $A$ of $\mathbb{E}$ to a JB-algebra $F(A)$. It remains to show that these JB-algebras are in fact JBW-algebras. But by assumption $[0,1]_{F(A)}$ is directed-complete and hence $F(A)$ is bounded directed-complete. Furthermore, we assume that every map is normal, and that the states separate the effects. Hence, there is a separating set of normal states. So $F(A)$ is indeed a JBW-algebra.
\end{proof}

\begin{remark}
  The assumptions we make in Definitions~\ref{def:sequential-effect-theory} and~\ref{def:sequential-complete-effect-theory} are not minimal. We have framed the definitions in this manner so that each point corresponds to one conceptual assumption. For instance, the first three axioms of an $\omega$-SEA from Definition~\ref{def:normal-SEA} are satisfied in any effect theory satisfying the first five axioms of Definition~\ref{def:sequential-effect-theory}, while the existence of images of pure maps (point 2 of Definition~\ref{def:sequential-effect-theory}) already follows from the other assumptions in a complete sequential $\omega$-effect-theory.
\end{remark}

\section{From effect theories to von Neumann algebras}\label{sec:neumann-from-effectus}

Recall that in Section~\ref{sec:opeffect} we used axioms from effectus theory to derive the structure of a Jordan algebra, and that we then added a tensor product in Section~\ref{sec:monoidalPETs} to restrict the Jordan algebras to just C$^*$-algebras. Analogously, in the previous section we used axioms from effectus theory to see that our systems had to be JBW-algebras, while in this section we will see what further restrictions the addition of a tensor product implies. The main result of this section will be that the existence of such a tensor product forces all JBW-algebras in the effect theory to be JW-algebras, and hence all the systems have an underlying von Neumann algebra.

For the remainder of this section we let $\mathbb{E}$ be a complete \emph{monoidal} sequential $\omega$-effect-theory with non-trivial irreducible scalars. Hence, Theorem~\ref{thm:JBW-embedding} applies and to all systems $A$ and $B$ of $\mathbb{E}$ we can associate JBW-algebras $V_A$ and $V_B$ such that $\eff(A)\cong [0,1]_{V_A}$ and $\eff(B)\cong [0,1]_{V_B}$. Furthermore, since $\mathbb{E}$ is assumed to be monoidal, we have a system $A\otimes B$ with a corresponding JBW-algebra $V_{A\otimes B}$. Recall that by the definition of a monoidal effect theory (Definition~\ref{def:monoidal-effect-theory}) we have $1\otimes 1 = 1$ and $(a\ovee b)\otimes c = (a\otimes c)\ovee (b\otimes c)$. By linearity, the tensor products of effects extends to a bilinear map $V_A\times V_B\rightarrow V_{A\otimes B}$ that we will denote with a tensor product symbol `$\otimes$' as well.

\begin{proposition}\label{prop:JBW-tensor-preserves-assert}
  Let $a\in \eff(A)$ and $b\in \eff(B)$. Then $\asrt_{a\otimes b} = \asrt_a\otimes \asrt_b$.
\end{proposition}
\begin{proof}
  Note $1\circ (\asrt_a\otimes \asrt_b) = (1\otimes 1)\circ (\asrt_a\otimes \asrt_b) = (1\circ \asrt_a)\otimes (1\circ \asrt_b) = a\otimes b$, so by uniqueness of $\dagger$-positive maps, it remains to show that $\asrt_a\otimes \asrt_b$ is $\dagger$-positive. But this follows because $(\asrt_{\sqrt{p}}\otimes \asrt_{\sqrt{q}})^2 = \asrt_{\sqrt{p}}^2 \otimes \asrt_{\sqrt{q}}^2 = \asrt_p\otimes \asrt_q$.
\end{proof}

\begin{corollary}
  Let $a\in \eff(a)$ and $1\in \eff(B)$. Then $\asrt_{a\otimes 1} = \asrt_a\otimes \id$.
\end{corollary}

\begin{corollary}
  Let $a\in \eff(A)$ and $b\in \eff(B)$. Then
    $a^2\otimes b^2 = (a\otimes b)^2$.
\end{corollary}
\begin{proof}
    $a^2\otimes b^2 = (a\circ \asrt_a)\otimes (b\circ \asrt_b) = (a\otimes b)\circ (\asrt_a\otimes \asrt_b) = (a\otimes b)\circ \asrt_{a\otimes b} = (a\otimes b)^2$.
\end{proof}

\begin{corollary}\label{cor:tensor-idempotents}
  Let $p\in \eff(A)$ and $q\in \eff(B)$ be sharp effects.
	Then $p\otimes q$ is sharp.
\end{corollary}

By Theorem~\ref{thm:unique-diamond-positivity}, the quadratic map $Q_{\sqrt{a}}$ for a positive $a$ is the unique $\diamond$-positive map satisfying $Q_{\sqrt{a}}(1) = a$. Since $\asrt_a$ is also $\diamond$-positive with $1\circ \asrt_a = a$ we must then have $\asrt_a = Q_{\sqrt{a}}$.
Denote by $*$ the Jordan product on $V_A$ and write $T_a(b) = a*b$ for the Jordan product map of $a$ (and similarly for $V_B$). $T_a$ is not a positive map and hence cannot be part of the effect theory. However,
recall that if $p$ is sharp (\ie~idempotent) that then $T_p = \frac12 (\id + Q_p - Q_{p^\perp})$ and hence that it is a linear combination of maps that do lie in the effect theory. Note that additionally $Q_p = \asrt_{p^2} = \asrt_p$.

\begin{proposition}
    Let $a\in V_A$ be arbitrary and $1\in V_B$, then $T_{a\otimes 1} = T_a\otimes \id$. Similarly, for $1\in V_A$ and $b\in V_B$ we have $T_{1\otimes b} = \id \otimes T_b$.
\end{proposition}
\begin{proof}
    We only show the first equation, as the second follows analogously. We prove the result for $a=p$ sharp. By the norm-continuity and linearity of the Jordan product in the first argument, this is sufficient as the sharp elements span a dense set.

    Note first that $(p\otimes 1)^\perp = p^\perp \mult 1$.
    We then calculate:
    \begin{align*}
    T_{p\otimes 1} &= \frac12 (\id\otimes \id + \asrt_{p\otimes 1} - \asrt_{(p\otimes 1)^\perp}) = \frac12 (\id \otimes \id + \asrt_p \otimes \id - \asrt_{p^\perp}\otimes \id) \\
    &= (\frac12 (\id +\asrt_p - \asrt_{p^\perp}))\otimes \id = T_p \otimes \id. \qedhere
    \end{align*}
\end{proof}

\begin{corollary}
    For all $a\in V_A$ and $b\in V_B$, $a\otimes 1$ and $1\otimes b$ operator commute.
\end{corollary}

\begin{corollary}\label{cor:tensor-is-Jordan-hom}
	The maps $a\mapsto a\otimes 1$ and $b\mapsto 1\otimes b$ are Jordan homomorphisms.
\end{corollary}

\begin{proposition}\label{prop:tensor-is-injective}
	The maps $a\mapsto a\otimes 1$ and $b\mapsto 1\otimes b$ are injective.
\end{proposition}
\begin{proof}
	We only show the first, as the second is analogous. Suppose $a\otimes 1 = a'\otimes 1$. Let $\omega$ be any state on the first system, and $\omega'$ any state on the second system. Then $\omega(a) = \omega(a)\omega'(1) = (\omega\otimes \omega')(a\otimes 1) = (\omega\otimes \omega')(a'\otimes 1) = \omega(a')$. Since states separate the effects, we then necessarily have $a=a'$.
\end{proof}

\begin{corollary}
	The maps $a\mapsto a\otimes 1$ and $b\mapsto 1\otimes b$ are normal.
\end{corollary}
\begin{proof}
	Since the maps are injective unital Jordan homomorphisms, the restriction to their domain is an order-isomorphism, and hence the maps must be normal.
\end{proof}

\begin{proposition}\label{prop:tensor-quadratic}
    Let $a\in V_A$ and $b\in V_B$ be arbitrary. Then $Q_{a\otimes b} = Q_a\otimes Q_b$.
\end{proposition}
\begin{proof}
    First suppose $a\in[0,1]_{V_A}$ and $b\in[0,1]_{V_A}$. Then $Q_a = \asrt_{a^2}$ and $Q_b=\asrt_{b^2}$, so that by Proposition~\ref{prop:JBW-tensor-preserves-assert} $Q_{a\otimes b} = \asrt_{(a\otimes b)^2} = \asrt_{a\otimes b}^2 = (\asrt_a\otimes \asrt_b)^2 = \asrt_{a^2}\otimes \asrt_{b^2} = Q_a\otimes Q_b$. Now for an arbitrary positive $a$ of course $Q_a = Q_{\norm{a} a/\norm{a}} = \norm{a}^2 Q_{a/\norm{a}}$, and hence the desired result also follows when $a\geq 0$ and $b\geq 0$.

    Now suppose $a=a_1+a_2$ where $a_1,a_2\geq 0$. Then $Q_a = Q_{a_1} + Q_{a_2} + 2Q_{a_1,a_2}$. We expand $Q_{a\otimes b}$ in two different ways. First we see that $Q_{a\otimes b} = Q_{a_1\otimes b + a_2\otimes b} = Q_{a_1\otimes b} + Q_{a_2\otimes b} + 2 Q_{a_1\otimes b, a_2\otimes b} = Q_{a_1}\otimes Q_b + Q_{a_2}\otimes Q_b + 2 Q_{a_1\otimes b, a_2\otimes b}$.
    Secondly, $Q_{a\otimes b} = Q_a\otimes Q_b = Q_{a_1}\otimes Q_b + Q_{a_2}\otimes Q_b + 2Q_{a_1,a_2}\otimes Q_b$. Comparing terms in both of these decompositions of $Q_{a\otimes b}$ we see that necessarily $Q_{a_1\otimes b, a_2\otimes b} = Q_{a_1,a_2}\otimes Q_b$.

    We can use this equation, and do a similar trick, but starting with $Q_{a_1\otimes b, a_2\otimes b}$ where $b=b_1 + b_2$ to give us the equation $2Q_{a_1,a_2}\otimes Q_{b_1,b_2} = Q_{a_1\otimes b_1, a_2\otimes b_2} + Q_{a_1\otimes b_2, a_2\otimes b_1}$.

    Finally, suppose $a$ and $b$ are arbitrary. Write $a=a^+ - a^-$ and $b=b^+-b^-$ where $a^+,a^-, b^+, b^- \geq 0$. Now if we expand both the expression $Q_{(a^+-a^-)\otimes (b^+-b^-)}$ and $Q_{a^+-a^-} \otimes Q_{b^+ - b^-}$ as much as possible using linearity and apply the previous rewrite rules, it is easily verified that these two expression are indeed equal.
\end{proof}

In order to proceed we need to use the concept of \emph{universal von Neumann algebras}.

\begin{theorem}[{\cite[Theorem~7.1.9]{hanche1984jordan}}]
	Let $V$ be a JBW-algebra. Then there exists an (up to isomorphism) unique von Neumann algebra $W^*(V)$ and a normal Jordan homomorphism $\psi: V\rightarrow W^*(V)_\sa$ such that $\psi(V)$ generates $W^*(V)$ as a von Neumann algebra and if $\mathfrak{B}$ is a von Neumann algebra with a normal Jordan homomorphism $\phi: V\rightarrow \mathfrak{B}_\sa$, then there is a normal *-homomorphism $\hat{\phi}: W^*(V)\rightarrow \mathfrak{B}$ such that $\hat{\phi}\circ \psi = \phi$.
\end{theorem}

\begin{corollary}\label{cor:JW-injective}
	A JBW-algebra $V$ is a JW-algebra if and only if $\psi: V\rightarrow W^*(V)$ is injective.
\end{corollary}
\begin{proof}
	If $\psi$ is injective, then $V$ is of course a JW-algebra. Conversely, if $V$ is a JW-algebra, then there must be an injective normal Jordan homomorphism $\phi:V\rightarrow \mathfrak{B}_\sa$ for some von Neumann algebra $\mathfrak{B}$, and hence by the universal property of $W^*(V)$, $\hat{\phi}\circ \psi = \phi$, which shows that $\psi$ must be injective.
\end{proof}

\begin{definition}
    Let $V$ be a JBW-algebra. We call $s\in V$ a \Define{symmetry} when $s^2 = 1$. Two idempotents $p,q \in V$ are \Define{exchangeable by a symmetry} if there exists a symmetry $s$ such that $Q_s p = q$.
\end{definition}

\begin{lemma}[{\cite[Lemma~4.4]{alfsen2012geometry}}]\label{lem:exchangeable-by-4-is-JW}
	Let $V$ be a JBW-algebra where the identity is the sum of at least 4 idempotents that are mutually exchangeable by a symmetry. Then $V$ is a JW-algebra.
\end{lemma}

\begin{lemma}\label{lem:symmetry-exceptional}
	Let $V\neq \{0\}$ be a purely exceptional JBW-algebra. Then the identity of $V$ is the sum of 3 orthogonal non-zero idempotents exchangeable by a symmetry.
\end{lemma}
\begin{proof}
	By Theorem~\ref{thm:purely-exceptional-char} we can write $V=C(X,E)$ where $E=M_3(\mathbb{O})_{\sa}$ for some hyperstonean space $X$. As $X$ is a type I$_3$ JBW-factor there exist orthogonal non-zero idempotents $q_1,q_2,q_3\in E$ mutually exchangeable by a symmetry such that $q_1+q_2+q_3 = 1_E$~\cite[Theorem 2.8.3]{hanche1984jordan}. Let $s_{ij}\in E$ for $i,j\in\{1,2,3\}$ be symmetries so that $Q_{s_{ij}} q_i = q_j$. Define then $f_i:X\rightarrow E$ as the constant function $f_i(x) = q_i$, and similarly $g_{ij}:X\rightarrow E$ by $g_{ij}(x) = s_{ij}$. Then indeed for every $x\in X: (Q_{g_{ij}} f_i)(x) = Q_{s_{ij}} q_i = q_j = f_j(x)$.
\end{proof}

\begin{lemma}\label{lem:tensor-symmetry}
  Let $p_1,q_1\in V_A$ be idempotents exchangeable by a symmetry $s_1\in V_A$, and let $p_2,q_2\in V_B$ be idempotents exchangeable by a symmetry $s_2\in V_B$. Then $p_1\otimes p_2$ and $q_1\otimes q_2$ are idempotents exchangeable by $s_1\otimes s_2$.
\end{lemma}
\begin{proof}
  That $p_1\otimes p_2$ and $q_1\otimes q_2$ are idempotents follows by Corollary~\ref{cor:tensor-idempotents}. That $s_1\otimes s_2$ is a symmetry follows by Proposition~\ref{prop:tensor-quadratic}, because $(s_1\otimes s_2)^2 = Q_{s_1\otimes s_2} 1 = (Q_{s_1}\otimes Q_{s_2})(1\otimes 1) = s_1^2\otimes s_2^2 = 1\otimes 1 = 1$. By the same proposition: $Q_{s_1\otimes s_2} (p_1\otimes p_2) = (Q_{s_1}\otimes Q_{s_2})(p_1\otimes p_2) = (Q_{s_1}p_1)\otimes (Q_{s_2}p_2) = q_1\otimes q_2$.
\end{proof}

\begin{proposition}
	$V_A$ is a JW-algebra.
\end{proposition}
\begin{proof}
	Since $V_A$ is a JBW-algebra we can write $V_A = V_1\oplus V_2$ where $V_1$ is a JW-algebra and $V_2$ is purely exceptional (Theorem~\ref{thm:JBW-decomposition}). We need to show that $V_2 =\{0\}$. Towards contradiction, suppose that $V_2\neq \{0\}$.

	Let $p\in V_A$ be the central idempotent corresponding to $V_2$. Then the compression system $C$ corresponding to $p$ has an associated JBW-algebra isomorphic to $V_2$. Let $q_1,q_2,q_3$ be a set of idempotents in $C$ exchangeable by symmetries $s_{ij}$ for $i,j\in \{1,2,3\}$, which exists by Lemma~\ref{lem:symmetry-exceptional}. Consider the system $C\otimes C$. By Lemma~\ref{lem:tensor-symmetry} $s_{ik}\otimes s_{jl}$ is a symmetry for all $i,k,j,l\in \{1,2,3\}$. This set of symmetries makes all nine idempotents $\{q_i\otimes q_j~;~i,j\in\{1,2,3\}\}$ in $C\otimes C$ mutually exchangeable by a symmetry.

  Hence, by Lemma~\ref{lem:exchangeable-by-4-is-JW}, $V_{C\otimes C}$ must be a JW-algebra. So then $V_{C\otimes C}$ embeds into $W^*(V_{C\otimes C})$ via an injective Jordan homomorphism (Corollary~\ref{cor:JW-injective}). 
  But we also have an injective Jordan homomorphism from $V_2$ to $V_{C\otimes C}$ given by $a\mapsto a\otimes 1$ (Corollary~\ref{cor:tensor-is-Jordan-hom} and Proposition~\ref{prop:tensor-is-injective}). Hence, $V_2$ embeds into $W^*(V_{B\otimes B})$. This contradicts the fact that $V_2$ is purely exceptional, so that we indeed must have had $V_2=\{0\}$.
\end{proof}

Let us denote by $\textbf{JW-alg}_{\text{npsu}}$ the full subcategory of $\textbf{JBW-alg}_{\text{npsu}}$ consisting of the JW-algebras. The above proposition shows that we indeed have the following.

\begin{theorem}
  Let $\mathbb{E}$ be a complete monoidal sequential $\omega$-effect theory with non-trivial irreducible scalars. Then there is a functor $F:\mathbb{E}\rightarrow \textbf{JW-alg}_{\text{npsu}}^\opp$ satisfying $F(\eff(A))\cong [0,1]_{F(A)}$. This functor is faithful if and only if $\mathbb{E}$ satisfies local tomography.
\end{theorem}

\begin{remark}
  Not all JW-algebras can exist as systems in $\mathbb{E}$. Indeed, adapting the results of Section~\ref{sec:monoidalPETs} we can show that there can be no quaternionic systems and that there cannot be both real and complex systems. The situation is however less clear for infinite-dimensional systems. We leave as a topic for future work what precise restrictions our assumptions imply on the allowed infinite-dimensional systems. It is also interesting to see what additional requirements are necessary to force all the JW-algebras in $\mathbb{E}$ to be of the form $\mathfrak{B}_{\sa}$ for some von Neumann algebra $\mathfrak{B}$, which would be the analogous infinite-dimensional version of Theorems~\ref{thm:seqprodlocalcomp} and~\ref{theor:compositealgebras}.
  Finally, note that the notion of `local tomography' in this theorem is that of Definition~\ref{def:tomographies}, and hence without knowing more about the functor, in particular whether it is strong monoidal, it is not a priori clear that this corresponds to the standard notion of local tomography.
\end{remark}



\if\ismain0 
  \ChapterOutsidePart
  \addtocontents{toc}{\protect\addvspace{2.25em}}
   \cleardoublepage
   \begingroup
	\phantomsection
	\emergencystretch=1em\relax
	\printbibliography[heading=bibintoc]
	\endgroup
   \cleardoublepage
   \phantomsection
   \printindex{math}{Abbreviations and mathematical notation}
   \cleardoublepage
   \phantomsection
   \printindex{default}{Index}
\fi 

\if\ismain0 

\setcounter{chapter}{5}
\setcounter{part}{1}
\ChapterOutsidePart
\pdfbookmark{\contentsname}{toc}
\microtypesetup{protrusion=false}
\tableofcontents
\microtypesetup{protrusion=true}
\ChapterInsidePart
\fi 

\part{Quantum Software from Diagrams}

\chapter{Calculating with diagrams}\label{chap:zxcalculus}

This introductory chapter is organised as follows. In Section~\ref{sec:intro-to-circuits} we give a brief introduction into the circuit model of quantum computing, recalling the most commonly used quantum gates and stating some important well-known results.

Then in Section~\ref{sec:ZX-diagrams} we introduce ZX-diagrams and show how to represent quantum circuits using them. Section~\ref{sec:ZX-calculus} recalls the graphical rewrite rules associated to the ZX-diagrams, collectively known as the ZX-calculus. We derive some simple consequences of these rules in Section~\ref{sec:simple-derivations}.

A concept that is tremendously useful for reasoning about ZX-diagrams is the \emph{phase gadget} that we will introduce in Section~\ref{sec:phasegadgets}. Section~\ref{sec:graph-like} defines a class of ZX-diagrams that are closer-aligned to simple graphs that we call \emph{graph-like}. We use this relation to graphs to define the graph operations of \emph{local complementation} and \emph{pivoting} on ZX-diagrams in Section~\ref{sec:local-complementation}. We demonstrate a first use of these rewrite rules in Section~\ref{sec:graph-theoretic-simp} by finding a diagrammatic proof of the Gottesman-Knill theorem.

\section{Quantum computation and quantum circuits}\label{sec:intro-to-circuits}

Before we proceed to the introduction of the ZX-calculus it will be helpful to establish the basic notions and definitions used in the circuit model of quantum computation.

In this thesis we will solely work with quantum computation based on qubits. A qubit is not any particular quantum system, but rather any physical quantum system that has two distinct orthogonal states. So a qubit could for instance be an ion in an ion trap, with the two states of the qubit corresponding to different hyperfine energy levels~\cite{iontrapoxford}, or a photon with the two states corresponding to different types of polarization~\cite{wang201818}. It can even be a more complicated arrangement of matter, such as Cooper pairs of electrons in superconducting circuits~\cite{superconductingdelft}, or a large collection of physical qubit systems that together form one \emph{logical} qubit in a quantum error correcting code~\cite{fowler2012surface}.

\indexd{Dirac notation} 
\index{math}{$\ket{\psi}$ (Dirac notation)}
We abstract away from these physical complications and simply represent a qubit by the 2-dimensional complex vector space $\C^2$. The possible states of a qubit then correspond to normalised vectors (up to global phase) of $\C^2$. We will write such states in \Define{Dirac notation} as $\ket{\psi}$. We use $\bra{\psi}$ to denote the Hermitian adjoint of the state $\ket{\psi}$. We denote the inner product of two states $\ket{\psi}$ and $\ket{\phi}$ by $\braket{\phi}{\psi}$. With a slight abuse of notation we then write $\ketbra{\psi}{\psi}$ for the linear map that projects onto the state $\ket{\psi}$, \ie $\ketbra{\psi}{\psi}(\ket{\phi}) = \braket{\psi}{\phi} \ket{\psi}$.

\index{math}{0@$\ket{0}$}\index{math}{1@$\ket{1}$}
\index{math}{+@$\ket{+}$}\index{math}{-@$\ket{-}$}
We refer to the states corresponding to the standard basis of $\C^2$ as \Define{computational basis states}\indexd{computational basis states} and denote them by $\ket{0}$ and $\ket{1}$. As these states form a basis for $\C^2$, we can write any other qubit state as a linear combination. I.e.~for any state $\ket{\psi}$ there exist complex numbers $a$ and $b$ such that $\ket{\psi} = a \ket{0} +b\ket{1}$. A different basis of quantum states that we will often use is the \Define{Hadamard basis} (sometimes also called the \Define{Fourier} basis)\indexd{Hadamard!--- basis}. This consists of states $\ket{+}$ and $\ket{-}$ defined as $\ket{+} := \frac{1}{\sqrt{2}} (\ket{0}+\ket{1})$ and $\ket{-} := \frac{1}{\sqrt{2}} (\ket{0}-\ket{1})$. Writing these states in regular vector notation we have:
\begin{equation*}
	\ket{0} \ = \ \begin{pmatrix}1\\ 0 \end{pmatrix} \quad\quad \ket{1} \ = \ \begin{pmatrix}0\\ 1 \end{pmatrix} \quad\quad \ket{+} \ = \ \frac{1}{\sqrt{2}}\begin{pmatrix}1\\ 1 \end{pmatrix} \quad\quad \ket{-} \ = \ \frac{1}{\sqrt{2}}\begin{pmatrix}1\\ -1 \end{pmatrix}
\end{equation*}

Because quantum states are normalised vectors, the numbers $a$ and $b$ must satisfy $\lvert a \rvert^2 + \lvert b \rvert^2 = 1$. Additionally, quantum states that differ only by a global phase $e^{i\theta}$ represent the same physical state. These two observations allow us to reduce the 4 real parameters present in $a$ and $b$ to just two: $\ket{\psi} = \cos \alpha \ket{0} + e^{i\beta} \sin \alpha \ket{1}$. This parametrisation allows us to present all the qubit states on the surface of a sphere, known as the \Define{Bloch sphere}\indexd{Bloch sphere}. On this sphere, $\ket{0}$ and $\ket{1}$ lie on respectively the north and south pole. The states $\ket{+}$ and $\ket{-}$ lie on opposing ends of the equator:
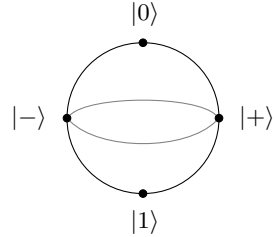
\ctikzfig{bloch-sphere}

A single qubit is of course not very useful. The power of quantum computing comes from how small systems combine into a composite system. A collection of $n$ qubits corresponds to the vector space $\C^2\otimes \C^2 \otimes \cdots \otimes \C^2 \cong \C^{2^n}$, and hence the dimension of the space grows exponentially with the number of qubits. This is an important reason for why quantum computation is more powerful than classical
computation\footnote{It is an open problem whether quantum computation is indeed more powerful than classical computation in the standard computational models, \ie~whether \textbf{BQP}$\neq$\textbf{P}. However, in certain other settings, such as Grover's oracle problem~\cite{grover1996fast}, or for circuits of limited depth~\cite{bravyi2018quantum}, there is a clear proven advantage for quantum computation.}. 
When we have a collection of qubits that are all in a computational basis state $\ket{x_i}$ for $x_i\in \{0,1\}$, we write the state of the full system as $\ket{x_0x_1\cdots x_n}$ as a shorthand for the tensor product $\ket{x_0}\otimes \ket{x_1}\otimes \cdots \otimes \ket{x_n}$.

The evolution through time of a physical quantum system depends on the energy present in the system and is governed by the Schr{\"o}dinger equation. This evolution acts on a state $\ket{\psi}\in \C^2$ by a unitary matrix. Carefully manipulating the environment of the system allows us to control the evolution of a quantum system. As a result we can construct ever more intricate quantum states and actually compute with the system. Different designs of quantum computers have different sets of evolutions that are possible to implement, and this gives rise to different \Define{gate sets}\indexd{gate set}: a collection of unitaries, usually acting on a small number of qubits, that can be implemented on (a subset of) the qubits of the computer.

For a single qubit, all the unitary matrices correspond to rotations around the Bloch sphere. In particular, the rotations corresponding to half-turns around the principal axes of the Bloch sphere are the \Define{Pauli matrices}\indexd{Pauli matrices}:
\begin{equation}
  \text{X} \ = \ \left(\begin{array}{cc}0 & 1\\1 & 0 \end{array}\right) \qquad
  \text{Y} \ = \ \left(\begin{array}{cc}0 & -i\\i & 0 \end{array}\right) \qquad
  \text{Z} \ = \ \left(\begin{array}{cc}1 & 0\\0 & -1 \end{array}\right)
\end{equation}
A rotation of an angle $\theta$ around the Z axis is then given by $R_Z(\theta) = \exp(-\frac12i\theta Z)$ (indeed, up to global phase $R_Z(\pi) = \text{Z}$). Similar expressions give $R_X(\theta)$ and $R_Y(\theta)$. We will refer to these gates as \Define{phase gates}\indexd{gate!phase ---}. Any rotation around a sphere can be decomposed as a series of 3 rotations around orthogonal axes, known as its \Define{Euler decomposition}\indexd{Euler decomposition}. Hence, for any single qubit unitary $U$ we can find angles $\alpha, \beta, \gamma$ such that (up to global phase) $U = R_Z(\alpha)R_X(\beta)R_Z(\gamma)$. 
Note that the Pauli X matrix interchanges the computational basis states: X$\ket{0} = \ket{1}$ and X$\ket{1} = \ket{0}$. In the context of computation it is therefore also sometimes referred to as the \Define{NOT gate}\indexd{gate!NOT ---}.
In addition to the Pauli matrices above, there are a couple other single qubit unitary gates that are widely used in the literature on quantum computation and carry special names. A rotation of $\pi/2$ around the Z-axis is known as an \Define{S gate}\indexd{gate!S ---}, while a rotation of $\pi/4$ is called a \Define{T gate}\indexd{gate!T ---}:
\begin{equation}
  \text{S} \ = \ \left(\begin{array}{cc}1 & 0\\0 & i \end{array}\right) \qquad\qquad T = \sqrt{\text{S}} \ =\ \left(\begin{array}{cc}1 & 0\\0 & e^{i\pi/4} \end{array}\right)
\end{equation}
As these are unitary matrices, their inverse is equal to the Hermitian adjoint. We will denote this by the dagger symbol $\dagger$. Hence T$^\dagger$ is the inverse of T.
Finally, there is the \Define{Hadamard gate} H.
\indexd{Hadamard!--- gate}
\index{math}{H (Hadamard gate)}
\indexd{gate!Hadamard ---} 
The Hadamard gate interchanges the computational basis with the Hadamard basis: $\text{H}\ket{0} = \ket{+}$, $\text{H}\ket{1} = \ket{-}$. As a matrix:
\begin{equation}
  \text{H} \ = \ \frac{1}{\sqrt{2}} \left(\begin{array}{cc}1 & 1\\1 & -1 \end{array}\right)
 \end{equation} 

For two qubits, there is a much wider variety of unitaries available. In particular, for any two single qubit unitaries $U_1$ and $U_2$ we can define their tensor product $U_1\otimes U_2$, and this itself is a unitary that acts on two qubits. These products however do not let the qubits interact. There are two specific gates that do act in an irreducible manner on two qubits that will be important to us: the \Define{controlled Z}\indexd{gate!CZ ---} and the \Define{controlled NOT}\indexd{gate!CNOT ---} gate (the latter is sometimes also called the controlled X gate). We shorten these names to just CZ and CNOT. These gates act on two qubits at once, and they apply a Z, respectively an X, gate on the second qubit if the first is in the $\ket{1}$ state. If instead the first qubit is in the $\ket{0}$ state, nothing happens. The first qubit hence \emph{controls} whether a gate gets applied. The matrices of these gates are the following:
\begin{equation}\label{eq:CNOT-matrix}
  \text{CZ} \ = \ \left(\begin{array}{cccc}1 & 0 & 0 & 0\\0 & 1 & 0 & 0 \\ 0&0&1&0 \\ 0&0&0&\text{-}1 \end{array}\right) \qquad\qquad \text{CNOT} \ = \ \left(\begin{array}{cccc}1 & 0 & 0 & 0\\0 & 1 & 0 & 0 \\ 0&0&0&1 \\ 0&0&1&0 \end{array}\right)
 \end{equation} 

The most widely used model of quantum computation is the \Define{circuit model}\indexd{circuit model}. In this model we take our input quantum state, apply a sequence of quantum gates to it, and finally measure (some of) the qubits to get some classical outcomes. The information about which gates to apply is usually presented as a \Define{quantum circuit}\indexd{quantum circuit}. For instance:
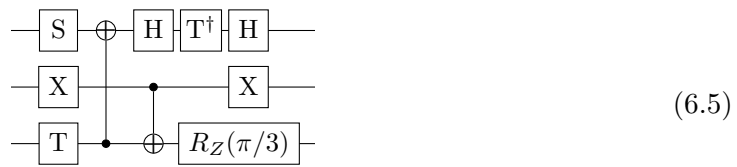
\begin{equation}\label{eq:example-circuit}
	\input{example-circuit.tikz}
\end{equation}
Here each horizontal line corresponds to a qubit (so this circuit contains 3 qubits). Each box corresponds to a gate that is applied to the qubit of the line that the box is on. Time flows from left to right, and hence the order of the boxes determines in which order the gates are applied to the qubits. The black dot connected to the white dot with a `+' represents a CNOT gate, where the black dot denotes the control qubit, and the white dot denotes the target qubit. The horizontal distance between each gate is immaterial, as the circuit only gives information about the order of application.

To achieve the full power of the circuit model one also needs \emph{ancillae} and \emph{classical control}. An \Define{ancilla}\indexd{ancilla} is simply a qubit that is always prepared in the same state (usually $\ket{0}$ or $\ket{+}$, or sometimes the $\ket{T} := T\ket{+}$ \Define{magic state}). It is hence not a free input of the circuit. When implementing a quantum computation one could choose to measure some of the qubits earlier than others. By using \Define{classical control}\indexd{classical control} we can let later parts of the circuit depend on these earlier measurement outcomes. In this thesis, unless otherwise specified, we will take `circuit' to mean \Define{unitary circuit}, \ie a circuit which does not contain ancillae or classical control and hence is fully defined by the gates that are in the circuit.\indexd{circuit!unitary ---}

Even though the dimension of the space of unitaries increases rapidly with the number of qubits, the set of single qubit gates together with either the CNOT or CZ gate is a \Define{universal gate set}~\cite{NielsenChuang}\indexd{gate set!universal ---}. This means that for any number of qubits $n$ and a unitary $U$ acting on $n$ qubits we can find a quantum circuit implementing $U$ such that all gates in the circuit are either single-qubit gates or CNOT/CZ gates.
Hence, by applying gates in the correct sequence in a quantum circuit we can construct any desired unitary, and thus any desired computation.

In some settings it will be useful or necessary to consider different gate sets. One of these gate sets that is particularly important is the set \Define{Clifford unitaries}\indexd{Clifford!--- unitary}. This is the group of unitaries that can be constructed out of quantum circuits consisting of Hadamard, S and CNOT gates. As S$^2 = $ Z, HZH $=$ X and XZ $=$ Y (up to global phase), this set contains all the Pauli matrices. It also contains the CZ gate as $(I\otimes \text{H})\text{CNOT}(I\otimes \text{H}) = \text{CZ}$. We will hence refer to all these gates as \Define{Clifford gates}\indexd{gate!Clifford ---}. Any quantum state that can be produced by applying a Clifford circuit to $\ket{0\cdots 0}$ is called a \Define{Clifford state}\indexd{Clifford!--- state}.
Note that Clifford circuits/states are sometimes also called \Define{stabiliser} circuits/states\indexd{Clifford!stabiliser}\indexd{stabiliser!see {Clifford}}. This name comes from the fact that the unitaries that map the group generated by the Pauli matrices to itself (up to global phase) are precisely the Clifford unitaries, and hence the Cliffords form the stabiliser of the Pauli group in the group of all unitaries. This definition leads to a generalisation known as the \emph{Clifford hierarchy} that we consider in more detail in Section~\ref{sec:clifhier}.
The Clifford unitaries are significant, because many well-known quantum protocols, such as teleportation, dense-coding, and quantum key distribution, can be performed using just Clifford unitaries. They are furthermore crucial in constructing quantum error correcting codes, and they display many inherently quantum features, such as entanglement and contextuality.

Even though Clifford circuits can perform many interesting tasks, they do not allow any computation that exceeds the speed of classical computers. The \Define{Gottesman-Knill theorem}\indexd{Gottesman-Knill theorem} states that any computation consisting just of Clifford operations can be efficiently simulated on a classical computer~\cite{aaronsongottesman2004}. Hence, in order to achieve speedups with a quantum computer, we will need to use additional quantum gates.

Interestingly, allowing just a single additional type of gate is sufficient to reproduce the full power of a quantum computer. If we supplement the Clifford gates with the {T gate}, we have a gate set that is \Define{approximately universal}\indexd{gate set!approximately universal ---}. This means that while we cannot exactly implement all unitaries, we can approximate any unitary to arbitrary precision (with just a relatively small overhead with respect to the universal gate set discussed above)~\cite{dawson2005solovay}. We will refer to this gate set as \Define{Clifford+T}\indexd{gate set!Clifford+T ---}. In the sense of computational complexity, computation with the Clifford+T gate set is just as powerful as computing with a universal gate set.

\section{ZX-diagrams}\label{sec:ZX-diagrams}

The ZX-calculus is a diagrammatic language introduced by Coecke and Duncan in 2008~\cite{CD1,CD2} that allows graphical manipulation of diagrams representing linear maps between qubits. These \emph{ZX-diagrams} can be seen as a generalisation of quantum circuit notation. Whereas a circuit consists of straight lines representing qubits and boxes representing quantum gates, a \Define{ZX-diagram}\indexd{ZX-diagram} consists of \emph{wires} and \emph{spiders}.  Wires entering the diagram from the left are \Define{inputs}; wires exiting to
the right are \Define{outputs}.

\Define{Spiders}\indexd{spider} are a special type of linear map which can have any number of inputs and outputs. Spiders come in two varieties, \Define{Z-spiders}\indexd{Z-spider} depicted as white dots,
\[\hfill \input{Zsp-a.tikz} \ \ :=\ \ \ketbra{0 \cdots 0}{0 \cdots 0} + e^{i \alpha} \ketbra{1 \cdots 1}{1 \cdots 1}, \hfill\]
and \Define{X-spiders}\indexd{X-spider} depicted as grey dots,
\[\hfill \input{Xsp-a.tikz} \ \ :=\ \ \ketbra{+ \cdots +}{+ \cdots +} + e^{i \alpha} \ketbra{- \cdots -}{- \cdots -}. \hfill\]

A special case is when spiders have a single input and output, in which case they form Z-phase and X-phase gates (up to a global phase):
\begin{align}\label{eq:zx-phase-gates}
\begin{tikzpicture}
	\begin{pgfonlayer}{nodelayer}
		\node [style=white phase dot] (0) at (0, 0) {$\alpha$};
		\node [style=none] (1) at (-1.25, 0) {};
		\node [style=none] (2) at (1.25, 0) {};
	\end{pgfonlayer}
	\begin{pgfonlayer}{edgelayer}
		\draw [style=simple] (1.center) to (0);
		\draw [style=simple] (0) to (2.center);
	\end{pgfonlayer}
\end{tikzpicture} \ \ &=\ \ \ketbra{0}{0} + e^{i \alpha} \ketbra{1}{1} = R_Z(\alpha), \\
\begin{tikzpicture}
	\begin{pgfonlayer}{nodelayer}
		\node [style=gray phase dot] (0) at (0, 0) {$\alpha$};
		\node [style=none] (1) at (-1.25, 0) {};
		\node [style=none] (2) at (1.25, 0) {};
	\end{pgfonlayer}
	\begin{pgfonlayer}{edgelayer}
		\draw [style=simple] (1.center) to (0);
		\draw [style=simple] (0) to (2.center);
	\end{pgfonlayer}
\end{tikzpicture} \ \ &=\ \ \ketbra{+}{+} + e^{i \alpha} \ketbra{-}{-} = R_X(\alpha). \nonumber
\end{align}
One can simply multiply out the matrices to verify that these diagrams are indeed equal to the definition given in the previous section. In particular, we have:
\begin{equation*}
	\text{S} \ = \ \begin{tikzpicture}
	\begin{pgfonlayer}{nodelayer}
		\node [style=none] (0) at (1.25, 0) {};
		\node [style=none] (2) at (-1.25, 0) {};
		\node [style=white phase dot] (5) at (0, 0) {$\frac\pi2$};
	\end{pgfonlayer}
	\begin{pgfonlayer}{edgelayer}
		\draw [style=simple] (5) to (2.center);
		\draw [style=simple] (5) to (0.center);
	\end{pgfonlayer}
\end{tikzpicture}
 \qquad \qquad \text{T} \ = \ \begin{tikzpicture}
	\begin{pgfonlayer}{nodelayer}
		\node [style=none] (0) at (1.25, 0) {};
		\node [style=none] (2) at (-1.25, 0) {};
		\node [style=white phase dot] (5) at (0, 0) {$\frac\pi4$};
	\end{pgfonlayer}
	\begin{pgfonlayer}{edgelayer}
		\draw [style=simple] (5) to (2.center);
		\draw [style=simple] (5) to (0.center);
	\end{pgfonlayer}
\end{tikzpicture}

\end{equation*}

Another special case is when $\alpha = 0$, in which case we omit the label:
\[\hfill \input{Zsp.tikz} \ \ :=\ \ \ketbra{0 \cdots 0}{0 \cdots 0} + \ketbra{1 \cdots 1}{1 \cdots 1} \hfill\]
\[\hfill \input{Xsp.tikz} \ \ :=\ \ \ketbra{+ \cdots +}{+ \cdots +} + \ketbra{- \cdots -}{- \cdots -} \hfill\]

In particular, when these \textbf{phaseless} spiders have a single input and output they are equal to the identity matrix:
\ctikzfig{zx-identity}

Phaseless spiders can be thought of as a generalisation of the \emph{GHZ state} to a linear map~\cite{GHZpaper}. Indeed the Z-spider with 0 inputs and 3 outputs is the usual \Define{GHZ state}\indexd{GHZ state} (up to normalisation):
\[\hfill \input{ghz.tikz} \ =\ \ket{000} + \ket{111} \hfill\]
Furthermore, the $\ket{0}$ and $\ket{+}$ states are just one-legged spiders:
\begin{equation}\label{eq:spider-basis1}
\hfill
\begin{tikzpicture}
	\begin{pgfonlayer}{nodelayer}
		\node [style=white dot] (0) at (-0.5, 0) {};
		\node [style=none] (1) at (0.5, 0) {};
	\end{pgfonlayer}
	\begin{pgfonlayer}{edgelayer}
		\draw [style=simple] (0) to (1.center);
	\end{pgfonlayer}
\end{tikzpicture}\ \ =\ \ \ket{0} + \ket{1} \ \approx \ket{+}
\qquad\qquad
\begin{tikzpicture}
	\begin{pgfonlayer}{nodelayer}
		\node [style=gray dot] (0) at (-0.5, 0) {};
		\node [style=none] (1) at (0.5, 0) {};
	\end{pgfonlayer}
	\begin{pgfonlayer}{edgelayer}
		\draw [style=simple] (0) to (1.center);
	\end{pgfonlayer}
\end{tikzpicture}\ \ =\ \ \ket{+} + \ket{-} \ \approx \ket{0}
\hfill
\end{equation}
Following Ref.~\cite{CKbook}, we use the symbol `$\approx$' to denote that the two sides are equal up to some non-zero complex number (in this case a factor of $\sqrt{2}$). 

Given two ZX-diagrams we can compose them either
by joining the outputs of the first to the inputs of the second giving the regular composition of linear maps, or by stacking the two diagrams to form the tensor product of the linear maps.

For instance, composing the states of Eqs.~\eqref{eq:spider-basis1} with respectively a Z = $R_Z(\pi)$ or X = $R_X(\pi)$ gate we get the $\ket{-}$ and $\ket{1}$ states:
\[\hfill \input{ket--.tikz}\ \ =\ \ \ket{0} + e^{i\pi} \ket{1} \ \approx \ket{-} \hfill\]
\[\hfill \input{ket-1.tikz}\ \ =\ \ \ket{+} + e^{i\pi} \ket{-} \ \approx \ket{1} \hfill\]

ZX-diagrams without any inputs or outputs are called \Define{scalars}\indexd{ZX-diagram!scalar ---}\indexd{scalar!ZX-diagram}. They are linear maps that are just complex numbers. In particular, we have:
\begin{equation}\label{eq:zx-scalars}
  \input{scalars.tikz}
\end{equation}
Interchanging the colours in these scalars preserves the scalar value.

Whenever useful, we will write a number instead of the scalar ZX-diagram that is equal to that number.
For instance, we can write the CNOT gate as follows:
\[
\text{CNOT}\ :=\ \sqrt{2}\ \ \input{cnot.tikz}
\]
We can write this vertical wire without ambiguity as it is easily verified by multiplying out the matrices that:
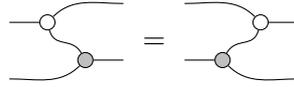
\[
  \input{cnot-left.tikz} \ =\ \input{cnot-right.tikz} 
\]

More generally, diagrams of spiders are less rigid than circuits. Much like diagrammatic depictions of tensor networks (e.g.~Ref.~\cite{Penrose}), any two diagrams of spiders with the same connectivity and ordering of inputs and outputs describe the same linear map (a property we will refer to as `only connectivity matters' in the next section).

We can represent arbitrary Z- and X-phase rotations using Eq.~\eqref{eq:zx-phase-gates}, and since we can also represent a CNOT gate, we have a universal set of gates in the ZX-calculus. In particular, we can represent the Hadamard gate.
As it is a single-qubit unitary, we can decompose it using its Euler angles. This gives a representation of the Hadamard gate in terms of Z- and X-spiders. The Hadamard gate will prove essential for reasoning with ZX-diagrams, so we introduce some special notation for it:
\begin{equation}\label{eq:had-def}
\input{had.tikz}\ \   =\ \frac{1}{\sqrt{2}}\left(\begin{array}{cc} 1 & 1 \\ 1 & -1 \end{array}\right)
\end{equation}

Using the composition rules of ZX-diagrams we can now represent arbitrary quantum circuits. For instance, the circuit~\eqref{eq:example-circuit} becomes:
\ctikzfig{example-circuit-ZX}

We will refer to the symbol in Eq.~\eqref{eq:had-def} as a \Define{Hadamard box}\indexd{Hadamard!--- box}.
A Hadamard box of course only has one input and one output, and in this sense it is more like an edge than like a vertex. To make this role as an edge clearer we introduce additional notation for spiders connected via a Hadamard box:
\begin{equation}\label{eq:def-had-edge}
\input{blue-edge-def.tikz}
\end{equation}
We will refer to such an edge as a \Define{Hadamard edge}\indexd{Hadamard!--- edge}. We can of course freely switch between the Hadamard edge notation and the Hadamard box notation, and we can always expand a Hadamard box into spiders using \eg Eq.~\eqref{eq:had-def}.

Sometimes we will need to explicitly distinguish between a diagram itself and the linear map it represents.
\begin{definition}\label{def:interpretation}
The \Define{interpretation}\indexd{ZX-diagram!interpretation} of a \zxdiagram $D$ is the linear map that such a diagram represents
and is written as $\intf{D}$.
We say two ZX-diagrams $D_1$ and $D_2$ are \Define{equivalent}\indexd{ZX-diagram!equivalence} when $\intf{D_1}=z\intf{D_2}$ for some non-zero complex number $z$.
\end{definition}
For more details regarding the interpretation of a ZX diagram see for instance~\cite{SimonCompleteness}.

\section{The ZX-calculus}\label{sec:ZX-calculus}
The power of using ZX-diagrams to represent linear maps comes from the set of rewrite rules associated to it, collectively known as the \Define{ZX-calculus}\indexd{ZX-calculus}. The ZX-calculus has a couple of variations. We will use only the most basic set of rewrite rules.

The first two rewrite rules are the \Define{spider fusion} rules\indexd{spider!--- fusion}\indexd{rule!spider fusion ---}, which say connected spiders of the same colour fuse together, and their phases add:
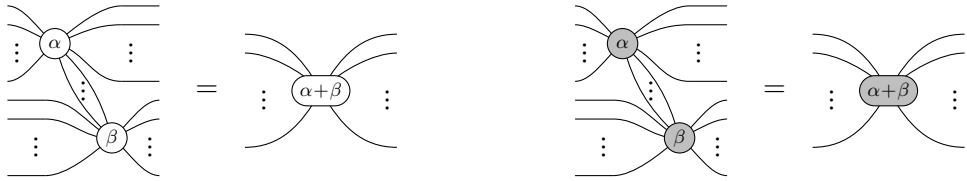
\[\hfill
\input{spider-fusion-Z.tikz} \qquad\qquad\qquad
\input{spider-fusion-X.tikz}
\hfill\]
As the phases on the spiders correspond to angles, we take this addition modulo $2\pi$. Note that these rules, like the others we will introduce, are \Define{sound}\indexd{ZX-calculus!soundness}\indexd{rule!soundness}, meaning that the two sides of the equation represent the same linear map.

As a straightforward application of these rules, we
see that Z- and X-phase gates commute through spiders of the same colour:
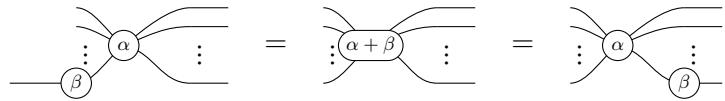
\ctikzfig{spider-fuse-commute}

The following two rules show that Pauli Z and X gates can be pushed through spiders of the opposite colour, changing the sign of the phase on the spider and introducing a global phase:
\begin{equation}\label{eq:x-stab}
\input{Zsp-stab2.tikz}\qquad\qquad\qquad\input{Xsp-stab2.tikz}
\end{equation}
The Hadamard box has some particularly elegant rewrite rules. The first rule for it expresses that it is self-inverse:
\[\hfill \input{had-self-inv.tikz} \hfill\]
For the second rule we recall that the Hadamard gate interchanges the computational and Hadamard bases. As a result, the Hadamard box acts as a `colour-changer' for spiders:\indexd{rule!colour change ---}
\ctikzfig{colour-ch2}
Using this rule we can get a simple expression for the CZ gate from the CNOT gate in the ZX-calculus:
\begin{equation}\label{eq:CZ-in-ZX}
\text{CZ} \ =\  (I\otimes \text{H})\,\text{CNOT}\,(I\otimes \text{H})\  =\  \input{cnot-to-cz.tikz}
\end{equation}

The final rewrite rules are based on a special property that the pair of the computational and Hadamard bases of the qubit have:
\Define{strong complementarity}\indexd{strong complementarity}~\cite[Section~9.3]{CKbook}. This is captured by the following two rules:
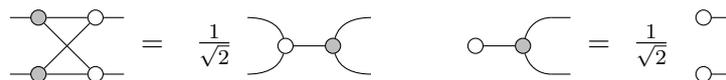
\ctikzfig{strong-compl}
We will refer to these two rules respectively as the \Define{exchange} rule\indexd{rule!exchange ---} and the \Define{copy} rule\indexd{rule!copy ---}.

Combining the copy rule with Eqs.~\eqref{eq:x-stab} and~\eqref{eq:zx-scalars} yields a more general version, for any $a \in \{0,1\}$ and $\alpha \in [0,2\pi)$:
\ctikzfig{gen-copy}

\begin{figure}[!t]
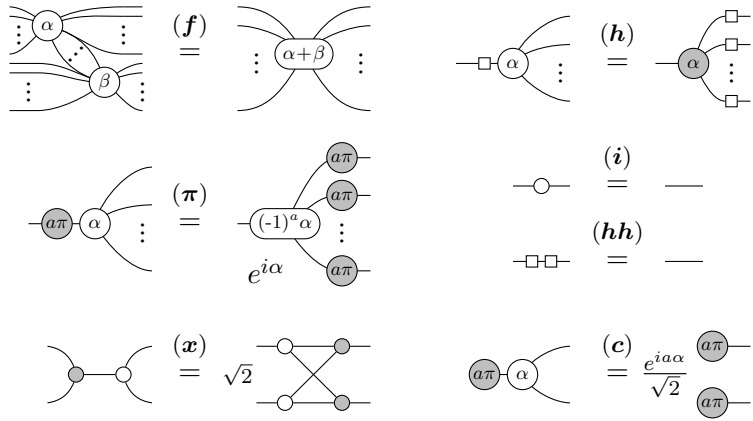

\ctikzfig{ZX-rules}
\caption{The rules of the ZX-calculus: spider-\SpiderRule{}usion, \HadamardRule{}adamard, \mbox{\PiRule{}-commutation}, \IdRule{}dentity, \HHRule{}-cancellation, e\HopfRule{}change and \CopyRule{}opy. These rules hold for all $\alpha, \beta \in [0, 2 \pi)$ and $a \in \{0,1\}$. Due to \HadamardRule and \HHRule they also hold with the colours interchanged.}\label{table:rewrite} \indexd{ZX-calculus!rules}
\end{figure}

We present all the rewrite rules discussed so far in Figure~\ref{table:rewrite}. 

Besides these concrete rewrite rules we also use the `meta-rule' that \Define{only connectivity matters} (OCM)\indexd{only connectivity matters}\index{math}{OCM (only connectivity matters)}. This says that two diagrams are equal when one can be continuously transformed into the other by moving spiders around.
In particular, the orientation of the wires is irrelevant so that we can apply the \Define{yanking equations}\indexd{yanking equations} whenever it proves useful:
\begin{equation}\label{eq:yank}
\input{yank-equations.tikz}
\end{equation}
Related to this are the symmetry equations of the spiders:\indexd{spider!symmetries}
\ctikzfig{spider-symmetries}
For each of the equalities here we of course mean that both sides have the same number of inputs and the same number of outputs.

\begin{remark}
    As a result of OCM, the orientation of the inputs and outputs in the rules of Figure~\ref{table:rewrite} is irrelevant, as we can pre-compose and post-compose each rule with `cups and caps' to change inputs to outputs and vice versa. For instance, for the `reverse' of the copy rule:
    \ctikzfig{gen-copy-reverse}
    Additionally, each rule also holds with the `colours interchanged', \ie~where we change every occurrence in the rewrite rule of a Z spider with an X spider and vice versa. This is because we can compose the rewrite rules with Hadamard gates both on the inputs and outputs, then use \HadamardRule to `push' the input Hadamard gate to the outputs, and finally cancel the paired Hadamards with \HHRule.
\end{remark}

\begin{remark}
    The rule-set of Figure~\ref{table:rewrite} is not minimal: we could remove or simplify some of the rules without losing any expressiveness. For instance, we only need \CopyRule for $a=0$, as the one with $a=1$ can be derived from the other rules. This presentation was chosen for expressiveness. We refer to Ref.~\cite{BackensSimplified} for a more minimal set of equivalent rules, and to Ref.~\cite[Definition~9.108]{CKbook} for a small set of rules that ignore the scalars.
\end{remark}

Finally, next to these diagrammatic rewrite rules we introduce a `meta-rule' that says that scalars can always be combined:\indexd{rule!scalars}
\begin{equation}\label{eq:zx-combine-scalar}
  \begin{tikzpicture}
	\begin{pgfonlayer}{nodelayer}
		\node [style=none] (0) at (-1.5, 0.5) {$z_1$};
		\node [style=none] (1) at (-0.75, -0.75) {$z_2$};
		\node [style=none] (2) at (1, 0) {=};
		\node [style=none] (3) at (3, 0) {$z_1z_2$};
	\end{pgfonlayer}
\end{tikzpicture}

\end{equation}
Here, the right-hand side denotes the product of the complex numbers $z_1$ and $z_2$.

\begin{remark}\label{rem:scalars}
As noted before, for some equations the exact scalar value is not important and we write `$\scalareq$' to denote both sides are only equal up to a non-zero scalar. In particular, in Chapters~\ref{chap:ms-mbqc}--\ref{chap:optimisation}, we will not need the exact value of the scalar at all, and hence we simply write `$=$' to denote equality up to non-zero scalar by, leaving scalars wholly implicit. 
The reason we present our rewrite rules here in a scalar-accurate way is because some of them (such as the rules of Section~\ref{sec:local-complementation}) have never been described before with accurate scalar values and because the scalars are important when using the ZX-calculus for circuit simulation; cf.~Sections~\ref{sec:graph-theoretic-simp} and~\ref{sec:simulation}.
\end{remark}

An important property for a set of rewrite rules is \Define{completeness}\indexd{ZX-calculus!completeness}. A set of rules is complete when they suffice to prove any equality that is true. In the setting of ZX-diagrams completeness means that two diagrams representing the same linear map can always be transformed into one another using the rewrite rules.

The set of rules we present in Figure~\ref{table:rewrite} is \emph{not} complete: there exist diagrams that represent the same linear map, but there is no way we can prove equality of these diagrams using just these rewrite rules~\cite{supplementarity}. In Section~\ref{sec:clifford-circuit-optimisation} we will see that when restricted to the Clifford fragment, \ie~where all spiders have a phase that is a multiple of $\frac\pi2$, the rules of Figure~\ref{table:rewrite} \emph{are} complete. Note that there do exist extended sets of rules that are complete for arbitrary ZX-diagrams~\cite{HarnyCompleteness,SimonCompleteness}. 

\begin{remark}
	As we are using the ZX-calculus as a calculational tool in this thesis, we are intentionally vague about what the actual underlying mathematical object is that defines a ZX-diagram and the set of rewrite rules. Formally, ZX-diagrams are morphisms in a \Define{PROP}\indexd{PROP}: a symmetric monoidal category which has as objects the natural numbers (corresponding to the number of qubits the diagram acts on). This category is freely generated by the spiders, monoidal structure, and \emph{compact closed} structure (\ie the yanking equations of Eq.~\eqref{eq:yank}). Closing the rewrite rules under tensor product, composition, and compact closed structure we can quotient the PROP by the rewrite rules so that two diagrams are identified when a series of rewrite rules maps one into the other. We refer the interested reader to Ref.~\cite{carette_completeness_2019} for more details.
\end{remark}

\section{Some simple derivations}\label{sec:simple-derivations}

In this section we will use the basic rules of the ZX-calculus presented in Figure~\ref{table:rewrite} to do some simple derivations.

The following lemma relates two ways to represent the Pauli Y eigenstates: as a $\pi/2$ rotation over the Z-axis or a $-\pi/2$ rotation over the X-axis.
\begin{lemma}
    The following holds in the ZX-calculus:
    \begin{equation}\label{eq:s-state-eq}
        \input{S-state-equality.tikz}
    \end{equation}
\end{lemma}
\begin{proof}
    \[\input{S-state-equality-pf.tikz} \qedhere\]
\end{proof}
Like with any other equation we derive, the same also holds with the colours of the spiders interchanged.

As an application of this rule, we can find a representation of the Hadamard gate into spiders that does not contain a global phase, which will make later derivations more elegant:
\begin{lemma}
    The following are also valid Euler angles decompositions of the Hadamard gate:
    \begin{equation}\label{eq:had-def3}
        \input{had-def3.tikz}
    \end{equation}
    Furthermore, the ZX-calculus proves the following representation:
    \begin{equation}\label{eq:had-def2}
        \input{had-def2.tikz}
    \end{equation}
\end{lemma}
\begin{proof}
    The first equation is just Eq.~\eqref{eq:had-def}, and the equality after that is the colour-swapped version.
    The first on the second line is proven by unfusing a $\pi$ phase and pushing it through:
    \ctikzfig{had-def3-pf}
    The second is again just a colour-swapped version.
    Now for the final equation~\eqref{eq:had-def2} we take the bottom right equation of Eq.~\eqref{eq:had-def3}, unfuse the grey $\pi/2$ and apply Eq.~\eqref{eq:s-state-eq} (with the colours swapped).
\end{proof}
These different forms for the Hadamard box reveal an additional symmetry present in our rewrite rules: every equation also holds with the sign of all the angles flipped. This was already clear for the rules of Figure~\ref{table:rewrite}, but now we see it also holds for the expansion of the Hadamard gate into spiders. The operation of flipping all the angles corresponds to taking the complex conjugate of the linear map.

The following rule will be essential for many of our arguments. It will allow us to get rid of multiple parallel edges between different spiders.
\begin{lemma}
  The following holds in the ZX-calculus:
  \begin{equation}\label{eq:hopf-law}
  \input{hopf-rule.tikz}
  \end{equation}
  \begin{proof}
    To prove this, we take advantage of the freedom to
    deform the diagram:
    \[\input{hopf-rule-pf.tikz} \qedhere\]
  \end{proof}
\end{lemma}

This rule can also be cast in terms of Hadamard edges:
\begin{lemma}
  The following holds in the ZX-calculus:
  \begin{equation}\label{eq:hopf-law-h}
  \input{par-edge-rem-scalar.tikz} 
  \end{equation}
  \begin{proof}
    \[\input{double-had-edge.tikz}\qedhere\]
  \end{proof}
\end{lemma}

Another useful simplification is the ability to get rid of self-loops.
\begin{lemma}
  The following equations holds in the ZX-calculus:
  \begin{equation}\label{eq:self-loops}
    \input{self-loop-rem.tikz} \qquad\qquad\ \ 
    \input{h-self-loop-rem.tikz}
  \end{equation}
\end{lemma}
\begin{proof}
  The first one follows by applying \IdentityRule from right to left and then using \SpiderRule. For the last one we do:
\begin{equation*}
\input{self-loop.tikz} \qedhere
\end{equation*}
\end{proof}

Finally, let us prove generalisations of the rules \HopfRule and \CopyRule to spiders of arbitrary arity.
\begin{lemma}
    The following holds in the ZX-calculus:
    \ctikzfig{bialgebra-arbitrary-arity}
    Here the right-hand side of the first equation is a complete bipartite graph.
\end{lemma}
\begin{proof}
    The second equation with $n=1$ is proved by \IdRule and for $n=2$ is proved using \CopyRule. The other cases are easily proven by induction by unfusing the `left-over' spiders with \SpiderRule and then applying \CopyRule.

    For the first equation we note that the $m=0$ case is covered by the left equation. Since $m=1$ is trivial, the first non-trivial case for the left equation is $m=2$. Keep $m$ fixed to this value and prove by induction on $n$ with the base case $n=2$ being \HopfRule. Now that we have the equation for all $n$ with $m=2$, we can do induction on $m$.
\end{proof}
In the remainder of this thesis we will refer to these generalisations of \HopfRule and \CopyRule in the same way as to the standard \HopfRule and \CopyRule.

\section{Phase gadgets}\label{sec:phasegadgets}

A concept that will be central in all the following chapters is the notion of a phase gadget. A \Define{phase gadget}\indexd{phase gadget} is simply an arity-1 spider with a phase $\alpha$, connected to another spider that has no phase, which, depending on the context, we will present in different ways:
\[
\input{phase-gadget.tikz}
\]
The first of these forms will come into play in Chapter~\ref{chap:ms-mbqc}, the second in Chapter~\ref{chap:optimisation} and the third in Chapter~\ref{chap:MBQC}. Note that we have already seen an example of the first type in the form of the Hadamard box in Eq.~\eqref{eq:had-def2}.

The ubiquity of phase gadgets comes from the simplicity of the linear map they implement\indexd{phase gadget!as linear map}. For example:
\begin{equation}\label{eq:phase-gadget-unitary}
  \input{phase-gadget-unitary.tikz} \ \  \scalareq \ \ U \text{ where }\  U
\ket{x_1, ..., x_n} =
e^{i \alpha (x_1 \oplus \ldots \oplus x_n)} \ket{x_1, ..., x_n}.
\end{equation}
Recall that we use the dotted lines in ZX-diagrams to represent Hadamard-edges (cf.~Eq.~\eqref{eq:def-had-edge}). The operation $\oplus$ denotes the XOR of the Boolean variables $x_i$. Hence, this diagram implements a diagonal unitary that applies a phase $e^{i\alpha}$ iff the parity of the input qubits is odd.

It can be shown that the above diagram is equal to a ladder of CNOT gates, followed by a single phase gate, followed by the reverse ladder of CNOT gates. For example, on 4 qubits:\indexd{phase gadget!CNOT ladder}
\begin{equation}\label{eq:phasegadget}
\input{phase-gadget-circ.tikz}
\end{equation}
This correspondence is easily proven for 2 qubits:
\ctikzfig{MS-simplify}
Here the last application of \IdRule removed the 2-ary X-spiders (recall from Figure~\ref{table:rewrite} that all the rules also hold with the colours interchanged).
The higher arity decompositions follow by induction. Note that while the presentation of a phase gadget is symmetric in the qubits, the circuit form of Eq.~\eqref{eq:phasegadget} is not. Hence, this circuit form obscures some of the structure that is present in a phase gadget.

The form of the unitary in Eq.~\eqref{eq:phase-gadget-unitary} implies that two phase gadgets with exactly the same connectivity should be able to fuse together. This is readily shown in the ZX-calculus:\indexd{phase gadget!--- fusion}
\begin{equation}\label{eq:fusing-gadgets}
    \scalebox{0.9}{\input{gf-proof.tikz}}
\end{equation}
This fact lies at the heart of the circuit optimisation strategy we will use in Chapter~\ref{chap:optimisation}.

Arbitrary diagonal unitaries, i.e. unitaries which are characterised by a $f : \{0,1\}^n \to \mathbb R$ via
$
U\ket{x_1, ..., x_n} = e^{i f(x_1, \ldots, x_n)} \ket{x_1, ..., x_n}
$
can always be expressed as a combination of phase gadgets. For example, for $f(x_1,x_2,x_3,x_4) = \frac \pi 4 x_1 \oplus x_4 + \frac \pi 8 x_1 \oplus x_2 - \frac \pi 4 x_1 \oplus x_3$ we get:
\begin{equation}\label{eq:phase-poly}
U \ \ \scalareq \ \ 
\input{phase-poly.tikz}
\end{equation}
The angles appearing in the phase gadgets for a function $f$ correspond to its \emph{semi-Boolean Fourier expansion} (see the appendix of Ref.~\cite{amy2018cnot} for more details). \indexd{Fourier expansion}This Fourier expansion allows us to write any $f : \{0,1\}^n \to \mathbb R$ as
\begin{equation}\label{eq:fourier}
  f(\vec x) = \alpha + \sum_{\vec y} \alpha_{\vec y} (x_1 y_1 \oplus \ldots \oplus x_n y_n),
\end{equation}
where $\vec x, \vec y \in \{ 0, 1 \}^n$ and $\alpha, \alpha_{\vec y} \in \mathbb R$. In the context of diagonal unitaries, $\alpha$ yields a global phase (which we can ignore), and each $\alpha_{\vec y}$ corresponds to a phase gadget.
The Fourier expansion of the semi-Boolean function corresponding to a diagonal unitary is called a \Define{phase polynomial}\indexd{phase polynomial}. In terms of ZX-diagrams, a phase polynomial is a unitary consisting of phase gadgets as in Eq.~\eqref{eq:phase-poly}. Phase polynomials play an important role in the optimisation of quantum circuits, cf.~Section~\ref{sec:optimisation-overview}.

As an example of the utility of phase polynomials, let us consider the case of the \Define{CCZ gate}\indexd{gate!CCZ ---}. This is a 3-qubit gate that applies a Z gate on the third qubit if the first two are in the $\ket{1}$ state. We can write down its action on the computational basis states as CCZ$\ket{x_1x_2x_3} = e^{i\pi x_1\cdot x_2\cdot x_3}\ket{x_1x_2x_3}$. In other words: it is a diagonal gate that applies a $e^{i\pi} = -1$ phase iff $x_1=x_2=x_3 = 1$. For values $x,y\in \{0,1\}$ it is easily verified that $2x\cdot y = x+y-2(x\oplus y)$. Applying this several times to $\pi x_1\cdot(x_2\cdot x_3)$ we see that it is equal to $\frac\pi4(x_1 + x_2 + x_3 - x_1\oplus x_2 - x_1\oplus x_3 - x_2\oplus x_3 + x_1\oplus x_2 \oplus x_3)$. This is the phase polynomial of the CCZ gate. We can use this expansion to write the CCZ gate as a collection phase gadgets as in Eq.~\eqref{eq:phase-poly}. Expanding the phase gadgets to a circuit using Eq.~\eqref{eq:phasegadget}, reveals a Clifford+T circuit that implements the CCZ gate.

\section{Graph-like diagrams}\label{sec:graph-like}
\indexd{ZX-diagram!graph-like ---}
In this section we introduce a special class of ZX-diagrams that are almost fully described as simple graphs.

\begin{definition}
\indexd{graph}\indexd{vertex!neighbours}\index{math}{vw@$v\sim w$ (connected vertices)}\index{math}{N(v)@$N(v)$ (set of neighbours)}
	A \Define{graph} $(V,E)$ consists of a set of \Define{vertices} $V$ and \Define{edges} $E$. Each edge $e\in E$ is a set with two elements $e=\{u,v\}$ where $u,v\in V$ are vertices. When $\{u,v\}\in E$ we say that $u$ and $v$ are \Define{connected} and are \Define{neighbours}. We write $u\sim v$ to denote that $u$ and $v$ are connected and $N(v)$ for the set of neighbours of $v$.
\end{definition}

\begin{remark}
	Note that our definition does not allow for directed edges, as we define an edge to be a set, not a tuple. Furthermore, since we explicitly require the set to contain 2 elements, it does not allow for self-loops. Finally, as $E$ is a set of edges, there is at most one edge between a pair of vertices. As a result there are no parallel edges. A graph with these restrictions is also called a \Define{simple graph}\indexd{graph!simple ---}.
\end{remark}

\begin{definition}\label{def:graph-form}
  A ZX-diagram is \Define{graph-like} when:
  \begin{enumerate}
    \item All spiders are Z-spiders.
    \item Z-spiders are only connected via Hadamard edges.
    \item There are no parallel Hadamard edges or self-loops.
    \item Every input and output wire is connected to a Z-spider and every Z-spider is connected to at most one input wire and at most one output wire.
  \end{enumerate}
\end{definition}

\begin{lemma}\label{lem:all-zx-are-graph-like}
  Every ZX-diagram is equal to a graph-like ZX-diagram.
\end{lemma}
\begin{proof} Starting with an arbitrary ZX-diagram, we apply \HadamardRule to turn all X-spiders into Z-spiders surrounded by Hadamard gates. We then remove excess Hadamards via \HCancelRule. Any non-Hadamard edge is removed by fusing the adjacent spiders with \SpiderRule. Self-loops are removed by applying Eq.~\eqref{eq:self-loops} while parallel Hadamard edges are removed by Eq.~\eqref{eq:hopf-law-h}.

At this point, the first 3 conditions are satisfied. To satisfy condition 4, we must deal with two special cases: (a) inputs/outputs not connected to any Z-spider, and (b) multiple inputs/outputs connected to the same Z-spider. For case (a), there are only two possibilities left: either an input and an output are directly connected (i.e. a `bare wire'), or they are connected to a Hadamard gate. These situations can both be removed by right-to-left applications of \IdentityRule and \HHRule as follows:
\ctikzfig{ident-graph-form-2}
For case (b), we can again use \IdentityRule and \HHRule to introduce `dummy' spiders until each input/output is connected to a single spider:
\ctikzfig{ident-graph-form}
Once this is done, the resulting ZX-diagram satisfies conditions 1-4.
\end{proof}

\noindent A useful feature of a graph-like ZX-diagram is that much of its structure is captured by its underlying \emph{open graph}.

\begin{definition}\label{def:open-graph}\indexd{graph!open ---}\indexd{open graph}
  An \Define{open graph} is a triple $(G,I,O)$ where $G = (V,E)$ is a graph, and $I \subseteq V$ is a set of \Define{inputs}\indexd{open graph!input} and $O \subseteq V$ a set of \Define{outputs}\indexd{open graph!output}. For a graph-like ZX-diagram $D$, the \Define{underlying open graph} $G(D)$\indexd{ZX-diagram!underlying open graph} is the open graph whose vertices are the spiders of $D$, whose edges correspond to Hadamard edges, and whose sets $I$ and $O$ are the subsets of spiders which have respectively input and output wires.
\end{definition}

\begin{figure}[!tb]
  \centering
  \scalebox{0.9}{\input{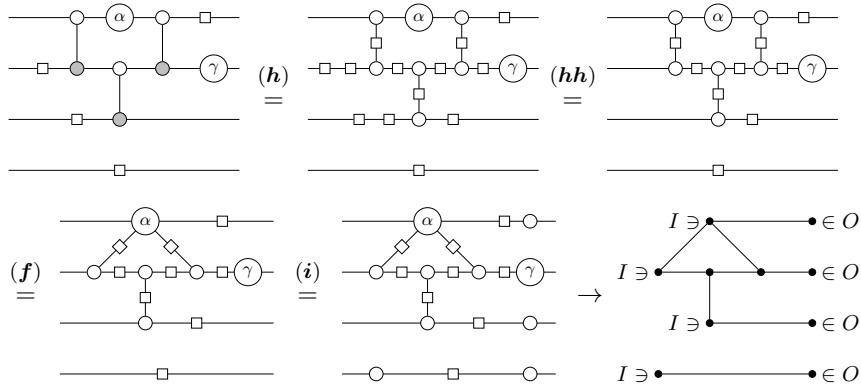}}
  \caption{A circuit, which is transformed into an equivalent graph-like ZX-diagram, and its underlying open graph.}
    \label{fig:underlying-graph}
\end{figure}

Indeed a graph-like ZX-diagram can be seen as an open graph with an assignment of angles to each of its vertices. Note that the sets $I$ and $O$ do not have to be disjoint. See Figure~\ref{fig:underlying-graph} for an example of a ZX-diagram transformed into a graph-like diagram.

It will be useful to introduce some further notation when dealing with (open) graphs.
\begin{definition}
  Let $(G,I,O)$ be an open graph. We write $\comp I:=V\setminus I$ and $\comp O:=V\setminus O$ for respectively the \Define{non-inputs}
  \indexd{vertex!non-input ---}\index{math}{O@$\comp O$ (non-input)} and \Define{non-outputs}
  \indexd{vertex!non-output ---}\index{math}{I@$\comp I$ (non-output)}. Vertices that are neither inputs nor outputs are called \Define{internal} vertices\indexd{vertex!internal ---}.
\end{definition}

Graph-like diagrams are a generalisation of \emph{graph states}~\cite{graphstates}, a well-studied class of quantum states. These states turn out to have a straightforward representation as ZX-diagrams.
\begin{definition}\label{def:graph-state}
	A \Define{graph state}\indexd{graph state} is a graph-like ZX-diagram that has no inputs, no internal vertices and no non-zero phases.
\end{definition}

Because a graph state has no internal vertices and no inputs, every spider must be connected to exactly one output. As no non-zero phases are allowed, a graph state is then completely specified by the underlying graph, and conversely any graph defines a graph state. For a given graph $G$ we write $\ket{G}$ for its corresponding graph state.

Being a graph state is quite restrictive, so we will often use a more relaxed notion:

\begin{definition}\label{def:GSLC}
	A \Define{graph state with local Cliffords} (GS-LC)\indexd{graph state!--- with local Cliffords}\index{math}{GS-LC (graph state with local Cliffords)} is a ZX-diagram that consists of a graph state composed with single qubit Clifford unitaries.
\end{definition}

Recall that a Clifford unitary is any unitary that can be constructed from a quantum circuit consisting of S, H and CNOT gates. Local Cliffords on the other hand are constructed only by using S and H gates. In the ZX-calculus a local Clifford unitary can always be represented by a sequence of arity 2 spiders with phases that are multiples of $\pi/2$.

\begin{example}
	On the left a graph state and on the right the same graph state composed with some local Cliffords to produce a diagram in GS-LC form.
	\ctikzfig{graph-state-ex}
\end{example}

The importance of graph states with local Cliffords comes from the following result:
\begin{theorem}[\cite{vandennest2004graphical}]\label{thm:clifford-state-GS-LC}
	Any Clifford state is equal to some GS-LC state.
\end{theorem}
We will actually reprove this result using the ZX-calculus in Section~\ref{sec:clifford-circuit-optimisation}. The main tool for doing so is an operation on graphs that is described in the next section.

\section{Local complementation and pivoting}\label{sec:local-complementation}

Local complementation is a graph transformation introduced by
Kotzig \cite{kotzig}.
\begin{definition}
  Let $G$ be a 
  graph and let $u$ be a vertex of $G$. 
  The \Define{local complementation of $G$ about $u$}\indexd{local complementation of $G$ about $u$}, 
  written as $G\star u$,\index{math}{Gstaru@$G\star u$ (local complementation)} is a graph which has the same vertices and edges as $G$, 
  except that all the neighbours $v,w$ of $u$ are connected in $G\star u$ if and only if they are not connected in $G$.
\end{definition}

\begin{example}\label{ex:local-complementation}
	An example of local complementations about the vertices $a$ and $b$ in the graph $G$.
\begin{equation*}
G\quad\input{graph1-lab.tikz}\qquad\qquad G\star a\quad\input{graph1-lab-1.tikz}\qquad\qquad (G\star a) \star b\quad\input{graph1-lab-2.tikz}
\end{equation*}
\end{example}

Interestingly, if we have a graph state $\ket{G}$, we can get a graph state $\ket{G\star u}$ just by applying some local Cliffords on the original graph state. This was originally shown in Ref.~\cite{vandennest2004graphical}. It was subsequently proved in the ZX-calculus in Ref.~\cite{DP1}. We will follow an adaptation of this proof provided in Ref.~\cite{CKbook}.

\begin{definition}
    We define $K_n$ to be the \Define{fully connected ZX-diagram}\indexd{ZX-diagram!fully connected ---}\index{math}{Kn@$K_n$(fully connected ZX-diagram)} on $n$ qubits, defined recursively as:
    \begin{equation}\label{eq:fully-connected-def}
        \input{fully-connected-def.tikz}
    \end{equation}
\end{definition}

When we fuse all the spiders in the $K_n$ we see that they indeed give totally-connected graphs of Hadamard edges:
\ctikzfig{fully-connected-ex}

Using this definition of $K_n$ we can state the equality that will allow us to do local complementations:\indexd{local complementation!in ZX-calculus}
\begin{lemma}[{\cite[Lemma~9.128]{CKbook}}]\label{lem:loc-comp-zx}
	The following holds in the ZX-calculus for all $n\in \N$:
    \begin{equation}\label{eq:lc-zx}
        \input{LC-zx.tikz}
    \end{equation}
\end{lemma}

Before we prove this, see Example~\ref{ex:local-complementation-zx} for a demonstration of how this is related to doing local complementations. The crucial point is that the introduction of a fully connected graph by Eq.~\eqref{eq:lc-zx} makes a parallel edge if there was already an edge present, which is then subsequently removed by Eq.~\eqref{eq:hopf-law-h}.

\begin{example}\label{ex:local-complementation-zx}
	Let us take the graph $G$ from Example~\ref{ex:local-complementation}, but now seen as the graph state $\ket{G}$.
	\ctikzfig{graph1-lab-zx}
	We indeed end up with $\ket{G\star a}$ (up to local Cliffords).
\end{example}

For the proof of Lemma~\ref{lem:loc-comp-zx} we will need the following base case.

\begin{lemma}[{\cite[Lemma~9.127]{CKbook}}]
	The following holds in the ZX-calculus:
	\begin{equation}\label{eq:local-comp2}
	   \input{local-comp2.tikz}
	   \end{equation}
\end{lemma}
\begin{proof}~
  \ctikzfig{local-comp2-pf1}
  \[\input{local-comp2-pf2.tikz} \qedhere\]
\end{proof}

\begin{proof}[Proof of Lemma~\ref{lem:loc-comp-zx}]
Note that for $n=0$ and $n=1$ this equation becomes:
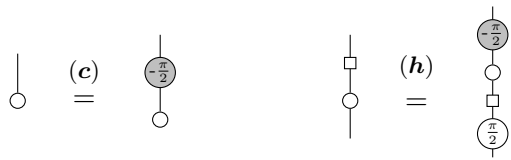
\ctikzfig{LC-zx-pf0}
The previous lemma shows $n=2$.
We now proceed by induction.
For our induction hypothesis, assume \eqref{eq:lc-zx} holds for some fixed $n\geq 2$, which we indicate as (ih) below. Then, for $n+1$ we calculate:  
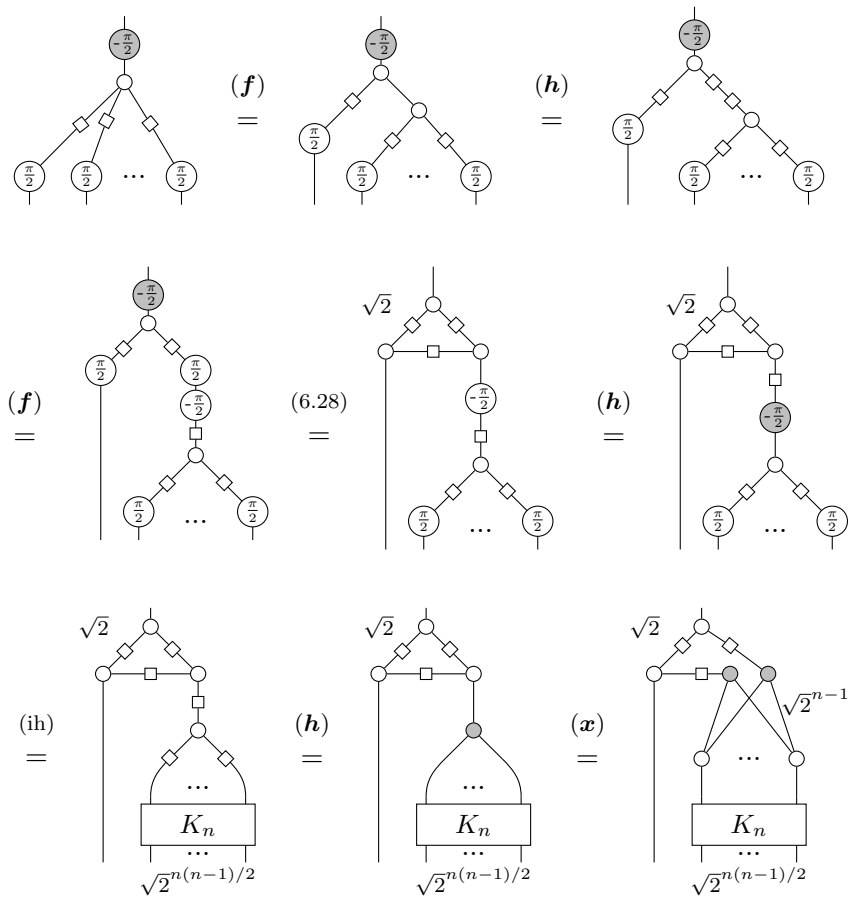
\ctikzfig{LC-zx-pf1}
\ctikzfig{LC-zx-pf2}
\ctikzfig{LC-zx-pf3}
\[\input{LC-zx-pf4.tikz} \qedhere\]
\end{proof}

Related to local complementation is the operation of pivoting.
\begin{definition}
  Let $G$ be a 
  graph, and let $u$ and $v$ be a pair of connected vertices in $G$. 
  The \indexd{pivot}\Define{pivot of $G$ along $uv$}, written as $G\wedge uv$, is the graph $G\wedge uv := G\star u \star v \star u$. \index{math}{Gwedgeuv@$G\wedge uv$}
\end{definition}
Although it is not obvious from the definition, because we have $G\star u\star v \star u = G\star v\star u\star v$, we do not care about the order of $u$ and $v$ when doing a pivot along the edge $uv$.

On a graph, pivoting consists in exchanging $u$ and $v$, and complementing the edges
between three particular subsets of the vertices: the common neighbourhood of $u$ and $v$ (i.e.~$N_G(u)\cap N_G(v)$), the exclusive neighbourhood of $u$ (i.e.~$N_G(u)\setminus (N_G(v)\cup \{v\})$), and exclusive neighbourhood of $v$ (i.e.~$N_G(v)\setminus (N_G(u)\cup \{u\})$):
\[G \quad\input{pivot-L.tikz}\qquad\qquad \quad G\wedge uv \quad\input{pivot-R.tikz}
\]
For a more concrete illustration of pivoting see Example~\ref{ex:pivot}.
\begin{example}\label{ex:pivot}
In the graph $G$ below, $\{a, b\}$ is in the neighbourhood of $u$ alone, $\{d\}$ is in the neighbourhood of $v$ alone, and $\{c\}$ is in the the neighbourhood of both. To perform the pivot along $uv$, we complement the edges connecting $\{a, b\}$ to $\{d\}$, $\{d\}$ to $\{c\}$ and $\{a, b\}$ to $\{c\}$. We then swap $u$ and $v$.
\[
G\quad\input{graph2.tikz} \qquad \quad G\wedge uv\quad\input{graph2-pivot.tikz}
\]
\end{example}

As a pivot is just three local complementations, a pivot can also be done on graph states using local Cliffords. It will however be useful to prove this fact more directly.

\begin{lemma}\indexd{pivot!in ZX-calculus}
	The following holds in the ZX-calculus:
	\begin{equation}\label{eq:pivot-desc}
	\input{pivot-desc.tikz}
	\end{equation}
	I.e.\ we can connect every vertex in the group of $n_1$ to those in $n_2$ and $n_3$ (and similarly for $n_2$), at the cost of introducing a swap and a Hadamard on the outputs of $u$ and $v$, and a $\pi$ phase on each of the vertices in $n_3$.
\end{lemma}
\begin{proof}
    First, we apply a colour change on $v$:
	\ctikzfig{pivot-desc-pf1}
    Now, comes the crucial step: we apply \HopfRule on $u$ and $v$:
	\ctikzfig{pivot-desc-pf2}
    This has resulted in many connected Z spiders which need to be fused again. For the spiders in the group of $n_3$ this will lead to self-loops and parallel edges that we deal with in the final step:
	\ctikzfig{pivot-desc-pf3}
	In this last step we removed $n_3$ self-loops, giving the term $(\frac{1}{\sqrt{2}})^{n_3}$, and we removed a parallel edge between every pair of vertices in $n_3$, meaning we have $n_3(n_3-1)/2$ parallel edges each giving a scalar of $\frac12$. It is easily checked that the final scalar multiplies out to the stated value in the lemma. 
\end{proof}

\section{Graph-theoretic simplification}\label{sec:graph-theoretic-simp}

The local complementation and pivoting rules of the previous section have no obvious directionality to them: it is not a priori clear in which order, left-to-right or right-to-left, the rules should be applied in order to make a diagram simpler. In this section we will introduce variations on these rules that \emph{do} have an obvious directionality. Versions of these rules will be used in Chapter~\ref{chap:MBQC} to simplify measurement patterns, but as an immediate consequence we will show here how they can be used to prove a version of the Gottesman-Knill theorem.\indexd{Gottesman-Knill theorem}

Our first rule uses local complementation. If the vertex we do the complementation about has a phase of $\pi/2$ or $-\pi/2$, and no output wire, then doing a complementation allows us to remove this vertex:
\begin{lemma}\indexd{local complementation!as simplification rule}~
 	\begin{equation}\label{eq:lc-simp}
 		\input{lc-simp.tikz}
 	\end{equation}
\end{lemma}
\begin{proof}
  We pull out all of the phases via \SpiderRule then apply the local complementation rule~\eqref{eq:lc-zx}:
  \ctikzfig{lc-simp-proof}
  Using Eq.~\eqref{eq:s-state-eq}, the topmost spider in the right-hand side above becomes an X-spider, with phase $\mp \pi/2$, which combines with the phase below it into an $a\pi$ phase, where $a=0$ if we started with $\pi/2$ and $a=1$ if we had started with $-\pi/2$. The resulting  X-spider copies and fuses with the neighbours:
  \ctikzfig{lc-simp-proof-2}
\end{proof}

The second rule uses pivoting. If we have a connected pair of vertices that each has a $0$ or $\pi$ phase, then doing a pivot about this pair allows us to remove them.
\begin{lemma}\indexd{pivot!as simplification rule}~
	\begin{equation}\label{eq:pivot-simp}
	\input{pivot-simp.tikz}
	\end{equation}
\end{lemma}
\begin{proof}
   We pull out all of the phases via \SpiderRule and apply the pivot rule~\eqref{eq:pivot-desc}:
   \ctikzfig{pivot-simp-proof}
   We then apply the colour-change rule to turn the Z-spiders with phases $j\pi$ and $k\pi$ into X-spiders. They can then be copied, colour-changed again and fused with their neighbours:
   \[\scalebox{0.95}{\input{pivot-simp-proof-2.tikz}}\]
   Note that the dangling scalar diagram appears because we copy twice and the vertices are connected. Using Eq.~\eqref{eq:zx-scalars} we see it is equal to $(-1)^{jk}\sqrt{2}$.
   It is straightforward to verify that the scalars multiply out as described.
\end{proof}

In the next chapters we will often refer to `applying' these rules (or similar ones) to certain vertices. By that we mean that the designated vertex (vertices) plays the role of the $\pi/2$ vertex (the $0$/$\pi$ vertices) in Eq.~\eqref{eq:lc-simp} (resp.~\eqref{eq:pivot-simp} and that we apply the rewrite rule from left to right.

Let us now demonstrate an immediate use-case for these rules.

Recall that the Gottesman-Knill theorem states that any quantum computation involving only Clifford operations can be efficiently simulated on a classical computer. We will show how this result can be rederived using the simplification rules above.\indexd{Gottesman-Knill theorem}

Specifically, we will show that given a Clifford circuit $C$ on $n$ qubits, we can efficiently find the amplitude $\bra{0\cdots 0}C\ket{0\cdots 0}$. We do this by writing that amplitude as a ZX-diagram, and then progressively simplifying this diagram until we can easily read of the scalar value. As the circuit is Clifford, the ZX-diagram will only contain phases that are multiples of $\pi/2$, which allows us to use these previous rules to maximal effect.\indexd{simulation!Clifford ---}

\begin{theorem}[Gottesman-Knill]
	Let $C$ be a Clifford circuit on $n$ qubits consisting of $k$ gates in the gate set $\{\text{CNOT},\text{H},\text{S}\}$. Then the amplitude $\bra{0\cdots 0}C\ket{0\cdots 0}$ can be exactly calculated classically using $O(k^3)$ elementary graph operations.
\end{theorem}
\begin{proof}
	Write the scalar ZX-diagram corresponding to $\bra{0\cdots 0}C\ket{0\cdots 0}$. Transform it into a graph-like diagram by applying Lemma~\ref{lem:all-zx-are-graph-like}. As there are no inputs and outputs, every spider is only connected to other spiders and always via a Hadamard edge. Keep applying Eq.~\eqref{eq:lc-simp} as long as there are spiders with a $\pm \pi/2$ phase to remove. The resulting diagram then must only have spiders with a $0$ or $\pi$ phase left. Isolated spiders can be easily written as scalars. If a spider is connected to another spider then we can remove the pair by applying Eq.~\eqref{eq:pivot-simp}. Doing such a pivot only changes phases by multiples of $\pi$ and hence does not introduce new spiders with phase $\pm \pi/2$. We conclude that the entire diagram can be reduced to a scalar, which gives the desired amplitude.

	For the time complexity, we note that constructing the ZX-diagram and writing it as a graph-like diagram takes time linear in the number of gates in the circuit, with the resulting graph having $O(k)$ vertices. Each local complementation or pivot could potentially change the connectivity of the entire graph and hence requires $O(k^2)$ elementary graph operations. As each pivot and local complementation removes a vertex we require at most $O(k)$ of these operations, and hence we see that we need at most $O(k^3)$ elementary graph operations.
\end{proof}

\begin{remark}
	The complexity $O(k^3)$ is pessimistic. Most graph operations will only involve a small part of the graph. This can be readily seen because a local complementation toggles the connectivity, and hence a highly connected subgraphs becomes sparse in the next step.
	Other versions of this result (such as Ref.~\cite{aaronsongottesman2004}) usually require linear time in the number of gates, and time $O(n^3)$ in the number of qubits. The reason we get a bound in terms of the number of gates is because we don't specify a strategy for doing the local complementations. We conjecture that if one does the rewrites in a more or less `chronological' order, that one gets the known time bounds.
\end{remark}

\begin{remark}
	We have actually not shown the full power of the Gottesman-Knill theorem: we have established that we can find single amplitudes, but we have not shown how to calculate marginal probabilities. For Clifford circuits this problem is easily addressed by using the \emph{CPM construction}, also known as \emph{doubling} the ZX-diagram. See for instance Ref.~\cite[Chapter~6]{CKbook}
\end{remark}

Using the ZX-calculus for circuit simulation can also be done for non-Clifford circuits. This is explored in more detail in Section~\ref{sec:simulation}.

\chapter{A simple model of computation}\label{chap:ms-mbqc}
\indexd{measurement-based quantum computation}\index{math}{MBQC (measurement-based quantum computation)}
This chapter and the next deal with \Define{measurement-based quantum computation} (MBQC). Unlike the circuit model of quantum computation, MBQC has no classical counterpart. 
There are a variety of different models of MBQC that work in different ways, but they all share some common aspects. They all start with some specific intricate resource state. The program that one wishes to execute consists of doing measurements on this state. Crucially, later choices of measurements may depend on previous measurement outcomes, a concept known as \Define{feed-forward}\indexd{feed-forward}. So while a computation in the circuit model is done by a series of unitary quantum gates, in MBQC it is done by implementing a specific series of outcome-dependent measurements on some resource state.

The most widely studied model of MBQC is called the \emph{one-way model}~\cite{MBQC1}. In this model the resource state is a graph state (cf.~Definition~\ref{def:graph-state}), and all the measurements are single qubit measurements restricted to specific planes of measurements. The one-way model is the topic of study of Chapter~\ref{chap:MBQC}. 
Because the resource state is a Clifford graph state, the power of universal quantum computation in the one-way model comes from doing measurements in a non-Clifford basis. In fact, any computation involving Pauli measurements can be done in a single time-step~\cite{MBQC2}.

It is natural to ask if we can invert this problem: is it possible to obtain universal computation by means of a non-Clifford resource state and just Pauli measurements? 
There are several ways to achieve this. For example, one could consider resource states which are prepared just like graph states, but with certain qubits prepared in a $\ket{T}$ magic state rather than the usual $\ket{+}$ state~\cite{danos2007pauli}. 
One can also consider \emph{hypergraph states}~\cite{gachechiladze2018changing}, a generalisation of graph states produced by multi-qubit $n$-controlled-$Z$ operations, represented graphically as hyper-edges. These were shown to admit a universal model of computation using Pauli measurements and feed-forward~\cite{takeuchi2018quantum}. A different approach was taken in \cite{miller2016hierarchy}, where a resource state was created that allowed non-deterministic approximately universal computation using just X, Y and Z measurements.

In this chapter, we introduce a new family of generalisations of graph states which admit universal deterministic computation using only Pauli X and Z measurements and feed-forward. We call these \Define{parity-phase} graph states, or P-graph states. Edges in P-graph states represent an application of the following \Define{parity-phase gate}, for some fixed angle $\alpha$:\indexd{parity phase gate}\indexd{gate!parity phase ---}
\[\hfill P(\alpha) = \exp(-i\frac\alpha2 Z\otimes Z) \hfill\]
We refer to this as a parity-phase gate because it introduces a relative phase of $\alpha$ between its even-parity eigenstates $\ket{00}, \ket{11}$ and its odd parity eigenstates $\ket{01}, \ket{10}$. We introduced such gates as phase gadgets in Section~\ref{sec:phasegadgets}.

These parity-phase gates are a popular primitive two-qubit entangling gate in certain hardware implementations of qubits, such as with ion trap qubits (via M\o{}lmer S\o{}rensen interactions~\cite{iontrapoxford,iontrapcolorado}), or with transmon-based superconducting qubits~\cite{superconductingdelft}. At the end of Section~\ref{sec:universal}, we comment briefly on near-future prospects of implementing this scheme using the latter.

When $\alpha = \frac\pi2$, we obtain resource states which are equivalent to graph states up to local Clifford operations. However, if $\alpha = \frac\pi4$, we can construct resources which are approximately universal for quantum computation using only single-qubit Pauli $X$ and $Z$ measurements and feed-forward. We call this the \Define{PPM model}\indexd{PPM model}, for \emph{parity-phase with Pauli measurements}.

A key feature which distinguishes P-graph states from standard graph states is that the entangling operation $P(\alpha)$, which is a CZ gate in the case of graph states, satisfies $P(\alpha)P(\alpha) \neq \id$ (except in the degenerate case where $\alpha = \pi$). Hence it is possible, and even desirable, to consider resource states described by graphs with multiple, parallel edges between nodes. For example:
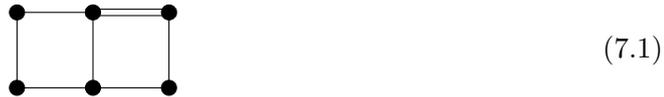
\begin{equation}\label{eq:first-ex}
    \input{ms-graph-state.tikz}
\end{equation}
These parallel edges correspond to multiple applications of the entangling gate $P(\alpha)$ to the adjacent qubits. For example, a doubled edge above indicates the application of $P(\alpha)^2 = P(2\alpha)$. In the graph theory literature, graphs such as \eqref{eq:first-ex} are sometimes referred to as \Define{undirected multigraphs}.

In the PPM model, doubled edges play a special role. Since $P(\frac\pi4)^2 = P(\frac\pi2)$ is equivalent, up to local Clifford operations, to a controlled-Z gate, subgraphs of a P-graph state containing only doubled edges behave in much the same way as traditional graph states. However, P-graph states additionally yield the ability to selectively inject $\pi/4$ phases into computations via nodes connected by single edges. One way to conceptualize this fact is to consider the two-qubit gates $P(\pi/4)$ as introducing `virtual' magic states between pairs of qubits. The phase data carried by this `virtual' magic state can either be destroyed or injected on to one of the neighbouring qubits, depending on the measurement choices, using a method similar to that of for instance~Ref.~\cite{bravyi2005universal}.

Notably, this dichotomy gives a clean separation of the efficiently simulable parts of the computation and the rest. In deriving a universal scheme for computation with P-graph states, we will note that feed-forward is only required in the vicinity of single edges. So, much like the case in the one-way model, the entire `Clifford part of the computation' can be done in a single time step.

In order to prove the correctness of our measurement patterns, we will use the ZX-calculus to reason about P-graphs. We will demonstrate the possibility of deterministic computation by `pushing' Pauli errors forward from measurements to qubits in future, which can be corrected. However, unlike previous work, we rely on the extra flexibility of ZX-diagrams to represent non-Clifford correlations between qubits and develop techniques for `pushing' errors through these edges using the ZX-calculus. 
As we shall see, the diagrams keep track of the extra (Clifford) errors introduced by propagating errors forward, and it enables us to derive a technique for performing Pauli and Clifford corrections purely by means of single-qubit measurement choices in the bases $\{ X, Z \}$.
This yields a measurement-based model which is very flexible. To give some evidence of this flexibility, we show in Section~\ref{sec:clifhier} how to generalise to P-graph states where a single edge denotes an application of $P(\frac{\pi}{2^{n-1}})$. This enables us to incorporate a familiar `trick' (see e.g.~\cite[Section III]{Gottesman}) into the model to deterministically implement any diagonal gate of the $n$-th level of the Clifford hierarchy.

To give a proof of universality, we introduce `hairy brickwork states', which are inspired by the brickwork states introduced in Ref.~\cite{brickworkuniversal} for universal computation in the one-way model. 

Alternatives to the one-way model have been considered, notably in a broad range of models by Gross \emph{et al.}~\cite{GrossEisartBeyond}, which include a variation on graph states called \emph{weighted graph states}, whose two-qubit interactions are equivalent to $P(\alpha)$ for values of $\alpha$ different from $\pi/2$, up to local unitaries.  However, our approach is different in two important ways. First, we use only very limited measurements, and second, our scheme is deterministic using feed-forward, eliminating the need for the `trial-until-success' strategies used by Ref.~\cite{GrossEisartBeyond}. In fact, our model was the first to offer an approximately universal model of MBQC using only Pauli X and Z measurements. A modification of the one-way model that allows for an additional operation to directly inject phases on qubits was considered in Ref.~\cite{danos2007pauli} where they showed this gives an approximately universal model requiring only Pauli X and Y measurements.

In Ref.~\cite{miller2016hierarchy} a model based on hypergraph states is constructed that only needs Pauli measurements to become universal, but its structure is more complex than ours and the protocols used are not deterministic. In later work, namely Refs.~\cite{takeuchi2018quantum} and \cite{gachechiladze2018changing} they do have a deterministic model using hypergraph states and Pauli measurements.
However, our protocol remains interesting for several reasons. First, parity-phase interactions are typically more primitive, in that they have simpler realisations within the gate sets of current hardware proposals. Second, the universal gate set we produce in our model is Clifford+T (or more generally, Clifford + arbitrary diagonal Clifford-hierarchy gates), as opposed to CCZ+Hadamard. While the latter is also universal, it requires extra overhead for encoding computations in a higher-dimensional space~\cite{ShiToffoliHadamard}. 
Third, and perhaps most importantly, we introduce a drastically different methodology to existing approaches. This yields a rather flexible family of models that enable us to explore a variety of multi-qubit interactions and graph topologies. Indeed it is a topic of active research to extend these techniques to hypergraph-based models, where the role of the ZX-calculus is played by the ZH-calculus~\cite{backens2018zhcalculus} (cf.~Section~\ref{sec:ZH-calculus}).

This chapter is structured as follows. In the next section we describe the PPM model. We give a description of this model in the ZX-calculus in Section~\ref{sec:ppm-zx}. Then we describe the implementation of several gates in the PPM model in Section~\ref{sec:ppm-patterns}. This leads to a proof of universality of our model in Section~\ref{sec:universal}. Then in Section~\ref{sec:clifhier} we discuss how our model can be generalised to allow gates from any level of the Clifford hierarchy. We conclude with some remarks in Section~\ref{sec:ppm-conclusion}.

\section{The PPM Model}
In this section we will give a full description of the PPM model. In Section~\ref{sec:ppm-zx} we will see how we can cast all the components of the model in the language of the ZX-calculus.

A \Define{P-graph state}\indexd{P-graph state} is described by an undirected multigraph. In practice, we only consider graphs that have just single and double edges:
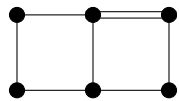
\ctikzfig{ms-graph-state}
A single edge describes the application of an $P(\pi/4)$ gate, whereas a double edge describes the application of an $P(\pi/2) = P(\pi/4)^2$ gate. 

A \Define{measurement pattern}\indexd{measurement pattern!in PPM model} is a P-graph state where each node is labelled by a \Define{measurement expression}\indexd{measurement expression} of the form ``$b \leftarrow \phi(a_1, \ldots, a_n)$'' where $b$ is a fresh variable called the \Define{output value} and $\phi$ is a classical function from boolean variables $a_1, \ldots, a_n$ to a single boolean value. In this case, we say for each $a_i$ that $b$ \Define{depends} on $a_i$. A pattern is \Define{well-founded} if there are no cyclic dependencies between variables, such as in the following pattern:
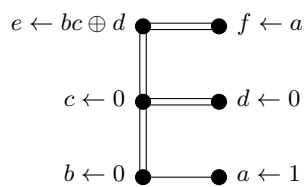
\ctikzfig{ms-pattern}

These expressions do not have any explicit time-ordering, but there are restrictions on the order in which measurements can be made, due to dependencies on prior outcomes. For instance, in the pattern above the qubit labelled $a\leftarrow 1$ needs to be measured before the qubit labelled $f\leftarrow a$, as the variable $a$ is introduced by the measuring of this first qubit. As a matter of convention, we will typically draw earlier measurements below later ones, i.e. `time' flows upward.

Computations are performed as follows:
\begin{enumerate}
  \item A qubit is initialised in the $\ket +$ state for each vertex in a P-graph state.
  \item For every edge in the graph, $P(\frac\pi4)$ is applied to the two qubits at its source and target. In particular, $P(\frac\pi2) = P(\frac\pi4)^2$ is applied to every pair of qubits connected by a double edge.
  \item For a qubit labelled ``$b \leftarrow \phi(a_1, \ldots, a_n)$'', where the values $a_1, \ldots, a_n$ are known, measure in the $X$-basis if $\phi(a_1, \ldots, a_n) = 0$ and the $Z$-basis otherwise. In either case, store the measurement result, either a $0$ or a $1$, in $b$.
  \item Optionally, perform some classical post-processing on the measurement results, in order to interpret the outcomes correctly.
\end{enumerate}
Note that, as with the one-way model, the two-qubit gates $P(\alpha)$ applied in step 2 all commute. The order of application is therefore irrelevant, and the undirected graph structure is indeed sufficient to describe the model.

To show that this model is universal, we will compose smaller patterns into larger ones. In order to do this, we give a notion of \Define{pattern fragment}\indexd{pattern fragment} analogous to the notion given for the one-way model. A pattern fragment is just like a measurement pattern, with the exception that we additionally identify two (not necessarily disjoint) subsets of vertices $I, O \subseteq V$ which respectively correspond to inputs and outputs. Inputs correspond to qubits that can be in an arbitrary state, rather than the fixed state $\ket{+}$. Outputs correspond to qubits which remain unmeasured after the application of the pattern-fragment.

Each vertex in $I$ is labelled with an \Define{input error expression}\indexd{error expression} of the form: ``$(z, x) \leftarrow \square$'' for fresh variables $z$ and $x$, which capture whether a Z or X error is being fed forward into this vertex. Unless the input vertex is also an output, the vertex will also be labelled with a measurement expression.
Measurement choice functions $\phi$ in the fragment are allowed to depend on the input errors as well as other measurement outcomes present in the fragment.

Each vertex in the output set $O$ is labelled by an \Define{output error expression} of the form: ``$\square \leftarrow (\zeta, \xi)$'' consisting of a pair of classical functions $\zeta, \xi$ which again can depend on the input errors and the results of measurements in the pattern fragment. These functions specify which $z$, respectively $x$, error is being fed forward. Vertices in $O$ are not measured so they do not contain a measurement expression.
\begin{example}
Consider the following pattern fragment:
\ctikzfig{ms-pattern-fragment}
The bottom left qubit is an input, and the top left qubit is an output. Note that the measurement plane of $e$ depends on the input $x$ error.
\end{example}

We say a pattern fragment \Define{implements}\indexd{pattern fragment!--- implementing a gate} a gate $G$ if, for any input state of the form: $X^x Z^z \ket\psi$, performing the pattern fragment yields $X^\xi Z^\zeta G \ket\psi$, i.e.~when regardless of the possible $X$ and/or $Z$ error that exists on the input state $\ket{\psi}$ the pattern fragment always transforms $\ket{\psi}$ to $G\ket{\psi}$, with the exception of some possible known error $X^\xi Z^\zeta$. This notion of implementing a gate extends in the natural way to gates with multiple input and output qubits:
\[\hfill
(X^{x_1}Z^{z_1} \otimes \ldots \otimes X^{x_n}Z^{z_n}) \ket\psi
\ \ \mapsto\ \ 
(X^{\xi_1}Z^{\zeta_1} \otimes \ldots \otimes X^{\xi_m}Z^{\zeta_m}) G \ket\psi
\hfill\]

Composing pattern fragments then results in the composition of their associated gates, hence it suffices for the sake of universality to show we can implement a universal set of quantum gates via pattern fragments.

\section{PPM in the ZX-calculus}\label{sec:ppm-zx}
\indexd{ZX-diagram!PPM model}
In order to express the PPM model in the ZX-calculus, we give graphical presentations for the parity-phase gate and for Pauli measurements. As discussed in Remark~\ref{rem:scalars} we will ignore global phases and global scalars in the ZX-diagrams, and we will use `$=$' to denote equality up to non-zero scalar (instead of writing `\scalareq').

It's not too hard to see that $P(\alpha) = \exp(-i\frac\alpha2 Z\otimes Z)$ is (up to a global phase) of the form of the unitary in Eq.~\eqref{eq:phase-gadget-unitary}. 
We then easily verify that:
\begin{equation}\label{eq:virtual-qubit}
  \hfill P(\alpha) \ =\ \input{zx-msgate.tikz}\hfill
\end{equation}

Pauli Z and X measurements non-deterministically introduce projections onto their respective eigenstates, namely $\{ \bra{0}, \bra{1} \}$ for $Z$-measurements and $\{ \bra{+}, \bra{-} \}$ for $X$-measurements. Hence, we can represent Pauli measurements in the ZX-calculus by:
\begin{equation}\label{eq:meas-effects}
\begin{split}
  \textit{\small Z-measure} & := \left\{ \begin{tikzpicture}
	\begin{pgfonlayer}{nodelayer}
		\node [style=none] (0) at (-0.5, 0) {};
		\node [style=gray phase dot] (1) at (0.25, 0) {$a \pi$};
	\end{pgfonlayer}
	\begin{pgfonlayer}{edgelayer}
		\draw [style=simple] (0.center) to (1);
	\end{pgfonlayer}
\end{tikzpicture}\right\}_{a\in\{0,1\}} \\
  \textit{\small X-measure} & := \left\{\begin{tikzpicture}
	\begin{pgfonlayer}{nodelayer}
		\node [style=none] (0) at (-0.5, 0) {};
		\node [style=white phase dot] (1) at (0.25, 0) {$a \pi$};
	\end{pgfonlayer}
	\begin{pgfonlayer}{edgelayer}
		\draw [style=simple] (0.center) to (1);
	\end{pgfonlayer}
\end{tikzpicture}\right\}_{a\in\{0,1\}}
\end{split}
\end{equation}

\begin{remark}
	Note that we represent a Z measurement with an X-spider, and vice versa. This is because spiders copy states of the opposite colour.
\end{remark}

There are two main ways in which we will use the $P(\alpha)$ gate.
First, the appearance of the `virtual qubit', i.e. the state being input between the two wires in equation \eqref{eq:virtual-qubit} suggests that we should be able to use it as a `magic state' that is waiting to be applied to one of its neighbouring (actual) qubits to introduce a phase.\indexd{magic state} More specifically, if we prepare the second qubit in the $\ket{+}$ state and then measure it in the Z basis, we obtain a Z-phase gate $R_Z(\pm \alpha)$, with the sign depending on the measurement outcome:
\begin{equation}\label{eq:zx-magic-injection}
	\input{zx-msgate-magic.tikz}
\end{equation}

Second, if we take $\alpha=\frac\pi2$ we can rewrite the expression of $P(\frac\pi2)$ in the ZX-calculus a bit further:
\begin{equation}\label{eq:mscx}
  \hfill\input{ms-cnot.tikz} \hfill
\end{equation}
Hence, this gate is equivalent, up to single-qubit Clifford unitaries, to a CZ gate.

Since $P(\alpha)P(\beta) = P(\alpha+\beta)$ we get $P(\frac\pi2) = P(\frac\pi4)^2$, so that the gate $P(\frac\pi4)$ is a $\sqrt{CZ}$ gate, up to local unitaries.

From this point, we will typically suppress explicit references to the spider-fusion rule \SpiderRule, and assume that spiders of the same colour are (un)fused as necessary. Similarly, we will suppress references to \IdRule and \HHRule and simply remove 2-legged spiders and pairs of $H$ gates as they appear.


We can now define the translation from P-graph states and patterns to ZX-diagrams, which is similar in spirit to the one given in Ref.~\cite{DP2} for graph states.
Qubits become white dots with a single output and single/double edges become edges decorated by the appropriate phases as follows:
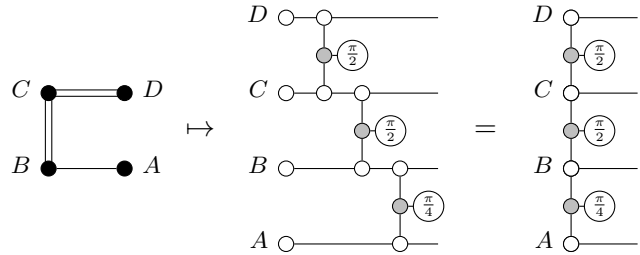
\ctikzfig{ms-translate}

It will be convenient to deform the right-hand side to match the topology of the associated P-graph state, in which case we can drop the qubit labels:
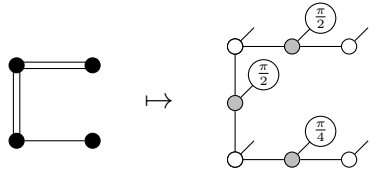
\ctikzfig{ms-translate2}
Note that all of the wires with a free end correspond to outputs, so there is no need to draw them exiting to the right of the diagram.

To compute the result of a measurement pattern, we post-compose with the appropriate effects, $\{ \bra{0}, \bra{1} \}$ for $Z$-measurements and $\{ \bra{+}, \bra{-} \}$ for $X$-measurements, using equation \eqref{eq:meas-effects}. This enables us to write patterns (without feed-forward) as a single ZX-diagram:
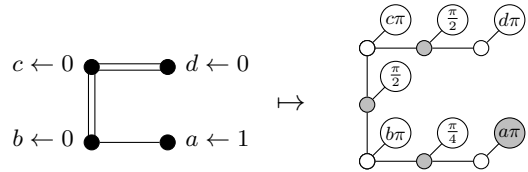
\ctikzfig{ms-translate3}
We could also represent the feed-forward within the diagram (\eg by conditionally applying Hadamard gates to outputs), but for our purposes it will be simpler just to do some simple case-distinctions.

Finally, pattern fragments can be expressed by not measuring outputs, and adding a new input wire for each input:
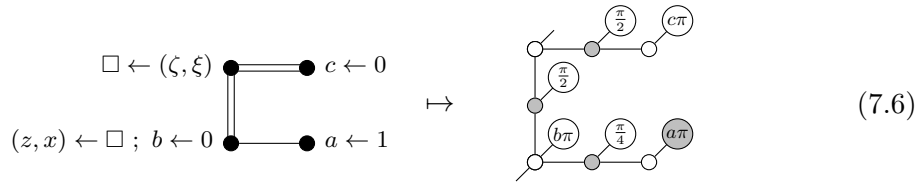
\begin{equation}\label{eq:pattern-translate}
  \hfill\input{ms-translate4.tikz}\hfill
\end{equation}

In order to implement a gate $G$, we should show that the right-hand side above, pre-composed with possible Pauli errors, implements $G$ followed by some possible Pauli errors. We can represent the possible Pauli errors as follows:
\begin{equation}\label{eq:errors}
  \hfill\input{error.tikz}\hfill
\end{equation}
Giving a deterministic implementation of a gate $G$ hence amounts to proving that there exist boolean functions $\zeta, \xi$ such that the following equation holds, for all values of the boolean variables $a, b, c, x, z$:
\ctikzfig{ms-translate5}

\begin{remark}
  Note that the colours play opposite roles in equations \eqref{eq:meas-effects} and \eqref{eq:errors}.
\end{remark}

\section{Measurement patterns for a universal set of gates}\label{sec:ppm-patterns}

\indexd{gate!in PPM model}
In this section, we will introduce several pattern fragments, and show that they deterministically implement certain quantum gates. We will start with a simple example, which uses one double-edge to implement an $X(\pi/2) = HSH$ gate. Following that, we find a different P-graph shape that can be used to selectively implement a CZ or $S \otimes S$. We end the section with a P-graph that, depending on the measurements we do on it, can implement $T$, $H$, or $S$ gates. These patterns will be used in Section~\ref{sec:universal} to establish universality of the PPM model.

The pattern for an $HSH$ gate is:
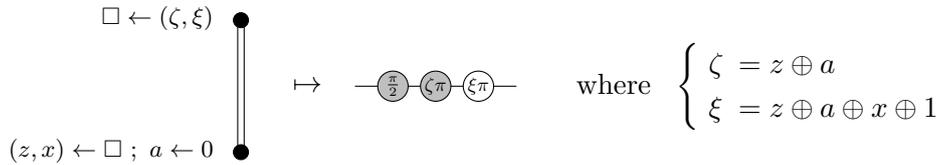
\begin{equation*}\label{eq:pat-HSH}
\input{ms-sgate.tikz}
\qquad\textrm{where}\ \ 
\begin{cases}
  \ \zeta & \!\!\!=  z\oplus a\\
  \ \xi & \!\!\!= z\oplus a\oplus x\oplus 1
\end{cases}
\end{equation*}

The bottom qubit is the input of the expression. We always measure it in the X basis ($a\leftarrow 0$) which gives us a measurement result $a$. We record the incoming $Z$ and $X$ error in the variables $z$ and $x$. The resulting $Z$ error at the end is now $z\oplus a$, and the X error is $z\oplus a\oplus x\oplus 1$. We can show the correctness of this fragment by performing translation \eqref{eq:pattern-translate} and reducing using the ZX-calculus:
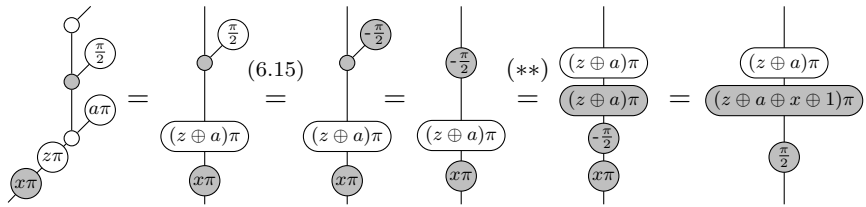
\ctikzfig{ms-sgate-zx}
Here equation $(**)$ is the standard Clifford commutation law of $S^\dagger X \scalareq X Z S^\dagger$, but with the bases interchanged. This commutation follows from \PiRule and the fact that, for $a \in \{0,1\}$, we have $(-1)^a \frac\pi2 = \frac\pi2 + a\pi$ (mod $2\pi$):
\begin{equation}\label{eq:S-commute}
    \hfill\input{S-commute.tikz}\hfill
\end{equation}

\indexd{PPM model!2-qubit gate}As the derivation of some of the other single qubit gate fragments, particularly that of the T gate, is quite tedious, let us first look at the following 2-qubit pattern that implements a CZ-gate:
\begin{equation*}\label{eq:pat-CZ}
\input{ms-cnot2.tikz} 
\qquad\textrm{where}\ \ \begin{cases}
\xi_i & \!\!\!= x_i \\
\zeta_1 & \!\!\!= z_1\oplus x_2\oplus a \oplus b \oplus 1 \\
\zeta_2 & \!\!\!= z_2\oplus x_1\oplus a \oplus b \oplus 1
\end{cases}
\end{equation*}
Note that the top and bottom qubits act as both inputs and outputs, so they are not measured. We measure $a$ in the Z-basis, and $b$ in the X-basis. Writing the resulting diagram out in the ZX-calculus (ignoring incoming Pauli errors for the moment) we get:
\ctikzfig{ms-zx-cnot}
\ctikzfig{ms-zx-cnot2}
Pauli errors propagate through a CZ in the following way:
\ctikzfig{zx-cx-propagate}
and analogously for errors on the other input. Putting the above derivation together with this error propagation gives the pattern fragment as specified above.

This pattern fragment implementing a CZ gate has the additional property that if we measure $b$ in the Z basis instead of the X basis, it \emph{disconnects}. It does not matter in which basis we measure $a$, so let us just choose the X basis for it. This pattern fragment is:
\begin{equation*}\label{eq:pat-SS}
\input{ms-cnot-disconnect.tikz}
\end{equation*}

\noindent This pattern implements an $S\otimes S$ gate:
\ctikzfig{ms-zx-cnot-disconnect}
Here at ($*$) we dropped the dangling scalar diagram. 
We hence see that the choice of measurement basis for $b$ `switches' the CZ-gate on or off.

\indexd{PPM model!T gate}We still need to find an implementation of some single-qubit gates. For this we introduce a more versatile P-graph, shaped like an `E', which can implement a variety of single-qubit gates. The first pattern fragment in the E-shape implements a $R_Z(\pi/4)$ gate, i.e. a $T$ gate:
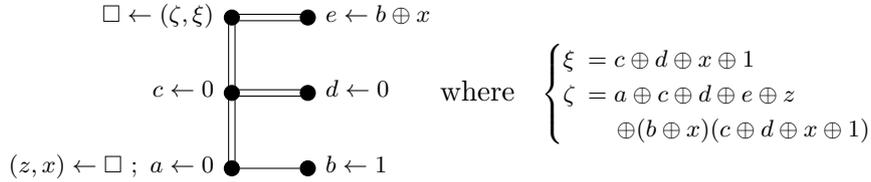
\begin{equation*}\label{eq:pat-T}
\input{ms-pattern-fragment.tikz}
\textrm{where \footnotesize $\ \ 
\begin{cases}
  \xi
  & \!\!\! = c \oplus d\oplus x \oplus 1 \\
  \zeta
  & \!\!\! = a \oplus c \oplus d \oplus e \oplus z \\
  & \ \ \oplus (b \oplus x)(c \oplus d \oplus x \oplus 1)
\end{cases}$}
\end{equation*}

For our analysis of this pattern, we will take the input errors to be $z=x=0$ for brevity. If there are errors present, we can perform a very similar analysis.

Note that the basis in which the qubit $e$ is measured depends on the incoming $X$ error and one of the measurement results in the pattern fragment itself. Let us translate first the P-graph into the ZX-calculus:
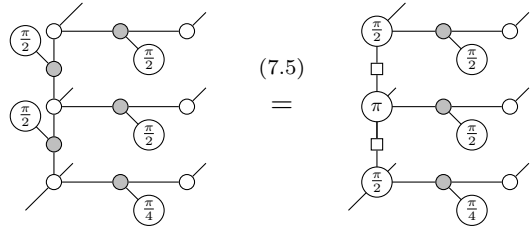
\ctikzfig{ms-zx-tgate1}
Before we incorporate the other measurements, we write down the diagram that results from measuring $a$, $c$ and $d$ in the $X$-basis and $b$ in the $Z$-basis, as these do not depend on any function:
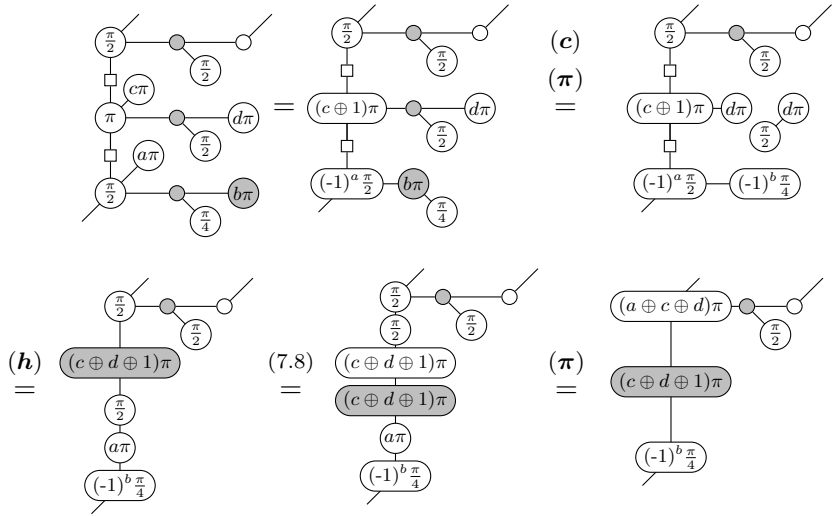
\ctikzfig{ms-zx-tgate2}
The goal of this pattern is to introduce a $\pi/4$ phase. We see that now we either have $\pi/4$ or $-\pi/4$ depending on $b$ (but additionally this value also depends on the incoming X-error that we ignoring for now). Now, if we measure $e$ in the X-basis, it gets cut off the main structure, while if we measure it in the Z-basis it introduces an extra $\pi/2$ phase. So, if we got $\pi/4$ (which is the case in the previous diagram when $b=0$), we measure $e$ in the X-basis:
\begin{equation}\label{eq:ms-zx-tgate3}
\input{ms-zx-tgate3.tikz}
\end{equation}
and otherwise we measure $e$ in the Z-basis:
\ctikzfig{ms-zx-tgate4}
We see that in both cases we indeed implement a $\pi/4$ Z-rotation with some Pauli X and Z error depending on the measurement outcomes. In the presence of a starting Pauli X and Z errors, the same procedure can be done and it will result in the error functions as stated in the pattern fragment. Note that classical control determines the sign of our rotation: if we decide to measure $e$ in the opposite basis (so in the Z basis when $b=0$ and the X basis when $b=1$), we implement a $T^\dagger$ gate. 

Using the same P-graph, but with a different set of measurements on the `hairs' of the fragment we can implement some different operators. For instance, the following pattern fragment gives a Hadamard gate:
\begin{equation*}\label{eq:pattern-had}
\input{ms-pattern-had.tikz}
\qquad\textrm{where} \ \ 
\begin{cases}
\xi & \!\!\!= a \oplus b \oplus c \oplus d \oplus 1\oplus z \\
\zeta & \!\!\!= c \oplus d \oplus e \oplus 1\oplus x
\end{cases}
\end{equation*}
Since there is no feed-forward, we can verify this in a single derivation (and for brevity we will again ignore incoming Z and X errors):
\ctikzfig{ms-zx-had}
In a similar way we can also produce an $S$-gate by measuring qubit $e$ in the Z basis and the rest in the X basis.

\section{Proof of universality}\label{sec:universal}

In the previous section, we have constructed a single P-graph that can implement a T gate and an H gate depending on the chosen measurements. Combining these, we can approximate any single-qubit unitary. We also presented a fragment that implements a CZ gate, and together with the H gate this allows us to make a CNOT, and hence we have an approximately universal set of gates.

\begin{figure}[!tb]
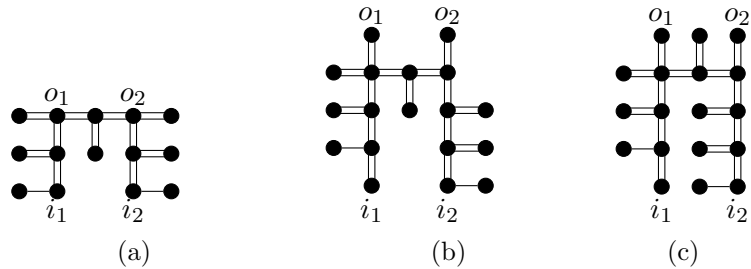

	\centering
	\subcaptionbox{\label{fig:block-tiles-1}}{\input{ms-brick-minimal.tikz} \qquad \qquad}
	\subcaptionbox{\label{fig:block-tiles-2}}{\input{ms-brick-pre.tikz} \qquad \qquad}
	\subcaptionbox{\label{fig:block-tiles-3}}{\input{ms-brick.tikz}}

	\caption{Variations on compositions of the `E'-shaped blocks and the CZ block. Figure a) shows a simple combination, in b) this is slightly expanded to make it asymmetric, and then in c) we reorganize the block so that it tiles better. The $i$'s denote inputs and $o$'s denote outputs. \label{fig:block-tiles}}
\end{figure}

It only remains to combine the fragments of these gates into a configuration that allows us to combine them arbitrarily. We will construct a fragment that is a combination of the simple blocks described in the previous section which fits neatly into a 2D square lattice. See Figure~\ref{fig:block-tiles}.
\indexd{brickwork state}

\begin{figure}[!tb]
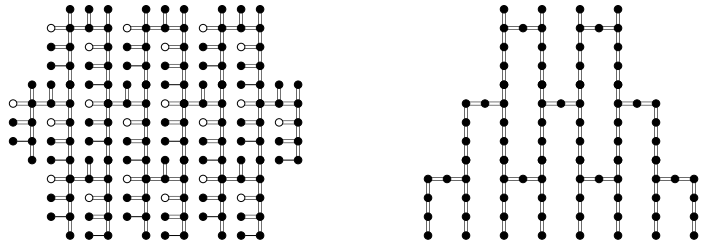

	\[\scalebox{0.50}{\input{ms-brickwork.tikz}} \qquad \qquad \scalebox{0.50}{\input{ms-brick-hairless.tikz}}\]
	\caption{On the left, a P-graph produced by tiling the brick of Figure~\ref{fig:block-tiles-3}. On the right, a pattern with the `hairs' shaved away. \label{fig:brick-tiling}}
\end{figure}

The `bricks' depicted in Figure~\ref{fig:block-tiles}.a) do not fit together in a square lattice, since there is no useful tiling we can produce without some qubits overlapping. We can solve this problem by considering a slightly larger brick, where the E-shape on the right is offset downward and extra double-edges are added to $i_1, i_2$ and $o_2$, giving rise to the pattern in b).
We then reposition some of the qubits on the outside in c) to make it more compact. 
We can picture the paths from $i_1$ to $o_1$ and $i_2$ to $o_2$ as two qubits passing through a circuit. The E-shapes on the left and the right can be used to apply S, T, or H gates depending on the choice of pattern. Similarly, the shape connecting the two qubits can be used to apply CZ or $\text{S} \otimes \text{S}$ to both qubits. The extra edges will always introduce HSH gates. By selecting which gates we actually want to execute, this single block can implement $3 \cdot 3 \cdot 2$ different two-qubit unitaries.
The asymmetry present in the brick allows us to efficiently tile them; see the left side of Figure~\ref{fig:brick-tiling}.

We see that all locations in a square grid are used. Qubits whose measurements could potentially depend on prior outcomes are shown in white. Of those, only the corrections forming part of the T pattern described rely on feed-forward. Hence \emph{all} of the other qubits can be measured simultaneously at the beginning of the computation.

This state can be viewed as consisting of lanes which carry the computation forward, attached to which are `hairs' which introduce extra phases. `Shaving off' all of the hairs reveals a state that has a somewhat similar structure to the brickwork state from Ref.~\cite{brickworkuniversal}; see the right side of Figure~\ref{fig:brick-tiling}.

Taking a more abstract viewpoint the pattern is of the following form:
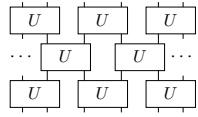
\[\scalebox{0.65}{\input{unitary-brick.tikz}}\]
Such a pattern can implement any Clifford+T circuit. Hence, similar to Ref.~\cite{brickworkuniversal}, we obtain
a model allowing for universal quantum computation.

While our primitive computational `brick' does not seem to be particularly canonical, the fact that it is missing some edges from a square lattice could have advantages when thinking about space-limited architectures. For instance, after dropping the extra `dummy' qubits $i_1$ and $o_1$ of Figure~\ref{fig:block-tiles-3} and a bit of folding, the resulting 16-qubit pattern fits into the 17-qubit `ninja star' design of the superconducting chip proposed in~\cite{superconductingdelft}, which is designed primarily for implementing a 17-qubit surface code: \\
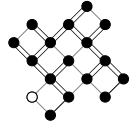
\vspace{-0.7cm}
\begin{equation*}
\scalebox{0.68}{\rotatebox{45}{\input{ms-brick-ninja.tikz}}}
\end{equation*}
A proof of concept for this computational model is hence potentially close at hand. Looking at the way the `E'-shape implements the T-gate we also see that the middle `hair' is actually not necessary. Removing this qubit allows us to fit a universal brick inside the superconducting chip of Ref.~\cite{otterbach2017unsupervised}.

\section{Climbing the Clifford hierarchy}\label{sec:clifhier}

The construction of a deterministic feed-forward strategy in the previous sections relied on the fact that a sign error in a T gate, i.e.\ a $\pi/4$ rotation, can be corrected by applying a $\pi/2$ rotation, and then correcting a sign error in a $\pi/2$ rotation by selectively applying a $\pi$ rotation. Since such a $\pi$ rotation can be commuted past all other gates, the resulting error can be handled at the end of the computation in a classical manner. 

We will now show that this works not only for $\pi/4$, but for all angles $\pi/2^n$ where $2 \leq n \in \mathbb N$. While the P-graphs of the previous sections had a single edge represent a $P(\frac\pi4)$ gate, we will now let a single edge represent $P(\frac{\pi}{2^n})$. If we have $k$ edges between two vertices then this represents a $P(\frac{\pi}{2^n})^k = P(\frac{k\pi}{2^n})$ gate. Now consider the following fragment:
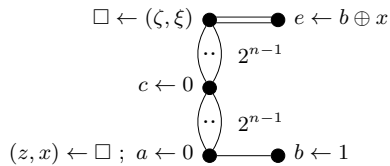
\begin{equation*}
\scalebox{0.9}{\input{ms-hierarchy.tikz}}
\end{equation*}
Here the $2^{n-1}$ refers to the number of wires between the vertices. I.e.~there is a $P(\pi/2)$ gate between the qubits there. Let us simplify the corresponding diagram in the ZX-calculus. We will not yet apply the measurement of the qubit $e$.
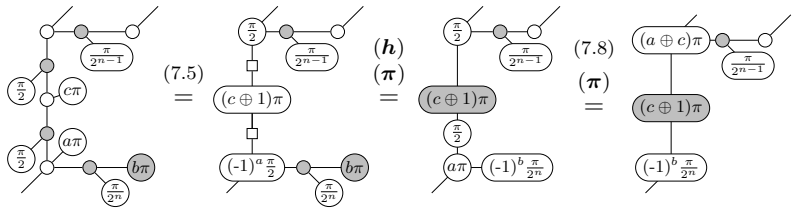
\begin{equation*}
\scalebox{0.9}{\input{ms-hierarchy-proof1.tikz}}
\end{equation*}

If $b=0$, then we have the rotation we want, so we can measure the remaining qubit $e$ in the $X$ basis, and we see that we are left with a $\pi/2^n$ rotation with some Pauli error. If $b=1$, however, we will measure $e$ in the Z-basis, and we calculate:
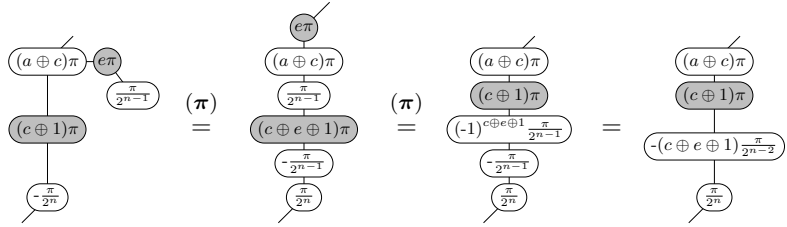
\begin{equation*}
\scalebox{0.9}{\input{ms-hierarchy-proof2.tikz}}
\end{equation*}

If $c\oplus e = 1$, then this is the desired computation. Otherwise, we are left with an unwanted $-\pi/2^{n-2}$ rotation. Since this undesired rotation is heralded by the outcome of a measurement, we can however decide to do a $\pi/2^{n-2}$ in its future to cancel out this rotation using exactly the same procedure. Trying to do this extra $\pi/2^{n-2}$ rotation could then introduce an unwanted $\pi/2^{n-3}$ rotation. After $n-2$ repetitions of this protocol we are therefore left with a $\pi$ error that can safely be incorporated into the classical feed-forward.

The relevant concept to understand this kind of iteration of rotations is that of the \Define{Clifford hierarchy}\indexd{Clifford!--- hierarchy}. The first level of this hierarchy $\mathcal{C}_1$ is defined to be the set of tensor products of the identity and the Pauli unitaries. 
The higher levels are then iteratively defined to be the multi-qubit unitaries that send the Pauli unitaries to a lower level of the hierarchy: $\mathcal{C}_n := \{U ~|~ \forall V\in \mathcal{C}_1.\  UVU^\dagger \in \mathcal{C}_{n-1}\}$. 
It turns out that the Clifford unitaries are precisely $\mathcal{C}_2$. While $\mathcal{C}_1$ and $\mathcal{C}_2$ are closed under composition, the higher levels no longer form groups. 
In fact, not much is known about the general structure of the higher sets in the Clifford hierarchy. 
What we do know however is that the \emph{diagonal} unitaries in $\mathcal{C}_n$ always form a group, and that for $n\geq 2$ each of these diagonal elements can be constructed using Clifford operations and the $\pi/2^{n-1}$ $Z$ rotation~\cite{cui2017diagonal}.
Using the above description of a deterministic implementation of a $\pi/2^{n-1}$ rotation we have therefore found a deterministic measurement-based model that can implement any diagonal $n$-th level Clifford operation using just Pauli measurements.

\section{Summary and outlook}\label{sec:ppm-conclusion}

We introduced a family of resource states that led to deterministic approximately universal quantum computation using measurements in just two bases. Furthermore, depending on the chosen parameters, diagonal gates of arbitrarily high levels of the Clifford hierarchy can be implemented.



This model highlights a link between the form of the parity-phase gate of equation~\eqref{eq:virtual-qubit} and quantum computing with magic states. It may be useful to consider if the representation of Ising-type interactions (i.e. parity-phase gates) as `virtual' magic states can be exploited for magic state distillation. For instance, on ion trap architectures, it is possible to introduce $O(n^2)$ parity-phase gates in a single time step using an $n$-qubit interaction~\cite{molmersorensen1999}. It would certainly be interesting so see whether this curious property can be used in the construction of fault-tolerant protocols.

The interactions needed to make the resource states described in this chapter are available `natively' in both ion trap and superconducting quantum computing hardware which means proofs of concept could be implemented in a short time frame. However, given long turnaround times for quantum measurements and feed-forward, it remains unclear if such a measurement-based scheme would yield benefits over the circuit model on such architectures. On the other hand, measurement-based schemes have already had some success in quantum optics~\cite{walther2005experimental}, where deterministic application of multi-qubit gates remains a significant challenge. While the scheme we described relied on resource states with a very specific structure, its likely that this could be relaxed using techniques similar to those employed in producing perfect cluster states from imperfect lattices~\cite{morley2017physical}. Furthermore, the use of multiple kinds of edges between qubits creates a possibility for more successful outcomes for non-deterministic entangling operations. That is, \textit{known} errors giving rise to non-maximal entanglement between pairs of qubits could still yield good resource states for universal deterministic computation. This could, for example, be exploited in models of universal quantum computation using linear optical devices and non-deterministic fusion gates~\cite{gimeno2015three}.

Finally, we note that the usage of the ZX-calculus for describing MBQC allows one to resort to tools such as Quantomatic~\cite{kissinger2015quantomatic} or the library PyZX described in Chapter~\ref{chap:optimisation} for verifying correctness of the calculations.\indexd{Quantomatic}

\chapter{Simplification of measurement patterns}\label{chap:MBQC}

In the previous chapter we described a concrete model of quantum computing based on a specific resource state that resulted in a deterministic method of computation. In this chapter we will take a more abstract viewpoint and consider an entire class of computational resource states. Specifically, we will consider a condition, the existence of \emph{gflow}, that ensures a deterministic computation is possible. We study how this property is preserved under changes, specifically simplifications, of the underlying resource state.

Throughout this chapter we will solely work with the \emph{one-way model}~\cite{MBQC1}. Unlike many other works studying the one-way model we will not restrict our measurements to a single plane, and instead we will allow measurements in all three of the principle axes of the Bloch sphere. This complicates some of the definitions, but ultimately simplifies our analysis, and will prove crucial for the results of Chapter~\ref{chap:optimisation}. 

This chapter contains two main results. The first regards the ability to remove qubits measured in a Clifford angle from a measurement pattern. It was already known that such qubits can be removed from a graph state~\cite{graphstates}, however we show that this can be done in such a way that the existence of gflow is preserved. Hence, we find that the number of qubits in a measurement pattern with gflow can be reduced to be proportional to the number of non-Clifford measurements, while still preserving deterministic realisability of the pattern.
The removal of Clifford vertices will form the basis of the circuit optimisation algorithm described in Section~\ref{sec:circuit-simplification}.

The second result is an efficient algorithm to extract a quantum circuit from a measurement pattern with gflow. It was already known how to extract a circuit from a pattern with measurements in a single plane (either involving ancillae and classical control~\cite{broadbent2009parallelizing,da2013compact} or without~\cite{miyazaki2015analysis}. Our algorithm in contrast can work with measurements in three planes, and results in a unitary circuit that is ancilla-free and measurement-free. By being able to work with measurements in three planes, we can use this algorithm to extract circuits from ZX-diagrams that contain phase gadgets (as these correspond in a natural way to YZ-plane measurements).

The chapter is structured as follows.
In Section~\ref{sec:one-way-model} we introduce the one-way model. Then in Section~\ref{sec:gflow} we recall the concept of gflow and the different types of determinism of a measurement pattern. In Section~\ref{sec:circuits-to-patterns} we see how any quantum circuit can be converted into a deterministic measurement pattern.

In Section~\ref{sec:gflow-rewrite} we see that graph operations like local complementation preserve the existence of gflow on a graph. We use these results in Sections~\ref{sec:pattern-local-complementation}--\ref{sec:further-optimisation} to simplify measurement patterns, particularly showing how to remove all qubits measured in a Clifford angle from the pattern.

Finally, in Section~\ref{sec:circuit-extraction} we tackle the converse problem of Section~\ref{sec:circuits-to-patterns}: transforming a measurement pattern into a circuit. We find an efficient algorithm that converts any measurement pattern with gflow into an equivalent unitary circuit.

\section{The one-way model}\label{sec:one-way-model}
\indexd{one-way model}
In this section we will describe the one-way model and its representation in the ZX-calculus. We will adopt the notation of the \emph{measurement calculus} from Ref.~\cite{danos2007measurement}.

The one-way model is in many ways similar to the PPM model of the previous chapter. The one-way model starts with the preparation of a specific graph state (cf.~Definition~\ref{def:graph-state}). The qubits in this graph state are then measured in some sequence, with the type of measurement possibly depending on previous measurement outcomes.

Instead of allowing arbitrary single-qubit measurements, measurements are usually restricted to three planes of measurement labelled \XY, \XZ, and \YZ.
For each plane, the state denoted $+$ is taken to be the desired outcome of the measurement and the state denoted $-$ is the undesired outcome, which will need to be adaptively corrected.
The allowed measurements are thus:
\begin{align*}
  \ket{+_{\XYm,\alpha}} &= \frac{1}{\sqrt{2}}\left(\ket{0} + e^{i\alpha} \ket{1} \right) &
  \ket{-_{\XYm,\alpha}} &= \frac{1}{\sqrt{2}}\left(\ket{0} - e^{i\alpha} \ket{1} \right) \\
  \ket{+_{\XZm,\alpha}} &= \cos\left(\frac{\alpha}{2}\right)\ket{0} + \sin\left(\frac{\alpha}{2}\right) \ket{1} &
  \ket{-_{\XZm,\alpha}} &= \cos\left(\frac{\alpha}{2}\right)\ket{0} - \sin\left(\frac{\alpha}{2}\right) \ket{1} \\
  \ket{+_{\YZm,\alpha}} &= \cos\left(\frac{\alpha}{2}\right)\ket{0} + i \sin\left(\frac{\alpha}{2}\right) \ket{1} &
  \ket{-_{\YZm,\alpha}} &= \cos\left(\frac{\alpha}{2}\right)\ket{0} - i \sin\left(\frac{\alpha}{2}\right) \ket{1}
\end{align*}
\noindent Here $0 \leq \alpha \leq 2\pi$. We present the positive outcomes of these measurement in the ZX-calculus in Table~\ref{tab:MBQC-to-ZX}.

\begin{table}[!t]
  \centering
  \renewcommand{\arraystretch}{2}
  \begin{tabular}{c||c|c|c|c|c}
   operator & $N_i$ & $E_{ij}$ & $\bra{+_{\XY,\alpha_i}}_i$ & $\bra{+_{\XZ,\alpha_i}}_i$ & $\bra{+_{\YZ,\alpha_i}}_i$ \\ \hline
   &&&&&\\
   diagram & \input{plus-state.tikz} & \input{cz.tikz} & \input{XY-effect.tikz} & \input{XZ-effect.tikz} & \input{YZ-effect.tikz}
  \end{tabular}
  \renewcommand{\arraystretch}{1}
  \caption{Translation from a measurement pattern to a \zxdiagram.\label{tab:MBQC-to-ZX}}
 \end{table}

Like in the PPM-model of the previous chapter, we describe a computation in the one-way model using a measurement pattern:
\begin{definition}[{\cite{danos2007measurement}}]\label{def:meas_pattern}
  A \Define{measurement pattern}\indexd{measurement pattern!in one-way model} consists of an $n$-qubit register $V$ with distinguished sets $I, O \sse V$ of input and output qubits and a sequence of commands consisting of the following operations:
  \begin{itemize}
    \item Preparations $N_i$, which initialise a qubit $i \notin I$ in the state $\ket{+}$.
    \item Entangling operators $E_{ij}$, which apply a CZ-gate to qubits $i$ and $j$.
    \item Measurements $M_i^{\ld, \alpha}$, which probabilistically project a qubit $i\notin O$ onto the orthonormal basis $\{\ket{+_{\ld,\alpha}},\ket{-_{\ld,\alpha}}\}$, where $\lambda \in \{ \XY, \XZ, \YZ \}$ is the measurement plane and $\alpha$ is the measurement angle.
    The projector $\ket{+_{\ld,\alpha}}\bra{+_{\ld,\alpha}}$ corresponds to outcome $0$ and $\ket{-_{\ld,\alpha}}\bra{-_{\ld,\alpha}}$
    corresponds to outcome $1$.
    \item Corrections $[X_i]^s$, which depend on a measurement outcome (or some Boolean function of measurement outcomes) $s\in\{0,1\}$ and act as the Pauli-$X$ operator on qubit $i$ if $s$ is $1$ and as the identity otherwise,
    \item Corrections $[Z_j]^t$, which depend on a measurement outcome (or a linear combination of measurement outcomes) $t\in\{0,1\}$ and act as the Pauli-$Z$ operators on qubit $j$ if $t$ is $1$ and as the identity otherwise.
  \end{itemize}
\end{definition}

Not every measurement pattern is actually physically possible. For this to be the case, the pattern needs to be runnable.
\begin{definition}[{\cite{danos2007measurement}}]\label{def:runnable_pattern}
A measurement pattern is \Define{runnable}\indexd{measurement pattern!runnable ---} if the following conditions hold.
\begin{itemize}
\item No correction depends on an outcome not yet measured.
\item All non-input qubits are prepared.
\item All non-output qubits are measured.
\item A command $C$ acts on a qubit $i$ only if $i$ has not already been measured, and one of (1)-(3) holds:
\begin{enumerate}[label=({\arabic*})]
\item $i$ is an input,
\item $i$ has been prepared and $C$ is not a preparation,
\item $i$ has not been prepared, $i$ is not an input, and $C$ is a preparation.
\end{enumerate}
\end{itemize}
\end{definition}
Runnable measurement patterns can be \Define{standardised}~\cite{danos2007measurement}, so that all preparations $N_i$ appear first, then all the edges $E_{ij}$, then the measurements $M_i^{\ld, \alpha}$ and finally the corrections.
The edges $E_{ij}$ used in a pattern characterize the \emph{\LOG} of the entangled resource state:

\begin{definition}[{cf.~\cite[p.5]{GFlow}}]\label{def:LOG}
  A \Define{\LOG}\indexd{labelled open graph}\indexd{graph!labelled open ---} is a tuple $\Gamma = (G,I,O, \lambda)$ where $(G,I,O)$ is an open graph (cf.~Definition~\ref{def:open-graph}), and $\lambda : \comp{O} \rightarrow \{ \XY, \YZ, \XZ\}$ assigns a measurement plane to each non-output qubit.\index{math}{lambda@$\lambda$ (assignment of measurement planes)}%
  \footnote{We adopt the notation of this from Ref.~\cite{GFlow} where their corresponding notion is called an `open graph state'. While they explicitly consider this to be a particular type of quantum state, our notion of a labelled open graph is just a type of graph.}
\end{definition}

A standardised measurement pattern is completely characterised by its \LOG together with an order of the measurements, measurement angles, and the necessary corrections.

In general, a single measurement pattern could implement many different linear maps depending on the specific outcomes of the measurements. In this chapter we will be primarily interested in patterns that implement the same linear map regardless of measurement outcome, \ie those that are deterministic. For such a pattern we can read of the linear map it implements from the pattern itself:

\begin{definition}\label{def:ogs-to-linear-map}
\indexd{linear map associated to \LOG}
\index{math}{Mgamma@$M_{\Gamma,\alpha}$ (linear map associated to \LOG)}
 Suppose $\Gamma=(G,I,O,\ld)$ is a \LOG corresponding to a runnable measurement pattern.
 Let $\alpha:\comp{O}\to [0,2\pi)$ be a set of measurement angles.
 The \Define{linear map associated to} $(\Gamma, \alpha)$, written as $M_{\Gamma,\alpha}$, is defined as follows:
 \[
  M_{\Gamma,\alpha} := \left( \prod_{i\in\comp{O}} \bra{+_{\ld(i),\alpha_i}}_i \right) E_G N_{\comp{I}}.
 \]
  Here $E_G := \prod_{i\sim j} E_{ij}$ and $N_{\comp{I}} := \prod_{i\in\comp{I}} N_i$.
\end{definition}
\begin{remark}
 Note that the projections $\bra{+_{\ld(i),\alpha_i}}_i$ on different qubits $i$ commute with each other.
 Similarly, the controlled-Z operations $E_{ij}$ commute even if they involve some of the same qubits.
 Finally, all the state preparations $N_i$ on different qubits commute.
 Thus, $M_{\Gamma,\alpha}$ is fully determined by $\Gamma$ and $\alpha$, and our definition is well-defined.
\end{remark}

Additionally, for such a measurement pattern we can also read of the ZX-diagram that represents the same linear map:

\begin{definition}\label{def:ogs-to-ZX}
 Suppose $\Gamma=(G,I,O,\ld)$ is a \LOG\ and $\alpha:\comp{O}\to [0,2\pi)$ is a set of measurement angles.
 Then its \Define{associated \zxdiagram}\indexd{labelled open graph!associated ZX-diagram} $D_{\Gamma,\alpha}$ is defined by translating the expression for $M_{\Gamma,\alpha}$ from Definition~\ref{def:ogs-to-linear-map} according to Table~\ref{tab:MBQC-to-ZX}, composing the elements in the obvious way, and then merging any sets of adjacent phase-free Z-spiders that are not measurement effects (i.e.\  fuse all the Z-spiders which come from the translation of a preparation or entangling command).
\end{definition}

\begin{example}
  The measurement pattern defined by the qubit register $V=\{ 1,2,3,4\}$ with $I=\{ 1,2 \}$ and $O = \{ 1,4 \}$, and the associated linear map
  $$ \bra{+_{\XYm,\frac{\pi}{2}}}_2 \bra{+_{\YZm,\frac\pi4}}_3 E_{14}E_{23}E_{24} E_{34} N_3 N_4$$
  is represented by the following \zxdiagram:
  \begin{equation*}
    \input{example-MBQC-translation.tikz}
  \end{equation*}
\end{example}

\begin{remark}\label{rem:MBQC-gadget}
	As can be seen in this example, qubits measured in the \YZ-plane are phase gadgets (cf.~Section~\ref{sec:phasegadgets}). The ZX-diagram associated to a measurement pattern that only has \XY-plane measurements is closely related to the graph-like diagrams of Section~\ref{sec:graph-like}. Indeed, if we fuse the spiders of the measurement effects to their corresponding vertices, the ZX-diagram is graph-like.
\end{remark}

The interpretation of the ZX-diagram associated to a measurement pattern is equal to its associated linear map.
\begin{lemma}\label{lem:zx-equals-linear-map}
 Suppose $\Gamma=(G,I,O,\ld)$ is a \LOG and $\alpha:\comp{O}\to [0,2\pi)$ is a set of measurement angles.
 Let $M_{\Gamma,\alpha}$ be the linear map specified in Definition~\ref{def:ogs-to-linear-map} and let $D_{\Gamma,\alpha}$ be the \zxdiagram constructed according to Definition~\ref{def:ogs-to-ZX}.
 Then $\intf{D_{\Gamma,\alpha}}=M_{\Gamma,\alpha}$.
\end{lemma}
\begin{proof}
 For each operator $M$ in Table~\ref{tab:MBQC-to-ZX} and its corresponding diagram $D_M$, it is straightforward to check that $\intf{D_M}=M$ (up to non-zero scalar).
 The result thus follows by the compositional properties of the interpretation $\intf{\cdot}$ and the fact that the rewriting of ZX-diagrams preserves semantics.
\end{proof}

We can also read of a measurement pattern from certain ZX-diagrams.
\begin{definition}\label{def:MBQC-form}
 A \zxdiagram is in \Define{MBQC form}\indexd{ZX-diagram!in MBQC form} if it consists of a `graph state' (cf.~Definition~\ref{def:graph-state}) in which each vertex of the graph may also be connected to:
 \begin{itemize}
  \item an input (in addition to its output), and
  \item a measurement effect (in one of the three measurement planes) instead of the output.
 \end{itemize}
\end{definition}

\begin{definition}\label{def:graph-of-diagram}
 Given a \zxdiagram $D$ in MBQC form, its \Define{underlying graph} $G(D)$ is the graph corresponding to the graph state part of $D$.\indexd{ZX-diagram!underlying labelled open graph}\index{math}{G(D)@$G(D)$ (underlying labelled open graph)}
\end{definition}

\begin{figure}
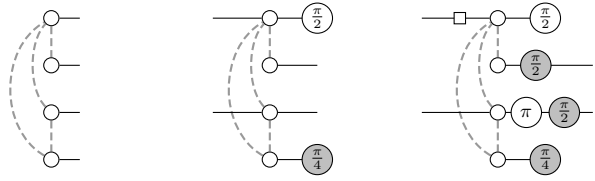

    \ctikzfig{example-MBQC-form}
    \caption{On the left, a graph state. In the middle, a diagram in MBQC form with the same underlying graph. On the right, an MBQC+LC form diagram with the same underlying labelled open graph.\label{fig:graph-state}} 
\end{figure}

See Figure~\ref{fig:graph-state} for an example of a graph state diagram and a diagram in MBQC form.

\begin{lemma}
 Let $\Gamma=(G,I,O,\ld)$ be a \LOG and let $\alpha:\comp{O}\to [0,2\pi)$ be a set of measurement angles.
 Then the \zxdiagram $D_{\Gamma,\alpha}$ constructed according to Definition~\ref{def:ogs-to-ZX} is in MBQC form.
\end{lemma}
\begin{proof}
 Consider performing the translation described in Definition~\ref{def:ogs-to-ZX} in two steps.
 The first step involves translating the preparation and entangling commands of the pattern $M_{\Gamma,\alpha}$ according to Table~\ref{tab:MBQC-to-ZX} and then merging any sets of adjacent green spiders.
 This yields a graph state diagram with some additional inputs.
 (The underlying graph is $G$).
 The second step is the translation of the measurement projections of $M_{\Gamma,\alpha}$.
 This yields measurement effects on some of the outputs of the graph state diagram.
 Thus, the resulting \zxdiagram is in MBQC form by Definition~\ref{def:MBQC-form}.
\end{proof}

The converse of this lemma also holds.

\begin{lemma}\label{lem:zx-to-pattern}
 Suppose $D$ is a \zxdiagram in MBQC form.
 Then there exists a \LOG $\Gamma=(G,I,O,\ld)$ and a set of measurement angles $\alpha:\comp{O}\to [0,2\pi)$ such that $\intf{D} = M_{\Gamma,\alpha}$.
\end{lemma}
\begin{proof}
 Let $G = G(D)$ be the underlying graph of $D$.
 Let $I\sse V$ be the set of vertices directly connected to input wires in $D$ and $O\sse V$ the set of vertices that are directly connected to an output wire in $D$.
 Fix $\ld:\comp{O}\to\{\XYm,\XZm,\YZm\}$ by using Table~\ref{tab:MBQC-to-ZX} in reverse to determine the measurement plane from the effect in the \zxdiagram.
 Let $\Gamma := (G,I,O,\ld)$.
 Finally, define $\alpha:\comp{O}\to [0,2\pi)$ to be the phase of the measurement effect connected to each non-output vertex in the \zxdiagram.
 Then $D = D_{\Gamma,\alpha}$ and thus the desired result follows from Lemma~\ref{lem:zx-equals-linear-map}.
\end{proof}

\begin{remark}
	We defined $G(D)$ to be a graph. Depending on context we will take it to be an open graph, with the inputs and outputs corresponding to the input and output vertices of $D$, or a labelled open graph, where the labels are given by the measurement effects present in $D$.
\end{remark}

\begin{remark}
 Given a fixed enumeration of the graph vertices the correspondence between an MBQC form diagram and tuples $(\Gamma,\alpha)$, where $\Gamma$ is a \LOG and $\alpha$ is a set of measurement angles, is one-to-one.
\end{remark}

It will turn out to be useful to consider a `relaxed' version of the MBQC form for \zxdiagrams.

\begin{definition}
  We say a \zxdiagram is in \Define{MBQC+LC}\indexd{MBQC+LC form}\indexd{ZX-diagram!in MBQC+LC form} form when it is in MBQC form (cf.~Definition~\ref{def:MBQC-form}) up to arbitrary single-qubit Clifford unitaries on the input and output wires (LC stands for `local Clifford').
  When considering the underlying graph of a \zxdiagram in MBQC+LC form, we ignore these single qubit Clifford unitaries.
\end{definition}
Note that an MBQC form diagram is an MBQC+LC form diagram with trivial single-qubit unitaries on its inputs and outputs. An example diagram in MBQC+LC form is given in Figure~\ref{fig:graph-state}.

The extra freedom of allowing local Cliffords on inputs and outputs makes it easier to do rewrites that stay in the `class' of MBQC+LC form diagrams. In particular, we will show that a variant of the local complementation and pivoting rewrites of Section~\ref{sec:local-complementation} transforms a MBQC+LC form diagrams into another MBQC+LC form diagram.

\section{Determinism and gflow}\label{sec:gflow}

Given a specific \LOG it will not always be possible to define a measurement pattern on it that implements a deterministic computation. In order to understand what structure is necessary in order to guarantee determinism, the notion of \emph{flow} (also called \emph{causal flow}) on open graphs was introduced by Danos and Kashefi~\cite{Danos2006Determinism-in-} as a sufficient condition on open graphs to distinguish those graphs which are capable of running a deterministic MBQC pattern. Causal flow is however not a necessary condition. That is, there are graphs that implement a deterministic pattern even though they do not have causal flow~\cite{GFlow,DP2}.
Hence, a \emph{generalised flow} was defined by Browne et al.~\cite{GFlow} in order to obtain a necessary condition. Before we give the (rather technical) definition of gflow, let us state the results that motivate its existence.

\begin{definition}[{\cite[p.5]{GFlow}}]
    The linear map implemented by a measurement pattern for a specific set of measurement outcomes is called a \Define{branch}\indexd{measurement pattern!branch} of the pattern.
    A pattern is \Define{deterministic}\indexd{measurement pattern!deterministic ---}\indexd{deterministic pattern} if all branches are equal up to a scalar.
	A pattern is \Define{strongly deterministic}\indexd{measurement pattern!strongly deterministic ---}\indexd{deterministic pattern!strongly ---} if all branches are equal up to a global phase.
	It is \Define{uniformly deterministic}\indexd{measurement pattern!uniformly deterministic ---}\indexd{deterministic pattern!uniformly ---} if it is deterministic for any choice of measurement angles.
	Finally, the pattern is \Define{stepwise deterministic}\indexd{measurement pattern!stepwise deterministic ---}\indexd{deterministic pattern!stepwise ---} if any intermediate pattern resulting from doing some measurements and their corresponding corrections is again deterministic.
\end{definition}

\begin{theorem}[{\cite[Theorem 2]{GFlow}}]\label{t-flow}
Let $\Gamma = (G,I,O,\ld)$ be a \LOG~with a gflow. Then for any choice of measurement angles $\alpha: \comp{O}\rightarrow [0,2\pi]$ there exists a measurement pattern $\mathfrak{P}$ which is runnable as well as uniformly, strongly and stepwise deterministic that realises the linear map $M_{\Gamma,\alpha}$ (cf.~Definition~\ref{def:ogs-to-linear-map}).
Conversely, if a pattern is stepwise, uniformly and strongly deterministic, then the underlying \LOG has a gflow.
\end{theorem}

Theorem~\ref{t-flow} shows that the existence of a gflow on a \LOG is sufficient and necessary for a particularly nice kind of determinism. As the name would suggest, gflow is a generalisation of causal flow:

\begin{definition}[{\cite{Danos2006Determinism-in-}}]\label{def:causal-flow}
Given an open graph $(G,I,O)$, a
\Define{causal flow} $(f,\prec)$\indexd{causal flow}\index{math}{$\prec$} on $G$  consists of a
function $f: \comp O \to \comp I$ and a partial order $\prec$ on the set
$V$ satisfying the following properties:
\begin{enumerate}
\item $f(v) \sim v$ \label{flowi}
\item $v \prec f(v)$ \label{flowii}
\item if $u \sim f(v)$ then $v \prec u$ \label{flowiii}
\end{enumerate}
\end{definition}
Recall that we write $v \sim u$ when $u$ and $v$ are adjacent in the graph.

The reason this definition talks about open graphs instead of \emph{labelled} open graphs is because causal flow is only defined for measurement patterns where all qubits are measured in the \XY plane, and hence we do not need the labelling.

The partial order in the causal flow tells us in which order the qubits should be measured (first the qubits at the bottom of the partial order, proceeding in the obvious way). When the wrong outcome is observed at a qubit $v$, a correction must be applied at the qubit $f(v)$. The conditions of causal flow then simply state that $f(v)$ must be a neighbour of $v$ that lies in $v$'s future and that any other qubit connected to $f(v)$ must also lie in $v$'s future, so that the correction applied at $f(v)$ only affects $v$ and qubits which have not yet been measured.

The notion of gflow differs from causal flow in two ways.
Firstly, $f(u)$ is taken to be a set of vertices instead of a single vertex, so that corrections can be applied to more than one vertex.
As a result of this change, the third condition of causal flow is now too strong: requiring that no element of $f(u)$ is adjacent to any vertex `in the past' of $u$ would be too restrictive.
When corrections are applied to an entire set of vertices, these corrections only affect qubits that are in the \emph{odd neighbourhood} of those vertices.
This motivates the second change in the definition of gflow, which is a parity condition: all vertices in the neighbourhood of $f(u)$ that lie in the past of $u$ are required to be in the even neighbourhood of $f(u)$.
As a result, either the effects of corrections do not propagate into the past or, if they do, an even number of corrections propagate so that it cancels out.

\begin{definition}[{\cite[p.7]{GFlow}}]\indexd{odd neighbourhood}\index{math}{OddG@$\odd{G}{A}$}
Let $G=(V,E)$ be a graph and $A\sse V$ any set of vertices. The \Define{odd neighbourhood} of $A$ in $G$ is $\odd{G}{A}= \{u\in V: \abs{N(u)\cap A}\equiv 1 \mod 2\}$, \ie the set of vertices that have an odd number of neighbours in $A$.
If the graph $G$ is clear from context, we simply write $\odd{}{A}$.
\end{definition}

We can now state the definition of gflow. We write $\pow{A}$ for the powerset of $A$.

\begin{definition}[{\cite[Definition~3]{GFlow}}]
\label{defGFlow}\indexd{gflow}
 A \LOG{} $(G,I,O,\ld)$ has \Define{generalised flow} (\Define{gflow} for short) if there exists a map $g:\comp{O}\to\pow{\comp{I}}$ and a partial order $\prec$ over $V$ such that for all $v\in \comp{O}$:
 \begin{enumerate}[label=({g}\theenumi), ref=(g\theenumi)]
  \item\label{it:g} If $w\in g(v)$ and $v\neq w$, then $v\prec w$.
  \item\label{it:odd} If $w\in\odd{}{g(v)}$ and $v\neq w$, then $v\prec w$.
  \item\label{it:XY} If $\ld(v)=\XYm$, then $v\notin g(v)$ and $v\in\odd{}{g(v)}$.
  \item\label{it:XZ} If $\ld(v)=\XZm$, then $v\in g(v)$ and $v\in\odd{}{g(v)}$.
  \item\label{it:YZ} If $\ld(v)=\YZm$, then $v\in g(v)$ and $v\notin\odd{}{g(v)}$.
 \end{enumerate}
 For a vertex $v$, we will refer to $g(v)$ as its \Define{correction set}\indexd{correction set}\index{math}{g(v)@$g(v)$}.
\end{definition}

\begin{remark}
If all qubits are measured in the \XY plane and every $g(u)$ is a singleton, then these conditions above become equivalent to those of a causal flow, and hence a causal flow is also a gflow with $g(v):=\{f(v)\}$.
Note that most of the literature that uses gflow focuses on patterns containing only \XY-plane measurements, and hence the definition of gflow is usually only given as conditions \ref{it:g}--\ref{it:XY}.\footnote{The original definition of gflow of Ref.~\cite{GFlow} states \ref{it:odd} in a different inequivalent way that turns out not to have the desirable properties. See Ref.~\cite{wetering-gflow} for an extended discussion on this.}
\end{remark}

In the case where all qubits are measured in the \XY plane (which allows us to ignore conditions \ref{it:XZ} and \ref{it:YZ}) we can give a game-like interpretation of gflow:
\begin{example}
Consider the following game. Suppose we have an open graph $G$ whose vertices are labelled with $0$'s and $1$'s, where a $1$ indicates the presence of an error. Define an operation $\textbf{flip}_v$, which flips all of the bits on the \textit{neighbours} of a given vertex $v$. Our goal is to propagate all of the errors present in $G$ to the outputs using only applications of the operation $\textbf{flip}_v$. For example:
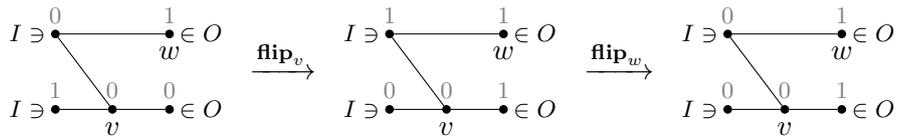
\begin{equation*}
\input{has-gflow.tikz}
\ \  \xrightarrow{\textbf{flip}_v} \ \ 
\input{has-gflow2.tikz}
\ \  \xrightarrow{\textbf{flip}_w} \ \ 
\input{has-gflow3.tikz}
\end{equation*}
For some open graphs and configurations of errors, this task might be impossible. For example, there is no solution for the following graph:
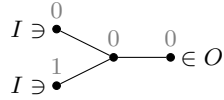
\ctikzfig{no-gflow}
However, we can always succeed if we are given the following data: an ordering $\prec$ of vertices which give a direction of `time' going from inputs to outputs, and, for each vertex, a correction set $g(v)$ of vertices in the future of $v$ (w.r.t. $\prec$) such that applying $\textbf{flip}_w$ for all $w \in g(v)$ flips the bit on $v$ without affecting any other bits, except for those lying in the future of $v$. By repeatedly finding the minimal vertex $v$ (w.r.t. $\prec$) with an error and applying $\textbf{flip}_w$ to all $w \in g(v)$, the procedure will eventually propagate all of the $1$'s to the outputs of $G$.
\end{example}

\begin{remark}
	Finding a gflow on a given \LOG can be done efficiently. See Theorem~\ref{thm:gflow-algorithm}.
\end{remark}

Gflow is a property that applies to \LOG{}s. For convenience we also define it for ZX-diagrams.
\begin{definition}\label{dfn:zx-gflow}
We say a \zxdiagram in MQBC(+LC) form has \Define{gflow}\indexd{gflow!on ZX-diagram} $(g,\prec)$ if its underlying \LOG\ $\Gamma$ has gflow $(g,\prec)$.
\end{definition}

\begin{remark}
    A measurement pattern specifies a \LOG, a set of measurement angles and a set of corrections. However, if the pattern is deterministic and the \LOG has gflow, then the corrections can be inferred from the gflow itself. Hence, we are warranted in conflating a measurement pattern with gflow with its \LOG and its measurement angles. This allows us to describe without ambiguity a measurement pattern with gflow as a MBQC form diagram where the underlying \LOG has gflow.
\end{remark}

\section{From circuits to measurement patterns}\label{sec:circuits-to-patterns}

Using the ZX-calculus we can transform any quantum circuit into a measurement pattern. This follows easily from the observation that any graph-like ZX-diagram (cf.~Definition~\ref{def:graph-form}) can be transformed into a ZX-diagram in MBQC form by unfusing all the phases into measurement effects of type \XY. E.g~using the graph-like diagram from Figure~\ref{fig:underlying-graph}:
\ctikzfig{graph-like-to-MBQC-form}
Note that we do not measure output qubits, and hence we cannot have phases on spiders connected to outputs. In the example above we therefore add some identity spiders onto the second qubit in order to disconnect the $\gamma$ from the output. This example is straightforwardly generalised:

\begin{lemma}\label{lem:zx-to-mbqc-form}
	Any ZX-diagram can be efficiently transformed into an equivalent ZX-diagram in MBQC form.
\end{lemma}
\begin{proof}
	First transform the ZX-diagram into a graph-like ZX-diagram using Lemma~\ref{lem:all-zx-are-graph-like}. Then disconnect every spider connected to an output which has a non-zero phase from the output by introducing identity spiders:
	\ctikzfig{disconnect-output-phase}
	Finally, unfuse every phase onto its own spider to make an \XY measurement effect. The resulting diagram is in MBQC form.
\end{proof}

Slightly more non-trivially, if we started with a quantum circuit, the resulting MBQC form diagram has a causal flow, and hence is deterministically implementable.
\begin{lemma}\label{lem:circuits-have-gflow}
	Let $D'$ be a ZX-diagram in MBQC form resulting from applying Lemma~\ref{lem:zx-to-mbqc-form} to a circuit $D$. Then the underlying open graph $G(D')$ has a causal flow.
\end{lemma}
\begin{proof}
  With every spider $v$ of the circuit $D$ we associate a number $q_v$ specifying on which `qubit-line' it appears.
  We also associate a `row-number' $r_v\geq 0$ specifying how `deep' in the circuit it appears.
  Suppose now that $v\sim w$ in $D$. If they are on the same qubit, so $q_v=q_w$, then necessarily $r_v\neq r_w$.
  Conversely, if they are on different qubits, $q_v\neq q_w$, then they must be part of a CZ or CNOT gate,
  and hence $r_v=r_w$.

  In the diagram resulting from Lemma~\ref{lem:all-zx-are-graph-like}, every spider arises from fusing together adjacent spiders on the same qubit line from the original diagram.
  For a spider $v$ in $D^\prime$ we can thus associate two numbers $s_v$, and $t_v$, 
  where $s_v$ is the lowest row-number of a spider fused into $v$, and $t_v$ is the highest.
  Spider fusion from $D$ only happens for spiders on the same qubit-line, and hence $v$ also inherits a $q_v$ from
  all the spiders that got fused into it. Any two spiders $v$ and $w$ in $D^\prime$ with $q_v=q_w$
  must have been produced by fusing together adjacent spiders on the same qubit-line,
  and hence we must have $t_v<s_w$ or $t_w<s_v$, depending on which of the spiders is most to the left.
  If instead $v\sim w$ and $q_v\neq q_w$, then their connection must have arisen from some CNOT or CZ gate in $D$, 
  and hence the intervals $[s_v,t_v]$ and $[s_w,t_w]$ must have overlap,
  so that necessarily $s_w\leq t_v$ and $s_v\leq t_w$.

  Now we let $O$ be the set of spiders connected to an output, 
  and $I$ the set of spiders connected to an input in $D^\prime$.
  For all $v\in \comp O$ we let $f(v)$ be the unique spider connected to $v$ on the right on the same qubit-line.
  We define a partial order as follows: $v\prec w$ if and only if $v=w$ or $t_v<t_w$. 
  It is straightforward to check that this is indeed a partial order.

  By construction $f(v)\sim v$ and the property $v \prec f(v)$, follows from $t_v < s_{f(v)}$ discussed above,
  so let us look at the third property of causal flow.

  Suppose $w\sim f(v)$. We need to show that $v\prec w$. If $v=w$ this is trivial so suppose $v\neq w$.
  First suppose that $q_w = q_{f(v)}$ (which is also equal to $q_v$). $f(v)$ has a maximum of two neighbours
  on the same qubit-line, and since one of them is $v$, this can only be if $w=f(f(v))$ and hence $v\prec f(v)\prec w$,
  and we are done.
  So suppose that $q_w\neq q_{f(v)}$. 
  By the discussion above, we must then have $s_w\leq t_{f(v)}$ and $s_{f(v)} \leq t_w$.
  Since we also have $t_v < s_{f(v)}$ we get $t_v < s_{f(v)} \leq t_w$ so that indeed $v\prec w$.
\end{proof}

\section{Graph operations on labelled open graphs}\label{sec:gflow-rewrite}

The property of having a gflow is important for \LOG{}s as it guarantees the possibility of doing deterministic computations on it. We might however want to change the \LOG somehow in order to accommodate other needs. In this section we will study some ways in which a \LOG with gflow can be changed so that the resulting \LOG still has a gflow. In particular, we will see that we can do a local complementation (cf.~Section~\ref{sec:local-complementation}) on the underlying graph while preserving the existence of a gflow.

Throughout this section, we will rely extensively on the symmetric difference of sets: ${A\symd B} := (A\cup B)\setminus (A\cap B)$. Note that $\symd$ is associative and commutative, so it extends to an $n$-ary operation in the obvious way. For $I:={1,\ldots,n}$ we have:
\[\Symdi{i \in I} A_i := A_1 \symd A_2 \symd \ldots \symd A_n \]
In particular, we have $a \in \Symdi{i \in I} A_i$ if and only if $a$ appears in an odd number of sets $A_i$. By convention, we assume $\Symdi{...}$ binds as far to the right as possible, i.e.
\[ \left( \Symdi{i \in I} A_i \symd B \right) := \left( \Symdi{i \in I} (A_i \symd B) \right) \]

The following lemma shows how the odd neighbourhood of a set evolves under local complementation:
\begin{lemma}\label{lem:oddneighbours}
Given a graph $G=(V,E)$, $A\subseteq V$ and $u \in V$, $$\odd {G\star u} A=\begin{cases}\odd G A \symd (N_G(u)\cap A) & \text{if $u\notin \odd G A$}\\ \odd G {A} \symd (N_G(u)\setminus A)&\text{if $u\in \odd G A$}\end{cases}$$
\end{lemma}
\begin{proof}
First notice that $\odd G .$ is linear: 
$$\forall A, B: \odd G {A\symd B} = \odd G A \symd \odd G B.$$ 
Moreover $\forall v, \odd G {\{v\}} = N_G(v)$, the neighbourhood of $v$ in $G$. 
As a consequence we see that for all subsets $A$ that $\odd G A  =\Symdi{v\in A} N_G (v)$. 

Local complementation acts as follows on the neighbourhood of an arbitrary vertex~$v$: 
$$N_{G\star u} (v) = \begin{cases} N_G(v) \symd N_G(u)\symd \{v\} &\text{if $v\in N_G(u)$}\\ N_G(v) &\text{otherwise} \end{cases}$$
As a consequence,
\begin{eqnarray*}
  \odd {G\star u} A&=&\Symdi{v\in A} N_{G\star u} (v)\\
  &=& \left(\Symdi{v\in A\cap N_G(u)} N_{G} (v)\symd N_G(u)\symd \{v\}\right) \symd  \left(\Symdi{v\in A\setminus  N_G(u)} N_{G} (v)\right)\\
  &=&\left(\Symdi{v\in A} N_{G} (v)\right) \symd \left(\Symdi{v\in A\cap N_G(u)}  N_G(u) \right) \symd  \left(\Symdi{v\in A\cap N_G(u)}  \{v\}\right)\\
  &=&\odd G A \symd \left(\Symdi{v\in A\cap N_G(u)}  N_G(u) \right) \symd (A\cap N_G(u))
\end{eqnarray*}
Notice that $|A\cap N_G(u)| \equiv 1 \bmod 2$ iff $u\in \odd G A$. Hence, if $u\notin \odd G A$, 
$
   \odd {G\star u} A  =\odd G A  \symd (A\cap N_G(u))
$.  Otherwise, if $u \in \odd G A$, then 
$    \odd {G\star u} A  =\odd G A \symd N_G(u)\symd (A\cap N_G(u)) =  \odd G A\symd (N_G(u)\setminus A)
$.
\end{proof}


We can now state the main lemma of this section that shows that a \LOG $(G,\lambda)$ with a gflow is transformed by local complementation into a \LOG $(G\star u,\lambda')$ that also has a gflow. Note that the proof is rather long due to the need to do several case distinctions.

\begin{lemma}\label{lem:lc_gflow}\indexd{local complementation!on labelled open graph}
 Let $(g,\prec)$ be a gflow for $(G,I,O,\ld)$ and let $u\in\comp{O}$. Then $(g',\prec)$ is a gflow for $(G\star u, I, O,\ld')$, where
 \[
  \ld'(u) := \begin{cases} \XZm &\text{if } \ld(u)=\XYm \\ \XYm &\text{if } \ld(u)=\XZm \\ \YZm &\text{if } \ld(u)=\YZm \end{cases}
 \]
 and for all $v\in \comp{O}\setminus\{u\}$
 \[
  \ld'(v) := \begin{cases} \YZm &\text{if } v\in N_G(u) \text{ and } \ld(v)=\XZm \\ \XZm &\text{if } v\in N_G(u) \text{ and } \ld(v)=\YZm \\ \ld(v) &\text{otherwise.} \end{cases}
 \]
 and the correction sets $g'$ are
 \[
  g'(u) := \begin{cases} g(u)\symd \{u\} &\text{if } \ld(u)\in\{\XYm,\XZm\} \\ g(u) &\text{if } \ld(u)=\YZm \end{cases}
 \]
 and for all $v\in \comp{O}\setminus\{u\}$,
 \[
  g'(v) := \begin{cases} g(v) &\text{if } u\notin\odd{G}{g(v)} \\ g(v)\symd g'(u) \symd \{u\} &\text{if } u\in\odd{G}{g(v)}. \end{cases}
 \]
\end{lemma}
\begin{proof}
We divide the proof in three parts, first proving that $g(u)$ satisfies \ref{it:g}--\ref{it:YZ}, then proving that $g(v)$ for $v\neq u$ satisfies \ref{it:g} and \ref{it:odd}, and finally $g(v)$ for $v\neq u$ satisfies \ref{it:XY}--\ref{it:YZ}. Each of these parts is further subdivided based on the value of $\lambda(u)$ and $\lambda(v)$.
  
\noindent \textbf{Part 1}: Note that condition~\ref{it:g} is trivially still satisfied by $g'(u)$. We will show that $g(u)$ satisfies the remaining conditions, subdividing into cases based on $\lambda(u)$.

 \noindent\textbf{Case 1a}: If $\ld(u)\in\{\XYm,\XZm\}$, then $u\in\odd{G}{g(u)}$ and $\ld'(u)\in\{\XYm,\XZm\}$ with $\ld'(u)\neq\ld(u)$.
 We have $\odd{G\star u}{\{u\}} = N_{G\star u}(u) = N_G(u)$.
 Thus
 \begin{align}
  \odd{G\star u}{g'(u)} &= \odd{G\star u}{g(u)\symd\{u\}} \nonumber \\
  &= \odd{G\star u}{g(u)} \symd \odd{G\star u}{\{u\}} \nonumber \\
  &= \odd{G}{g(u)} \symd (N_G(u)\setminus g(u)) \symd N_G(u) \nonumber \\
  &= \odd{G}{g(u)} \symd (N_G(u)\cap g(u)) \label{eq:ld_XYXZ}
 \end{align}
 which implies
 \begin{itemize}
  \item $u\in\odd{G\star u}{g'(u)}$ since $u\in\odd{G}{g(u)}$ and $u\notin N_G(u)$,
  \item $u\in g'(u)$ if and only if $u\notin g(u)$ (this is desired because $\ld'(u)\neq \ld(u)$); together with the previous item this yields~\ref{it:XY} or \ref{it:XZ}, as appropriate,
  \item if $v\in\odd{G\star u}{g'(u)}$, then either $v\in\odd{G}{g(u)}$ or $v\in g(u)$, so either $u=v$ or $u\prec v$; this is~\ref{it:odd}.
 \end{itemize}
 So $g'(u)$ indeed satisfies all the required conditions.

 \noindent\textbf{Case 1b}: If $\ld(u)=\YZ$, then $u\notin\odd{G}{g(u)}$, hence
 \begin{equation}\label{eq:ld_YZ}
  \odd{G\star u}{g'(u)} = \odd{G\star u}{g(u)} = \odd{G}{g(u)} \symd (N_G(u)\cap g(u)).
 \end{equation}
 This implies
 \begin{itemize}
  \item $u\notin\odd{G\star u}{g'(u)}$ because $u\notin\odd{G}{g(u)}$ and $u\notin N_G(u)$,
  \item $u\in g'(u)$ because $u\in g(u)$; together with the previous item this yields Condition~\ref{it:YZ},
  \item if $v\in\odd{G\star u}{g'(u)}$, then either $v\in\odd{G}{g(u)}$ or $v\in g(u)$, so either $u=v$ or $u\prec v$; this is Condition~\ref{it:odd}.
 \end{itemize}
 So $g'(u)$ indeed satisfies all the required conditions.

 \noindent\textbf{Part 2}:
We will show that the correction set $g(v)$, where $v\neq u$,
satisfies conditions~\ref{it:g} and \ref{it:odd} of gflow,
splitting into subcases depending on whether  $u\in\odd{G}{g(v)}$ or $u\notin\odd{G}{g(v)}$.

 \noindent\textbf{Case 2a}:
 If $u\notin\odd{G}{g(v)}$, first note that
 \[
  \odd{G\star u}{g'(v)} = \odd{G\star u}{g(v)} = \odd{G}{g(v)} \symd (N_G(u)\cap g(v)).
 \]
 Hence if $w\in\odd{G\star u}{g(v)}$, then either $w\in\odd{G}{g(v)}$ or $w\in g(v)$, so $v=w$ or $v\prec w$; this is condition~\ref{it:odd}.
 Condition~\ref{it:g} is trivially still satisfied.

 \noindent\textbf{Case 2b}:
 If $u\in\odd{G}{g(v)}$, first note that $v\prec u$.
 If $w\in g'(v)$, then $w\in g(v)$ or $w\in g'(u)$ or $w=u$ and in each case $w=v$ or $v\prec w$, this is condition~\ref{it:g}.
 Furthermore, we have
 \begin{align*}
  \odd{G\star u}{g'(v)} &= \odd{G\star u}{g(v) \symd g'(u) \symd \{u\}} \\
  &= \odd{G\star u}{g(v)} \symd \odd{G\star u}{g'(u)} \symd N_G(u) \\
  &= \odd{G}{g(v)} \symd (N_G(u)\setminus g(v)) \symd \odd{G}{g(u)} \symd (N_G(u)\cap g(u)) \symd N_G(u) \\
  &= \odd{G}{g(v)} \symd (N_G(u)\cap g(v)) \symd \odd{G}{g(u)} \symd (N_G(u)\cap g(u))
 \end{align*}
 where the third step uses the property that 
 $$\odd{G\star u}{g'(u)} = \odd{G}{g(u)} \symd (N_G(u)\cap g(u))$$
for any $\ld(u)$, which follows from combining \eqref{eq:ld_XYXZ} and \eqref{eq:ld_YZ}.
 Hence, if ${w\in\odd{G\star u}{g'(v)}}$, then at least one of the following holds:
 \begin{itemize}
  \item $w\in\odd{G}{g(v)}$, so $v=w$ or $v\prec w$.
  \item $w\in g(v)$ so $v=w$ or $v\prec w$.
  \item $w\in\odd{G}{g(u)}$, so $u=w$ or $u\prec w$; in both cases $v\prec w$ since $v\prec u$.
  \item $w\in g(u)$, so $u=w$ or $u\prec w$; in both cases $v\prec w$ since $v\prec u$.
 \end{itemize}
 In each case, Condition~\ref{it:odd} is satisfied.

 \noindent\textbf{Part 3}:
 Finally we show that the correction set $g(v)$, where $v\neq u$,
satisfies conditions~\ref{it:XY}, \ref{it:XZ} or \ref{it:YZ} of gflow,
splitting into subcases depending on whether $v\in N_G(u)$ or $v\notin N_G(u)$, and also $\lambda(v)$.

 \noindent\textbf{Case 3a}: Suppose $v\in N_G(u)$ and distinguish cases according to $\ld(v)$.
 \begin{itemize}
  \item Suppose $\ld'(v)=\XYm$. Then $\ld(v)=\XYm$, and hence $v\notin g(v)$ and $v\in\odd{G}{g(v)}$.
  We have
   \begin{itemize}
    \item $v\in\odd{G\star u}{g'(v)}$ since $v\in\odd{G}{g(v)}$ and $v\notin g(v)$,
    \item $v\notin g'(v)$,
   \end{itemize}
   which together give condition~\ref{it:XY}.
  \item Suppose $\ld'(v) = \XZm$. Then $\ld(v)=\YZm$, and hence $v\in g(v)$ and $v\notin\odd{G}{g(v)}$.
   We have
    \begin{itemize}
     \item $v\in\odd{G\star u}{g'(v)}$ since $v\notin\odd{G}{g(v)}$ and $v\in N_G(u)\cap g(v)$,
     \item $v\in g'(v)$,
    \end{itemize}
   which together give condition~\ref{it:XZ}.
   \item Suppose $\ld'(v) = \YZm$. Then $\ld(v)=\XZm$, and hence $v\in g(v)$ and $v\in\odd{G}{g(v)}$.
   We have
    \begin{itemize}
     \item $v\notin\odd{G\star u}{g'(v)}$ since $v\in\odd{G}{g(v)}$ and $v\in N_G(u)\cap g(v)$,
     \item $v\in g'(v)$,
    \end{itemize}
   which together give condition~\ref{it:YZ}.
 \end{itemize}

 \noindent\textbf{Case 3b}: Suppose $v\notin N_G(u)$ and distinguish cases according to $\ld(v)$.
 \begin{itemize}
  \item Suppose $\ld(v)=\XYm$, this case is analogous to the corresponding one in Subcase~3a, so condition~\ref{it:XY} is satisfied.
  \item Suppose $\ld(v)=\XZm$, then $v\in g(v)$ and $v\in\odd{G}{g(v)}$.
   Furthermore, $\ld'(v) = \XZm$.
   We have
    \begin{itemize}
     \item $v\in\odd{G\star u}{g'(v)}$ since $v\in\odd{G}{g(v)}$ and $v\notin N_G(u)\cap g(v)$,
     \item $v\in g'(v)$,
    \end{itemize}
   which together give condition~\ref{it:XZ}.
  \item Suppose $\ld(v)=\YZm$, then $v\in g(v)$ and $v\notin\odd{G}{g(v)}$.
   Furthermore, $\ld'(v) = \YZm$.
   We have
    \begin{itemize}
     \item $v\notin\odd{G\star u}{g'(v)}$ since $v\notin\odd{G}{g(v)}$ and $v\notin N_G(u)\cap g(v)$,
     \item $v\in g'(v)$,
    \end{itemize}
   which together give condition~\ref{it:YZ}. \qedhere
 \end{itemize}
\end{proof}

The previous result required the vertex that was complemented on to not be an output. A similar result holds when it \emph{is} an output.
\begin{lemma}
 Let $(g,\prec)$ be a gflow for $(G,I,O,\ld)$ and let $u\in O$. Then $(g',\prec)$ is a gflow for $(G\star u, I, O,\ld')$, where for all $v\in \comp{O}:$
 \[
  \ld'(v) := \begin{cases} \YZm &\text{if } v\in N_G(u) \text{ and } \ld(v)=\XZm \\ \XZm &\text{if } v\in N_G(u) \text{ and } \ld(v)=\YZm \\ \ld(v) &\text{otherwise.} \end{cases}
 \]
 Furthermore, for all $v\in \comp{O}$:
 \[
  g'(v) := \begin{cases} g(v) &\text{if } u\notin\odd{G}{g(v)} \\ g(v) \symd \{u\} &\text{if } u\in\odd{G}{g(v)}. \end{cases}
 \]
\end{lemma}
\begin{proof}
 The proof is basically the same as that of Lemma~\ref{lem:lc_gflow} if we take $g(u)$ and $g'(u)$ to be empty.
 The output vertex has no label, so its label does not need to be updated.
\end{proof}

As a pivot is just a series of three local complementations, doing a pivot also preserves the existence of a gflow on a \LOG. Let us explicitly state the effect this has on the measurement type of vertices.
\begin{corollary}\label{cor:pivot_gflow}
	Let $(G,I,O,\ld)$ be a \LOG with a gflow and let $u,v\in\comp{O}$ be connected by an edge. Then the \LOG $(G\land uv, I, O,\ld')$ has a gflow, where
	 \[
	  \ld'(a) = \begin{cases} \YZm &\text{if } \ld(a)=\XYm \\
	                                     \XZm &\text{if } \ld(a)=\XZm \\
	                                     \XYm &\text{if } \ld(a)=\YZm \end{cases}
	 \]
	 for $a\in\{u,v\}$, and $\ld'(w)=\ld(w)$ for all $w\in \comp{O}\setminus\{u,v\}$
\end{corollary}
\begin{proof}
    $G\land uv = G\star u\star v \star u$ so it suffices to check that $\ld'$ as specified here is indeed what you get if you apply Lemma~\ref{lem:lc_gflow} first for $u$, then for $v$ and then for $u$ again. For instance, if $\lambda(u)=\XYm$, then after the first application of Lemma~\ref{lem:lc_gflow} we have $\lambda_1(u) = \XZm$. Then after the local complementation on $v$ we get $\lambda_2(u) = \YZm$ (as $u$ and $v$ are neighbours), and then applying the final local complementation on $u$ again we get $\lambda'(u) = \lambda_3(u) = \YZm$ as desired. The other cases are checked similarly.
\end{proof}

Perhaps somewhat surprisingly, the deletion of some types of vertices also preserves the existence of gflow.

\begin{lemma}\label{lem:deletepreservegflow}
    Let $(g,\prec)$ be a gflow for $(G,I,O,\ld)$ and let $u\in \comp{O}$ with $\ld(u) \neq \XYm$. Then $(g',\prec)$ is a gflow for $(G\setminus\{u\},I,O,\ld)$ where $\forall v\in V, v\neq u$:
    \[g'(v) := \begin{cases} g(v) &\text{if } u\not \in g(v)\\ g(v)\symd g(u) &\text{if } u \in g(v) \end{cases}\]
\end{lemma}
\begin{proof}
    Observe that $u\in g(u)$ as $\ld(u)\neq \XYm$ and hence $u\not \in g'(v)$ for both cases in the definition. Hence, $g'$ is indeed a function on the graph $G\backslash \{u\}$.

    To check that $g'$ is indeed a gflow we check the necessary conditions for all $v\in G\setminus\{u\}$. First, if $u\not \in g(v)$, then $g'(v) = g(v)$ and hence we are done. If $u \in g(v)$, then $v\prec u$ and hence also $v\prec w$ where $w \in g(u)$ or $w\in \odd{G}{g(u)}$. Since $g'(v) = g(v)\symd g(u)$ we see that then conditions \ref{it:g} and \ref{it:odd} are indeed satisfied. For conditions \ref{it:XY}--\ref{it:YZ} we note that we cannot have $v \in g(u)$ or $v\in \odd{G}{g(u)}$. As a result $v\in g'(v) \iff v\in g(v)$ and $v\in \odd{G\setminus\{u\}}{g'(v)} \iff v \in \odd{G}{g(v)}$. Since the labels of all the vertices stay the same, \ref{it:XY}--\ref{it:YZ} stay satisfied.
\end{proof}

\begin{remark}
    The condition in the previous lemma that $\ld(u) \neq \XYm$ is necessary. Removing a vertex with label \XY will, in general, create a graph which does not allow a gflow. For instance consider the following open graph:
    \ctikzfig{line-graph}
    Here the first two vertices both have label \XY. This graph has a gflow specified by $I\prec u \prec O$ and $g(I) = \{u\}$, $g(u) = \{O\}$, but removing $u$ will disconnect the graph and hence the resulting graph does not have a gflow. Note that if we were to have the same graph, but with $u$ in a different measurement plane, then the graph will \emph{not} have a gflow to start with (because if it did, then we would need $u\in g(I)$, so that $I\prec u$ but also $u\in g(u)$ so that $I\in \odd{}{g(u)}$ giving $u\prec I$), and hence this does not contradict Lemma~\ref{lem:deletepreservegflow}.
\end{remark}

In general, removing a qubit from a measurement pattern (\ie a vertex from the underlying \LOG) will change the semantics of the computation. In Section~\ref{sec:removing-clifford-vertices} we will see that there are however some semantics preserving operations that come down to deleting a vertex.

The final gflow-preserving graph-operation we will consider is adding vertices to inputs and outputs. This is sometimes necessary to unfuse a phase on an input or output (cf.~Lemma~\ref{lem:zx-to-mbqc-form}).

\begin{lemma}\label{lem:gflow-add-output}
Let $\Gamma=(G,I,O,\ld)$ be a \LOG\ with $G=(V,E)$. Let $\Gamma'$ be the \LOG\ that results from converting an output $u\in O$ into a vertex measured in the \XY-plane and adding a new output vertex $u'$ after it:
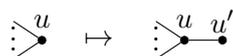
\ctikzfig{gflow-add-output}
I.e.\ let $\Gamma'=(G',I,O',\ld')$, where $G'=(V',E')$ with $V'=V\cup\{u'\}$ and $E'=E\cup\{u\sim u'\}$, $O'=(O\setminus\{u\})\cup\{u'\}$, and $\ld'(v)$ is the extension of $\ld$ to domain $V'\setminus O'$ with $\ld'(u)=\XYm$. Then if $\Gamma$ has a gflow, $\Gamma'$ also has a gflow.
\end{lemma}
\begin{proof}
 Suppose $\Gamma$ has a gflow $(g,\prec)$.
 Let $g'$ be the extension of $g$ to domain $V'\setminus O'$ which satisfies $g'(u)=\{u'\}$, and let $\prec'$ be the transitive closure of $\prec\cup\{(u,u')\}$.

 The tuple $(g',\prec')$ inherits \ref{it:g} and \ref{it:XY}--\ref{it:YZ} for all $v\in V\setminus O$ because the correction sets have not changed for any of the original vertices.
 Furthermore, $u'\in\odd{G'}{g'(v)}$ for any $v$ implies $u\in g'(v)$ as $u$ is the only neighbour of $u'$; hence $u'\in\odd{G'}{g'(v)}$ implies $v\prec' u \prec' u'$.
 Therefore \ref{it:odd} continues to be satisfied for all $v\in V\setminus O$.

 Now, for $u$, \ref{it:g} holds because $u\prec' u'$ by definition, \ref{it:odd} holds because $\odd{G'}{g'(u)}=\{u\}$, and \ref{it:XY} can easily be seen to hold.
 Thus, $(g',\prec')$ is a gflow for $\Gamma'$.
\end{proof}

\begin{lemma}\label{lem:gflow-add-input}
Let $\Gamma=(G,I,O,\ld)$ be a \LOG\ with $G=(V,E)$.
 Let $\Gamma'$ be the \LOG\ that results from converting an input $u\in I$ into a vertex measured in the \XY-plane and adding a new input vertex $u'$ in the \XY-plane before it:
 \ctikzfig{gflow-add-input}
 I.e.\ let $\Gamma'=(G',I',O,\ld')$, where $G'=(V',E')$ with $V'=V\cup\{u'\}$ and $E'=E\cup\{u\sim u'\}$, $I'=(I\setminus\{u\})\cup\{u'\}$, and $\ld'(v)$ is the extension of $\ld$ to domain $V'\setminus O$ which satisfies $\ld'(u')=\XYm$. Then if $\Gamma$ has gflow, $\Gamma'$ also has gflow.
\end{lemma}
\begin{proof}
 Suppose $\Gamma$ has a gflow $(g,\prec)$.
 Let $g'$ be the extension of $g$ to domain $V'\setminus O$ which satisfies $g'(u')=\{u\}$, and let $\prec'$ be the transitive closure of $\prec\cup\{(u',w):w\in N_G(u)\cup\{u\}\}$.

 The tuple $(g',\prec')$ inherits the gflow properties for all $v\in V\setminus O$ because the correction sets have not changed for any of the original vertices and because the additional inequalities in $\prec'$ do not affect the gflow properties for any $v\in V\setminus O$.
 The latter is because
 \begin{itemize}
  \item $u'\notin g'(v)=g(v)$ for any $v\in V\setminus O$, and
  \item $u'\notin\odd{G'}{g'(v)}=\odd{G'}{g(v)}$ for any $v\in V\setminus O$ since its only neighbour $u$ was an input in $\Gamma$ and thus satisfies $u\notin g(v)$ for any $v\in V\setminus O$.
 \end{itemize}
 Now, for $u'$, \ref{it:g} holds by the definition of $\prec'$.
 Note that $\odd{G'}{g(u')}=N_{G'}(u)$, so \ref{it:odd} also holds by the definition of $\prec'$.
 Finally, \ref{it:XY} holds because $u'\notin g(u')$ and $u'\in\odd{G'}{g(u')}=N_{G'}(u)$.
 Thus, $(g',\prec')$ is a gflow for $\Gamma'$.
\end{proof}

\section{Graph operations on measurement patterns}\label{sec:pattern-local-complementation}

The previous section detailed how to change a \LOG while preserving the existence of a gflow. A measurement pattern however also has measurement angles associated to it. So in order to do these transformations in a way that preserves the linear map we are implementing, we need to incorporate the changes to these measurement angles. That is what we will do in this section, by representing measurement patterns as MBQC-form diagrams, and rewriting these using the ZX-calculus.

First, let us see how to rewrite MBQC+LC diagrams to MBQC-form diagrams by incorporating the local Cliffords into the measurement pattern.

\begin{lemma}\label{lem:SQU-to-MBQC-form}
 Any \zxdiagram $D$ which is in MBQC+LC form can be brought into MBQC form.
 Moreover, if the MBQC-form part of $D$ involves $n$ qubits, of which $p$ are inputs and $q$ are outputs, then the resulting MBQC-form diagram contains at most $n+2p+4q$ qubits. If the underlying \LOG of $D$ has a gflow, then so does the resulting MBQC-form diagram.
\end{lemma}
\begin{proof}
 Any single-qubit Clifford unitary can be expressed as a composite of three phase gates~\cite[Lemma~3]{BackensCompleteness}.
 Note that this result holds with either choice of spider, i.e.\ any single-qubit Clifford unitary can be expressed as \input{SQC-red.tikz} or \input{SQC-green.tikz}.

 Now, with the Z-X-Z version, for any Clifford operator on an input, we can `push' the final Z-phase gate through the graph state part onto the outgoing wire.
 There, it will either merge with the measurement effect or with the output Clifford unitary:
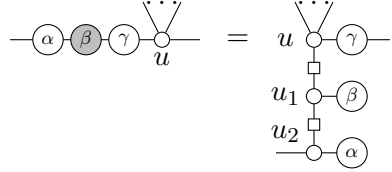
\ctikzfig{SQC-in-replacement}
 If $\gamma\in\{0,\pi\}$, merging the phase shift with a measurement effect may change the angle but not the phase label, e.g.\ if $\gamma=\pi$:
 \begin{center}
  \input{pivot-pi-phases-XY.tikz} \qquad \input{pivot-pi-phases-XZ.tikz} \qquad \input{pivot-pi-phases-YZ.tikz}
 \end{center}
 If $\gamma\in\{\frac\pi2,-\frac\pi2\}$, merging the phase shift with a measurement effect will flip the phase labels \XZ and \YZ, e.g.\ if $\gamma=-\frac\pi2$:
 \begin{center}
  \input{lc-N-XY.tikz} \qquad \input{lc-N-XZ.tikz} \qquad \input{lc-N-YZ.tikz}
 \end{center}
 Thus we need to add at most two new qubits to the MBQC-form part when removing a Clifford unitary on the input.

 For a Clifford unitary on the output, we have
  \ctikzfig{SQU-out-replacement}
 Thus we add at most four new qubits.

 Combining these properties, we find that rewriting to MBQC form adds at most $2p+4q$ new qubits to the pattern.

 The underlying \LOG has changed by adding new vertices before inputs and vertices after outputs. Hence, Lemmas~\ref{lem:gflow-add-output} and~\ref{lem:gflow-add-input} ensure that the resulting MBQC-form diagram still has a gflow.
\end{proof}

Eqs.~\eqref{eq:lc-zx} and~\eqref{eq:pivot-desc} showed how to apply a local complementation and pivot on a ZX-diagram by introducing some local Clifford spiders, while in Section~\ref{sec:gflow-rewrite} we saw what the effect is of these operation on the gflow of a \LOG. Now we combine these to modify an MBQC form diagram while preserving the existence of a gflow.

\begin{lemma}\label{lem:lc-MBQC-form-non-input}\indexd{local complementation!on measurement pattern}
 Let $D$ be an MBQC+LC diagram and let $u\in G(D)$ be a non-input vertex.
 Then the diagram resulting from applying Eq.~\eqref{eq:lc-zx} on $u$ (\ie a local complementation on $u$), can be transformed into an equivalent MBQC+LC diagram $D'$ with $G(D')=G(D)\star u$. If $D$ has a gflow, then so does $D'$.
\end{lemma}
\begin{proof}
 Suppose $D$ is an MBQC+LC diagram, $\Gamma=(G,I,O,\ld)$ the corresponding \LOG, and $\alpha:\comp{O}\to[0,2\pi)$ the associated measurement angles.
 By assumption, $u\notin I$, so -- with the exception of the output wire or the edge to the measurement effect -- all edges incident on $u$ connect to neighbouring vertices in the graph.
 The input wires on the other qubits can be safely ignored.
 To get back an MBQC+LC diagram after Eq.~\eqref{eq:lc-zx} is applied to $u$, we only need to rewrite the measurement effects, and hence we need to construct new $\lambda'$ and $\alpha'$ for these measurement effects. We do that as follows.

 First of all, there are no changes to the measurement effects on vertices $v\not\in N(u)\cup\{u\}$, and hence for those vertices we have $\lambda'(v)=\lambda(v)$ and $\alpha'(v)=\alpha(v)$.

 The vertex $u$ gets a $\frac\pi2$ X-phase from the application of Eq.~\eqref{eq:lc-zx}. If $u\in O$, then it has no associated measurement plane or angle. In this case, this red $\frac\pi2$ simply stays on the output wire, as allowed in an MBQC+LC diagram. When $u\notin O$, there are three possibilities, depending on $\ld(u)$:
 \begin{itemize}
  \item If $\ld(u)=\XYm$, then the new measurement effect is
   \ctikzfig{lc-u-XY}
   i.e.\ $\ld'(u)=\XZm$ and $\alpha'(u)=\frac{\pi}{2}-\alpha(u)$.
  \item If $\ld(u)=\XZm$, then the new measurement effect is
   \ctikzfig{lc-u-XZ}
   i.e.\ $\ld'(u)=\XYm$ and $\alpha'(u)=\alpha(u)-\frac{\pi}{2}$.
  \item If $\ld(u)=\YZm$, then the new measurement effect is
   \ctikzfig{lc-u-YZ}
   i.e.\ $\ld'(u)=\YZm$ and $\alpha'(u)=\alpha(u)+\frac{\pi}{2}$.
 \end{itemize}
 The vertices $v$ that are neighbours of $u$ get a $-\frac\pi2$ Z-phase. Again, if such a $v$ is an output, this phase can be put as a local Clifford on the output. If it is not an output, then there are also three possibilities depending on $\ld(v)$:
 \begin{itemize}
  \item If $\ld(v)=\XYm$, then the new measurement effect is
   \ctikzfig{lc-N-XY}
   i.e.\ $\ld'(v)=\XYm$ and $\alpha'(v)=\alpha(v)-\frac{\pi}{2}$.
  \item If $\ld(v)=\XZm$, then the new measurement effect is
   \ctikzfig{lc-N-XZ}
   i.e.\ $\ld'(v)=\YZm$ and $\alpha'(v)=\alpha(v)$.
  \item If $\ld(v)=\YZm$, then the new measurement effect is
   \ctikzfig{lc-N-YZ}
   i.e.\ $\ld'(v)=\XZm$ and $\alpha'(v)=-\alpha(v)$.
 \end{itemize}
 With these changes, we see that the resulting diagram $D'$ is indeed in MBQC+LC form. The underlying graph $G(D')$ results from the local complementation about $u$ of the original graph $G(D)$. Furthermore, the measurement planes changed in the same way as in Lemma~\ref{lem:lc_gflow}, and hence if $D$ has a gflow, then so will $D'$.
\end{proof}

\begin{proposition}\label{prop:MBQC-lc-MBQC}\indexd{pivot!on measurement pattern}
 Let $D$ be an MBQC+LC diagram and let $u\in G(D)$.
 Then the diagram resulting from applying Eq.~\eqref{eq:lc-zx} on $u$ (\ie a local complementation on $u$), can be transformed into an equivalent MBQC+LC diagram $D'$. If $D$ has a gflow, then so does $D'$.
\end{proposition}
\begin{proof}
 If $u$ is not an input vertex, the result is immediate from Lemma~\ref{lem:lc-MBQC-form-non-input}.

 If instead $u$ is an input vertex, we modify $D$ by replacing the input wire incident on $u$ by an additional graph vertex $u'$ measured in the \XY-plane at angle 0, and a Hadamard unitary on the input wire:
 \ctikzfig{input-replacement}
 Throughout this process, the measurement effect on $u$ (if any) does not change, so it is left out of the above equation.
 In the resulting diagram $D'$, $u$ is no longer an input.
 Furthermore, $D'$ is an MBQC+LC diagram.
 Thus, the desired result follows by applying Lemma~\ref{lem:lc-MBQC-form-non-input} to $D'$. That the resulting diagram has a gflow follows by combining Lemmas~\ref{lem:gflow-add-input} and~\ref{lem:lc_gflow}.
\end{proof}

A pivot is just a sequence of three local complementations.
Thus, the previous lemma already implies that an MBQC+LC diagram that has been pivoted on can also be brought back into MBQC+LC form. Nevertheless, it will be useful to explicitly write out how the measurement planes and angles of the vertices change.

\begin{lemma}\label{lem:pivot-MBQC-form-non-input}
 Let $D$ be an MBQC+LC diagram and let $u,v\in G(D)$ be neighbouring non-input vertices.
 Then the diagram resulting from applying Eq.~\eqref{eq:pivot-desc} to $u$ and $v$ (\ie a pivot about $u\sim v$) can be transformed into an equivalent MBQC+LC form diagram $D'$ satisfying $G(D') = G(D)\wedge uv$. If $D$ has a gflow, then so will $D'$.
\end{lemma}
\begin{proof}
  Let $\Gamma=(G,I,O,\ld)$ be the \LOG underlying $D$ and let $\alpha:\comp{O}\to[0,2\pi)$ be the associated measurement angles.
  We will denote the measurement planes after pivoting by $\ld':\comp{O}\to\{\XYm,\XZm,\YZm\}$ and the measurement angles after pivoting by $\alpha':\comp{O}\to[0,2\pi)$.
  Let $a\in\{u,v\}$, then:
  \begin{itemize}
    \item If $a$ is an output, we consider the Hadamard resulting from the pivot operation as a Clifford operator on the output.
    \item If $\ld(a)=\XYm$ then $\ld'(a) = \YZm$ and if $\ld(a)=\YZm$ then $\ld'(a) = \XYm$:
   \ctikzfig{pivot-u-XY}
   In both cases, the measurement angle stays the same: $\alpha'(a) = \alpha(a)$.
   \item If $\ld(a)=\XZm$, then
   \ctikzfig{pivot-u-XZ}
   \ie $\ld'(a) = \XZm$ and $\alpha'(a) = \frac\pi2 - \alpha(a)$.
  \end{itemize}

  The only other changes on measurement effects are $\pi$ Z-phases on vertices $w\in N(u)\cap N(v)$.
  For measured (i.e.\ non-output) vertices, these preserve the measurement plane and are absorbed into the measurement angle in all three cases:
   \begin{align*}
  (\ld'(w), \alpha'(w)) =
  \begin{cases}
  (\XYm, \alpha(w) + \pi) & \text{if } \ld(w) = \XYm \mspace{-1.5mu} \quad \input{pivot-pi-phases-XY.tikz} \\
  (\YZm, -\alpha(w)) & \text{if } \ld(w) = \YZm \quad  \input{pivot-pi-phases-YZ.tikz} \\
  (\XZm, -\alpha(w)) & \text{if } \ld(w) = \XZm \quad  \input{pivot-pi-phases-XZ.tikz}
  \end{cases}
  \end{align*}
  If instead $w$ is an output vertex, the $\pi$ Z-phase becomes a Clifford gate on the output wire.
  The measurement planes and the graph change exactly like in Corollary~\ref{cor:pivot_gflow} and hence $D'$ has a gflow when $D$ does.
\end{proof}

\section{Removing Clifford vertices}\label{sec:removing-clifford-vertices}

In this section, we will see that we can rewrite measurement patterns so that less qubits are involved in the pattern while preserving the computation performed and the presence of a gflow.

\begin{definition}\label{dfn:internal-boundary-Clifford}
  Let $D$ be a \zxdiagram in MBQC+LC form, with underlying \LOG $(G,I,O,\ld)$ and corresponding set of measurement angles $\alpha:\comp{O}\to [0,2\pi)$. We say a measured vertex $u\in G$ is \Define{Clifford} when $\alpha(u) = k\frac\pi2$ for some $k$.\indexd{Clifford!--- vertex} Otherwise the vertex is \Define{non-Clifford}.
\end{definition}

Our goal will be to remove as many internal Clifford vertices as possible. Recall that in Section~\ref{sec:graph-theoretic-simp} we managed to use local complementation and pivoting rules to delete spiders with a specific phase. In this section we will do something similar, but slightly more general.

We make a key observation for our simplification scheme: an \YZ or \XZ measurement with a $0$ or $\pi$ phase can be removed from the pattern by modifying its neighbours in a simple manner:

\begin{lemma}\label{lem:ZX-remove-YZ-Pauli}
Let $D$ be a ZX-diagram in MBQC+LC form, and let $u\in G(D)$ be a non-input vertex with $\lambda(u) \neq \XY$ and $\alpha(u) = a\pi$ where $a=0$ or $a=1$. Then we can efficiently find an equivalent diagram $D'$ with $G(D')=G(D)\backslash \{u\}$. If $D$ has a gflow, then so does $D'$.
\end{lemma}
\begin{proof}
	Since $\lambda(u) \neq \XY$, its $a\pi$ measurement angle is an X-spider.
 It is then straightforward to show using the axioms of the ZX-calculus, that:
  \ctikzfig{remove-YZ-measurement}
  These $a\pi$ Z-phases on the right-hand side can be absorbed into the measurement of the neighbouring vertices (or for output vertices, added as a local Clifford). This does not change the plane of the measurement, only the angle. The resulting diagram $D'$ is then also in MBQC+LC form. Lemma~\ref{lem:deletepreservegflow} shows that the existence of a gflow is preserved.
\end{proof}

We can now state a version of Eq.~\eqref{eq:lc-simp}, but in the context of measurement patterns.

\begin{lemma}\label{lem:lc-simp}
  Let $D$ be a ZX-diagram in MBQC+LC form with vertices $V$, and let $u\in V$ be a non-input vertex with $\lambda(u) = \YZ$ or $\XY$ and $\alpha(u) = \pm\frac\pi2$. Then we can efficiently find an equivalent diagram $D'$ with vertices $V\setminus \{u\}$. If $D$ has a gflow, then so does $D'$.
\end{lemma}
\begin{proof}
  Apply a local complementation about $u$ and reduce the diagram to MBQC+LC form with Lemma~\ref{lem:lc-MBQC-form-non-input}. This lemma shows that the presence of a gflow is preserved.
  As can be seen from Lemma~\ref{lem:lc-MBQC-form-non-input}, if $u$ was in the \XY plane, then it will be transformed to the \XZ plane and will have a measurement angle of $\frac\pi2 \mp\frac\pi2$. As a result, its measurement angle is of the form $a\pi$ for $a\in\{0,1\}$.
  If instead it was in the \YZ plane, then it stays in the \YZ plane, but its angle is transformed to $\frac\pi2 \pm\frac\pi2$ in which case it will also be of the form $a\pi$ for $a\in\{0,1\}$.
  In both cases we can remove the vertex $u$ using Lemma~\ref{lem:ZX-remove-YZ-Pauli} while preserving the presence of a gflow.
\end{proof}

Analogously, the following can be seen as a generalisation of Eq.~\eqref{eq:pivot-simp}.

\begin{lemma}\label{lem:pivot-simp}
  Let $D$ be a ZX-diagram in MBQC+LC form with vertices $V$, and let $u,v \in V$ be neighbouring non-input measured vertices.
  Suppose that either $\ld(u)=\XYm$ with $\alpha(u) = a\pi$ for $a\in \{0,1\}$ or $\ld(u) = \XZm$ with $\alpha(u) = (-1)^a\frac\pi2$.
  Then we can efficiently find an equivalent diagram $D'$ with vertices $V\setminus \{u\}$. If $D$ has a gflow, then so does $D'$.
\end{lemma}
\begin{proof}
  We apply a pivot about $uv$ and reduce the diagram to MBQC+LC form with Lemma~\ref{lem:pivot-MBQC-form-non-input}. This preserves the presence of gflow.
  As can be seen from Lemma~\ref{lem:pivot-MBQC-form-non-input}, if $\lambda(u) = \XYm$ then $\lambda'(u) = \YZ$ with $\alpha'(u)=\alpha(u) = a\pi$. If instead we had $\lambda(u) = \XZm$ (and thus $\alpha(u) = (-1)^a\frac\pi2$), then $\lambda'(u) = \XZm$, but $\alpha'(u) = \frac\pi2 - \alpha(u) = \frac\pi2 - (-1)^a \frac\pi2 = a\pi$. In both cases we can remove the vertex $u$ using Lemma~\ref{lem:ZX-remove-YZ-Pauli} while preserving the presence of a gflow.
\end{proof}

Combining the previous lemmas we can remove any internal Clifford vertex, except for some internal Clifford vertices that are only connected to boundary vertices. While it might in general not be possible to remove such vertices, when the diagram has a gflow, we can always find an equivalent smaller diagram.

\begin{lemma}\label{lem:removeboundaryPauli}
  Let $D$ be a ZX-diagram in MBQC+LC form which has a gflow. Denote its vertices by $V$. Let $u\in V$ be an internal vertex that is only connected to input and output vertices. Suppose that either $\ld(u)=\XYm$ with $\alpha(u) = a\pi$ for $a\in \{0,1\}$ or $\ld(u) = \XZm$ with $\alpha(u) = (-1)^a\frac\pi2$. Then we can efficiently find an equivalent diagram $D'$ with gflow and vertices $V\backslash\{u\}$.
\end{lemma}
\begin{proof}
  We prove the result for $\ld(u)=\XYm$ and $\alpha(u) = a\pi$. The other case is similar.

  We claim that $u$ is connected to at least one output that is itself not an input. Suppose it is not. Then the diagram looks like the following:
  \ctikzfig{ZX-Pauli-projector}
  Here `LC' denotes that there are local Cliffords on the inputs.
  Since $D$ has gflow, the entire diagram must be (proportional to) an isometry, and hence it must still be an isometry if we remove the local Cliffords on the inputs. But we note that we then have the map
  \ctikzfig{ZX-Pauli-projector2}
  before the rest of the diagram. This map is not invertible. This is a contradiction, as the entire diagram cannot then be an isometry.

  So $u$ is connected to some output vertex $v$ which is not an input. We can then do a pivot on $uv$ using Lemma~\ref{lem:pivot-MBQC-form-non-input}. This adds a Hadamard gate beyond $v$, and changes the label of $u$ to \YZ. We can then remove $u$ using Lemma~\ref{lem:ZX-remove-YZ-Pauli}. The resulting diagram then still has gflow.
\end{proof}

\begin{theorem}\label{thm:simplifiedZXdiagram}
  Let $D$ be a ZX-diagram in MBQC+LC form that has a gflow. Then we can efficiently find an equivalent ZX-diagram $D'$ in MBQC+LC form, which also has gflow and which contains no non-input Clifford vertices.
\end{theorem}
\begin{proof}
  Starting with $D$ we simplify the diagram step by step using the following algorithm:
  \begin{enumerate}
  \item Using Lemma~\ref{lem:lc-simp} repeatedly, remove any non-input vertex measured in the \YZ or \XY plane which has a $\pm \frac\pi2$ phase.
  \item Using Lemma~\ref{lem:ZX-remove-YZ-Pauli} repeatedly, remove any non-input vertex measured in the \YZ or \XZ plane with angle $a\pi$.
  \item Using Lemma~\ref{lem:pivot-simp} repeatedly, remove any non-input vertex which is connected to any other internal vertex and is either measured in the \XY plane with an $a\pi$ phase or is measured in the \XZ plane with a $\pm \frac\pi2$ phase. If any have been removed, go back to step 1.
  \item If there are non-input measured Clifford vertices that are only connected to boundary vertices, use Lemma~\ref{lem:removeboundaryPauli} to remove them. Then go back to step 1. Otherwise we are done.
\end{enumerate}
By construction there are no internal Clifford vertices left at the end. Every step preserves the existence of a gflow, so the resulting diagram still has a gflow.
As every step removes a vertex, this process terminates in at most $n$ steps, where $n$ is the number of vertices in $D$. Each of the steps possibly requires doing a pivot or local complementation requiring $O(n^2)$ elementary graph operations. Hence, the algorithm requires at most $O(n^3)$ elementary graph operations.
\end{proof}

We can reformulate this theorem in terms of measurement patterns. This shows that any qubit measured in a Clifford angle can be removed from a pattern by modifying it in a suitable manner.

\begin{theorem}\label{thm:simplifiedMBQCpattern}
  Let $(G,I,O,\ld,\alpha)$ represent a uniformly, strongly, and stepwise deterministic measurement pattern with $q$ inputs and outputs and $n$ non-Clifford measured qubits.
  Then we can efficiently find a uniformly, strongly and stepwise deterministic measurement pattern that implements the same linear map and uses at most $(n+8q)$ measurements.
\end{theorem}
\begin{proof}
  Let $D$ be the ZX-diagram in MBQC form from Lemma~\ref{lem:zx-equals-linear-map} that implements the same linear map as the measurement pattern $\pat:=(G,I,O,\ld,\alpha)$.
  As $\pat$ is uniformly, strongly and stepwise deterministic, it has a gflow by Theorem~\ref{t-flow}, and hence $D$ also has gflow by Definition~\ref{dfn:zx-gflow}.
  Let $D'$ be the ZX-diagram in MBQC+LC form produced by Theorem~\ref{thm:simplifiedZXdiagram}.
  Since $D'$ has no internal Clifford vertices, its MBQC-form part can have at most $n$ internal vertices.
  It may still have boundary Clifford vertices, and by assumption $\abs{O}=\abs{I}=q$, so the MBQC-form part contains at most $(n+2q)$ vertices.

  Denote by $D''$ the MBQC-form diagram produced by applying Lemma~\ref{lem:SQU-to-MBQC-form} to $D'$.
  Then $D''$ has at most $((n+2q)+6q)$ vertices.

  We can construct a labelled open graph $\Gamma'$ and measurement angles $\alpha'$ from $D''$ using Lemma~\ref{lem:zx-to-pattern}. As $D''$ has a gflow, $(\Gamma', \alpha')$ represents a uniformly, strongly and stepwise deterministic measurement pattern.
  This new pattern involves at most $(n+8q)$ qubits.
\end{proof}

\begin{remark}
    While it has been known that qubits measured in a Clifford angle can be removed from a graph state~\cite{graphstates}, as far as the author is aware, the result that this can be done on a measurement pattern while preserving determinism is new.
\end{remark}

\section{Further pattern optimisations}\label{sec:further-optimisation}

In this section we continue the process of removing qubits from measurement patterns started in the previous section by finding a few more simplification rules. Before we do that however, we will show that a measurement pattern with measurements in all three planes can always be reduced to one with measurements in two planes, namely \XY and \YZ, in a straightforward manner using pivoting and local complementation.

\begin{definition}\label{def:pseudonormalform}
 A \LOG is in \Define{phase-gadget form}\indexd{phase-gadget form} if
 \begin{itemize}
  \item there does not exist any $v\in\comp{O}$ such that $\ld(v) = \XZm$, and
  \item there do not exist any neighbours $v,w\in\comp{O}$ such that $\ld(v)=\ld(w)=\YZm$.
 \end{itemize}
 We say an MBQC+LC form diagram or a measurement pattern is in phase-gadget form when its underlying \LOG is.
\end{definition}

\begin{remark}
We call this a phase-gadget form since, as discussed in Remark~\ref{rem:MBQC-gadget}, the \XY-type vertices form a graph-like ZX-diagram, while the \YZ-type vertices are phase gadgets that are connected to this graph-like diagram of \XY-type vertices.
\end{remark}

\begin{proposition}\label{prop:ZXtopseudonormalform}
Let $D$ be a ZX-diagram in MBQC+LC form which has a gflow.
Then we can efficiently find an equivalent ZX-diagram $D'$ in MBQC+LC form with the same number of vertices that has a gflow and is in phase-gadget form.
\end{proposition}
\begin{proof}
Set $D_0:=D$ and iteratively construct the diagram $D_{k+1}$ based on $D_k$.

\begin{itemize}
\item
  Suppose the diagram $D_k$ contains a pair of vertices $u \sim v$ that are both measured in the \YZ-plane.
  Note that any input vertex $w$ has $\ld(w) = \XYm$, as
  otherwise $w \in g(w)$ contradicting the definition of the co-domain of $g$ as given in Definition~\ref{defGFlow}.
  Therefore $u,v \notin I$.
  Let $D_{k+1}$ be the MBQC+LC diagram that results from pivoting on that edge $u \sim v$ (Lemma~\ref{lem:pivot-MBQC-form-non-input}).
  This changes the measurement plane for $u$ and $v$ from \YZ to \XY
  and it does not affect the measurement planes for any other vertices:
  \ctikzfig{rm-adj-red}

\item
  Otherwise, if there is no such connected pair but there is some vertex $u$ that is measured in the \XZ-plane (which is again necessarily not an input) we let $D_{k+1}$ be the MBQC+LC diagram that results from applying a local complementation on $u$ (Lemma~\ref{lem:lc-MBQC-form-non-input}):
  \ctikzfig{rm-adj-red2}

  As can be seen from Lemma~\ref{lem:lc-MBQC-form-non-input},
  this process changes the measurement on $u$ from \XZ to \YZ
  and it does not affect the label of any vertices that are measured in the \XY-plane.
\item
  If there is no such connected pair nor any vertex that is measured in the \XZ-plane
  then $D_k$ is already in the desired form, so we are finished.
\end{itemize}

The number of vertices not measured in the \XY-plane decreases with each step,
and no vertices are added, so this process terminates.
Since a pivot is just a sequence of local complementations,
$D_{k+1}$ has gflow if $D_k$ had gflow
(Corollary~\ref{cor:pivot_gflow}).
Finally every step preserves equivalence, so $D_{k+1}$ is equivalent to $D_k$.
\end{proof}

Now let us proceed with two new rewrite rules that allow us to further remove some measured qubits.

\begin{lemma}\label{lem:removeidvertex}
    Let $D$ be an MBQC+LC diagram with an internal vertex $u$ measured in the \YZ plane, and suppose it has a unique neighbour $v$ measured in the XY plane. Then there is an equivalent MBQC+LC diagram $D'$ with $G(D') = G(D)\backslash \{u\}$. If $D$ had gflow, then $D'$ also has gflow.
\end{lemma}
\begin{proof}
    We do the following rewrite:
    \ctikzfig{id-simp-1}
    The resulting diagram is then again an MBQC+LC diagram. The change to the \LOG\ comes down to deleting a YZ vertex. By Lemma~\ref{lem:deletepreservegflow} this preserves gflow.
\end{proof}

Since vertices measured in the \YZ axis are basically phase gadgets, we can fuse them together when they have the same set of neighbours (cf.~Eq.~\eqref{eq:fusing-gadgets}).
\begin{lemma}\label{lem:removepairedgadgets}
    Let $D$ be an MBQC+LC diagram with internal vertices $u$ and $v$ both measured in the YZ plane and with $N(u) = N(v)$. Then there is an equivalent diagram $D'$ with $G(D') = G(D)\backslash\{u\}$. If $D$ had gflow, then $D'$ also has gflow.
\end{lemma}
\begin{proof}
    We do the following rewrite:
    \ctikzfig{gadget-simp}
    This was shown to be sound in Eq.~\eqref{eq:fusing-gadgets}.
    The new diagram is still an MBQC+LC diagram, and the \LOG\ has only changed by deleting a YZ vertex. Hence, by Lemma~\ref{lem:deletepreservegflow} this preserves gflow.
\end{proof}

While the simplifications outlined in Section~\ref{sec:removing-clifford-vertices} remove Clifford vertices, the rewrites of this section succeed in removing vertices that have a non-Clifford measurement angle. Furthermore, since the measurement angles are added together, this might result in additional Clifford vertices that can then be removed in turn.
This iterative simplification process will prove crucial in the next chapter where we will use it for optimising the T-count of quantum circuits (cf.~Section~\ref{sec:clifford-T-optimisation}).

\section{Circuit extraction}\label{sec:circuit-extraction}

We saw in Section~\ref{sec:circuits-to-patterns} how to convert a circuit into a measurement pattern.
In this section we will do the converse and find a way to convert a measurement pattern with gflow back into a circuit. We call this problem of converting a measurement pattern, or more generally any ZX-diagram, into a circuit the \Define{circuit extraction problem}\indexd{circuit extraction}.
In Refs.~\cite{broadbent2009parallelizing,da2013compact} it was shown how to convert a measurement pattern into a circuit involving a number of ancillae proportional to the number of measured qubits in the pattern, while also requiring measurements and classical control in the resulting circuit. In Ref.~\cite{miyazaki2015analysis} an algorithm was found that results in ancilla-free circuits, but which also only worked with measurements in a single plane.
Our algorithm can deal with measurements in all three planes and results in an ancilla-free circuit.

Before we describe our algorithm we will require some additional results regarding gflow.

\subsection{Maximally delayed and focused gflow}

First, to perform our circuit extraction we will require a more specific type of gflow. In particular, we require a gflow that is \emph{maximally delayed} and \emph{focused}.
These ideas were originally introduced in Ref.~\cite{MP08-icalp,mhalla2011graph} for patterns with measurements in a single plane. We extend them here to work with measurements in three planes.

Intuitively, a gflow is maximally delayed when the corrections for each vertex are applied as late as possible. In order to figure out how delayed a gflow is, we need to `stratify' the vertices into layers of vertices that can be corrected at the same time.

\begin{definition}[{Generalisation of~\cite[Definition~4]{MP08-icalp} to multiple measurement planes}]
\label{defVk}
 For a labelled open graph $(G,I,O,\ld)$ and a gflow $(g,\prec)$ of $(G,I,O,\ld)$, let
 \[
  V_k^\prec = \begin{cases} \max_\prec (V) &\text{if } k= 0 \\ \max_\prec (V\setminus(\bigcup_{i<k} V_i^\prec)) &\text{if } k > 0 \end{cases}
 \]
 be layers of \emph{anti-chains} (i.e. incomparable elements in the partial order $\prec$) where $\max_\prec(X) := \{u\in X~;~\forall v\in X, \neg(u\prec v)\}$ is the set of the \Define{maximal elements} of $X$.\indexd{maximal element (poset)}
\end{definition}

Note that $V=\bigcup_k V_k^\prec$ and that there is some $N\in \N$ such that $V_n^\prec=\emptyset$ for all $n\geq N$.

\begin{definition}[{Generalisation of~\cite[Definition~5]{MP08-icalp} to multiple measurement planes}]
\label{defMoreDelayed}
 For a labelled open graph $(G,I,O,\ld)$ and two gflows $(g,\prec)$ and $(g',\prec')$ of $(G,I,O,\ld)$, we say
 $(g,\prec)$ is \Define{more delayed} than $(g',\prec')$ if for all $k$,
 \[
  \abs{\bigcup_{i=0}^k V_i^\prec} \geq \abs{\bigcup_{i=0}^k V_i^{\prec'}}
 \]
 and there exists a $k$ such that the inequality is strict.
 A gflow $(g,\prec)$ is \Define{maximally delayed} if there exists no gflow of the same \LOG that is more delayed.\indexd{gflow!maximally delayed ---}
\end{definition}

\begin{theorem}[{cf.~\cite{MP08-icalp} and \cite[Appendix C]{wetering-gflow}}]\label{thm:gflow-algorithm}
\label{thmGFlowAlgo}
There exists an efficient algorithm that decides whether a given
\LOG has a gflow, and that outputs a gflow if one exists.
Moreover the gflow this algorithm finds is maximally delayed.
\end{theorem}

\begin{corollary}\label{cor:gflow-max-delayed}
	A \LOG has a gflow if and only if it has a maximally delayed gflow.
\end{corollary}

\begin{remark}\label{rem:maxdelayedgflow}
    It's not too hard to see that if $(g,\prec)$ is a maximally delayed gflow for $(G,I,O,\ld)$, that necessarily $V_0^\prec = 0$. As a consequence we have for any $v\in V_1^\prec$ that $g(v)\sse O\cup \{v\}$ and $\odd{}{g(v)}\sse O\cup \{v\}$.
\end{remark}

The second property we need for our gflows is that of being \Define{focused}\indexd{gflow!focused ---}. This notion was originally introduced for patterns where all measurements are in the \XY-plane~\cite{mhalla2011graph}, and intuitively says that corrections only affect the vertex they are meant to correct and no others. This notion can be extended to patterns with measurements in all planes, but is particularly nice for patterns in phase-gadget form. In that setting all vertices that are in a correction set $g(v)$ of a focused gflow must be of \XY type, while all vertices in $\odd{}{g(v)}$, \ie~the vertices being corrected, must be either $v$ itself or of type \YZ.

\begin{lemma}\label{lem:successor-gflow}
 Let $(G,I,O,\ld)$ be a \LOG which has gflow $(g,\prec)$.
 Suppose there exist $v,w\in\comp{O}$ such that $v\prec w$.
 Define $g'(v):=g(v)\symd g(w)$ and $g'(u):=g(u)$ for all $u\in\comp{O}\setminus\{v\}$, then $(g',\prec)$ is a gflow.
\end{lemma}
\begin{proof}
 As the correction set only changes for $v$, the gflow properties remain satisfied for all other vertices.
 Now, suppose $w'\in g'(v)$, then $w'\in g(v)$ or  $w'\in g(w)$.
 In the former case, $v\prec w'$, and in the latter case, $v\prec w\prec w'$, since $(g,\prec)$ is a gflow. So \ref{it:g} holds.
 Similarly, suppose $w'\in\odd{}{g'(v)}$, then by linearity of $\odd{}{\cdot}$ we have $w'\in\odd{}{g(v)}$ or $w'\in\odd{}{g(w)}$.
 Again, this implies $v\prec w'$ or $v\prec w\prec w'$ since $(g,\prec)$ is a gflow. So \ref{it:odd} holds.
 Finally, $v\prec w$ implies $v\notin g(w)$ and $v\notin\odd{}{g(w)}$.
 Therefore $v\in g'(v) \Longleftrightarrow v\in g(v)$ and $v\in\odd{}{g'(v)}\Longleftrightarrow v\in\odd{}{g(v)}$.
 Thus \ref{it:XY}--\ref{it:YZ} hold and $(g',\prec)$ is a gflow.
\end{proof}

\begin{lemma}\label{lem:focus-single-vertex}
 Let $(G,I,O,\ld)$ be a \LOG with a gflow $(g,\prec)$. Let $v\in\comp{O}$.
 Then there exists $g':\comp{O}\to\pow{\comp{I}}$ such that
 \begin{enumerate}
  \item $(g',\prec)$ is a gflow and $g'(w)=g(w)$ for all $w\in\comp{O}$ with $w\neq v$,
  \item for all $w\in g'(v)\cap\comp{O}$, either $v=w$ or $\ld(w) = \XYm$,
  \item for all $w\in \odd{}{g'(v)}\cap\comp{O}$, either $v=w$ or $\ld(w)\neq \XYm$.
 \end{enumerate}
\end{lemma}
\begin{proof}
 Let $g_0:=g$, we will modify the function in successive steps to $g_1,g_2$, and so on.
 For each non-negative integer $k$ we define
 \begin{align*}
  S_{k,\XYm} &:= \{u\in (\odd{}{g_k(v)}\cap\comp{O}) \setminus\{v\} : \ld(u)=\XYm\}, \\
  S_{k,\XZm} &:= \{u\in (g_k(v)\cap\comp{O}) \setminus\{v\} : \ld(u)=\XZm\}, \\
  S_{k,\YZm} &:= \{u\in (g_k(v)\cap\comp{O}) \setminus\{v\} : \ld(u)=\YZm\},
 \end{align*}
 and set $S_k := S_{k,\XYm} \cup S_{k,\XZm} \cup S_{k,\YZm}$.
 If $S_k=\emptyset$, let $g':=g_k$ and stop.
 Otherwise, choose $w_k\in S_k$ among the elements minimal in $\prec$, and define
 \[
  g_{k+1}(u) := \begin{cases} g_k(v)\symd g_k(w_k) &\text{if } u=v \\ g_k(u) &\text{otherwise.} \end{cases}
 \]
 Note $w_k\in S_k$ implies $w_k\neq v$, as well as either $w_k\in g_k(v)$ or $w_k\in\odd{}{g_k(v)}$.
 Thus if $(g_k,\prec)$ is a gflow, then $v\prec w_k$, and hence by Lemma~\ref{lem:successor-gflow}, $(g_{k+1},\prec)$ is also a gflow.
 Since $(g_0,\prec)$ is a gflow, this means $(g_k,\prec)$ is a gflow for all $k$.

 Now, if $w_k\in S_{k,\XYm}$, then $w_k\in\odd{}{g_k(w_k)}$ by \ref{it:XY}.
 This implies $w_k\notin\odd{}{g_{k+1}(v)}$, and thus $w_k\notin S_{k+1}$.
 Similarly, if $w_k\in S_{k,\XZm} \cup S_{k,\YZm}$, then $w_k\in g_k(w_k)$ by \ref{it:XZ} or \ref{it:YZ}.
 This implies $w_k\notin g_{k+1}(v)$, and thus $w_k\notin S_{k+1}$.
 Hence, in each step $S_{k+1}$ has at least one less minimal element than $S_k$.

 Suppose there exists $w'\in S_{k+1}\setminus S_k$, then either $w'\in g_k(w_k)$ or $w'\in\odd{}{g_k(w_k)}$ and thus in either case $w_k\prec w'$.
 In other words, we always remove a minimal element from the set and add only elements that come strictly later in the partial order.
 Therefore, the process terminates after $n\leq\abs{V}$ steps, at which point $S_n=\emptyset$.
 Then the function $g'=g_n$ has the desired properties: (1) holds because we never modify the value of the function on inputs other than $v$ and every step results in a gflow, and (2) and (3) hold because $S_n=\emptyset$.
\end{proof}

Based on these lemmas, we can now show the focusing property.
These results state basically that correction sets can be taken to only contain measurements in the \XY plane, while the effects of these corrections, apart from the qubit where they should have an effect, are only felt on qubits measured in a different plane.

\begin{proposition}\label{prop:focused-gflow}
 Let $(G,I,O,\ld)$ be a \LOG which has a gflow.
 Then $(G,I,O,\ld)$ has a maximally delayed gflow $(g,\prec)$ with the following properties for all $v\in V$:
 \begin{itemize}
  \item for all $w\in g(v)\cap\comp{O}$, either $v=w$ or $\ld(w)= \XYm$, and
  \item for all $w\in \odd{}{g(v)}\cap\comp{O}$, either $v=w$ or $\ld(w)\neq \XYm$.
 \end{itemize}
 We call a gflow with these properties \Define{focused}.
\end{proposition}
\begin{proof}
 Let $(g_0,\prec)$ be a maximally delayed gflow of $(G,I,O,\ld)$, which exists by Corollary~\ref{cor:gflow-max-delayed}.
 Set $n:=\abs{V}$ and consider the vertices in some order $v_1,\ldots,v_n$.
 For each $k=1,\ldots,n$, let $g_k$ be the function that results from applying Lemma~\ref{lem:focus-single-vertex} to the gflow $(g_{k-1},\prec)$ and the vertex $v_k$.
 Then $g_k$ satisfies the two properties for the vertex $v_k$.
 The function $g_k$ also equals $g_{k-1}$ on all inputs other than $v_k$, so in fact $g_k$ satisfies the two properties for all vertices $v_1,\ldots,v_k$.
 Thus, $g_n$ satisfies the two properties for all vertices.
 Moreover, the partial order does not change, so $(g_n,\prec)$ is as delayed as $(g_0,\prec)$; i.e.\ it is maximally delayed.
 Hence if $g:=g_n$, then $(g,\prec)$ has all the desired properties.
\end{proof}

Finally, using this combination of maximally delayed and focused gflow, we can show that we can always find a maximal vertex that is connected to an output, a fact that is crucial for the extraction algorithm.

\begin{lemma}\label{lem:maxdelayednotempty}
 Let $(G,I,O,\ld)$ be a labelled open graph in phase-gadget form, which furthermore satisfies $\comp{O}\neq\emptyset$.
 Suppose $(G,I,O,\ld)$ has a gflow.
 Then there exists a maximally delayed gflow $(g,\prec)$ such that $N_G(V_1^\prec)\cap O \neq \emptyset$. In other words, there exists a maximal vertex connected to an output.
\end{lemma}
\begin{proof}
    By Proposition~\ref{prop:focused-gflow}, there exists a maximally delayed gflow of $(G,I,O,\ld)$ such that no element of a correction set (other than possibly the vertex being corrected) is measured in the $\YZ$ plane.
    Let $(g,\prec)$ be this gflow.

    Since by assumption the open graph does not consist solely of outputs, we have $V_1^\prec\neq \emptyset$ (c.f.~Remark~\ref{rem:maxdelayedgflow}). Hence, the following arguments are non-trivial.
    Again by Remark~\ref{rem:maxdelayedgflow}, we have $g(v)\subseteq O\cup \{v\}$ for any $v\in V_1^\prec$.
    Now if there is a $v\in V_1^\prec$ with $\ld(v) =  \XYm$, then $v\in\odd{}{g(v)}$.
    Hence this $v$ must be connected to at least one output, and we are done.
    As the graph is in phase-gadget form, there are no vertices labelled \XZ and hence we may now assume that $\ld(v)=\YZm$ for all $v\in V_1^\prec$.

    Suppose first that $V_2^\prec = \emptyset$, so that the only non-output vertices are in $V_1^\prec$. The vertices in $V_1^\prec$ are all labelled \YZ and thus appear in their own correction sets; this means they cannot be inputs because inputs do not appear in correction sets. The vertices in $V_1^\prec$ are not outputs either, so each of them must have at least one neighbour (since otherwise it would just be a scalar).
    Being in phase-gadget form implies that two vertices labelled \YZ cannot be adjacent, and all vertices in $V_1^\prec$ are labelled \YZ.
    Thus by process of elimination, any vertex $v\in V_1^\prec$ must have a neighbour in $O$, and we are done.

    So now suppose there is some vertex $w\in V_2^\prec$. Then, regardless of $\ld(w)$, we have $g(w) \sse V_1^\prec\cup O\cup\{w\}$ and $\odd{}{g(w)} \sse V_1^\prec\cup O\cup\{w\}$.
    By the focusing property, for any $v\neq w$ with $\ld(v)=\YZm$ we have $v\notin g(w)$. Since all vertices in $V_1^\prec$ have label \YZ we must therefore have $g(w)\cap V_1^\prec = \emptyset$ and hence $g(w)\sse O\cup \{w\}$.
    
    We claim that we must now necessarily have $\odd{}{g(w)}\cap V_1^\prec \neq \emptyset$, \ie that there is a $v\in V_1^\prec$ that is in $\odd{}{g(w)}$. Suppose this is not the case. Then we define a new partial order $\prec' \ :=\  \prec \setminus \{(w,u): u\in V_1^\prec\}$ which results in a gflow $(g,\prec')$: dropping the given inequalities from the partial order does not affect the gflow properties since $u\in V_1^\prec$ implies $w\notin g(u)$ and $w\notin \odd{}{g(u)}$.
    But furthermore we see that $(g,\prec')$ is more delayed than $(g,\prec)$, because $w$ (and potentially some of its predecessors) move to an earlier layer, contradicting the assumption that $(g,\prec)$ is maximally delayed.
    Hence, there is indeed a $v\in V_1^\prec$ such that $v\in \odd{}{g(w)}$.
    Now if $\ld(w)=\XYm$, we have $w\not\in g(w)$ and hence $g(w)\sse O$ so that there must be some $o\in O$ that is connected to $v$ and we are done.
    Otherwise, if $\ld(w)=\YZm$, then $w\in g(w)$, but since both $v$ and $w$ are measured in the \YZ plane they are not neighbours, and hence there still must be an $o\in O$ that is connected to $v$.
    Thus, the gflow $(g,\prec)$ has the desired property.
\end{proof}

\subsection{Extracting a circuit}\label{sec:extracting-circuit-details}
\indexd{circuit extraction!overview}

In this section we will find a way to extract a (unitary) circuit from a measurement pattern. This involves some new results that we will prove along the way. In Section~\ref{sec:extraction-algorithm} we will condense the procedure into a short and efficient algorithm.
As we wish to extract a circuit we will assume the measurement pattern has the same number of inputs and outputs. Our algorithm will crucially rely on the existence of a gflow, and hence we will assume the pattern has a gflow. For convenience we represent the pattern by a MBQC+LC diagram, and describe the algorithm in terms of ZX-diagrams. 

The algorithm will consist of making sequential changes to the \zxdiagram so that the diagram looks progressively more like a circuit. During the process, there will be a `frontier': a set of Z-spiders such that everything to their right looks like a circuit, while everything to their left (and including the frontier vertices themselves) is an MBQC-form \zxdiagram equipped with a gflow.
We will refer to the MBQC-form diagram on the left as the \Define{unextracted} part of the diagram, and to the circuit on the right as the \Define{extracted} part of the diagram.
For example:
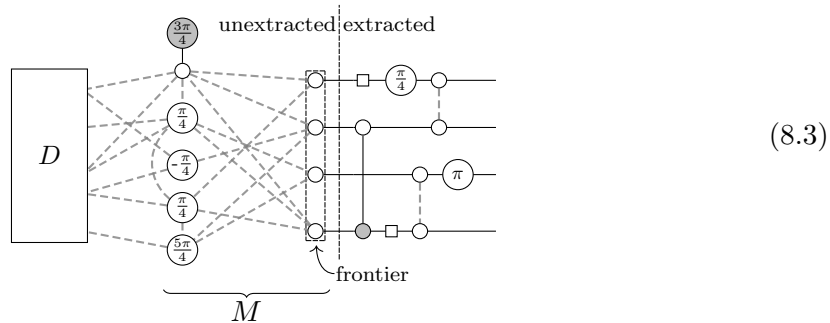
\begin{equation}\label{ex:frontier-example}
\scalebox{1.0}{\input{frontier-example.tikz}}
\end{equation}
In this diagram, we have merged the \XY measurement effects with their adjacent vertices, in order to present a tidier picture.
The matrix $M$ is the biadjacency matrix between the vertices on the frontier and all their neighbours to the left of the frontier.
For the purposes of the algorithm below, we consider the extracted circuit as no longer being part of the diagram, and hence when we refer to `outputs' below, we mean the vertices on the frontier.

The general idea now is to start at the end of the diagram, find a suitable vertex connected to a frontier vertex, and transform the diagram in such a way that we can `consume' this vertex. We then keep repeating this procedure until all vertices are consumed.

\textbf{Step 0}: First, we transform the pattern into phase-gadget form using Proposition~\ref{prop:ZXtopseudonormalform}, ensuring that all vertices are measured in the \XY or \YZ planes, and that vertices measured in the \YZ plane are only connected to vertices measured in the \XY plane. This can be done efficiently, and preserves the interpretation of the diagram. Furthermore, the resulting diagram still has a gflow.

\textbf{Step 1}: 
We unfuse any connection between output vertices as a CZ gate into the extracted circuit, and we consider any local Clifford operator on the output vertices as part of the extracted circuit. For example:
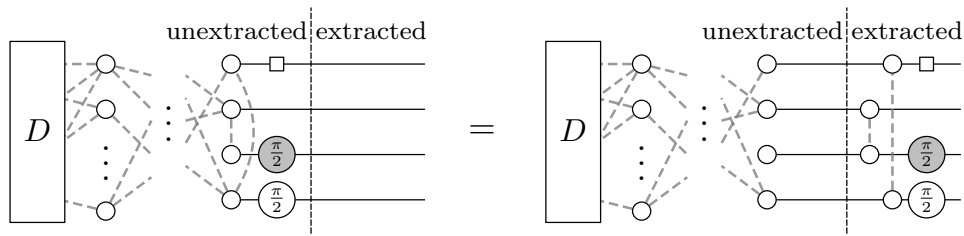
\[\scalebox{1.2}{\input{example-unfuse-gates.tikz}}\]
This process changes the unextracted diagram in two ways. The first simply removes local Clifford operators, which does not affect the underlying \LOG. The second removes connections between the frontier vertices. While this changes the \LOG, it preserves the existence of gflow:
\begin{lemma}\label{lem:remove-output-edges-preserves-gflow}
	Let $(G,I,O,\lambda)$ be a \LOG{} with gflow. Then $(G',I,O,\lambda)$ where $G'$ is equal to $G$ but with all edges between the vertices in $O$ removed also has gflow.
\end{lemma}
\begin{proof}
	We claim that if $(g,\prec)$ is a gflow for $G$, then it is also a gflow for $G'$. Note that $\odd{G'}{g(v)}\cap \comp{O} = \odd{G}{g(v)}\cap \comp{O}$ as the only changes to neighbourhoods are between the output vertices. It is then easily checked that all properties of Definition~\ref{defGFlow} are still satisfied.
\end{proof}
Thus, the resulting unextracted diagram continues to be in MBQC+LC form and it continues to have a gflow.
If the only unextracted vertices are on the frontier, go to step~5, otherwise continue to step~2.

\textbf{Step 2}:
The unextracted diagram is in phase-gadget form and has a gflow.
Thus, by Lemma~\ref{lem:maxdelayednotempty}, it has a maximally delayed gflow $(g,\prec)$ such that $N_G(V_1^\prec)\cap O \neq \emptyset$, where $V_1^\prec$ is the `most delayed' layer before the outputs (see Definition~\ref{defVk}).
Such a gflow can be determined efficiently by first finding any maximally delayed gflow using the algorithm of Theorem~\ref{thmGFlowAlgo} and then following the procedure outlined in the proof of Lemma~\ref{lem:maxdelayednotempty}.

Now, if any of the vertices in $V_1^\prec$ are labelled \XY, pick one of these vertices and go to step~3. Otherwise, all the maximal non-output vertices (with respect to $\prec$) must have label \YZ; go to step~4.

\textbf{Step 3}: This step is where the non-trivial part of the extraction 
happens. 
We have a maximal non-output vertex $v$ labelled \XY, which we want to extract. Since it is maximal in $\prec$, we know that $g(v)\sse O$ by Remark~\ref{rem:maxdelayedgflow}.
As the gflow is maximally delayed, we have $\odd{}{g(v)}\cap \comp O = \{v\}$. Let us consider an example. In the following diagram, we have indicated the vertex $v$ and its correction set $g(v)$:
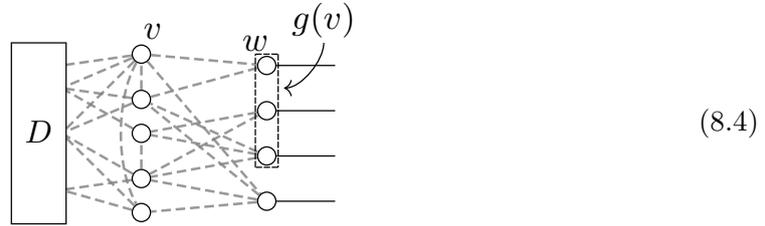
\begin{equation}\label{eq:example-extracted-vertex}
\scalebox{1.2}{\input{example-extracted-vertex.tikz}}
\end{equation}
Note that we are ignoring the measurement effects on the left hand-side spiders for clarity.
In the above example, the biadjacency matrix of the bipartite graph between the vertices of $g(v)$ on the one hand, and their neighbours in the unextracted part on the other hand, is
\begin{equation}\label{eq:biadjacency-example}
	\begin{pmatrix}
		1&1&0&0\\
		0&0&1&1\\
		0&1&1&1
	\end{pmatrix}
\end{equation}
where the rows correspond to vertices of $g(v)$, and vertices are ordered top-to-bottom. We do not include the bottom-most output in the biadjacency matrix, as it is not part of $g(v)$, and we do not include the bottom left spider, as it is not connected to any vertex in $g(v)$.

The property that $\odd{}{g(v)}\cap \comp O = \{v\}$ now corresponds precisely to the following fact: if we sum up all the rows of this biadjacency matrix modulo 2, the resulting row vector contains a single 1 corresponding to the vertex $v$ and it has zeros everywhere else.
It is straightforward to see that this is indeed the case for the matrix of \eqref{eq:biadjacency-example}.

This fact will allow us to extract the vertex $v$. However, in order to use these row operations, we need to see how these can be applied to a ZX-diagram.

\begin{lemma}\label{lem:cnotgflow}
  The following equation holds.
  \begin{equation}
  \input{cnot-pivot.tikz} 
  \end{equation}
  Here $M$ describes the biadjacency matrix of the vertices on the right to the vertices on the left, and $M^\prime$ is the matrix produced by adding row 1 to row 2 (modulo 2) in $M$. Furthermore, if the diagram on the left-hand side has a gflow, then so does the one on the right-hand side.
\end{lemma}
\begin{proof}
  We will show how to transform the first diagram into the second in such a way that gflow and equality is preserved at every step. For clarity we will not draw the entire diagram, but instead focus on the relevant part. First of all, we note that we can add CNOTs in the following way while preserving equality:
  \ctikzfig{cnot-pivot3}
  As we are only adding vertices at the outputs, it should be clear how the gflow can be extended to incorporate these new vertices (cf.~Lemma~\ref{lem:gflow-add-output}).

  Now let $A$ denote the set of vertices connected to the top vertex, but not to the vertex beneath it, $B$ the set of vertices connected to both, and $C$ the vertices connected only to the bottom one. Further restricting our view of the diagram to just the neighbourhood of these two spiders, we see that we can apply a pivot-and-delete rewrite as in Eq.~\eqref{eq:pivot-simp}:
  \ctikzfig{cnot-pivot4}
  Looking at the connectivity, it is straightforward to see that the matrix $M$ has now been changed in exactly the way described.
  The underlying \LOG still has a gflow because it has changed according to Corollary~\ref{cor:pivot_gflow}, followed by two deletions of vertices as in Lemma~\ref{lem:deletepreservegflow}. 
\end{proof}

Pick any output $w\in g(v)$. 
Lemma~\ref{lem:cnotgflow} shows that the application of a CNOT to two outputs corresponds to a row operation on the biadjacency matrix, which adds the row corresponding to the target to the row corresponding to the control. Hence if, for each $w'\in g(v)\setminus\{w\}$, we apply a CNOT with control $w$ and target $w'$, the effect is to add all the rows in $g(v)$ to that of $w$:
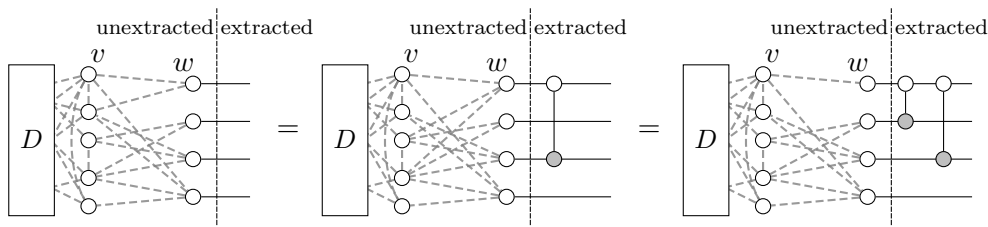
\[\scalebox{1}{\input{example-extracted-vertex-cnots.tikz}}\]
As a result, $w$ is now only connected to $v$, but $v$ may still be connected to other vertices in $O\setminus g(v)$ (which is indeed the case in our example). For each such vertex $u$, applying a CNOT with control $u$ and target $w$ removes the connection between $u$ and~$v$:
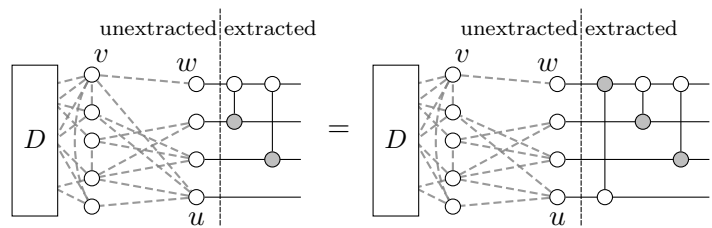
\[\scalebox{1}{\input{example-extracted-vertex-cnots2.tikz}}\]
Now we can extract $v$ by removing $w$ from the diagram, adding a Hadamard to the circuit (this comes from the Hadamard edge between $v$ and $w$), adding the measurement angle of $v$ to the circuit as a Z-phase gate, and adding $v$ to the set of outputs of the graph (i.e.\ the frontier):
\begin{equation}\label{eq:extract-vertex}
\scalebox{1}{\input{extract-vertex.tikz}}
\end{equation}
On the underlying \LOG this corresponds to removing $w$ and adding $v$ to the list of outputs. Since $w$ was only connected to $v$, and no other output vertex is connected to $v$, it must be that $w$ is not part of any correction set $g(v')$ for any other $v'$ in the graph (because if it were, then the odd neighbourhood necessarily contained $v$, which contradicts the gflow being focused). So the resulting \LOG still has a gflow.

As the vertex $w$ has been removed, the number of vertices in the unextracted part of the diagram is reduced by 1. We now go back to step 1.

\textbf{Step 4}: All the maximal vertices are labelled \YZ. Since we chose our gflow according to Lemma~\ref{lem:maxdelayednotempty}, we know that at least one of these vertices is connected to an output. Pick such a vertex $v$, and pick a $w\in O\cap N_G(v)$ (this set is non-empty). Pivot about $vw$ using Eq.~\eqref{eq:pivot-desc} and reduce the resulting diagram to MBQC form with Lemma~\ref{lem:pivot-MBQC-form-non-input}. Afterwards, $v$ has label \XY and $w$ has a new Hadamard gate on its output (which will be dealt with in the next step 1).

We have changed one vertex label in the unextracted part of the diagram from \YZ to \XY. Since no step introduces new \YZ vertices, step~4 can only happen as many times as there are \YZ vertices at the start of the algorithm. Go back to step 1.

\textbf{Step 5:} At this point, there are no unextracted vertices other than the frontier vertices, all of which have degree 2 and can be removed using rule $(\bm{i1})$.
Yet the outputs might be connected to the inputs in some permuted manner and the inputs might carry some local Cliffords:
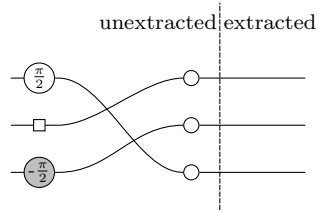
\ctikzfig{example-permutation}
This is easily taken care of by decomposing the permutation into a series of SWAP gates, at which point the entire diagram is in circuit form.

Since step 3 removes a vertex from the unextracted diagram, and step 4 changes a measurement plane from \YZ to \XY (and no step changes measurement planes in the other direction), this algorithm terminates. All steps correspond to ZX-diagram rewrites, so the resulting diagram is a circuit that implements the same linear map as the original diagram.

\subsection{An efficient circuit-extraction algorithm}\label{sec:extraction-algorithm}
\indexd{circuit extraction!algorithm}

Now that we have a procedure for extracting a circuit from a pattern, we can simplify some of the steps involved.

In step 2, instead of using the gflow to find a maximal vertex, we do the following: Write down the biadjacency matrix of the bipartite graph consisting of outputs on one side and all their neighbours on the other side. For example, diagram~\eqref{eq:example-extracted-vertex} would give the matrix:
\begin{equation}\label{eq:matrix2}
	\begin{pmatrix}
		1&1&0&0&0\\
		0&0&1&1&0\\
		0&1&1&1&0\\
		1&1&0&1&1
	\end{pmatrix}
\end{equation}
Now perform a full Gaussian elimination on this $\mathbb{Z}_2$ matrix. In the above case, this results in the matrix:
\begin{equation}\label{eq:matrix_after_elim}
	\begin{pmatrix}
		1&0&0&0&0\\
		0&1&0&0&0\\
		0&0&1&0&1\\
		0&0&0&1&1
	\end{pmatrix}
\end{equation}
Any row in this matrix containing a single 1 necessarily corresponds to a maximal vertex with label \XY. For instance, in the matrix in \eqref{eq:matrix_after_elim}, the first row has a single 1 in the first column, and hence the top-left spider of~\eqref{eq:example-extracted-vertex} is maximal. Similarly, the second row has a single 1, appearing in column 2, and hence the second spider from the top on the left in~\eqref{eq:example-extracted-vertex} is maximal.

If we found at least one maximal vertex labelled \XY with this method, we implement the row operations corresponding to the Gaussian elimination procedure as a set of CNOT gates using Lemma~\ref{lem:cnotgflow}. Doing this with the diagram~\eqref{eq:example-extracted-vertex} gives:
\begin{equation}
	\scalebox{1.0}{\input{example-extracted-gauss.tikz}}
\end{equation}

We see that every row which had a single 1 now corresponds to a frontier spider with a single neighbour, and hence we can extract vertices using the technique of \eqref{eq:extract-vertex}:
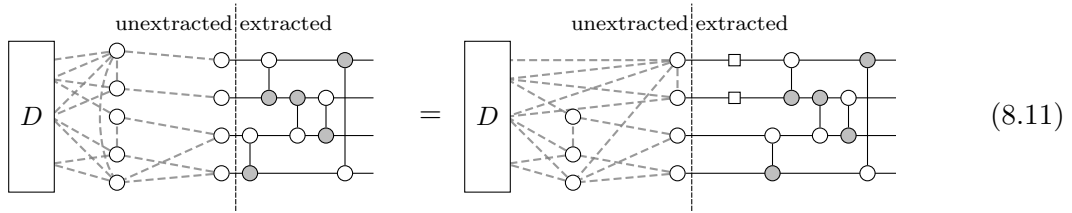
\begin{equation}\label{eq:example-extracted-3}
	\scalebox{1.0}{\input{example-extracted-3.tikz}}
\end{equation}

As we now extract multiple vertices at a time, there could be connections between the new frontier vertices (for instance between the top two frontier spiders in \eqref{eq:example-extracted-3}). These turn into CZ gates the next time step 1 is done.

If the Gaussian elimination does not reveal a row with a single 1, then we are in the situation of step 4. We perform pivots involving a vertex with label \YZ and an adjacent output vertex until there is no vertex with a label \YZ which is connected to an output. We then go back to step 1.

Interestingly, by using these shortcuts we can extract a circuit without having a gflow explicitly calculated. The fact that there \emph{is} a gflow is crucial for the correctness of the algorithm though: without it there is no guarantee that our Gaussian elimination approach will succeed, and in fact when the diagram does not have gflow it will in general not succeed.

\begin{algorithm}[!b]
\caption{Circuit Extraction}\label{alg:extraction}
\begin{algorithmic}[1]
\Procedure{Extract}{$D$}\Comment{input is MBQC+LC diagram $D$}
  \State Init empty circuit $C$
  \State $G,I,O\gets $ Graph$(D)$\Comment{get the underlying graph of $D$}

  \State $D,C \gets$ ProcessOutputs($D,O,C$)

  \While{$\exists v\in D\backslash O$}\Comment{there are still vertices to be processed}
    \State $D,O,C \gets $ ExtractVertices$(D,O,C)$ \Comment{See Algorithm~\ref{alg:extract-vertex}}
  \EndWhile
  \For{$v\in O$} \Comment{the only vertices still in $D$ are in $O$}
    \If{$v$ connected to input has Clifford}
      \State $C\gets$ Cliffords(Qubit($v$))
    \EndIf
  \EndFor
  \State Perm $\gets$ Permutation from inputs to outputs \Comment{step 5 of Section~\ref{sec:extracting-circuit-details}}
  \For{swap$(q_1,q_2)$ in PermutationAsSwaps(Perm)}
    \State $C\gets$ swap$(q_1,q_2)$
  \EndFor
  \State \textbf{return} $C$
\EndProcedure

\Procedure{ProcessOutputs}{$D,O,C$} \Comment{Corresponds to step 1 of Section~\ref{sec:extracting-circuit-details}}
    \For{$v\in O$}
        \If{$v$ has local Cliffords}
          \State $C\gets $ Cliffords(Qubit($v$))
          \State Remove Cliffords from $v$ on output wire
        \EndIf
      \EndFor
      \For{edge between $v$ and $w$ in $O$}
        \State $C\gets $ CZ(Qubit($v$), Qubit($w$))
        \State Remove edge between $v$ and $w$
  \EndFor
  \State \textbf{return} $D,C$
\EndProcedure
\end{algorithmic}
\end{algorithm}

\begin{algorithm}[!b]
\caption{Extracting a vertex in the circuit extraction algorithm}\label{alg:extract-vertex}
\begin{algorithmic}[1]
\Procedure{ExtractVertices}{$D,O,C$} \Comment{See Section~\ref{sec:extraction-algorithm}}
  \While{There is \YZ vertex connected to $O$}
      \State $v \gets$ \YZ vertex connected to $O$
      \State $w \gets$ a neighbour of $w$ in $O$
      \State $D\gets $ Pivot($D$,$v$,$w$)
  \EndWhile
  \If{any \YZ vertex found}
    \State $D,C \gets$ ProcessOutputs$(D,O,C)$ \Comment{See Algorithm~\ref{alg:extraction}}
    \State \textbf{return} $D,O,C$
  \EndIf
  \State $N\gets $ Neighbours$(O)$
  \State $M \gets$ Biadjacency$(O,N)$ 
  \State $M^\prime \gets $ GaussReduce$(M)$ 
  \State Init $vs$ \Comment{initialise empty set $vs$}
  \For{row $r$ in $M^\prime$}
    \If{sum$(r) == 1$}\Comment{there is a single 1 on row $r$}
      \State Set $v$ to vertex corresponding to nonzero column of $r$
      \State Add $v$ to $vs$ \Comment{$v$ will be part of the new frontier}
    \EndIf
  \EndFor
  \State $M \gets $  Biadjacency($O,ws$) \Comment{smaller biadjacency matrix}
  \For{$(r_1,r_2) \in $ GaussRowOperations$(M)$}
    \State $C\gets $ CNOT(Qubit$(r_1)$, Qubit$(r_2)$)
    \State Update $D$ based on row operation
  \EndFor
  \For{$v\in vs$} \Comment{all $v$ now have a unique neighbour in $O$}
    \State $w \gets $ Unique neighbour of $v$ in $O$
    \State $C\gets $ Hadamard(Qubit$(w)$)
    \State $C\gets $ Phase-gate$(Phase(v),Qubit($w$))$
    \State Remove $w$ from $D$ and $O$
    \State Add $v$ to $O$
  \EndFor
  \State $D,C \gets$ ProcessFrontier$(D,O,C)$
  \State \textbf{return} $D,O,C$
\EndProcedure
\end{algorithmic}
\end{algorithm}

For a pseudocode description of the extraction algorithm see Algorithms~\ref{alg:extraction} and~\ref{alg:extract-vertex}. The observations we have made allow us to state the following theorem:

\begin{theorem}\label{thm:extraction-algorithm}
    Let $\pat$ be a measurement pattern with gflow and $n$ inputs and outputs containing a total of $k$ qubits. Then there is an algorithm running in time $O(n^2k^2 + k^3)$ that converts $\pat$ into an equivalent $n$-qubit circuit that contains no ancillae. The number of non-Clifford gates in the circuit is equal to the number of qubits in $\pat$ measured in a non-Clifford angle.
 \end{theorem}
 \begin{proof}
The runtime for the extraction algorithm is dominated by Gaussian elimination of the biadjacency matrices which has complexity $O(n^2m)$, where $n$ is the number of rows, corresponding to the number of outputs, and $m$ is the number of columns, corresponding to the neighbours of the outputs. In principle $m$ could be as large as the number of vertices in the graph and hence could be as large as $k$.
Doing a pivot requires in the worst case to toggle the connectivity of almost the entire graph, which requires $k^2$ elementary graph operations.
Since for every vertex in the graph we might have to do a pivot and a Gaussian elimination the complexity for the entire algorithm is upper-bounded by $O(k(n^2k + k^2)) = O(n^2k^2 + k^3)$.
\end{proof}

Note that if $k\geq O(n^2)$, which will be the case for most useful computations, the bound in the theorem becomes $O(k^3)$. We expect this bound to be rather pessimistic as it would require the diagram to be almost fully connected in every step of the computation. As pivots toggle connectivity, this seems rather unlikely, and indeed empirical data seems to suggest that the extraction is much more sensitive to the number of outputs rather than the total number of vertices. We will discuss several ways in which this circuit extraction algorithm can be improved and generalised in Chapter~\ref{chap:future}. In the next chapter we will use the algorithm to optimise circuits.

\chapter{Optimisation of quantum circuits}\label{chap:optimisation}

In this chapter we will apply the results of Chapters~\ref{chap:zxcalculus} and~\ref{chap:MBQC} to two problems: quantum circuit optimisation, and verification of equality of quantum circuits. 

To understand the context of our results we start with an overview of existing approaches to circuit optimisation in Section~\ref{sec:optimisation-overview}. Then we will present our main ZX-calculus based simplification routine in Section~\ref{sec:circuit-simplification}. This is based on the rewrite rules of measurement patterns described in Theorem~\ref{thm:simplifiedZXdiagram} extended with the additional simplifying rewrite rules of Section~\ref{sec:further-optimisation}.

In order to apply our optimisation routine to circuits of a realistic size, we made a software library named \emph{PyZX}. The architecture and implementation of PyZX is discussed in Section~\ref{sec:pyzx}.

In Section~\ref{sec:clifford-circuit-optimisation} we will apply our simplification method to Clifford circuits and see that it reduces them to a pseudo-normal form that has several desirable features: it has an optimal number of free parameters, and in the linear nearest neighbour connectivity model its 2-qubit gate depth is less then any other existing normal form for Clifford circuits.

Then in Section~\ref{sec:clifford-T-optimisation} we apply our method to Clifford+T circuits with the goal of minimising the number of T gates present in the circuit. We will see that our method outperforms or matches all of the other existing ancilla-free T-count optimisers.

In order to get a circuit-to-circuit T-count optimiser, we need to use the circuit extraction algorithm of the previous chapter. This has the drawback that it is relatively slow and that for many circuits it actually increases the number of 2-qubit gates in the circuit. In order to get around these issues we introduce the method of \emph{phase teleportation} in Section~\ref{sec:phase-teleportation}, which allows us to bypass the need for circuit extraction. With phase teleportation we can reduce the T-count without changing any other feature of the circuit.

For any software implementation it is of course crucial to know that it has been implemented correctly. We discuss in Section~\ref{sec:circuit-verification} how our optimisation procedure is essentially self-checking, and can produce a certificate of equality for any of our optimised circuits.

We end with some concluding remarks in Section~\ref{sec:optimisation-conclusion}

\section{Introduction to quantum circuit optimisation}\label{sec:optimisation-overview}

The goal of quantum circuit optimisation is to take a given circuit $C$ that implements some unitary $U$ and transform it into a circuit $C'$ that still implements $U$, but is better than $C$ in some chosen metric. Before we get into the work that has been done on this topic, we will outline the most important metrics that are being considered.

An obvious one is \Define{total gate count}\indexd{gate count}. A circuit which contains more gates is of course going to be harder to implement, and hence it makes sense to want to minimise the number of gates in a circuit.
A related metric is that of \Define{2-qubit gate count}\indexd{2-qubit gate count} where the goal is to minimise the number of gates that act on 2 (or more) qubits, such as CNOT and CZ gates. This is an important metric because in many types of quantum computers two-qubit gates take more time or introduce more noise into the system than single qubit gates.

In most types of quantum computers operations acting on disjoint sets of qubits can be performed in parallel, and hence the total time required to implement a circuit is not related to the number of gates needed, but rather the number of parallel layers of gates that are needed. This number is referred to as the \Define{depth}\indexd{depth (of a circuit)} of the circuit. This gives the metric of gate depth, but since 2-qubit gates generally take more time to implement than single qubit gates, another metric in use is \Define{2-qubit gate depth} that only measures the number of 2-qubit gate layers needed.

The above-mentioned metrics (in combination with \emph{routing} discussed below) are generally the most important when it comes to near-term quantum computation. The situation is different when it comes to computation at the logical level of a fault-tolerant quantum computer. In this setting one can only easily perform gates that interact well with the error-correcting code used. This usually includes Clifford gates, but excludes non-Clifford ones. A gate like the T gate must then be implemented using \emph{magic state injection}\indexd{magic state!--- injection} where a magic state is prepared and combined with the circuit in a particular manner to implement a T gate (similar to how this is done in Eq.~\eqref{eq:zx-magic-injection}). The problem is that such magic states will be prepared at the noisy physical level and must be \emph{distilled}\indexd{magic state!--- distillation} in order to be used in a fault-tolerant manner. This is generally very costly. For instance, in Ref.~\cite[Appendix M]{fowler2012surface} they work through an example of Shor's algorithm run on a surface code\indexd{surface code} (a leading candidate for fault-tolerant quantum computation) where about 95\% of the total number of qubits, and almost all the execution time is required for magic state distillation\footnote{Recent work in optimising the design and layout of magic state distillation factories has greatly improved the resource cost of implementing T gates in the surface code~\cite{Gidney2019efficientmagicstate,Litinski2019gameofsurfacecodes}. Nevertheless, they still account for most of the cost of implementing a circuit on the surface code.}. In the surface code, a T gate has been estimated to be about 50 times~\cite{fowler2012bridge} to 300 times~\cite{gorman2017quantum} more costly to implement than a CNOT gate in the surface code. A relevant metric for fault-tolerant computation is therefore the \Define{T-count} of a circuit: the number of T gates required to implement the circuit.\indexd{T-count}

A final consideration is that in most types of quantum computers it is not possible for every pair of qubits to interact with each other, since they might for instance be physically far apart. When optimising circuits we might therefore also want to make sure the resulting circuit only contains 2-qubit gates between qubits that can actually interact with one another. We will refer to this problem as \Define{routing} the circuit\indexd{circuit!--- routing}. A related problem is \Define{qubit mapping}\indexd{qubit mapping} where the qubit lines of a circuit must be mapped to the (physical) qubits of the quantum computer.

Before we continue with an overview of the field of quantum circuit optimisation let us remark that the general problem of optimisation is unlikely to have an efficient solution.
The complexity class \textbf{QMA}\index{math}{QMA@\textbf{QMA} (complexity class)} is the quantum analogue to \textbf{NP}, and hence contains problems that are believed to be hard. A complete problem for \textbf{QMA} is determining whether a circuit implements the identity~\cite{bookatz2012qmacomplete}. As a consequence, the general problem of determining what the optimal number of gates (or 2-qubit gates or T gates) is to implement a given unitary is probably intractable: if we could determine the optimal gate count of a circuit, then we can easily test whether a circuit implements the identity by checking whether its optimal gate-count is equal to zero. This explains why all known results that find optimal implementations of circuits take exponential time in the number of qubits.

There are basically four branches of quantum circuit optimisation: approximate synthesis, exact synthesis, heuristic optimisation and circuit routing. We will now give a short overview of previous results and methods used in the first three, with a focus on heuristic optimisation. Circuit routing will not be relevant for this thesis (although we discuss it briefly in Section~\ref{sec:circuit-routing}). For the purpose of this section we will define `efficient' to mean `taking time polynomial in the relevant parameters'. 

\Define{Synthesis}\indexd{circuit!--- synthesis} constructs a circuit directly from a description of the unitary. The input is thus not a circuit, but instead, for instance, a matrix.\indexd{Solovay-Kitaev algorithm} The seminal result in approximate synthesis is the \Define{Solovay-Kitaev algorithm}~\cite{dawson2005solovay}\indexd{Solovay-Kitaev algorithm} that gives a way to arbitrarily closely approximate a unitary on any number of qubits with any approximately universal gate set. The runtime for this algorithm is not surprisingly exponential in the number of qubits, and furthermore is not optimal in the number of gates used.
Other synthesis results focus on a restricted number of qubits. Refs.~\cite{bocharov2012resource,ross2014optimal} showed how to efficiently synthesise a single qubit unitary with an optimal number of T gates, while Ref.~\cite{glaudell2020optimal} does the same for 2-qubit unitaries with an optimal number of controlled-S gates. 

In contrast to approximate synthesis, exact synthesis focuses on unitaries that can be exactly represented by a circuit in some chosen gate set.
For instance, in Ref.~\cite{vidal2004universal} they find that any 2-qubit unitary can be implemented with at most 3 CNOT gates when arbitrary single qubit unitaries are allowed.
In Ref.~\cite{meet-in-the-middle2013} they implement a smart brute-force approach to finding optimal depth Clifford+T circuits of small size. In particular they verified that the 3-qubit \Define{Toffoli gate}\indexd{gate!Toffoli ---}, \ie the controlled-CNOT gate, has a optimal T-count of 7.\footnote{This only holds when restricted to unitary circuits. When ancillae and classical control are allowed the T-count can be reduced to 4~\cite{jones2013low}, or even 2 for certain pairs of Toffoli gates~\cite{Gidney2018halvingcostof}.} Recall that in Section~\ref{sec:phasegadgets} we gave a decomposition of the CCZ gate into phase gadgets that also had a T-count of 7. The Toffoli gate is simply the CCZ gate with the third qubit surrounded by Hadamard gates, so our decomposition was indeed optimal.
The brute-force approach was improved in Ref.~\cite{di2016parallelizing} by exploiting parallelisation which allowed them to find optimal T-counts for 4 qubit circuits.
A brute force approach was also used to find optimal implementations (in terms of total gate count, CNOT count and depth) of Clifford circuits up to 5 qubits~\cite{PhysRevA.88.052307}.

For circuits and unitaries on a larger number of qubits it quickly becomes impractical to synthesise optimal implementations. This has lead to a burgeoning field of \Define{heuristic optimisation}\indexd{circuit optimisation!heuristic ---} algorithms. These take in an existing quantum circuit, apply a set of rules and transformations and output a new circuit that implements the same unitary, but (hopefully) performs better on the chosen metric. For these approaches there is usually no guarantee that the resulting circuit is optimal in any sense.

Perhaps the most straightforward set of heuristics involve what we will refer to as \Define{peephole optimisation}\indexd{circuit optimisation!peephole ---}, but which is sometimes also referred to as \emph{template matching}. Peephole optimisation works by replacing small `chunks' of a circuit by more optimal counterparts. The process is repeated until no more chunks can be optimised. The power of this method derives from the variety of chunks that can be matched. The most widely used peephole optimisations consist of combining adjacent gates. For instance, canceling adjacent CNOT gates and Hadamard gates and combining adjacent Z-phase gates:
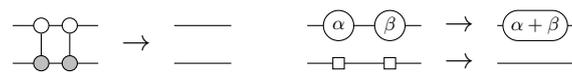
\ctikzfig{example-optimisations}
A more thorough approach is applied in for instance Ref.~\cite{PhysRevA.88.052307} where every 4-qubit Clifford sub-circuit is replaced by an optimal implementation of the same circuit. Peephole optimisations are usually combined with a representation of the circuit that includes information on which gates can be commuted past one another. This allows gates to be `moved out of the way' so that more matches can be found. An often-used representation for this is a \emph{directed acyclic graph} that contains information concerning which gates necessarily have to follow other gates. For a systematic study on the use and complexity of peephole optimisation in quantum circuits we refer to Ref.~\cite{iten2019efficient}.

Another important subfield where peephole optimisation is widely used is in \Define{reversible logic} synthesis\indexd{reversible logic synthesis}. Important components of many quantum circuits are essentially classical logical circuits implementing for instance adders or multipliers. Since any classical reversible logic circuit can be implemented using just Toffoli, CNOT and NOT gates, the goal of synthesising these circuits is usually minimising the number of Toffoli gates. The many results in this field, see e.g.~Refs.~\cite{Adriano-Barenco:1995qy,Maslov2005Toffoli-network,wille2010effect}, will not be of importance to us, but we will highlight one key result. A circuit that consists fully of CNOT gates is known as a \Define{linear reversible circuit}\indexd{circuit!linear reversible ---}. In Ref.~\cite{markov2008optimal} they found that such a circuit on $n$ qubits can be represented by an $n\times n$ matrix of zeros and ones, and that it can be efficiently synthesised using a modified Gaussian elimination algorithm into a circuit containing an asymptotically optimal number of CNOT gates.

There is one final heuristic approach we will discuss that is especially important for T-count optimisation. For any $n$-qubit unitary $U$ that can be implemented by a circuit consisting of just diagonal gates and CNOT gates we can find a \emph{phase polynomial}\indexd{phase polynomial} $f: \{0,1\}^n\rightarrow \R$ (cf.~Section~\ref{sec:phasegadgets}) and an invertible $n\times n$ matrix $A$ over $\mathbb{Z}_2$ such that $U\ket{\vec{x}} = e^{i f(\vec{x})} \ket{A \vec{x}}$ for all $\vec{x} \in \{0,1\}^n$~\cite{amy2014polynomial}. The action of the matrix $A$ is a linear reversible circuit and hence can be implemented solely using CNOT gates. All of the non-Clifford information is then captured in the phase polynomial $f$. There exist efficient methods to synthesise a circuit from a phase polynomial $f$. We gave a simple one in Section~\ref{sec:phasegadgets}, but there are also more intricate synthesis methods~\cite{amy2014polynomial,amy2018cnot}. 
By representing every sub-circuit consisting of CNOTs and diagonal gates as a phase polynomial and resynthesising we can already significantly reduce the T-count of the full circuit. As shown in Eq.~\eqref{eq:phasegadget}, a phase gadget, which is just one term of a phase polynomial, can be implemented as a circuit using a `ladder' of CNOT gates. Since a phase gadget is symmetric in the qubits, the orientation of the ladder is irrelevant. By representing the circuit as a phase polynomial this `irrelevant' information is modded out, allowing us to do simplifications that are hard to see in a circuit description. For instance\footnote{This particular example could also be simplified by a simple rule that says that phase gates can be commuted through a pair of CNOT gates with opposite target and control such as is done in Ref.~\cite{nam2018automated}, but for complicated configurations, for instance if we had phase gadgets on 3 qubits, this would be increasingly cumbersome.}:
\ctikzfig{phase-poly-example}
The first to take this approach was Ref.~\cite{amy2014polynomial}. This was improved upon by Ref.~\cite{abdessaied2014quantum} by applying some peephole optimisations that reduce the number of Hadamard gates that act as a barrier to phase polynomial approaches. Finally, it was combined with an even larger set of peephole optimisations in Ref.~\cite{nam2018automated} that produced circuits with a significantly lower CNOT count.

\indexd{circuit optimisation!phase polynomial ---}Interestingly, different phase polynomials can represent the same unitary. Finding equivalent phase polynomials that have less terms, and thus require less T gates to implement, is done by dedicated \Define{phase polynomial optimisers}. The first to use such an optimiser was Ref.~\cite{amy2016t} where they showed the problem of finding optimal equivalent phase polynomials to be related to the problem of \emph{Reed-Muller decoding}\indexd{Reed-Muller decoding}, which is believed to be a hard problem. The results of Ref.~\cite{amy2016t} were improved upon by Ref.~\cite{heyfron2018efficient} where they exploited a relation to the problem of \emph{3-tensor factorisation}. They call their approach \emph{third order duplicate and destroy}, abbreviated to TODD\indexd{circuit optimisation!TODD}\indexd{TODD}. In Section~\ref{sec:clifford-T-optimisation} we will use TODD in combination with our ZX-calculus-based approach. Finally, let us mention Ref.~\cite{deBeaudrap2020Techniques} where they introduce a general framework for finding optimisations to phase polynomials.

Recently some work has been done on generalising the phase polynomial approach so that it can also deal with Hadamard gates. This has lead to a representation of quantum circuits as a series of \Define{Exponentiated Pauli operators}~\cite{Litinski2019gameofsurfacecodes}. This was used in Ref.~\cite{zhang2019optimizing} to get T-counts similar to the method we will introduce.

\section{Simplification of ZX-diagrams}\label{sec:circuit-simplification}

In this section we will describe the ZX-diagram simplification strategy we will use to do circuit optimisation. This strategy uses the rules from Section~\ref{sec:graph-theoretic-simp} in combination with those of Section~\ref{sec:further-optimisation}.

Our starting point will be a circuit given in the form of a ZX-diagram. If the circuit we want to simplify contains gates that are not native to the ZX-calculus, such as Toffoli gates, we decompose these gates in a way that allows us to represent them directly in the ZX-calculus.

Before we describe our rewrite strategy let us remark that whenever a rewrite introduces parallel edges or self-loops that we immediately implicitly remove these using Eqs.~\eqref{eq:hopf-law-h} and~\eqref{eq:self-loops}), \ie:
\begin{align*}
&\quad\input{par-edge-rem.tikz} \\ 
	&\input{self-loop-rem.tikz}\qquad \qquad\qquad\ \ 
    \input{h-self-loop-rem.tikz}
\end{align*}

\begin{figure}[!t]
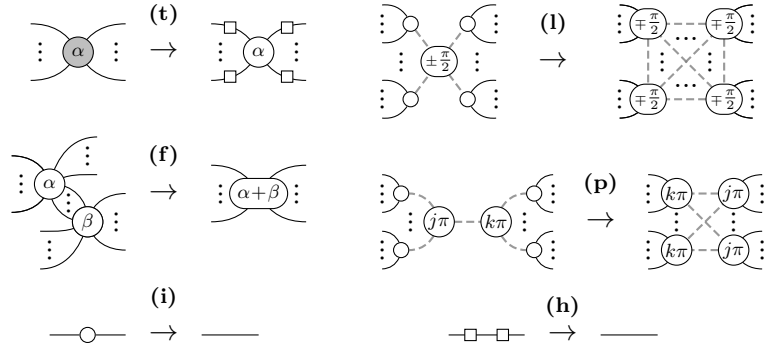
\indexd{local complementation!as simplification rule}\indexd{pivot!as simplification rule}
  \centering
  \ctikzfig{zx-clifford-axioms-simple4}
  \caption{Main set of simplification rules (scalar factors not included). The rules \textbf{(l)} and \textbf{(p)} are not presented in full generality (for that see Eqs.~\eqref{eq:lc-simp} and~\eqref{eq:pivot-simp}). In particular all the Z-spiders are allowed to have phases and in \textbf{(p)} the two main spiders are also allowed to share neighbours. If an application of one of these rules would add an edge between a pair of spiders that are already connected, then instead this connection is removed (cf.~Eq.~\eqref{eq:hopf-law-h}).\label{fig:simplification-rules-basic}}
\end{figure}

The main rewrite rules are shown in Figure~\ref{fig:simplification-rules-basic}. 
The first steps consist in making the diagram graph-like (cf.~Definition~\ref{def:graph-form}):
\begin{enumerate}
  \item First, we change the \textbf{(t)}ype of all the X-spiders to Z-spiders by introducing Hadamard boxes. This means that all other rules only need to deal with Z spiders.
  \item As the previous step potentially introduced many Hadamard boxes, we now proceed with canceling double \textbf{(h)}adamard boxes.
  \item Now \textbf{(f)}use all the neighbouring Z spiders.
  \item Remove \textbf{(i)}dentities. This step might produce more pairs of adjacent Hadamard boxes and hence we go back to step 2) to remove these.    
\end{enumerate}

\begin{figure}[!htb]
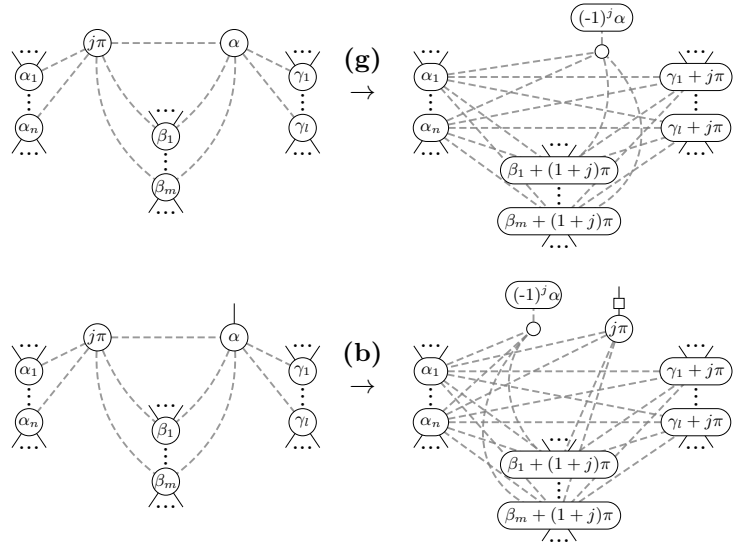

    \centering
    \scalebox{0.9}{\input{pivot-simp-gadget.tikz}}\\
    \vspace{0.4cm}
    \scalebox{0.9}{\input{pivot-simp-boundary-gadget.tikz}}
    \caption{Gadgetisation rules based on the pivot rule~\eqref{eq:pivot-desc}. As in that equation, the right-hand side contains fully connected bipartite graphs between the 3 sets of vertices. \label{fig:gadget-rules}}
\end{figure}

We repeat these steps until a fixed point is reached.
This procedure is very similar to that described in Lemma~\ref{lem:all-zx-are-graph-like}. Recall that a graph-like ZX-diagram is essentially the same thing as a ZX-diagram in MBQC form where every measured spider is measured in the \XY-plane (cf.~Lemma~\ref{lem:zx-to-mbqc-form}). Since we started with a circuit, the diagram, when viewed as an MBQC form diagram, has gflow (cf.~Lemma~\ref{lem:circuits-have-gflow}). The other rules we will introduce in this section similarly preserve the existence of a gflow on the diagram.

Now we will apply more involved rules in order to get rid of Clifford vertices as described in Sections~\ref{sec:graph-theoretic-simp} and~\ref{sec:removing-clifford-vertices}.

\begin{enumerate}[resume]
  \item Apply a \textbf{(l)}ocal complementation to every internal vertex with a $\pm\pi/2$ phase in order to remove it. See also Eq.~\eqref{eq:lc-simp}.
  \item Apply a \textbf{(p)}ivot to every pair of internal vertices with a $0$ or $\pi$ phase in order to remove the pair. See also Eq.~\eqref{eq:pivot-simp}.
\end{enumerate}

These rewrite rules will remove many of the internal Clifford spiders, but not all of them. In particular, we could still have internal vertices left that have a $0$ or $\pi$ phase that are connected solely to boundary spiders and internal spiders with a non-Clifford phase.
In order to get rid of these spiders we require two generalisations of the \textbf{(p)}ivoting rule that we already implicitly used in Lemmas~\ref{lem:pivot-simp} and~\ref{lem:removeboundaryPauli}. See Figure~\ref{fig:gadget-rules}. 

\indexd{phase gadget!gadgetisation}We call these `gadgetisation' rules, as they produce phase gadgets. The proof of correctness of these rules follows very similarly to that of Eq.~\eqref{eq:pivot-simp}. For \textbf{(g)} we unfuse the $j\pi$ and $\alpha$ spiders, apply the pivot rule Eq.~\eqref{eq:pivot-desc}, and then copy the $j\pi$ through. For \textbf{(b)} we do the same, except we unfuse the $\alpha$ like so:
\ctikzfig{unfuse-alpha}
We apply these rules with the following strategy:
\begin{enumerate}[resume]
    \item Apply a \textbf{(b)}oundary gadgetisation for every internal vertex with a $0$ or $\pi$ phase that is connected only to boundary spiders. If it is connected to multiple boundaries, give a preference to boundary spiders that have a Clifford phase, as the resulting phase gadgets can be easily simplified away later.
    \item Apply a \textbf{(g)}adgetisation to every remaining internal spider with a $0$ or $\pi$ phase. Such a spider is necessarily connected to some other internal spider (since otherwise it would not be connected to anything else making it a scalar).
\end{enumerate}
\begin{remark}
    For the purpose of these rewrite rules, we consider a phase gadget as a different type of vertex that is not available for further gadgetisation. This is necessary to prevent the previous step being an infinite loop where the `base' of a phase gadget keeps getting used in a gadgetisation.
\end{remark}
\begin{remark}
    The rewrite rule \textbf{(b)} only works if the boundary is connected to exactly one input or output. If the internal vertex is not connected to any such boundary, then spiders can be unfused to make the boundary have the right shape. This only introduces additional boundary spiders, not any internal ones.
\end{remark}

After these steps, any internal spider either has a non-Clifford phase or is part of a phase gadget. Note furthermore that the phase gadget produced by rules \textbf{(g)} and \textbf{(b)} is connected to exactly what the $j\pi$ spider was connected to. As all the phase gadgets in our diagram are produced by applying these rules to spiders with a $0$ or $\pi$ phase that is not connected to any other spider with a $0$ or a $\pi$ phase, each phase gadget is then necessarily not connected to any other phase gadget.
This means that the resulting diagram fits the description of a MBQC+LC diagram in phase-gadget form (cf.~Definition~\ref{def:pseudonormalform}), where the phase gadgets are interpreted as \YZ measurements, and the `regular' spiders are interpreted as qubits measured in the \XY-plane.

\begin{figure}[!htb]
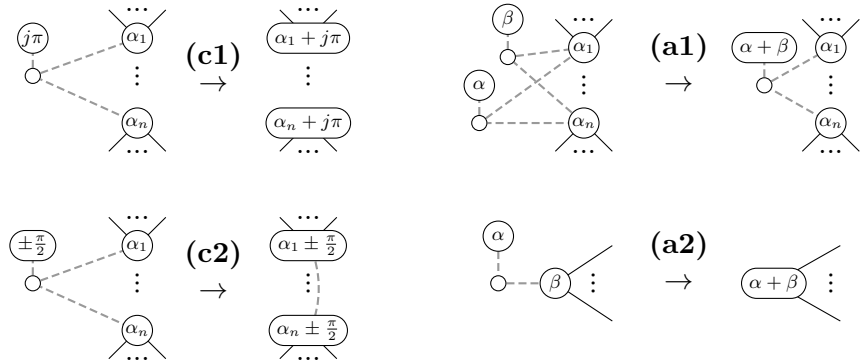

    \centering
    \ctikzfig{fusion-rules}
    \caption{Copy and fusion rules for phase gadgets. The right-hand side of \textbf{(c2)} represents a fully connected graph between all the neighbours of the removed phase gadget. \label{fig:fusion-rules}}
\end{figure}

We present now the final set of rewrite rules. See Figure~\ref{fig:fusion-rules}. The \textbf{(c)}opy rules allow us to remove the phase gadgets with a Clifford phase. The first is easily proven with \CopyRule, while the second uses Eq.~\eqref{eq:s-state-eq} and then \textbf{(l)}. The two other rules allow us to \textbf{(a)}dd together the phases of particular configurations of spiders. The first is proven in Eq.~\eqref{eq:fusing-gadgets} while the second follows easily from \IdRule and \HHRule.

Now we can present the final steps of our simplification procedure:
\begin{enumerate}[resume]
    \item Apply \textbf{(c1)} and \textbf{(c2)} to remove all Clifford phase gadgets.
    \item Apply \textbf{(a1)} to any pair of phase gadgets that have the same set of neighbours. If this introduces new Clifford phase gadgets then go back to the previous step.
    \item Apply \textbf{(a2)} to any phase gadget which has a single neighbour. If this introduces any new internal Clifford spiders, go back to step 4. Otherwise we are done.
\end{enumerate}

If step 11 introduced Clifford vertices, steps 4-6 can change the connectivity of the diagram so that phase gadgets can potentially become connected to one another. In this case, the pivoting of step 6 turns these phase gadgets into regular vertices. These steps might result in new internal spiders with a $0$ or $\pi$ phase which are only connected to non-Clifford spiders. Steps 7 and 8 can again remove these. As noted below steps 7 and 8, we do not consider the base of a phase gadget as available for gadgetisation.

Most of the power of this algorithm, especially when dealing with T-count optimisation, comes from the latter steps dealing with gadgetisation. We will hence refer to the full simplification routine of steps 1-11 as \gadgetsimp.\indexd{gadget-simp@\gadgetsimp}

\section{PyZX}\label{sec:pyzx}

The simplification procedure described above would of course be tedious to implement by hand on large diagrams. In order to apply this procedure to diagrams of a useful size, we implemented the routine in software.

PyZX (pronounced like `physics' without the `h') is a Python-based library designed for reasoning with large quantum circuits and ZX-diagrams. PyZX is Free and Open Source Software, licensed under GPLv3. The project is hosted on GitHub and available at:
\begin{center}
  \href{https://github.com/Quantomatic/pyzx}{\texttt{https://github.com/Quantomatic/pyzx}}
\end{center}
It allows users to efficiently rewrite ZX-diagrams using built-in simplification strategies. In this section we aim to give a short overview of the general architecture of PyZX and some of the utilities it offers. 

PyZX is written in pure Python with the only dependencies being NumPy~\cite{numpy} and Matplotlib~\cite{matplotlib}.

There are two main data-structures present in PyZX: {\circuit}s and {\graph}s. A \circuit is basically a wrapper around a list of gates, while a \graph represents a ZX-diagram. 

The \circuit class is the entry-point for importing and exporting circuits to and from PyZX. It also provides methods to do gate-level operations, such as converting a Toffoli-circuit into a Clifford+T circuit or taking the adjoint of all its gates. There is also a variety of circuit optimisation schemes that act directly on the \circuit class which can be found in the \texttt{optimize} sub-module. 

A \circuit consists of a list of {\gate}s, which in turn are small classes containing some information about the gate and how to convert it into various representations, such as ZX-diagrams or the \emph{QASM} format~\cite{qasmpaper}.

\indexd{ZX-diagram!in PyZX}The \graph class is more interesting. The graphs in PyZX are simple, undirected graphs with typed vertices and edges. Vertices come in three types: boundaries, Z-spiders and X-spiders. Each vertex can be labelled by a phase that is stored as a fraction representing a rational multiple of $\pi$. For instance, a label of $\frac12$ corresponds to a phase of $\frac\pi2$. 

The edges come in two types, which correspond respectively to a regular connection between spiders, and a Hadamard-edge. As ZX-diagrams allow parallel edges between spiders and self-loops, we need a way to deal with these in PyZX graphs. Adding an edge where there is already one present will simply replace it. Often, it is more convenient to use the rules of the ZX-calculus to resolve parallel edges and self-loops whenever a new edge is added. How this should be done depends on the types of edges and vertices involved:
\begin{equation}\label{eq:edge-cases}
\input{edge-cases.tikz}
\end{equation}
Note that these rules as presented here are not scalar-accurate. In PyZX the scalar is stored separately from the graph, as a complex number. This number is automatically updated in order to preserve the correct scalar value of the diagram.

The \graph class abstracts away some of the details of the underlying representation of the graph. As a result, a different implementation of a graph can simply be a subclass of the \texttt{BaseGraph} class in order to work with all the other functionality of PyZX. 

The default pure Python implementation of a graph in PyZX stores its connectivity as a dictionary of dictionaries, where the first level has vertices as keys (identified by an integer), with the values being another dictionary containing all the neighbours of this vertex. The values of these dictionaries in turn specify by which type of edge the vertices are connected, regular or Hadamard. Phases are stored in another dictionary. We have experimented with other graph back-ends, but we found that this simple implementation is fast enough to handle and simplify diagrams with hundreds of thousands of vertices in a reasonable time-frame.

Usually you will not want to manually create a ZX-diagram. There are various ways to import ZX-diagrams into PyZX. Directly importing ZX-diagrams is possible using the Quantomatic format~\cite{kissinger2015quantomatic}, but PyZX can also read files describing quantum circuits in a variety of languages. It currently supports QASM~\cite{qasmpaper}, the Quipper ASCII format~\cite{green2013quipper,quipper}, the QC/TFC format used by the Reversible Circuit Benchmarks page~\cite{maslovreversible} and the QSIM format used by the quantum supremacy circuits of Ref.~\cite{arute2019quantum}.

The simplification strategies for ZX-diagrams in PyZX are built in a hierarchical manner. 

At the bottom level there are \Define{rules} that consist of a \Define{matcher} and a \Define{rewriter}. A matcher loops over all the vertices or edges of a graph (which one depends on the rule) and tries to find as many non-overlapping sections of the graph that are suitable for the rule to be applied. Once it finishes, it returns a list of matches (what such a match object looks like differs per rule). The rewriter takes this list of matches and figures out which changes to the graph need to be made. Since all the matches are non-overlapping, this can be done for each match separately. If these changes involve adding vertices or changing phases, then this is done immediately, but in order to reduce overhead, other changes (adding edges or removing edges and vertices) are collected until the rewriter has processed all matches, and are then implemented simultaneously. 

For a simple example of a rule, let us consider \Define{identity removal}:
\begin{equation}\label{eq:id-simp}
\input{id-simp.tikz}
\end{equation}
The matcher loops over vertices and tries to find those vertices that have a phase of zero, and exactly two neighbours. When it has found such a vertex, it removes its neighbours from the list of candidates (in order to prevent overlapping applications), and it makes it into a match that contains the vertex, its neighbours, and the type of the edge that should be made between these neighbours in the rewritten graph (in the case above, since both edges to the middle vertex were Hadamard edges, the resulting edge is a regular edge, cf.~\HHRule). The rewriter takes this list of matches, builds a list of vertices to be removed, and edges to be added, and applies this all at once. As the neighbours of the removed vertex could already have been connected in the original graph, this could lead to a self-loop that is handled as in Eq.~\eqref{eq:edge-cases}.

The next stage in the hierarchy of simplification strategies are the \Define{basic simplifiers}. These are built on a single rule. A simplifier keeps applying the matcher and rewriter of a single rule, until the matcher finds no more matches. Because a matcher only finds non-overlapping rules, and thereby might miss possible applications for the rule the first time around, and because the rewritten diagram might produce new sections of the graph that are suitable for application of the rule, the simplifier might need to do the process of matching and rewriting many times before no new matches are found. In order for the simplifier to not get stuck in an infinite loop, it is important that the rule actually simplifies the diagram in some manner, i.e.\ that some kind of metric on the graph is reduced. In the case of the basic simplifiers used in PyZX, this metric is usually that the number of vertices in the graph is reduced, but it could also be something more complex such as trading one type of vertex for another one.

Let us consider the identity removal simplifier (named \texttt{id\_simp} in PyZX). Its simplifier removes all the identity spiders in the diagram as in \eqref{eq:id-simp}. The rule never generates new arity-2 zero-phase spiders, hence the only way in which the simplifier has to run multiple times is when there were multiple identities in a row, i.e.\ when the rule applications had overlap. Since every application of the rule removes a vertex, this simplifier indeed terminates.

At the top level of the hierarchy are the \Define{compound simplifiers}. These simply combine other simplifiers into a more complicated simplification strategy, by applying other simplifiers in some particular order, potentially looping over this combination until none of the simplifiers finds any more reductions.

For instance, PyZX also implements a basic simplifier that fuses spiders together (cf.~\SpiderRule), called \texttt{spider\_simp}. We could combine \texttt{id\_simp} and \texttt{spider\_simp} together in a compound simplifier as follows:
{
\begin{python}
def fuse_simp(g):
    i = 0
    while True:
        i1 = id_simp(g)
        i2 = spider_simp(g) 
        if i1 == 0 and i2 == 0: break  # No matches found
        i += 1
    return i
\end{python}
}
Here \texttt{i1} and \texttt{i2} contain the number of iterations the simplifiers had to go through before no more matches were found.
This simplifier simply keeps removing identities and fusing spiders until this is no longer possible. Note that removing an identity might make more possibilities for fusing spiders, and fusing spiders might make a zero-phase arity-two spider. Hence, we indeed need the loop in this function to make full use of these simplifiers.

Some simplification rules might disconnect part of the graph, in particular making dangling scalar diagrams consisting of 1 or 2 spiders (like happens in for instance Eq.~\eqref{eq:ms-zx-tgate3}). These scalar diagrams are automatically converted into a complex number and combined with the scalar of the \graph.

\section{Clifford circuit optimisation}\label{sec:clifford-circuit-optimisation}
\indexd{circuit optimisation!Clifford ---}
As a first demonstration of the power of \gadgetsimp, let us apply it to Clifford circuits.
Recall that the simplification routine of Section~\ref{sec:graph-theoretic-simp} is a weaker version of \gadgetsimp. This routine was already capable of completely reducing a scalar Clifford diagram to a single number. Similarly, \gadgetsimp is able to reduce any Clifford diagram to a pseudo-normal form.

For a Clifford diagram any spider has a phase in the set $\{0,\pi/2,\pi,-\pi/2\}$. Local complementation removes any internal spider with a phase of $\pm\pi/2$. The pivoting steps remove most of the internal $\pi$ spiders. The only ones that remain are those that are only connected to boundary spiders (as there are no remaining internal non-Clifford spiders they could be connected to). Hence, the boundary gadgetisation in step 7 makes these into phase gadgets. These phase gadgets are all removed in step 9 using \textbf{(c1)}. Hence, the resulting diagram has no internal spiders left. The resulting diagram will then look something like the following:
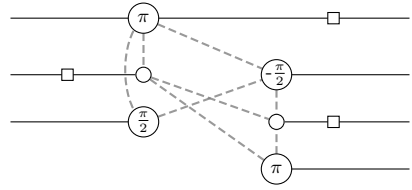
\ctikzfig{gslc-general-form}
I.e.~each spider is connected to an input or output wire, possibly both (assuming that we are representing solitary scalar spiders as a complex number). If the diagram had no inputs, then we recognize this form as a graph state with local Clifford gates (cf.~Definition~\ref{def:GSLC}). We have thus reproved Theorem~\ref{thm:clifford-state-GS-LC}, which said that every Clifford state is equal to a GS-LC state. Hence, \gadgetsimp is a general procedure to bring Clifford diagrams to a pseudo-normal form that is closely related to the GS-LC form of states, and hence we will refer to these diagrams as being in GS-LC normal form.

\indexd{Clifford!--- normal-form}This form is especially interesting when the input diagram is a Clifford circuit. The resulting diagram will look like the following:
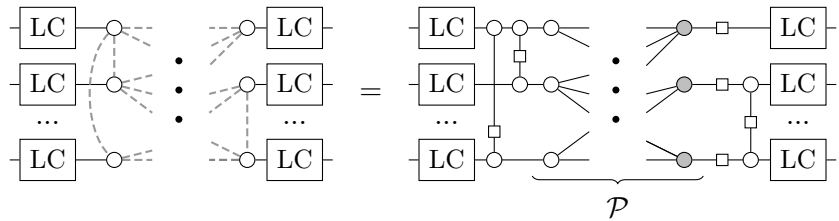
\ctikzfig{gslc-extract1prime}
Here each box of `LC' stands for a local Clifford that consists of a possible Hadamard gate followed or preceded by some power of an S gate. In the left diagram, each spider can be connected to any other spider. To get to the right-hand diagram we simply unfuse some spiders to make CZ gates out of spiders connected on the same side, and converted some Z-spiders into X-spiders to create the middle part of the diagram labelled $\mathcal{P}$.
We can transform this part of the diagram into a circuit using the Gaussian elimination procedure of Section~\ref{sec:extracting-circuit-details}.
This works because $\mathcal{P}$ has the form of a linear reversible circuit. That is, it is a permutation of computational basis states where the output state is given in terms of parities of the inputs, e.g.
$ \ket{x_1, x_2, x_3, x_4} \mapsto \ket{x_1 \oplus x_2, x_1 \oplus x_3, x_4, x_3}.$
Such a unitary can always be realised via CNOT gates. This was first observed in Ref.~\cite{markov2008optimal}.

With this decomposition of $\mathcal{P}$ into CNOT gates, we obtain a Clifford circuit with 6 layers of gates:
\begin{equation}\label{eq:clifford-normal-form}
  \textrm{Local Clifford + CZ + CNOT + H + CZ + Local Clifford}
\end{equation}
Here each local Clifford layer can be further decomposed in a layer of Hadamard gates and a layer of S$^k$ gates, giving an 8-layer pseudo-normal form for Clifford circuits.

There are a variety of pseudo-normal forms for Clifford circuits in the literature, starting with the 11-layer form given by Aaronson and Gottesman~\cite{aaronsongottesman2004} and the coarser-grained 5-layer form of Dehaene and De Moore~\cite{dehaene2003clifford}, which led to improved versions by Maslov and Roetteler~\cite{maslov2018shorter} and van den Nest~\cite{nest2010clifford}, respectively. While there are some superficial similarities between the normal form of Eq.~\eqref{eq:clifford-normal-form} and these earlier ones, there is at least one notable difference. All of the forms mentioned above require at least two distinct CNOT layers, but (with the exception of Ref.~\cite{aaronsongottesman2004}) require just a single later of Hadamard gates. On the other hand, our normal form has just a single CNOT layer, at the cost of multiple Hadamard layers. We will now see that this trade-off has some positive consequences.

In Ref.~\cite{maslov2018shorter} the authors argued that since there are $2^{2n^2 +O(n)}$ Clifford unitaries on $n$ qubits, one needs at least $2n^2 + O(n)$ Boolean degrees of freedom to specify all the $n$ qubit Clifford unitaries, and furthermore they found a normal form which has the same number of degrees of freedom and hence is asymptotically optimal in this sense. In that same work they also found a different Clifford pseudo-normal form that has a lower 2-qubit gate depth when restricted to a `linear nearest neighbour architecture` where 2-qubit gates are only allowed between adjacent qubits put on a line. The 2-qubit gate depth of this form in that architecture is bounded by $14n-4$. Our normal form improves on this bound, while at the same time also satisfying the $2n^2 + O(n)$ asymptotically optimal gate count.

\begin{theorem}
    The GS-LC pseudo-normal form on $n$ qubits has an asymptotically optimal number of degrees of freedom $2n^2 + O(n)$. Furthermore, any Clifford unitary in this normal form can be mapped to a linear nearest neighbour architecture with a 2-qubit gate depth of $9n-2$.
\end{theorem}
\begin{proof}
    The argument follows closely the one given in Ref.~\cite{maslov2018shorter}, but for a different normal form. Our normal form has 5 layers of single qubit Clifford gates. The Hadamard layers each add at most $n$ gates, while the S-phase layers add at most $3n$ gates, hence these layers only add a linear number of degrees of freedom. Each CZ layer adds $n^2/2$ degrees of freedom, while a CNOT layer adds $n^2$ degrees of freedom~\cite[Section I]{maslov2018shorter}. Hence the total degrees of freedom is given by $3n+2 \cdot 3n + 2 \cdot n^2/2 + n^2 = 2n^2 + O(n)$.

    For the 2-qubit gate depth, we note that any CNOT circuit can be implemented on a linear nearest neighbour architecture in depth $5n$~\cite{kutin2007computation}. A CZ circuit followed or preceded by a series of SWAP gates that reverses the qubit order can be implemented in depth $2n+2$ on a linear nearest neighbour architecture~\cite[Thm.~6]{maslov2018shorter}. But by~\cite[Cor. 7]{maslov2018shorter}, when we have two of these CZ circuits, possibly separated by some other gates, then this pair of CZ circuits can be implemented in 2-qubit gate depth $4n-2$. As the only layers in our pseudo-normal form that contribute to the 2-qubit gate depth are two CZ layers and a CNOT layer we then indeed have a total depth of $5n + 4n-2 = 9n - 2$.
\end{proof}

\subsection{On completeness of the ZX-calculus}
\indexd{ZX-calculus!completeness}
An important aspect of the ZX-calculus that we have not yet discussed in detail is \Define{completeness}, \ie~whether any two ZX-diagrams that represent equal linear maps can be transformed into one another using the graphical rewrite rules of the ZX-calculus. The set of rules we use is known to not be complete for the set of all ZX-diagrams, or even just those where all phases are multiples of $\pi/4$~\cite{supplementarity}. 
Backens showed in 2013 that the rule-set we use (cf.~Figure~\ref{table:rewrite}) is however complete for Clifford diagrams~\cite{BackensCompleteness}. Their proof consists of two stages. 
The first shows that any Clifford ZX-diagram can be converted to the GS-LC pseudo-normal form. 
The second step builds on the work of Ref.~\cite{elliott2008graphical} to show that two GS-LC normal forms representing the same linear map can be transformed into one another by a sequence of local complementations. 
The first step, that any Clifford diagram can be brought to GS-LC pseudo-normal form, is proved by showing how all ZX-diagrams consisting of a GS-LC normal form composed with any generator of the ZX-calculus can be brought back into GS-LC form, a process that requires many case distinctions. The results we have discussed above give a different proof of this first step: starting from any Clifford ZX-diagram, apply \gadgetsimp to convert it into GS-LC normal form. This strategy is preferable for multiple reasons: it requires less case distinctions, it builds on rewrite rules (local complementation and pivoting) that are needed in the second step regardless, every rewrite step consists of an actual simplification of the diagram, and the process can even be applied if the diagram is not fully Clifford.

\section{Clifford+T circuit optimisation}\label{sec:clifford-T-optimisation}

We have seen that \gadgetsimp is capable of normalising Clifford circuits. In this section we will see how we can use \gadgetsimp to produce a circuit-to-circuit optimisation routine that can reduce the number of non-Clifford gates needed to implement a circuit. We will benchmark an implementation of this routine and see that it matches or outperforms all other existing methods that have the same goal.

When \gadgetsimp is applied to a Clifford circuit, it eliminates all the internal spiders, leaving a compact diagram. This is no longer the case when the circuit being simplified has non-Clifford gates. In that case, each non-Clifford gate can result in additional internal spiders, resulting in a diagram with a potentially intricate internal structure. The main problem then if we want to use \gadgetsimp as a circuit-to-circuit routine, is to transform the resulting ZX-diagram back into a circuit.

Fortunately, we have in fact already solved this problem in Chapter~\ref{chap:MBQC}. The graph-like diagrams produced by \gadgetsimp can be seen as MBQC+LC diagrams where the phase gadgets correspond to vertices measured in the \YZ-plane, and every other vertex is measured in the \XY-plane. As noted in Section~\ref{sec:circuit-simplification}, each of the rewrite rules of \gadgetsimp corresponds to a rewrite rule of Chapter~\ref{chap:MBQC} and preserves the existence of gflow. Hence, using the algorithm of Section~\ref{sec:extraction-algorithm} we can convert the diagram back into a circuit. As described in Theorem~\ref{thm:extraction-algorithm} the number of non-Clifford gates in the resulting circuit matches the number of non-Clifford vertices in the diagram.

We have then the following circuit-to-circuit optimisation routine:
\begin{enumerate}
    \item Write your circuit in the Clifford+T gate set and interpret it as a ZX-diagram.
    \item Simplify the diagram with \gadgetsimp.
    \item Extract a new circuit from the simplified diagram using the results of Section~\ref{sec:extraction-algorithm}.
\end{enumerate}

Note that the procedure actually works for a gate set that includes all Z rotation gate, and not just T gates. We will however focus on Clifford+T, because there is a wider variety of benchmark circuits available for this gate set, and more research has been done in optimising these circuits.

We have implemented the routine in PyZX and applied it to a set of benchmark circuits. Specifically, we used all of the Clifford+T benchmark circuits from Refs.~\cite{amy2014polynomial,nam2018automated} (except for some of the larger members of the \texttt{gf($2^n$)-mult} family).  Most of these circuits include Toffoli gates. We have decomposed these into the Clifford+T gate set using a standard decomposition. These benchmark circuits are widely used in other approaches to quantum circuit optimisation (in addition to the aforementioned, also in Refs.~\cite{amy2016t,heyfron2018efficient}) and include components that are likely to be of interest to quantum algorithms, such as adders or Grover oracles. See Table~\ref{fig:T-benchmark} for the list of results.

\newcommand{\better}{\rowfont{\bfseries}}
\newcommand{\worse}{\rowfont{\itshape}}
\newcommand{\plustoddbetter}{\rowfont{\upshape\bfseries}}
\newcommand{\regular}{\rowfont{}}

\begin{table}
\centering
\scalebox{1.0}{
\begin{tabular}{LCCCCCC}
Circuit & $n$& T& Best & Method & PyZX & \parbox{1.2cm}{PyZX\\+TODD} \\[0.2cm]
\hline
\better adder$_8$ & 24 & 399 & 213 & RM$_m$ & 173 & \plustoddbetter 167 \regular \\
Adder8 & 23 & 266 & 56 & NRSCM & 56 & 56 \\
Adder16 & 47 & 602 & 120 & NRSCM & 120 & 120 \\
Adder32 & 95 & 1274 & 248 & NRSCM & 248 & 248 \\
Adder64 & 191 & 2618 & 504 & NRSCM & 504 & 504 \\
barenco-tof$3$  & 5 & 28 & 16 & Tpar & 16 & 16 \\
barenco-tof$4$  & 7 & 56 & 28 & Tpar & 28 & 28 \\
barenco-tof$5$  & 9 & 84 & 40 & Tpar & 40 & 40 \\
barenco-tof$10$  & 19 & 224 & 100 & Tpar & 100 & 100 \\
tof$_3$ & 5 & 21 & 15 & Tpar & 15 & 15 \\
tof$_4$ & 7 & 35 & 23 & Tpar & 23 & 23 \\
tof$_5$ & 9 & 49 & 31 & Tpar & 31 & 31 \\
tof$_{10}$ & 19 & 119 & 71 & Tpar & 71 & 71 \\
\worse csla-mux$_3$  & 15 & 70 & 58 & RM$_r$ & 62 & \plustoddbetter 45 \regular \\
\worse csum-mux$_9$  & 30 & 196 & 76 & RM$_r$ & 84 & \plustoddbetter 72 \regular \\
\better cycle17$_3$ & 35 & 4739 & 1944 & RM$_m$ & 1797 & 1797 \regular \\
\worse gf($2^4$)-mult & 12 & 112 & 56 & TODD & 68 & \plustoddbetter 52 \regular \\
\worse gf($2^5$)-mult & 15 & 175 & 90 & TODD & 115 & \plustoddbetter 86 \regular \\
\worse gf($2^6$)-mult  & 18 & 252 & 132 & TODD & 150 & \plustoddbetter 122 \regular  \\
\worse gf($2^7$)-mult & 21 & 343 & 185 & TODD & 217 & \plustoddbetter 173 \regular \\
\worse gf($2^8$)-mult & 24 & 448 & 216 & TODD & 264 & \plustoddbetter 214 \regular \\
ham15-low & 17 & 161 & 97 & Tpar & 97 & 97 \\
\better ham15-med & 17 & 574 & 230 & Tpar & 212 & 212 \regular \\
ham15-high & 20 & 2457 & 1019 & Tpar & 1019 & \plustoddbetter 1013 \\
hwb$_6$ & 7 & 105 & 75 & Tpar & 75 & \plustoddbetter 72 \\
\better hwb$_8$ & 12 & 5887 & 3531 & RM$_{m\&r}$ & 3517 & \plustoddbetter 3501 \regular \\
\worse mod-mult-55 & 9 & 49 & 28 & TODD & 35 & \plustoddbetter 20 \regular \\
mod-red-21 & 11 & 119 & 73 & Tpar & 73 & 73 \\
\better mod5$_4$ & 5 & 28 & 16 & Tpar & 8 & \plustoddbetter 7 \regular \\
\better nth-prime$_6$ & 9 & 567 & 400 & RM$_{m\&r}$ & 279 & 279 \regular \\
\worse nth-prime$_8$ & 12 & 6671 & 4045 & RM$_{m\&r}$ & 4047 & \plustoddbetter 3958 \regular \\
qcla-adder$_{10}$ & 36 & 589 & 162 & Tpar & 162 & \plustoddbetter 158 \\
\worse qcla-com$_7$ & 24 & 203 & 94 & RM$_m$ & 95 & \plustoddbetter 91 \regular \\
qcla-mod$_7$ & 26 & 413 & 235${}^\textrm{a}$ & NRSCM & 237 & \plustoddbetter 216 \\
rc-adder$_6$ & 14 & 77 & 47 & RM$_{m\&r}$ & 47 & 47 \\
vbe-adder$_3$ & 10 & 70 & 24 & Tpar & 24 & 24
\end{tabular}
}
\caption{Benchmark circuits from \cite{AmyGithub} and \cite{NRSCMGithub}. The columns \emph{$n$} and \emph{T} contain the number of qubits and T gates in the original circuit. \emph{Best} is the previous best-known ancilla-free T-count for that circuit and \emph{Method} specifies which method was used: \emph{RM$_m$} and \emph{RM$_r$} refer to the \emph{maximum} and \emph{recursive} Reed-Muller decoder of Ref.~\cite{amy2016t}, \emph{Tpar} is Ref.~\cite{amy2014polynomial}, \emph{TODD} is Ref.~\cite{heyfron2018efficient} and \emph{NRSCM} is Ref.~\cite{nam2018automated}. \emph{PyZX} and \emph{PyZX+TODD} specify the T-counts produced by respectively our method, and our method combined with TODD. Numbers shown in bold are better than previous best, and italics are worse. The superscript (a) indicates an error in a previously reported T-count.
\label{fig:T-benchmark}}
\end{table}

Of the 36 benchmark circuits, we are at or improving upon the best previously known ancilla-free T-count for 26 circuits ($\sim$72\%), and we improve on 6 ($\sim$17\%). If we apply some simple post-processing afterwards (in the form of peephole optimisation) and feed the resulting circuit into the TODD phase polynomial optimiser~\cite{heyfron2018efficient}, we improve on the state of the art for 20 circuits ($\sim$56\%). These two methods seem to complement each other well in the ancilla-free regime, obtaining significantly better numbers than either of the two methods alone, and matching or beating all other methods for every circuit tested.

For 20 of the 36 circuits, we exactly match the best previously known result, which is interesting, since the methods we use are quite different in nature from previous methods.
The circuits where PyZX seems to do considerably better are ones that contain many Hadamard gates.
The fact that PyZX achieves improvements when there are many Hadamard gates is as expected, as most other successful methods employ a dedicated phase-polynomial optimiser~\cite{amy2014polynomial,amy2016t,nam2018automated,heyfron2018efficient} that is hampered by the existence of Hadamard gates. On the other hand, the only circuits where phase polynomial techniques significantly out-perform our methods are in the \texttt{gf($2^n$)-mult} family. After some simple preprocessing, these circuits have almost no Hadamard gates, hence they are very well-suited to phase polynomial techniques.

As noted before, our optimisation routine is agnostic to the values of the non-Clifford phases. We have also tested our routine on the quantum Fourier transform circuits of Ref.~\cite{nam2018automated} that include more general non-Clifford phases, and in each case found that our non-Clifford gate count exactly matched their results.

\section{Phase teleportation}\label{sec:phase-teleportation}

An important detail we glossed over in the previous section is whether our optimisation routine preserves or optimises other metrics of interest. For circuits meant to be run on the logical level of a fault-tolerant quantum computer, the most important metric, after the number of non-Clifford gates, is the number of 2-qubit gates. Unfortunately, in this regard the method described above behaves rather poorly. For most of the benchmark circuits tested, the extracted circuit contains (many) more 2-qubit gates then the original circuit. In Chapter~\ref{chap:future} we will discuss ways in which the extraction algorithm can be improved in order to reduce the number of 2-qubit gates in the resulting circuit. In this section we will describe a way in which the extraction stage of the algorithm can be skipped, bypassing the issue completely.

The method, that we call \Define{phase teleportation}, \indexd{phase teleportation} relies on the observation that \gadgetsimp is completely parametric in the values of its non-Clifford phases: the decision of which rewrite rule to apply or how it should be applied never depends on the exact value of the non-Clifford phase. 

We continue to use \gadgetsimp, but instead of working with concrete phases, we begin by replacing every non-Clifford phase in our starting circuit $C$ with a fresh variable name $\alpha_1, \ldots, \alpha_n$ resulting in a parametric circuit $C[\alpha_1,\ldots, \alpha_n]$. We store the concrete angles in a separate table $\tau : \{1, \ldots, n\} \to \mathbb R$. To get back the original circuit we simply insert these angles back into the variables: $C=C[\tau]$.

Next, we perform \gadgetsimp on $C[\alpha_1,\ldots,\alpha_n]$ \emph{symbolically}. That is, we work on a \zxdiagram whose spiders are labelled not just with phase angles, but with polynomials over the variables $(\alpha_1, \ldots, \alpha_n)$.

The interesting step happens when two variables are added together by the \textbf{(a1)} or \textbf{(a2)} rule of Figure~\ref{fig:fusion-rules}. One of two things can occur: $(a)$ the two variables have the same sign or $(b)$ they have different signs:
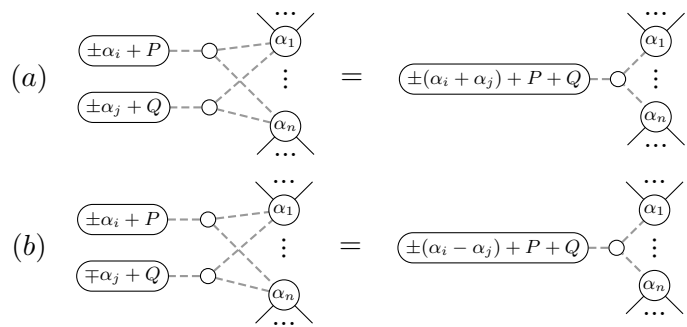
\[
(a) \ \ \ \input{gf-symbolic.tikz}
\]
\[
(b) \ \ \ \input{gf-symbol-diff.tikz}
\]
Since none of our simplifications will copy any of the variables we started with, these are the only occurrences of $\alpha_i$ and $\alpha_j$ in the \zxdiagram. Hence, in the case $(a)$, if we replace $\alpha_i$ with $\alpha_i + \alpha_j$ and $\alpha_j$ with $0$, we get an equivalent diagram.

Put another way, in case $(a)$, we can update our table $\tau$ by setting $\tau'(i) := \tau(i) + \tau(j)$, $\tau'(j) := 0$, and $\tau'(k) := \tau(k)$ for $k \notin \{i,j\}$. As the ZX-diagrams described by the tables of phases $\tau'$ and $\tau$ are the same, we see that $C[\tau]$ and $C[\tau']$ must also describe equivalent ZX-diagrams, and hence implement the same unitary. Crucially, $C[\tau']$ now contains fewer non-Clifford phases, since $\alpha_j$ has `teleported' to combine with $\alpha_j$.
Case $(b)$ is similar, except we should set $\tau'(i) := \tau(i) - \tau(j)$.

This observation yields Algorithm~\ref{alg:phase-teleportation}.\indexd{phase teleportation!algorithm}
\begin{algorithm}
    \caption{Phase teleportation\label{alg:phase-teleportation}}
    Starting with a circuit, do the following:
  \begin{enumerate}
    \item Choose unique variables $\alpha_1, \ldots, \alpha_n$ for each non-Clifford phase and store the pair $(C, \tau)$, where $C$ is the parametrised circuit and $\tau : \{1, \ldots, n\} \to \mathbb R$ assigns each variable to its phase.
    \item Interpret $C$ as a \zxdiagram and apply \gadgetsimp to it while doing the following:
    \begin{quote}
      Whenever \textbf{(f)}, \textbf{(a1)}, or \textbf{(a2)} is applied to a pair of vertices or phase-gadgets containing variables $\alpha_i$ and $\alpha_j$, update the phase table $\tau$ as described for cases $(a)$ and $(b)$ above.
    \end{quote}
    \item When \gadgetsimp is done, $C[\tau]$ still describes the same unitary, but now contains possibly fewer non-Clifford phases.
  \end{enumerate}
\end{algorithm}

\begin{remark}
    When two variables $\alpha_i$ and $\alpha_j$ are added together so that $\tau(i)+\tau(j)$ is a multiple of $\pi/2$ (or similarly for $\tau(i)-\tau(j)$), the spider will be treated as a Clifford spider for the remainder of the application of \gadgetsimp. The reason this works is because in the updated $\tau'$, both $\tau'(i)$ and $\tau'(j)$ now carry a Clifford phase, and hence we can treat the values $\alpha_i$ and $\alpha_j$ as if they were Clifford to start with.
\end{remark}

By construction, the phase teleportation algorithm results in a circuit with the same number of non-Clifford gates as the one based on circuit extraction described in the previous section. But in contrast to that algorithm, phase teleportation does not change anything else in the circuit. In particular, it does not change the number or location of the 2-qubit gates, and hence for a majority of the benchmark circuits results in smaller circuits. This makes phase teleportation ideally suited as a first step in a compound simplification step, where later steps can optimise, for instance, 2-qubit gate count. These methods can then perform better since there will in general be less `obstructions' in the form of non-Clifford gates. 

It should also be noted that in our implementation, the application of \gadgetsimp, or equivalently, that of phase teleportation, will usually not take more than a few seconds, even for circuits containing tens of thousands of gates. In contrast, extracting a circuit from the resulting diagram can take multiple minutes. Hence, phase teleportation allows the optimisation to be done much faster, as it skips this expensive extraction stage.

Phase teleportation could also be directly used on parametric circuits such as the quantum variational eigensolver~\cite{peruzzo2014variational}. In this setting phase teleportation will combine together redundant free parameters.

\section{Verification of equality}\label{sec:circuit-verification}

It is of course crucial that the implementation of an optimisation routine does not change the unitary the circuit implements, \ie that it preserves the semantics of the circuit. An `easy' way to verify that the original circuit is equal to the optimised one is to directly calculate the matrix of the unitary. This unfortunately takes memory exponential in the number of qubits. The general problem of determining whether two quantum circuits implement the same unitary is complete for the complexity class \textbf{QMA}~\cite{bookatz2012qmacomplete}, which is the quantum analogue of \textbf{NP}, and hence we do not expect there to be a much more efficient procedure that will work for all pairs of circuits.

Using \gadgetsimp we can construct an equality verification scheme: given two circuits $C_1$ and $C_2$ (or more generally, any two ZX-diagrams with the same number of inputs and outputs), we make the circuit $C=C_1\circ C_2^\dagger$ and apply \gadgetsimp to $C$ (the dagger $\dagger$ represents the adjoint of the circuit). If the resulting diagram is the identity, \ie where all the spiders have been simplified away, we conclude that $C_1$ and $C_2$ implement the same unitary, and otherwise our procedure has failed in providing an answer.

\indexd{circuit!adjoint}Note that the adjoint of a circuit is easily constructed by reversing the order of the gates, and taking the adjoint of each individual gate. Because $C_1$ and $C_2$ implement unitaries,  if they indeed implement the same unitary, then $C$ will implements the identity unitary. It is our hope then that \gadgetsimp succeeds in finding this reduction. If it indeed does, then the set of rewrites that simplifies $C$ to the identity forms a certificate of equality of $C_1$ and $C_2$. If the simplification however does not succeed in fully reducing the diagram, then we cannot conclude anything either way: it might be that the circuits are not equal, and that that is why the simplification failed, or it could be that our simplification schema was not powerful enough to find the right reduction. The utility of this verification is therefore completely determined by the set of equalities it can verify.

We used this validation scheme to verify correctness of all the optimised benchmark circuits of Ref.~\cite{nam2018automated}, except for \texttt{qcla-mod$_7$} on which it failed. Using the Feynman tool~\cite{AmyVerification} we then showed that this optimised circuit indeed contained an error.

We can also use the validation scheme to verify equality of our own optimisation routine. This might seem counter-intuitive as you would not expect a simplification routine to be able to verify its own correctness. The reason why the success of our validation schema can still be seen as strong evidence of the correctness of our implementation, is due to the specific nature of \gadgetsimp. It applies all possible occurrences of a rewrite before moving on to the next rewrite. Hence, after just a few rewrite steps, $C := C_1\circ C_2^\dagger$ where $C_2$ is the optimised version of $C_1$ no longer looks like a concatenation of $C_1$ and $C_2^\dagger$. The set of rewrites that are done to $C$ will be vastly different to those used in optimising $C_1$. It is then unlikely that an error in the implementation will cancel itself out. We used this validation scheme to verify correctness of all the optimised circuits in Table~\ref{fig:T-benchmark}.

It is unclear exactly which pairs of circuits this method is able to verify equality of. When both circuits are Clifford, the method will always succeed, but beyond that it is hard to say anything concrete. We conjecture that if two circuits can be transformed into one another using the rewrite rules of Figure~\ref{table:rewrite}, that the method should be able to verify this equality. The intuition behind this conjecture is that these rewrite rules only concern the Clifford structure of the circuit (in the sense that there is no rewrite rule that uses any phases other than multiples of $\pi/2$). As \gadgetsimp removes all the Clifford spiders, it stands to reason that the different representations of a diagram under these rewrite rules are `modded out'.

\section{Conclusion}\label{sec:optimisation-conclusion}

In this chapter we introduced a new approach to quantum circuit optimisation using the ZX-calculus. We found a single simplification algorithm that is at the same time capable of reducing Clifford circuits to a beneficial normal form, while also matching or surpassing the state-of-the-art in ancilla-free T-count optimisation when combined with the TODD phase polynomial optimiser.
We found we could extract circuits from the simplified ZX-diagrams using the algorithm described in Section~\ref{sec:extracting-circuit-details}, or by using the new method of phase teleportation. Finally, we saw that the optimisation method is essentially self-checking, being powerful enough to produce a certificate of equality for its optimised circuits.

Shortly after the results shown in Table~\ref{fig:T-benchmark} appeared online, Ref.~\cite{zhang2019optimizing} appeared on the arXiv, which used the technique of exponentiated Pauli's discussed in Section~\ref{sec:optimisation-overview}. On every circuit that was benchmarked both in their paper as well as in Table~\ref{fig:T-benchmark}, the same T-count was found. This is rather surprising as our method and theirs seem very different. This implies one of two things. Either both methods are capable of finding some kind of canonical local optimum, or both methods are actually doing the same thing when viewed through the right lens.
We conjecture that it might be the second option. If this is correct then it might be possible to describe ZX-calculus rewrite rules using the language of exponentiated Pauli's.

The results of this chapter cover just a few ways in which diagrammatic reasoning can help with optimising quantum circuits. In the next chapter we discuss several avenues in which research could be carried forward.

\chapter{Future applications of diagrammatic reasoning}\label{chap:future}

In this thesis we have seen that the ZX-calculus can be successfully applied to the study of measurement-based quantum computation, circuit optimisation and circuit verification. There are however many more aspects of quantum computation that we believe can benefit from the usage of the ZX-calculus and related graphical calculi.

In this chapter we will discuss several ways in which our optimisation strategy can be improved and generalised. Specifically, we discuss how the circuit extraction algorithm can be improved to reduce the CNOT count, how the presence of gflow can possibly be used to facilitate optimisation involving ancillae and classical control, and how our methods can be applied to circuit routing and optimisation of Toffoli circuits.
Finally, we will discuss some preliminary research into using ZX-diagrams to do circuit simulation based on the stabiliser decomposition method.

Although the results we discuss in this chapter are cause for hope that the ZX-calculus will indeed be useful in a wide variety of fields, the reader is advised to bear in mind that the research discussed in this chapter is still in an early stage, and hence will contain a lot of speculation.

\section{Improving circuit extraction}\label{sec:extraction-improvements}
\indexd{circuit optimisation!CNOT ---}
\indexd{circuit extraction!improvements}
As remarked in Section~\ref{sec:phase-teleportation}, the application of our optimisation scheme can result in a higher number of 2-qubit gates than what we started out with. The solution we proposed there was to use phase teleportation. While this solves the problem, it disregards much of the structure we have derived about the circuit in the process of simplifying its ZX-diagram. In this section we will take another look at circuit extraction, and remark on a couple of ways in which it can be improved. 

But first let us remark on the differences in performance between some classes of circuits. The worst increases in 2-qubit gate count are seen when optimising circuits that implement classical reversible functions. In that case most of the gates in the circuit arise from decompositions of Toffoli gates. The decomposition of a Toffoli gate into the Clifford+T gate set is already highly optimised. The reason we then see an increase in the CNOT count is because we basically `forget' this information about this optimised placement of CNOT gates in the Toffoli gates when we extract a circuit. The Gaussian elimination algorithm we use for extraction is necessarily a heuristic, and hence we cannot expect it to perform as well as the optimal CNOT placements in the original decomposition of a Toffoli gate. In this setting we would expect a more conservative peephole optimiser to perform better.

A class of circuits where our algorithm actually succeeds in producing 2-qubit gate counts that are \emph{better} than the state-of-the-art (at the moment of writing) are the `quantum chemistry' circuits benchmarked in Ref.~\cite{phaseGadgetSynth}. Those circuits essentially consist of many phase gadgets, each of which is surrounded by local Clifford gates. When such a circuit is presented as a series of gates, some kind of order of placement of the `CNOT-ladders' (cf.~Eq.~\eqref{eq:phasegadget}) must be chosen. While work is being done on finding good heuristics for placing these ladders (such as in Ref.~\cite{phaseGadgetSynth}), our algorithm essentially gets rid of this problem by actually representing the phase gadgets as phase gadgets, and only outputting a circuit using extraction after the optimisation is done.

There are several ways in which the 2-qubit gate count of our extraction algorithm can be improved. First, we observe that for many circuits it only takes a few row operations on the biadjacency matrix of its simplified ZX-diagram to produce a row that corresponds to an extractable vertex. When this is the case, instead of applying a Gaussian elimination algorithm, we can simply brute-force through all possible combinations of row operations until we find a combination that results in an extractable vertex. For $n$ qubits, there are $n(n-1)/2$ ways to add rows together, $n(n-1)(n-2)/6$ to add 3 rows together, and so on in a exponential manner for more rows. Hence, this kind of brute-forcing is only possible when a small number of row operations suffice (and when the number of qubits is not too high). 
We implemented this brute-force method in PyZX. To limit runtime, we instituted a cut-off for when to stop searching and apply the Gaussian elimination algorithm instead. We see that in practice 1 or 2 row operations suffice for most vertices in most circuits (although some circuits resist this brute-force approach almost entirely).

The second improvement is more conceptual in nature. The Gaussian elimination lets us discover which vertices can be extracted. However, not every row operation that we did in the Gaussian elimination necessarily contributes to the extractability of vertices. Intuitively, we can halt the Gaussian elimination algorithm `early' so that row operations that do not assist in extraction are not implemented as CNOT gates. For instance, suppose we did a Gaussian elimination which needed 14 row operations to fully reduce the matrix, and which resulted in 3 extractable vertices. We can then walk through these 14 row operations step by step, and after every step check how many vertices are extractable. Perhaps it takes 4 row operations for the first vertex to become extractable, 7 for the second one and 10 for the third one. Since we already know that 3 vertices is the maximum number of extractable vertices, we no longer need the remaining $14-10=4$ row additions, saving 4 CNOT gates. Once those row operations have been filtered out, we can also check through the remaining operations to see which ones do not actually influence the extractability of the vertices and remove those as well.

The final improvement to CNOT count of the extraction algorithm we will consider is also conceptual in nature. The columns of the matrix that is being eliminated correspond to unextracted vertices. These vertices have no inherent order to them. We can hence arbitrarily permute columns in the matrix to our benefit. Ideally we would use a Gaussian elimination algorithm that eliminates a matrix to the `nearest permutation' instead of to the identity, but we are not aware of such an algorithm existing. A good heuristic however is simply picking some beneficial order of the columns. Instead of picking this order at the start and letting the elimination run its course, we can do a `lazy' choice of column permutation where we update our choice after every individual column elimination. 

Let us describe a heuristic that uses this method and seems to work quite well.
Before we start the Gaussian elimination, we look among all columns for one that has a 1 in the first row. If there are multiple, then we pick the column that has the lowest number of 1's.  We now permute the columns to make this column the first one. We then let our Gaussian elimination algorithm eliminate this first column. Since there is a 1 in the first row, the algorithm will not have to do a pivot to this row, and because we picked a column with as few 1's as possible, the algorithm will require less row operations to eliminate this column. With the first column cleared, we pick the second column. Here we again pick the column that has the fewest 1's, but now must have a 1 on the second row. We do a column permutation to make this the second column and again let our elimination algorithm clear out the second column. We proceed, picking columns with a 1 on the diagonal while containing as few 1's as possible until we have exhausted all the rows.

By using these modifications---the brute-force approach, the early stopping, and the column swapping---we get significantly better CNOT counts. In general this however does not seem to change the quantitative picture that our algorithm does worse on classical reversible circuits, and performs well on the quantum chemistry circuits.

We have now discussed ways to improve the extraction for ZX-diagrams from which we can already extract a circuit. A complementary question is whether the class of ZX-diagrams we can turn into a circuit can be increased in size. 
The results of this thesis show that a circuit can be extracted from a ZX-diagram as long as the diagram has a gflow in a suitable way. It is then an interesting question whether this condition is necessary. Is there some algorithm that extracts a circuit from any ZX-diagram? 
As a ZX-diagram does not necessarily represent a unitary, and a (deterministic) circuit is always unitary, extraction is not always possible. If we have the promise that the ZX-diagram represents a unitary then we can always extract a circuit in exponential time by simply calculating the linear map the ZX-diagram represents and using existing unitary synthesis methods. 

We conjecture that there is no method that \emph{efficiently} extracts a circuit from a ZX-diagram with a promise of unitarity. This is motivated by the following argument. It is known that the computational power of quantum computation increases drastically when post-selection on measurement outcomes is allowed~\cite{aaronson2005quantum}. It is conceivable that we have a post-selected quantum circuit doing some hard calculation, that just happens to be proportional to a unitary. We can easily represent such a post-selected circuit as a ZX-diagram, and if we had an efficient method for extracting a circuit, then we could convert this post-selected circuit into a regular one, bringing the power of post-selection to regular quantum computation. This would have drastic consequences like the collapse of the polynomial hierarchy.

\section{Circuit routing}\label{sec:circuit-routing}
\indexd{circuit!--- routing}
An issue that we chose to gloss over in Chapter~\ref{chap:optimisation} is circuit routing: ensuring that 2-qubit gates only happen between pairs of qubits that are actually connected in a given qubit architecture. There is however an interesting way in which this requirement can be incorporated in our optimisation routine.

The ZX-diagram simplification algorithm proceeds entirely the same: in order to conform to a given qubit architecture we only need to change the circuit extraction algorithm. Note that there are two places in this algorithm where 2-qubit gates appear. CZ gates appear when connections between vertices on the boundary are removed during extraction. We currently do not know of any intelligent way to make these conform to the architecture. So for now we propose simply using existing methods to route the CZ gates (for instance~Ref.~\cite{cowtan2019qubit}).

The interesting part is the Gaussian elimination stage that introduces CNOT gates. Since a row addition step between rows $r_1$ and $r_2$ adds a CNOT between the qubits corresponding to the rows $r_1$ and $r_2$, making CNOT gates conform to the architecture reduces to finding a way to do Gaussian elimination on a matrix using a restricted set of row additions. A method to do this kind of Gaussian elimination using \emph{Steiner trees} has been developed in Ref.~\cite{KissingerCNOT2019} (and independently in Ref.~\cite{nash2019quantum})\indexd{Steiner tree}. These methods can immediately be applied to our circuit extraction algorithm to route the CNOT gates. Unfortunately, this usually seems to result in worse results than straightforward circuit routing algorithms. The results can be improved by applying the techniques from the previous sections that improve the CNOT count of the extraction, but currently not enough to compete with existing methods.

Recall that in the previous section, an improvement to the Gaussian elimination was made by permuting the columns of the matrix in a specific manner. We speculate that it should be possible to improve the Steiner tree algorithm in circuit extraction by finding a different heuristic for picking column permutations that uses information from the given circuit architecture.

\section{Optimisation with ancillae}
\indexd{circuit optimisation!--- with ancillae}
Our circuit optimisation results focused on optimising quantum circuits without the use of ancillae. It is known that using ancillae can greatly improve several metrics, such as T-count~\cite{Gidney2018halvingcostof,heyfron2018efficient} and circuit depth~\cite{dias2013global}. It then seems like a worthwhile pursuit to find a method to do optimisation with ancillae using the ZX-calculus.

When given a ZX-diagram, we can easily turn it into a circuit with ancillae, as long as we allow \Define{post-selection}\indexd{post-selection}, i.e.~a measurement where we fix the outcome to a particular value. For instance, consider the following graph-like ZX-diagram (that implements a 3-control Toffoli gate):
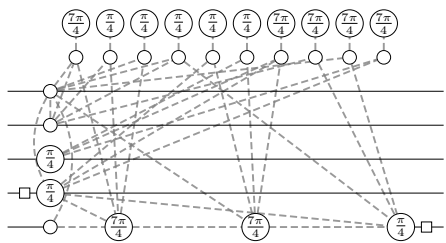
\[\scalebox{0.9}{\input{tof3-zx-opt.tikz}}\]
We can now unfuse the internal spiders (but leaving the phase gadgets intact) to get the following diagram:
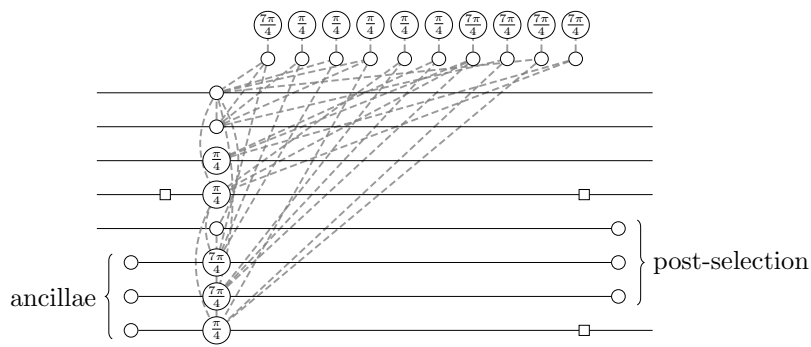
\[\scalebox{0.9}{\input{zx-opt-aspect.tikz}}\]
Here we indeed see ancillae prepared in the $\ket{+}$ state, and some post-selections to the $\bra{+}$ effect.
We could implement this circuit by repeatedly executing it, performing measurements instead of post-selections, until we get a run where the measurements get the outcome specified by the post-selections. However, the expected number of runs to do this increases exponentially with the number of post-selections. This method for introducing ancillae in ZX-diagrams is thus not one that scales.

In order to get a scalable method, we need some kind of way to correct for the wrong measurement outcomes. In Chapter~\ref{chap:MBQC} we saw that this problem was solved in the one-way model using gflow. As our extraction algorithm also requires the existence of a gflow, this suggests that there should be some way to modify our extraction algorithm to make it also introduce ancillae and measurements that can be corrected. While we have been able to do circuit extraction with ancillae and classical control on some small circuits by hand, it is still unclear how to do this in a generic manner.

\section{Toffoli circuit optimisation with the ZH-calculus}\label{sec:ZH-calculus}

In this thesis we focused solely on the ZX-calculus. There are however also other graphical calculi for linear maps between qubits. A well-studied alternative to the ZX-calculus, is the \Define{ZW}-calculus~\cite{CoeckeKissinger2010compositional,hadzihasanovic2015diagrammatic}\indexd{ZW-calculus}. This calculus also includes the Z-spider, but instead of Hadamard gates and X-spiders, the ZW-calculus takes the \Define{W-spider} to be fundamental, which is a generalisation of the \Define{W-state} $\ket{W}:= \ket{001}+\ket{010}+\ket{100}$\indexd{W-state}. The ZW-calculus has been useful in proving completeness (it was the first graphical calculus for qubits to be proven complete, and it served as the basis for the first completeness proofs of the ZX-calculus~\cite{HarnyCompleteness,SimonCompleteness}). It even allowed for a complete axiomatisation of Fermionic quantum computation~\cite{hadzihasanovic2018diagrammatic}\indexd{Fermionic quantum computation}.
The ZW-calculus has however not been used in any more practical tasks (so far). This could stem from the fact that the W-state that is fundamental to the ZW-calculus is simply not used as much in quantum computation protocols.

In 2018, a new calculus was developed by Backens and Kissinger called the \Define{ZH-calculus}\indexd{ZH-calculus}~\cite{backens2018zhcalculus}. This calculus generalises the arity-2 Hadamard box used in the ZX-calculus to an H-box of arbitrary arity:
\[\input{H-spider-free.tikz}\ =\ \sum (-1)^{i_1\ldots i_m j_1\ldots j_n} \ket{j_1\ldots j_n}\bra{i_1\ldots i_m}\]
The sum in this equation is over all $i_1,\ldots, i_m, j_1,\ldots, j_n\in\{0,1\}$ so that an H-box represents a matrix with all entries equal to 1, except for the bottom right element which is equal to $-1$. We see then that a single-input single-output H-box is a Hadamard box (up to normalisation).
The benefit of the ZH-calculus comes from the easy representation of CCZ gates (and thus also Toffoli gates), which is a straightforward extension of the representation of a CZ gate in the ZX-calculus (cf.~Eq.~\eqref{eq:CZ-in-ZX}):
\[CCZ\ = \ \input{CCZ-H.tikz}\]
In a similar manner we can write down $n$-controlled Z gates.
Because these more complicated gates have such a simple structure in the ZH-calculus we can more easily see and apply simplifications on circuits containing Toffoli gates.

We were able to use the ZH-calculus to rederive some well-known simplifications involving Toffoli gates~\cite{GraphicalFourier2019}, such as the 4 T-count implementation of the `Toffoli$^*$' gate of Selinger~\cite{selinger2013quantum}, the 4 T-count implementation of the Toffoli gate using an ancilla and classical control of Jones~\cite{jones2013low}, and the 4 T-count implementation of a compute-uncompute pair of Toffoli gates of Gidney~\cite{Gidney2018halvingcostof}. By using the ZH-calculus we unified the derivation of all these rules. It would be interesting to see if we can use the ZH-calculus to derive more sophisticated Toffoli rewriting tricks that would be hard to find using just the circuit model.

We conclude this section with a promising approach for using the ZH-calculus to systematically optimise circuits based on Toffoli gates. Recall that we viewed the Hadamard box in the ZX-calculus as an edge, resulting in a description of a ZX-diagram as a simple graph. We were then able to remove all internal Clifford spiders using the graph-theoretic operations of local complementation and pivoting. 
Analogously, in the ZH-calculus we can view H-boxes as \Define{hyper-edges}\indexd{hyper-edge} connecting any number of vertices together. This gives a description of a ZH-diagram as a \Define{hypergraph}\indexd{hypergraph}. It turns out that the hypergraph operations of hyper-local-complementation and hyper-pivoting have analogues in the ZH-calculus, and they also allow for the removal of some type of Clifford spiders~\cite{Lemonnier2020hypergraph}. Interestingly, these operations turn out to have a somewhat equivalent description in the path-sum approach of Amy~\cite{AmyVerification}, and point towards a connection between the ZH-calculus and the path-sum of a circuit.

For most of the benchmark Toffoli circuits shown in Table~\ref{fig:T-benchmark}, the application of these hypergraph rules reduces them to a kind of normal form where there are no internal spiders left, similar to how we were able to reduce Clifford circuits to a normal form (cf.~Section~\ref{sec:clifford-circuit-optimisation}). While intriguing, it is not yet clear whether this normal form is usable for some task.

\section{Circuit simulation}\label{sec:simulation}

The final topic we will discuss in this thesis is circuit simulation. We already saw in Section~\ref{sec:graph-theoretic-simp} that we can use the ZX-calculus to completely reduce a scalar Clifford diagram, and in this way efficiently classically simulate the application of a Clifford circuit to a Clifford input state. It turns out that this method can be generalised to work for simulation of Clifford+T circuits using the method of \emph{stabiliser decompositions}~\cite{bravyi2016trading}. Let us first explain the basics of this method.

Suppose we wish to calculate the amplitude $A = \bra{0\cdots 0}C\ket{0\cdots 0}$ where $C$ is a given Clifford+T circuit. We can use the magic state injection method (cf.~Eq.~\eqref{eq:zx-magic-injection}) for every T gate to turn this T gate into a $\ket{T}$ magic state ancilla so that $C$ will just be a post-selected purely Clifford circuit $C'$ with some $\ket{T}$ ancilla inputs. Hence, the problem of calculating $\bra{0\cdots 0}C\ket{0\cdots 0}$ can be reduced to calculating an amplitude $A = \bra{0\cdots 0 T\cdots T}C'\ket{0\cdots 0}$ for a Clifford circuit $C'$, so that the only non-Clifford part of this computation is captured in these T magic state ancillae.

We know that we can easily calculate this type of amplitude if the input state and output effect are Clifford. Hence, if we simply expand each $\ket{T}$ as a sum of Clifford states like $\ket{T} = \ket{0} + e^{i\pi/4}\ket{1}$, then we can decompose the calculation of $A$ into a sum 
$$A = \sum_{x_1x_2\cdots x_t \in \{0,1\}}e^{i\frac\pi4(x_1+\ldots+x_t)} \bra{0\cdots 0 x_1\cdots x_t}C'\ket{0\cdots 0}.$$
Each of the terms is now the amplitude corresponding to a Clifford computation and hence can be calculated relatively efficiently.
Note that a `standard' simulation method that uses a representation of the input state and sequentially updates it based on the gates in the circuit necessarily scales exponentially with the number of qubits. In contrast, this method scales only polynomially with the number of qubits, but exponentially with the number of T gates. Indeed, if the circuit contained $t$ T gates, then the above decomposition of $A$ contains $2^t$ terms.
This method then seems to be promising for large circuits that contain only a small number of T gates.

The exponent $2^t$ can be improved significantly. A \Define{stabiliser decomposition}\indexd{stabiliser decomposition} of an arbitrary state $\ket{\psi}$ is a representation $\ket{\psi} = \sum_{i=1}^k \lambda_i \ket{\phi_i}$ where $\lambda_i \in \C$ and all the $\ket{\phi_i}$ are Clifford states. Since the Clifford states span the space of all states, such a decomposition always exists (although it cannot necessarily be found efficiently if $\ket{\psi}$ is arbitrary, because the dimension of the space is exponential). The \Define{stabiliser rank}\indexd{stabiliser rank} of $\ket{\psi}$ is defined as the minimal number of terms $k$ needed to write $\ket{\psi}$ as a sum of Clifford states. 

In the example above we used the decomposition $\ket{T} = \ket{0} + e^{i\pi/4}\ket{1}$ as the basis for a stabiliser decomposition of $\ket{T}^{\otimes t}$ (\ie $t$ copies of the $\ket{T}$ state) that contains $2^t$ terms. For example, taking $t=2$ we had $\ket{TT} = \ket{00} + e^{i\pi/4} \ket{01} + e^{i\pi/4} \ket{10} + e^{i\pi/2} \ket{11}$. But we can in fact group these terms in a more clever way: 
\begin{equation}\label{eq:TT-decomposition}
\ket{TT} = (\ket{00} + e^{i\pi/2} \ket{11}) + e^{i\pi/4} (\ket{01}+\ket{10}).
\end{equation}
This is a stabiliser decomposition of $\ket{TT}$ of rank 2 (and in fact, since $\ket{TT}$ is not Clifford itself, this is the lowest possible).
By grouping $\ket{T}^{\otimes t}$ as $(\ket{TT})^{\otimes t/2}$ (assuming that $t$ is even) we get a stabiliser decomposition of $\ket{T}^{\otimes t}$ that only requires $2^{t/2}$ terms.

The currently best-known decomposition of $\ket{T}^{\otimes t}$ for large $t$ relies on a decomposition of $\ket{T}^{\otimes 6}$ into 7 terms~\cite{bravyi2016trading}. Using this decomposition we can find a stabiliser decomposition of $\ket{T}^{\otimes t}$ requiring $7^{t/6} = 2^{\alpha t}$ terms where $\alpha = \log_2(7)/6 \approx 0.468$. This is hence a bit better then the exponent of $0.5$ when using the decomposition of Eq.~\eqref{eq:TT-decomposition}.

The current best implementation of a variant of this simulation algorithm using the decomposition described above is given in Ref.~\cite{Bravyi2019simulationofquantum} and can simulate 40 to 50 qubit circuits on a desktop computer containing more than 60 non-Clifford gates.

Because the simulation cost scales exponentially in the number of T gates, it makes sense to apply a T-count optimiser on the circuit before you start simulating it. Even just a modest 10\% reduction in the T-count could make the simulation vastly cheaper if the T-count is significant.
This presents an opportunity for the usage of the ZX-calculus as it is not just bound to optimising circuits, but can instead optimise the number of non-Clifford spiders on any ZX-diagram.

To use the ZX-calculus for this simulation method we make a few changes to the algorithm. Our goal is still to calculate $A=\bra{0\cdots 0}C\ket{0\cdots 0}$. Instead of using magic state injection to rewrite $C$ to a Clifford circuit with ancillae, we just simplify the scalar ZX-diagram corresponding to $A$ using the procedure outlined in Section~\ref{sec:circuit-simplification}. As this is a scalar diagram, the resulting diagram will have no Clifford vertices left: every spider will either carry a non-Clifford phase, or be part of a phase gadget with a non-Clifford phase.

The next step consists of replacing a set of spiders by a stabiliser decomposition. We can cast the decompositions $\ket{T} = \ket{0}+e^{i\pi/4}\ket{1}$ and the one of $\ket{TT}$ presented in Eq.~\eqref{eq:TT-decomposition} in terms of ZX-diagrams as follows:
\ctikzfig{T-gate-decomposition}
With a little more work, the `6-to-7' decomposition of Ref.~\cite{bravyi2016trading} can be shown to correspond to the following equality:
\begin{equation}\label{eq:magic-state-decomposition}
\scalebox{0.85}{\input{magic-state-decomposition.tikz}}
\end{equation}
To use this equality on a scalar graph-like ZX-diagram we pick 6 spiders carrying a phase that is an odd multiple of $\frac\pi4$, and hence are non-Clifford. We then unfuse a $\frac\pi4$ phase from each of them to create the left-hand side of Eq.~\eqref{eq:magic-state-decomposition}. For example:
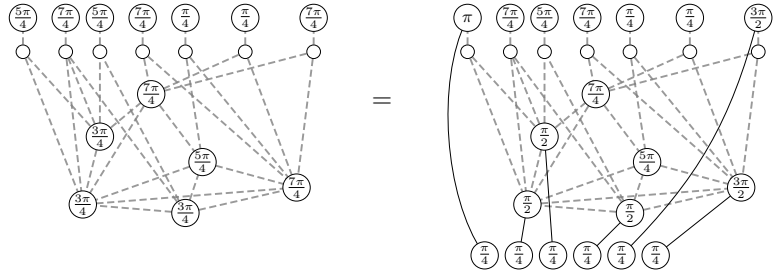
\[\scalebox{0.9}{\input{example-scalar-diagram.tikz}}\]
By applying Eq.~\eqref{eq:magic-state-decomposition} to the right-hand side of the above diagram we get a sum of 7 ZX-diagrams. Crucially, each of the resulting ZX-diagrams has 6 fewer non-Clifford spiders, and hence we can further simplify each of these diagrams until these spiders are removed. This simplification might in turn cancel out some of the non-Clifford spiders. It is this last potential cancellation of additional non-Clifford spiders where the benefit lies of using the ZX-calculus for this simulation method. This is best illustrated with an example.

Consider the 7-qubit circuit \texttt{hwb6} from the benchmark circuits of Figure~\ref{fig:T-benchmark}. Of course this circuit can easily be directly simulated by calculating its matrix, but we will use it to demonstrate the potential usefulness of the ZX-calculus for the stabiliser decomposition simulation method. The circuit has 105 T gates. While this number is too high to simulate it, when we apply the circuit optimisation routine of Section~\ref{sec:clifford-T-optimisation} it reduces to 75, which starts to become feasible to simulate. Now we apply the state $\ket{++---+-}$ to this circuit and we post-select for the effect $\bra{+011-1-}$ (chosen as to result in a final diagram that is as large as possible). We then simplify the resulting scalar ZX-diagram as described. The resulting diagram has 33 non-Clifford spiders. If we were to use the 6-to-7 method on all these 33 spiders, this would result in 67228 Clifford terms. 
However, if we apply it to 6 spiders at a time, simplifying after each step and canceling additional spiders, we only need 8865 terms (on this particular run). This number is very sensitive to which 6 spiders are chosen at every step. A slightly more complicated procedure that tries different choices for each step and picks the best one was able to fully decompose the diagram using just 49 terms, 3 orders of magnitude better than the naive approach.

While this certainly looks very promising, and could potentially result in a large improvement over existing stabiliser decomposition simulation methods, our current implementation is still too slow to compete with established methods. It also remains to be seen whether this improvement in the number of terms needed continues to be present when simulating larger circuits. 

\if\ismain0 
	\ChapterOutsidePart
  	\addtocontents{toc}{\protect\addvspace{2.25em}}
   \cleardoublepage
   \phantomsection
   \addcontentsline{toc}{chapter}{Bibliography}
   \printbibliography[heading=bibintoc]
   \cleardoublepage
   \phantomsection
   \printindex{math}{Abbreviations and mathematical notation}
   \cleardoublepage
   \phantomsection
   \printindex{default}{Index}
\fi 

\backmatter
\ChapterOutsidePart
\addtocontents{toc}{\protect\addvspace{2.25em}}

\cleardoublepage
\begingroup
\phantomsection
\emergencystretch=1em\relax
\printbibliography[heading=bibintoc]
\endgroup


\cleardoublepage
\begingroup
\phantomsection
\chaptermark{}
\printindex{math}{Abbreviations and mathematical notation}
\cleardoublepage
\phantomsection
\printindex{default}{Index}
\endgroup

\chapter{Research data management}

This thesis research has been carried out under the institute policy of the Institute for Computing and Information Sciences of the Radboud University Nijmegen. Information regarding this research data management policy can be found at \url{https://www.ru.nl/icis/research-data-management/policy-protocol/} (last accessed March 13th 2020).

The results of Chapter~\ref{chap:optimisation}, specifically the data of Table~\ref{fig:T-benchmark}, was produced using PyZX. This software developed as part of this thesis research is licensed under the Apache license and is available at \url{https://github.com/Quantomatic/pyzx}.

\end{document}